\documentclass[amsmath,amssymb]{book}

\usepackage{graphicx}
\usepackage{amsfonts}
\usepackage{amssymb}
\usepackage{amsmath}
\usepackage{amsthm}

\begin{document}

\newtheorem{defi}{Definition}[chapter]
\newtheorem{assu}{Assumption}[chapter]
\newtheorem{state}{Statement}[chapter]
\newtheorem{hypo}{Hypothesis}[chapter]

\frontmatter

\title{\huge{A STRICT EPISTEMIC APPROACH TO PHYSICS}}

\author{Per \"{O}stborn$^{1,2}$\\\\\tiny{$^1$Division of Mathematical Physics, Lund University, Sweden}\\\tiny{per.ostborn@matfys.lth.se}\\\tiny{$^2$Department of Archaeology and Ancient History, Lund University, Sweden}\\\tiny{per.ostborn@klass.lu.se}}


\maketitle

\newpage
\pagestyle{empty}
.
\vspace{60mm}
\begin{figure}[h!]
\begin{center}
\includegraphics[width=25mm,clip=true]{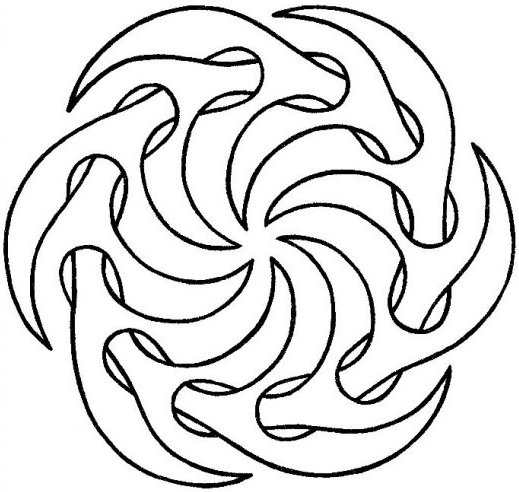}
\end{center}
\end{figure}

\newpage
\pagestyle{empty}
.
\vspace{40mm}
\begin{quote}
\emph{Indeed, when we say that we do not care for philosophy, what we are likely to do is substitute an implicit, hence immature and uncontrolled philosophy for the explicit one.}
\end{quote}
\begin{flushright}
Mario Bunge \cite{bunge}
\end{flushright}
\vspace{5mm}
\begin{quote}
\emph{[N]o content can be grasped without a formal frame and [...] any form, however useful it has hitherto proved, may be found to be too narrow to comprehend new experience.}
\end{quote}
\begin{flushright}
Niels Bohr \cite{bohr0}
\end{flushright}
\vspace{5mm}
\begin{quote}
\emph{[O]nly by renouncing an explanation of life in the ordinary sense do we gain a possibility of taking into account its characteristics.}
\end{quote}
\begin{flushright}
Niels Bohr \cite{bohr1}
\end{flushright}

\newpage
\pagestyle{empty}
.
\vspace{40mm}
\begin{center}
A word of warning
\end{center}
\vspace{10mm}
\begin{quote}
\emph{The frontiers of our research are lost in dazzling light. Plutarch, writing of the fountain-heads of history, says that when we push our investigations to extremes, they all fall into vagueness, rather like maps where the margins of known lands are filled in with marshes, deep forests, deserts and uninhabitable places. That explains why the most gross and puerile of rhapsodies are to be found among thinkers who penetrate most deeply into the highest matters: they are engulfed by their curiosity and their arrogance.\\
The beginnings and the ends of our knowledge are equally marked by an animal-like stupor: witness Plato's soarings aloft in clouds of poetry and the babble of the gods to be found in his works. Whatever was he thinking about when he defined Man as an animate creature with two legs and no feathers? He furnished those who wanted to laugh at him with an amusing opportunity for doing so. For, having plucked a live capon, they went about calling it `Plato's man'.}
\end{quote}
\begin{flushright}
Michel de Montaigne \cite{montaigne}
\end{flushright}

\newpage
\section*{\normalfont{SHORT SUMMARY}}
\label{abstract}

The general view is that all fundamental physical laws should be formulated within the framework given by quantum mechanics (QM). In a sense, QM therefore has the character of a metaphysical theory. Consequently, if it is possible to derive QM from more basic principles, these principles should be of general, philosophical nature. Here, we derive the formalism of QM from well-motivated epistemic principles. A key assumption is that a physical theory that relies on entities or distinctions that are unknowable in principle gives rise to wrong predictions. First, an epistemic formalism is developed, using concepts like knowledge and potential knowledge, where the physical state $S$ corresponds to the potential knowledge of the physical world. It is demonstrated that QM emerges from this formalism. However, Hilbert spaces, wave functions and probabilities are defined in certain well-defined observational contexts only. This means that the epistemic formalism is broader than QM. In the fundamental layer of description, the physical state $S$ is a subset of a state space $\mathcal{S}=\{Z\}$, such that $S$ always contains several elements $Z$. These elements correspond to unattainable states of complete knowledge of the world. The evolution of $S$ cannot be determined in terms of the individual evolution of the elements $Z$, unlike the evolution of an ensemble in classical phase space. The evolution of $S$ is described in terms of sequential time $n\in \mathbf{\mathbb{N}}$, which is updated according to $n\rightarrow n+1$ each time an event occurs, each time potential knowledge changes. Sequential time $n$ can be separated from relational time $t$, which describes distances between events in space-time. There is an entire space-time associated with each $n$, in which $t$ represents the knowledge at sequential time $n$ about the temporal relations between present and past events.

When a wave function $\Psi$ is defined, its evolution can be parametrized with a continuous evolution parameter $\sigma$ that interpolates between two sequential times $n$ and $n+m$ according to $\frac{d}{d\sigma}\Psi(\mathbf{r}_{4},\sigma)=\frac{ic^{2}\hbar}{2\langle E\rangle}\Box\Psi(\mathbf{r}_{4},\sigma)$, where $\mathbf{r}_{4}=(\mathbf{r},ict)$, if the observed object is known to be free. The parametrization of the evolution is chosen to be `natural' in the sense that $\frac{d}{d\sigma}\langle t\rangle=1$, where $\langle t\rangle$ is the expected time distance between the events that define $n$ and $n+m$. The squared rest mass $m_{0}^{2}$ is proportional to an eigenvalue that describes a stationary state of this equation. The Dirac equation follows as a `square root' of the stationary state equation from the condition that $m_{0}^{2}$ has to be non-negative, which in turn is related to the directed nature of sequential time $n$. The introduction of $n$ releases $t$, so that it can be treated as an observable with an uncertainty $\Delta t$, just like $\mathbf{r}$. The full symmetry between the space-time of four-positions $\mathbf{r}_{4}$ and the reciprocal space of four-momenta $\mathbf{p}_{4}$ is unleashed. As a consequence, the spectrum of space-time may have both discrete and continuous parts, just like the spectrum of momentum-energy.

The suggested epistemic formalism also sheds new light on physical concepts and principles such as entropy, Pauli's exclusion principle, the spin-statistics theorem, and the gauge principle. A longer summary is found in Section \ref{summaryconclusions}.

\newpage
\section*{\normalfont{PREFACE}}
\label{preface}
The present work is the result of a growing feeling that there has to be some clear-cut philosophical perspective from which we can see the structure of physical law clearly. In a sense, it is a return to the perspective of Bohr, Heisenberg and Pauli. My aim has been to use their epistemic interpretation of quantum mechanics as a starting point, and to continue as far as possible along this road.

The other clear-cut philosophical perspective is the purely ontic one. To follow that road further is to continue the search for realistic physical models. However, the hints that Nature gave us in the twentieth century clearly point in the opposite direction. To construct models based on naive realism has become like combing the hair the wrong way, against experimental facts. Consequently, it seems to me, recent attempts in this direction have been contrived or vague - or both.

Is there a third road? I cannot see one. Nevertheless, most people seems to be hesitating to embark wholeheartedly on the epistemic road. They remain where experiment left them, accepting the unavoidable, such as Heisenberg's uncertainty relations, but are unwilling to go any further.

Let me give an example. The acceptance of Heisenberg's relations means that the practical inability to measure position and momentum simultanesously is transcended to the principle that we should reject models in which elementary particles have well-defined positions and momenta at the same time. If the position is well-defined, momentum is described as a set of probability amplitudes, one amplitude for each momentum value. However, it is practically impossible to determine the momentum of an elementary particle except in certain experimental contexts. Nevertheless, most physicists accept a universally valid wave function, in which there is always a little probability amplitude flag attached to each momentum value. To continue wholeheartedly along the epistemic road would mean to reject models that use probabilities in the mathematical formalism in situations where the quantity to which the probability refer cannot be observed by anyone. It is only justified to say that God plays dice if we can see the dice in his hand and make statistics of the numbers that turn up. This may seem like an innocent observation, but it throws the Hilbert space out the window as a candidate for a fundamental playground for physical law.

What does it actually mean to "follow the epistemic road"? To me, it means to use the basic distinctions of our perceptions as a foundation when physical models are constructed. We can distinguish darkness from light, sight from sound, two objects from each other, and a logically valid conclusion from one that isn't. We can also distinguish the past from the future, object from subject, and me from you.

Physics is life, as my senior high school physics teacher Karin Sj\"{o}holm used to say. To follow the epistemic road is to take that exclamation seriously. Let me quote the beginning of Violeta Parra's classic song \emph{Gracias a la vida}
\footnote{Translated by Joan Robertson (URL = http://jveronr.blogspot.se/2013/08/gracias-la-vida-lovely-poem-and-song-by.html). The original recording of the song was released in Violeta Parra's album \emph{Las \'{U}ltimas Composiciones} (RCA V\'{i}ctor, Chile, 1966).}:

\begin{quote}
\emph{Thank you, Life, for giving me so much\\
You gave me two eyes and when I open them\\
I can distinguish between black and white\\
And the starry background of the sky above\\
And in the multitudes, the man I love.\\
\\
Thank you, Life, for giving me so much\\
You gave me my hearing with all its power\\
To record, night and day, crickets and canaries\\
Hammers, turbines, barks, rain showers\\
And the tender voice of my beloved.}
\end{quote}

Some physicists seem to deny that the perceived categories of life have anything to do with fundamental physics, trying to make them emerge from something else. Some of them emphasize the quest for `unification' so much that they seem to forget that we cannot get distinctions as an output from a model without distinctions as input. The problem is just to decide which distinctions are primary and which are secondary. Those who reject the physical significance of the distinctions that are primary from a subjective point of view, such as that between past and future, on the basis that they are just `mental constructs' seem to disregard that what they do, in effect, is to replace them with other, more convoluted mental constructs, whose components often depend on the primary subjective distinctions they set out to get rid of. We may imagine one of them writing to a colleague: "Last night I realized that time is not a necessary ingredient in my physical model, today I am working out the mathematical details, and tomorrow I will present my calculations at a seminar." To me, this is just nuts. We are stuck in the existence we are born into, and to understand it we must use the categories of perception that we are given, including the categories of thought.

Perhaps the clearest example of this predicament is the paradoxical nature of all attempts to deny the existence of subjective experience. All these attempts rely on the quality that is denied. As Peter Hankins puts it, as a motto for his wonderful blog \footnote{P. Hankins, Weblog \emph{Conscious Entitites} (URL = http://www.consciousentities.com/)} about the philosophy of mind: "If the conscious self is an illusion - who is it that's being fooled?"

It is not enough to point out the fundamental role of the subjective categories of perception. We must find principles that give form to physical law, that limit the possibilites. We can turn to epistemics for this purpose also, and I dare say that all successful principles of this kind found in the twentieth century have an epistemic root. For example, special relativity stems in part from the criterion that the observations of all subjects are equally valid, regardless their relative state of motion. Physical law is invariant to a change of perspective from one to the other. Just as it does not play any role whose knowledge we use to determine physical law, the overall amount of knowledge that we possess should not play any role either. We do not have to know everything to be able to say something true. Einstein's equivalence principle can be seen in this light. If knowledge about the surroundings of an elevator that accelerates upwards is blanked out, the observations from within should still be valid. Then there is no falling, just an apparent force that pushes you to the ground. A beam of light is seemingly bent in response to this force. Thus gravity bends light and space-time is curved. Another epistemic principle that we rely on is that physical models that presupposes knowledge that is unattainable in principle, distinctions that cannot be made, should spit out wrong answers. If we treat the interchange of two identical elementary particles as a physical operaton that gives us a new state, we get the wrong answer to problems in statistical mechanics; we have to use Fermi-Dirac or Bose-Einstein statistics rather than Maxwell-Boltzmann statistics. Here, I will use this principle to argue that we must get some other result in a double-slit experiment than that that predicted by a classical model in which the particles pass one slit or the other, if we have prepared the experimental setup so that it is impossible forever to judge which slit the particle actually passed. I will argue that the only consistent alternative to such a classical model is interference and Born's rule. 

I cannot help the feeling that it has been very fruitful to follow the epistemic road at some length. So much insight comes out easily and naturally, it seems to me. However, even an epistemic approach to physics needs a coherent ontology, the existence of something that transcends our own perceptions. The ontology that emerges is sketched in section \ref{summaryconclusions}. I know that many people will not feel comfortable with this ontology, since it does not contain `little billiard balls' with well-defined positions that exist regardless whether anyone observes them or not. I ask these people to consider the apparent effectiveness of the present approach, at least.

In a certain sense, I have walked the epistemic road in the opposite direction as compared to the fathers of the Copenhagen interpretation of quantum mechanics. They wanted to understand a given physical formalism, and arrived at an epistemic interpretation. I start with a set of philosophical assumptions of epistemic nature and try to use them to derive (or at least carefully motivate) the physical formalism. The advantage of this reverse approach is that the conceptually well-defined starting point enables a better understanding of the components of the formalism, and its domain of validity. Another advantage with the reverse approach is that it makes it possible to understand better not only the meaning of quantum mechanics, but also some other physical concepts and principles, such as Pauli's exclusion principle, the gauge principle and entropy.

All of these claims depend on the validity of my arguments, of course. If a sufficient number of them hold water, then the suggested approach provides more than just another interpretation of quantum mechanics. Such interpretations tend to be sterile in the sense that they cannot be distinguished experimentally, and they offer no directions for further research that may lead to new predictions. As a consequence, the adherents of different interpretations sometimes engage in philosophical discussions that cannot be resolved scientifically. Instead, I am challenging opponents to the present approach to physics to show that their own approach is even more effective, making it possible to motivate an even larger chunk of physics using fewer or more natural assumptions. I also encourage readers to find logical or mathematical flaws in my reasoning. That is, I am hoping for technical rather than philosophical criticism. 

The present text is a rather unusual animal, a mixed breed dog, or rather a puppy. Judging the puppy by its size, it must have a Great Dane among its forefathers. With its long legs, which are slightly out of control, the puppy wants to explore the whole world at the same time, stumbling into things along the way, breaking some of them. Leaving the metaphors aside, I feel the need to say something about the aim, development, and final form of the present text. Otherwise the reader may not know what to make of it.

First of all, despite its length, this text should be seen as a sketch - a detailed sketch at some places, less detailed at others. I have worked on and off with this material since early fall 2011. At that time, my friend Mi Lennhag was about to go to Lithuania, Kaliningrad and Poland to make interviews for a PhD-project about informal economy and everyday corruption in Post-Soviet Eastern Europe. I embraced the idea to come along as a driver. I had decided some months earlier to take seriously the growing feeling mentioned in the first sentence of this preface, and to reserve time to think intensely about these matters. This trip would give me the opportunity, undisturbed as I would be by duties and distractions at home. My inner romantic loved the idea to walk the same streets as Kant in Kaliningrad, thinking about what we can know and what we can't, and what that tells us about the world.

With me on our roadtrip I took the insight from childhood that subject and object are equally fundamental aspects of the world, and that any proper model must acknowledge this fact - not only at the philosophical level, but also at the physical and mathematical one. (I have always had a hard time understanding how people can think differently, but I know, of course, that many do. It is such a primary insight that it is impossible to argue about it - either you agree, or you don't.) With me I also carried a problem that had nagged me ever since I first came across quantum mechanics. In all accounts I read, spatially extended wave functions were described as `unreal' or as `spooky superpositions', which meant that only perfectly precise values were considered `real'. In such an interpretation, the wave function has to collapse to a perfect delta spike as soon as an observation of a continuous quantity is made. How could it be that an observation with our imperfect senses, or with detectors with finite resolution, could cause the value of a quantity to be defined with inifinte precision? As I third item in my luggage I carried the idea, discussed above, that physical models that presuppose knowledge that is unattainable in principle should be in conflict with experiment.

In Kaliningrad, I used this idea to motivate Born's rule in a sketchy fashion. I also got the idea to regard the physical state $S$ as an extended set consisting of all states of complete knowledge of the world that are not excluded by the present incomplete knowledge. This idea was complemented with the image that a set of alternative outcomes of an experiment can be viewed as a set of slices of this set $S$, a set of distinct subsets, like slices of a bread. This image got rid of the troublesome interpretation that a wave function has to collapse all the way down to a delta spike to be judged `real'. All that is required is that the bread collapses to one of its slices. The bread and the slice are equally real, all that happens in the observation is that the knowledge about the observed quantity becomes more precise - but not infinitely precise!

With these simple ideas as a starting point, I aimed at writing an inspirational paper, maybe 20-30 pages long, discussing an approach to understand quantum mechanics better, without too much details, finishing it within a couple of months. I got off the ground during that trip, and I thank Mi for offering me to join her, good companionship, and her sharing with me in the evenings the experiences of the people she interviewed.

However, the more I worked with my paper, the more loose ends turned up, together with more and more ideas how the basic ideas might be applied in other areas of physics. The paper grew. It is not until now I feel that I have something self-contained, something with enough detail to make it possible for others to decide whether there is substance to the approach.

To make it easier for the reader to pass judgement in this respect, I have highlighted crucial assumptions, definitions and statements. I felt it was necessary to make the skeleton of the discussion visible in this way because of the large amount of material. This skeleton hopefully makes it easier to spot mistakes and unclear points, to suggest alterations and improvements. I want to emphasize that none of the presented assumptions, definitions or statements aspire at mathematical or logical precision. They are not final, neither in form, nor in content. Their only purpose is to help the reader to get through the text, and to form an opinion about it.

The reader will notice that I make a number of definite statements, but no numerical predictions or postdictions. For example, I state that the approach leads to a positive cosmological constant, but I do not try to estimate its value. I suggest some new equations, but I do not try to solve them to extract numbers that can be compared with experiment. There are two reasons for this. The first is that I have worked for so long with this material on my own that I felt it was time to present something, in order to get input from others, as soon as I had substantiated my ideas in each area in which I saw the potential to apply these ideas. To go to the next level and do real physics in each area before I present something would take too much time for me. The second reason is related to the first. Since my analytical abilities are very limited, I thought it would be better, and much more time-efficient, if I could inspire others to attack these kinds of problems.

Even though I offer no numbers at this stage, I put forward at least two predictions. First, all elementary fermions should be massive. Second, there should be no pairing of elementary fermions and bosons, as in supersymmetric theories. They emerge from the analysis as conceptually different kinds of entities. I make the last statement without any detailed knowledge about supersymmetry. As a physicist, I have worked only with classical non-linear dynamics, complex systems, and with the application of these theories in different areas. I have no more than undergraduate training in fundamental, modern physics. The reader will probably notice this fact in my style of presentation, in the lack of conventional notation. This lack of knowledge has made it impossible for me to answer the question that I find most pressing, the question whose answer may falsify my entire construction: can the picture of interactions and transformations of elementary particles I paint be made consistent with the vast body of experimental facts that are so well explained by quantum field theory? The rest mass concept appear very naturally in the present formalism, I think, as the quantity in four-momentum space that corresponds to the Lorentz distance in space-time. I have some hope that rest masses of elementary particles can be extracted from eigenvalue equations, as discussed in Section \ref{evconsequences}. How does all of this go together with the Higgs mechanism?

\tableofcontents

\mainmatter
\pagestyle{headings}

\chapter{\normalfont{IDEAS AND CONCEPTS}}
\label{concepts}

\section{Background and aim}
\label{background}

The underlying meaning of quantum mechanics (QM) has been subject to debate since the birth of the theory ninety years ago. A wide spectrum of interpretations have been proposed, with radically different perspectives \cite{interpretations,stanfordphilo}.
Some theorists think that QM is incomplete or approximate. The suggested changes may involve the introduction of hidden variables \cite{hidden,leggett}, mechanisms for objective state reduction \cite{OR}, or other nonlinearities in the evolution \cite{nonlinear}. Another view is that there is redundancy in the standard postulates of QM \cite{neumann}; the most radical example is the 'many-worlds' interpretation of Everett and DeWitt \cite{manyworlds,dewitt}, according to which linear evolution of superposed states is all there is. Zurek tries to derive some postulates from the others \cite{zurek,zurek2}. In recent years, several attempts have been made to derive the Hilbert space formalism of QM from other principles, which are easier to interpret physically \cite{axiomatic,dakic,masanes}. One approach is to use as a foundation the concept of information \cite{infoderivation,goyal,aaronson}.

Already Bohr, Heisenberg and colleagues focused on information, or rather knowledge. The Copenhagen interpretation stresses that the quantum state encapsulates what can be known about a system, and that it is meaningless to ask for anything else. This epistemic perspective has gained renewed interest. Caves, Fuchs and Schack have introduced an interpretation of QM in which the collapse of the wave function is an update of subjective, bayesian probabilities. Fuchs and Schack have given the name Qbism to this approach \cite{fuchs}. In Anton Zeilinger's eyes "the reduction of the wave packet is just a reflection of the fact that the representation of our information has to change whenever the information itself changes" \cite{zeilinger}. This seems to be the only position to take in order to understand some consequences of QM that have been confirmed in recent experiments. For instance, the work by Dopfer \cite{dopfer,zeilinger3} made it clear that the wave function of one part of an entangled system collapses even if no measurement is made on the other part until a moment later. The \emph{potential} to gain knowledge of the state of the other part is enough to collapse the wave function. Nature does not risk future contradictions. The non-local character of QM becomes less mysterious if the non-local character of knowledge itself is considered: if it is known that two particles have opposite momenta, and the momentum of one of them is measured,  the momentum of the other is immediately known. Distance has nothing to do with it. Time has nothing to do with it either, considering the possibility to use memory and deduction to gain knowledge of the future or the past.

Knowledge has an inevitable subjective side to it: someone has knowledge about something. The association of quantum mechanical states with states of knowledge therefore suggests that knowing subjects play a fundamental role in the modern scientific world view. Nevertheless, the common drive behind many attempts to understand or alter QM has been to explain away or suppress this feature.

My aim is to confront the subjective aspect of knowledge face to face, turn such an epistemology into symbolic form, and show that the formalism that results provides a simple and coherent way to understand QM. No artificial distinction will be needed between a quantum microscopic world and a classical macroscopic world, between system and apparatus, or between system and environment. I will also try to demonstrate that the epistemic formalism provides a fruitful perspective in order to gain better understanding other aspects of modern physics than just the core principles of QM.

\section{The subjective and the objective}

The traditional scientific perspective is to treat the objective, material world as primary, and the subjective world as illusory or secondary - emergent from the material world. This approach has had immense success. To do away with gods, spirits, souls, and intent as basic components of, and active agents in, the world was necessary for progress. However, I think that this one-sidedness makes itself felt and prevents further development as science evolves and aims to provide a complete world view.

Some would say that broadening the perspective would cause science to deteriorate into obscurity and mysticism. In my view, the mysticism lies in the method rather than the subject. I am convinced that it is possible to explore the subjective aspect of the world, and its relation to the objective aspect, with the scientific methods of experiment, logic and mathematics.

As a qualitative starting point, I formulate the following assumption.  

\begin{assu}[\textbf{Intertwined dualism}]
Material objects and aware subjects are both fundamental components of the world. They emerge from each other and cannot be considered in isolation from each other.
\label{intertwined}
\end{assu}

Consequently, any proper attempt to describe the world in a systematic, scientific way should acknowledge this fact by treating the subjective and the objective on equal footing.

A materialistic person often argue as follows: vision is nothing more than photons interacting with the rods and cones in the retina, causing electric signals to be transmitted to the visual cortex, where they are interpreted, with the help of many different areas of the brain, as images containing distinct objects. All the other senses, thoughts and feelings can be explained in a similar way, as they all correspond to electrical and chemical processes in biological structures built, ultimately, by elementary particles.

This reasoning confuses correspondence and identification. The statement `vision is nothing more than...' must be replaced by `vision corresponds to...'. But when this replacement is done, the subjective sensations have not been reduced to physical processes, just matched with them. In response to this, some people maintain that the subjective part of the correspondence is non-existent or illusory. In that case, there is only one element left in the correspondence, and one might say that it reduces to an identification, as claimed. At this basic level, it is hard to come up with arguments. Either it appears self evident that there are subjective sensations, or it does not.

\begin{figure}[tp]
\begin{center}
\includegraphics[width=80mm,clip=true]{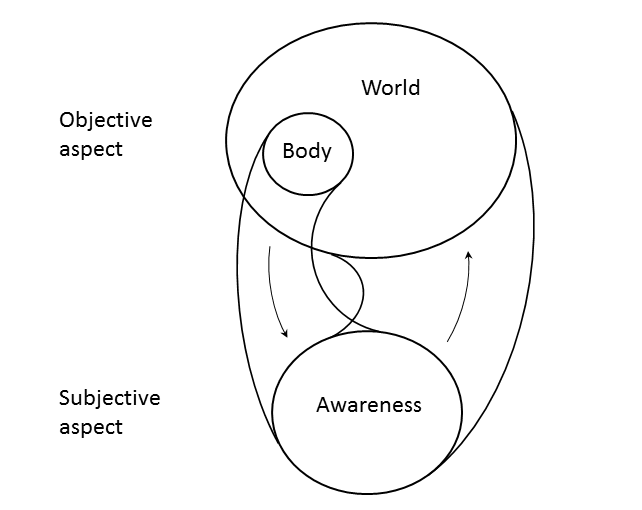}
\end{center}
\caption{Schematic illustration of the outlined world view. There is nothing objective independent of the subjective, and the subjective is completely embedded in the objective. Since the body (including the brain) is regarded as part of the world, fantasies and illusions are also considered as awareness of the world. However, the \emph{outside} world emerges from just a subset of awareness, just as awareness emerges from just a subset of the world.}
\label{Fig1}
\end{figure}

Nevertheless, let me try to illustrate why the latter viewpoint is self-contradictory. If one maintains that physical processes in the brain are all there is, one must realize that this viewpoint involves a mental image of small particles running around in the brain. They must necessarily be assigned attributes such as color and size, and one may add effects such as flashes when electric discharges take place. The point is that a mental image must be used to rule out the primary existence of mental images. If one tries to avoid visualization, one have to use a more abstract mental picture nevertheless.

This argument may give the impression that the subjective aspect of the world should be regarded as primary, and that the objective, material aspect is illusory or emergent. This is equally misleading. First of all: just as it is contradictory to imagine a world without an imaginative observer, it is impossible to imagine an observer being aware, without being aware of \emph{something}. Awareness in itself is a meaningless concept.

Still, it is possible argue that this objective aspect of the world is secondary - nothing more than a mental image. To avoid a purely semantic discussion, we must give an operational meaning to such a statement. To me, the proper meaning is the hypothesis that we can choose freely what to be aware of, that there are no physical laws, independent of our will, that limit the freedom of choice. Apart from the fact that such a world view is in conflict with everyday experience, you can argue against it in several ways.

For one thing, it implies solipsism. Assume that there are two aware beings in such a world. From the perspective of being A, being B must be an object, and consequently A is in total control over B, and can make her behave in any way she wants. But if we turn the perspective around, we conclude that B must be in total control over A, and we have a contradiction. Thus, in any non-solipsist model of the world, there must be something `out there', at least in the form of physical law independent from both A and B, that limits the freedom of perception.

Even if you accept solipsism, a world without physical law indepdendent of your will would be absurd. Then you have to wish everything that happens into existence. You cannot decide to go to a stream to look at the dancing water - you have to wish every movement of each water drop into existence, and before that, for each step you take in your promenade, you must wish the foot to bounce back from the ground, and so on. An exhausting world indeed.

If you give up and wish things to happen by themselves, you introduce independent law. You may wish to change the rules at a later time, but in the meantime you have handed the world over to something else, something `out there', which is then implicitly assumed to exist. If you insist on omnipotence, but assume that some things are harder to make happen than others, you assume independent physical law in the form of a `resistance table' telling which things are hard to make happen in a given circumstance.\footnote{Note that the humoristic argument against the omnipotency of God is inapplicable, but in any case superfluous: in the question "Can God make a stone so heavy that he cannot lift it?", the existence of physical law in the form of weight - resistance to the wish to lift, independent of the will of God - is already assumed.}

\begin{assu}[\textbf{Existence of physical law}]
The freedom of the evolution of awareness is limited by rules independent of our will and of our conception of such rules.
\label{thereislaw}
\end{assu}

The independence of the rules from our conception of them is necessary to avoid a contradiction of the same kind as the one concerning omnipotent will. In effect, what we have is an assumption of absolute truth.

\begin{assu}[\textbf{Independent truth}]
There is a concept of truth independent of our perception of truth.
\label{thereistruth}
\end{assu}

In my attempts to express this truth, I will make extensive use of the word \emph{object}. Let me therefore specify the meaning of this word, as used in the following discussions.

\begin{defi}[\textbf{Object}]
In composite states of awareness, each distinguishable part of the experience corresponds to a separate object.
\label{objectdef}
\end{defi}

In this way, two simultaneous emotions or two components of a composite smell become two objects in the same way as two objects that can be visually separated by the eye.

\begin{defi}[\textbf{Material object}]
An object whose appearance, disappearance or evolution is governed by physical law, according to Assumption \ref{thereislaw}, at least in part.
\label{mobjectdefi}
\end{defi}

Note that this definition does not separate `real' material objects from imagined ones. Flawed perception of the outside world is as deeply rooted in the material world as proper perception; the difference is that the former correspond mostly to the state of material objects in the brain. To this end, we assume the following hypothesis, supported by modern science:

\begin{assu}[\textbf{Detailed materialism}]
Different states of subjective awareness correspond to different states of material objects, and vice versa.
\label{localmaterialism}
\end{assu}

Loosely speaking, the assumption is that mental states cannot evolve independently from material states; they are firmly rooted in the material world. In a sense, this assumption follows from the assumption of intertwined dualism, which forbids separation between the subjective and the objective world. However, the materialistic assumption is stronger in the sense that it provides a local correspondence in addition to a global one - each detail of a mental state corresponds to a detail of the physical state.

If we accept Assumption \ref{localmaterialism}, we have to conclude the following.

\begin{state}[\textbf{All objects are material objects}]
\label{allmobjects}
\end{state}

Therefore we drop the word `material' when we speak about objects in all discussions that follow.

Even if the assumption of detailed materialism may not appear as self-evident as the previous assumptions, it is very natural. The conclusion that there is subject-independent physical law is, in an operational sense, equivalent to the statement that there is an objective world. The subject must then be regarded as placed \emph{in} this objective world, since its subjective experiences depend on it. There must then be an interface between the subject an the world. If the subjective experiences are composite, the interface must be composite. This is pure logic and corresponds to the existence of an extended body. Since the body is extended and placed in the objective world, it must be seen as an extended object itself, ruled by the same physical law as the objects around it. Experiences have to correspond to interactions across the interface - between the body and the outside world - so that each subjective experience should correspond to a physical process or state.\footnote{Note that it is not the surface of the body that shuld be considered the interface between the subjective and the objective, but the body as a whole, and that it is the necessary extended nature of the body that enables the close correspondence between its states and the spectrum of subjective experiences.}

It would be very strange if this correspondence were limited in depth, so that the materialism got lost at a given stage when we probe finer and finer details of states of awareness and states of the extended body - ultimately the brain. If we ask Nature more and more detailed questions about this correspondence, it should continue to give answers, if it gives some answers to start with.

The above assumptions and definitions concerning the objective world and material objects are quite allowing, but nevertheless the narrowest operational ones I can think of. Any aware entity capable of interacting with the environment must follow the basic rules in this philosophical game - even hypothetical ghosts or spirits. Therefore they must be regarded as material in the same sense as any other aware being, even if the physical rules they operate under might be of an unusual nature, giving them an elusive or transparent appearance.

In the same way, the suggestion that life, as we perceive it, is an illusion or simulation created by some external operators (like in the movie The Matrix), falls within the framework of intertwined duality. The only difference is that our relation to the proper objective aspect of the world becomes more complex; it becomes harder to unravel the proper physical law. The external operators themselves manipulate the `true' objective world to create the illusion. If there are no such external operators fooling us, there is no operational way to distinguish the presumed illusion from interaction with a proper objective world, as defined above, and the word illusion becomes nothing more than a label. The concept of an illusion requires the existence of something real, and the concept of a simulation requires an agent that performs the simulation with real tools, such as a computer.

I have already used quite freely the words awareness, subjective experience, perception, and consciousness. The common meaning of these words is described in the most allowing definition of them all.

\begin{defi}[\textbf{Awareness}]
The existence of any subjective experience.
\label{awareness}
\end{defi}

The word consciousness could have been chosen instead of awareness. Regardless the choice of word, the concept is meant to cover the entire subjective aspect of the intertwined duality.

\begin{defi}[\textbf{States of awareness}]
Subjective experiences that are subjectively distinguishable.
\label{awarestates}
\end{defi}

Thus, awareness of a ball is one state of awareness, awareness that the ball is made of leather is a another state, awareness that one is aware of the ball is third, awareness that the preceding two states are distinct is a forth state of awareness, awareness that this was a wrong conclusion is a fifth state, the insight that it is right after all is a sixth, and so on.

\section{Knowledge and potential knowledge}
\label{knowledge}

By knowledge, I mean properly interpreted awareness. This interpretation \emph{may} have a component of deduction, using logic or physical law, in order to gain knowledge about present, past or future states of the objective world, and also about distant places outside the field of direct perception.

\begin{defi}[\textbf{Knowledge}]
Awareness with proper interpretation.
\label{knowledgedef}
\end{defi}

\begin{figure}[tp]
\begin{center}
\includegraphics[width=80mm,clip=true]{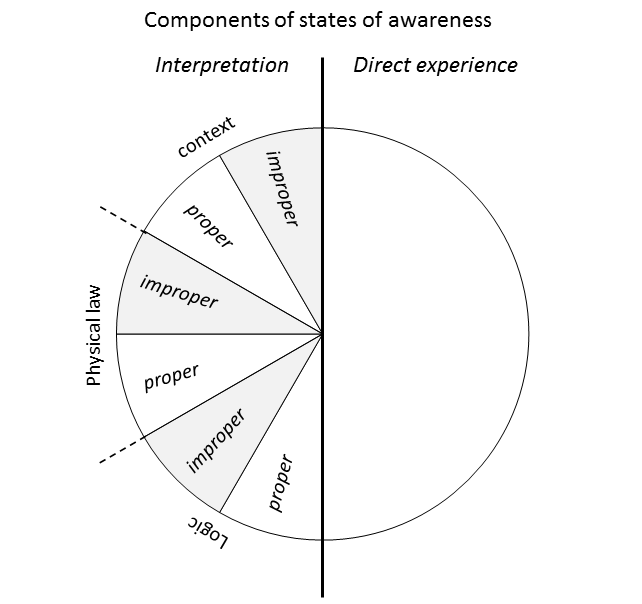}
\end{center}
\caption{Knowledge is direct experience in combination with at least one of the three components of interpretation. There must be no improper such component involved. The component of interpretation called `context' may, for instance, be the distinction between external and internal experiences, or between present and past experiences.}
\label{Fig2}
\end{figure}

Even if all states of awareness are `proper' in the sense that they correspond to actual states of the objective world, they may be improper if an interpretation "I see something in front of me" is added to the state of awareness, when, in fact, the vision is all in your head. The mistake is simply that the material objects that correspond to your state of awareness are not the ones you thought. The assumed independent truth and physical law is the judge. From these assumptions it follows that logic or physical law used in deductions can also be proper or improper. The situation is illustrated in Fig. \ref{Fig2}.

\begin{defi}[\textbf{State of knowledge}]
State of awareness consisting of two other states of awareness: 1) a state of direct experience, and 2) a state of proper interpretation.
\label{knowledgestate}
\end{defi}

The aim of the present text is to propose a way to identify epistemic facts with physical law and physical states. But it would be impossible to use states of knowledge as the only basis for such an enterprise. States of knowledge are fleeting and momentary, and we know much more than that of which we are momentarily aware. Therefore I introduce the concept of potential knowledge.

\begin{defi}[\textbf{Potential knowledge}]
The set of all candidates of knowledge that may become knowledge, that is, may come into mind.
\label{potknow}
\end{defi}

This definition is vague. To sharpen the idea, it is necessary to introduce time and temporal order.

\begin{defi}[\textbf{State of potential knowledge at time t}]
The knowledge obtainable in principle, at time $t$ or any later time, from the possible subjective experiences at time $t$.
\label{potknowstate}
\end{defi}

The possible subjective experiences at time $t$ are all physical processes which take place in the body at time $t$, and which may correspond to an aware state. It is often hard to decide whether you are aware of something or not; there is a gradual scale in clarity of subjective experiences. They may become clearer at a later time, as memories, or they will never be realized in your mind. All such potential subjective experiences are seen as the basis of potential knowledge. Even if you are clearly aware of something, interpretation may be delayed - you may suddenly put things together in your mind - so that the transition from awareness to knowledge is delayed. New information arriving at a later time may also make a better interpretation possible, increasing the knowledge about time $t$ at a later time.

In contrast, if information about time $t$ reach you later, at time $t'$, but this information cannot be associated with any (potentially) aware memories of $t$, this knowledge is \emph{not} part of your potential knowledge at $t$. To exemplify, imagine that you find an old newspaper a time $t'$, from a time $t$ when you were a small child, and reads about a blizzard. If the act of reading stirs up (proper) memories of the snowfall, this proves that the blizzard is part of your potential knowledge at $t$. On the other hand, if no memories come into mind, and you furthermore realize that you lived in distant town at time $t$, then you can be quite sure that the blizzard is not part of your potential knowledge at $t$. This suspicion would become certainty if you suddenly realize that the article was an account of a blizzard that took place before you were born. 

What status should such deduced knowledge have in terms of potential knowledge? Obviously, the aware physical act of reading the newspaper is part of potential knowledge at time $t'$. But the information you get refers back to time $t$. However, without personal experience, you can never be sure that the newspaper properly reports the event. Nevertheless, each observation, such as reading, give \emph{some} rudimentary information of the past, by the use of logic and physical law alone. Some alternatives are ruled out. In the same way, physical law gives information about the future, given the present state of potential knowledge. Retrodictions at time $t'$ about a past time $t$, and predictions about a future time $t''$ are not considered part of potential knowledge at the target times $t$ and $t''$, simply because that would introduce redundancy in the description. We have assumed the independent existence of physical law, and by definition that law specifies exactly what retrodictions and predictions can be made, for any state of potential knowledge. 

This does not mean that deductions are never part of potential knowledge. But there must be an element of subjective experience involved. For instance, imagine that you walk across a floor in your stockinged feet. Suddently, at time $t$, something thorns your socket, but you hurry on. Later, at time $t'$, you return and investigate the floor, finding just one nail sticking out. You conclude that this nail did the job. Then it is part of the potential knowledge at time $t$ that that particular nail torn the socket.

It must be stressed that the exclusion of pure deductions in the state of potential knowledge has only formal importance. Strictly speaking, only the directly experienced outcome of a physical experiment is part of potential knowledge, such as a number on a display, a pointer that moves, and so on. The quantities that are aimed to be measured are just necessary consequences, given the known experimental setup and physical law. But nothing is risked by being sloppy on this point, saying that the deduced information, such as the chemical composition of a sample, the distance to a star, and so on, is also part of potential knowledge.

The above discussions mostly concern the potential knowledge of a given person. It is not specified in the definition of potential knowledge whether it belongs to one subject or many. These matters are dealt with in section \ref{collective}.

Even if it is virtually impossible to determine the limits of the potential knowledge, it seems clear that such limits exist - a collection of finite beings cannot potentially be aware of everything in the universe at any given time $t$ (see Section \ref{limits}). Therefore potential knowledge appears to me to be a well-defined, non-trivial concept.

The basic correspondence I want to establish in this text is that between the potential knowledge at a given time and the physical state at that time.

\begin{defi}[\textbf{Physical state representation at time} $t$]
A symbolic expression that is a representation of the potential knowledge at $t$. The physical state representation determines, via physical law that acts on the symbols, all that can be said about the evolution of the world from time $t$.
\label{firststatedef}
\end{defi}

Actually, this definition contains an assumption:

\begin{assu}[\textbf{A physical state representation exists}]
It is possible to construct a physical state representation, as defined above.
\label{stateexists}
\end{assu}

The interpretational ability to distinguish between the present and the past is crucial in the definition of potential knowledge, and therefore becomes fundamental in a physical theory based on the above definition of physical state.

\begin{defi}[\textbf{Determinism}]
The world is deterministic if and only if, given the potential knowledge at time $t$, physical law uniquely determines the potential knowledge at any later time.
\label{determinism}
\end{defi}

People supporting the realistic, deterministic world view often see knowledge as a function of the physical state, but not vice versa. In other words, potential knowledge is often still regarded to be limited, so that this knowledge is not sufficient to determine the physical state, upon which the future depends. The above definition of determinism will not be fulfilled, and is therefore too narrow.

However, this is a semantic problem. People taking this perspective could redefine words and say that the possibility in principle to determine future physical states perfectly corresponds to unlimited potential knowledge. Then the above definition of determinism is saved. Such an implicit epistemic perspective was taken by Laplace in his famous discussion of determinism, where he imagined an intelligence with perfect knowledge of the physical state and of physical law \cite{laplace}.

\begin{figure}[tp]
\begin{center}
\includegraphics[width=80mm,clip=true]{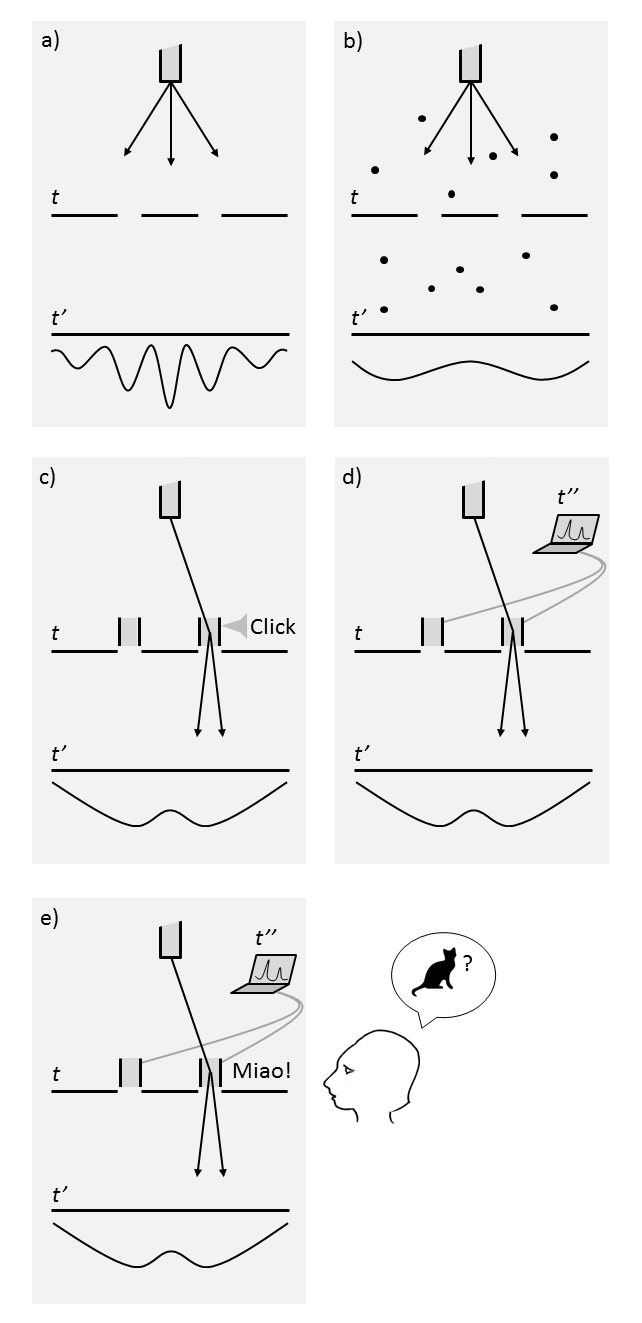}
\end{center}
\caption{Potential knowledge in the double-slit experiment. The particle passes the slits at time $t$ and hits the screen at time $t'$. a) Interference pattern and no path information. b) Perturbation that destroys interference pattern but gives no path information. c) Detector that gives direct response the moment the particle passes, providing path information at time $t$. d) Detector that gives path information only when checked at a later time $t''$. e) Detector that the observer thinks is the same as in d), but actually sounds like a cat the instant $t$ the particle passes.}
\label{Fig2b}
\end{figure}

To illustrate the possible relation between potential knowledge and physics, consider a double-slit experiment which gives rise to an interference pattern on a screen [Fig. \ref{Fig2b}(a)]. In such a situation, the information which slit the particle passes must be considered for ever outside potential knowledge. Otherwise there is a risk of contradiction - interference is inconsistent with path information. This is an example of the requirement of epistemic consistency (Assumption \ref{epconsistency}). A perturbation that destroys the interference pattern may or may not mean that there is a chance that path information becomes potential knowledge. In other words, even if the possibility to gain path information necessarily destroys interference patterns, the destruction of interference patterns does not necessarily mean that there is a possibility to gain path information.

Consider a perturbation in the form of a gas in the experimental chamber that may interact with the particle [Fig. \ref{Fig2b}(b)]. It is unlikely that the state of the gas after the passing of the particle encodes path information that leaves a trace in the observers's body that corresponds to a specific aware state. Nevertheless, the particle becomes entangled with the gas, leading to decoherence. The loss of the interference pattern means that we no longer have any practical means to decide that path information is forever outside potential knowledge. 

Assume, on the other hand, that the perturbation takes the form of a detector at one of the slits. Now there are two alternatives. The detector may produce a macroscopic physical signal the instant $t$ the particle passes (like a Geiger counter), a signal that may be recorded by the senses of the experimenter [Fig. \ref{Fig2b}(c)]. In this case, path information evidently becomes part of potential knowledge at time $t$. Alternatively, the output of the detector is shielded and becomes available only if the experimenter chooses to examine the detector at a later time [Fig. \ref{Fig2b}(d)]. If the examination takes place at time $t''$, path information becomes part of potential knowledge at time $t''$. Note that if $t'<t''$, it is possible to gain partial path knowledge in advance, if the experimenter looks where the particle hits the detector screen, in the form of increased probability that it passed one of the slits. In any case, it is part of potential knowledge at time $t$ that the experimental setup is such that it is possible to gain path information in the future. Therefore, to ensure that no contradiction follows at time $t''$, there can be no interference pattern.

What if the observer is ignorant about the experimental setup, so that she cannot interpret what she experiences in the lab? For example, the detector may be such that it sounds like a cat each time a particle passes [Fig. \ref{Fig2b}(e)], while she thinks that she has to login to a computer to obtain path information, as in Fig. \ref{Fig2b}(d). She hears the `miao', thinks that a cat has sneaked in, and then reads the computer record at time $t''$, concluding that a particle passed at time $t$. Is knowledge of the passing of the particle part of potential knowledge at time $t''$ referring back to $t$, or part of potential knowledge at time $t$ referring to the present? Since there is a potential for the observer to read the manual of the strange apparatus at a still later time $t'''$, and then recall the `miao' she heard at time $t$, we must conclude that the knowledge belongs to the potential knowledge at $t$. Otherwise we get a contradiction: we cannot both lack path information about the particle at time $t$ and have it (acquired at a later time).

\begin{assu}[\textbf{Epistemic consistency}]
The world cannot be properly described by two different physical states at the same time. A proper retrodiction at time $t''$ about $t<t''$ must be consistent with the physical state at time $t$, as properly remembered at time $t''$.
\label{epconsistency}
\end{assu}

This assumption can be seen as the physical counterpart to the assumption of consistency of mathematics. The principle can be seen as a weaker form of causality, and may mask lack of determinism: the potential knowledge at time $t$ constrains the state of potential knowledge at time $t''$, excluding everything that may lead to contradiction. So, when the observer who heard the 'miao' log into the computer at time $t''$ to read the detector record, it is certain that it will show that a particle passed at $t$, even if she is unaware of it herself. A state of potential knowledge just before time $t''$ without path information about the particle would mean that it is possible that the detector record showed at time $t''$ that the particle was \emph{not} detected at time $t$, which would be inconsistent with the memory of the 'miao'.

Note that if the world were deterministic, there would be no need to require epistemic consistency in this sense. The strict causality in such a world means that there is no need to eliminate the risk of unexpected events that cause contradictions.

Epistemic consistency is discussed in more detail in Section \ref{structureknowledge}.

\section{Composite knowledge}
\label{composite}

We know many things. The logical constant \emph{and} must be used to describe most states of knowledge.

In other words, knowledge usually consists of awareness of several objects at the same time. We have defined the word \emph{object} in a very general way (Definition \ref{objectdef}). But objects are not just objects, they have attributes making it possible to distinguish objects with different qualities. This is the usual way to talk about knowledge: Someone (the subject) knows something (attributes) about something (the object). It seems impossible to talk about knowledge in any other way.

Each attribute has a set of possible values. The attribute \emph{color} can take the values corresponding to all the colors in the visible spectrum. The crucial aspect of attribute values is that they can be subjectively ordered.

\begin{defi}[\textbf{Attributes and their values}]
Values of attributes are a set of qualities that can be associated with each other and ordered. Such a set defines an attribute.
\label{attributevalues}
\end{defi}

Thus, the word `value' does not denote a numerical value in this context, but a subjective quality that conforms with Definition \ref{attributevalues}. However, the possibility to order the attribute values means that they can be represented by numbers. Of course, this is what motivates the choice of term.

The ordering of values may be sequential, like the set of possible colors, or multi-dimensional, like the set of possible flavors. Here, the axes are defined by the basic flavors sweet, sour, salt, bitter, and umami. In the case of multi-dimensional ordering, a set of attributes (the five flavors) can be associated with each other to form a meta-attribute (flavor).

The statement that values can be ordered means that given an arbitrary pair of values, it is possible to tell whether a third value is placed between the first two values or not. Note that this conception of ordering says nothing about \emph{direction}, about which value comes after another. If an attribute has only one or two values, the `ordering' may be established as a matter of definition. 

\begin{figure}[tp]
\begin{center}
\includegraphics[width=80mm,clip=true]{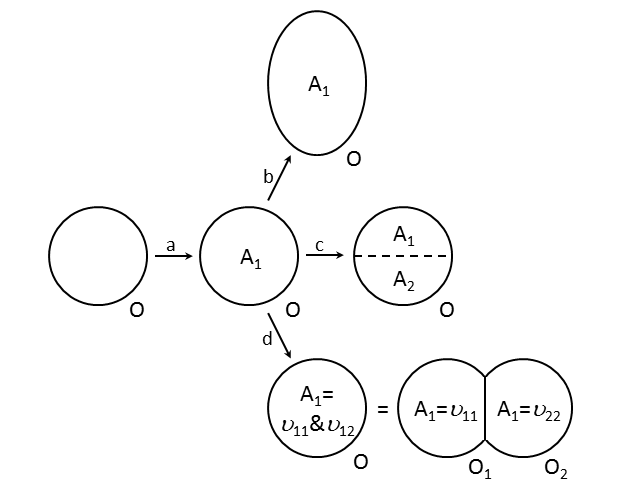}
\end{center}
\caption{Growth of knowledge via increased knowledge about internal attributes $A$ of objects $O$. Knowledge may increase in different ways along the arrows. a) Knowledge that there is an attribute $A_{1}$ of object $O$, b) increased knowledge of the value of $A_{1}$, c) knowledge that there is an an additional attribute $A_{2}$, d) knowledge that $A_{1}$ has two values $\upsilon_{11}$ and $\upsilon_{12}$, which means that $O$ can be divided into two objects $O_{1}$ and $O_{2}$.}
\label{Fig4}
\end{figure}

\begin{figure}[tp]
\begin{center}
\includegraphics[width=80mm,clip=true]{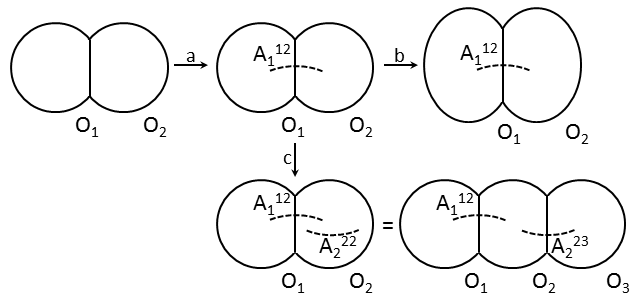}
\end{center}
\caption{Growth of knowledge via increased knowledge about relational attributes $A$ connecting different objects. Knowledge may increase in different ways along the arrows. a) Knowledge that there is an attribute $A_{1}^{12}$ relating objects $O_{1}$ and $O_{2}$. It is assumed that the objects have different values of some internal attribute, so that they can be distinguished. b) increased knowledge of the value of $A_{1}^{12}$, c) knowledge that there is an an additional (possibly different) attribute $A_{2}^{22}$, relating object $O_{2}$ with itself. This means that $O_{2}$ can be divided into two objects $O_{2}$ and $O_{3}$, related by the relabeled attribute $A_{2}^{23}$.}
\label{Fig5}
\end{figure}

Attributes may be of two kinds. They may be \emph{internal}, such as color or rest mass, referring to the object itself, or \emph{relational}, such as distance or angle, relating two or more objects. The angular momentum of an object is a relational attribute, since its direction relates the object to other objects. In contrast, its magnitude can be seen as an internal attribute. Increase of knowledge can always be described as increase of knowledge about internal or relational attributes, as described in Figs. \ref{Fig4} and \ref{Fig5}.

Knowledge increase often means that one object is realized to be composite. For example, tasting wine, new flavors may reveal themselves after some time, so that the one attribute (flavor) of the object (wine) turn out to have several values. Then the object can be divided into several objects with one attribute each. These new `smaller' objects of course may share the values of other attributes of the original `larger' object, such as temperature and color in the case of wine.

\begin{defi}[\textbf{Division of an object}]
An object which turns out to have two different values of a given internal attribute, or a relational attribute referring to the object itself, can be seen as two different objects.
\label{objectdivi}
\end{defi}

\begin{defi}[\textbf{Composite object}]
An object that can be divided.
\label{compobject}
\end{defi}

Among conceivable relational attributes, distance is the primary example. If we approach a pointlike object, we may appreciate that it actually is extended. This means that it has a size, or a distance referring to the object itself. Then it can be divided into smaller objects.

\begin{defi}[\textbf{Elementary object}]
An object which is impossible to divide in the above sense.
\label{elementaryobject}
\end{defi}

For such an elementary object, the number of attributes must be fixed, and so must the values of the internal attributes. If two values of a given attribute were allowed, and more knowledge was gained about which of these applied for a particular elementary object, it could turn out that both applied. Then it could be divided and would not be elementary.

It is meaningless to talk about the size of elementary objects embedded in physical space, since it is only defined if the relational attribute distance is referring to the object itself, which is forbidden by definition.

\begin{figure}[tp]
\begin{center}
\includegraphics[width=80mm,clip=true]{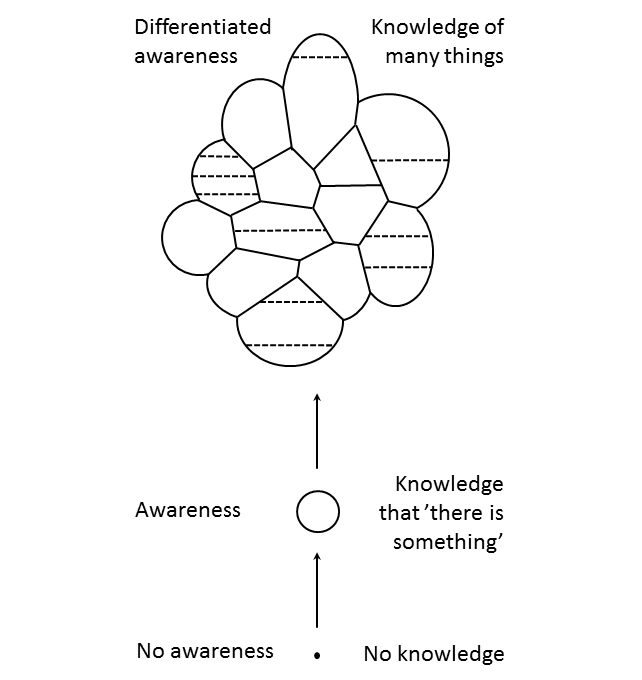}
\end{center}
\caption{The foam picture of knowledge. There is a discontinuous jump from the state of no awareness and no knowledge to `naked' awareness, corresponding to the basic state of knowledge `there is something'. This basic element of knowledge can then grow by division to knowledge of many objects with different attributes. The identification of the state of potential knowledge with the physical state implies that this state should have a discrete structure of the same kind.}
\label{Fig5b}
\end{figure}

To speak about a composite world, it is inevitable to define its constituent parts or objects by the use of the concept of \emph{distinction}. We make a distinction between this and that. A distinction always has two ends - never $\sqrt{2}$ ends. Such a process of distinguishing corresponds to object division, or distinguishing between different attributes of a given object. Increase of knowledge can thus be seen as a stepwise process in which objects may divide into integer numbers of 'smaller' objects, for which an integer number of attributes can be distinguished. In this way a `foam' of knowledge is successively built (Fig. \ref{Fig5b}). This line of reasoning can be summarized in the following assumption.

\begin{assu}[\textbf{Construction of knowledge}]
Any state of knowledge can be constructed in a countable number of steps in which distinctions are made, starting from the basic element of knowledge (Fig. \ref{Fig5b}).
\label{constructknow}
\end{assu}

The following statement follows.

\begin{state}[\textbf{Countability of objects}]
There are at most countably many objects, and none of these object can be divided into more than countably many objects.
\label{countobjects}
\end{state}

This definition allows division of an object ad infinitum. We adopt a stronger atomistic assumption.

\begin{assu}[\textbf{The depth of knowledge is finite}]
No object can be divided more than a finite number of times.
\label{finitedepth1}
\end{assu}

The picture of knowledge painted above is discrete in the sense that it always consists of objects and attributes that can be indexed. This fact does \emph{not} mean, however, that the \emph{values} of the attributes have to be discrete. For example, the value of the internal attribute mass may in principle take any real value, just as the relational attributes distance and angle. The latter observation means that physical space may very well be described as continuous.

The point is that the concept of continuity can be defined by a sequence of mathematical propositions that consist of discrete set of symbols representing distinct concepts - that is by a state of internal composite knowledge having the structure described above.

The assumption that (potential) knowledge is in one-to-one correspondence with the physical state of the world is fundamental in this text. Therefore it becomes a basic hypothesis that the foam-like, partial discreteness of knowledge is refleceted in the structure of the physical world.

There is a certain similarity between the construction of allowed states of knowledge and the construction of allowed sets in the Zermelo-Fraenkel axiomatic set theory \cite{ZFC,fraenkel}. To exclude improper candidates, the proper ones are created from already accepted sets via simple rules. Of course, set theory has much richer structure, partly because it does not require that all sets are built from the same basic set (c.f. Fig. \ref{Fig5b}). One set candidate excluded by the procedure to create sets successively from simpler ones is that of Russell's paradox: the set of all sets that do not have themselves as elements. Since states of knowledge should correspond to physical states, such self-contradictory states must be excluded. One example is the state of knowledge of Socrates, who knew that he knew nothing.
 
\section{Incomplete knowledge}
\label{incomplete}

Most often we must use the logical constant \emph{or} to specify the state of (potential) knowledge, apart from \emph{and}. For example, we may know that the value of attribute $A_{1}$ of some object is $\upsilon_{11}$ or $\upsilon_{12}$ or $\upsilon_{13}$. The increase of knowledge about $A_{1}$ along arrow b in Fig. \ref{Fig4} means that the set of possible values shrinks. As another example, we may know that an object we are aware of can be divided into ten or more smaller objects. That is, the number of constituent objects is ten or eleven or twelve or ...

\begin{defi}[\textbf{Defocused knowledge}]
Potential knowledge about an object that can increase without causing object division.
\label{defocus}
\end{defi}

In other words, an knowledge of an object is defocused when it is not known how many attributes it has, when the values of its attributes are incompletely known, or when it is not known how many objects the given object can be divided into. Note that both composite and elementary objects may be defocused. 

\begin{defi}[\textbf{Conditional knowledge}]
In a defocused composite object, gain of knowledge about one constituent object may imply gain of knowledge about another, and vice versa. Such knowledge is conditional.
\label{conditional}
\end{defi}

For example, consider a sealed box of wine bottles. If you know that the wine in all bottles is taken from the same barrel in the winery, then you also know that if the wine in one bottle turns out to be sour, so will the wine in all the other bottles.

\section{Individual and collective knowledge}
\label{collective}

An object that is the body of a subject must have properties that distinguish it from other objects, according to the requirement that all concepts that are intended to be part of physical theory shall be operationally defined.

The only such property I can imagine is the following: if particles interact with the senses of an object that corresponds to a subject, this fact immediatly becomes part of potential knowledge. As long as we only consider the body of \emph{one} subject, this is just a repetition of the ideas already presented. But if we explicitly consider several distinct subjects, new elements are added to the picture.

\begin{figure}[tp]
\begin{center}
\includegraphics[width=80mm,clip=true]{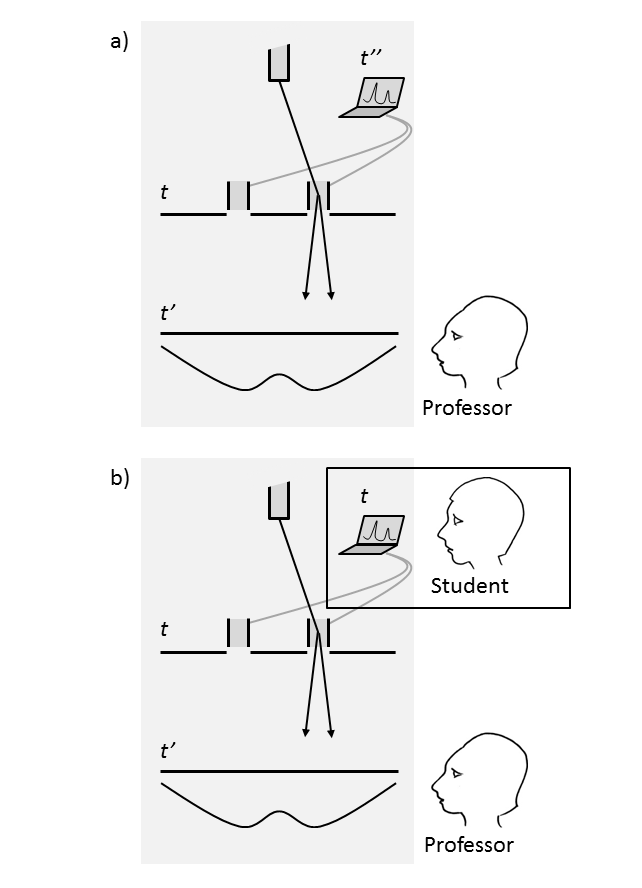}
\end{center}
\caption{a) If the detector output is shielded from the experimenter until time $t''$, path information becomes potential knowledge at time $t''$ (Fig. \ref{Fig2b}). b) If a subject monitors the detector, it becomes potential knowledge already at time $t$, even if the experimenter is shielded from the output.}
\label{Fig4b}
\end{figure}

Consider two subjects, let us call them the professor and the student. Figure \ref{Fig4b}(a) shows the same experimental setup as in Fig. \ref{Fig2b}(d). The professor goes to the computer at time $t''$ and learns that a particle passed the right slit at the earlier time $t$. This fact becomes potential knowledge at time $t''$ referring back to $t$. Consider now Fig. \ref{Fig4b}(b), where the student sits in a room with the computer and reads in real time possible detections of particles passing the right slit.

Again, at time $t''$ the professor knocks on the door and asks if the particle passed at time $t$. If the body of the student is assumed to be an object just like any other object, the room with the student and the computer is a detector with sheilded output, just like in Figure \ref{Fig4b}(a). Consequently, the fact that the particle passed the right slit again becomes potential knowledge at time $t''$. In contrast, if the body of the student is on equal footing with that of the professor, the passing of the particle becomes potential knowledge already at time $t$, since the student reads detections in real time.

Clearly, the former assumption corresponds to solipsism. In that case, there is no operational difference between dead matter and other individuals, from ones one perspective. And from the epistemic viewpoint this means that there are no other subjects. Instead, we assume is the other possibility.

\begin{assu}[\textbf{Many subjects}]
Physcical law allows more than one subject.
\label{manysubjects}
\end{assu}

Apart from the fact that this assumption reflects our na\"{i}ve ideas about life and matter, it is implicitly suggested by the principles of epistemic invariance, as discussed in section \ref{epinvariance}. These are just reformulations of Einstein's relativity principle and equivalence principle, which can be seen as expressions of observer democracy; there is no special `I' that has the final word when it comes to proper perception of physical law.

As discussed above, the operational way to express the existence of many subjects is the following.

\begin{assu}[\textbf{Potential knowledge is collective}]
The potential knowledge that corresponds to the physical state is the union of the potential knowledge of all subjects.
\label{collectiveknow}
\end{assu}

\begin{figure}[tp]
\begin{center}
\includegraphics[width=80mm,clip=true]{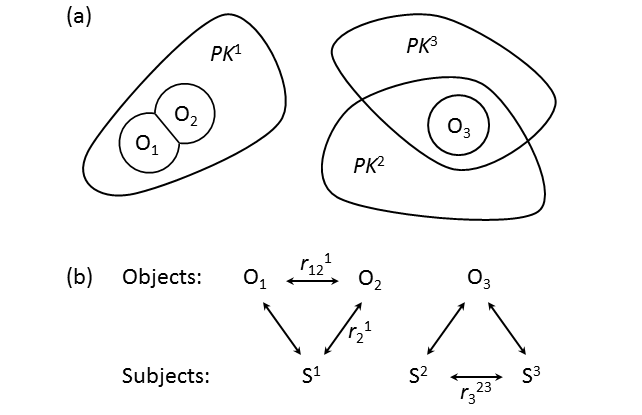}
\end{center}
\caption{a) Potential knowledge $PK$ is the union of individual potential knowledge $PK^{k}$ of subjects $S^{k}$. Individual potential knowledge may overlap, which means knowledge of at least one common identifiable object $O_{j}$. b) The fact that subject $S^{k}$ knows a identifiable object $O_{j}$ defines a relation $r_{j}^{k}$ between them. That one subject knows two objects defines a relation $r_{jj'}^{k}$ between the objects. If two subjects know the same object, a relation $r_{j}^{kk'}$ between the subjects is defined.}
\label{Fig7b}
\end{figure}

The assumption is illustrated in Fig. \ref{Fig7b}. Natural as it may seem, it has strange consequences. The student may be millions of light years away. The observations that she makes immediately affects the range of possible observations the professor can make, since the observations of the student change the state of potential knowledge, which is in one-to-one correspondence with the physical state of the world. This state controls the evolution of awareness of the professor as well as that of the student.

We run into contradictions if we want to keep the idea that there are many subjects, but relax the idea that individual observations immediately affect the common pool of potential knowledge. The operational meaning of the latter relaxation would be that before the student and the professor meet, the observations of the student do not limit the possible observations of the professor, and vice versa. But then they may tell each other contradictory memories when they actually meet, in opposition to the assumption of epistemic consistency (Assumption \ref{epconsistency}).

As discussed in relation to identifiable objects (Section \ref{limits}), two subjects may be potentially aware of the same object, so that the potential knowledge of individuals often overlap (Fig. \ref{Fig7b}). Speculating that primitive organisms have a small degree of awareness, it may well be so that the corresponding knowledge is so tiny that it is contained by the knowledge of other beings, and does not contribute at all to the collective potential knowledge that specify the state of the world.

There is a deeper question hiding in these considerations: is there a fundamental difference between different subjects and different aspects of the same subject? Put differently: is there a fundamental difference between different bodies and different parts of the same body? What is the fundamental difference between ants in an ant colony and the cells in your body? If there is no fundamental difference, then this section is superfluous. However, in Section \ref{individualsubjects} we will discuss the capability to make independent choices as a possible identifier of separate subjects.

\section{Limits of knowledge}
\label{limits}

In this section, possible limits to potential knowledge will be discussed. To say that there is something we lack potential knowledge about, there must be a way to identify this `something'. It must be an object which we know exists, but which we can learn more about.

To say that we learn something new about an object, or lose knowledge about something else, we must be able to decide that the two states of (potential) knowledge point to the same object, that it is possible to track. In other words, there has to be a relation between a given observer and a given identifiable object, that is maintaned during a period of time [Fig. \ref{Fig7b}(b)]. As a consequence, it is possible to define a relation between two observers by the fact that both are related to the same identifiable object.

To be able to track a specific object among all the objects in a state of knowledge, at least one of three conditions must be fulfilled: 1) the objects are related by relational attributes that stay constant or change continuously with time. 2) they have different values of a given internal attribute, and these values stay constant or change continuously. 3) they have different attributes.

\begin{figure}[tp]
\begin{center}
\includegraphics[width=80mm,clip=true]{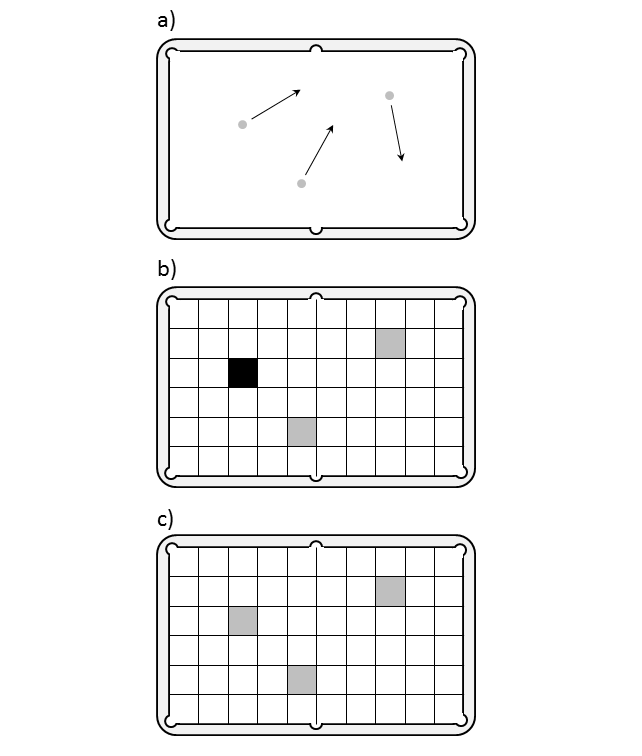}
\end{center}
\caption{Identifiable objects must have a) attributes that change continuously (like relative distance), or b) have internal attributes that differ, and are known to stay constant (like colour). If none of these conditions is fulfilled, as in c), it is impossible to keep individual track of them.}
\label{Fig9}
\end{figure}

If none of these condistions is fulfilled, there is no operational way to identify with certainty two objects observed at different times as being \emph{the same} (Fig. \ref{Fig9}).

For example, playing billiards with identical grey balls, to identify the ball you hit with the queue when all the balls start bouncing about requires that their relational attribute \emph{distance} changes continuously (criterion 1). However, if you paint the ball you hit black, you may identify it as times goes even if the game takes place on a strange billiard table forming a discrete lattice - assuming knowledge that the color of a ball does not change with time (criterion 2).

If you listen to harmony singing during the game, the pitch of the voices must change continuously to keep track of which voice belongs to which individual (criterion 2). However, you can always distinguish the voices from the balls. The two groups of objects have different internal attributes (criterion 3).

More formally, two objects have different internal attributes if there is an attribute $A_{1}$ of object 1, and it is part of potential knowledge that object 2 does \emph{not} have attribute $A_{1}$. In practice, such knowledge arises if you observe attribute $A_{2}$ of object 2, and there is a physical law telling that if an object has attribute $A_{2}$, it does not have attribute $A_{1}$.

\begin{defi}[\textbf{Identifiable objects}]
Two objects are identifiable during the time interval $\Delta t$ if and only if they fulfil at least one of the continuity conditions 1) or 2), or there is an attribute $A$ such that one of the objects have attribute $A$ while the other does not. One of these three facts must be part of potential knowledge during $\Delta t$.
\label{identifiableobjects}
\end{defi}

If the identifiability is due to condition 1) or 2), then the exact value of the attribute in question must belong to potential knowledge during the time interval. Otherwise it is not possible to decide that it is constant or continuously changing. These matters are discussed further in section \ref{identifiability}.

\begin{defi}[\textbf{Independent objects}]
An identifiable object is independent of another identifiable object if and only if physical law allows at least one of its attributes to take several values even if the attributes of the other object are completely known.
\label{indobjects}
\end{defi}

\begin{defi}[\textbf{Independent attributes}]
An attribute of an independent object is independent if and only if physical law allows it to take several values even if all other attributes of the given object are completely known.
\label{indattributes}
\end{defi}

Naively, the independent objects are the `real' objects - we can keep track of them and they possess their own degrees of freedom; they do not just mirror the behavior of other objects.

\begin{assu}[\textbf{Existence of independent objects}]
There are independent objects, as defined above.
\label{indexit}
\end{assu}

Having come so far, we may speak of those things we lack potential knowledge about.

\begin{defi}[\textbf{The currently unknowable}]
The currently unknowable is the attributes of those independent objects about which the potential knowledge may increase in the future.
\label{currunknown}
\end{defi}

As potential knowledge changes with time, knowledge of some attributes may decrease as knowledge of others increases. Increased knowledge may mean that the currently known objects can be further divided. If there is something currently unknowable, the logical constant \emph{or} has to be used to specify the state of current potential knowledge. In other words, potential knowledge is incomplete.

\begin{state}[\textbf{Potential knowledge is incomplete}]
There is something currently unknowable.
\label{incompleteknowledge}
\end{state}

This statement can be motivated by contradiction. Assume that there is nothing currently unknowable. That is, the potential knowledge includes perfect knowledge about which independent objects can be divided, and the knowledge about none of these objects is defocused - all attributes are perfectly known. According to our epistemic approach, the state of the world is then perfectly known.

The body is an identifiable object separate from the objects in the outside world, since we assume the interpretational ability to properly distinguish internal from external experiences (Fig. \ref{Fig2}). Referring directly to the definition of identifiability, the known attribute $A$ that the body has, but the objects in the outside world have not, is simply `this is the body of a subject'. To see that the body and the objects of the outside world have to be regarded as independent objects, it is enough to contemplate the fact that we must leave room for internal processes, such as interpretation, that do not just mirror the state of the outside world. We conclude that the independent objects belonging to the body are a proper subset of all independent objects in the world.

\begin{figure}[tp]
\begin{center}
\includegraphics[width=80mm,clip=true]{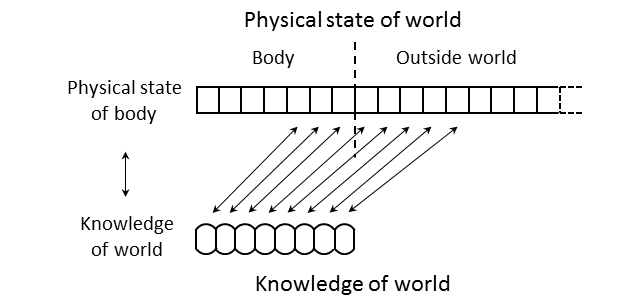}
\end{center}
\caption{Knowledge of the world in terms of perceived objects, and the physical state of the world in terms of mathematical objects. The tilted arrows indicate the one-to-one correspondence between these (Assumption \ref{pkscorrespond}). The knowledge of the world is encoded in the physical state of the body according to detailed materialism (Assumption \ref{localmaterialism}), so that there is also a vertical correspondece. This correspondence is not object-to-object, since one object of perception usually corresponds to many objects in the body. The vision of a black dot involves the eye, the visual nerve and the visual cortex. The unmatched mathematical objects in the top row illustrates our incomplete knowledge of the body and of the world.}
\label{Fig8}
\end{figure}

According to the principle of detailed materialism (Assumption \ref{localmaterialism}), the assumed perfect potential knowledge of all objects in the world has to be a function of the physical state of the body, which is in one-to-one correspondence to the assumed perfectly known objects of the body. Therefore, the perfect potential knowledge becomes a proper subset of itself.

Actually, to arrive at this conclusion, there is a need to clarify what is meant by this one-to-one correspondence.

\begin{assu}[\textbf{Correspondence between knowledge and physical state}]
Potential knowledge of object $O_{k}$ corresponds to a mathematical object $\bar{O}_{k}$ present in the representation of the physical state.
\label{pkscorrespond}
\end{assu}

This correspondence is further explored in section \ref{state}.

The body consists of a finite number of objects according to Assumption \ref{finitedepth1}, and we concluded that the objects of the body is a proper subset of all the objects in the world. Thus it is impossible to put the objects of the body into one-to-one correspondence with all objects in the world (Fig. \ref{Fig8}). We have a contradiction, and thus there is something currently unkowable.

Note that we have implicitly used the observation that the body is an \emph{independent} object in the world. If the physical state of the body was a function of the physical state of the outside world there would be no contradiction.

Also, there would be no contradiction if we allow panpsychism. That is to say, we have to exclude the possibility that \emph{all} objects in the world belong to the body of some perceiving subject. In the above motivation of Statement \ref{incompleteknowledge}, we used the fact that the body of \emph{one} subject is always a proper subset of the world. The natural world-view is that the union of all bodies is also a proper subset of all the objects in the world. In that case, the argument goes through even if we allow an arbitrary number of subjects. If the world is infinite in extension, we may allow infinitely many subjects without destroying the argument. Problems arise only if all the external objects that a given subject observe belong to the body of another subject. Then the potential knowledge can indeed be complete. We could help each other to complete the knowledge according to the principle "I scratch your back if you scratch mine".

\begin{assu}[\textbf{No panpsychism}]
The union of the bodies of all subjects in the world is a proper subset of the set of all objects in the world.
\label{nopanpsychism}
\end{assu}    

To summarize, we know that there is something we cannot know anything about. At any time, it is part of potential knowledge that there is something outside potential knowledge. This means that any state of potential knowledge is defocused (Definition \ref{defocus}), and possibly conditional (Definition \ref{conditional}). This fact must somehow be represented in the mathematical expression of the physical state. 

\begin{figure}[tp]
\begin{center}
\includegraphics[width=80mm,clip=true]{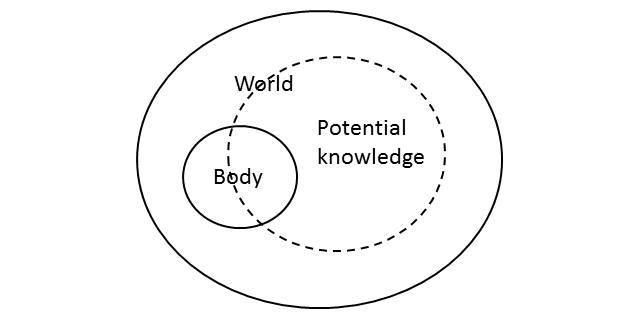}
\end{center}
\caption{The assumptions of intertwined duality and detailed materialism imply that some parts of the body and some parts of the outside world are currently unknowable. Compare Fig. \ref{Fig1}.}
\label{Fig6}
\end{figure}

We can go a step further and conclude that potential knowledge of the external world and of the body must both be incomplete; they mirror each other (Fig. \ref{Fig6}). Assume that potential knowledge of the external world is incomplete. This incompleteness is mapped via detailed materialism to a `fuzzy' physical state of the body where the number of objects it contains or their values are not precisely determined. Thus potential knowledge of the body is also incomplete. Going the other way around, assume that potential knowledge of the body is incomplete. The corresponding fuzzy physical state clearly cannot represent a perfectly perceived external world.

\begin{figure}[tp]
\begin{center}
\includegraphics[width=80mm,clip=true]{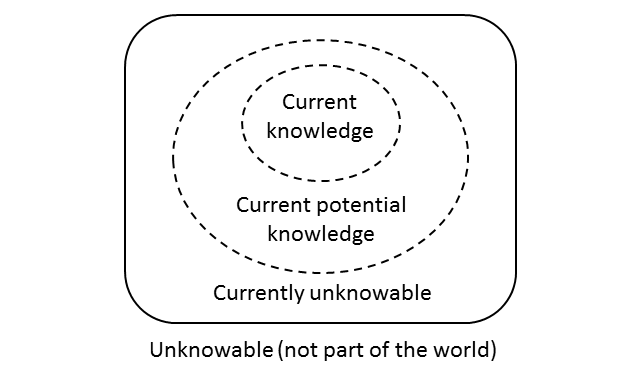}
\end{center}
\caption{An onion of knowledge. The current knowledge corresponds to the properly interpreted current state of awareness. The potential knowledge represents those things that are knowable in principle at a given moment. It is part of potential knowledge that there is something currently unknowable. The solid boundary of the outermost layer of knowability corresponds to boundary of the world. There is such a boundary since the existence of proper and improper interpretation of awareness is assumed (Fig. \ref{Fig2}). In other words, not everything that is conceivable is part of the world.}
\label{Fig3}
\end{figure}

The unmatched objects in the physical state in Fig. \ref{Fig8} correspond to the currently unknowable, but knowable in principle (Fig \ref{Fig3}). The assumption of detailed materialism (Assumption \ref{localmaterialism}) implies that all subjectively perceived objects correspond to physical states of objects in the body. These bodily objects should themselves all be perceivable \emph{in principle}. Each change of subjective perception should in principle be possible to distinguish by mapping the activity of objects in the brain, e. g. in a PET scan. In this sense, the vertical correspondence between perceived objects of knowledge and physical objects of the body is assumed to be perfect (left side of Fig. \ref{Fig8}).

A related conclusion, which follows from the motivation of Statement \ref{incompleteknowledge}, is that if there is \emph{some} potential knowledge of the outside world, the potential knowledge of the body (including the brain) is imperfect, that is, some parts of it belong to the currently unknowable (Fig. \ref{Fig4}).

Statement \ref{incompleteknowledge} can be seen as a rephrased Heisenberg's uncertainty principle. The motivation here is static, while the motivation is most often dynamic: if an observer, a limited being embedded in the world, reach out for new knowledge, she necessarily mess things up by the movement, losing some of her old knowledge.

The epistemic world view presented so far can be summarized by the onion of knowledge shown in Fig. \ref{Fig3}.

Let me return to the question of determinism.

\begin{state}[\textbf{Conditions for determinism}]
a) If there is something currently unknowable, then the world is not deterministic. b) If there is nothing currently unknowable, the world may or may not be deterministic.
\label{detcond}
\end{state}

To establish a), it is enough to notice that the premise means that there are attributes whose values $\upsilon$ are outside potential knowledge at current time $t$ (incomplete knowledge), and these can vary independently from the values $\upsilon'$ of the known attributes. Since we can learn to know $\upsilon$ at a later time $t'$ (they are just \emph{currently} unknowable), and there is no way these can be deduced from the potential knowledge at $t$ represented by $\upsilon'$ (independent objects), the world is not deterministic. Statement a) can of course be turned around: if the world is deterministic, then there is nothing currently unknowable.

To motivate b), imagine a world governed by Newton's laws where a stochastic term with white noise is added. This term cannot represent states of hidden independent objects that we could follow through time and learn more about, because of the lack of temporal correlations. In other words, it cannot represent something currently unknowable.

From statements 2 and 3 we conclude the following.

\begin{state}[\textbf{Absence of determinism}]
The world is indeterministic. No group of subjects can ever determine its future exactly, even in principle.
\label{nodeterminism}
\end{state}

The motivation of this statement was based on the existence of the currently unknowable. A reasonable assumption is that all apparent randomness in nature is due to this unavoidable lack of knowledge, rather than the existence of stochastic noise at a fundamental level of description.

\begin{assu}[\textbf{Noiseless physical law}]
Physical law is such that if knowledge is complete, evolution is deterministic. The number of objects and their attributes will be completely known at all times.
\label{noiseless}
\end{assu}

This assumption requires that \emph{all} objects are identifiable. To pinpoint the form of physical law more exactly, the following assumption is helpful.

\begin{assu}[\textbf{Mathematical physical law}]
The evolution of objects with time, as dictated by physical law, can be expressed as mathematical operations on number schemata that describe the potential knowledge of these objects.
\label{mathlaw}
\end{assu}

This potential knowledge includes lists of objects, and lists of attributes and numerical representations of their values. It may also include lists of probabilities that a given attribute value will be found if knowledge about it increases.

In relation to physical law and the incompleteness of knowledge (Statement \ref{incompleteknowledge}), let us return to elementary objects (Definition \ref{elementaryobject}). It is natural to consider the possibility that such objects exist. However, to be certain that an object is elementary, potential knowledge of this object must be complete. Given the unavoidable incompleteness of knowledge, we can never be sure that a candidate of an elementary object is indeed elementary. Their existence can neither be proved, nor disproved, in any operational sense. From a strict epistemic perspective, we should therefore not refer to elementary objects in expressions of physical law. More precisely, this follows from the assumption of epistemic minimalism (Section \ref{minimalism}).

The explicit form of epistemic minimalism (Assumption \ref{explicitepmin}) says that the introduction in a physical model of elements that cannot be verified or refuted should give rise to wrong predictions. If we introduce elementary objects as the basic building blocks in a physical model of the world, then the world functions \emph{as if} there were a fixed number of such elements present. Each perceived object could be divided into a given set of these elementary objects. But since we cannot know exactly how many elementary objects there are in a macroscopic object, and we cannot verify or refute the hypothesis that their number stays fixed, the world cannot function \emph{as if} this was the case, according to the principle of explicit epistemic minimalism.

Nevertheless, it is possible to introduce a weaker form of elementarity that is epistemically acceptable.

\begin{defi}[\textbf{Minimal set} $\mathcal{M}$ \textbf{of objects}]
Let $\mathcal{M}$ be a set of distinguishable independent objects defined exclusively by their internal attributes. Then $\mathcal{M}$ is a minimal set if and only if it fulfils the following conditions. 1) The number of objects in $\mathcal{M}$ is finite. 2) Division of any object in $\mathcal{M}$ gives rise to objects that are elements in $\mathcal{M}$ themselves. 3) There is no subset of $\mathcal{M}$ that fulfils condition 2).
\label{minimalset}
\end{defi}

Each object in $\mathcal{M}$ has to be defined by the same set of attributes, and the sets of allowed attribute values for each object must be distinct from the sets of values that belong to the other objects in $\mathcal{M}$.

\begin{defi}[\textbf{Minimal object} $M_{l}$]
An object that belongs to a minimal set $\mathcal{M}$ of objects.
\label{minimalobject}
\end{defi}

Just like elementary objects, minimal objects form a basic object layer, but in contrast to elementary objects, they can be divided. Needless to say, these minimal objects correspond to elementary particles, as perceived in current high energy physics, particles that may transform according to the allowed Feynman diagrams, creating other elementary particles in the process.

Just like for elementary objects, it is meaningless to assign a size to a minimal object $M_{l}\in \mathcal{M}$. A minimal object $M_{l}$ with a size attribute must always be considered to consist of at least two other objects $M_{l'}$ and $M_{l''}$, since size requires a measure of distance, and distance relates two objects. According to property 2), we must have $M_{l'}\in \mathcal{M}$ and $O_{l''}\in \mathcal{M}$. Then it is sufficient to consider these two objects as minimal, not $M_{l}$. Thati is, to fulfil property 3), $M_{l}$ must be excluded from $\mathcal{M}$.

Let us reformulate Assumption \ref{finitedepth1} using the concept of minimal objects.

\begin{assu}[\textbf{The depth of knowledge is finite}]
Any object can be divided into minimal objects in a finite number of steps.
\label{finitedepth}
\end{assu}

If the world $\Omega$ as a whole is regarded to be an object, then it can only contain a finite number of minimal objects. This is too restrictive. We want to allow the world to be infinite in size, so that we can come across a (countably) infinite number of objects if we travel along a straight line forever. To make this possible we have to make a distinction between the division of an object and the appearance of new objects among those already perceived. Even if the depth of knowledge is finite, we do not want to put boundaries to the scope of knowledge by assumption.

\begin{defi}[\textbf{The complement to an object}]
Let $O$ be an object. Then the world $\Omega$ consists of $O$ and the complement $\Omega_{O}$ to $O$, where $\Omega_{O}$ may or may not contain a countably infinite number of minimal objects.
\label{complement}
\end{defi}

Thus the complement to an object is not an object itself. The world contains at least one object. The complement to this object is defined, and the option that this complement can be divided into infinitely many minimal objects is left open. This means that the world may contain infinitely many objects.

There is another sense in which the world cannot be regarded as an object. By definition, the world lacks a complement. In contrast, all objects have a complement which contains at least one other object. Each perceived object corresponds to another object in the body of the perceiving subject according to the assumption of detailed materialism (Assumption \ref{localmaterialism}).

\begin{state}[\textbf{All objects have a complement}]
To each object $O$ is associated a complement $\Omega_{O}$ that contains at least one object.
\label{objectcomplement}
\end{state}

\begin{state}[\textbf{The world is not an object}]
The world $\Omega$ is allowed to contain a countably infinite number of minimal objects. We cannot associate any complement to $\Omega$ that contains any objects.
\label{worldnoobject}
\end{state}

Let me give a second motivatation of the first part of this statement as follows. To say that the world is an object is to state \emph{a priori} that it contains a finite number of objects. But we can never be sure about that. We cannot exclude all hypothetical states of the world that contains an infinite number of objects; such states do not necessarily contradict our potential knowledge. To be able to acommodate such states, the world itself must be described as something else than an object.

In contrast, our actual \emph{potential knowledge} of the world \emph{may} correspond to a (composite) object. This knowledge corresponds to an object if and only if the number of aware subjects is finite. This is so since each of the finite bodies of these subjects can only perceive a finite number of objects.

\begin{state}[\textbf{The potential knowledge of the world may correspond to an object}]
The potential knowledge $PK_{\Omega}$ corresponds to an object $O_{\Omega}$ if and only if the number of subjects is finite.
\label{kworldmayobject}
\end{state}

Of course, any individual subject is unable to decide whether there are infinitely many other subjects or not. Therefore the question whether the knowledge of the world represents an object is probably undecidable.

\section{Epistemic invariance}
\label{epinvariance}

Previously, we assumed that physical law is independent of our conception of physical law (Assumption \ref{thereislaw}). Now, we assume that it is independent of our knowledge.

\begin{assu}[\textbf{Epistemic invariance}]
The evolution of a set of objects, as given by physical law, is independent of the potential knowledge of these
objects and their surroundings.
\label{epistemicinvariance}
\end{assu}

Physical law acts on the objects and their attributes; there is nothing else to act on. The attributes may take different values, and they may be more or less known. The meaning of Assumption \ref{epistemicinvariance} is thus twofold: the evolution given by physical law depends neither on the \emph{content} of potential knowledge, nor on the \emph{amount} of potential knowledge.

\begin{figure}[tp]
\begin{center}
\includegraphics[width=80mm,clip=true]{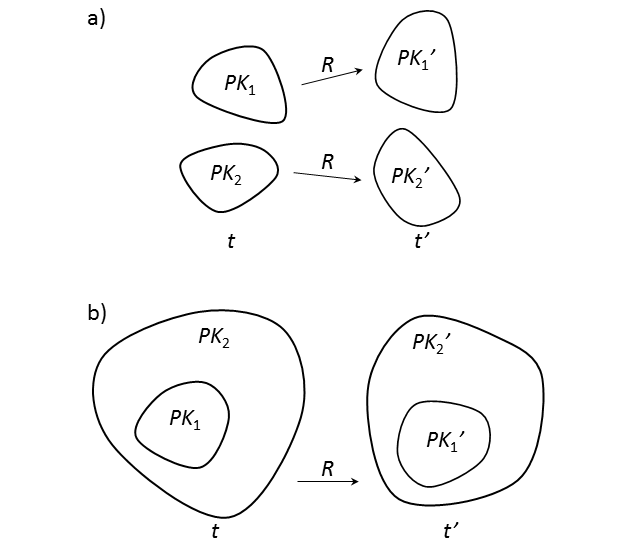}
\end{center}
\caption{Epistemic invariance with respect to a) content and b) amount of potential knowledge. a) The same evolution rule $\mathcal{R}$ applies to all states $PK$. b) This rule is such that if one state of potential knowledge is contained in another, then this fact stays true at all subsequent times. This is an alternative formulation of the principle of reductionism.}
\label{Fig12}
\end{figure}

Generally speaking, the existence of physical law means that for each state of potential knowledge $PK_{j}$, there is a rule $\mathcal{R}_{j}$ that limits its evolution in time [Fig. \ref{Fig12}(a)]. To formulate a universally valid law, in a worst case scenario, it would be necessary to list the rules $(\mathcal{R}_{1},\mathcal{R}_{2},\ldots)$ for each of the potentially infinitely many allowed states $(PK_{1},PK_{2}, \ldots)$.

The assumption that physical law is independent of the content of knowledge means that it is possible to contract this list to a single rule $\mathcal{R}$, consisting of a finite number of logical and mathematical symbols. This corresponds to the Newtonian paradigm that there is a clear division between physical state and physical law. The law acts on the state to make it change with time, whereas the law itself is independent of state and time.

Lee Smolin \cite{smolin} has challenged this paradigm from a cosmological perspective, proposing a model where the state and a slowly varying law is merged into a \emph{metastate}, acted upon by a \emph{metalaw}. This metalaw must itself be independent of the metastate to avoid infinite regress, but it is argued that the metalaw can be much simpler than the slowly varying law. In the formalism developed here, these considerations make no difference, since we make no specific assumptions about the \emph{form} of physical law $\mathcal{R}$ that distinguish law from metalaw. If a metalaw exists, $\mathcal{R}$ will correspond to the combined law and metalaw, acting on the state $PK$.

A more traditional way to allow for laws that change with time is to regard time as a parameter whose value is part of the state. We merge state and time, instead of state and law. However, such explicit time dependence is in opposition to the strict epistemic perspective I strive to adopt. There is no operational way to distinguish two points in time if the states of potential knowledge at the two instants are identical, including identical potential memories of the past. We arrive at the following conclusion.

\begin{state}[\textbf{Physical law is independent of time}]
Physical law is invariant under time translations.
\label{timeindependence}
\end{state}

Let us turn to the second part of the meaning of Assumption \ref{epistemicinvariance}. A state of potential knowledge $PK_{1}$ is a subset of another state $PK_{2}$ when at most the same independent objects are known in $PK_{1}$ as in $PK_{2}$, and there is at most the same knowledge of \emph{all} the attributes of these objects in $PK_{1}$ as there are in $PK_{2}$.

The assumption is then that if $PK_{1}$ is a subset of $PK_{2}$ at some time $t$, then the same is true for the corresponding evolved states of knowledge at any subsequent time $t'$ [Fig. \ref{Fig12}(b)].

There is a potential problem with this way to describe the assumption that physical law is independent of the amount of potential knowledge. If subjects can search for new knowledge in such a way that their choices are not entirely dictated by physical law, then the state of affairs expressed in Fig. \ref{Fig12}(b) is not waterproof. The subjects possessing potential knowledge $PK_{1}$ may then choose to look for new knowledge that is not part of $PK_{2}'$, and find before time $t'$. If this freedom of choice is real, the qualification that no such choices are made in the time interval $[t,t']$ must be added.

The time independent relation that $PK_{1}$ is a subset of $PK_{2}$ corresponds to the principle of reductionism. The statement `the behaviour of the whole can always be understood in terms of the behaviour of its parts' can be translated into our language as `the evolution of a state of less potential knowledge is always consistent with the evolution of states of more potential knowledge'. The reductionistic principle follows since the additional knowledge may, of course, be gained by object division, that is, by identification of the parts of a system. In other words, according to Assumption \ref{epistemicinvariance}, the world is not holistic in the sense that the avoidance of additional knowledge enables new behaviour.

\begin{figure}[tp]
\begin{center}
\includegraphics[width=80mm,clip=true]{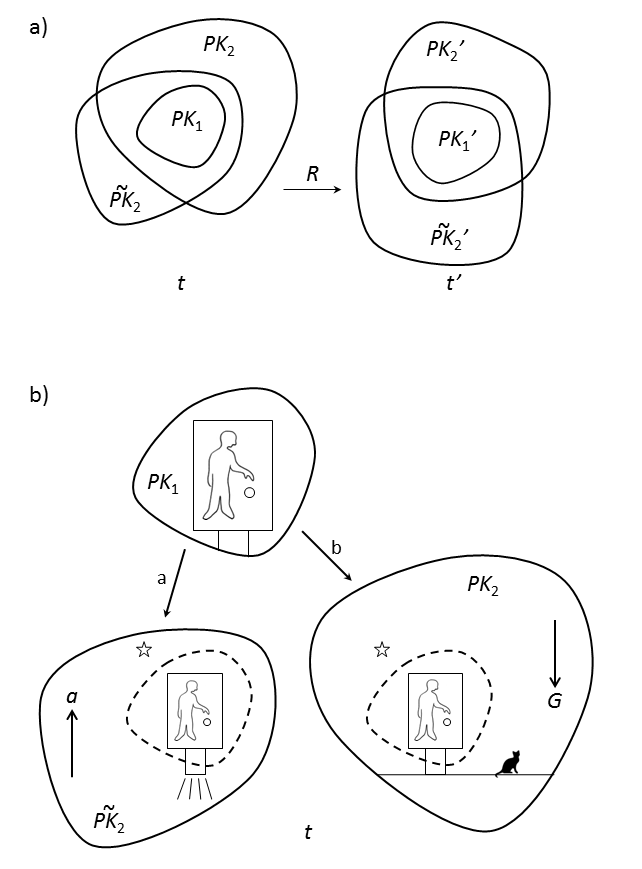}
\end{center}
\caption{a) A corollary of the epistemic invariance expressed in Fig. \ref{Fig12} is that if two states of potential knowledge $PK_{2}$ and $\tilde{PK}_{2}$ overlap, they do so at all subsequent times. b) The idea is illustrated by Einstein's equivalence principle. Potential knowledge grows along arrows a and b, just as in Figs. \ref{Fig4} and \ref{Fig5}.}
\label{Fig13}
\end{figure}

Any state $PK_{2}$ is embedded in the currently unknowable (Fig. \ref{Fig3}), so that it is always possible to choose another state $\tilde{PK}_{2}$ of which $PK_{1}$ is a subset [Fig. \ref{Fig13}(a)]. Thus, according to Assumption \ref{epistemicinvariance}, the evolved state $PK_{1}'$ must stay a subset of both the evolved states $PK_{2}'$ and $\tilde{PK}_{2}'$. It follows that if two states of potential knowledge contain common knowledge, they will continue to do so.

Einstein's equivalence principle provides a neat illustration [Fig. \ref{Fig13}(b)]. Let $PK_{1}$ correspond to a state of potential knowledge when all observers are stuck inside an elevator, and assume that its walls do not let any physical information from the outside reach the senses of the observers. Let $PK_{2}$ be a state where there are outside observers noting that the elevator is standing on a gravitating body, and let $\tilde{PK}_{2}$ be a state where there are outside observers noting that the elevator is accelerating in free space. The evolution of $PK_{1}$ must be consistent with the evolution of both these states of enlarged knowledge. In other words, if the evolution rule $\mathcal{R}$ is such that it is impossible to tell from the inside whether the elevator is accelerating or affected by gravity (or a combination of both) at a given time, it will remain impossible forever.

\begin{figure}[tp]
\begin{center}
\includegraphics[width=80mm,clip=true]{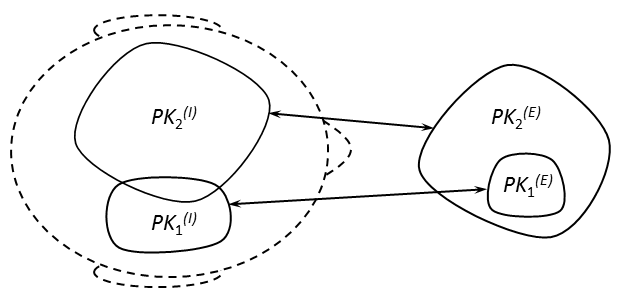}
\end{center}
\caption{Illustration of Statement \ref{sizecontent}. If potential knowledge $PK^{(E)}$ of the external objects shrinks from $PK_{2}^{(E)}$ to $PK_{1}^{(E)}$, the potential knowledge $PK^{(I)}$ of the internal objects of the observing subjects changes in such a way that $PK_{1}^{(I)}$ is not a subset of $PK_{2}^{(I)}$. The state of the body becomes knowably different. It takes a different state of the brain to encode a fuzzier visual image, not just less knowledge of the state that encoded the sharper image. The state of potential knowledge $PK$ is the union of $PK^{(E)}$ and $PK^{(I)}$. Therefore two states $PK$ and $PK'$ are never subsets of each other in practice.}
\label{Fig13b}
\end{figure}

It must be stressed that the assumption that the physical evolution rule does not depend on the amount of potential knowledge is a purely theoretical statement about the structure of physical law. It is not possible in practice to have two states of potential knowledge such that one is a subset of the other (Fig. \ref{Fig13b}). It is, of course, possible to have different amounts of knowledge of a given set of \emph{external} objects. But such a difference must be encoded as different states of the \emph{internal} objects in the bodies of the knowing subjects, simply because it corresponds to a different perception. This conclusion follows from detailed materialism (Assumption \ref{localmaterialism}). The corresponding states of knowledge of the internal objects therefore have different content, to some extent, and are not subsets of each other. Since the state of potential knowledge is the knowledge of \emph{all} objects, both external and internal, two such states are never subsets of each other. If our eyesight becomes fuzzier, we know someting we did not know before: either we have changed accomodation, our eyes have deteriorated, or we have lost our glasses.

\begin{state}[\textbf{If potential knowledge changes size, it changes content}]
If the state of potential knowledge is $PK_{a}$ at time $t'$, there is no past time $t$ or future time $t''$ such that the state of potential knowledge $PK_{b}$ at that time fulfils $PK_{a}\subset PK_{b}$ or $PK_{b}\subset PK_{a}$.
\label{sizecontent}
\end{state}

Just as the evolution rule is assumed to be the same regardless the amount and content of potential knowledge, we also assume that it is the same for all subjects.

\begin{assu}[\textbf{Individual epistemic invariance}]
The evolution of a set of objects, as given by physical law, is independent of who posseses the potential knowledge of these objects and their surroundings.
\label{indepistemicinvariance}
\end{assu}

According to Assumption \ref{collectiveknow}, potential knowledge is the union of the individual potential knowledge of all subjects. The elements in a union commute, so that there is complete democracy among subjects. No `master subject' exists. The attributes of the objects in the state of potential knowledge are the attributes that the individual subjects perceive. Different subjects may assign different values to these attributes, for example spatio-temporal distances. Therefore, since the evolution rule $\mathcal{R}$ is assumed to be common to all subjects, and since it acts upon these attributes, we run into contradiction if it acts upon them in such a way that the evolved state becomes different depending on whose individually perceived attribute values we let it act on. (We need to bother only about identifiable, independent objects according to Definition \ref{indobjects}. It is only for such objects it is meaningful to say that two subjects refer to the same object, and agree or disagree about its attributes. Only then can a relation $r_{j}^{kk'}$ between two subjects $k$ and $k'$ be defined according to Fig. \ref{Fig7b}.)

\begin{state}[\textbf{Individually invariant evolution rule}]
Let $\mathcal{R}$ be the common evolution rule for each state of potential knowledge and each subject according to Assumptions \ref{epistemicinvariance} and \ref{indepistemicinvariance}. Then $\mathcal{R}$ acts upon individually perceived attribute values in such a way that the evolved state of potential knowledge is the same regardless whose perceived attribute values we let $\mathcal{R}$ act upon. If $\{\upsilon_{i}^{k}\}$ is a set of attribute values perceived by subject $k$ and $\{\upsilon_{i}^{k'}\}=T\{\upsilon_{i}^{k}\}$ is the corresponding set perceived by $k'$, then we must have $R\{\upsilon_{i}^{k'}\}=TR\{\upsilon_{i}^{k}\}$, that is $[R,T]=0$.
\label{individualepinv}
\end{state}

Another way to express the same conclusion, without referring to attributes, is the following.

\begin{state}[\textbf{Individually invariant evolution rule 2}]
Let $\mathcal{R}$ be the common evolution rule for each state of potential knowledge and each subject according to Assumptions \ref{epistemicinvariance} and \ref{indepistemicinvariance}. Then $\mathcal{R}(PK^{k}\cap PK^{k'})=\mathcal{R}PK^{k}\cap \mathcal{R}PK^{k'}$, where $PK^{k}$ and $PK^{k'}$ are the states of potential knowledge of subjects $k$ and $k'$, respectively.
\end{state}

In words, the evolution of the common part of the potential knowledge of two subjects is the same regardless which of them use physical law to determine this evolution - it is objective in the sense that it transcends the individual.  

\section{Epistemic closure}
\label{minimalism}

History of physics teaches that it is a dead end to introduce objects that are epistemically unreachable, such as the aether, or make use of attributes that cannot be operationally defined or measured, such as positions in absolute space. Physical law is epistemically picky. This principle may be called \emph{epistemic minimalism}. We will see below that it is possible to identify two levels of this principle: implicit and explicit minimalism.

Even though it is a standard conclusion, it may be instructive to argue, using the vocabulary introduced above, that the concept of absolute speed is epistemically empty. To speak of absolute speed, there has to be at least one observable point in space with special status, in relation to which absolute speeds can be defined. With our general definition of an object, this point becomes an object $O_{o}$, and its special status as reference becomes one of its internal attributes. The absolute speeds of all other objects becomes relational attributes relating these to $O_{o}$. However, in such a state of knowledge, the speed $v$ of $O_{o}$ remains undefined. To define it, there has to be another object $O_{o'}$ in relation to which the speed of $O_{o}$ can be determined. We end up in infinite regress.

In the same way, it is impossible to define absolute acceleration epistemically. In other words, the perspectives where an object accelerates in relation to a background of other objects, and where the background accelerates in the opposite direction in relation to the given object, correspond to the same state of knowledge of spatial and temporal relational attributes. It is an elementary fact, however, that we can \emph{feel} acceleration, but this subjective feeling cannot be used to \emph{define} absolute acceleration from the spatio-temporal attributes alone.

\begin{defi}[\textbf{Implicit epistemic minimalism}]
Physical models can be expressed without the introduction of distinctions that cannot be subjectively perceived as such, or deduced from such perceptions. This is true, in particular, when it comes to the introduction of objects, attributes and attribute values. It is also true with regard to discriminations between attribute values in a model of the physical state.
\label{implicitepmin}
\end{defi}

The primary example is Galilean invariance. There is no need to use the attribute \emph{absolute speed} to formulate physical law: it is invariant under the transformation $x\rightarrow x+vt$ for any constant $v$.

Even if physical law does not need the idea of absolute speed, it is nevertheless compatible with Galilean invariance. Therefore Newton could uphold the idea of absolute space. This is the reason I call this level of epistemic minimalism \emph{implicit}.

A second, important, application of implicit epistemic minimalism is that physical law should not depend on the concept of elementary objects (Definition \ref{elementaryobject}), as discussed in section \ref{limits}, since it is in principle impossible to determine whether an object is elementary, given the fundamental incompleteness of knowledge (Statement \ref{incompleteknowledge}). Their role may instead be played by minimal objects (Definition \ref{minimalobject}), which are epistemically more appropriate.

We argued above that the attribute \emph{absolute acceleration} is impossible to define epistemically, just as \emph{absolute speed}. Thus, according to epistemic minimalism, it should be possible to formulate physical law without referring explicitly to the acceleration of a given object. To this end, there has to be another physical circumstance that gives rise to the same \emph{feeling}, so that the physical law that accounts for this feeling does not need to refer explicitly to acceleration. This other circumstance is gravity, of course, and we again arrive at the equivalence principle.

Going from the equivalence principle to general relativity, we need special relativity, and to arrive at special relativity, we need the stronger version of epistemic minimalism.

\begin{assu}[\textbf{Explicit epistemic minimalism}]
The introduction in a physical model of entities or distinction that cannot be subjectively perceived, or deduced from such perceptions, leads to conflict with physical law. The model gives rise to wrong predictions. This is true, in particular, when it comes to the introduction of objects, attributes and attribute values. It is also true with regard to discriminations between attribute values in a model of the physical state.
\label{explicitepmin}
\end{assu}

This principle constrains the form of physical law further. It means that Nature explicitly answers "yes" or "no" if we ask her whether a proposed object or attribute has epistemic meaning.

To obey explicit epistemic minimalism, physical law must be inconsistent with the notion of absolute speed, since we concluded above that this concept is epistemically empty. The notion of absolute speed implies the addition law for velocities, and whenever the addition law holds it is possible to uphold the notion of absolute speed. A bit more generally, we may say that the notion of absolute space and time can be kept alive if and only if the addition law for velocities always holds.

The addition law may expressed as follows. Consider any three objects $O_{1}$, $O_{2}$ and $O_{3}$. Let $u_{12}$ be the relative velocity of $O_{1}$ and $O_{2}$, as judged in the rest frame of $O_{1}$. In the same way, let $u_{23}$ be the relative velocity of $O_{2}$ and $O_{3}$, as judged in the rest frame of $O_{2}$. Then the relative velocity of $O_{1}$ and $O_{3}$ is

\begin{equation}
u_{13}=u_{12}+u_{23},
\label{additionlaw}
\end{equation}
as judged in the rest frame of $O_{1}$.

To rule out the notion of absolute speed, it is thus necessary and sufficient that physical law sometimes break Eq. (\ref{additionlaw}). One way to do this is to introduce a maximum speed $c$ that no object ever exceeds. To give such a concept epistemic meaning, all subjects must agree that a given type of object, under given circumstances that all agree upon, always travels at speed $c$ in their own reference frame. Of course, this ansatz leads to the Lorentz transformation. In the limit $c\rightarrow \infty$ we get $\sqrt{1-(v/c)^{2}}\rightarrow 1$, and we get back the Galileo transformation, for which Eq. (\ref{additionlaw}) holds. Thus, the introduction of a maximum speed is the only possible way to break Eq. (\ref{additionlaw}). In other words, the finite speed of light and special relativity can be seen as an expression of explicit epistemic minimalism.

To reach this conclusion, we made some crucial assumptions, though. To say that the Lorentz transformation contradicts Eq. (\ref{additionlaw}), we allowed the bodies of any two subjects $1$ and $2$ to constitute two of the three objects, say $O_{1}$ and $O_{2}$. Further, the relative speed $u_{13}$ is taken to be the speed of object $O_{3}$ relative to subject $1$, whereas the relative speed $u_{23}$ is taken to be the relative speed of object $O_{3}$ relative to the other subject $2$. If, instead, $u_{23}$ is taken to mean the relative velocity of object $O_{3}$ relative object $O_{2}$ as seen by subject 1, Eq. (\ref{additionlaw}) would still hold for all three objects in the reference frame of subject 1. In effect, we have assumed that the addition law for velocities should hold regardless which subjective reference frames we choose to measure the three relatives velocities.

This is an application of individual epistemic invariance (Statement \ref{individualepinv}). Thus, the finite speed of light and the Lorentz invariance of physical law, can be seen as consequences of explicit epistemic minimalism \emph{and} epistemic invariance. Conversely, the well-established fact that physical law indeed has these properties gives a strong hint that the idea of an epistemically invariant `subject democracy' and collective potential knowledge is correct. In a solipsistic world-view, these properties of physical law would not be necessary.  

The statistics of identical particles is another example of explicit epistemic minimalism. Since it is epistemically meaningless to treat permutations of identical particles as different states, the principle implies that it must lead to wrong statistics if such permutations are included as distinct states in the calcultation of statistical weights. And, of course, it does, since it gives rise to Maxwell-Boltzmann statistics, which in many experimental situations is physically very different from the correct Bose-Einstein and Fermi-Dirac statistics of bosons and fermions, respectively. In contrast, if the minimalism would have been implicit, it would have made no physical difference whether permutations of identical particles were included or not.

We will argue that all minimal objects that are building blocks to perceivable objects are fermions and obey the Pauli exclusion principle. (This picture is developed gradually, and is summarized in section \ref{fermbos}). The simple reason is that it does not make epistemic sense to say that two objects are found in the same state. To be able to determine in practice that we are dealing with two objects, they must be divided; there must be some knowledge that tells them apart, be it that their spatio-temporal positions are different or that their internal attributes differ. From this perspective, the fact that electrons and other building blocks of matter follow Fermi-Dirac rather than Bose-Einstein statistics is an expression of explicit epistemic minimalism. 

As a fourth example of this principle, we may take the fact the orbital angular momentum of a possibly rotating object has to be set to zero if we have no potential knowledge at all where in its orbit the object is positioned at a given time, that is, if the probability distribution of its position is spherically symmetric. Allowing for non-zero angular momentum in such a case gives rise to erroneous physical predictions, making the minimalism explicit.

\begin{figure}[tp]
\begin{center}
\includegraphics[width=80mm,clip=true]{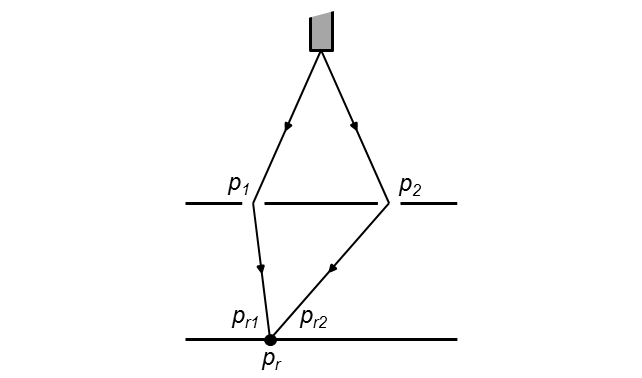}
\end{center}
\caption{Probabilities that can be used to illuminate the difference between implicit and explicit epistemic minimalism in the double-slit experiment. See text for further explanation.}
\label{Fig14}
\end{figure}

A central idea in the upcoming attempt to motivate quantum mechanics from epistemic principles is that the double-slit experiment should be looked at in the light of explicit epistemic minimalism. Consider Fig. \ref{Fig14}, and assume that it is forever outside potential knowledge which slit the particle actually passes. However, there is (potential) knowledge that it passes slit 1 with probability $p_{1}$ and slit 2 with probability $p_{2}$. Implicit epistemic minimalism would mean that it does not matter for the evolution of the system whether we assume that it actually takes one of the paths, even if we can never know which. The only option is then to combine probabilities as if events 1 and 2 are mutually exclusive. That is, for any pair of probabilities $p_{k}$ and $p_{rk}$ we must have

\begin{equation}
p_{r}=p_{1}p_{r1}+p_{2}p_{r2},
\label{normalprob}
\end{equation}
where $p_{r}$ is the probability that the particle finally hits the point $r$ on the detector screen, $p_{k}$ is the probability that it passes slit $k$, and $p_{rk}$ is the probability that it hits $r$ given that it has passed slit $k$.

In contrast, explicit epistemic minimalism means that physical law must contradict the possibility that there is (unknowable) path information. The only way to get the message through is to let

\begin{equation}
p_{r}\neq p_{1}p_{r1}+p_{2}p_{r2}
\label{quantprob}
\end{equation}
for some choice of probabilities. In Section \ref{bornrules} we discuss how this condition leads to Born's rule and contributes to the fact that physical states can be seen as elements in a Hilbert space.

Loosely speaking, what explicit minimalism does in this case is to let Nature give a clear answer "no" to the question "If a tree falls in the forest and no one sees it, does it then fall?". The question "Does this mean that it does not fall?" would also get a negative answer. The tree neither falls or does not; the question is ill-posed, it has no knowable answer. In contrast, in a world described by classical mechanics, Nature would be unable to answer anything at all.

The previous discussion concerned the fact that physical law seems to make use of \emph{no more} than what can be perceived and distinguished in principle. Turning the perspective around, we may argue that physical law should make use of \emph{everything} than can be perceived and distinguished from something else. Loosely speaking, if two things can be distinguished, there will be a corresponding distinction in physical law.

\begin{state}[\textbf{Epistemic completeness}]
All subjectively perceived distinctions, or distinctions deduced from such perceptions, correspond to distinctions in proper models of physical law and of the physical state. This is true, in particular, when it comes to objects, attributes and attribute values.
\label{epcompleteness}
\end{state}

Comparing Assumption \ref{explicitepmin} and Statement \ref{epcompleteness}, the former basically says `everything physical is epistemic', while the latter says `everything epistemic is physical'. Statement \ref{epcompleteness} can be seen as a consequence of detailed materialism (Assumption \ref{localmaterialism}), the hypothesis that every detail of subjective perception has a physical description.

Since we can deduce that there is a distinction between potential knowledge and the currently unknowable (Fig. \ref{Fig3}), a corresponding distinction should be made in physical law. This is accomplished by the distinction between Eqs. [\ref{normalprob}] and [\ref{quantprob}].

We concluded above that gravity is envitable given the possibility to \emph{feel} acceleration, since there has to be another interpretation of this feeling, according to implicit epistemic minimalism. Actually, we jumped to conclusion. We also have to assume epistemic completeness to be sure that the subjective distinction between \emph{feeling} and \emph{no feeling} of acceleration corresponds to a distinction in physical law. This distinction is expressed by the fact that the evolution of the physical state depends on whether or not there is acceleration (or gravity). The acceleration dependence is formulated in Newton's second law.

Another application of epistemic completeness is time. In the next section the distinction between relational and sequential time is elaborated. The former aspect of time is an attribute relating two objects, just as distance, whereas the latter aspect corresponds to our ability to order objects or events into the past, the present and the future. The distinction between the two is at the core of our perception of time. Statement \ref{epcompleteness} thus dictates that this distinction should be respected in equations that express physical law. Such an equation is suggested in Section \ref{eveq}.

\begin{figure}[tp]
\begin{center}
\includegraphics[width=80mm,clip=true]{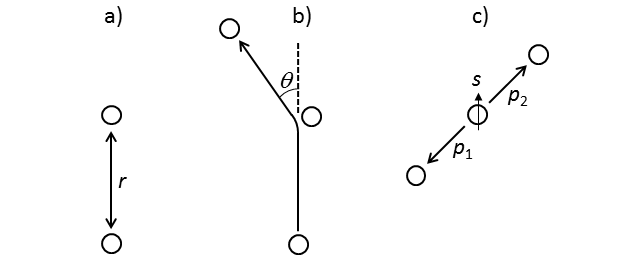}
\end{center}
\caption{Relations between objects in physical space can be described by distances $r$, angles $\theta$, and orientation. Our ability to perceive such a space is reflected in physical law as dependencies on a) distance and b) angle, and the fact c) that parity violation occurs.}
\label{Fig16}
\end{figure}

The ability to construct and give meaning to mathematical concepts is often imagined to be independent of physical law. However, if detailed materialism is taken seriously, the perceived mathematical world cannot `hover' above the physical world, but must be related to it in every detail. For example, it should be possible to relate all aspects of the mathematical representation of physical space to the behaviour of attributes of objects. In other words, it must be possible to relate the defining properties of $\mathbf{\mathbb{R}}^{3}$ and Riemannian manifolds to distinctions made by physical law.

The ability to distinguish different distances from each other is reflected by the distance-dependence present in physical law (Fig. \ref{Fig16}). The strength and the time delay of the interaction between two objects $O_{i}$ and $O_{j}$ both depend on a variable $r\geq 0$ that we interpret as distance, and so does wave diffraction. Using these dependencies to determine the distances $\{r_{ij}\}$ in a collection of objects, it is \emph{a priori} impossible to exclude any set $\{r_{ij}\}$ as a possible outcome.

However, it is a basic fact that the triangle inequality is always fulfilled. This condition is reflected by physical law if we assume that interactions between $O_{i}$ and $O_{j}$ depend on the \emph{shortest} distance between the two objects, and that it is this distance that we denote by $r_{ij}$. Introducing the concept of shortest path implies that there are other possible paths. The only meaning that can be attached to this statement is that there are paths from $O_{i}$ to $O_{j}$ via other objects $O_{k}$. Breaking the triangle inequality means that there is an $O_{k}$ such that $r_{ik}+r_{jk}<r_{ij}$, i.e. that the path via $O_{k}$ is shorter than $d_{ij}$, and we have a contradiction.

Of course, the set of distances fulfill a much more restrictive constraint than the triangle inequality. It is an experimental fact that distances seem always to be related in such a way that locally, the objects and distances can be embedded in a Euclidean, three-dimensional space.

\begin{figure}[tp]
\begin{center}
\includegraphics[width=80mm,clip=true]{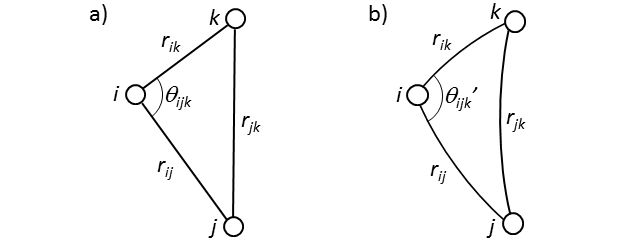}
\end{center}
\caption{If curvature is allowed, it is necessary to define angles to specify the space. Even if all distances are equal, the angles may differ.}
\label{Fig16b}
\end{figure}

Does the distance-dependence of physical law provide enough information to construct physical space? The angle $\theta_{ijk}$ is uniquely defined by the three distances $r_{ij}$, $r_{ik}$ and $r_{jk}$, given that space is Euclidean and three-dimensional (Fig. \ref{Fig16b}). However, as Einstein pointed out, the equivalence principle implies that the apparent trajectory of a light beam is bent by gravity, like the trajectory of any other object. Therefore, since the shortest path between two objects must be operationally defined by some interaction of objects, it cannot be described by a straight line (in the Euclidean sense) in the presence of gravity. Then the angles $\theta_{ijk}$ are \emph{not} uniquely given by the distances (except in the limit where the distances go to zero), but must be determined independently. To do so, and thus to define space unambiguously, there have to be physical processes that depend explicitly on angle (Fig. \ref{Fig16}). And of course there are such processes, such as scattering. 

Still, one basic ingredient is missing before we can define space as we know it, namely orientation. The concept of reflection cannot be introduced using the relational attributes distance and angle only. The reason why we appear to be able to construct the mirror images of the triangles in Fig. \ref{Fig16b} is just that they are embedded in a larger oriented space, our own. 

\begin{figure}[tp]
\begin{center}
\includegraphics[width=80mm,clip=true]{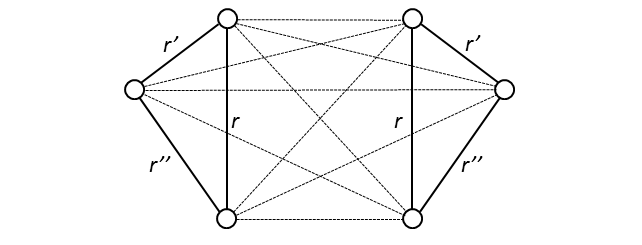}
\end{center}
\caption{It is possible to define mirror images in a minimal space defined just by distance, even if the objects are identical. But the space nevertheless lacks orientation.}
\label{Fig17}
\end{figure}

Nevertheless, it is possible to define mutual mirror images. The corresponding objects in the mirror triangles in Fig. \ref{Fig17} can be identified since these have identical distance relations to the other two objects in the triangle (solid lines). Proper mirror images can be distinguished from rotated images, since in the first case it is impossible to make them coincide by continuously decreasing all distances relating the two images to zero (dotted lines), without violating the constraint imposed by the dimensionality of the space.

To be able to define orientation, it is necessary to introduce an \emph{internal} attribute that is coupled to distance, in the sense that it defines a direction in relation to the distance between two objects. In addition, there has to be a physical law that depends on this direction. Spin and spin-dependent weak interactions fulfil these tasks. In other words, the possibility to distinguish left from right correponds to the existence of parity violation in physical law.

Mathematically speaking, spin and parity violation makes it possible to describe a Euclidean space as a vector space, since for each vector $\mathbf{v}$ it makes the reflected vector $-\mathbf{v}$ a meaningful concept. It becomes appropriate to assign a position vector to each object, to embed the space in a coordinate system. Of course, this procedure introduces redundancy in the representation, since any translation, rotation or motion of the coordinate system relative to the set of objects and attributes that constitutes the physical state cannot be defined from within this state itself. Keeping this in mind, the representation provides a great simplification as compared to working explicitly with the distances $\{r_{ij}\}$ and spins $s_{i}$.

\begin{figure}[tp]
\begin{center}
\includegraphics[width=80mm,clip=true]{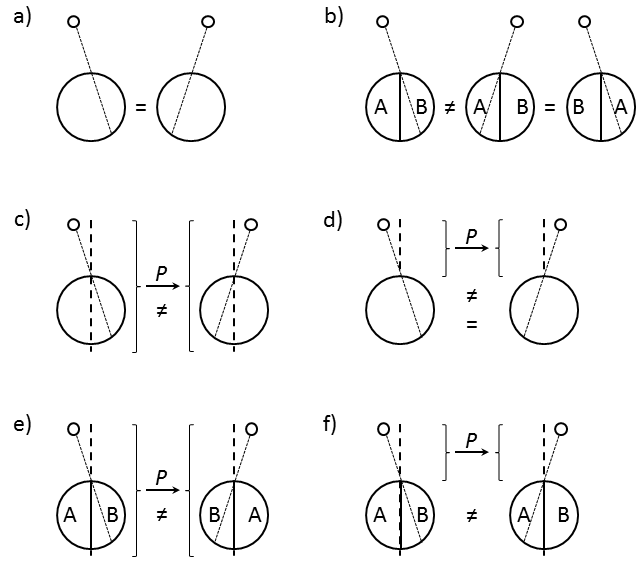}
\end{center}
\caption{Distinction between left and right. A subject (large circle) views an object (small circle). a) If the body halves are mirror images (Fig. \ref{Fig17}), the state where the object is seen to the left is identical to the state where it is moved to the right. No distinction between left and right is possible. b) If the body halves are asymmetric, subjective distinction between left and right is possible even if the space in which subjects live lacks orientation. c) Parity violation makes the space oriented and defines a distinct mirror image. d) If the parity operation $P$ is applied to the perceived object but not to the body, subjective distinction is possible if and only if the particular object violates parity. e)-f) If the body halves are asymmetric and there is parity violation, distinction between left and right is always possible.}
\label{Fig18}
\end{figure}

This story is too good to be the true, or at least to be the whole truth. It is possible to enable subjective distinction between left and right by means of less fundamental physical facts than parity violation (Fig. \ref{Fig18}). Namely, since the two hemispheres of the brain are different, the physical state of the brain as it processes a sensory input will be different depending on whether the input comes from the left or the right [Fig. \ref{Fig18}(b)]. This physical difference enables a difference in state of awareness, a subjective sense of left and right, according to the assumption of detailed materialism.\footnote{Similarly, the ability to decide whether an image is upside down in the absence of reference points must be attributed to the top-down asymmetry of the brain. Or, more properly, to asymmetry with respect to the corresponding plane defined by the projection of the image on the visual cortex.} 

But why has the left-right asymmetry of the brain (and other internal organs) evolved in the first place? There seems to be no consensus on this matter \cite{levin}. Of course, the ability to distinguish left from right, possibly with the help of distinct hemispheres, has significant evolutionary value. It makes it much easier to repeat a path that leads to a certain goal, based on the memory of this path. It also makes it possible to find or avoid hidden objects in an apparently symmetric environment based on the memory of the location of these objects. One may speculate that the fact that sensory input from left is always processed in the far right of the brain, and vice versa, has evolved as a way to magnify the difference between these physical states. A nerve signal travelling through the entire brain has the potential to make clear footprints in the distinct hemispheres, and in turn be affected by them in distinct ways. In this manner, the subjective distinction between left and right might become more robust.

The appearance of back-front and top-bottom asymmetries is easier to understand. The evolutionary drive to develop a back-front asymmetry is the need for an animal to move in order to find food. Thus its velocity vector corresponds to the vector defined by the alimentary canal. The top-down asymmetry is an adaptation to life on the ground of earth. (For stationary plants, this is the only inherent asymmetry axis.)

There have to be seeds to develop these asymmetries. The attachment of the embryo to the placenta (the birth cord) defines a back-front axis, and gravity defines a top-bottom axis. However, the seed for the development of left-right asymmetry is poorly understood. It might be the result of spontaneous symmetry breaking from a random fluctuation, like the direction of magnetization in the absence of external fields \cite{edlund}. But since almost all bodies of a given species are oriented in the same way \cite{levin}, there must be a preferred direction.

Such a seed of chirality is provided by the fact that only L-amino acids and D-sugars are present in the biological world. Several hypotheses have been put forward to explain this fact. Some involve polarized cosmic radiation or parity dependent binding energies. Thus spin and parity violation might be at the root of matters after all. However, since the parity-dependent energy shifts are minuscule, there has to be some magnifying process involved \cite{kondepudi,fitz}.  

It is important to note that regardless the biological basis for left-right asymmetry, only parity violation can motivate the use of an oriented space with position coordinates to describe the entire physical world of external objects \emph{and} bodies of subjects.   

Let me conclude this long section with its moral. Explicit epistemic minimalism means that any entity or distinction introduced in a physical model should correspond to a knowable entity or distinction. Epistemic completeness means that any knowable entity or distinction chould have a counterpart in a physical model. We get a one-to-one correpondence.

\begin{assu}[\textbf{Epistemic closure}]
There is a one-to-one correspondence between knowable entities and distinctions, and entities and distinctions in proper physical models.
\label{closure}
\end{assu}

\section{Time}
\label{time}

The concept of \emph{distinction} was used to argue that any given state of potential knowledge consists of a discrete set of objects. In the same way, from the epistemic perspective, to say that time has passed, it must be possible to make a distinction between now and then. Something must have changed subjectively. Hence the evolution of potential knowledge can be described as a discrete sequence $\{PK(n), PK(n+1), PK(n+2), \ldots\}$.

\begin{figure}[tp]
\begin{center}
\includegraphics[width=80mm,clip=true]{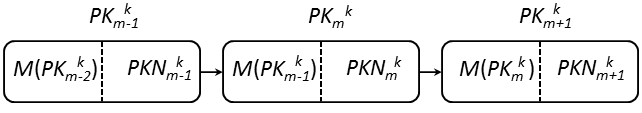}
\end{center}
\caption{The evolution of potential knowledge of a given subject $k$. At time $m$, the potential knowledge $PK_{m}^{k}$ consists of knowledge $PKN_{m}^{k}$ of the present, and knowledge $M(PK_{m-1}^{k})$ of the past, in the form of potential memories $M$. This part of $PK_{m}^{k}$ points to its predecessor, and makes it possible to order the states of knowledge into a sequence corresponding to the flow of time.}
\label{Fig20}
\end{figure}

Any perceivable change defines a new state of potential knowledge. It may be a bird appearing in the sky, or the appearance of the thought `nothing has changed in the sky' in the head of the skywatcher. In the latter case an internal object has appeared, corresponding to a changed physical state of the brain.

The characteristic aspect of time is that it flows. Time is directed, the ordering of states of knowledge is essential. The two sets $\{PK(n), PK(n+1), PK(n+2), \ldots\}$ and $\{PK(n), PK(n+2), PK(n+1), \ldots\}$ are different.

This primary fact corresponds to the interpretational ability to distinguish the present from the past (Fig. \ref{Fig2}). More precisely, most states of knowledge consist both of objects that correspond to (potential) memories, and of objects that correspond to the present. The ability to tell which objects are which has to be assumed. In this way, for a given subject, a predecessor of each state of individual potential knowledge is defined, and the ordering is established (Fig. \ref{Fig20}). This procedure may be compared to the Peano axioms, where each natural number is assumed to have a successor, making the natural numbers an ordered set.

We cannot take for granted that memory is perfect; it is not certain that $M(PK_{m-1}^{k})$ equals $PK_{m-1}^{k}$, provided the proper interpretation, indicated by $M$, that the former state corresponds to the immediate past is removed. If this would be the case, potential knowledge would grow without bound as time passes. This seems to contradict the fact that potential knowledge is limited. The matter is further discussed in Section \ref{entropy}.

In fact, we cannot be sure even that knowledge of the temporal ordering of memories is always preserved. A preserved ordering means that we can write $M(PK_{m-1}^{k})=M(\{M(PK_{m-2}^{k}),PKN_{m-1}^{k}\})=\{M(M(PK_{m-2}^{k})),M(PKN_{m-1}^{k})\}$. If this relation always applies, the memories $PKN_{m-u}^{k}$ of time $m-u$ enter the state of potential knowledge as $M^{u}(PKN_{m-u}^{k})$, where the position of $PKN_{m-u}^{k}$ in the chain of memories is identified by the superscript $u$ in $M^{u}$.

\begin{figure}[tp]
\begin{center}
\includegraphics[width=80mm,clip=true]{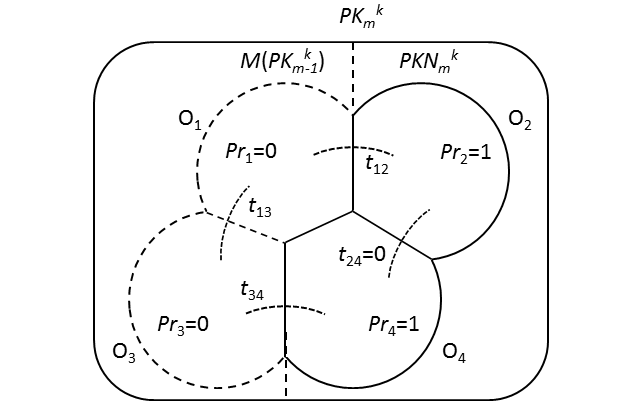}
\end{center}
\caption{The part $PKN^{k}_{m}$ of personal potential knowledge $PK^{k}_{m}$ that corresponds to the present, and the part $M(PK^{k}_{m-1})$ that corresponds to the past, each consist of a number of objects (c.f. Fig. \ref{Fig20}). Present objects are assigned `presentness' attribute $Pr=1$, and past objects are assigned $Pr=0$. The relational attribute $t$ can be defined between any two objects. It must be chosen such that the numbers sum to zero when added in a circle, e.g. $t_{13}+t_{12}+t_{24}+t_{34}=0$. If $Pr_{i}=1$ and $Pr_{j}=0$, the only available knowledge of $t_{ij}$ is that it is positive. Its magnitude is completely unknown.}
\label{Fig21}
\end{figure}

Nevertheless, the basic ability to order experiences in present and past is assumed. The fact that an object in a state of knowledge belongs to the past, to the set of memories, can be thought of as an internal attribute of that object. We may define a `presentness' attribute $Pr$, with possible values $Pr=0$ (past) and $Pr=1$ (present).

\begin{defi}[\textbf{The presentness attribute} $Pr$]
The internal attribute $Pr$ is defined for any object $O$. It has two possible values, zero and one. We have $Pr_{O}(n)=1$ if and only if $O\subseteq PKN(n)$, and $Pr_{O}(n)=0$ if and only if $O\subseteq M(PK(n-1))$. 
\label{presentness}
\end{defi}

In an individual state of potential knowledge $PK^{k}$, the distance in time $t_{ij}$ between any two objects $i$ and $j$ can be defined as a relational attribute that fulfils $t_{ij}\geq 0$ whenever $Pr_{i}=1$, and $t_{ij}=-t_{ji}$ (Fig. \ref{Fig21}). These relations reflect the directed nature of time, and can be generalized to the statement that the sum of time differences in a closed loop of objects is zero. They make the time distances different from spatial distances $r_{ij}$, which are non-negative and invariant under index exchange.

Knowledge of the temporal attributes $t$ may be incomplete (defocused or conditional) in the same way as knowledge of any other attribute. In fact, if $Pr_{i}=1$ and $Pr_{j}=0$, the only knowledge we have about $t_{ij}$ is that it is positive. The reason is the following. Given any two objects $O_{i}$ and $O_{j}$, the realization that they are related by the number $t_{ij}$ is a perceived change of potential knowledge. Thus it corresponds to a temporal update $n\rightarrow n+1$. Even if $Pr_{j}(n)=1$, we have $Pr_{j}(n+1)=0$ by definition. Since we do not have any knowledge of the magnitude of $t_{ij}$ until time $n+1$, we conclude that such knowledge can only exist in states for which $Pr_{i}=Pr_{j}=0$. (A trivial exception is the case where $Pr_{i}=Pr_{j}=1$ and $t_{ij}=0$ by definition.)

If knowledge of $t_{ij}$ is perfect for all objects belonging to the past, the relations described above mean that these objects can be represented as points on a directed time-axis (where the location of the origin is arbitrary). If knowledge of some $t_{ij}$ is defocused, temporal knowledge necessarily becomes conditional, due to the rule that time differences added in a loop must be zero. They cannot be independently chosen; if knowledge of one temporal attribute increases, so does knowledge of the others. Actually, the same is true for spatial distances $r_{ij}$. The empirical fact that they can be embedded in a three-dimensional space means that they cannot be independently chosen either. (Numerical knowledge of $r_{ij}$ is only available if $Pr_{i}=Pr_{j}=0$, for the same reason as for $t_{ij}$.)

In the context of relativity, objects described in this way are called events. Relations between these objects or events are described by the attribute pair $(r,t)$, which makes it possible to embed them in Minkowski space-time. Each state of potential knowledge $PK(n)$ can therefore be (partially) represented by a map of objects spread across space-time, a map that is updated each time potential knowledge is updated. This picture resolves the apparent contradiction between a `frozen' space-time and a flowing time. What we get is a sequence of frozen space-times.

This twofold nature of time should be manifest in evolution equations. In fact, it is well known that we run into problems when the sequential and relational aspects of time are described by a single variable $t$. In the field of quantum gravity, it is difficult to formulate evolution equations in a frozen, curved spacetime. Due to the general covariance, there is no way to single out a temporal direction in this manifold. This is known as `the problem of time'.

\begin{figure}[tp]
\begin{center}
\includegraphics[width=80mm,clip=true]{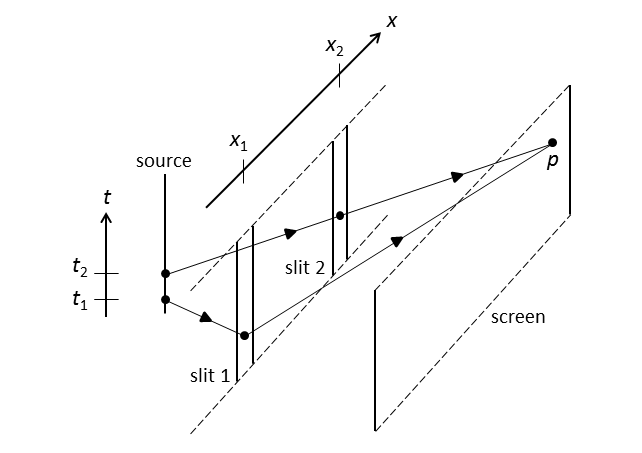}
\end{center}
\caption{The double-slit experiment not only demonstrates interference of spatially separated paths, passing the slits at $x_{1}$ or $x_{2}$, but also interference of temporally separated paths, starting from the source at two different times $t_{1}$ and $t_{2}$.}
\label{Fig21b}
\end{figure}

The need to separate the two aspects of time can be seen already in the double-slit experiment. Let us add a vertical time axis to the standard picture (Fig. \ref{Fig21b}). Assume 1) that a single object hits the detector screen at a point $p$ off the symmetry axis of the experimental setup (to the left or to the right), 2) that the speed of the object on its path from the source to the screen is known (such as the speed of light in the case of photons), and 3) that information about which slit the object passed is outside potential knowledge. Then, according to Eq. \ref{quantprob}, there is interference between the two alternative paths. But the two paths correspond to two different departure times from the source. Thus there is not only spatial interference between paths departing from the two slits located at positions $x_{1}$ and $x_{2}$, but also temporal interference between paths departing from the source at times $t_{1}$ and $t_{2}$. Furthermore, the Lorentz transformation partially transforms the distances $t_{21} = t_{2}-t_{1}$ and $x_{21} = x_{2}-x_{1}$ into each other when the state of motion of the observer changes.

Temporal and spatial interference must therefore be treated on an equal footing in proper evolution equations. In each state, different relational attributes $t$ must be allowed. This is not the case, for instance, in the Schr\"odinger equation, where the wave function $\Psi(x,t)$ is distributed in space, allowing an uncertainty $\Delta x > 0$ of spatial position, but is perfectly localized in time (as a delta spike), corresponding to a perfectly known relational time $\Delta t = 0$.

This is not a problem if the state described by the Schr\"odinger equation is stationary. In the double-slit experiment this corresponds to a periodic wave source, or a time-independent probability of object emission from the source. Then the detection probability densities at the screen are also independent of time. From an epistemic perspective, this corresponds to a complete lack of knowledge when the object is emitted from the source, implying that $t_{12}$ can take any value so that $\Delta t=\infty$.

In any physically and epistemically relevant situation, however, there is some knowledge of the timing of the particle emission. For instance, it cannot take place before the experiment starts. In other words, the state is never stationary. Then, to calculate the time dependent detection probabilities at the screen, the probability amplitudes of emission at all times $t_{12}$ must be known, due to temporal interference, in the same way as the spatial part of the wave function must be completely known to calculate its evolution in the Schr\"odinger equation. In other words, at each sequential time instant we must allow a $\Psi(x,t)$ with both $\Delta x > 0$ and $\Delta t > 0$ to be able to determine the physical state at the next time instant.

\begin{figure}[tp]
\begin{center}
\includegraphics[width=80mm,clip=true]{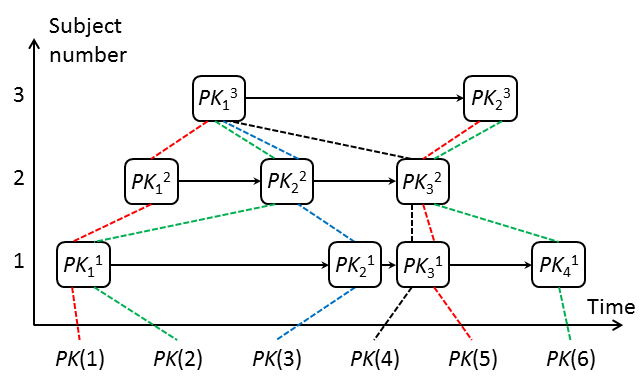}
\end{center}
\caption{The potential knowledge $PK(n)$ is updated each time any of the individual states of knowledge $PK_{m_{k}}^{k}$ is updated. For example, $PK(2)$ is the union of $PK_{1}^{1}$, $PK_{2}^{2}$, and $PK_{1}^{3}$. This state is updated to $PK(3)$ when $PK_{1}^{1}\rightarrow PK_{2}^{1}$. If two individual updates have space-like separation, they must be considered to occur simultaneosly, as the updates $PK_{2}^{1}\rightarrow PK_{3}^{1}$ and $PK_{2}^{2}\rightarrow PK_{3}^{2}$. Overlaps between individual states of potential knowledge typically occur (Fig. \ref{Fig7b}), but this is not shown here for clarity. }
\label{Fig22}
\end{figure}

Let us return to the sequential aspect of time, and the problem how to treat the existence of several subjects. Since the potential knowledge $PK(n)$ is the union of the states of individual potential knowledge $PK_{m_{k}}^{k}$ of subjects $k$, the time indicator $n$ is updated each time any of the individual time indicators $m_{k}$ are updated (Fig. \ref{Fig22}). In such a schema, it is necessary to be able to tell in which temporal order $m_{k}$ and $m_{k'}$ are updated. Otherwise the ordering in the sequence $\{PK(n), PK(n+1), PK(n+2), \ldots\}$ becomes ill-defined, which means that the evolution of the physcial state becomes ill-defined.

In case the updates $m_{k}\rightarrow m_{k}+1$ and $m_{k'}\rightarrow m_{k'}+1$ correspond to updates of two objects with time-like separation, the ordering is unambiguous. If the separation is space-like, no ordering that all subjects are certain to agree on can be assigned. The resolution to this problem is to say that the events occur `at the same time'. This means that the updates of the individual potential knowledge of subjects $k$ and $k'$ that are reflected in the updates $m_{k}\rightarrow m_{k}+1$ and $m_{k'}\rightarrow m_{k'}+1$, correpond to a single update of potential knowledge $PK(n)$, reflected in a single update $n\rightarrow n+1$. An example is shown in Fig. \ref{Fig22}, in the update $PK(3)\rightarrow PK(4)$.

Note that we always consider updates of objects that are placed along the world line of some subject who actually observes this object. That is, we may equally well say that we are trying to order all subjective changes of perception. These subjective changes correspond to objects placed at the same spatial location as the subject itself.

\begin{defi}[\textbf{Temporal updates}]
If a subject $k$ observes the change of an object $O$ in its immediate vicinity, this event belongs to time $n+1$ if and only if it is place inside the light cone of another event that belongs to time $n$, and it is located along the world-line of $k$, or of another subject $k'$.
\label{temporalupdates}
\end{defi}

\begin{figure}[tp]
\begin{center}
\includegraphics[width=80mm,clip=true]{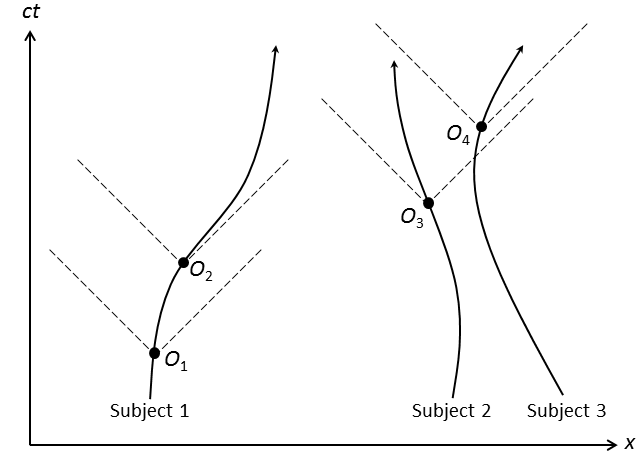}
\end{center}
\caption{Illustration of the rule that governs temporal updates in the presence of several subjects (Definition \ref{temporalupdates}). We define events as perceived appearances or changes of objects $O$ that are placed along wold-lines of subjects. Object $O_{2}$ perceived by subject $1$ defines a temporal update since it is located within the lightcone of object $O_{1}$ perceived by the same subject. Object $O_{3}$, on the other hand, does not correspond to a temporal update. Object $O_{4}$ again defines a temporal update. We may say that the set of events $\{O_{1}\}$ belongs to time $n$, the set $\{O_{2},O_{3}\}$ to time $n+1$, and the set $\{O_{4}\}$ to time $n+2$.}
\label{Fig96}
\end{figure}

This definition is illustrated in Fig. \ref{Fig96}. It means that a subjectively perceived event defines a temporal update if and only if an immediately preceding event of the same kind can influence it causally.

Definition \ref{temporalupdates} avoids the problem that the ordering of two objects or events with space-like separation is ambiguous depending on the state of motion of the observer. In that case we presuppose observers that move with different speeds in relation to the two objects or events. They are not placed where the two events actually happen. Therefore they have to deduce the timing upon which they disagree. We can disregard their deductions since they does not correspond to direct perceptions. (This point is further discussed in section \ref{boundstates}).

To summarize, in the present description, the perceived flow of time is represented by a discrete sequence of states of potential knowledge. The continuous time parameter $t$ loses its global, absolute status, and are replaced by continuous relational attributes $t_{ij}$, connecting pairs of objects $O_{i}$ and $O_{j}$ in any given state of knowledge. In this manner, relational time enters the physical description in the same way as distance, as attributes relating object pairs.

\begin{assu}[\textbf{The concept of time has two components}]
1) A sequential ordering of states of potential knowledge $PK(n),PK(n+1),PK(n+2),\ldots$, and 2) a relational attribute $t_{ij}$ that relates any pair of objects $O_{i}$ and $O_{j}$ in each state $PK(n)$. These objects in $PK(n)$ may belong to the present, $PKN(n)$, or past, $M(PK(n-1))$. If $O_{i}$ belong to the present and $O_{j}$ to the past, then $t_{ij}=-t_{ji}>0$ If $O_{i}$ and $O_{j}$ both belong to the present, then $t_{ij}=0$.
\label{timeconcept}
\end{assu}

Physical law becomes a rule that relates $PK(n)$ and $PK(n+1)$. More precisely, given $PK(n)$ a rule is assumed to exist that constrains the set $\{r_{ij},t_{ij}\}$, as well as other attributes, that are perceivable in any updated state of potential knowledge $PK(n+1)$.

In the vocabulary of this section, the assumption that physical law only depends on independent (identifiable) objects, means that it only depends on objects that can belong both to the present and past part of a given state of potential knowledge. To elaborate on this point, any object $O\subseteq PKN(n)$ that we need to consider must have a chance to persist in the updated state of knowledge: $O\subseteq PKN(n+1)$. This statement only have meaning if there is also a memory of $O$ from time $n$ in $PK(n+1)$, that is $O\subseteq M(PK(n))$. In other words, apart from $O\subseteq PK(n+1)$ with $Pr_{O}=1$, there must be a related object $O'=M(O)\subseteq PK(n+1)$ with $Pr_{O'}=0$. Simply put, any relevant object is possible to trace in some sequence of states $PK(n)$.

To me, there is one situation in which the present two-fold picture of time appears to be the only reasonable one. It is when we listen to music. The appreciation of harmonies, and the emotional response they give rise to in the present, depends crucially on memories of sounds in the immediate past, to the extent that the music would cease to exist without these memories. That is, each present state of the listener contains both the present and the past; each fleeting `now' can be unfolded to an entire space-time. At the basic level of physical desciption, the very perception of a sound relies on memories of the past, since an extended period of relational time $t$ is needed to determine the frequencies that define the sound that we hear at a given moment. 

\section{Interaction between subject and object}
\label{twoways}

The traditional scientific view is that the aware state is a function of the physical state, and that the evolution of the physical state is determined by the physical state itself. Consequently, the outcome of a subjective choice is regarded to be a function of the physical state of the body just before the choice is made. This means that the subjective aspect of the world is a slave under the objective aspect (Fig. \ref{Fig1}). Philosophers have given the name \emph{epiphenomenalism} to the hypothesis that the subjective aspect of the world exists, but that it just mirrors the objective one.

\begin{figure}[tp]
\begin{center}
\includegraphics[width=80mm,clip=true]{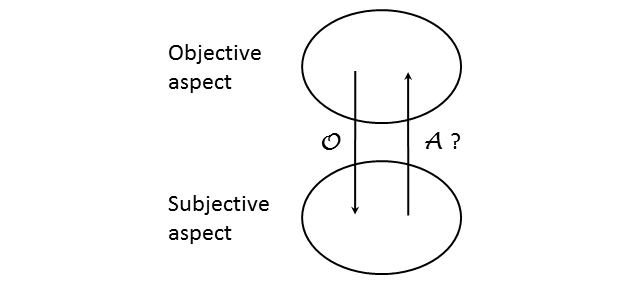}
\end{center}
\caption{If the physical state determines the aware state, but the aware state cannot affect the physical state, the objective aspect of the world becomes more fundamental, contrary to the spirit of intertwined dualism. $\mathcal{O}$ stands for the observation of events that are determined by physical law, and $\mathcal{A}$ stands for an action that is not determined by physical law, but may be constrained by it. Compare Figs. \ref{Fig1} and \ref{Fig8}.}
\label{Fig24}
\end{figure}

I want to conclude this introductory section by noting that from a purely aesthetic perspective, this hypothesis is unsatisfactory if intertwined dualism is accepted as a proper basis for scientific understanding (Assumption \ref{intertwined}). In this picture, the subjective and objective aspects emerge from each other, and one of them cannot be regarded as more fundamental than the other. In contrast, in the epiphenomenalistic picture the objective aspect becomes more fundamental. A more symmetric picture is that the subject can passively `observe' the evolution of objects, but that it can also actively affect them by some `action' $\mathcal{A}$ (Fig. \ref{Fig24}). These matters are further discussed in Sections \ref{individualsubjects} and \ref{probabilities}.

The main point is that there are situations in which the conditions that determine the update $PK(n)\rightarrow PK(n+1)$ cannot be completely described in terms of causation and probability. This fact consititutes a `white hole' in physical law, in which there is playground for the action $\mathcal{A}$. However, such a free action can only be defined in negative terms, as an event that is neither the result of a deterministic law, nor a probabilistic one. Furthermore, the hole has limits, meaning that physical law constrains which free actions are possible.

Most often the lack of determinism of quantum mechanics is discussed in terms of the probabilities encoded in the amplitudes of the wave function. Can these probabilities be described in terms of the deterministic evolution of hidden variables, or are they fundamental? However, there is an even deeper indeterminism, which is less often discussed, for some strange reason.

To define the probabilities, we must first choose which experiment we are going to do, which variable we are going to observe. The formalism tells us nothing about how this choice comes about, not even in a probabilistic sense. Nevertheless the choice makes a clear physical difference when it comes to the future state of the observed system, and the future state of the world at large. This is an example of the `white hole'.

To circumvent the problem we might argue that there is a meta-experiment going on in our brain, in which the outcome is the decision to do this or that experiment, and that there are well-defined probabilities in a meta-wavefunction associated with each possible experiment. However, this idea just moves the white hole to the level of the meta-experiment. We end up in infinite regress.

\chapter{\normalfont{EPISTEMIC FORMALISM}}
\label{formalism}

In this chapter, the ideas and conclusions presented above are given a tighter symbolic form. I aim to present a formal skeleton that can be transfomed into a mathematical formalism of an extended quantum mechanics. The reader is referred back to the previous sections to add more conceptual flesh to the formal bones.

\section{The structure of knowledge}
\label{structureknowledge}

We let $K_{m}^{k}$ be the aware knowledge of subject $k\in\mathbf{\mathbb{N}}$ at individual time instant $m\in\mathbf{\mathbb{N}}$. We let $PK_{m}^{k}$ be the potential knowledge of $k$ at individual time instant $m$, that is, all candidates of knowledge that may become knowledge $K_{m'}^{k}$ at any later time $m'>m$, properly interpreted by $k$ as her own memories of time $m$.

$PK_{m}^{k}$ can be expressed as the union of the potential knowledge $PKN_{m}^{k}$ of potentially perceived objects belonging to the present (time instant $m$), and the knowledge $M(PK_{m-1}^{k})$ of objects that are memories of all past times $m''<m$, potentially perceivable at time $m$:

\begin{equation}
PK_{m}^{k}=PKN_{m}^{k}\cup M(PK_{m-1}^{k}).
\label{presentpast1}
\end{equation}

The state of potential knowledge that corresponds to the physical state is the union of the potential knowledge of all subjects. The sequential time $n$ is updated each time one of the individual times $m$ is updated, corresponding to a subjective change in subject $k$ (Fig. \ref{Fig22}).

\begin{equation}\begin{array}{l}
PK(n)=\bigcup_{k} PK_{m_{k}}^{k}\\
m_{k}\rightarrow m_{k}+1 \Rightarrow n\rightarrow n+1
\end{array}\end{equation}
Individual potential knowledge often overlap, corresponding to different subjects being potentially aware of the same objects (Fig. \ref{Fig7b}). Dividing $PK(n)$ into present and past parts as in Eq. \ref{presentpast1}, we may write

\begin{equation}\begin{array}{lll}
PKN(n) & = & \bigcup_{k} PKN_{m_{k}}^{k}\\
M(PK(n-1)) & = & \bigcup_{k} M(PK_{m_{k}-1}^{k})\\
\end{array},\end{equation}
leading to

\begin{equation}
PK(n)=PKN(n)\cup M(PK(n-1)).
\label{presentpast2}
\end{equation}

Instead of using the letter $M$ to label memories, we may assign a presentness attribute $Pr\in\{0,1\}$ to any object $O$ potentially perceived by any subject, with $Pr=1$ if it belongs to the present, and $Pr=0$ if it belongs to the past. Then we may write

\begin{equation}
PK(n)=\bigcup_{l}O_{l},
\end{equation}
A description of the knowledge of all attributes of $O_{l}$ (including $Pr$) is assumed to be contained in the symbol $O$:

\begin{equation}
O_{l}=
\left[\begin{array}{c}
\{pk_{l}(A_{i})\}_{i}\\
\{pk_{l,l' }(A_{j})\}_{j,l'}\\
\{pk_{l,l',l''}(A_{k})\}_{k,l',l''}\\
\vdots\\
\end{array}\right].
\label{objectcontent}
\end{equation}
Here, $A_{i}$ is the $i$:th internal attribute of object $O_{l}$ (such as presentness or spin). The ordering of the attributes is arbitrary but fixed. $A_{j}$ is the $j$:th relational attribute relating two objects $O_{l}$ and $O_{l'}$ (such as distance or time). $A_{k}$ is the $k$:th relational attribute relating three objects $O_{l}$, $O_{l'}$, and $O_{l''}$ (such as angle). As far as I am aware, no fundamental relational attribute relating four or more objects is needed in today's physics, but the option must be left open. $pk_{l}(A_{i})$ is the potential knowledge of the internal attribute $A_{i}$, and correspondingly for $pk_{l,l'}(A_{j})$ and $pk_{l,l',l'''}(A_{k})$. Defocused and conditional knowledge must be allowed for.

Beside direct experiences of the present or memories of the past, deduced knowledge also have an important role to play (Fig. \ref{Fig2}). Many objects used in scientific descriptions are deduced. For instance, analyzing the chemical composition of a sample, we observe objects in the form of digits appearing on a display of a measuring apparatus, and we remember the experimental setup. But the actual chemical compounds are just deduced objects, arrived at with the help of logic and physical law from memories $O_{1}$ belonging to $M(PK(n-1))$, and observations $O_{2}$ belonging to $PKN(n)$.

\begin{defi}[\textbf{Quasiobject} $\tilde{O}$]
An object that is not directly perceived, but deduced via physical law.
\label{quasiobjectdefi}
\end{defi}

To each quasiobject $\tilde{O}$ corresponds exactly one real object $O\subset PKN$ namely the object of insight about $\tilde{O}$. For a general deduction we may write

\begin{equation}
\begin{array}{c}
O_{1}\subset M(PK(n-1))\cup O_{2}\subset PKN(n)\\
\Downarrow\\
O_{3}\subset PKN(n+1)\leftrightarrow\tilde{O}_{3},
\end{array}
\label{deduction1}
\end{equation}
where $\tilde{O}_{3}$ is the deduced quasiobject. Note that time is updated from $n$ to $n+1$ when the deduction is made; we can distinguish the states before and after we have had an insight.

\begin{figure}[tp]
\begin{center}
\includegraphics[width=80mm,clip=true]{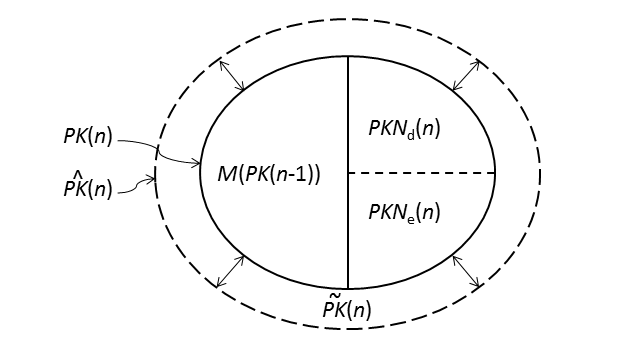}
\end{center}
\caption{The present part of potential knowledge $PKN(n)$ can be divided into the present experiences $PKN_{e}(n)$ and the presently made deductions $PKN_{d}(n)$, that use the memory $M(PK(n-1))$ of the previous state of potential knowledge as premise. Each object of deductional insight in $PKN_{d}(n)$ point to a quasiobject, an object that is deduced to exist at some time, but of which we may have direct experience. $\tilde{PK}(n)$ is the union of all quasiobjects. The extended potential knowledge $\hat{PK}(n)=PK(n)\cup \tilde{PK}(n)$ is a function of $PK(n)$ and it is inappropriate to say that it is larger than $PK(n)$. Compare Figs. \ref{Fig3}, \ref{Fig20}, and \ref{Fig21}.}
\label{Fig24b}
\end{figure}

Just as the ability to distinguish the present from the past is taken to be a basic fact, treated like an assumption, we assume the interpretational ability to distinguish direct experiences from deductions (Fig. \ref{Fig2}):

\begin{equation}
PKN(n)=PKN_{e}(n)\cup PKN_{d}(n),
\end{equation}
where the subscripts $e$ and $d$ stand for experience and deduction, respectively (Fig. \ref{Fig24b}). We may therefore rewrite Eq. \ref{deduction1} as

\begin{equation}
\begin{array}{c}
O_{1}\subset M(PK(n-1))\cup O_{2}\subset PKN_{e}(n)\cup O_{3}\subset PKN_{d}(n)\\
\Downarrow\\
O_{4}\subset PKN_{d}(n+1)\leftrightarrow\tilde{O}_{4}.
\end{array}
\label{deduction2}
\end{equation}
where an old deduction $O_{3}\leftrightarrow\tilde{O}_{3}$ is allowed as input for a new one, $O_{4}\leftrightarrow\tilde{O}_{4}$ , beside memories $O_{1}$ and observations $O_{2}$.

A given identifiable object (Definition \ref{identifiableobjects}) may be a quasiobject at one time $n$, but be directly perceived at another time. A classic example is the sun: it does not cease to exist after sunset, we know that it must still be there. But if there is only one subject on earth, or if all subjects live on the same hemisphere, no one has the potential to see it directly. Some quasiobjects are expected to be quasibjects at all times, however. It is hard to imagine aware beings that can ever perceive electrons or other elementary particles directly.

\begin{state}[\textbf{There may be temporary quasiobjects}]
An identifiable object $O$ may be a quasiobject at time $n$, that is, $O\subset PKN_{d}(n)\leftrightarrow\tilde{O}$, but be directly perceived at another time $n'$, that is, $O\subset PKN_{e}(n')$.
\label{temporalquasi}
\end{state}

The quasiobjects may refer to the past (retrodictions), to the present or to the future (predictions). This mean that the presentness attribute $Pr$ take a third value, say $2$, for quasiobjects, corresponding to a predicted object. The range of $Pr$ is thus $\{0,1,2\}$ for quasiobjects and $\{0,1\}$ for objects.

Let $\tilde{PK}(n)$ be the state of potential `quasiknowledge', the state of all possible deductions from $PK(n-1)$. Since there corresponds one object of potential insight to each potential quasiobject, we may write

\begin{equation}
\tilde{PK}(n)\leftrightarrow PKN_{d}(n).
\label{deductioncorr}
\end{equation}
This relation expresses the fact that since $\tilde{PK}(n)$ exhausts the possible deductions from $PK(n-1)$, it is a function of $PK(n)$. Therefore, it is redundant to include it as part of $PK(n)$. But in some cases it is convenient to include it in an extended state $\hat{PK}(n)$ of potential knowledge (Fig. \ref{Fig24b}). We have

\begin{equation}
\hat{PK}(n)=PKN(n)\cup M(PK(n-1))\cup\tilde{PK}(n).
\label{extendedknowledge}
\end{equation}

\begin{figure}[tp]
\begin{center}
\includegraphics[width=80mm,clip=true]{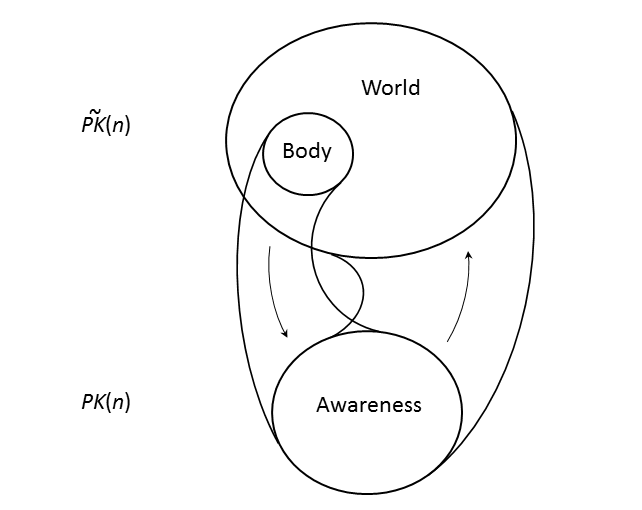}
\end{center}
\caption{The same illustration of the idea behind `intertwined dualism' as in Fig. \ref{Fig1}, where potential knowledge $PK$ is identified with the subjective aspect, and potential `quasiknowledge' $\tilde{PK}$ consisiting of deduced quasiobjects is identified with the objective aspect. Compare also Fig. \ref{Fig8}.}
\label{Fig24c}
\end{figure}

In the picture of extended potential knowledge (Fig. \ref{Fig24b}), we may lift out the `belt' of quasiobjects in $\tilde{PK}$ and place it at safe distance, as an opposing pole that represent `physical objects', as opposed to the percevived objects, representing `subjective experiences'. It must be remembered, however, that the quasiobject represent the `objective world' only to the extent that they are necessary objects in the mathematical representation of the physical law that governs the evolution of the subjective experiences. In this way we get back the original picture of intertwined duality (Fig. \ref{Fig1}) with renamed aspects (Fig. \ref{Fig24c}).

Potential knowledge about the past in the form of memories can never increase as time goes. However, extended potential knowledge including quasiobjects may increase. We may have experiences today that were not precisely dictated by physical law (no determinism), which make it possible to deduce something new about the past, something that was not part of potential knowledge back then. We have to require, though, that the new knowledge about the past does not contradict the previous knowledge about the past, and does not lead to predictions about the present that contradicts what we observe now.

For example, astronomers may construct a telescope with wich they observe a distant interstellar gas cloud. Due to the finite speed of light they observe its properties as they were a long time ago. There were no aware subjects in the gas cloud at that time who formed memories of its properties, and before the telescope was built, the radiation from the cloud that reached the earth was too faint to leave traces in the bodies of terrestial subjects that have the potential to become aware experiences. Therefore, we may observe all kinds of properties of the gas cloud that were not fixed before the observation, without running the risk of contradiction.

This discussion about extended potential knowledge has brought us back to the question of epistemic consistency (Assumption \ref{epconsistency}). This is natural, since consistency is all about deduction. We divide the consistency requirement into two parts.

First, the present is not allowed to contradict the past, given physical law. (In contrast, one cannot say that two objects that both are part of the present contradict each other.)

\begin{assu}[\textbf{Epistemic consistency 1}]
Assume that $O$ is an identifiable object (Definition \ref{identifiableobjects}) that can be both directly perceived and deduced (Definition \ref{temporalquasi}). Physical law must be such that for any state of potential knowledge $PK(n)$, $\tilde{PK}(n)$ does not contradict $M(PK(n-1))$ in the sense that the properly deduced quasiobject $\tilde{O}\in\tilde{PK}(n)$ is never described by deduced attribute values that are knowably different from those of the same remembered object $O\in M(PK(n-1))$.
\label{epconsistency1}
\end{assu}

One may want to add the condition that memories experienced at time $n'$ that refer back to time $n$ must be consistent with $PKN(n)$. However, this is included in the notion of `proper memories' that is implicit in the state $M(PK(n'-1))$.

Second, the present must also conform with the past in the following sense.

\begin{assu}[\textbf{Epistemic consistency 2}]
Let $\tilde{PK}_{new}(n;n')$ be the new potential knowledge about time $n$ that physical law make it possible to deduce at a later time $n'>n$. That the deduced knowledge is new means that $\tilde{PK}_{new}(n;n')\cap \hat{PK}(n)=\varnothing$. Let $\hat{PK}_{new}(n)=\hat{PK}(n)\cup \tilde{PK}_{new}(n;n')$. Then, for any time $n''>n$, we have $\hat{PK}_{new}(n'')=\hat{PK}(n'')$ for any (extended) potential knowledge $\hat{PK}_{new}(n'')$ that may follow from $\hat{PK}_{new}(n)$ via physical law.
\label{epconsistency2}
\end{assu}

\begin{figure}[tp]
\begin{center}
\includegraphics[width=80mm,clip=true]{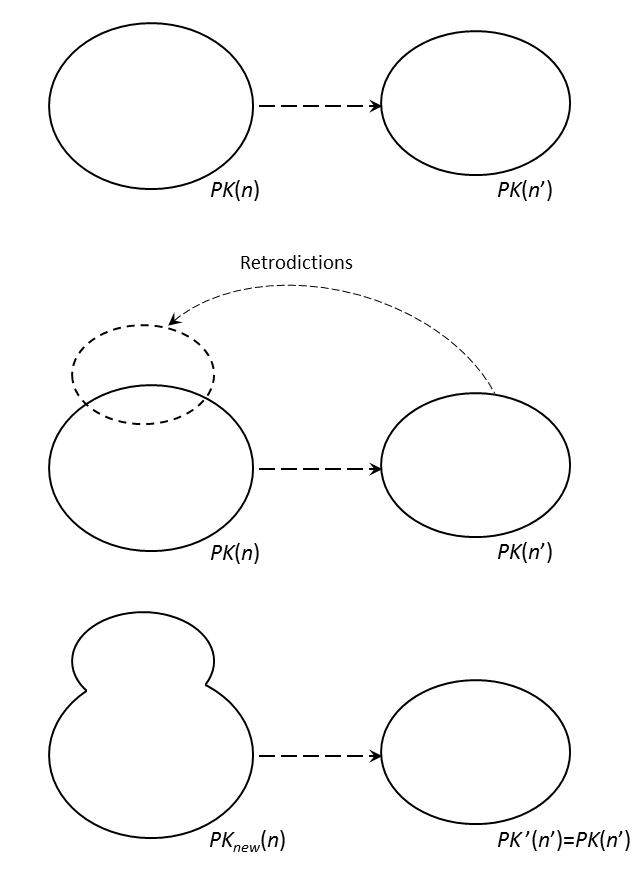}
\end{center}
\caption{Illustration of Assumption \ref{epconsistency2}. Anything that can be deduced about time $n$ from any possible future state of potential knowledge $PK(n')$ must be insignificant enough, so that it does not cause a future state $PK'(n')$ that is different in any way from the state $PK(n')$ we actually see. Such a forbidden difference might be an outright contradiction, but it might also amount to more (or less) focused knowledge, or more or less conditional knowledge.}
\label{Fig25c}
\end{figure}

The idea is illustrated in Fig. \ref{Fig25c}. This more subtle kind of consistency is the one we discussed in relation to the interstellar gas cloud. The same line of reasoning was used in Section \ref{knowledge}, in connection with the discussion of the double slit experiment, and led to the first attempt to formulate the meaning of epistemic consistency (Assumption \ref{epconsistency}). Suppose that we see an interference pattern. If it were possible after that to gain knowledge that make it possible to retrodict which slit the particle actually passed, then the future state that would follow if this knowledge were there at the time of passing would contradict the state that we actually perceived and remember now. We would get no interference pattern. This is an expression of the well-known quantum mechanical rule that interference patterns only appear when it is impossible in principle to gain path information at a later time, be it via regained memories, later deduction from memories, or deduction from knowledge acquired later.

If we recall Eq. \ref{deductioncorr}, and refer to Fig. \ref{Fig24b}, then we see that Assumption \ref{epconsistency2} can be seen as a requirement of self-consistency for $PK(n)$. The retrodictions that are possible to make must point to quasiobjects $\tilde{PK}(n)$ that correspond to real objects of a past physical state that can evolve, together with objects corresponding to potential memories of this past time, to the same physical state we started out with, with its retrodictions (Fig. \ref{Fig25c}). The circle must close.

The actual deductions made by any group of subjects is of course a much smaller state of quasiknowledge than $\tilde{PK}(n)$, due to lack of intelligence, lack of knowledge of physical law, or simply because the analytical mind is resting. Therefore the above strict requirement of consistency is never put to a complete test, and one may argue that it might be loosened. But because of the lack of determinism (Statement \ref{nodeterminism}), Nature can never be sure about the future intellectual capabilities and activities of its creatures, and cannot risk any contradiction by underestimating their intelligence.




The notation $O_{l}\subset PK$ has already been used for an object being part of a state of potential knowledge. As expressed in Eq. \ref{objectcontent}, the state of knowledge of the attributes is implicit in the symbol $O_{l}$. More generally, we may write

\begin{equation}
PK'\subseteq PK
\end{equation}
if $PK$ contains at least the same objects as $PK'$, and the potential knowledge of the attributes of the shared objects is at least as great in $PK$ as in $PK'$. This means that at least the same attributes are known, and that the knowledge of the value of each attribute is at least as exact.

\section{The physical state}
\label{state}

According to the tentative Definition \ref{firststatedef}, the physical state representation is a specification in symbolic form of the state of potential knowledge $PK$. Let us call such a schema of symbols $\bar{S}$. The fact that $\bar{S}$ represents $PK$ may be expressed as

\begin{equation}
\bar{S}\hookrightarrow PK.
\end{equation}

We require that a given representation $\bar{S}$ corresponds to at most one state of potential knowledge $PK$, but we do not exclude the possibility that several representations $\bar{S}$ may correspond to the same state $PK$:

\begin{equation}\begin{array}{l}
\bar{S}\hookrightarrow PK\\
\bar{S}'\hookrightarrow PK.
\end{array}
\label{elbowroom}
\end{equation}

In other words, we allow the symbolic representation of $PK$ to have redundant degrees of `symbolic freedom' that do not reflect degrees of `epistemic freedom' in $PK$.

The hypothesis that all kinds of knowledge can be described by sets of objects, with attributes and attribute values (Assumptions \ref{stateexists} and \ref{constructknow}), suggests that it is indeed possible to represent $PK$ as a symbolic schema $\bar{S}$. But it is not self-evident that there is a sensible way to represent it uniquely. This is why we allow the symbolic elbow-room expressed in Eq. [\ref{elbowroom}].

\begin{defi}[\textbf{Exact knowledge representations} $\bar{Z}$]
The symbolic schema $\bar{Z}$ is exact if and only if it represents a state of knowledge $\kappa$ for which the potential knowledge of all independent objects and attributes is complete.
\label{exactrep}
\end{defi}

Symbolically,

\begin{equation}
\bar{Z}\hookrightarrow\kappa
\end{equation}

In order to obtain a clear definition of a physical state, and of physical law that acts on such states, we would like to define physical states $S$ and exact states $Z$ so that they do not refer to specific symbolic representations. Later on, the distinction between the physical state $S$ and a particular representation

\begin{equation}
\bar{S}\hookrightarrow S
\end{equation}
of this state will be essential.

The incompleteness of any actual state of knowledge $PK$ is expressed by the use of the logical constant \emph{or}, as discussed in section \ref{incomplete}. This or that object may exist, and that attribute may have this or that value, possibly conditional on the value of other attributes or the existence of other objects.

Let $\kappa_{i}$ be a state of complete knowledge. Then there is a set $\{\kappa_{i}\}_{PK}$ such that $PK$ could turn out to be $\kappa_{i}$ if it were possible to remove the incompleteness of knowledge inherent in $PK$, as if you put on perfect glasses. We may write

\begin{equation}
PK=\kappa_{1} \;\mathrm{or}\; \kappa_{2} \;\mathrm{or} \;\kappa_{3} \;\mathrm{or}\; \ldots,\ \ \ \kappa_{i}\in\{\kappa_{i}\}_{PK}.
\label{kappaset}
\end{equation}

Even conditional knowledge is contained in this simple description, in the form of exclusion of states $\kappa_{i}$ that do not fulfil the relevant conditions. (For simplicity we use an index $i$ to distinguish different states of complete knowledge even if we have not decided whether they form a countable set or not.)

We may simply identify $\kappa_{i}$ with a point $Z_{i}$ in a state space $\mathcal{S}$.

\begin{defi}[\textbf{The state space} $\mathcal{S}$]
To each $\kappa_{i}$ we associate exactly one `point' $Z_{i}\in\mathcal{S}$. The state space $\mathcal{S}$ is the set of all possible exact states $Z_{i}$.
\label{statespacedef}
\end{defi}

The space of possibilities is limited by a given a set of independent attributes that describe any object, a given range of possible values of each attribute, and a lower bound on the number of objects, given by the necessary existence of an observer who observes at least one object. The properties of state space are further discussed in section \ref{statespaces}.

The  physical state $S$ can then be defined as the union of all exact states $Z_{i}$ that correspond to a state of complete knowledge $\kappa_{i}$ that is not excluded by the actual potential knowledge $PK$.

\begin{defi}[\textbf{The physical state} $S$]
We have $S=\bigcup_{i}Z_{i}$, where $Z_{i}\leftrightarrow\kappa_{i}\in\{\kappa_{i}\}_{PK}$ for each $i$. The set $\{\kappa_{i}\}_{PK}$ is defined by Eq. [\ref{kappaset}].
\label{statedef}
\end{defi}

Note that even if the physical state $S$ is never given by an exact state $Z$ (Statement \ref{nodeterminism}), these states are nevertheless well-defined, since the concept of complete knowledge is needed to define our actual incomplete knowledge.

Since $Z$ has no specific symbolic form, $S$ has no specific symbolic form either. In this way we get rid of any redundancy of representation. The relations between $Z$ and $\kappa$, and between $S$ and $PK$, become one-to-one correspondences: 

\begin{equation}\begin{array}{lll}
Z & \leftrightarrow & \kappa\\
S & \leftrightarrow & PK.
\end{array}\end{equation}

\begin{figure}[tp]
\begin{center}
\includegraphics[width=80mm,clip=true]{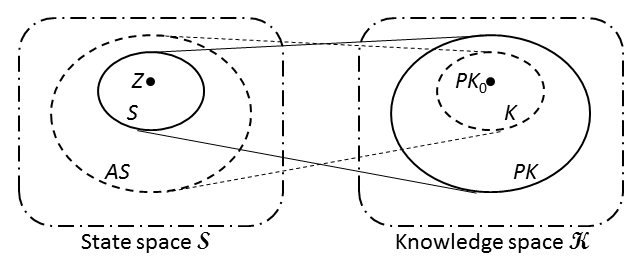}
\end{center}
\caption{The state space $\mathcal{S}$ has exact states $Z$ as elements. Larger states of knowledge $PK$ are mapped to smaller physical states $S$ and vice versa. In particular, the aware knowledge $K$ is contained in the potential knowledge $PK$, whereas the corresponding `aware physical state' $AS$ contains the physical state $S$. Since knowledge is always incomplete (Statement \ref{incompleteknowledge}), $S$ never shrinks to a point $Z$. Compare Figs. \ref{Fig3} and \ref{Fig154}.}
\label{Fig25}
\end{figure}

In relation to Definition \ref{statespacedef} we noted that the extent of state space $\mathcal{S}$ was limited by the number of independent attributes, the number of possible values of these attributes, and a minimum number of objects. These limitations define an outer boundary of $\mathcal{S}$. The minimal state of potential knowledge $PK_{0}$ is mapped to this boundary (dash-dotted closed curve in the left panel of Fig. \ref{Fig25}).

\begin{equation}
\mathcal{S}\leftrightarrow PK_{0}
\end{equation}

Note that there is only one such state $PK_{0}$, corresponding to the `naked' awareness `I am' or `there is something' (Fig. \ref{Fig5b}).

At the other side of the coin, the extent of knowledge space $\mathcal{K}$ is limited by the outer boundary of the currently unknowable (Figs. \ref{Fig3} and left panel of Fig. \ref{Fig25}). Attributes or values of attributes that cannot be perceived or determined even in principle are not part of any state $\kappa$ of complete knowledge. We may say that each such state $\kappa_{i}$ fills the entire knowledge space $\mathcal{K}$ consisting of the potential knowledge and the currently unknowable (but knowable in principle). Consequently any exact state $Z_{i}$ is mapped to the entire knowledge space

\begin{equation}
\forall i:\ \mathcal{Z}_{i}\leftrightarrow \mathcal{K}
\end{equation}

More generally,

\begin{equation}
PK'\subset PK \Leftrightarrow S'\supset S.
\label{kommunicerande}
\end{equation}
For the aware state of knowledge $K$ we have $K\subseteq PK$. Let $AS\leftrightarrow K$ be the corresponding `aware physical state'. Then $AS\supseteq S$, according to Eq. \ref{kommunicerande}. This relation is illustrated in Fig. \ref{Fig25}.

The sizes of the knowledge state and the physical state behave like information content and entropy, respectively. Larger information content means smaller entropy, and vice versa. The entropy concept is discussed further in section \ref{entropy}.

We may say that the knowledge states $PK$ correspond to the subjective component of the world, whereas the physical states correspond to the objective component. Compare Figs. \ref{Fig1} and \ref{Fig8}.

Any actual state of potential knowledge is larger than $PK_{0}$ (Fig. \ref{Fig25}). There are always some states $\kappa_{i}$ of complete knowledge that contradict our knowledge. If all we perceive is a grayish haze and a humming noise, we can exclude all $\kappa_{i}$ corresponding to a clear blue sky, all $\kappa_{i}$ corresponding to a barking dog being in the neighborhood, and so on.

This means that for any physical state $S$ there are exact states $Z_{i}$ such that $Z_{i}\notin \ S$. The maximal set of such states form a non-empty complement $S^{C}$ to $S$ such that

\begin{equation}\begin{array}{rcl}
S\cup S^{c} & = & \mathcal{S}\\
S\cap S^{c} & = & \varnothing
\end{array}\end{equation}

\begin{figure}[tp]
\begin{center}
\includegraphics[width=80mm,clip=true]{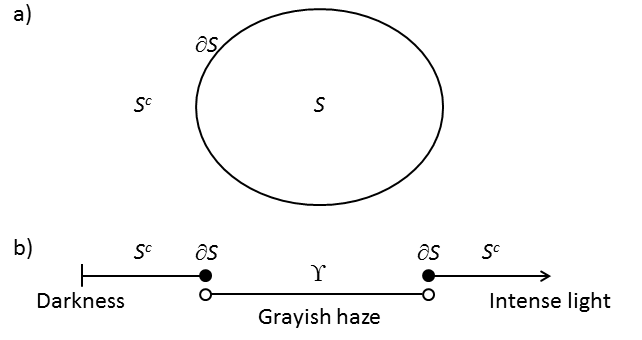}
\end{center}
\caption{(a) The physical state $S$, its boundary $\partial S$, and its complement $S^{c}$, consisting of all exact states $Z$ that contradict the potential knowledge $PK$ corresponding to $S$. By convention, $S^{c}$ is a closed set with $\partial S\subseteq S^{c}$. (b) It is easier to determine $S^{c}$ by exclusion of impossible alternatives, than to find all $Z$ in $S$ that are consistent with $PK$. If a grayish haze is perceived, all $Z$ with too high or too low light intensity can be excluded. Light intensity is an attribute $A$, and the potential knowledge $pk(A)$ of its value corresponds to the middle interval $\Upsilon$ of the line.}
\label{Fig27}
\end{figure}

If the elements in $\mathcal{S}$ are ordered in a predefined manner, e.g. according to the values of their attributes, a boundary $\partial S$ between $S$ and $S^{c}$ can always be defined. (One of the sets must be defined to be open, the other closed.) If light intensities are ordered along a line, both extreme light and almost no light contradict the perception of the grayish haze (Fig. \ref{Fig27}).

Each individual set $S$, $S^{c}$ or $\partial S$ can be used as a description of the physical state. From an epistemic point of view, it is preferable to define it with the help of $S^{c}$. Then we don't have to refer to individual exact states $Z$, which we can never pinpoint since potential knowledge is always incomplete (Statement \ref{incompleteknowledge}). In the grayish haze, we can immediatley exclude \emph{all} exact states with too low or too high light intensity from $S$, but we cannot include all exact states with light intensity in the allowed middle interval. Some of them may have values of other attributes that are forbidden, or may contain forbidden attributes or objects. For this reason, I choose to define $S^{s}$ to be the closed set, so that $\partial S\subseteq S^{c}$. From a mathematical point of view, this has no significance.

\begin{figure}[tp]
\begin{center}
\includegraphics[width=80mm,clip=true]{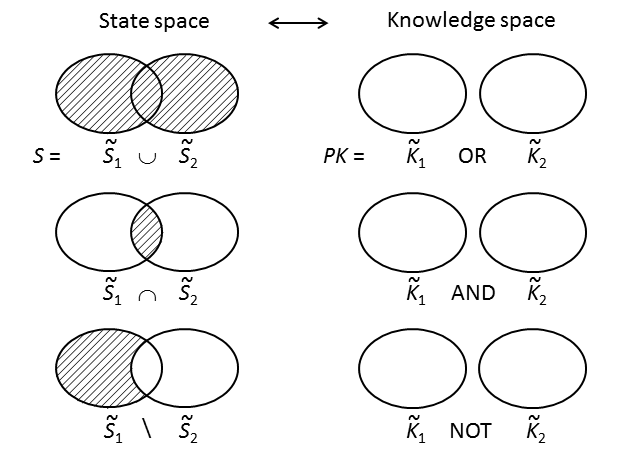}
\end{center}
\caption{The relation between composite states of knowledge and the physical state. The knowledge element $\tilde{K}_{i}$ corresponds to the physical state $\tilde{S}_{i}$. Given knowledge $\tilde{K}_{1}$, the addition `OR $\tilde{K}_{2}$' reduces the amount of knowledge and makes the physical state grow (becoming more defocused). The additions `AND $\tilde{K}_{2}$' and `NOT $\tilde{K}_{2}$' increases the amount of knowledge and make the physical state shrink.}
\label{Fig28}
\end{figure}

Figure \ref{Fig28} shows how composite states of knowledge relate to composite physical states. The logical constants OR, AND and NOT are replaced by the set theoretical binary operations $\cup$, $\cap$, and $\setminus$. The two states of knowledge $\tilde{K}_{1}$ and $\tilde{K}_{2}$ may refer to the individual potential knowledge $PK^{1}$ and $PK^{2}$ of two subjects, to two objects $O_{1}$ and $O_{2}$, to the knowledge $pk(A_{1})$ and $pk(A_{2})$ of two attributes of the same object, or to two elements of knowledge about the value of a single attribute, or any other pair of knowledge elements.

The knowledge elements $\tilde{K}_{1}$ and $\tilde{K}_{2}$ are shown as non-overlapping. This is always true for a pair of objects or a pair of attributes, which are distinct by definition. It may not be true for a pair of individual states of knowledge (Fig. \ref{Fig7b}), or a pair of knowledge elements about the value of some attribute. However, this does not affect the symbolic translation to the physical state.

As long as $\tilde{K}_{1}$ and $\tilde{K}_{2}$ do not contradict each other, $\tilde{S}_{1}$ and $\tilde{S}_{2}$ will overlap. If they would contradict each other, $\tilde{K}_{1}$ OR $\tilde{K}_{2}$ and $\tilde{K}_{1}$ AND $\tilde{K}_{2}$ would be improper states of knowledge. The state of knowledge $\tilde{K}_{1}$ NOT $\tilde{K}_{2}$ would be improper if $\tilde{K}_{1}$ and $\tilde{K}_{2}$ overlapped.

Let us complete the dictionary translating elements of $PK$ to symbols that can be used to specify $S$:
\begin{enumerate}
  \item Object $O$ $\leftrightarrow$ $S_{O}$.
  \item Potential knowledge $pk(A)$ of attribute $A$ $\leftrightarrow$ $\Upsilon$.
  \item OR $\leftrightarrow$ $\cup$
  \item AND $\leftrightarrow$ $\cap$
  \item NOT $\leftrightarrow$ $\setminus$
\end{enumerate} 

Here $S_{O}$ is the state of an object, and $\Upsilon$ is a set of values of $A$ not excluded by the potential knowledge $pk(A)$ (Fig. \ref{Fig27}). Let us elaborate on the object state $S_{O}$, since we will use this concept frequently in what follows. The concept is meaningful when we consider a physical state $S$ that corresponds to potential knowledge $PK$ of at least two objects. If we choose one object $O$ among these, $S_{O}$ is the physical state that would result if the knowledge about all the other objects was erased.

\begin{defi}[\textbf{The state} $S_{O}$ \textbf{of object} $O$]
$S_{O}$ is the union of all exact states $Z$ in state space that do not contradict the fact that $O$ exists, or the potential knowledge of its internal attributes.
\label{objectstate}
\end{defi}

Let the corresponding state of potential knowledge be $PK_{O}$. We have defined $\Omega_{O}$ as the complement to $O$ (Definition \ref{complement}). Generally, $S\subseteq S_{O}\cap S_{\Omega_{O}}$, referring to the middle panel in Fig. \ref{Fig28}.

\begin{defi}[\textbf{The state of the environment} $S_{\Omega_{O}}$ \textbf{to object} $O$]
$S_{\Omega_{O}}$ is the union of all exact states $Z$ in state space that do not contradict the existence of any of the perceived objects in the complement $\Omega_{O}$ to $O$, or the potential knowledge of the attributes internal to this complement.
\label{environmentstate}
\end{defi}

If there would be no knowledge $PK_{R}$ about the relational attributes that relate $O$ to its environment $\Omega_{O}$, and if there would be no conditional knowledge $PK_{C}$ that relates $O$ and $\Omega_{O}$, then we would have $PK=PK_{O}\cup PK_{\Omega_{O}}$. This is the same as to say $S=S_{O}\cap S_{\Omega_{O}}$. However, whenever $S_{O}$ is defined, there is also some knowledge about the relation between $O$ and its environment. We must therefore always write $PK=PK_{O}\cup PK_{\Omega_{O}}\cup PK_{R}$. This means that $PK$ is larger than $PK_{O}\cup PK_{\Omega_{O}}$, or that $S\subset S_{O}\cap S_{\Omega_{O}}$.

\begin{state}[\textbf{Any object is related to its environment}]
For any perceived object $O$ we have $S\subset S_{O}\cap S_{\Omega_{O}}$.
\label{relobjectstate}
\end{state}

There may also be conditional knowledge relating object $O$ and its complement $\Omega_{O}$. In that case we should write  

\begin{equation}
PK=PK_{O}\cup PK_{\Omega_{O}}\cup PK_{R}\cup PK_{C}.
\label{generalpk}
\end{equation}

In the scientific modelling of the behavior of an object we often assume that it is isolated. In our terminology this approximation corresponds to the assumption that $S=S_{O}\cap S_{\Omega_{O}}$. We know, of course, that this is never quite true, as expressed in Statement \ref{relobjectstate}. It is impossible in principle to make $S$ fill the entire intersection between $S_{O}$ and $S_{\Omega_{O}}$, even if we may come close.

\begin{defi}[\textbf{An isolated object}]
An object $O$ is isolated if and only if $S=S_{O}\cap S_{\Omega_{O}}$.
\label{isoobjectstate}
\end{defi}

\begin{figure}[tp]
\begin{center}
\includegraphics[width=80mm,clip=true]{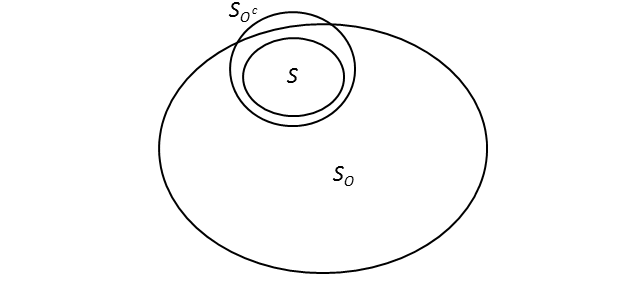}
\end{center}
\caption{$S_{O}$ is the state corresponding to knowledge of object $O$. $S_{\Omega_{O}}$ is the state corresponding to all other objects in a state $S$ of composite knowledge. $S_{O}$ is shown large since it corresponds to `small' knowledge of a single object (c.f. Fig. \ref{Fig25}). The state $S$ is a proper subset of the intersection between $S_{O}$ and the environment $S_{\Omega_{O}}$ whenever $O$ is related to its environment.}
\label{Fig28b}
\end{figure}

The set-theoretic relations expressed in Statement \ref{relobjectstate} and Defintion \ref{isoobjectstate} are illustrated in Fig. \ref{Fig28b}. 

If we do not involve quasiobjects in the description of $PK$, then the number of objects, and the number of attributes of each object, is always precisely known at each sequential time $n$, given by the distinctions that are possible to make in the potential perception. If we involve quasiobjects such as elementary particles in the description, the number $N$ of objects in the specification of the state is typically not precisely known, since many values of $N$ may be consistent with the potential knowledge of the system under consideration. In this case we may re-express \ref{generalpk} as

\begin{equation}
PK=\left(\bigcup_{j=1}^{J}PK_{Oj}\right)\cup PK_{R}\cup PK_{C}.
\label{generalpk2}
\end{equation}

In terms of the physical state we get

\begin{equation}
S=\left(\bigcap_{j=1}^{J}S_{Oj}\right)\cap \Sigma_{R}\cap \Sigma_{C}.
\label{generalstate}
\end{equation}

Let us discuss the role of conditional knowledge in this kind of expression. It is possible to divide this knowledge into $K$ distinct conditions $C_{k}$, so that we may write

\begin{equation}
\Sigma_{C}=\bigcap_{k=1}^{K} \Sigma_{Ck},
\label{generalcond}
\end{equation}
where $\Sigma_{Ck}$ is the set of exact states $Z$ for which condition $C_{k}$ is fulfilled. Often, it is more straightforward to remove the exact states for which the conditions are \emph{not} fulfilled, than to check for consistency with all the conditions, as will be exemplified below. From this perspective,

\begin{equation}
S=\left(\bigcap_{j=1}^{J}S_{Oj}\right)\cap \Sigma_{R}\setminus \Sigma^{c}_{C1}\setminus \Sigma^{c}_{C2}\setminus\ldots\setminus \Sigma^{c}_{CK},
\end{equation}
where $\Sigma^{c}_{Ck}$ is the set of exact states $Z$ for which a given condition $C_{k}$ is not fulfilled.

It may be instructive to be a litte more concrete. To this end we note first that $S_{Oj}$ is just an array of the internal attributes $A_{ij}$ of object $O_{j}$, which can be specified as a list of the sets $\Upsilon_{ij}$ of allowed values:

\begin{equation}
S_{Oj}=
\left[\begin{array}{c}
\Upsilon_{1j}\\
\Upsilon_{2j}\\
\vdots\\
\Upsilon_{nj}
\end{array}\right],
\label{statearray}
\end{equation}
where we may have $n=n(j)$.

Conditional knowledge is a set of conditions that relate the values of attributes - either different attributes of the same object, or attributes of different objects. In the simplest case,

\begin{equation}
C_{k}:\ \ \upsilon_{ij}\in\Delta_{ij}\Rightarrow \upsilon_{i'j'}\in\Delta_{i'j'},
\end{equation}
where $\Delta_{ij}\subset\Upsilon_{ij}$, and $\Delta_{i'j'}\subset\Upsilon_{i'j'}$. In words, if the value $\upsilon_{ij}$ of the attribute $A_{i}$ of object $O_{j}$ belongs to some subset $\Delta_{ij}$ of the allowed values $\Upsilon_{ij}$, then the value $\upsilon_{i'j'}$ of the attribute $A_{i'}$ of object $O_{j'}$ belongs to a subset $\Delta_{i'j'}$ of the allowed values of this attribute. More complicated conditions can of course be formulated, for example $\upsilon_{ij}\in\Delta_{ij}\; \mathrm{OR}\; \upsilon_{i'j'}\in\Delta_{i'j'} \Rightarrow \upsilon_{i''j''}\in\Delta_{i''j''}$.

To illustrate, consider a state consisting of two objects with two attributes each, and one condition $C$:

\begin{equation}\begin{array}{l}
S=\left[
\begin{array}{c}
\Upsilon_{11}\\
\Upsilon_{21}\end{array}\right]\cap
\left[
\begin{array}{c}
\Upsilon_{12}\\
\Upsilon_{22}\end{array}\right]\cap \Sigma_{R}\setminus \Sigma_{C}^{c}\\\\
C: \ \ \upsilon_{21}\in\Delta_{21} \Rightarrow \upsilon_{12}\in\Delta_{12}
\end{array}
\end{equation}
Writing $\Upsilon_{21}=\Delta_{21}\cup\Delta_{21}^{c}$, and $\Upsilon_{12}=\Delta_{12}\cup\Delta_{12}^{c}$, we get

\begin{equation}\begin{array}{lll}
S & = & \left[
\begin{array}{c}
\Upsilon_{11}\\
\Delta_{21}\cup\Delta_{21}^{c}\end{array}\right]\cap
\left[
\begin{array}{c}
\Delta_{12}\cup\Delta_{12}^{c}\\
\Upsilon_{22}\end{array}\right]\cap \Sigma_{R}\setminus \Sigma_{C}^{c}\\\\
& = & \left(\left[
\begin{array}{c}
\Upsilon_{11}\\
\Delta_{21}\end{array}\right]\cap
\left[\begin{array}{c}
\Delta_{12}\\
\Upsilon_{22}\end{array}\right]\right)\cup\\\\
&   & \left(\left[
\begin{array}{c}
\Upsilon_{11}\\
\Delta_{21}^{c}\end{array}\right]\cap
\left[\begin{array}{c}
\Delta_{12}\\
\Upsilon_{22}\end{array}\right]\right)\cup\\\\
&   & \left(\left[
\begin{array}{c}
\Upsilon_{11}\\
\Delta_{21}\end{array}\right]\cap
\left[\begin{array}{c}
\Delta_{12}^{c}\\
\Upsilon_{22}\end{array}\right]\right)\cup\\\\
&   & \left(\left[
\begin{array}{c}
\Upsilon_{11}\\
\Delta_{21}^{c}\end{array}\right]\cap
\left[\begin{array}{c}
\Delta_{12}^{c}\\
\Upsilon_{22}\end{array}\right]\right)\cap \Sigma_{R}\setminus \Sigma_{C}^{c}\\\\
& = & \left(\left[
\begin{array}{c}
\Upsilon_{11}\\
\Delta_{21}\end{array}\right]\cap
\left[\begin{array}{c}
\Delta_{12}\\
\Upsilon_{22}\end{array}\right]\right)\cup\\\\
&   & \left(\left[
\begin{array}{c}
\Upsilon_{11}\\
\Delta_{21}^{c}\end{array}\right]\cap
\left[\begin{array}{c}
\Delta_{12}\\
\Upsilon_{22}\end{array}\right]\right)\cup\\\\
&   & \left(\left[
\begin{array}{c}
\Upsilon_{11}\\
\Delta_{21}^{c}\end{array}\right]\cap
\left[\begin{array}{c}
\Delta_{12}^{c}\\
\Upsilon_{22}\end{array}\right]\right)\cap \Sigma_{R}.
\end{array}\end{equation}
The third set within curved brackets on the right hand side of the second equality equals $\Sigma_{C}^{c}$, and is simply removed in the last equality. The presence of a condition means that it is no longer possible to separate all objects and write $S=S_{O1}\cap S_{O2}\cap \ldots S_{OJ}\cap \Sigma_{R}$, even if the sets $\Upsilon_{ij}$ of allowed values of the attributes are modified.

\begin{state}[\textbf{Conditional knowledge leads to non-separable states}]
Suppose that $\Sigma_{C}^{c}\neq\varnothing$. Then $S\neq \left(\bigcap_{j}S_{Oj}'\right)\cap \Sigma_{R}$ whenever $S_{Oj}'$ is a state that is defined exclusively in terms of the internal attributes of object $O_{j}$. 
\label{nonseparable}
\end{state}

If quasiobjects are allowed in the specification of the state, their number $N$ is not be fixed by $PK$. Then conditions must be allowed that relate the number of quasiobjects with values of attributes, or the number of quasiobjects of different types with each other.

\begin{figure}[tp]
\begin{center}
\includegraphics[width=80mm,clip=true]{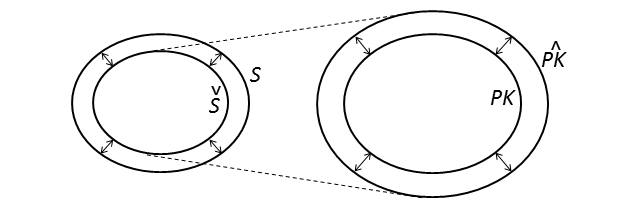}
\end{center}
\caption{The extended potential knowledge $\hat{PK}$ includes deduced quasiobjects (Fig. \ref{Fig24b}). The corresponding physical state $\check{S}$ may be called the reduced state, and is a function of $S$.}
\label{Fig28c}
\end{figure}

We introduced the extended state of potential knowledge $\hat{PK}$ that includes quasiobjects in Fig. \ref{Fig24b}. The corresponding physical state $\check{S}$ may be called the reduced state (Fig. \ref{Fig28c}). We my write

\begin{equation}
\hat{PK}\leftrightarrow\check{S}.
\end{equation}

Regarding the reduced state for a particular object $O$, or quasiobject $\tilde{O}$, we may correspondingly write

\begin{equation}\begin{array}{rcl}
\hat{PK}_{O}\leftrightarrow\check{S}_{O}\\
\hat{PK}_{\tilde{O}}\leftrightarrow\check{S}_{\tilde{O}}
\end{array}\end{equation}

Since $S$ and $\check{S}$ are basically equivalent, it is not always necessary to be strict and repeatedly use the word `reduced' or the diacritic sign above the symbol $S$ when we speak about a state that involves quasiobjects.

\section{Knowability of the physical state}
\label{knowstate}

We have argued that the exact state $Z$ of the world is not knowable. Accordingly, we describe the physical state as a set $S$ that contains all those $Z$ that are not excluded by the actual knowledge. But is the state $S$ itself exactly knowable?

\begin{figure}[tp]
\begin{center}
\includegraphics[width=80mm,clip=true]{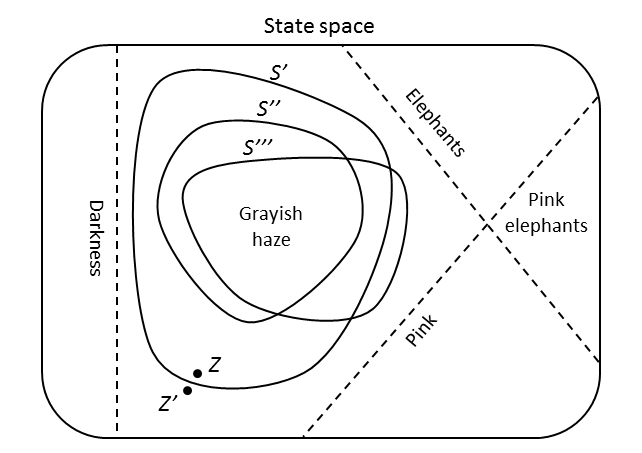}
\end{center}
\caption{The visual perception of nothing but grayish haze exludes those regions of state space in which everything is pink, those in which there are elephants, and those in which everything is dark. Accordingly, the state $S$ must be located in the complement to these regions, defined by the dashed lines. But we cannot say which set $S'$, $S''$ or $S'''$ is the actual physical state $S$. To determine $S$ exactly, we need to discriminate between two arbitrarily close exact states $Z$ and $Z'$ at either side of $\partial S$. This is not possible when knowledge is incomplete. Compare Fig. \ref{Fig27}.}
\label{Fig28d}
\end{figure}

In any state of knowledge, there is something we can exclude as being in conflict with our perception. To perceive something always means that there is something else that we do not perceive. Referring to Fig. \ref{Fig27}, if we see just a grayish haze, we know that we do not perceive a pink haze, or an elephant. This means that for any physical state $S$ (or state of any object $S_{O}$), there are always some parts of state space that we know do not belong to $S$ (or $S_{O}$). The situation is illustrated in Fig. \ref{Fig28d}, where the regions bounded by the staright dashed lines are assumed to be knowably outside $S$. However, this leaves considerable freedom to choose a state $S$ within these limits, as illustrated by the three sets $S'$, $S''$ and $S'''$.

To determine $S$ exactly means to locate its boundary $\partial S$ exactly. To do this, we have to be able to compare pairs of exact states $(Z,Z')$ on either side of $\partial S$ that are arbitrarily close in state space, and judge, based on our knowledge, that one of them belong to $S$ while the other one does not. In a sense, this corresponds to a precision of knowledge equivalent of the knowledge of the exact state $Z$ itself. Since the incompleteness of knowledge (Statement \ref{incompleteknowledge}) implies that we don't have such a precision, the boundary $\partial S$ is not exactly knowable. Admittedly, this is heuristic reasoning. We nevertheless formulate the following statement.

\begin{state}[\textbf{The physical state} $S$ \textbf{is not exactly knowable}]
Assume that some object attributes can take values that are dense. Then the state of potential knowledge $PK$ is not sufficient to determine the boundary $\partial S$ exactly.
\label{unknownstate}
\end{state}

We could try to introduce a `metastate' $MS$ that expresses the actual knowledge of the physical state $S$ in the sense that it is a collection of all states $S', S'', S''', \ldots$ that we cannot exclude as candidates:

\begin{equation}
MS=\{S',S'',S''',\ldots\}.
\end{equation}

But $MS$ is not precisely knowable, since we cannot locate $\partial MS$ exactly for the same reason as we cannot locate $\partial S$ exactly. In other words, we cannot decide exactly which states $S',S'',S''',\ldots$ we should include in $MS$. Introducing a `meta-metastate' $MMS$ does not help either. We end up in infinite regress if we proceed further along this road.

To pinpoint the knowledge of the physical state, it does not help, either, to try to define a `maximal state' $S_{\max}$ that exludes only those regions of state space that we know contradict our perceptions (e.g. those containing pink elephants). The boundary $\partial S_{\max}$ defined by these regions is not exactly knowable for the same reason as $\partial S$ is not exactly knowable.

There is a minimum size of $S$ - it cannot be allowed to shrink towards an exact state $Z$. This allows us to talk about the `minimal state' $S_{\min}$, just as we talked about the maximal state $S_{\max}$. But again, it is impossible to determine the boundary $\partial S_{\min}$. Fuzziness is omnipresent. All we can say is that there exist two non-trivial sets $S_{\min}$ and $S_{\max}$ such that the physical state $S$ fulfils

\begin{equation}
S_{\min}\subseteq S \subseteq S_{\max}.
\end{equation}
We say that the two sets are non-trivial since $S_{\min}$ contains several exact states $Z$, and $S_{\max}$ is a proper subset of state space.

\begin{figure}[tp]
\begin{center}
\includegraphics[width=80mm,clip=true]{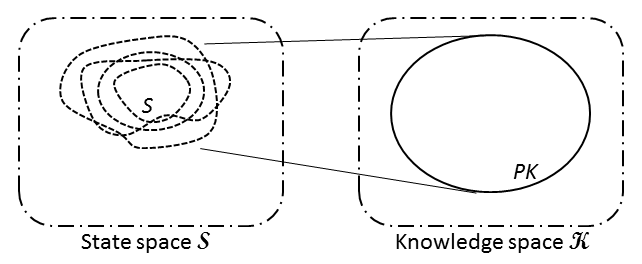}
\end{center}
\caption{The state of potential knowledge $PK$ does not determine the physical state $S$ completely. This means that there are always exact states $Z$ such that it is not possible to decide whether $Z$ is consistent with $PK$ or not. In other words, the boundary $\partial S$ is fuzzy. In contrast, it has no meaning to say that it is impossible to decide whether an element of knowledge is part of $PK$ or not. Either we potentially know someting, or we do not. The boundary $\partial PK$ is precisely defined. Compare Fig. \ref{Fig25}.}
\label{Fig154}
\end{figure}

The `known physical state' thus appears to be an inherently fuzzy concept, escaping our hands like a slippery soap. However, the physical state itself is well-defined, since an exact state $Z$ is either consistent with our potential knowledge $PK$, or it is not. There are no alternatives in between, no fuzziness. The state $S$ is just the set of all $Z$ that are consistent with $PK$. Therefore we have to distinguish between knowability and existence in the case of physical states. They are epistemically well-defined, even if they are unknowable in their details (Fig. \ref{Fig154}).

The distinction between the physical state and the `known physical state' plays little role in what follows. We will come back to it, however, in section \ref{evolutionparameter}, where we introduce the continuous evolution parameter $\sigma$, and also in section \ref{entropy}, where we discuss entropy.

\section{Physical law}
\label{law}

Any physical law allows us to define an evolution operator $u$ that acts on any $S$ and gives another physical state $uS$ such that the next physical state is a subset of $uS$: 

\begin{equation}
S(n+1)\subseteq uS(n).
\label{defu}
\end{equation}

In other words, given the present physical state, any physical law limits the possibilities of the future.

\begin{defi}[\textbf{The evolution} $u_{1}$]
The stepwise evolution operator $u_{1}$ is defined by the condition that $u_{1}S(n)$ is the smallest possible set $C\subseteq\mathcal{S}$ such physical law dictates that $S(n+1)\subseteq C$.
\label{evolutionu1}
\end{defi}

The above definition is very general - it should hold for any conceivable evolution operator $u_{1}$. We will also make use of the more specific properties of $u_{1}$.

\begin{assu}[\textbf{Evolution} $u_{1}$ \textbf{is unique and invertible}]
Two states $S$ and $S'$ are distinct if and only if $u_{1}S$ and $u_{1}S'$ are distinct: $S\cap S'\neq\varnothing\Leftrightarrow u_{1}S\cap u_{1}S'\neq\varnothing$.
\label{uniqueu1}
\end{assu}

\begin{figure}[tp]
\begin{center}
\includegraphics[width=80mm,clip=true]{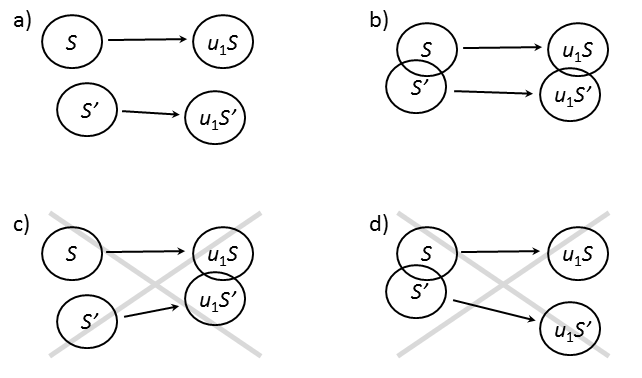}
\end{center}
\caption{Physical law as expressed by $u_{1}$ is analogous to an invertible function. Evolution of two states $S$ and $S'$ according to cases a) and b) are allowed by Assumption \ref{uniqueu1}, whereas evolution according to c) and d) are forbidden.}
\label{Fig31b}
\end{figure}

That $S$ and $S'$ are distinct means that they do not overlap. If that is the case, then $u_{1}S$ and $u_{1}S'$ does not overlap either (Fig. \ref{Fig31b}). On the other hand, if two states $S$ and $S'$ do overlap, so do $u_{1}S$ and $u_{1}S'$. The former statement can be regarded as an assumption of `subjective' invertibility: two subjectively distinct states cannot evolve into states which are subjectively indistinguishable. The latter statement is an assumption that the evolution $u_{1}$ is a function in a subjective sense: two states that are subjectively the same cannot evolve into states that are subjectively different.

\begin{state}[\textbf{The evolution} $u_{1}$ \textbf{is a mapping from the power set} $\mathcal{P}(\mathcal{S})$ \textbf{of state space to itself}]
We may write $u_{1}:\mathcal{P}(\mathcal{S})\rightarrow \mathcal{P}(\mathcal{S})$, where the domain $\mathcal{D}_{u}\subset\mathcal{P}(\mathcal{S})$ of the mapping is the set $\{\Sigma\}$ of those sets $\Sigma\subset\mathcal{S}$ that may correspond to a physical state $S(n)$ or a physical state $S_{O}$ of an object $O$.
\label{evodef}
\end{state}

We know that $\mathcal{D}_{u}$ is a proper subset of $\mathcal{P}(\mathcal{S})$ since exact states $Z$ are elements of both $\mathcal{S}$ and $\mathcal{P}$, and these can never correspond to physical states $S(n)$ or $S_{O}$, due to the incompleteness of knowledge (Statement \ref{incompleteknowledge}). The range $\mathcal{R}_{u}$ of the mapping must also be a proper subset of $\mathcal{P}(\mathcal{S})$, since we cannot have $u_{1}S(n)=Z$ for some exact state $Z$ at the same time as we fulfil the defining property $S(n+1)\subseteq u_{1}S(n)$ of $u_{1}$. We do not attempt, here, to determine the relation between the domain $\mathcal{D}_{u}$ and the range $\mathcal{R}_{u}$ of $u_{1}$. We cannot be sure that the set $u_{1}\Sigma$ corresponds to a possible physical state $S$ or $S_{O}$ just because $\Sigma$ does. All we know is that there is a subset of $u_{1}\Sigma$ that corresponds to such a physical state whenever $\Sigma$ does. Therefore we cannot be sure that $\mathcal{R}_{u}\subseteq\mathcal{D}_{u}$. On the other hand, we cannot be sure that $\mathcal{R}_{u}\supseteq\mathcal{D}_{u}$ at this stage either.

\begin{figure}[tp]
\begin{center}
\includegraphics[width=80mm,clip=true]{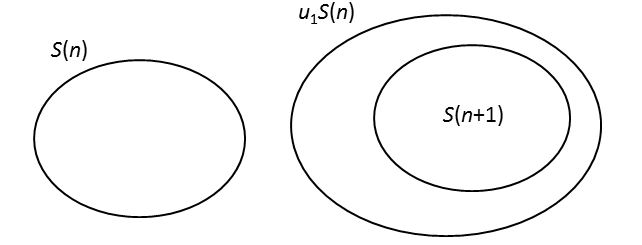}
\end{center}
\caption{Since $S(n+1)\subseteq u_{1}S(n)$ must be distinct from $S(n)$, by definition of successive states, we can define the evolution operator $u_{1}$ such that $S(n)$ and $u_{1}S(n)$ are always distinct.}
\label{Fig29}
\end{figure}

The sequence of time instants is defined by the fact that we can tell the corresponding states apart. They are manifestly distinct; it is not possible for one exact state $Z$ to be consistent with both $S(n)$ and $S(n+1)$:

\begin{equation}
S(n+1)\cap S(n)=\varnothing.
\end{equation}
It follows that $S(n)\cap u_{1}S(n)=\varnothing$ (Fig \ref{Fig29}).

\begin{defi}[\textbf{Determinism}]
Physical law is deterministic if and only if $S(n+1)=u_{1}S(n)$ for all states $S(n)$.
\label{determinism2}
\end{defi}

\begin{defi}[\textbf{State determinism}]
Let $u_{2}\equiv u_{1}u_{1}$, $u_{3}\equiv u_{1}u_{1}u_{1}$, and so on. The evolution of the state $S(n)$ is deterministic if and only if $S(n+m)=u_{m}S(n)$ for all $m\geq 1$.
\label{determinism3}
\end{defi}

\begin{state}[\textbf{Exact states evolve deterministically}]
The evolution of a state $S(n)$ is deterministic if and only if $S(n)$ is exact, that is, $S(n)=Z$ for some exact state $Z$.
\label{exactdeterminism}
\end{state}

To motivate Statement \ref{exactdeterminism}, recall that $u_{1}$ is assumed to be noiseless (Assumption \ref{noiseless}); it depends on independent objects, and independent attributes of these (Definitions \ref{indobjects} and \ref{indattributes}), but on nothing else. If the evolution is not deterministic for an exact state $Z$, there must be a stochastic term in $u_{1}$, contrary to Assumption \ref{noiseless}. Conversely, if the evolution is deterministic for an inexact state $S$, the lack of knowledge of the attributes in this inexact state would not affect their evolution, and hence the unknown attributes cannot vary independently. 

This means that in the absence of noise, complete knowledge and deterministic evolution goes hand in hand. Therefore we could equally well \emph{define} an exact state $Z$ as a state that evolves deterministically, according to Definition \ref{determinism3}.

\begin{defi}[\textbf{An exact state} $Z$]
A state $S$ is exact if and only if it evolves deterministically. 
\label{exactstate}
\end{defi}

This definition is more restrictive than to say that $S$ is exact if and only if knowledge is complete, since it rules out exact states in a world with noisy physical law.

Since knowledge is incomplete, the true state $S(n)$ is never exact. Therefore there is a time $n+m$ such that $S(n+m)\subset u_{m}S(n)$.

\begin{defi}[\textbf{State reduction}]
A state reduction occurs at time $n+1$ if and only if $S(n+1)\subset u_{1}S(n)$.
\label{statereduction}
\end{defi}

\begin{state}[\textbf{State reductions do occur}]
There are times $n$ such that a state reduction occurs at time $n+1$.
\label{reductionsoccur}
\end{state}

After these considerations, we can state what we mean by physical law.

\begin{defi}[\textbf{Physical law}]
The evolution $u_{1}$ together with a rule that tells which states $S(n+1)\subseteq u_{1}S(n)$ are allowed, given $S(n)$, define what can be said about physical law.
\label{physicallaw}
\end{defi}

Note that we do not, in this definition, refer to probabilities for different outcomes $S(n+1)\subset u_{1}S(n)$ after a state reduction. Probabilities will be treated as deduced quantitites rather than fundamental components of physical law, as discussed in section \ref{probabilities}.

Given any inexact state $S$ we can imagine a set of alternative states that result if additional knowledge is gained. For instance, if we see a raptor in the sky but cannot decide which kind, an exhaustive set of alternatives consists of all species of raptors that live in our country. Any such set of alternatives divides $S$ into distinct subsets $S_{j}$ such that

\begin{equation}\begin{array}{l}
S=\bigcup_{j}S_{j},\\
S_{j}\cap S_{j'}=\varnothing \;\mathrm{for}\;\mathrm{all}\; j\neq j'.
\label{division}
\end{array}\end{equation}

If the list of alternatives is not exhaustive to begin with, it can trivially be completed by a last alternative `not any of the above'. For instance, the bird may be `a golden eagle, a white-tailed eagle, or some other raptor'. Formally, an incomplete set $\{S_{j}\}$ is completed by adding $S\setminus \bigcup_{j}S_{j}$ as the last alternative.

\begin{figure}[tp]
\begin{center}
\includegraphics[width=80mm,clip=true]{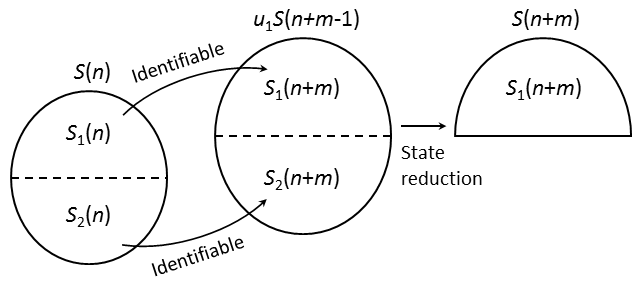}
\end{center}
\caption{A process where the first of two alternatives $S_{1}$ and $S_{2}$ conceivable at time $n$ turns out to be true at time $n+m$. The distinction separating $S_{1}$ and $S_{2}$ must correspond to objects that are identifiable in the time interval $[n,n+m]$.}
\label{Fig30}
\end{figure}

A partition of the state into alternatives may represent knowledge that can be gained at a later time, given the present state $S(n)$, or it may be an imagined list of alternatives, where it is forever impossible to decide which is true. In the first case, let $S(n)=\bigcup_{j}S_{j}(n)$, and assume that one alternative turn out to be true at some later time $n+m$. This statement is only meaningful if the objects corresponding to the alternatives are identifiable, possible to trace from time $n$ to time $n+m$ (Definition \ref{identifiableobjects}). This is no restriction since we have assumed that all objects relevant to physics are identifiable (Assumption \ref{noiseless}). The process is illustrated in Fig. \ref{Fig30}.

\begin{figure}[tp]
\begin{center}
\includegraphics[width=80mm,clip=true]{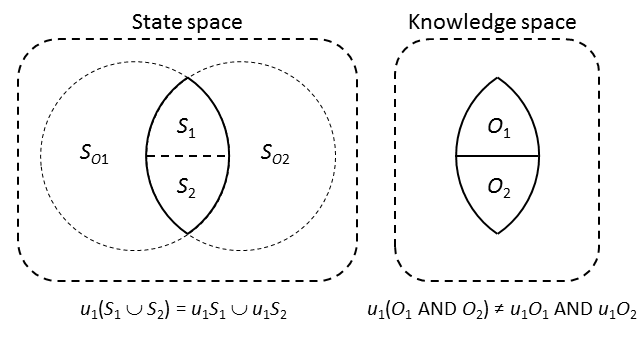}
\end{center}
\caption{The evolution of alternatives $S_{1}$ and $S_{2}$ in state space is independent. The evolution of objects $O_{1}$ and $O_{2}$ in knowledge space is not independent in general, since objects are interacting. $S_{O}$ is basically the state that corresponds to knowledge $PK_{O}$ of $O$ only. Ignoring relational and conditional knowledge, $S=S_{O_{1}}\cap S_{O_{2}}$. Realization of one alternative $S_{j}$ means that knowledge of $O_{1}$ or $O_{2}$ increases.}
\label{Fig31}
\end{figure}

For any such partition into alternatives, we have

\begin{equation}
u_{1}S=u_{1}\left(\bigcup_{j}S_{j}\right)\supseteq\bigcup_{j}u_{1}S_{j}
\label{linear1}
\end{equation}
whenever the evolution $u_{1}S_{j}$ is defined for each $S_{j}$, that is, $S_{j}\in \mathcal{D}_{u}$ according to Definition \ref{evodef}. This relation must hold since the sets $S_{j}$ represent \emph{alternatives}. If it would not hold, the alternatives $S_{j}$ would behave as interacting \emph{objects}. This would represent a confusion of the state space with the knowledge space (Fig. \ref{Fig31}). The interaction of objects can be expressed in state space as

\begin{equation}
u_{1}\left(\bigcap_{l}S_{O_{l}}
\right)\not\supseteq\bigcap_{l}u_{1}S_{O_{l}}.
\end{equation}

Equation \ref{linear1} can also be seen as an expression of epistemic invariance (Assumption \ref{epistemicinvariance}). Physical law depends neither on the content, nor the amount of potential knowledge. In state space, these conditions can be expressed as

\begin{assu}[\textbf{Epistemic invariance}]
$u_{1}$ is independent of $S$, meaning that it can be represented in a single closed form $\bar{u}_{1}$ that applies to all state representations $\bar{S}$. Also, $S'\subseteq S\Rightarrow u_{1}S'\subseteq u_{1}S$ for any two states $S$ and $S'$.
\label{epistemicinvariance2}
\end{assu}

As emphasized in section \ref{epinvariance}, this is a purely theoretical statement of the structure of physical law. In practice, no pair of states $S(n)$ and $S(m)$ can be subsets of each other. We express this below as the `irreducibility of physical state' (Statement \ref{irreduciblestate}).

It is important to note that any set $\{S_{j}\}$ that correspond to alternatives that can be realized constitute a discrete set. This is true regardless the structure of the underlying exact states $Z$. The concept of distinction is again essential; the alternatives must be possible to distinguish subjectively from each other. This fact in itself implies that the set contains a countable number of elements. This conceptual reason for discreteness is mirrored by a physical reason. Consider, for example, an experiment that measures the distance between two objects. Even if distance in itself may be a continuous attribute, the result always depends on an apparatus with finite precision. That apparatus may be the eye, with its discrete array of rods and cones.

It is not possible to let the division [\ref{division}] into alternatives be a division all the way down to the exact states $Z$, and still write down Eq. (\ref{linear1}). Since we know that we can never know the exact states (Statement \ref{incompleteknowledge}), physical law should not refer to $Z$; we should not be able to write $u_{1}Z$. In fact, explicit epistemic minimalism (Assumption \ref{explicitepmin}) requires that we fail if we try to express physical law as if it were acting on the exact states. We conclude the following.

\begin{state}[\textbf{Physical law is irreducible}]
Let $S=\cup_{j} Z_{j}$. It is not possible to define an exact evolution operator $u_{Z}$ such that $u_{1}S=\cup_{j}u_{Z}Z_{j}$ holds true for all states $S$.
\label{irreduciblelaw}
\end{state}

This statement can also be related to Statement \ref{sizecontent}. The fact that two states of potential knowledge are never subsets of each other is equivalent to the fact that two physical states $S(n)$ and $S(n+m)$ are never subsets of each other. The state $S(n)$ is what it is, it cannot be `reduced' by physical action to be better focused on some $Z\in S(n)$.

\begin{state}[\textbf{Physical state is irreducible}]
Given a physical state $S(n)$, we never have $S(m)\subset S(n)$ at some earlier or later time $m$.
\label{irreduciblestate}
\end{state}

This statement can be strengthened. Not only are two subsequent states never subsets of each other, they cannot even overlap.

\begin{state}[\textbf{No partial recurrence}]
Given a physical state $S(n)$, we never have $S(m)\cap S(n)\neq\varnothing$ at some earlier or later time $m$.
\label{norecurrence}
\end{state}

The motivation is a bit different than for Statement \ref{irreduciblestate}, and less fundamental. That two states overlap means that they cannot be subjectvely distinguished. To say that the state at some time $m>n$ cannot be distinguished from the state at time $n$ means that sequential time breaks down. The very statement $m>n$ loses its meaning, since the set of memories associated with $S(m)$ cannot be distinguished from those associated with $S(n)$. No temporal comparisons can be made. This fact is discussed in a cosmological context in Section \ref{entropy}, in relation to Fig. \ref{Fig129}.

Even if two states $S(n)$ and $S(m)$ can never overlap in physical practice, we can still imagine overlapping states in the formulation of theoretical properties of these states. It may concern the action on these states of the evolution operator $u_{1}$, like the invertibility expressed in Assumption \ref{uniqueu1}, or the unknowability of the details of the boundary $\partial S$, as discussed in Section \ref{knowstate}.  

The irreducibility of physical law means that the state $S$ does not evolve as en ensemble of states in the phase space of classical mechanics, or as an ensemble of states in the Hilbert space of quantum mechanics. The evolution is not pointwise. The interior of $S$ has no structure, there is no measure defined on it; it is just a set of exact states not excluded by potential knowledge. Therefore, the only way to avoid pointwise evolution is to let it depend in a non-local way on the boundary $\partial S$.

\begin{figure}[tp]
\begin{center}
\includegraphics[width=80mm,clip=true]{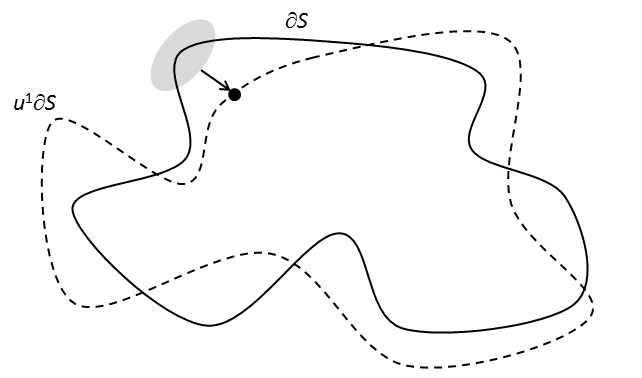}
\end{center}
\caption{The state boundary $\partial S$ as a membrane. The evolution of the `center of mass' of the state is subtracted to emphasize the vibrations of $\partial S$. The exact states $Z\in\partial S$ do not evolve like independent points in an ensemble, just like the evolution of the points $x$ on a membrane do not evolve independently. In that case, the evolution of $x$ depends on an entire neighbourhood of $x$, defining the tension at $x$.}
\label{Fig55}
\end{figure}

This reminds us of a membrane, where the evolution of the position of one point on the membrane does not only depend on its present position, but also on the present position of neighbouring points, defining tension by spatial derivatives (Fig. \ref{Fig55}).   

\begin{state}[\textbf{The state boundary} $\partial S$ \textbf{is a membrane}]
We may write $u_{1}S =u_{1}\partial S = f(\partial S)$, but there is no operator $g_{1}$ acting on exact states $Z_{\partial}\in\partial S$ such that $\partial S=\{Z_{\partial}\}$ and $u_{1}\partial S=\{g_{1}Z_{\partial}\}$.
\label{membrane}
\end{state}

In practice, we only need to consider the evolution of actual physical states $S$, or actual states of objects $S_{O}$. The evolution of exact states $Z$ may, at the most, have a theoretical significance as a basis for expressions of physical law in symbolic or mathematical form. But the irreducibility of physical law means that such expressions are not possible. Therefore, we may restrict the domain $\mathcal{D}_{u}$ of $u_{1}$ to states $S$ that may correspond to actual potential knowledge $PK$, or states of objects $S_{O}$ that may correspond to actual potential knowledge of an object $O$.

\begin{state}[\textbf{The domain of the evolution operator is the set of observable states}]
The domain $\mathcal{D}_{u}$ of the evolution operator $u_{1}$ is the set of all states $S$ that may correspond to an actual state of potential knowledge $PK$ together with the set of all object states $S_{O}$ that may correpond to a state of potential knowledge $PK_{O}$ about an object $O$.
\label{evolutiondomain}
\end{state}

Note that we have to distinguish between states $S$ and states of objects $S_{O}$. Total states $S$ necessarily involve a subject that has knowledge, and an object $O$ that corresponds the body of this subject. A state $S_{O}$ may correspond to a tiny part of knowledge, say of a grain of sand $O$, without the explicit involvement of a perceiving subject. In reality, the evolution of an object $O$ depends on its surroundings, but we nevertheless let $u_{1}$ be \emph{defined} for the states of hypothetical isolated object $O$.

Since we have introduced the notion of quasiobjects $\tilde{O}$, such as electrons or atoms, we have to be able to speak about the evolution of the state $\check{S}_{O}$ of such deduced objects, given that the state corresponds to extended potential knowledge $\hat{PK}_{O}$ that is attainable in principle (Fig. \ref{Fig28c}). Since the use of deduction via physical law introduces no new knowledge, we have $\check{S}_{\{\tilde{O}\}}\leftrightarrow S_{O}$, where the object $O$ is deduced to consist of a set $\{\tilde{O}\}$ of quasiobjects.

\begin{defi}[\textbf{The evolution} $\check{u}_{1}$ \textbf{of quasiobjects}]
Given that $S'=u_{1}S$ and that $\check{S}\leftrightarrow S$ and $\check{S}'\leftrightarrow S'$, we define $\check{u}_{1}$ by the relation $\check{S}'=\check{u}_{1}\check{S}$. A corresponding statement holds if we apply $\check{u}_{1}$ to a state $\check{S}_{\tilde{O}}$ of a quasiobject $\tilde{O}$.
\label{qevolution}
\end{defi}

It is assumed to be possible to divide any object into minimal objects (Assumption \ref{finitedepth}). We will discuss below why at least some of these minimal objects have to be quasiobjects (Statement \ref{alwaysquasi}). Epistemic invariance (Assumptions \ref{epistemicinvariance} and \ref{epistemicinvariance2}) implies that increased potential knowledge in the form of object division does not change the evolution of the system (Fig. \ref{Fig12}). As discussed in section \ref{epinvariance} this corresponds to reductionism. It follows that the evolution of any state $S$ or $S_{O}$ can be expressed in terms of the evolution of the corresponding states $\check{S}_{M}$ or $\check{S}_{OM}$ expressed in terms of minimal objects $O_{M}$ only.

\begin{state}[\textbf{The evolution} $u_{1}$ \textbf{expresses reductionism}]
Suppose that $S'=u_{1}S$, and that $S\leftrightarrow\check{S}_{M}$. Then $S'\leftrightarrow \check{u}_{1}\check{S}_{M}$. A corresponding statement holds for states of objects $S_{O}\leftrightarrow\check{S}_{OM}$.
\label{evred}
\end{state}

What form does the state $\check{S}_{M}$ take? Let

\begin{equation}
\check{S}_{j}=\left(\bigcap_{j}\check{S}_{M_{j}}\right)\cap \check{\Sigma}_{R}\cap \check{\Sigma}_{C}.
\end{equation}
This is a state of a set of minimal objects $M_{l}$ with states $\check{S}_{Oj}$ related by a web of relational attributes corresponding to the set $\check{\Sigma}_{R}\subset\mathcal{S}$, and by a set of conditions corresponding to the set $\check{\Sigma}_{C}\subset\mathcal{S}$ (compare Eq. [\ref{generalstate}]). It is chosen such that $\check{S}_{j}$ is consistent with $S$. In other words, the minimal objects present in $\check{S}_{j}$, with their states and relations, can account for $S$. Since the minimal objects may be quasiobjects, several such arrangements, containing a varying number of minimal objects, may be consistent with or account for $S$. We have to write

\begin{equation}
\check{S}_{M}=\bigcup_{j}\check{S}_{j}.
\end{equation}
This is a partition of the state $S\leftrightarrow\check{S}$ into alternatives just like in Eq. [\ref{division}], evolving according to Eq. [\ref{linear1}]. It should be noted, however, that these alternatives are just imagined, in the sense that it can never be decided by means of an observation which is true.

\begin{figure}[tp]
\begin{center}
\includegraphics[width=80mm,clip=true]{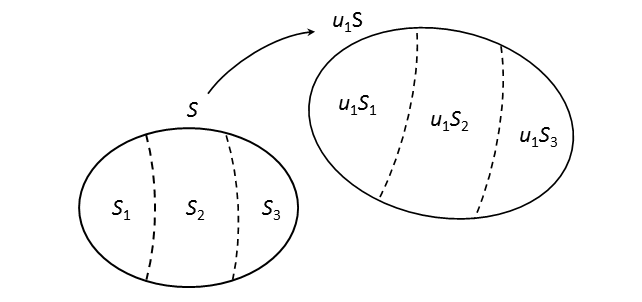}
\end{center}
\caption{Illustration of Statement \ref{linearev} and Eq [\ref{linear}]. The evolution of the whole is the union of the evolution of the parts. This holds for all partitions $\{S_{j}\}$ for which the evolution $u_{1}$ can be applied tp each $S_{j}$. Since $u_{1}$ cannot be applied to exact states $Z$, the irreducibility of physical law is not violated (Statement \ref{irreduciblelaw}).}
\label{Fig31c}
\end{figure}

Let us clarify the evolution [\ref{linear1}] further (Fig. \ref{Fig31c}).

\begin{state}[\textbf{Linear evolution} $u_{1}$]
Assume that $S=\bigcup_{j}S_{j}$, and that $S_{j}\cap S_{j'}=\varnothing$ for all $j\neq j'$. Whenever $u_{1}S_{j}$ is defined for each $j$, the following equation holds. The same equation holds for quasiobjects, if we put check-marks above all symbols.
\label{linearev}
\end{state}

\begin{equation}\begin{array}{l}
u_{1}S=\bigcup_{j}u_{1}S_{j}\\
u_{1}S_{j}\cap u_{1}S_{j'}=\varnothing \;\mathrm{for}\;\mathrm{all}\; j\neq j'.
\label{linear}
\end{array}\end{equation}

To motivate this statement, we start by referring back to Eq. [\ref{division}]. We have already motivated Eq. \ref{linear1}. The invertibility of evolution (Assumption \ref{uniqueu1}) further implies that $u_{1}S_{j}\cap u_{1}S_{j'}=\varnothing$. Thus we may write $u_{1}S=\left(\bigcup_{j}u_{1}S_{j}\right)\cup \Sigma_{rest}$, where all sets on the right hand side are disjoint, and $\Sigma_{rest}$ is a hypothetical `rest set'. Suppose that $\Sigma_{rest}$ corresponds to the evolution of an observable state. Then the state $\Sigma_{rest}^{-}\equiv u_{1}^{-1}S_{R}$ is defined according to the invertibility of $u_{1}$. We must then have $\Sigma_{rest}^{-}\cap S_{j}=\varnothing$ for each $j$ according to the uniqueness of $u_{1}$. This is the same as to say $\Sigma_{rest}^{-}\cap S=\varnothing$. But then invertibility demands that $u_{1}\Sigma_{rest}^{-}\cap u_{1}S=\varnothing$, that is, $\Sigma_{rest}\cap u_{1}S=\varnothing$. We have a contradiction if $\Sigma_{R}$ is not the empty set. Thus, if $\Sigma_{R}$ is non-empty, then it cannot be the evolution of a state for which $u_{1}$ can be defined. Since $u_{1}$ can be defined for all observable states of objects, $\Sigma_{R}$ must be an unobservable state, which can be disregarded since it lacks meaning from the epistemic point of view.

Since the reduced evolution $\check{u}_{1}$ is defined for states $\check{S}_{O}$ of minimal objects that are observable in principle, we may use Statement \ref{linearev} to write

\begin{equation}
\check{u}_{1}\check{S}_{M}=\bigcup_{j}\check{u}_{1}\check{S}_{j}.
\end{equation}

Let us try to classify levels of possiblity for realization of the alternatives in a given set (\ref{division}). Note first that, by definition of potential knowledge, it can never be known which alternative is true in the present state $S(n)$, the very state we use to define the alternatives. If it could, by direct perception or later deduction, the present state would not be $S(n)$ but one of the alternatives $S_{j}\subset S(n)$. This leaves us three levels of knowability of the alternatives, as described in Table \ref{levels}.

\begin{table}
	\centering
		\begin{tabular}{|l|l|}
		\hline
		  &\\
		1 & The alternatives can never be realized, they are not observable states. \\
		  &\\
		2 & There is a time $\hat{n}>n$ such that it is possible that one alternative $S_{j}$\\
		  & is realized at some time $n'\geq \hat{n}$, i.e. $S(n')=S_{j}(n')$, but it is not \\
		  & dictated by physical law that this will happen. We let $\hat{n}$ be the\\
		  & smallest possible such time.\\
		  &\\
		3 & There is also a time $\check{n}>n$ such that physical law dictates that one\\
		  & alternative $S_{j}$ will be realized at some time $n'\leq \check{n}$, i.e. $S(n')=S_{j}(n')$.\\
		  & We let $\check{n}$ be the smallest possible such `deadline for decision'.\\
		  &\\
		\hline
		\end{tabular}
	\caption{Three knowability levels of alternatives.}
	\label{levels}
\end{table}

The present state $S(n)$ and the nature of the set $\{S_{j}\}$ determine which level applies. It has to assumed that the objects involved in the alternatives are identifiable during the relevant time span (Fig \ref{Fig30}). At level 2, each $S_{j}(n)$ must have a chance to be identifiable at least in the time interval $[n,\hat{n}]$. At level 3, each $S_{j}(n)$ \emph{has to be} identifiable at least in $[n,\check{n}]$. If these conditions are not met, the level 2 candidate is degraded to level 1, and the level 3 candidate is degraded to level 2.

As an example of alternatives at level 1, the question whether the electron in your liver with the highest speed at time $n$ has spin pointing towards the head or the feet at some later time $n'>n$, will most probably be impossible ever to answer.

The question whether a raptor in the sky is a golden eagle, a white-tailed eagle or some other species corresponds to alternatives at level 2. The question can clearly be answered later, but not necessarily so. It requires that a nearby friend has binoculars at hand, or that you track it until you come close enough to be able to resolve its features with your naked eye. It is necessary to track it continuously to be sure that it is the same bird you finally speciate as you observed to begin with. This is the condition of identifiability.

Alternatives at level 3 typically arise in prepared states, designed to answer specific questions. An experimental setup where the experiment has started is one example. Nature is challenged to show its cards. In some cases the necessity to get an answer follows trivially from the question. At time $n$ you may ask whether it will rain at time $\check{n}$. The alternative $S_{1}(n)$ is the union of the weather conditions at time $n$ that will lead to rain at time $\check{n}$, and $S_{2}(n)$ is the complementary set. At time $\check{n}$ we will know whether we live in state $S_{1}(\check{n})$ or $S_{2}(\check{n})$. We might even know it a short time before that. The two states are identifiable by defininition.

\begin{figure}[tp]
\begin{center}
\includegraphics[width=80mm,clip=true]{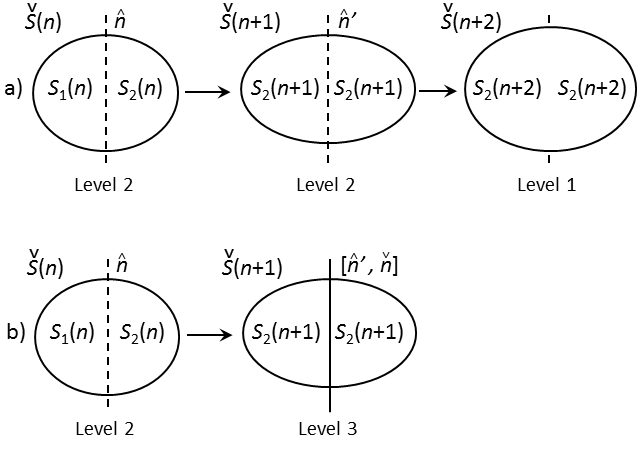}
\end{center}
\caption{Sequences of states in which changes occur in the knowability level, or among the numbers $\hat{n}$ and $\check{n}$ that are associated with alternatives at knowability levels 2 or 3. A dashed vertical line that separates the alternatives represents level 2, an absent line represents level 1, and a solid line level 3. The set $S$, the alternatives $\{S_{j}\}$, the knowability level and possibly the numbers $\hat{n}$ and $\check{n}$ together represent the reduced state $\check{S}$. See text for further discussion.}
\label{Fig32b}
\end{figure}

The question whether a given partition [\ref{division}] belongs to knowability level 1, 2 or 3 is a function of the state $S$, and so are the numbers $\hat{n}$ and $\check{n}$. These numbers and the knowability level can be said to be attribute values of quasiobjects that are part of the extended potential knowledge $\hat{PK}$, contributing to the reduced state $\check{S}$ (Fig. \ref{Fig28c}). These attribute values may change with time, just like any other attribute value.

This point is illustrated in Fig. \ref{Fig32b}. In panel a) we see a sequence of states in which the knowability level is 2 to begin with. The associated number $\hat{n}$ changes in the first time step. In the second step the knowability level changes to 1. In panel b) the knowability level changes from 2 to 3. Note that the knowability level cannot change once it has reached level 1 or 3. However, the pair of numbers $[\hat{n},\check{n}]$ associated with knowability level 3) may change. In plain language, we may delay the inevitable, or get it over with.

Note also that it is meaningless to speak about alternatives at level 2 or 3 that intersect, so that $S_{j}\cap S_{j'}\neq\varnothing$ for some $j\neq j'$. Then it would be impossible to decide which alternative is picked. Accordingly, if physical law includes a rule that specifies how alternatives are picked, then this rule needs to be defined for distinct alternatives only.

\section{State spaces}
\label{statespaces}

In this section we characterize the state space $\mathcal{S}$ in which physical states $S$ and object states $S_{O}$ live. We also introduce the concept of an object state space $\mathcal{S}_{O}$, in which we can embed object states $S_{O}$, but not the entire physical state $S$. Which of these two state spaces is more useful depends on the problem at hand. Further, we introduce the volume measure $V$ in these spaces. It will be essential in the following discussions about probability (Section \ref{probabilities}) and entropy (Section \ref{entropy}).

Our epistemic formalism is based on the concept of identifiable objects, which are characterized by a set of independent attributes $\{A_{i}\}$, each of which may take values $\upsilon_{ij}$ from a given range $\Upsilon_{i}$. The characteristic quality of attribute values is that they can be ordered (Definition \ref{attributevalues}).

These basic facts give some structure to the state space. An object state $S_{O}$ is defined as those states of the world that are not excluded by the knowledge about object $O$. This knowledge can be expressed as a set of intervals $\{\Delta_{i}\}\subseteq\Upsilon_{i}$ of values $\upsilon_{ij}$ of attribute $A_{i}$ that are not excluded by this knowledge. We can therefore define one axis in state space for each attribute $A_{i}$, and points on these axes that correspond to possible attribute values.

To represent the knowledge that correspond to the entire physical state $S$, we will have to add one set of such axes for each object in the world, representing the set of internal attributes of these objects. We also have to add axes that correspond to the relational attributes needed to describe all interrelations between the objects.

We see that it is meaningful to speak of a dimension of the state space that equals the total number of these axes. However, it is not possible to use the basic epistemic concepts in order to define angles between these axes, or numerical distances between points. There is no inner product, and no inherent metric. Take the spatial distance $r_{12}$ between two objects $O_{1}$ and $O_{2}$ as an example. Knowledge of this distance may be part of the state of potential knowledge $PK$. Thus it is a property of the corresponding \emph{physical state} $S$ that contains $O_{1}$ and $O_{2}$, rather than a property of the underlying \emph{state space} $\mathcal{S}$. A physical state $S$ that contains a distance $r_{12}$ also contains reference objects $\{O_{r12}\}$ which are used to measure this distance by means of comparison. If such reference objects are lacking in $S$, then the distance $r_{12}$ is simply not defined - even if $O_{1}$ and $O_{2}$ are still there.

Let us try to formalize the above discussion about attributes, attribute values and distances between these values.

\begin{defi}[\textbf{Ordered attribute}]
Suppose that the set $\Upsilon_{i}=\{\upsilon_{ij}\}$ defines the possible values of attribute $A_{i}$. Let $(\upsilon_{ij},\upsilon_{il})$ be a pair of different values such that  $\upsilon_{ij}\in\Upsilon_{i}$ and $\upsilon_{il}\in\Upsilon_{i}$. The set $\Upsilon_{i}$ is ordered if and only if, for each possible pair $(\upsilon_{ij},\upsilon_{il})$, all values $\upsilon_{ik}\in\Upsilon_{i}$ such that $\upsilon_{ik}\neq \upsilon_{ij}$ and $\upsilon_{ik}\neq \upsilon_{il}$ has the quality of either being \emph{in between} the members of $(\upsilon_{ij},\upsilon_{il})$ or not.
\label{orderedvalues}
\end{defi}

\begin{defi}[\textbf{Betweenness}]
Betweenness is a relation between three different values $\upsilon_{ij}$, $\upsilon_{ik}$ and $\upsilon_{il}$ of the same attribute $A_{i}$. If $\upsilon_{ik}$ is between the members of the pair $(\upsilon_{ij},\upsilon_{il})$ we write $\upsilon_{ij}\succ \upsilon_{ik}\prec \upsilon_{il}$. Otherwise we write $\upsilon_{ij}\not\succ \upsilon_{ik}\not\prec \upsilon_{il}$.

When the values are permuted in these formal expressions, the following rules hold. We have $\upsilon_{ij}\succ \upsilon_{ik}\prec \upsilon_{il}$ if and only if $\upsilon_{il}\succ \upsilon_{ik}\prec \upsilon_{ij}$. If $\upsilon_{ij} \not\succ \upsilon_{ik}\not\prec \upsilon_{il}$, then $\upsilon_{ik}\succ \upsilon_{ij}\prec \upsilon_{il}$ or $\upsilon_{ij}\succ \upsilon_{il}\prec \upsilon_{ik}$.

Suppose that there are more than three different values of $A_{i}$, and that the betweenness relation is defined for each triplet picked from the quadruplet $\{\upsilon_{ij},\upsilon_{ik},\upsilon_{il},\upsilon_{im}\}$. Then the following transitivity rules hold. If $\upsilon_{ij}\succ \upsilon_{ik}\prec \upsilon_{il}$ and $\upsilon_{ik}\succ \upsilon_{il}\prec \upsilon_{im}$, then $\upsilon_{ij}\succ \upsilon_{ik}\prec \upsilon_{im}$ and $\upsilon_{ij}\succ \upsilon_{il}\prec \upsilon_{im}$. If $\upsilon_{ij}\succ \upsilon_{ik}\prec \upsilon_{il}$ and $\upsilon_{ik}\succ \upsilon_{im}\prec \upsilon_{il}$, then $\upsilon_{ij}\succ \upsilon_{ik}\prec \upsilon_{im}$.
\label{betweenness}
\end{defi}

\begin{figure}[tp]
\begin{center}
\includegraphics[width=80mm,clip=true]{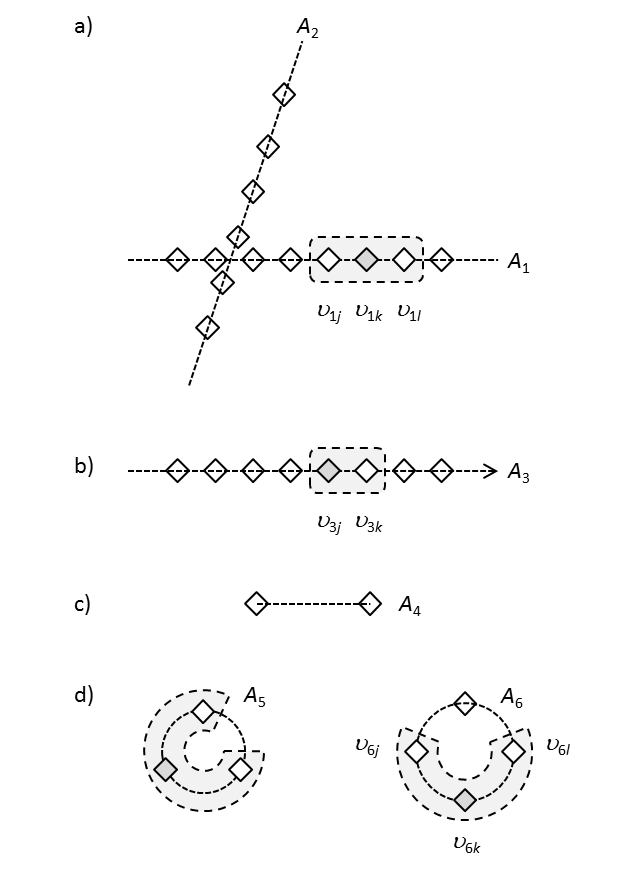}
\end{center}
\caption{The attributes $A_{i}$ can be regarded as axes that span state space. a) The \emph{ordering} of the attribute values along a given axis is determined from the concept of \emph{betweenness}. We assume that it is always possible to decide whether a value $\upsilon_{1k}$ is placed between another pair of values $(\upsilon_{1j},\upsilon_{1l})$ or not. b) The values of sequential time has an additional structure, namely \emph{direction}. For such an attribute $A_{3}$, it is always possible to decide which value $\upsilon_{3k}$ \emph{succeeds} another value $\upsilon_{3j}$. c) For attributes with only two possible values (like electron spin projections), the betweenness quality cannot be defined. d) Circular attributes are defined by the property that for any pair $(\upsilon_{6j},\upsilon_{6l})$, all other values $\upsilon_{6k}$ are placed between the members of this pair.}
\label{Fig138}
\end{figure}

In other words, the attribute values are ordered if and only if they are all equipped with a binary `betweenness' quality with respect to any other pair of values. The idea is illustrated in Fig. \ref{Fig138}(a). To be able to apply this definition, there has to be at least three attribute values $\upsilon_{ij}$ in the set $\Upsilon_{i}$.

The concept of ordering says nothing about direction. We have not added any arrows to the dashed attribute axes in Fig. \ref{Fig138}(a). If we have a preconceived notion about direction, we can say that Definition \ref{orderedvalues} is symmetric with respect to a direction reversal. However, to make this statement clear, we should first define explicitly what we mean by directedness and its relation to ordering.

\begin{defi}[\textbf{Directed attribute}]
Suppose that the set $\Upsilon_{i}=\{\upsilon_{ij}\}$ defines the possible values of attribute $A_{i}$. The set $\Upsilon_{i}$ is directed if and only if, for each $\upsilon_{ij}\in\Upsilon_{i}$, all values $\upsilon_{ik}\in\Upsilon_{i}$ such that $\upsilon_{ik}\neq \upsilon_{ij}$ has the quality of either being a successor to $\upsilon_{ij}$ or not.
\label{directededvalues}
\end{defi}

\begin{defi}[\textbf{Succession}]
Succession is a relation between two different values $\upsilon_{ij}$ and $\upsilon_{ik}$ of the same attribute $A_{i}$. If $\upsilon_{ik}$ is a successor to $\upsilon_{ij}$ we write $\upsilon_{ij}<\upsilon_{ik}$. Otherwise we write $\upsilon_{ij}>\upsilon_{ik}$.

The rule for value permutation in this formal expression is $\upsilon_{ij}<\upsilon_{ik}$ if and only if $\upsilon_{ik}>\upsilon_{ij}$.

Suppose that there are more than two different values of $A_{i}$, and that the succession relation is defined for each pair picked from the triplet $\{\upsilon_{ij},\upsilon_{ik},\upsilon_{il}\}$. Then the following transitivity rules hold. If $\upsilon_{ij}<\upsilon_{ik}$ and $\upsilon_{ik}<\upsilon_{il}$, then $\upsilon_{ij}<\upsilon_{il}$.
\label{succession}
\end{defi}

In plain language, the attribute values are ordered if and only if they are all equipped with a binary `succession' quality with respect to any other value. The idea is illustrated in Fig. \ref{Fig138}(b). To be able to apply this definition, it is sufficient that there are two attribute values $\upsilon_{ij}$ in the set $\Upsilon_{i}$.

\begin{defi}[\textbf{The ordering of directed attribute values}]
Suppose that $\Upsilon_{i}$ is ordered and directed. Then $\upsilon_{ij}\succ\upsilon_{ik}\prec\upsilon_{il}$ if and only if 1) $\upsilon_{ij}<\upsilon_{ik}$ and $\upsilon_{ik}<\upsilon_{il}$, or 2) $\upsilon_{ij}>\upsilon_{ik}$ and $\upsilon_{ik}>\upsilon_{il}$.
\label{orderdirected}
\end{defi}

The inclusion of both alternatives 1) and 2) in this definition provides the symmetry of the ordering with respect to a change of direction that we mentioned above.

The following statement follows directly from the above definitions.

\begin{state}[\textbf{Directed attributes are ordered}]
All directed sets $\Upsilon_{i}$ are ordered, but an ordered set $\Upsilon_{i}$ does not have to be directed.
\label{directedimpliesordered}
\end{state}

We have introduced one inherently directed attribute, namely sequential time $n$ (Section \ref{time}). However, we do not require that attribute values are directed, only that they are ordered. Since the definition of an ordered attribute (Definition \ref{orderedvalues}) requires three possible values, we have to make an exception to allow for attributes with only two possible values, like the spin direction of an electron [Fig. \ref{Fig138}(c)].

\begin{defi}[\textbf{Attribute}]
An attribute $A_{i}$ is defined as a set of values $\Upsilon_{i}$ which can be subjectively associated with each other. If $\Upsilon_{i}$ contains more than two elements, it must be an ordered set according to Definition \ref{orderedvalues}.
\label{attridefi}
\end{defi}

To make possible a symbolic and numerical representation of the physical state and of physical law within an epistemic approach to physics, we assume the following.

\begin{assu}[\textbf{All knowledge can be expressed in terms of objects, attributes and their values.}]
To specify the physical state $S$, it is sufficient to specify which sets of objects, and which sets of attribute values that describe and relate them, are not excluded by potential knowledge.
\label{knowattri}
\end{assu}

This assumption implies the following statement about state space $\mathcal{S}$.

\begin{state}[\textbf{The attributes span state space.}]
All physical states $S$ can be embedded in a state space $\mathcal{S}$ spanned by axes defined by attributes, according to Fig. \ref{Fig138}. Each object that may be present in $S$ must be allowed its own set of attribute-axes.
\label{spans}
\end{state}

How does the color attribute of quarks fit into the above description of attributes and their values? There is clearly no direction of its three possible values `green', `red', and `blue'. The concept of `betweenness' becomes kind of degenerate if there are only three possible values, since each value must be considered to be placed between the other two. Each value has the betweenness quality `yes' with respect to every possible pair of other values. There is no betweenness quality `no' around. Nevertheless, this is not required in the definition of an ordered set of attribute values. We can even turn this degeneracy vice into a virtue by regarding the color attribute as an example of an interesting type of ordered attribute, namely the \emph{circular} ones. 

\begin{defi}[\textbf{Circular attribute}]
An ordered set $\Upsilon_{i}=\{\upsilon_{ij}\}$ of attribute values is circular if and only if, for any pair $(\upsilon_{ij},\upsilon_{il})$ of different attribute values such that $\upsilon_{ij}\in\Upsilon_{i}$ and $\upsilon_{il}\in\Upsilon_{i}$, \emph{all} the other attribute values $\upsilon_{ik}$ have the quality of being in between the members of $(\upsilon_{ij},\upsilon_{il})$, that is, $\upsilon_{ij}\succ\upsilon_{ik}\prec\upsilon_{il}$.
\label{circularvalues}
\end{defi}

The idea behind such circular attributes is illustrated in Fig. \ref{Fig138}(d). They are naturally represented by complex numbers of modulus one. Since the defintion of a circular attribute relies on that of an ordered attibute, it has to have at least three possible values. The following statement is easily realized.

\begin{state}[\textbf{Circular attributes cannot be directed}]
No set of attribute values $\Upsilon_{i}$ can fulfil Definitions \ref{orderedvalues} and \ref{directededvalues} at the same time.
\label{circularundirected}
\end{state}

An ordered attribute that is not circular may be called linear. It is possible to introduce the notions of discrete and continuous linear attributes with the tools that we have introduced. To be able to do the same with circular attributes, we need to define new tools. Let us therefore start with the linear case. 

\begin{defi}[\textbf{Discrete linear attribute}]
Suppose that the set $\Upsilon_{i}=\{\upsilon_{ij}\}$ defines the possible values of the linear attribute $A_{i}$.
The set $\Upsilon_{i}$ is discrete if and only if, for each value $\upsilon_{ij}\in\Upsilon_{i}$, there is another value $\upsilon_{il}\in\Upsilon_{i}$ such that there is no value $\upsilon_{ik}\in\Upsilon_{i}$ that has the quality of beeing in between the members of the pair $(\upsilon_{ij},\upsilon_{il})$.
\label{discreteattribute}
\end{defi}

\begin{defi}[\textbf{Continuous linear attribute}]
Suppose that the set $\Upsilon_{i}=\{\upsilon_{ij}\}$ defines the possible values of the linear attribute $A_{i}$.
The set $\Upsilon_{i}$ is continuous if and only if there is no value $\upsilon_{ij}\in\Upsilon_{i}$ for which there is another value $\upsilon_{il}\in\Upsilon_{i}$ such that there is no value $\upsilon_{ik}\in\Upsilon_{i}$ that has the quality of beeing in between the members of the pair $(\upsilon_{ij},\upsilon_{il})$.
\label{continuousattribute}
\end{defi}

A linear attribute that is netiher discrete nor continuous may be called mixed.

\begin{defi}[\textbf{Mixed linear attribute}]
Suppose that the set $\Upsilon_{i}=\{\upsilon_{ij}\}$ defines the possible values of a linear attribute $A_{i}$.
The set $\Upsilon_{i}$ is mixed if, for some but not for all values $\upsilon_{ij}\in\Upsilon_{i}$, there is another value $\upsilon_{il}\in\Upsilon_{i}$ such that there is no value $\upsilon_{ik}\in\Upsilon_{i}$ that has the quality of beeing in between the members of the pair $(\upsilon_{ij},\upsilon_{il})$.
\label{mixedattribute}
\end{defi}

The above three definitions are formal. At the same time, their aim is to capture the essence of different kinds of \emph{physical} attributes. Therefore, in an epistemic approach to physics, there must be clearcut criteria to distinguish between these three kinds of attributes by means of observation. The definitions are only menaingful if an attribute can be \emph{knowably} discrete, \emph{knowably} continuous, or \emph{knowably} mixed.

\begin{defi}[\textbf{Knowably discrete linear attribute}]
Referring to the terminology in Definition \ref{discreteattribute}, suppose that the state of potential knowledge $PK$ contains an object $O_{j}$ with value $\upsilon_{ij}$ of attribute $A_{i}$ and another object $O_{l}$ with value $\upsilon_{il}$ of the same attribute. Then $A_{i}$ is knowably discrete if for each such object $O_{j}$ there is a value $\upsilon_{il}$ for which there is no such corresponding object $O_{l}$ for which physical law allows the presence of a third object $O_{k}$ in $PK$ with value $\upsilon_{ik}$ of $A_{i}$ such that $\upsilon_{ik}$ has the quality of being in between the members of the pair $(\upsilon_{ij},\upsilon_{il})$. The above is true also if some or all of the abovementioned objects are quasiobjects, and the corresponding attribute values are deduced from the attribute values of other, directly perceived objects.
\label{knowdiscreteattribute}
\end{defi}

\begin{defi}[\textbf{Knowably continuous linear attribute}]
Referring to the terminology in Definition \ref{continuousattribute}, suppose that the state of potential knowledge $PK$ contains an object $O_{j}$ with value $\upsilon_{ij}$ of attribute $A_{i}$ and an object $O_{l}$ with value $\upsilon_{il}$ of the same attribute. Then $A_{i}$ is knowably continuous if for each such pair of objects, knowable physical law cannot exlude the presence of a third object $O_{k}$ with value $\upsilon_{ik}$ of $A_{i}$ that has the quality of beeing in between the members of the pair $(\upsilon_{ij},\upsilon_{il})$. The above is true at least if we allow all of these three objects to be quasiobjects for which the corresponding attribute values are deduced from the attribute values of other, directly perceived objects.
\label{knowcontinuousattribute}
\end{defi}

The definition of a knowably mixed attribute is analogous. We may say that an attribute is judged to be continuous whenever we cannot confirm by means of experiment and theory that it is discrete (at least partially). Put differently, it is continuous whenever we cannot \emph{exclude} the possibility that it is indeed continuous. This acceptance of everything that cannot be excluded is typical of the epistemic approach used in this treatise. For example, the physical state is defined as the set of all hypothetical exact states of the world that are not excluded by the incomplete potential knowledge.

With the above definition of a continuous attribute, we cannot rule out the possibility that the ordered sequence of continuous attribute values can be represented by a sequence of rational numbers. That is, from a mathematical point of view we should rather call this type of attribute \emph{dense}. However, from a physical point of view, we can never distinguish a dense attribute from a truly continuous one, whose numerical representation requires irrational numbers. 

Let us turn to circular attributes. It is clear that Defintion \ref{discreteattribute}, which specifies the meaning of discreteness, cannot be fulfilled by a circular attribute (Definition \ref{circularvalues}). Using the concept of betweenness, Definition \ref{discreteattribute} introduces the notion that two attribute values are close, that they are nearest neighbors. This approach is impossible for circular attributes. Therefore we have to introduce such a notion of \emph{closeness} explicitly, as an additional primitive relation between attribute values.

\begin{defi}[\textbf{Closeness}]
Closeness is a relation between two different values $\upsilon_{ij}$, $\upsilon_{ik}$ of the same circular attribute $A_{i}$. If $\upsilon_{ik}$ is the nearest neighbor to $\upsilon_{ij}$ we write $\upsilon_{ij}\bowtie \upsilon_{ik}$, otherwise we write $\upsilon_{ij}\not\bowtie \upsilon_{ik}$. We have $\upsilon_{ik}\bowtie \upsilon_{ij}$ if and only if $\upsilon_{ij}\bowtie \upsilon_{ik}$. If $\upsilon_{ij}\bowtie \upsilon_{ik}$ and $\upsilon_{ik}\bowtie \upsilon_{il}$, then $\upsilon_{ij}\not\bowtie \upsilon_{il}$.
\label{closeness}
\end{defi}

\begin{defi}[\textbf{Discrete circular attribute}]
Suppose that the set $\Upsilon_{i}=\{\upsilon_{ij}\}$ defines the possible values of the circular attribute $A_{i}$, and that the binary closeness quality is defined for each pair of different values  $(\upsilon_{ij},\upsilon_{ik})$.
Then the set $\Upsilon_{i}$ is discrete if and only if for each value $\upsilon_{ij}$ there is exactly one pair of values $(\upsilon_{ik},\upsilon_{il})$ such that $\upsilon_{ij}\bowtie \upsilon_{ik}$ and $\upsilon_{ij}\bowtie \upsilon_{il}$.
\label{continuouscattribute}
\end{defi}

Definition \ref{closeness} is not quite satisfactory, since it is \emph{ad hoc}. It makes it possible to speak of discrete and continuous circular attributes, but it cannot be used for anything else. In particular, I cannot see how it can be used to defined a mixed circular attribute in analogy with a mixed linear attribute (Definition \ref{mixedattribute}).

\begin{defi}[\textbf{Continuous circular attribute}]
Suppose that the set $\Upsilon_{i}=\{\upsilon_{ij}\}$ defines the possible values of the circular attribute $A_{i}$, and that the binary closeness quality is defined for each pair of different values  $(\upsilon_{ij},\upsilon_{ik})$.
Then the set $\Upsilon_{i}$ is continuous if and only if $\upsilon_{ij}\not\bowtie \upsilon_{ik}$ for each such pair.
\label{discretecattribute}
\end{defi}

We argued above that there is no inherent metric in state space. Since state space is assumed to be spanned by the attributes (Statement \ref{spans}), it would be sufficient that such a metric tells us the distance between any pair of attribute values. If all attributes were discrete, such a distance could actually be defined. We could simply count the number of attribute values that have the quality of being in between those in the pair (Definition \ref{betweenness}). This procedure is not possible for continuous attribute. We always get the answer infinity.

\begin{state}[\textbf{State spaces are not metrical}]
It is impossible to assign a metric to a state space spanned by a set of attributes among which at least one is continuous. 
\label{nometrics}
\end{state}

\begin{figure}[tp]
\begin{center}
\includegraphics[width=80mm,clip=true]{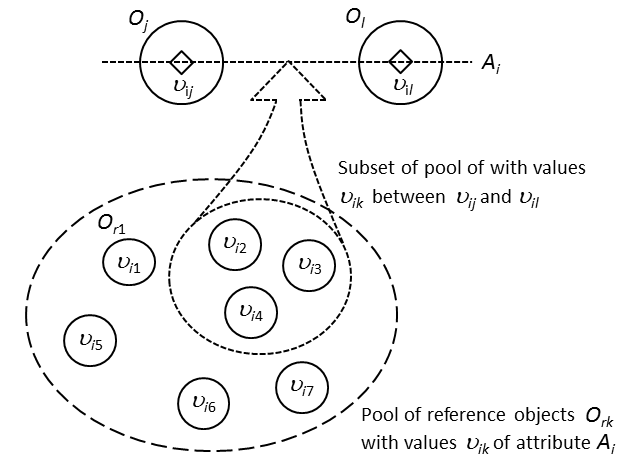}
\end{center}
\caption{The pool of reference objects can be seen as a ruler chopped into unit pieces, and then reassembled between the object $O_{j}$ and $O_{l}$. In this example, the distance between the values $\upsilon_{j}$ and $\upsilon_{l}$ of these objects is three, in the unit defined by the pool.}
\label{Fig139}
\end{figure}

\vspace{5mm}
\begin{center}
$\maltese$
\end{center}
\paragraph{}

Allow me to make a digression at this point, to describe in the above vocabulary what it means in practice to assign a distance between two values of the same attribute. We assume that we have a state $S$ that contains two objects $O_{j}$ and $O_{l}$ with values $\upsilon_{ij}$ and $\upsilon_{il}$ of attribute $A_{i}$, respectively (Fig. \ref{Fig139}). To be able to say that the distance between $\upsilon_{ij}$ and $\upsilon_{il}$ is $d_{jl}$, the state $S$ must also contain a pool of $N$ distinct reference objects $\{O_{rk}\}$ with values $\{\upsilon_{ik}\}$ of $A_{i}$ that are all different. Then $d_{jl}\leq N$ is the number of objects in the pool that have a value $\upsilon_{rk}$ between $\upsilon_{ij}$ and $\upsilon_{il}$. The pool of reference objects acts as a ruler and their number $N$ and values $\{\upsilon_{ik}\}$ define the unit in which the distance $d_{jl}$ is measured.

To be generally valid, the above sketch of a definition of a state $S$ that contains a numerical value of a distance should allow values $\upsilon_{i}$ that are not perfectly specified, but are known to be found within intervals $\Delta_{ij}$ and $\Delta_{il}$ for the objects $O_{j}$ and $O_{l}$, and within the intervals $\{\Delta_{ik}\}$ for the reference objects $\{O_{rk}\}$. The value intervals of the reference objects should be disjoint, that is $\{\Delta_{ik}\}\cap \{\Delta_{ik}\}=\varnothing$. Regarding the distance $d_{jl}$, we may say that the knowledge encoded in such a state $S$ is that $N\geq d_{jl}\leq N+1$ in the unit defined by $N$ and $\{\Delta_{ik}\}$.

\begin{defi}[\textbf{Interval betweenness}]
Interval betweeness is a relation between three disjoint sets of values $\Delta_{ij}$, $\Delta_{ik}$ and $\Delta_{il}$ of the same attribute $A_{i}$. The set $\Delta_{ik}$ is between $\Delta_{ij}$ and $\Delta_{il}$ if and only if $\upsilon_{ij}\succ \upsilon_{ik}\prec \upsilon_{il}$ for any value triplet $(\upsilon_{ij},\upsilon_{ik},\upsilon_{il})$ such that $\upsilon_{ij}\in\Delta_{ij}$, $\upsilon_{ik}\in\Delta_{ik}$, and $\upsilon_{il}\in\Delta_{il}$. We write $\Delta_{ij}\succ \Delta_{ik}\prec \Delta_{il}$. The set $\Delta_{ik}$ is not between $\Delta_{ij}$ and $\Delta_{il}$ if and only if $\upsilon_{ij}\not\succ \upsilon_{ik}\not\prec \upsilon_{il}$ for any such value triplet. We write $\Delta_{ij}\not\succ \Delta_{ik}\not\prec \Delta_{il}$. If none of the two above conditions is fulfilled, then the interval betweenness quality is not defined. $\Delta_{ij}$ does not possess the interval betweenness quality in relation to the members of the pair $(\Delta_{ij},\Delta_{il})$.
\label{ibetweenness}
\end{defi}

\begin{defi}[\textbf{Distance between attribute values}]
Let $\Delta_{ij}$ be the set of values of attribute $A_{i}$ not excluded by the potential knowledge about object $O_{j}$. The distance $d_{jl}$ between the disjoint set of values $\Delta_{ij}$ and $\Delta_{il}$ of attribute $A_{i}$ of two objects $O_{j}$ and $O_{l}$ that is contained in the physical state $S$ is defined in the following circumstances. The state $S$ contains a pool of reference objects $\{O_{rk}\}$ with disjoint sets of allowed values $\{\Delta_{ik}\}$. The betweenness quality of each set $\Delta_{ik}$ is defined with respect to the members of the pair of sets $(\Delta_{ij},\Delta_{il})$. Then $d_{jl}$ is the number of reference objects $O_{rk}$ for which $\Delta_{ij}\succ \Delta_{ik}\prec \Delta_{il}$.
\label{distance}
\end{defi}

Note that knowledge about $d_{jl}$ can be incomplete, just like knowledge about the value $\upsilon_{ij}$ of $A_{i}$. Incomplete knowledge about $d_{jl}$ means that we are not sure about how many reference objects can be fitted between $O_{j}$ and $O_{l}$. We have to define a set $D_{jl}$ of possible values of $d_{jl}$, just like we have defined a set $\Delta_{ij}$ of possible values $\upsilon_{ij}$. If $D_{jl}$ is a connected interval without holes (which is easy to define using the concept of betweenness), we may write

\begin{equation}
D_{jl}=[d_{jl}-\frac{\Delta d_{jl}}{2},d_{jl}+\frac{\Delta d_{jl}}{2}]
\label{uncertaininterval}
\end{equation}
for some interval of uncertainty $\Delta d_{jl}$.

As an example of a distance $d_{jl}$, we may take the relational time $t_{jl}$ passed between two events or objects $O_{j}$ and $O_{l}$. The pool of reference object used to determine this time interval may be a set of $N$ heartbeats of a subject who observes both $O_{j}$ and $O_{l}$. If she can fit in $T$ heartbeats between $O_{j}$ and $O_{l}$, then the time passed is $T$, expressed in the unit defined by the activity of her own heart. These heartbeats may come at irregular intervals, as measured by \emph{another} pool of reference objects, but in terms of the original reference objects $\{O_{rk}\}$ themselves, there is nothing with which to judge whether the unit of measurement is uniform or not. All we can say is that $\{O_{rk}\}$ provides a unit.

We have defined sequential time $n$ to be a collective attribute that can be used to characterize the physical state $S$ of the entire world. Therefore, all subjects agree on which objects or events $O_{k}$ occur between any pair $O_{j}$ and $O_{l}$ of other events. In other words, everyone  agrees whether $t_{j}\succ t_{k}\prec t_{l}$ or $t_{j}\not\succ t_{k}\not\prec t_{l}$. This means that everone should agree on the measured time $t_{jl}$ as long as they use the same set of reference events. We can try to generalize this quality of time to a general statement about all attributes.

\begin{assu}[\textbf{Universality of betweenness}]
Suppose that two subjects agree that the range of values $\Delta_{ik}$ of each reference object $O_{rk}$ in a pool $\{O_{rk}\}$ possesses the binary interval betweenness quality in relation to the corresponding ranges of objects $O_{j}$ and $O_{l}$. Then, for each $O_{rk}$, they agree whether $\Delta_{ij}\succ \Delta_{ik}\prec \Delta_{il}$ or $\Delta_{ij}\not\succ \Delta_{ik}\not\prec \Delta_{il}$.
\label{ubetween}
\end{assu}

Without this assumption, the entire construction of a generally valid state space spanned by ordered attributes would collapse, since its structure would become subjective.

\begin{state}[\textbf{Agreement on measured distances}]
Suppose that two observers use the same pool $\{O_{rk}\}$ of reference objects to measure the distance of attribute $A_{i}$ between objects $O_{j}$ and $O_{l}$. Then they arrive at the same answer provided they agree that the range of values $\Delta_{ik}$ of each reference object $O_{rk}$ possesses the interval betweenness quality in relation to the corresponding value ranges of $O_{j}$ and $O_{l}$.
\label{dagreement}
\end{state}

How do these considerations go together with special relativity, which implies that two subjects may measure different time intervals $t_{jl}$ and $t_{jl}'$ between the same pair of events? There is no contradiction, since these two subjects use \emph{different} pools of reference objects to obtain the different distances $t_{jl}$ and $t_{jl}'$. In a typical example, each subject uses a pool of reference objects (a clock) which is at rest in her own rest frame. Consider the twin paradox. Let $O_{j}$ be the event when one twin boards a space ship which takes her at high speed into deep space, and let $O_{l}$ be the event when she returns. If she measures the number $t_{jl}$ of her own heartbeats, and compares this number with the number $t_{jl}'$ of heartbeats that her earthbound twin can fit in between the two events, then they find that $t_{jl}<t_{jl}'$. But if they both decide to use the heartbeats of the earthbound twin to measure the time interval, they will agree on the result: $t_{jl}=t_{jl}'$. Of course, the same is true if they both decide to use the heartbeats of the space traveller as a pool of reference objects.

\vspace{5mm}
\begin{center}
$\maltese$
\end{center}
\paragraph{}

We argued above that the number of heartbeats between two events is the measure of the relational time passed between these events in the chosen unit regardless whether these heartbeats come at regular intervals or not. Nevertheless, we need to discuss more thouroghly what meaning, if any, can be assigned in the conceptual framework introduced so far to the statement that two intervals are equal, or to the statement that one is larger than the other.

We certainly have the subjective ability to decide such issues, even if the judgement may not be perfectly precise. We feel if a hearbeat is premature even if we do not explicitly check it with a clock. An hour feels longer than a minute even if we do not count heartbeats. We can decide roughly whether two people are equally tall, and if not, who is taller. If two men and one woman talk, it is most often evident that the pitch of the two male voices are more similar than the pitches of the female voice and one of the male voices. In short, we have an intuitive sense of scale. The question is whether the above formal framework is sufficient to account for this sense of scale, or if it have to be introduced explicitly.

\begin{figure}[tp]
\begin{center}
\includegraphics[width=80mm,clip=true]{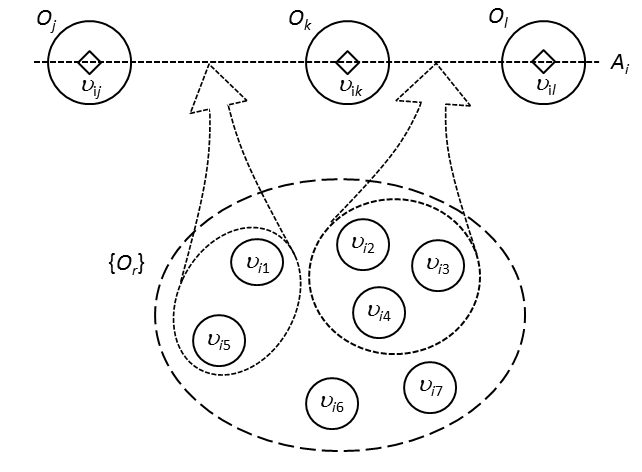}
\end{center}
\caption{A single pool $\{O_{r}\}$ of reference objects can be used to compare the distances $d_{jk}$ and $d_{kl}$ between the two pairs of attribute values $(\upsilon_{j},\upsilon_{k})$ and $(\upsilon_{k},\upsilon_{l})$. We get $d_{jk}=2$ and $d_{kl}=3$ so that $d_{jk}<d_{kl}$. However, the outcome of this comparison depends on the pool chosen, so that this method cannot be used to compare distances objectively. Compare Fig. \ref{Fig139}.}
\label{Fig202}
\end{figure}

To compare two distances, the only formal possibility we have at our disposal at this stage is to use a pool of reference objects, some of which `fit' between the object pair used to define the first distance, and some other fit between the objects pair used to define the second distance (Fig. \ref{Fig202}). The two distances are equal if and only if the number of reference objects in the first group equals the number in the second group. Using this method, we conclude in the example shown in Fig. \ref{Fig202} that $d_{jk}<d_{kl}$.

The problem with this mode of comparison is that the outcome depends on the the pool of reference objects we use, the unit we choose. The result is not an inherent property of the attribute values of the two pairs of objects that define the two distances. It is not a property of the world we observe, but depends on the mode of observation.

If the attribute $A_{i}$ is discrete we can nevertheless measure and compare distances objectively. A properly measured distance $d_{jk}$ can be operationally defined as the \emph{maximum} number of reference objects whose attribute values can be fitted between those of the object pair $(O_{j},O_{k})$. This is impossible if the attribute is continuous. Then infinitely many reference objects can in principle be fitted between any such object pair whenever $\upsilon_{j}\neq\upsilon_{k}$.

Even if we cannot measure a single distance in the continuous case and get an objective pure number as a result, we might get the idea to choose equidistant attribute values of all the reference objects in the pool (Fig. \ref{Fig202}) in order to \emph{compare} two distances $d_{ij}$ and $d_{kl}$ objectively. But this is circular reasoning, of course. We make use of equidistance in order to define it.

The only way to account for the subjective ability to compare distances between values of continuous attributes is therefore to introduce scale as a primary quality of state space.

\begin{defi}[\textbf{Relative distance}]
Relative distance is a primary relation defined between any two pairs of values $(\upsilon_{j},\upsilon_{k})$ and $(\upsilon_{l},\upsilon_{m})$ of the same continuous attribute. This means that the distance $d_{jk}$ between $\upsilon_{j}$ and $\upsilon_{k}$ and the distance $d_{lm}$ between $\upsilon_{l}$ and $\upsilon_{m}$ are always defined in the sense that exactly one of the following three relations always hold: $d_{jk}<d_{lm}$, $d_{jk}=d_{lm}$ or $d_{jk}>d_{lm}$. If $(\upsilon_{j},\upsilon_{k})=(\upsilon_{l},\upsilon_{m})$ then $d_{jk}=d_{lm}$. These relations are not defined in the collective potential knowledge $PK$, only in the personal potential knowledge $PK^{m}$ of a subject $m$.
\label{reldistance}
\end{defi}

The reason why we have to say that distance comparisons are personal in general is that they are indeed personal in the case of spatio-temporal distances. Two subjects may judge relative spatial or temporal distances differently if they are accelerating in relation to each other. For example, in the twin paradox, the first time distance $t_{ij}$ may be the time passed between the decision to test Einstein's prediction and the actual departure of the space ship. The second time distance $t_{jl}$ is the time passed after that until the twins reunite after the journey. The earthbound twin may have to wait a very long time for her sibling to return, so that she judges that $t_{jl}>t_{ij}$, whereas the space-travelling twin finds that his trip was very brief, so that $t_{jl}'<t_{ij}'$ from his perspective (where $t_{ij}'\approx t_{ij}$).

In contrast, if we would only take inertial relative motion of different subjects into account, the judgement of relative spatio-temporal distances would collective or universal, since the Lorentz transformation is linear and therefore preserves such relations.

\vspace{5mm}
\begin{center}
$\maltese$
\end{center}
\paragraph{}

In order to apply the formal machinery introduced above to spatial relational attributes, we first need to assign a spatial position $\mathbf{r}_{j}$ to a given object $O_{j}$, and call that position the spatial attribute value of $O_{j}$. We cannot start with spatial distances $r_{jl}$ between two objects $O_{j}$ and $O_{l}$, since we describe distances between attribute values of two objects as secondary to the values themselves. On the other hand, we cannot presuppose a coordinate system in which $\mathbf{r}_{j}$ is defined by a numerical value. That would amount to an added content in the physical state $S$ that cannot be part of state space $\mathcal{S}$ itself, just like numerical distances cannot be considered as a primary part of $\mathcal{S}$, as discussed above.

\begin{figure}[tp]
\begin{center}
\includegraphics[width=80mm,clip=true]{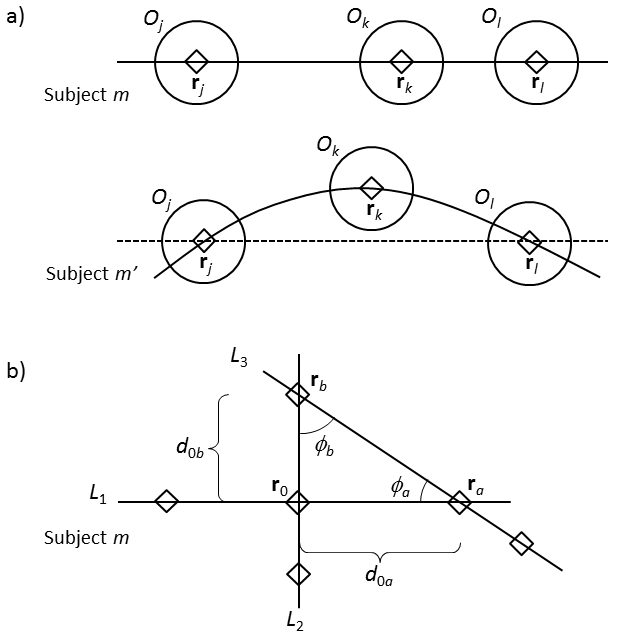}
\end{center}
\caption{a) The question whether the positions of three objects are placed along a straight line may have different answers depending on which subject you ask. If someone judges that they lie on a straight line, the question which position is between the other two has a definitive answer. b) The combination of three such straight lines makes it possible to define a triangle. If two of the lines are orthogonal, and if the personal physical state of subject $m$ is such that the distances $d_{0a}$ and $d_{0b}$ are defined with the same kind of pool of reference objects (ruler), then the angles $\phi_{a}$ and $\phi_{b}$ can be defined in the ususal way. Compare Fig. \ref{Fig139}.}
\label{Fig140}
\end{figure}

If numerical values are forbidden, how can we possibly define $\mathbf{r}_{j}$? First, we have to take special relativity into account, so that we should actually speak of the spatio-temporal four-position $\mathbf{r}_{4}$.(Below, we will sometimes suppress the index "4" for notational simplicity, and we hope that this will not cause confusion.) We have based our approach to attribute values on the idea of betweenness. However, since spatio-temporal positions are four-dimensional, it is not possible in general to assign a betweenness quality $\mathbf{r}_{j}\succ \mathbf{r}_{k}\prec \mathbf{r}_{l}$ or $\mathbf{r}_{j}\not\succ \mathbf{r}_{k}\not\prec \mathbf{r}_{l}$ to the position $\mathbf{r}_{j}$ in relation to an arbitrary pair of positions $(\mathbf{r}_{j},\mathbf{r}_{l})$. We are forced to add another primary quality to physical space, namely \emph{straightness} (Fig. \ref{Fig140}).

\begin{defi}[\textbf{Straightness}]
Straightness is a relation between three distinct objects $O_{j}$, $O_{k}$, and $O_{l}$, for which three different spatio-temporal positions $\mathbf{r}_{j}$, $\mathbf{r}_{k}$, and $\mathbf{r}_{l}$ can be defined. The quality is not defined in the collective potential knowledge $PK$, only in the personal potential knowledge $PK^{m}$ of a subject $m$. If $\mathbf{r}_{j}$, $\mathbf{r}_{k}$, and $\mathbf{r}_{l}$ are placed along a straight line as judged by $m$, we write $\mathbf{r}_{j}-\mathbf{r}_{k}-\mathbf{r}_{l}$, otherwise $PK_{m}:\mathbf{r}_{j}\not-\mathbf{r}_{k}\not-\mathbf{r}_{l}$. The straightness quality does not depend on the ordering in these expressions.
\label{straightness}
\end{defi}

The personal aspect of this quality means that we do not exlude the possibility that different subjects $m$ judge the straightness of the same triplet of objects differently. Just like \emph{betweenness} and \emph{succession}, \emph{straightness} is a basic epistemic quality that we use to assign structure to attribute values, and therefore to state space.

The existence of the quality of straightness will be used to derive the evolution equation in Section \ref{eveq}. A free specimen whose evolution we want to determine is defined as an object which travels along a straight trajectory. The fact that straightness is not a collectively or universally defined quality means that physical law must be insensitive to it. This fact will be taken into account in Sections \ref{eveqi} and \ref{gaugeprinciple}.

Is it really possible to judge whether a \emph{trajectory} in space-time is straight without any measurements? It is clear that we can judge in such primary way whether a \emph{line} is straight. But this judgement is immediate, it concerns a purely spatial straightness. A trajectory also has a temporal component. We resort to Einstein's elevator (Fig. \ref{Fig13}), epistemic invariance, and the primary \emph{feeling} of acceleration or gravitation.

\begin{defi}[\textbf{Straight trajectory}]
The trajectory $\mathbf{r}_{4}^{m}(n)$ of a subject $m$ in space-time is straight, as judged from within her own state $PK^{m}$ of personal potential knowledge, if one of the following two conditions is met. 1) She can match her trajectory to a straight spatial line with the help of her memory of previous positions along this line. She experiences no feeling of acceleration in the direction of this line in so doing. 2) She has no outside reference points that can define a straight line along which she moves (she is at rest as far as she knows), and experience a constant feeling of acceleration or gravitation.
\label{straj}
\end{defi}

With the help of condition 1) she can also determine whether the trajectory of an object that she observes is straight or not, by moving along with it.

We may say that the betweenness quality of $\mathbf{r}_{j}$ in relation to the spatio-temporal positions in the pair $(\mathbf{r}_{k},\mathbf{r}_{l})$ is defined if and only if $\mathbf{r}_{j}-\mathbf{r}_{k}-\mathbf{r}_{l}$. In this way we can assign an ordered structure to spatio-temporal positions, just like we have done for any other attribute.

This approach means that the betweenness of positions, and their ordering, becomes a subjective relation of three objects, since it relies on the subjective straightness quality. This fact does not contradict Assumption \ref{ubetween}, however, since the presupposed collective agreement on the answer to the question whether an attribute value is placed between two other values or not, relies on the assumption that the betweenness quality is defined for these three values. In short, for spatio-temporal positions, the subjectivity resides in the straightness, not in the betweenness.

To make the presentation more comprehensive, let us account for incomplete knowlege about spatio-temporal positions by introducing the concept of interval straightness, just like we introduced interval betweenness in Definition \ref{ibetweenness}.

\begin{defi}[\textbf{Interval straightness}]
Interval straightness is a relation between three disjoint sets of values $\mathbf{R}{j}$, $\mathbf{R}_{k}$ and $\mathbf{R}_{l}$ of the spatio-temporal position $\mathbf{r}_{4}$. The quality is not defined in the collective potential knowledge $PK$, only in the personal potential knowledge $PK^{m}$ of a subject $m$. The sets $\mathbf{R}_{j}$, $\mathbf{R}_{k}$, and $\mathbf{R}_{l}$ are placed along a straight line as judged by $m$ if and only if $\mathbf{r}_{j}-\mathbf{r}_{k}-\mathbf{r}_{l}$ for at least one position triplet $(\mathbf{r}_{j},\mathbf{r}_{k},\mathbf{r}_{l})$ such that $\mathbf{r}_{j}\in\mathbf{R}_{j}$, $\mathbf{r}_{k}\in\mathbf{R}_{k}$, and $\mathbf{r}_{l}\in\mathbf{R}_{l}$. In that case we write $\mathbf{R}_{j}-\mathbf{R}_{k}-\mathbf{R}_{l}$.
\label{istraightness}
\end{defi}

\begin{defi}[\textbf{Interval betweenness of spatio-temporal positions}]
Interval betweenness of spatio-temporal positions is a relation between three sets $\mathbf{R}_{j}$, $\mathbf{R}_{k}$ and $\mathbf{R}_{l}$ of the spatio-temporal position $\mathbf{r}_{4}$ which can be defined in the potential knowledge $PK^{m}$ of subject $m$ if and only if $\mathbf{R}_{j}-\mathbf{R}_{k}-\mathbf{R}_{l}$, as judged by $m$. It is indeed defined in this case if and only if we can write $\mathbf{R}_{j}\succ\mathbf{R}_{k}\prec\mathbf{R}_{l}$ or $\mathbf{R}_{j}\not\succ\mathbf{R}_{k}\not\prec\mathbf{R}_{l}$ according to Definition \ref{ibetweenness}.
\label{ipbetweenness}
\end{defi}

In order to make it possible to measure angles, we introduce orthogonality as a sixth and final primary epistemic quality that gives structure to sets of attribute values. All these qualities are listed in Table \ref{attributestructure}. Angles are seen as functions of measured distances $r_{jl}$ between a triplet of spatial positions $(\mathbf{r}_{j}, \mathbf{r}_{k},\mathbf{r}_{l})$. As such they are part of specific types of states $S$ with appropriate pools of reference objects rather than part of the structure of state space itself, as discussed above.

Note that I speak about spatial positions here, not spatio-temporal four-positions. As far as I understand, the primitive, subjective judgement whether a pair of straight lines are orthogonal or not has to be instantaneous, so that it lacks the temporal component. Another way to put it is to say that we have to use different pools of reference objects to measure spatial and temporal distances, different units. Therefore a presupposed orthogonality in space-time becomes dependent on the choice of reference objects. It cannot correspond to a primitive structure of attribute values in state space, since it depends on the content of the state $S$ that is embedded in state space.

\begin{defi}[\textbf{Orthogonality}]
Orthogonality is a relation between five spatial positions, which we may call $\mathbf{r}_{\pm 1}$, $\mathbf{r}_{0}$ and $\mathbf{r}_{\pm 2}$. It is defined in the potential knowledge $PK^{m}$ of subject $m$ if and only if we can write $\mathbf{r}_{-1}-\mathbf{r}_{0}-\mathbf{r}_{1}$ and $\mathbf{r}_{-2}-\mathbf{r}_{0}-\mathbf{r}_{2}$. The two straight lines defined by these relations are either orthogonal or not. If they are, we write $(\mathbf{r}_{-1},\mathbf{r}_{0},\mathbf{r}_{1})\bot (\mathbf{r}_{-2},\mathbf{r}_{0},\mathbf{r}_{2})$, otherwise we write $(\mathbf{r}_{-1},\mathbf{r}_{0},\mathbf{r}_{1})\not\bot (\mathbf{r}_{-2},\mathbf{r}_{0},\mathbf{r}_{2})$.
\label{orthogonal}
\end{defi}

We may now define angles by the construction of triangles from three straight lines, two of which are orthogonal (Fig \ref{Fig140}). Since the defintion relies on straightness, angles are well-defined only for individual subjects. Therefore they cannot be used to specify physical law in any fundamental representation of these laws. We skip the definition of the `interval orthogonality' that should be used when the potential knowledge about the spatial positions is incomplete. The reader should have grasped the idea already from the definition of the `interval betweenness' and the `interval straightness'. If these ideas are pursued, we get a set of angles $\Phi$ allowed by our knowledge, rather than the single angle $\phi$ provided by the definition below.

\begin{table}
	\centering
		\begin{tabular}{|l||c|c|c|}
		\hline
		Quality & Range of validity & Applies to & Definition \\
		\hline
		Betweenness & Collective & All attributes & \ref{betweenness} \\
		Succession & Collective & Sequential time & \ref{succession}\\
		closeness & Collective & circular attributes & \ref{closeness}\\
		relative distance & Personal & continuous attributes & \ref{reldistance}\\
		Straightness & Personal & Spatio-temporal positions & \ref{straightness}\\
		Orthogonality & Personal & Spatial positions & \ref{orthogonal}\\
		\hline
		\end{tabular}
	\caption{The epistemic qualities that we use as starting points to define the structure of state space.}
	\label{attributestructure}
\end{table}

\begin{defi}[\textbf{Angles}]
Consider Fig. \ref{Fig140}. The line $L_{1}$ is defined by distinct objects with three spatial positions $(\mathbf{r}_{-1},\mathbf{r}_{0},\mathbf{r}_{1})$ such that $\mathbf{r}_{-1}-\mathbf{r}_{0}-\mathbf{r}_{1}$ as judged by subject $m$, and $\mathbf{r}_{-1}\succ\mathbf{r}_{0}\prec\mathbf{r}_{1}$. The line $L_{2}$ is defined by distinct objects with three spatial positions $(\mathbf{r}_{-2},\mathbf{r}_{0},\mathbf{r}_{2})$ such that $\mathbf{r}_{-2}-\mathbf{r}_{0}-\mathbf{r}_{2}$, $\mathbf{r}_{-2}\succ\mathbf{r}_{0}\prec\mathbf{r}_{2}$ and $(\mathbf{r}_{-1},\mathbf{r}_{0},\mathbf{r}_{1})\bot (\mathbf{r}_{-2},\mathbf{r}_{0},\mathbf{r}_{2})$, as judged by the same subject $m$. The line L3 is defined by three positions $(\mathbf{r}_{a},\mathbf{r}_{b},\mathbf{r}_{c})$ such that $\mathbf{r}_{a}-\mathbf{r}_{b}-\mathbf{r}_c$. This triplet contains exactly one position from L1 and exactly one position from L2. None of the three positions that define L3 is $\mathbf{r}_{0}$. This means that we need six objects to define the triangle enclosed by $L_{1}$, $L_{2}$ and $L_{3}$. Suppose that $S$ is such that the two distances $r_{0a}$ and $r_{0b}$ are defined for subject $m$ according to Definition \ref{distance}. The only knowable difference between the two pools of reference objects (rulers) used to measure $r_{0a}$ and $r_{0b}$ should be the spatial positions of each corresponding object pair. Then we can introduce the angles $\phi_{a}\equiv\tan^{-1}(d_{0b}/d_{0a})$, $\phi_{b}\equiv\tan^{-1}(d_{0a}/d_{0b})$, and $\phi_{0}=\pi$ (or some other constant, depending of the choice of angular unit).
\label{angles}
\end{defi}

\vspace{5mm}
\begin{center}
$\maltese$
\end{center}
\paragraph{}

Our assumptions made it possible to conclude that the physical state $S$ can always be represented by a set of minimal objects when physical law is applied to it (Statement \ref{evred}). To determine the dimension $D[\mathcal{S}]$ of state space, it is therefore sufficient to take into account the maximum number $N$ of minimal objects in the world, the number $N_{Ai}$ of independent internal attributes that are needed to specify each minimal object in an exhaustive set $\{M_{l}\}$, and also the number $N_{Ar}$ of independent relational attributes that is need to specify the relations between $N$ minimal objects. We have argued in Section \ref{time} that relational time $t$ is an attribute that can vary independently among the objects in the state $S(n)$. Therefore there are four independent relational spatio-temporal attributes which have to be specified to be able to deteremine the evolution of $S(n)$. In Section \ref{eveq} we will see that these four attributes are matched by the four relational attributes associated with reciprocal space-time, namely momentum and energy. Thus $N_{Ar}=8N$. All in all we get

\begin{equation}
D[\mathcal{S}]=8N\times N_{Ai}.
\label{statespacedimension}
\end{equation}

In a personal state of knowledge $PK^{m}$, there is a notion of orthogonality (Definition \ref{orthogonal}), so that the values of the $8N$ relational attributes can be arranged in an orthogonal coordinate system. If the universe is infinite, or if the answer to the question whether it is infinite belongs to the unknowable, then we cannot exclude that $N=\infty$, so that we get $D[\mathcal{S}]=\infty$. If it is knowable in principle that the universe is finite, on the other hand, we must have a finite state space dimension according to Eq. [\ref{statespacedimension}], because of the assumption that the depth of knowledge is finite (Assumption \ref{finitedepth}).

\begin{figure}[tp]
\begin{center}
\includegraphics[width=80mm,clip=true]{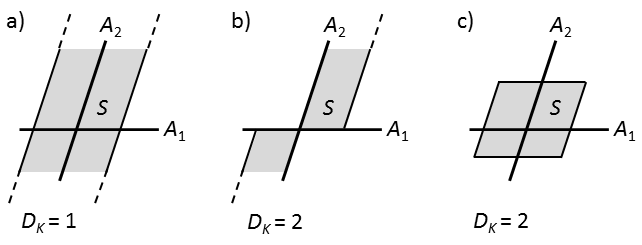}
\end{center}
\caption{The knowledge dimension $D_{K}$ in a two-dimensional state space, spanned by attributes $A_{1}$ and $A_{2}$. There is knowledge associated with $A_{1}$ in all three cases, meaning that some of its values can be excluded, so that $D_{K}\geq 1$. a) There is no knowledge associated with $A_{2}$, so that $D_{K}=1$. b) There is conditional knowledge associated with $A_{2}$, so that $D_{K}=2$. c) There is knowledge associated with $A_{2}$, so that $D_{K}=2$.}
\label{Fig141}
\end{figure}

As a matter of principle, we may ask whether we can exclude the existence of something we do not know anything about at the moment. Potential knowledge can increase in the future. New objects may come into our field of vision, and we may learn more about the objects we already see. Therefore it is questionable whether we should assign a dimension $D[\mathcal{S}]$ to state space \emph{a priori}, regardless whether this dimension is finite or infinite. The epistemic approach is to focus on what we know. In this spirit we introduce the following concept.

\begin{defi}[\textbf{Knowledge dimension of state space}]
The knowledge dimension $D_{K}[\mathcal{S}]$ of state space $\mathcal{S}$ equals the number of distinctions in the state of knowledge $PK$ that corresponds to the physical state $S$. Only distinctions pertaining to directly perceived objects $O$ count. Quasiobjects $\tilde{O}$ are disregarded.
\label{knowdim}
\end{defi}

That $S$ contains $N$ different objects may be said to correspond to $N$ distinctions. Referring to the foam picture of knowledge (Fig. \ref{Fig5b}), the addition of one object $O$ correponds to the distinction of this new object from all the objects that existed previously.

To each known attribute $A_{i}$ of $O$ corresponds the distinction between the values of this attribute which conform with the knowledge about $A_{i}$, and those which do not. Let $A_{i}$ be color. If we see that $O$ is grey, this perception means that we can exclude some other colors. We have made a distinction. We conclulde that we can associate one distinction to each attribute $A_{i}$ of $O$ for which we can exclude some values. Turning the argument around, if we cannot exclude any values of $A_{i}$, then this attribute of $O$ is not perceived at all.

Even so, we may have conditional knowledge pertaining to $A_{i}$. Say that we know that the value of $A_{1}$ is $\upsilon_{11}$ or $\upsilon_{12}$, but that we know nothing about the value of $A_{2}$. Nevertheless, we know that $A_{1}=\upsilon_{11}\Rightarrow A_{2}=\upsilon_{21}$. Already such indirect knowledge about $A_{2}$ lifts it from the darkness of complete ignorance. We may therefore regard a condition involving $A_{i}$ as a distinction associated with $A_{i}$, since a condition picks an implication $A\Rightarrow B$ and drops the alternative $A\not\Rightarrow B$. 

\begin{state}[\textbf{Specification of the knowledge dimension}]
We may write $D_{K}[\mathcal{S}](n)=\prod_{l=1}^{N_{O}(n)}NA_{Ol}(n)$, where $N_{O}(n)$ is the number of distinct objects $O$ contained in $S(n)$, and $NA_{Ol}(n)$ is the number of distinct attributes that can be associated with $O_{l}$ at sequential time $n$. An attribute $A_{i}$ of $O_{l}$ is distinct if and only if at least one of the following two condistions is fulfilled: 1) $A_{i}$ is directly observed by some subject, or 2) there is conditional potential knowledge that involves $A_{i}$.
\label{knowdimnum}
\end{state}

This statement is illustrated in Fig. \ref{Fig141}. Clearly,
 
\begin{equation}
D[\mathcal{S}]\geq D_{K}[\mathcal{S}](n),
\end{equation}
at each sequential time $n$.

We argue that even if $D[\mathcal{S}]$ is undetermined, this does not matter much. In a sense, the important part of $\mathcal{S}$ is spanned by the $D_{K}[\mathcal{S}]$ attributes we know anything about. In fact, we claim that the evolved state $u_{1}S$ is completely determined by the projection of $S$ onto part spanned by the $D_{K}[\mathcal{S}]$ known attributes.

\begin{state}[\textbf{The known parts of state space determine the evolution}]
Let $S_{K}\equiv\Pi_{K}S$ be the projection of the state $S$ onto the subset $\mathcal{S}_{K}$ of state space $\mathcal{S}$ spanned by the $D_{K}[\mathcal{S}]$ known attributes specified in Statement \ref{knowdimnum}. Then $u_{1}\Pi_{K}S=\Pi_{K}u_{1}S$.
\label{knowevo}
\end{state}

This statement is almost self-evident, when you think about it for a minute. We have identified the physical state which we are going to evolve with a state of knowledge. Thus the evolution $u_{1}$ must depend on this knowledge, and on nothing else. We know nothing about the attributes in the part $\mathcal{S}\setminus \mathcal{S}_{K}$ of state space. Therefore $u_{1}$ cannot depend on these parts. It must depend on something known, it cannot depend on nothing.

To be more concrete, consider Fig. \ref{Fig141} again. In panel a) there is no knowledge associated with attribute $A_{2}$. We can specfy the state $S$ completely by saying $\upsilon_{1}^{\min}<\upsilon_{1}<\upsilon_{1}^{\max}$, where $\upsilon_{1}$ is the value of attribute $A_{1}$. This fact is the only thing the evolution $u_{1}$ can depend upon. In relation to Statement \ref{membrane} and Fig. \ref{Fig55} we put forward the idea that we can identify the state boundary $\partial S$ as the essential part of $S$, treat it like a membrane and write $u_{1}S=u_{1}\partial S$. The state boundary in Fig. \ref{Fig141}(a) has two parts $\partial S_{1}$ and $\partial S_{2}$ which are specified by $\partial S_{1}:\upsilon_{1}=\upsilon_{1}^{\min}$ and $\partial S_{2}:\upsilon_{1}=\upsilon_{1}^{\max}$. Again, we see that the values $\upsilon_{2}$ of $A_{2}$ play no role.

The situation is different in Fig. \ref{Fig141}(b). Even if we have no knowledge of the value $\upsilon_{2}$ of $A_{2}$, there is conditional knowledge that relates the values of $A_{1}$ and $A_{2}$. To describe $S$ and its evolution we need to take $A_{2}$ into account. This is the same as to say that $\partial S$ depends on $A_{2}$ as well as on $A_{1}$ The boundary has two horizontal parts $\partial S_{h1}: \upsilon_{2}=0, \upsilon_{1}^{\min}\leq\upsilon_{1}\leq0$ and $\partial S_{h2}: \upsilon_{2}=0, 0\leq\upsilon_{1}\leq\upsilon_{1}^{\max}$.

In Fig. \ref{Fig141}(c), we have direct knowledge about the valus of both $A_{1}$ and $A_{2}$. Clearly, to specify $S$ or $\partial S$ we need to involve both $A_{1}$ and $A_{2}$. Again there are horizontal parts of $\partial S$. In a heuristic notation, we may say that the essential quality of $S$ in panels b) and c) that forces us to take $A_{2}$ into account in the evolution is that there are points on $\partial S$ at which $d(\partial S)/d\upsilon_{2}\neq 0$.

Recall that Definition \ref{knowdim} and Statement \ref{knowdimnum} refer to directly perceived objects, and to those attributes of these objects that we actually observe. This direct knowledge is represented by the physical state $S$. We can model $S$ with deduced quasiobjects like minimal objects, and create a reduced state $\check{S}$. This state will typically have more than $D_{K}[\mathcal{S}]$ degrees of freedom, making it overdetermined. Since $S$ and any properly constructed reduced state $\check{S}$ are equivalent and represent the same potential knowledge, $\check{S}$ cannot contain more than $D_{K}[\mathcal{S}]$ observables upon which its evolution depends. This means, for example, that we should not associate probability amplitudes to the state of each elementary particle in a quantum mechanical state representation in which different such states are superposed. Matters related to this fact are discussed in section \ref{compspec}.

\vspace{5mm}
\begin{center}
$\maltese$
\end{center}
\paragraph{}

The distinction between the full state space $\mathcal{S}$ and the known part $\mathcal{S}_{K}$ of this state space might be interesting from a conceptual point of view, but we will not make much use of it in what follows. In fact, we will not be concerned much about the dimensionality of state space at all, and will continue to represent states $S$ as sets in state space without referring to any axes.

Let us introduce another distinction between two types of state spaces that we will actually make us of. We will use one or the other depending on the problem at hand, depending on what point we wish to make. The distinction is that between the full state space $\mathcal{S}$ and the object state space $\mathcal{S}_{O}$.

We introduced the physical state $S(n)$ as the union of all exact states $Z$ that do not contradict the potential knowledge $PK(n)$ at time $n$ (Definition \ref{statedef}). The state space $\mathcal{S}$ is the set of all possible exact states $Z$ of the world (definition \ref{statespacedef}). We further introduced the object state $S_{O}$ as the subset of $\mathcal{S}$ that consists of the union of all $Z$ that do not contradict the existence of object $O$, with its known attributes (Definition \ref{objectstate}).

We may, however, also consider the space $\mathcal{S}_{O}$ of all possible exact states $Z_{O}$ of an \emph{object} rather than of the entire \emph{world}. (Recall that we do not consider the world to be an object, according to Statement \ref{worldnoobject}.) Then we may embed the object state in this object state space $\mathcal{S}_{O}$. Since this object state $S_{OO}$ is a subset of another space, we have to give it a different name than $S_{O}$. If we describe $S_{O}$ as all exact states of the object $O$ and the surrounding world $\Omega_{O}$ that is consistent with the knowledge about $O$, we may describe $S_{OO}$ as all exact states of $O$ that is consistent with the knowledge about $O$, ignoring the rest of the world.

\begin{figure}[tp]
\begin{center}
\includegraphics[width=80mm,clip=true]{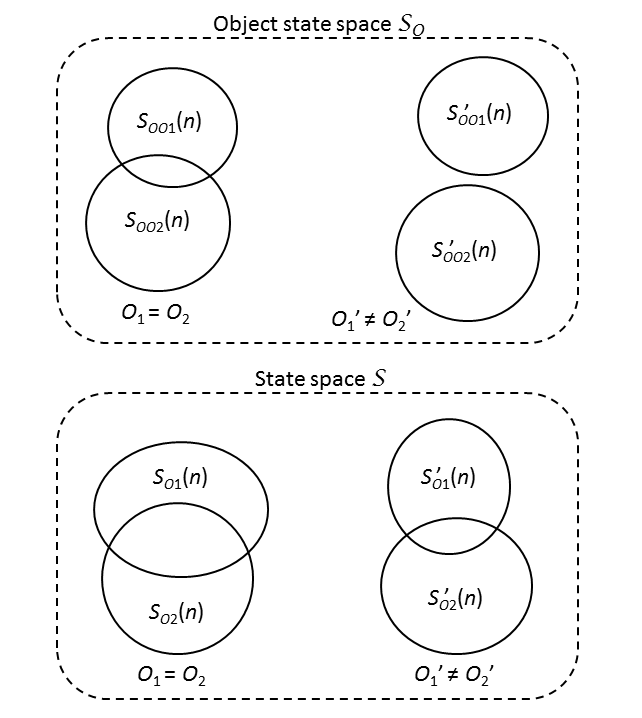}
\end{center}
\caption{The object state space $\mathcal{S}_{\mathcal{O}}$ and the state space $\mathcal{S}$. If two object states $S_{OO1}$ and $S_{OO2}$ overlap in $\mathcal{S}_{\mathcal{O}}$, then they represent the same object. Otherwise they represent different objects. In contrast, states of objects $S_{O}$ perceived at the same time always overlap, even if they are known to be distinct.}
\label{Fig90b}
\end{figure}

What difference does it make? Consider two distinct objects $O_{1}$ and $O_{2}$ that are observed at the same time $n$. Clearly there is a physical state $S(n)$ that is consistent with the simultaneous existence of both these objects, and therefore there are exact states $Z$ that are consistent with these objects. In other words, $S_{O1}\cap S_{O2}\neq\varnothing$. On the other hand, there is no exact object state $Z_{O}$ that is consistent with both these objects. If there were, we would be unable to distinguish them. This means that $S_{OO1}\cap S_{OO2}=\varnothing$. These relations are illustrated in Fig. \ref{Fig90b}. The represention of object states in object state space will be useful when we discuss object division (Section \ref{divideconserve}) and the Pauli exclusion principle (Section \ref{spinstatistics}). In short, it is useful to illustrate discussions about the distinguishability of objects.

\begin{defi}[\textbf{The object state space} $\mathcal{S}_{\mathcal{O}}$]
$\mathcal{S}_{\mathcal{O}}$ is the set of all exact object states $Z_{O}$ allowed by physical law.
\label{objectstatespace}
\end{defi}

\begin{defi}[\textbf{The state} $S_{OO}$ \textbf{in object state space}]
$S_{OO}$ is the set of exact object states $Z_{O}\in \mathcal{S}_{\mathcal{O}}$ that are not excluded by the potential knowledge of the attributes of object $O$.
\label{stateobjectspace}
\end{defi}

\vspace{5mm}
\begin{center}
$\maltese$
\end{center}
\paragraph{}

We have argued that there is no inherent metric in state space. It is nevertheless possible to define a volume measure. It will be essential in our discussions about probability and entropy (Sections \ref{probabilities} and \ref{entropy}).

\begin{defi}[\textbf{Attribute value space} $\mathcal{S}(A,\upsilon)$]
Let $A$ be an independent attribute according to Definition \ref{indattributes}. $\mathcal{S}(A,\upsilon)\subseteq \mathcal{S}$ is the set of exact states $Z$ for which there is at least one object for which $A$ is defined, and for which the value of $A$ is $\upsilon$. The attribute value space in object state space $\mathcal{S}_{O}(A,\upsilon)\subseteq \mathcal{S}_{O}$ is the set of exact object states $Z_{O}$ for which $A$ is defined, and for which the value of $A$ is $\upsilon$.
\label{valuespacedef}
\end{defi}

\begin{defi}[\textbf{State space volume}]
The measure $V[S]\geq 0$ is defined for any state $S\in\mathcal{S}$, and is such that $V[\mathcal{S}(A,\upsilon)]=V[\mathcal{S}(A,\upsilon')]$ for any independent attribute $A$, and any pair of values $(\upsilon,\upsilon')$ of $A$ allowed by physical law. Also, $V[S_{1}\cup S_{2}]=V[S_{1}]+V[S_{2}]-V[S_{1}\cap S_{2}]$ for any two states $S_{1}$ and $S_{2}$. For any exact state $Z$ we have $V[Z]=1$. The volume $V[S_{O}]\geq 0$ of an object state $S_{O}\subseteq \mathcal{S}$ is defined in the same way. The measure $V_{O}[S_{OO}]\geq 0$ is defined analogously for any object state $S_{OO}\subseteq\mathcal{S}_{O}$ in object state space, replacing $Z$ with $Z_{O}$.
\label{voldef}
\end{defi}

The condition $V[S(A,\upsilon)]=V[S(A,\upsilon')]$ can be interpreted as a statement that for each exact state $Z$ for which the value of $A$ is $\upsilon$ there is exactly one exact state $Z'$ for which the value is $\upsilon'$. We thus compare state space volumes in the same way as we compare the sizes of two sets $\Sigma_{1}$ and $\Sigma_{2}$ by putting elements of $\Sigma_{1}$ into one-to-one correspondence with elements of $\Sigma_{2}$. Nevertheless, we avoid reference to the individual elements $Z$ of the space $\mathcal{S}$ in this condition. We do so because they lack physical or epistemic meaning if considered one by one. We know, however, that they are there as a collective, since our knowledge is incomplete (Statement \ref{incompleteknowledge}). Such a statement requires an unknowable completion of knowledge, which is provided by the shadowy exact states $Z$. `We know that there is something about which we cannot know anything.'

Nevertheless, we let $V[Z]=1$ to indicate that a state consistent with at least one exact state has positive volume, and that the exact state is a well-defined concept. As such, it can be assigned a fixed unit volume. But this assignment cannot be used to calculate the volume of actal physical states; the link between $V[Z]$ and $V[S]$ is very weak. Instead we have to compare the volumes of different states. Such a relative volume specification will be sufficient for our purposes. 

Note that we do not say anything in Definition \ref{voldef} about the structure of the set of possible values $\upsilon$. These values may be continuous or discrete (Definitions \ref{discreteattribute} and \ref{continuousattribute}). In the case of continuous values one may object that two sets $\Sigma_{1}$ and $\Sigma_{2}$ with continous elements may have different length, area or volume even if each element $\upsilon$ of $\Sigma_{1}$ can be put into one-to-one correspondence with each element $\upsilon'$ of $\Sigma_{2}$.

An example is the mapping of the entire complex plane onto the Riemann sphere. The area of the first set is infinite while that of the other is finite. However, this statement depends on \emph{a priori} knowledge that allows us to calculate areas without referring to each individual complex element. It depends on a predefined two-dimensional grid of integers that is used as a standard to compare the size of the plane to that of the sphere. Some elements are marked as special, and all the others are interpreted to be placed between these special elements.

However, in our case the value $\upsilon$ is just an arbitrary numerical encoding or labelling of the `epistemic value' of attribute $A$ (section \ref{state}). There is no justification \emph{a priori} to regard some epistemic values as special and to label them with integer numerical values. Therefore this measurement should not appear in any proper definition of state space volume. It should be defined from scratch.

Another way to put it is to say that the definition of a coordinate system in state space requires that we place reference objects with a known distance from each other, and measure other objects in relation to the reference objects. But then we are using object states and observations to define state space (Definition \ref{distance} and Fig. \ref{Fig139}). We have defined it the other way around, and describe all objects and observations with the help of the state space.

\begin{defi}[\textbf{Continuous attribute interval space} $S(A,\Delta\upsilon)$]
Let $A$ be a continuous independent attribute according to Definitions \ref{continuousattribute} and \ref{indattributes}, and let $\Delta\upsilon$ be an interval of values $\upsilon$. Then $S(A,\Delta\upsilon)$ is the set of exact states $Z$ for which there is at least one object for which $A$ is defined, and for which the value of $A$ is an element of $\Delta\upsilon$.
\label{valueintspace}
\end{defi}

The elements of two such sets or intervals $\Delta\upsilon$ and $\Delta\upsilon'$ can be put in one-to-one correspondence to each other. The condition $V[S(A,\upsilon)]=V[S(A,\upsilon')]$ in the definition of state space volume means that the associated volume elements in this one-to-one correspondece are equal. Summing them up, we conclude the following.

\begin{state}[\textbf{Equipartition of continuous attribute value intervals}]
For any continuous independent attribute $A$, and for any two intervals $\Delta\upsilon$ and $\Delta\upsilon'$ according to Definition \ref{valueintspace}, we have $V[S(A,\Delta\upsilon)]=V[S(A,\Delta\upsilon')]$.
\label{equivalue}
\end{state}

Some consequences of these considerations are discussed in section \ref{contextrep} (Statement \ref{equiarea} and Fig. \ref{Fig69f}).

Despite the equality relation in Statement \ref{equivalue}, the introduction of the attribute value space $S(A,\upsilon)$ in Definition \ref{valuespacedef} makes it possible to define the volume so that we respect some intuitive notions about relative size, such as the statement that a two-dimensional plane is `bigger' than a one-dimensional line. This is not possible if we compare the size of sets using just one-to-one correspondence of elements. To illustrate this point, consider a state space spanned by the two attributes $A_{1}$ and $A_{2}$. The points on the $A_{1}$-axis may be said to correspond to the set $S(A_{2},\upsilon_{2}^{0})$, whereas the entire state space can be identified with $\bigcup_{\Sigma(\upsilon_{2})}S(A_{2},\upsilon_{2})$, where $\Sigma(\upsilon_{2})$ is the set of all possible values of $A_{2}$, so that  $\upsilon_{2}^{0}\in\Sigma(\upsilon_{2})$. From Definition \ref{voldef} we get $V[\bigcup_{\Sigma(\upsilon_{2})}S(A_{2},\upsilon_{2})]=\sum_{\Sigma(\upsilon_{2})}V[S(A_{2},\upsilon_{2})]>V[S(A_{2},\upsilon_{2}^{0})]$ whenever $A_{2}$ can take more than one value.

\section{Properties and property spaces}
\label{propspaces}

\begin{figure}[tp]
\begin{center}
\includegraphics[width=80mm,clip=true]{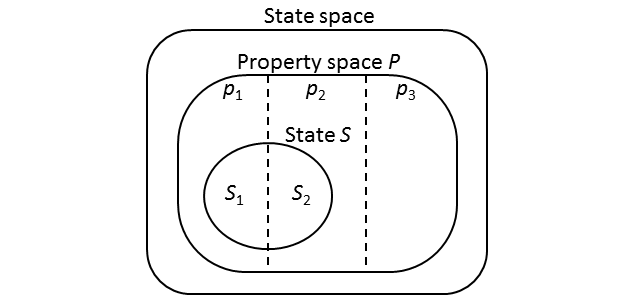}
\end{center}
\caption{The property space $\mathbf{\mathcal{P}}$ of a property $P$ is the union of all states for which there are objects such that $P$ is defined. $\mathbf{\mathcal{P}}=\bigcup_{k}\mathbf{\mathcal{P}}_{j}$, where $\mathbf{\mathcal{P}}_{k}$ is the union of states for which $P$ has value $p_{j}$.}
\label{Fig33}
\end{figure}

Let us return to the concept of altervatives, discussed at some length in Section \ref{law}. Consider the property $P$ that defines the set of alternatives $\{S_{j}\}$. This property defines an abstract set $\mathcal{P}$ (Fig. \ref{Fig33}).

\begin{defi}[\textbf{Property space}]
The property space $\mathcal{P}$ is the union of all exact states $Z$ for which there are objects such that the value $p$ of the property $P$ can be defined. Each possible value $p_{j}$ defines a set $\mathcal{P}_{j}$ as the union of those $Z\in \mathcal{P}$ for which the value of $P$ is $p_{j}$. We get $\mathcal{P}=\bigcup_{j}\mathcal{P}_{j}$.
\label{propspace}
\end{defi}

A property $P$ is a statement about attributes of objects. It may concern one attribute of one object, or several attributes of several objects. In general,

\begin{equation}
p=f(\{\upsilon_{il}\}),
\label{pstatement}
\end{equation}
where $\upsilon_{il}$ is the value of attribute $A_{i}$ of object $O_{l}$. Clearly, any attribute is a property, but the opposite is not necessarily true. By definition, the values $\upsilon_{i}$ of a given attribute $A_{i}$ are always possible to order. In contrast, the values $p_{j}$ of a property $P$ cannot always be ordered. Ordering is impossible when $p_{j}$ is a function of attribute values $\upsilon_{il}$ and $\upsilon_{i'l}$ belonging to different attributes $A_{i}$ and $A_{i'}$. For example, the color of the tail feathers of a bird can be ordered according to the spectrum. However, letting $P$ represent bird species, the different species $p_{j}$ cannot be ordered, since the classification depends on other attributes than feather colors.

The following statement is a reformulation of the incompleteness of knowledge (Statement \ref{incompleteknowledge}):

\begin{state}[\textbf{Simultaneously knowable properties}]
The values of all properties cannot be known at the same time.
\label{simknowprop}
\end{state}

In other words, there are pairs of properties such that knowledge of the value of one makes it impossible to know the value of the other at the same time.

If $P$ is the species of birds, then $\mathcal{P}$ is the union of all $Z$ for which there is a bird. We have $S\subset\mathcal{P}$ for any actual state $S$ with birds, since in any situation we know more than just `there is a bird'. We know the landscape in which we see the bird, we have self-awareness, and so on.

Referring to Fig. \ref{Fig33}, we may let $p_{1}$ represent a golden eagle, $p_{2}$ represent a raptor that is not a golden eagle, and $p_{3}$ a bird that is not a raptor. If we are sure that we see some kind of raptor in the sky, then $\mathcal{P}_{3}\cap S =\varnothing$.

To give a physical example, consider the spin quantun number of electrons. For the total spin there is only one possible value $1/2$. $\mathcal{P}$ is the union of all $Z$ for which there is at least one electron, and $\mathcal{P}_{1}=\mathcal{P}$ is the set of all states where $P_{1}=1/2$. Obviously, in this case, $\mathcal{P}$ is almost as big as state space $\mathcal{S}$ itself. If it is assumed that electrons are necessary components in the body of any subject, $\mathcal{P}=\mathcal{S}$, since subjects are necessary ingredients in any state of knowledge, and thus in any physical state (Definition \ref{firststatedef}). 

For the spin component in a given direction, $\mathcal{P}$ is the union of all $Z$ which contain electrons and one apparatus that defines the direction, and is capable of measuring the spin in this direction. We have $p_{1}=1/2$, $p_{2}=-1/2$, and $\mathcal{P}=\mathcal{P}_{1}\cup\mathcal{P}_{2}$. The word `apparatus' must be understood in a general sense, as a part of the physical state that makes it possible to deduce the spin direction from the corresponding state of knowledge as an attribute of a quasiobject.

The property space $\mathcal{P}$ is a function of the property $P$ only; it does not depend on the present state, on time, or on anything else. All these details are `summed over', or rather `taken the union over'. Note also that for any well-defined property $P$, it is possible to define the set $\mathcal{P}$.

\section{Present and future alternatives}
\label{presfut}

\begin{figure}[tp]
\begin{center}
\includegraphics[width=80mm,clip=true]{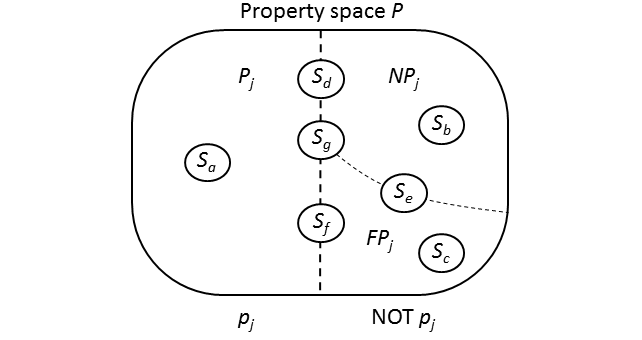}
\end{center}
\caption{Present and possible future properties. Each property value $p_{j}$ spans the property space $\mathcal{P}$ in itself if we replace all other values by the complement `not $p_{j}$'. The region $\mathcal{FP}_{j}$ is the union of those states $S$ which do not have property value $p_{j}$ at present, but will have it at some future time. The region $\mathcal{NP}_{j}$ is the union of those states for which the property value is not, and will never be $p_{j}$. The state $S$ may relate to the three regions in seven different ways.}
\label{Fig33b}
\end{figure}

We may use the concept of properties to describe and classify more clearly the division of the state $S$ into a set of alternatives $\{S_{j}\}$, as expressed in Eq. [\ref{division}]. One should, for instance, distinguish between states that may have a property now, at time $n$, say with value $p_{j}$, and states that may have it if we investigate the matter at some future time $n+m$.

We may divide property space $\mathcal{P}$ into two parts: $\mathcal{P}_{j}$ and $\mathcal{P}_{j}^{c}$, where $\mathcal{P}_{j}^{c}$ is the region in which the value of property $P$ is not $p_{j}$ (Fig. \ref{Fig33b}). For states $S$ that are embedded in one of these regions, we know for sure whether the property value at present is $p_{j}$ or not. For states that overlap both regions, we do not know.

We may define the region $\mathcal{FP}_{j}$ as the union $\bigcup_{j}S_{j}(n)$ of those states $S_{j}(n)\subset\mathcal{P}_{j}^{c}$ for which there is a time $n+m$ such that $u_{m}S_{j}(n)\subset\mathcal{P}_{j}$. That is, $\mathcal{FP}_{j}$ is the region of property space consisting of states that we know will have property value $p_{j}$ at some future time, but which do not have it now. Further, we may define $\mathcal{NP}_{j}=\mathcal{P}_{j}^{c}/\mathcal{FP}_{j}$ as the union of those states for which we know that the property value is not $p_{1}$ now, and that it will never be.

A state $S$ may overlap these three regions in various ways, as expressed in Fig. \ref{Fig33b}. The state $S_{b}$ will, by definition, be a subset of $\mathcal{NP}_{j}$ forever. The evolved states $u_{m}S_{d}$, $u_{m}S_{e}$ and $u_{m}S_{g}$ will forever (for any $m\geq 1$) at least partially belong to $\mathcal{NP}_{j}$. However, physical law may or may not allow that there is a time $n+m$ such that $S(n+m)\subset\mathcal{P}_{j}$ given that $S(n)$ belongs to one of the classes $S_{d}$, $S_{e}$ or $S_{g}$. If it is allowed, then property $p_{j}$ is \emph{realizable}.

\begin{defi}[\textbf{Realizable property}]
The property value $p_{j}$ is realizable if and only if $S(n)$ belongs to class $S_{d}$, $S_{e}$ or $S_{g}$ (Fig. \ref{Fig33b}), and physical law allows that $S(n+m)\subset\mathcal{P}_{j}$ for some $m\geq 1$.
\label{realizableprop}
\end{defi}

In this definition, we exclude the trivial cases when the state already have property value $p_{j}$, or will turn out to have it in the future with necessity.

Let

\begin{equation}
\tilde{\mathcal{P}}_{j}=\mathcal{P}_{j}\cup\mathcal{FP}_{j}
\label{pfp}
\end{equation}
be the region in which states have property value $p_{j}$ now, or will have it at some future time. Even if two regions $\mathcal{P}_{j}$ and $\mathcal{P}_{j'}$ never overlap, two regions $\tilde{\mathcal{P}}_{j}$ and $\tilde{\mathcal{P}}_{j'}$ may or may not overlap. If they do not overlap, the property values are mutually exclusive. By this we mean not only that the two property values cannot occur at the same time - they cannot occur in succession either.

\begin{defi}[\textbf{Mutually exclusive properties}]
Two property values $p_{j}$ and $p_{j'}$ are mutually exclusive if and only if $\tilde{\mathcal{P}}_{j}\cap \tilde{\mathcal{P}}_{j'}=\varnothing$.
\label{mexprop}
\end{defi}

Note again that even if two property values are \emph{not} mutually exclusive, we still have $\mathcal{P}_{j}\cap \mathcal{P}_{j'}=\varnothing$.

To examplify, the property values $p_{1}$ and $p_{2}$ that a given particle in a given double slit experiment passes slit 1 and 2, respectively, are mutually exclusive. The setup is such that if the particle passes one of the slits, it cannot pass the other at a later time. In contrast, if $p_{1}$ corresponds to the fact that the distance between two objects is $x_{1}$ and $p_{2}$ corresponds to the fact the distance is $x_{2}$, the property values can occur one after the other if the objects are moving, even if they, of course, cannot occur simultaneously.

\begin{defi}[\textbf{Present alternative} $S_{j}$]
Let $S_{j}\equiv S(n)\cap\mathcal{P}_{j}$ and $u_{1}S_{j}\equiv u_{1}S(n)\cap\mathcal{P}_{j}$. The alternative $S_{j}$ is a present alternative if and only if $u_{1}S_{j}\neq\varnothing$ and $u_{1}S_{j}\subset u_{1}S(n)$. Further, physical law must allow that $S(n+1)\subseteq u_{1}S_{j}$.
\label{presentalt}
\end{defi}

A present alternative can be immediately realized. This means that the change of knowledge that corresponds to the realization that alternative $S_{j}$ is true may define the next temporal update $n\rightarrow n+1$. Two present alternatives $S_{j}$ and $S_{j'}$ may or may not correspond to mutually exclusive properties. Even if they are not mutually exclusive, we still have $u_{1}S_{j}\cap u_{1}S_{j'}=\varnothing$ from the invertibility of physical law (Assumption \ref{uniqueu1}). A present alternative always corresponds to a realizable property.

The condition that $u_{1}S_{j}$ is a proper subset of $u_{1}S(n)$ means that it should be uncertain whether alternative $S_{j}$ is or will ever become realized. We exclude the trivial cases in Fig. \ref{Fig33b} where $S(n)$ belongs to class $S_{a}$, so that the corresponding property is already realized, or to classes $S_{c}$ or $S_{f}$, so that the property is or will be realized by necessity.

\begin{defi}[\textbf{A complete set of present alternatives}]
A set of present alternatives $\{S_{j}\}$ is complete if and only if $S(n)=\bigcup_{j}S_{j}$.
\label{setpresentalt}
\end{defi}

\begin{figure}[tp]
\begin{center}
\includegraphics[width=80mm,clip=true]{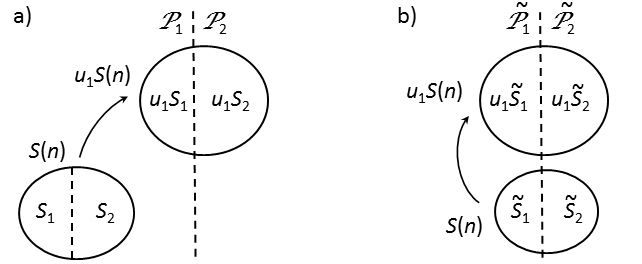}
\end{center}
\caption{Present and future alternatives. a) $\{S_{1},S_{2}\}$ is a complete set of present alternatives at time $n$. At the next moment $n+1$, the state may turn out to have property value $p_{1}$ or $p_{2}$. b) $\{\tilde{S}_{1},\tilde{S}_{2}\}$ is a complete set of future alternatives. At some time $n+m$, the state may turn out to have property value $p_{1}$ or $p_{2}$. When one of the property values is realized, the other property value is excluded forever if they are mutually exclusive. By definition of the sets $\tilde{\mathcal{P}}_{j}$, the partition $\{\tilde{S}_{j}\}$ of the state $S$ is then `locked onto' these sets. In contrast, the possible property values of the present alternatives may `float' from value to value, like the position of a ball rolling on the ground.}
\label{Fig33c}
\end{figure}

For example, if you are about to step up on the bathroom scale, a subset of the range of weights that can be displayed make up a complete set of present alternatives (if you are not too heavy). The alternatives are not mutually exclusive, however, since you can gain or lose weight until the next time you step up on the scale. It follows from the definition of a present alternative that a complete set of such alternatives always contains more than one element (Fig. \ref{Fig33c}a).

Let us turn to alternatives that may be realized further into the future.

\begin{defi}[\textbf{Future alternative} $\tilde{S}_{j}$]
Let $\tilde{S}_{j}=S(n)\cap\tilde{\mathcal{P}}_{j}$ and $u_{1}\tilde{S}_{j}\equiv u_{1}S(n)\cap\tilde{\mathcal{P}}_{j}$. Then $\tilde{S}_{j}$ is a future alternative if and only if $\tilde{S}_{j}\neq\varnothing$ and $\tilde{S}_{j}\subset S(n)$. Further, physical law must allow that $S(n+m)\subseteq \tilde{S}_{j}$ for some $m\geq 2$.
\label{futurealt}
\end{defi}

Just as a present alternative, a future alternative corresponds to a realizable property. We want to define a complete set of future alternatives analogous to a complete set of present alternatives. To do so, we must require that the sets $\tilde{S}_{j}$ do not overlap. Otherwise the quality of completeness would be hard to use effectively.

If the corresponding property values $p_{j}$ are mutually exclusive (Definition \ref{mexprop}), the future alternatives $\tilde{S}_{j}$ are automatically disjoint. We may, however, define disjoint alternatives $\tilde{S}_{j}$ even if this is not the case. Consider again the property \emph{distance}. Two different distances may be found in succession, so that the property values are not mutually exclusive. But in a physical setup prepared to measure the distance between the objects at a given time, the possible outcomes nevertheless define disjoint future alternatives, defined as `the first outcome of the measurement that we have prepared'.

\begin{defi}[\textbf{A complete set of future alternatives}]
A set of future alternatives $\{\tilde{S}_{j}\}$ is complete if and only if $S(n)=\bigcup_{j}\tilde{S}_{j}$ and $\tilde{S}_{j}\cap \tilde{S}_{j'}=\varnothing$ for all $j\neq j'$.
\label{setfuturealt}
\end{defi}

Just as for present alternatives, a complete set of future alternatives always contains more than one element (Fig. \ref{Fig33c}b).

A future alternative may be said to be invariant in time. We have $\tilde{S}_{j}\subseteq\tilde{\mathcal{P}}_{j}$. By definition of the region $\tilde{\mathcal{P}}_{j}$ we also have $u_{1}\tilde{S}_{j}\subseteq\tilde{\mathcal{P}}_{j}$. If no state reduction occurs at time $n+1$, meaning that $S(n+1)=u_{1}S(n)$, then the set $\{u_{1}\tilde{S}_{j}\}$ is also a complete set of future alternatives.

\begin{state}[\textbf{The regions} $\tilde{\mathcal{P}}_{j}$ \textbf{are invariant under evolution} $u_{1}$]
If $\tilde{S}_{j}\subseteq\tilde{\mathcal{P}}_{j}$, then $u_{1}\tilde{S}_{j}\subseteq\tilde{\mathcal{P}}_{j}$ .
\label{invariantregions}
\end{state}

Of course, when the value $p_{j}$ is actually observed there is a state reduction, and after that the regions $\tilde{\mathcal{P}}_{j}$ do not need to be invariant in time.

A complete set of present or future alternative belongs to knowability level 2 or 3 in the table given in Section \ref{law}. Two alternatives $S_{j}$ and $S_{j'}$ (or $\tilde{S}_{j}$ and $\tilde{S}_{j'}$) in a complete set are subjectively distinguishable by definition. More than that, once one alternative is realized, all future states are subjectively distinguishable from those that would follow if another alternative was realized. This follows from the invertibility of physical law (Assumption \ref{uniqueu1}), which means that $S_{j}\cap  S_{j'}=\varnothing\Rightarrow u_{m} S_{j}\cap u_{m} S_{j'}=\varnothing$.

\begin{state}[\textbf{A choice between alternatives is definitive}]
When one present or future alternative is realized, each future state is subjectively distinct from all future states that would have followed if another alternative from the same complete set was realized.
\label{definitechoice}
\end{state}

The main concepts introduced in this section are present and future alternatives. To summarize their meaning in plain language, present alternatives are things that `may be about to happen', whereas future alternatives are possible outcomes of an investigation.

\section{Realization of alternatives}
\label{graphical}

By definition, present and future alternatives corresponds to the observation of a realizable property (Definition \ref{realizableprop}). It is useful to be able to refer to such alternatives as realizable.

\begin{defi}[\textbf{Realizable alternative}]
An alternative is realizable if and only if it corresponds to a present alternative $S_{j}$ (Definition \ref{presentalt}) or a future alternative $\tilde{S}_{j}$ (Definition \ref{futurealt}).
\label{realizablealt}
\end{defi}

Note that alternatives can be realizable only in an indeterministic world with incomplete knowledge. An element of contingency is presupposed in the definition of present and future alternatives.

In the defintions of present and future alternatives, we required that it is possible that $S(n+1)\subseteq S_{j}$ and $S(n+m)\subseteq \tilde{S}_{j}$, respectively.  In practice, we must always have equalitites in these relations. To say that $S(n+1)\subset S_{j}$ is the same as saying that we gain some more knowledge at time $n+1$ than the knowledge that property $P$ has value $p_{j}$. This possibility is excluded by definition, since each temporal update $n\rightarrow n+1$ is associated with a given piece of new knowledge, a given change of perception.

\begin{state}[\textbf{The size of the physical state when an alternative is realized}]
Suppose that a present alternative $S_{j}$ is realized at time $n+1$. Then $S(n+1)=u_{1}S_{j}$. Suppose that a future alternative $\tilde{S}_{j}$ is realized at time $n+m$. Then $S(n+m)=u_{m}\tilde{S}_{j}$.
\label{realizedsize}
\end{state}

Before a realizable alternative $S_{j}$ or $\tilde{S}_{j}$ is actually realized, its evolution is defined according to the prescriptions in Definitions \ref{presentalt} and \ref{presentalt}. If no realizable alternative pertaining to property $P$ is ever realized, its evolution is defined forever in the same way. If value $p_{j}$ is finally observed, the alternative becomes identical to the physical state, whose evolution is always defined. If another value $p_{j}'$ of $P$ is finally observed, the alternative ceases to exist, so that the question whether its evolution is defined is no longer meaningful. We conclude that the evolution operator $u_{1}$ can be applied an arbitrary number of times to any realizable alternative.

Suppose that a present alternative $S_{j}$ is realized at time $n+1$. Then we can track-back its origin as an alternative, since the invertibility of the evolution gives physical meaning to the expression $S_{j}=u_{1}^{-1}S(n+1)$. In the same way, if a future alternative is defined at time $n$ and finally realized at time $n+m$, then we can reconstruct its entire history in a physical sense according to $\tilde{S}_{j}=u_{m}^{-1}S(n+1)$, $u_{1}\tilde{S}_{j}=u_{m-1}^{-1}S(n+1)$, $u_{2}\tilde{S}_{j}=u_{m-2}^{-1}S(n+1)$, and so on.

We may say that it is meaningful to talk about the evolution of a realizable alternative since it has the potential to be physically track-backed (if the alternative is realized). On the other hand, if an alternative is \emph{not} realizable, then such track-back can never be performed. The evolution of such an alternative cannot be given any physical meaning.

\begin{state}[\textbf{Physical states and realizable alternatives are the domain of the evolution operator}]
Let $\Sigma_{j}\equiv S(n)\cap\mathcal{P}_{j}$ or $\Sigma_{j}\equiv S(n)\cap\tilde{\mathcal{P}}_{j}$. The expression $u_{m}\Sigma_{j}$ is defined for any $m\geq 1$ if and only if $\Sigma_{j}=S(n)$ or $\Sigma_{j}$ corresponds to a realizable alternative $S_{j}$ or $\tilde{S}_{j}$.
\label{realizabledomain}
\end{state}

\begin{figure}[tp]
\begin{center}
\includegraphics[width=80mm,clip=true]{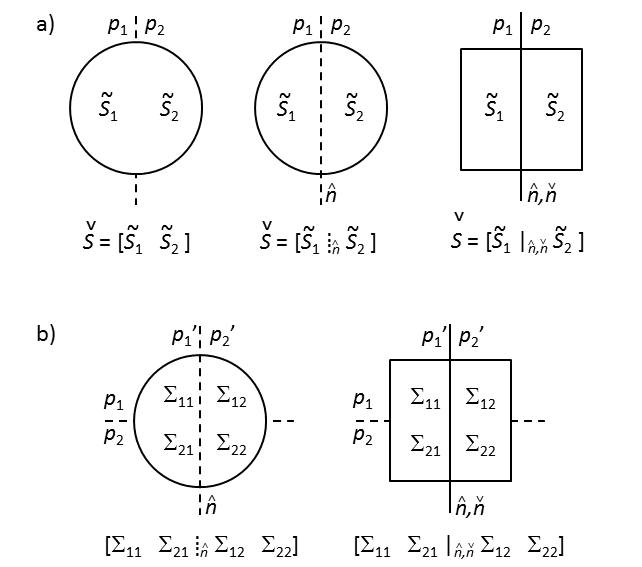}
\end{center}
\caption{(a) Alternatives $\tilde{S}_{1}$ and $\tilde{S}_{2}$ with three levels of knowability. The alternatives correspond to two values $p_{1}$ and $p_{2}$ of property $P$. No vertical line means that the alternatives can never be realized (level 1), a dashed line means that they may be realized (level 2), and a solid line means that one of them will be realized (level 3). States at level 3 are given square shape to emphasize that they are `prepared'. (b) Combinations of alternatives that correspond to two different properties $P$ and $P'$, meaning that these properties are observed in succession. Compare Fig. \ref{Fig32b}}
\label{Fig34}
\end{figure}

In the remainder of this section, we will get aquainted with one graphical and one symbolic way to illustrate realizable alternatives at different knowability levels (Table \ref{levels}), and what happens when they are actually realized. We have already discussed the realization of one alternative in a single complete set of such alternatives, corresponding to the observation of a single property $P$. It is more interesting to consider the sequential observation of several properties $P,P',P'',\ldots$. To get the main points, we discuss the simplest case where we observe in succession two properties $P$ and $P'$ with two values each. We denote these values $(p_{1},p_{2})$ and $(p_{1}',p_{2}')$, respectively. Curious situations may occur when there is \emph{a priori} conditional knowledge that relates the values of $P$ and $P'$.

Figure \ref{Fig34} develops the graphic notation introduced in Fig. \ref{Fig32b}, and also introduces a corresponding symbolic notation. Alternatives at level 3 are divided by a solid line, level 2 alternatives are divided by a dashed line, whereas alternatives at level 1 are not divided by any line. States at level 3 are given square shape to express that they are `prepared' or `designed'. Despite the choice of words, this does not have to mean anything teleological. At knowability levels 2 and 3, the lower time limit $\hat{n}$ for decision is marked, and at level 3 the upper time limit $\check{n}$ is also marked.

The sets $\Sigma_{ij}$ represent the regions in state space in which $P$ has value $p_{i}$ and $P'$ has value $p_{j}$. In the examples shown in Fig. \ref{Fig34}, these sets are not realizable alternative since property $P$ is never observed, so that the alternatives that correpond to the values $p_{1}$ and $p_{2}$ have knowability level 1.

To make things simpler, we focus on complete sets of future alternatives $\tilde{S}_{j}$ that extend into one or more property value spaces $\tilde{\mathcal{P}}_{j}$ (Definition \ref{invariantregions} and Eq. [\ref{pfp}]), which corresponds to mutually exclusive property values (Definition \ref{mexprop}). This is convenient since it makes the description time independent, as long as no observation is made. The reason is the invariance under evolution that is expressed in statement \ref{invariantregions}. This invariance prevents the states and alternatives from floating around in state space. Instead they are `nailed' by the property value division lines $\partial\tilde{\mathcal{P}}_{j}$.

Let us turn the attention to the corresponding symbolic notation. The right state in Fig. \ref{Fig34}(b) corresponds to an experiment in which property $P$ is defined but unknowable, and in which property $P'$ will be observed with certainty sooner or later. We write

\begin{equation}
\check{S}=\left[\begin{array}{cc}
\Sigma_{11} & \Sigma_{21}
\end{array}\right|_{\hat{n}',\check{n}'}
\left.\begin{array}{cc}
\Sigma_{12} & \Sigma_{22}
\end{array}\right].
\label{binaryex}
\end{equation}

\begin{figure}[tp]
\begin{center}
\includegraphics[width=80mm,clip=true]{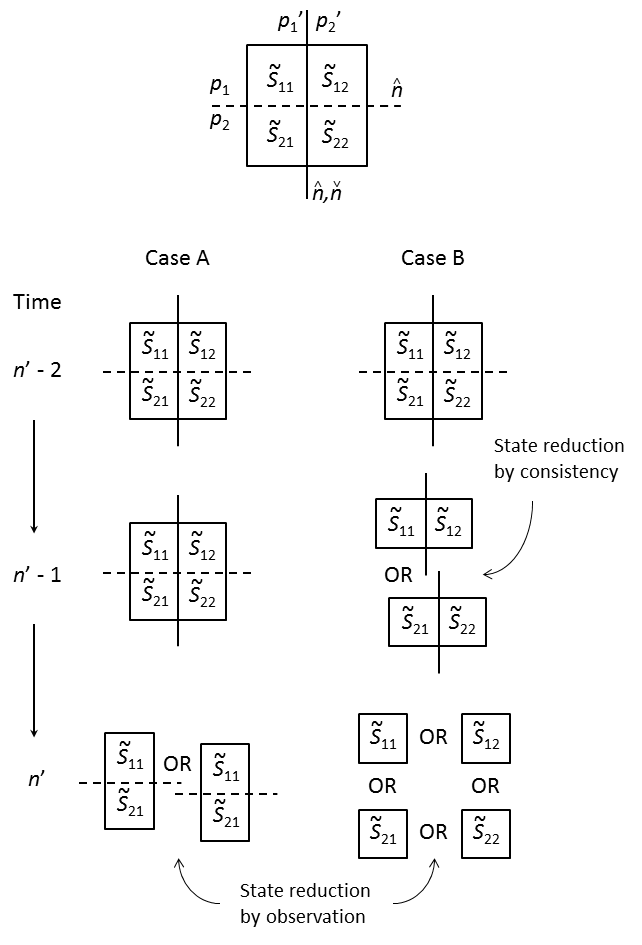}
\end{center}
\caption{An experiment in which two properties $P$ and $P'$ are observed, and we have conditional knowledge so that the value of $P'$ depends on the value of $P$. When the alternative values of property $P$ are at level 2, and the values of property $P'$ are at level 3, the evolution of the state depends on whether $P$ refers to a time before $P' $ is observed. If no (Case A), the state is reduced when the value of $P'$ is observed at time $n'$. If yes (Case B), the state is further reduced at an earlier time. This takes place either by direct observation of the value of $P$, or, if no observation is made, at time $n'-1$ by the requirement of epistemic consistency.}
\label{Fig34b}
\end{figure}

Assume instead that $P$ may or may not be observed in the experiment, so that the alternatives that correspond to the values $p_{1}$ and $p_{2}$ have knowability level 2. This situation is shown in the left panel of Fig. \ref{Fig34b}). Assume further that there is conditional knowledge such that the value of $P'$ depends to some degree on the value of $P$. Either knowledge of the value of $P$ excludes some values of $P'$, or knowledge of $P$ affects the probability to observe different values of $P'$. This situation may occur, for example, if the $P$ is associated with an event that is known to precede the event that is associated to $P$, and these two events are causally connected (have a time-like separation).

Let $\hat{n}$ be the lower time limit for the observation of $P$, and let $\hat{n}'$ and $\check{n}'$ be the lower and upper time limits for the observation of $P'$, respectively. Also, let $n'$ be the time when the value of property $P'$ is actually observed ($\hat{n}'\leq n'\leq\check{n}'$). We must treat the cases $\hat{n}\geq n'$ and $\hat{n}<n'$ separately. Let us call them cases A and B. In case A, it is impossible to know the value $p_{j}$ of $P$ before the value of $P'$ has been observed at time $n'$. Thus the influence of $P$ on $P'$ cannot make itself felt, so that this case is analogous to that represented in Eq. \ref{binaryex}. For $n<n'$ the state is

\begin{equation}
\check{S}(n)=\left[\begin{array}{ccc}
\tilde{S}_{11}(n) & \vdots_{\hat{n}} & \tilde{S}_{21}(n)
\end{array}\right|_{\hat{n}',\check{n}'}
\left.\begin{array}{ccc}
\tilde{S}_{12}(n) & \vdots_{\hat{n}} & \tilde{S}_{22}(n)
\end{array}\right].
\label{doublestate}
\end{equation}

At time $n'$ the state reduces to

\begin{equation}\begin{array}{clll}
u_{1}\check{S}(n'-1)\\
\downarrow\\
\check{S}(n') & = & \left[\begin{array}{ccc}
\tilde{S}_{11}(n') & \vdots_{\hat{n}} & \tilde{S}_{21}(n')
\end{array}\right] & \mathrm{OR}\\\\ 
& & \left[\begin{array}{ccc}
\tilde{S}_{12}(n') & \vdots_{\hat{n}} & \tilde{S}_{22}(n')
\end{array}\right],
\end{array}
\label{singlestates1}
\end{equation}
where $\check{S}(n'-1)$ has the form (\ref{doublestate}).

In case B, it \emph{may} be possible to learn the value $p_{j}$ before time $n'$. The time when we \emph{actually} get to know $p_{j}$ does not matter; it may be before or after time $n'$. If knowledge is gained before time $n'$, the state is reduced at this time to one of the halves shown in Fig. \ref{Fig34b}, corresponding to $p_{1}$ or $p_{2}$. However, even if no such observation of $P$ is made, epistemic consistency (Assumptions \ref{epconsistency}, \ref{epconsistency1} and \ref{epconsistency2}) requires that $\check{S}$ is nevertheless reduced at time $n'-1$ from a state of the form (\ref{doublestate}) to a state of the following form.

\begin{equation}\begin{array}{clll}
u_{1}\check{S}(n'-2)\\
\downarrow\\
\check{S}(n'-1) & = & \left[\begin{array}{c}
\tilde{S}_{11}(n'-1)
\end{array}\right|_{\hat{n}',\check{n}'}
\left.\begin{array}{c}
\tilde{S}_{12}(n'-1)
\end{array}\right] & \mathrm{OR}\\\\ 
& & \left[\begin{array}{c}
\tilde{S}_{21}(n'-1)
\end{array}\right|_{\hat{n}',\check{n}'}
\left.\begin{array}{c}
\tilde{S}_{22}(n'-1)
\end{array}\right].
\end{array}
\label{singlestates}
\end{equation}
Otherwise, if knowledge about $p_{j}$ is gained at some later time $n''>n'$, it would be possible to deduce at time $n''$ that $\check{S}(n'-1)$ had the above form. We would have a knowable contradiction, since the evolution to time $n'$ from this reduced state may be different than it actually was.

In the absence of direct observation of the value $p_{j}$ before time $n'$, there is no epistemic reason to say that the reduction takes place earlier than at time $n'-1$. No observation and no consistency argument will ever be able to tell whether it has happened earlier. Therefore such a `reduction by consistency' can be defined to occur at time $n'-1$.

At time $n'$, the state (\ref{singlestates}) is reduced further, to

\begin{equation}\begin{array}{clllll}
u_{1}\check{S}(n'-1)\\
\downarrow\\
\check{S}(n') & = & \left[\tilde{S}_{11}(n')\right] & \mathrm{OR} &\left[\tilde{S}_{12}(n')\right] & \mathrm{OR}\\
&&&&&\\
& & \left[\tilde{S}_{21}(n')\right] & \mathrm{OR} & \left[\tilde{S}_{22}(n')\right].
\end{array}
\label{choppedstates}
\end{equation}

\begin{figure}[tp]
\begin{center}
\includegraphics[width=80mm,clip=true]{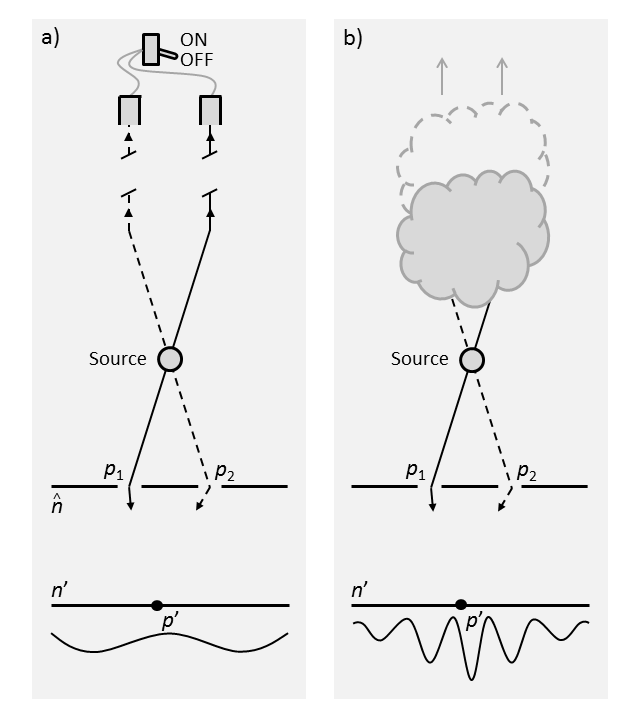}
\end{center}
\caption{Gedankenexperiment to illustrate Case B in Fig. \ref{Fig34b}. The source emits two particles in opposite directions. One of these particles passes a slit. Property $P$ is the slit the particle passes. Property $P'$ is the position at the screen hit by the particle that passes a slit. a) If there are two detectors that can detect the backward-moving twin particle, the state reduces at time $n'-1$, just before the particle hits the screen at time $n'$. It does not matter whether the detectors are turned on or off. b) If path information is erased in a `cloud', the values of $P$ are degraded to knowability level 1, and the evolution corresponds to that in Case A in Fig. \ref{Fig34b}. The cloud has to be close enough, so that information is erased before the particle hits the screen.}
\label{Fig36}
\end{figure}

To make case B more concrete, let us consider an example. Imagine a double-slit experiment where the source emits two (entangled) particles in opposite directions (Fig. \ref{Fig36}). Property $P$ represents the slit the particle has passed. By definition, the property values $p_{1}$ and $p_{2}$ cannot be known before the time $\hat{n}$ of passage. Property $P'$ is the position on the detector screen behind the slits. The values of $P'$ cannot be known before a time $\hat{n}'$ given by $\hat{n}$ plus the shortest distance from the slits to the screen divided by the fastest possible speed of the particle. Assume that the particle hits the screen at time $n'\geq\hat{n}'$.

Imagine first that the twin of the particle that passes a slit is deflected into a detector [Fig. \ref{Fig36}(a)]. There is one detector for each slit, so that detection of the twin particle determines which slit the original particle passed. The two detectors may be placed far away, so that the twin particle reaches them long after the original particle hits the screen. Regardless whether the two detectors are turned on or off at the time the experiment starts, path information may or may not be gained, since the status of the detectors can be changed afterwards. Therefore the alternatives corresponding to property values $p_{1}$ or $p_{2}$ are at knowability level 2. Regardless the time of detection of the twin particle, the path information refers to a time \emph{before} the original particle hits the screen. Thus, to respect epistemic consistency, Nature must choose a state corresponding to the passage through one slit or the other already at time $n'-1$, even if no observation of the path is made at this time, or even if no observation will ever be made. If Nature would not make such a choice, we would have a superposition of the two alternative values of $P$, which would give rise to an interference pattern on the screen. This pattern would contradict the knowledge about the value of $P$ that \emph{might} be gained after the interference pattern is observed.

In contrast, if it is known that path information is erased before time $n'$, there is no risk of inconsistency. Such an eraser is represented in Fig. \ref{Fig36}(b) as a cloud in which the twin particle disappears before the original particle passes through one of the slits. The alternatives corresponding to property values $p_{1}$ or $p_{2}$ are degraded to level 1.

However, if the eraser is successively moved away, the possibility to gain path information finally survives after time $n'$, and we are back in a situation where Nature has to choose path at time $n'-1$ to make sure that no contradiction will occur. Nature is about to paint itself into a logical corner, so to say, and must jump out of it to preserve consistency.

This choice corresponds to a sudden increase of potential knowledge at time $n'-1$. It is not clear to me how this event, dictated by logic and consistency, relates to ordinary increase of potential knowledge by direct observation. The latter event is associated with the potential for the subjective perception of something new. But what about the former type of event?

\begin{state}[\textbf{Potential knowledge may increase in two ways}]
A state reduction $u_{1}S(n)\rightarrow S(n+1)\subset u_{1}S(n)$ may occur either as the result of subjective perception, or as the result of the requirement of epistemic consistency (Assumptions \ref{epconsistency}, \ref{epconsistency1} and \ref{epconsistency2}).
\label{twowaysred}
\end{state}

Of course, there are more possible combinations of alternatives than those shown in Figs. \ref{Fig34} and \ref{Fig34b}. However, these are sufficient to introduce the graphic and symbolic notation and to illustrate state reduction by observation or by epistemic consistency.

\begin{figure}[tp]
\begin{center}
\includegraphics[width=80mm,clip=true]{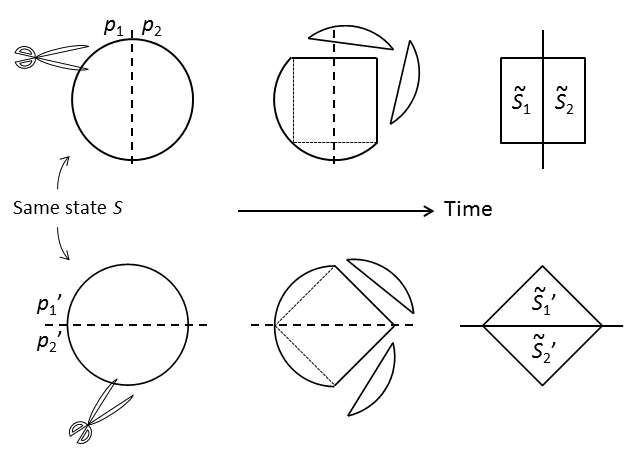}
\end{center}
\caption{Preparation of a state. Because of the lack of determinism, any given state can be prepared to different expermiental setups, designed to observe different properties $P$ and $P'$. The scissor cuts correspond to state reductions.}
\label{Fig35}
\end{figure}

To strengthen the intuition for the graphic notation, Fig. \ref{Fig35} shows how a single state can be prepared in different ways to become two different experiments to determine the values of properties $P$ and $P'$, respectively. Of course, this is possible because of the lack of determinism. This indeterminism makes it possible to apply different sequences of `scissor cuts' to the same original state. Each of these cuts represent a state reduction.

\section{Minimal objects}
\label{minimal}

In this section characetrize such objects more exactly than was done in Section \ref{limits} (Definitions \ref{minimalset} and \ref{minimalobject}).

Consider a set $\mathcal{M}$ of object species $M_{l}$ such that each $M_{l}\in \mathcal{M}$ is defined by the same set of internal attributes $\{A_{1},\ldots,A_{m}\}$. Each $M_{l}$ is specified by a set

\begin{equation}
\Upsilon_{(l)}=\{\Upsilon_{1l},\ldots,\Upsilon_{ml}\}
\label{minimalvalueset}
\end{equation}

of allowed values of these attributes that is distinct from the corresponding sets $\Upsilon_{(l')}$ of all other objects $M_{(l')}\in \mathcal{M}$. That these sets are distinct means that for each pair of object species $M_{l}\in \mathcal{M}$ and $M_{l'}\in \mathcal{M}$,

\begin{equation}
\exists A_{i}:\ \Upsilon_{il}\cap\Upsilon_{il'}=\varnothing.
\label{distinctobjects}
\end{equation}
This condition means that each object species $M_{l}\in \mathcal{M}$ is inherently different from all the others. To make this statement meaningful, we must require that each set $\Upsilon_{il}$ is constant in time. Otherwise the value of attribute $A_{i}$ in Eq. (\ref{distinctobjects}) of minimal object $M_{l}$ may be observed to be $\upsilon_{1}\in\Upsilon_{il}$ at one time, and $\upsilon_{2}\in\Upsilon_{il'}$ the next time. If no object division has taken place, such blending of attribute values means that $M_{l}$ and $M_{l'}$ must be considered to be the same object species $M_{l''}$, with $\Upsilon_{il''}=\Upsilon_{il}\cup\Upsilon_{il'}$. More formally: if no object division takes place,
\begin{equation}
\forall i,l:\ u_{1}\Upsilon_{il}=\Upsilon_{il}.
\label{constantobjects}
\end{equation}

If $\mathcal{M}$ furthermore fulfils Definition \ref{minimalset}, it is a minimal set of objects.

To relate the vocabulary used here to well-known quantities, let us consider the internal attributes of elementary fermions. These are, in arbitrary order:

\begin{equation}\begin{array}{ll}
A_{1}: & total\;spin\\
A_{2}: & generation\\
A_{3}: & baryon\;number\\
A_{4}: & lepton\;number\\
A_{5}: & electric\;charge\\
A_{6}: & colour\;charge.
\end{array}
\label{attributelist}\end{equation}

Spin components, parity, positions and momentum are all relational attributes, since they refer to other objects that define a spatio-temporal reference frame. Regarding the problem how to treat rest mass, we will argue in section \ref{evconsequences} that it is a derived internal attribute, that the set of masses of the minimal objects is a function of the other internal attributes in the above list. In other words, the rest mass is not considered to be an independent attribute (Definition \ref{indattributes}). Therefore it is not necessary to include it in a specification of the members $M_{l}$ of a minimal set $\mathcal{M}$.

The sets of possible values of these six attributes are:

\begin{equation}\begin{array}{lll}
\Upsilon_{1} & = & \frac{1}{2}\\
\Upsilon_{2} & = & \{1,2,3\}\\
\Upsilon_{3} & = & \{-\frac{1}{3},0,\frac{1}{3}\}\\
\Upsilon_{4} & = & \{-1,0,1\}\\
\Upsilon_{5} & = & \{-1,-\frac{2}{3},-\frac{1}{3},0,\frac{1}{3},\frac{2}{3},1\}\\
\Upsilon_{6} & = & \{0,1,\mathrm{exp}(i\frac{2\pi}{3}),\mathrm{exp}(i\frac{4\pi}{3})\}.
\end{array}\end{equation}

In this list we have included the possible attributes of elementary particles as well as anti-particles. For the colour charges, we set $r\leftrightarrow 1$, $g\leftrightarrow\mathrm{exp}(i\frac{2\pi}{3})$, and $b\leftrightarrow\mathrm{exp}(i\frac{2\pi}{3})$. The three non-zero values of color charges are thus seen as a circular attribute (Definition \ref{circularvalues} and Fig. \ref{Fig138}). 

As an example, let us specify the sets $\Upsilon_{(l)}$ of allowed attribute values (Eq. [\ref{minimalvalueset}]) that define the electron and the down-quark.

\begin{equation}\begin{array}{lll}
\Upsilon_{(electron)} & = & \left\{1/2,1,0,1,-1,0\right\}\\
\Upsilon_{(d-quark)} & = & \left\{1/2,1,1/3,0,-1/3,\{1,\mathrm{exp}(i\frac{2\pi}{3}),\mathrm{exp}(i\frac{4\pi}{3})\}\right\}
\label{fermionvaluesets}
\end{array}
\end{equation}

The fact that the set $\Upsilon_{6,d-quark}=\{1,\mathrm{exp}(i\frac{2\pi}{3}),\mathrm{exp}(i\frac{4\pi}{3})\}$ that corresponds to color charge contains more than one value reflects the fact that we may identify a minimal object without having to assume complete potential knowledge of all the internal attribute values. The color does not stay constant in the sense of Eq. [\ref{circularvalues}], and therefore a single color cannot constitute a set $\Upsilon_{il}$. Put differently, the reason why quarks with different colors are not considered to be different elementary fermions is that it is not possible to observe the color of an individual quark and track this colored object until it divides. This makes such a differentiation meaningless from the epistemic point of view. In contrast, it is possible to deduce which set of quarks are contained in a composite fermion. The members of this deduced set of quarks stay the same until the composite fermion divides. Each of them can be tracked by means of deduction. This makes quarks valid quasiobjects even though they cannot be observed individually.

We may express the discussion about color charge in the following alternative way. Let $\Upsilon_{i}$ be the set of possible value of attribute $A_{i}$. Then the vector

\begin{equation}
\upsilon=(\upsilon_{1},\upsilon_{2},\ldots,\upsilon_{6}),
\label{completeminimal}
\end{equation}
where $\upsilon_{i}\in\Upsilon_{i}$, represents a state of complete knowledge about the internal attributes of some minimal object. The fact that there may be several such vectors $\upsilon$, $\upsilon'$, $\ldots$ that are identified with the same object $M_{l}$ can be expressed as a degeneracy

\begin{equation}\begin{array}{lcl}
\upsilon & \leftrightarrow & M_{l}\\
\upsilon' & \leftrightarrow & M_{l}.\\
& \vdots &
\end{array}
\end{equation}
We may also write $M_{l}\leftrightarrow\{\upsilon,\upsilon',\ldots\}$, where each element can be described as an `exact internal state' $Z_{I}$ of the minimal object $M_{l}$. The color charge illustrates the fact that knowledge of the internal attributes is never complete for some minimal objects (the quarks). In other words, the internal state of a quark is never exact.

\section{Identifiability}
\label{identifiability}

The reader may have noted that I stress the discreteness of alternatives, and the discreteness of sequential time, but still crucially refer to the concept of identifiable objects, which is defined with the help of continuity (Definition \ref{identifiableobjects}). Clearly, this concept must be redefined in an epistemically sound way.

The intuition is the following: if an object in a state of knowledge $PK(n)$ cannot be distinguished from an object in the next state $PK(n+1)$, then these two objects must be seen as the same object. It is impossible to tell them apart, and it is epistemically unsound to give them different names. This one and the same object is therefore possible to track from time $n$ to time $n+1$, and we may say that it is identifiable at time $n$.

\begin{figure}[tp]
\begin{center}
\includegraphics[width=80mm,clip=true]{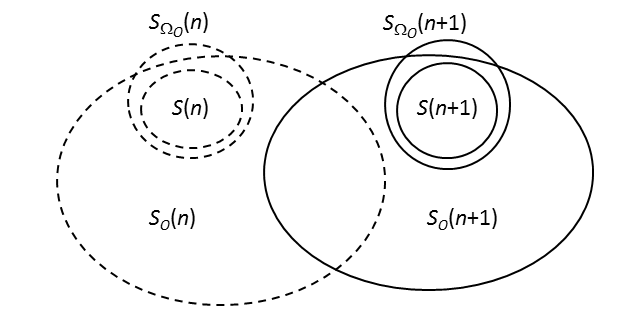}
\end{center}
\caption{That object $O$ is identifiable at time $n$ means that $S_{OO}(n)$ and $S_{OO}(n+1)$ overlap. This is true also for the two object states $S_{O}(n)$ and $S_{O}(n+1)$ shown here. In contrast, $S_{\Omega_{O}}(n)$ and $S_{\Omega_{O}}(n+1)$ must be disjoint, since \emph{some} objects must be subjectively distinguishable in the states $S(n)$ and $S(n+1)$. Graphically, it becomes clear that the smaller part of knowledge that is encapsulated in $O$, the larger chance that $O$ is identifiable. It becomes more probable that one of all the other objects change in the temporal update $n\rightarrow n+1$. Compare Fig. \ref{Fig28b}.}
\label{Fig39}
\end{figure}

Of course, there must be other objects in $PK(n+1)$ that \emph{can} be distinguished from objects in $PK(n)$, by definition of sequential time. Time is updated when a subjective distinction can be made between now and then. These other objects can collectively be seen as a clock that ticks each time sequential time is updated (Fig. \ref{Fig39}). These other objects constitute the environment or complement $\Omega_{O}$ to $O$, with state $S_{\Omega_{O}}$ (Definition \ref{environmentstate}). The identifiable object `floats' through time in the sense that one cannot say for sure that its attributes have changed between two subsequent tickings of the clock. In contrast, it may be possible to perceive a change of the identifiable object at a later time $n+m$, as compared to time $n$, where $m>1$.

Let us formalize these ideas. Recall the Definition \ref{objectstate} of the state $S_{O}$ of an object $O$ expressed in state space $\mathcal{S}$, and the Definition \ref{stateobjectspace} of the state $S_{OO}$ of the same object expressed in object state space $\mathcal{S}_{O}$. Figure \ref{Fig90b} tries to poinpoint the essential difference between these two ways to express the object state.

\begin{defi}[\textbf{Identifiable object at time} $n$]
An object $O$ is identifiable at time $n$ if and only if $S_{OO}(n)\cap S_{OO}(n+1)\neq\varnothing$.
\label{identifiable}
\end{defi}

\begin{figure}[tp]
\begin{center}
\includegraphics[width=80mm,clip=true]{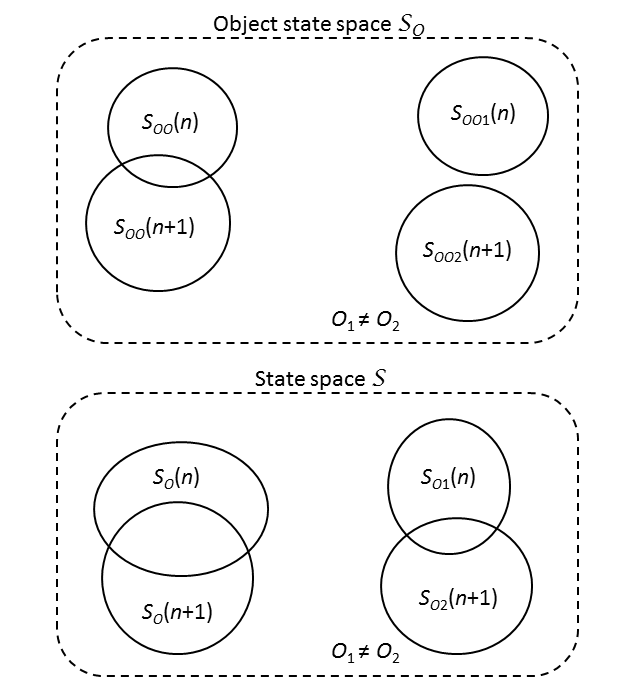}
\end{center}
\caption{Tracking an object through time in object state space $\mathcal{S}_{\mathcal{O}}$ and in state space $\mathcal{S}$. If two object states $S_{OO}(n)$ and $S_{OO}(n+1)$ overlap in $\mathcal{S}_{\mathcal{O}}$, then they represent the same identifiable object $O$. Otherwise they represent different objects $O_{1}$ and $O_{2}$. When two object states are represented in the full state space $\mathcal{S}$, they overlap at subsequent times regardless whether they correspond to a single identifiable object $O$, or two different objects $O_{1}$ and $O_{2}$. Thus the overlap of subsequent object states can be used to define a single identifiable object only if we represent these states in $\mathcal{S}_{O}$. Compare Fig. \ref{Fig90b}.}
\label{Fig90c}
\end{figure}

This condition means that there are exact object states $Z_{O}$ that are compatible with both $S_{OO}(n)$ and $S_{OO}(n+1)$ (Fig. \ref{Fig39}). In other words, there is no potential knowledge that can tell the objects at subsequent times apart. If these two object states do not overlap, one the other hand, we have to say that they are the states of different objects. To reach this conclusion, we have to represent the object states in object state space, as illustrated in Fig. \ref{Fig90c}.

It may be possible to track an object during extended periods of time:

\begin{defi}[\textbf{Identifiable object during a time interval}]
An object that is identifiable at all times $n, n+1,\ldots,n+m$ is identifiable in the time interval $[n,n+m]$.
\label{identifiableinterval}
\end{defi}

\begin{figure}[tp]
\begin{center}
\includegraphics[width=80mm,clip=true]{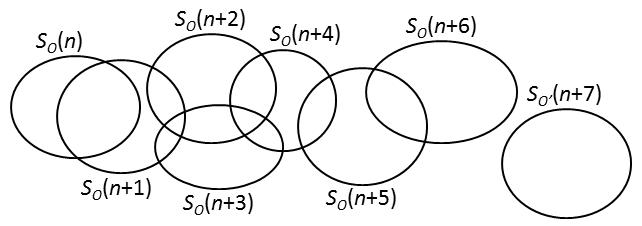}
\end{center}
\caption{The evolution of the state $S_{OO}$ of an object $O$ that is identifiable in the time interval $[n,n+6]$. The object $O'$ at time $n+7$ cannot be identified with object $O$ at time $n+6$, and is therefore given a separate name.}
\label{Fig40}
\end{figure}

This definition is illustrated in Fig. \ref{Fig40}. Even if the object is identifiable at every time instant in the above sequence, it may be possible to distinguish the object at some time $n+\mu$ from the object at time $n$, that is, we may have $S_{OO}(n)\cap S_{OO}(n+\mu)=\varnothing$, where $1<\mu\leq m$. Due to its identifiability at each instant, the object nevertheless preserves its identity.

When one object changes and time is updated, there are other objects that apparently stay the same. A leave may fall from the tree, but the stem of the tree does not move. Not everything changes at the same time. This fact can be related to the feeling that states of knowledge at different times belong to the same world, even though they are different. Even if everything around us seem to change at once, some internal objects may stay the same, such as mood and memories.

\begin{defi}[\textbf{Identity of a world}]
A world specified by a sequence of states $S(n)$ preserves its identity during time interval $[n,n+m]$ if and only if there is an identifiable object at all times $n, n+1,\ldots,n+m-1$.
\label{worldidentity}
\end{defi}

The object that is identifiable may change from one instant to the next. When the leave falls, the immobility of the stem preserves the identity of the world. But when the leave has fallen, the tree may be cut down, and the leave resting on the ground preserves the identity of the world. In this way, the evolution of a given, identifiable world may be compared to walking: when one foot is lifted, the other is resting on the ground.

The notion of minimal objects (Section \ref{minimal}), the assumptions of finite depth of knowledge and noiseless physical law (Assumptions \ref{finitedepth} and \ref{noiseless}), together with the assumption of epistemic invariance (Assumptions \ref{epconsistency}, \ref{epconsistency1} and \ref{epconsistency2}), imply that any physical state $S(n)$ can be specified in terms of a finite set of identifiable minimal objects, upon which the evolution $u_{1}$ acts (Definition \ref{evolutionu1}). Consequently, to specify $u_{1}$ completely, it is sufficient to specify the action of $u_{1}$ on identifiable minimal objects. Thus it is essential to define identifiability of minimal objects.

\begin{defi}[\textbf{Identifiable minimal object}]
A minimal object $O_{M}$ is identifiable if and only if it is quasi-identifiable at all times. A minimal object is quasi-identifiable if and only if, given the evolution of potential knowledge, it cannot be excluded that $O_{M}$ evolves along a continuous trajectory, in accordance with the `naive' notion of identifiability (Definition \ref{identifiableobjects}).
\label{minimalidentity}
\end{defi}

To actually confirm that a minimal object travels along a continuous trajectory would require that the time interval $t(n)$ between the sequential time instants $n-1$ and $n$ at which we observe it could be made arbitrarily small for each $n$. This cannot be taken for granted, and it is not necessary according to the present notion of identifiability, based on overlapping sequential object states. In fact, arbitrarily small $t(n)$ are forbidden according to the discussion in section \ref{boundstates}.

\begin{state}[\textbf{Minimal objects are identifiable}]
The evolution of any object is always consistent with a model in which it is composed of identifiable minimal objects.
\label{allminimalidentity}
\end{state}

Statement \ref{allminimalidentity} follows from the assumption of noiseless physical law (Assumption \ref{noiseless}). The identifiability of minimal objects means that we are never forced into a physical model where objects pop out of nowhere, or suddenly disappear.

The fact that minimal objects are identifiable gives meaning the concept of the \emph{evolution} of a minimal object. That is, we are allowed to use a model in which predict the trajectory of \emph{a particular} minimal object $O_{M}$ given the present knowledge of its state.

\begin{defi}[\textbf{The evolution of a minimal object}]
Consider a particular minimal object $O_{M}$ at time $n$ with state $S_{O_{M}}(n)$. Its evolution $u_{1}(S(n))S_{O_{M}}(n)$ is the smallest set such that we always have $S_{O_{M}}(n+1)\subseteq u_{1}(S)S_{O_{M}}(n)$. 
\label{minievolution}
\end{defi}

If minimal objects were not identifiable, it would not make any sense to single out their evolution in the evolution $u_{1}S(n)$ of the entire world. Note that we have to let the evolution operator $u_{1}$ depend on $S(n)$ when we apply it to $S_{O_{M}}(n)$, since $O_{M}$ may interact with its environment.

\begin{figure}[tp]
\begin{center}
\includegraphics[width=80mm,clip=true]{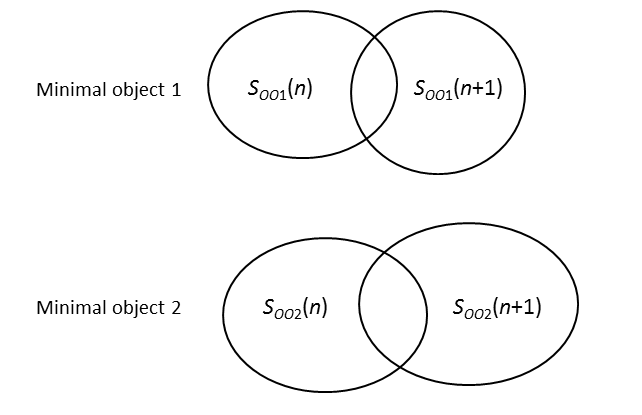}
\end{center}
\caption{If all minimal objects in the physical state $S$ were directly perceived and were still identifiable, then there would be no object $O_{C}$ that could act as a clock and make it possible to distinguish time $n$ from time $n+1$ in the sense that $S(n)\cap S(n+1)=\varnothing$. For such a clock we must require $S_{OOC}(n)\cap S_{OOC}(n+1)=\varnothing$. In this example the world consists of two directly perceived minimal objects, and there is no such clock. We conclude that some minimal objects must be deduced quasiobjects.}
\label{Fig41}
\end{figure}

How can distinct time instants appear if all minimal objects are identifiable? If none of the minimal objects can be told apart at times $n$ and $n+1$, how can the states $S(n)$ and $S(n+1)$ they build collectively be told apart? Some of the minimal objects must be deduced quasiobjects, as illustrated in Fig. \ref{Fig41}. If the state of potential knowledge $PK(n)$ consisted of directly perceivable minimal objects only, there would be no object that could act as a clock, that could define the distinction between $S(n)$ and $S(n+1)$. (Compare the discussion in the caption to Fig. \ref{Fig39}).

\begin{state}[\textbf{There are always minimal objects that cannot be individually perceived}]
At all sequential time instants $n$, there are some minimal objects that are quasiobjects.
\label{alwaysquasi}
\end{state}

The existence of minimal quasiobjects makes it possible to speak about identifiable objects that are not directly perceived at each time instant at which it is identifiable. If we look at the moon, close our eyes, and then look at it again, we want to be able to say that it is the same moon we are looking at the second time.

\begin{defi}[\textbf{Quasi-identifiable object during a time interval}]
An object $O$ is quasi-identifiable during the time interval $[n,n+m]$ if and only if it can be described in terms of the same set of identifiable minimal objects for any $n'\in[n,n+m]$, and these minimal objects evolve according physical law.
\label{quasiidentifiable}
\end{defi}

This definition makes it possible to introduce the \emph{evolution} of a quasi-identifiable object in the same way as we introduced the evolution of a minimal object (Definition \ref{minievolution}).

\begin{defi}[\textbf{The evolution of a quasi-identifiable object}]
Consider an object $O$ which is quasi-identifiable at time $n$, having state $S_{O}(n)$. Its evolution $u_{1}(S(n))S_{O}(n)$ is the smallest set such that we always have $S_{O}(n+1)\subseteq u_{1}(S)S_{O}(n)$. 
\label{objectevolution}
\end{defi}

We see that an identifiable object is always quasi-identifiable, but a quasi-identifiable object does not need to be identifiable. We may also say that we can \emph{be sure} that an identifiable object stays the same throughout the period during which it is identifiable, whereas we cannot \emph{exclude} that a quasi-identifiable object stays the same throughout the same period of time, given physical law. Needless to say, it is this weaker form of identifiability that we use in everyday life. We do not need to stare constantly at a flower in the kitchen window to say that it is the same flower we see each morning as we drink our coffee.

\section{States of the body and of the world}
\label{bwstates}

It is not altogether clear how to make the distinction between the external world and the bodies of aware subjects. Does my toenail belong to my body or to the outside world? What if I cut it off? What about the retina, which is essential for visual preception, and thus for subjective knowledge?

\begin{figure}[tp]
\begin{center}
\includegraphics[width=80mm,clip=true]{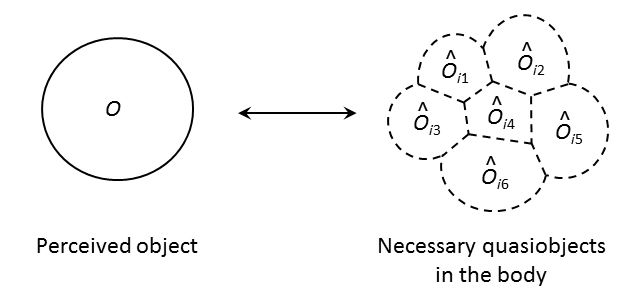}
\end{center}
\caption{According to the assumption of detailed materialism, each perceived object $O$ correspond to a set of internal quasiobjects $\hat{O}_{ik}$ in the body. There is a one-to-one correspondence between the perceived state $S_{O}$ of $O$ and the (reduced) state $\check{S}_{i}$ of these quasiobjects.}
\label{Fig41a1}
\end{figure}

Let us try to clear things out. Each perception has two ends. It consists of the perceived object $O$, and those objects in the body that correspond to the subjective perception according to detailed materialism (Assumption \ref{localmaterialism}). These objects in the body are quasiobjects; they are the objects that can be deduced from physical law as necessary for the perception at hand (Fig. \ref{Fig41a1}). For example, any visual perception requires a lens $O_{i1}$, a retina $O_{i2}$, a visual nerve $O_{i3}$ and a visual cortex $O_{i4}$. (The subscript $i$ indicates that we are talking about an internal object of the body.) Even if we do not normally perceive these objects directly, we know that they must be there in our heads - otherwise we would have seen nothing.

Let $\check{S}_{i}$ be the state consisting of all exact states of a body of a preceiving subject that cannot be excluded by the perception of $O$. Since bodily objects $\hat{O}_{ik}$ are deduced quasiobjects, the state $\check{S}_{i}$ of these objects is classified as a reduced state (Fig. \ref{Fig28c}). We have a one-to-one correspondence

\begin{equation}
S_{O}\leftrightarrow\check{S}_{i}.
\end{equation}

let $O^{k}(n)$ be the totality of the objects potentially perceived by subject $k$ at time $n$. Then the corresponding reduced state of the body of $k$ can be written $\check{S}_{i}^{k}(n)$.

\begin{defi}[\textbf{The body of a subject}]The body of subject $k$ at time $n$ is the set $\mathcal{B}^{k}(n)$ of quasiobjects that are part of all exact states $Z\in \check{S}_{i}^{k}(n)$.
\label{body}
\end{defi}

This definition identifies a minimal group of necessary quasiobjects: none of them can be excluded in a body that is able to account for the present state of individual potential knowledge $PK^{k}(n)$. The body becomes a dynamical object, where the constituent objects may change from one time to the next.

Even if the perceived object $O^{k}(n)$, of course, is necessary to account for the perception, we exclude it from the body - only the corresponding quasiobjects count (Fig. \ref{Fig41a1}). Otherwise we would have to consider a flower that we see on the ground to be part of our body.

\begin{figure}[tp]
\begin{center}
\includegraphics[width=80mm,clip=true]{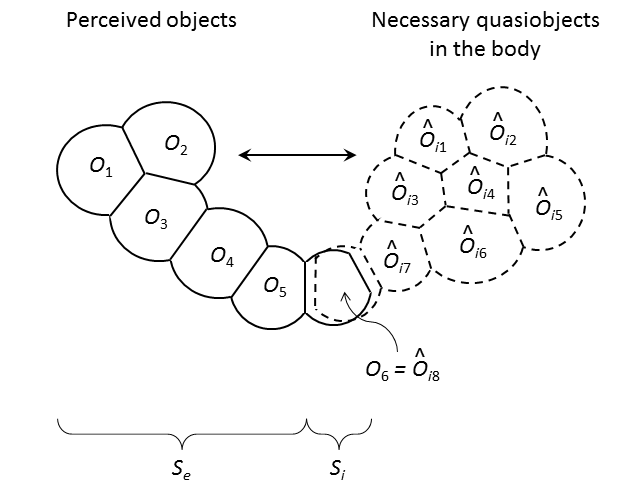}
\end{center}
\caption{Sometimes we perceive part of our own body. Some of the perceived objects $O_{l}$ can be identified with some of the quasiobjects $\hat{O}_{il'}$ that defines the body. If we snap our fingers, the sensation of snapping necessitates the same fingers (quasiobject $\hat{O}_{i8}$) as we see directly in front of our head (object $O_{6}$).}
\label{Fig41a1b}
\end{figure}

It often happens, however, that some of the perceived objects indeed are part of the body (Fig. \ref{Fig41a1b}). If we let our two hands touch, the direct sensation of touch necessitates as quasiobjects the very same skin with nerve endings as we also perceive directly. Some of the perceived objects $O_{l}$ can be identified with some of the necessary quasiobjects $\hat{O}_{il'}$. In this situation, the states $S_{O_{l}}$ and $\check{S}_{\hat{O}_{il'}}$ of the objects we identify must overlap, to respect epistemic consistency. If we look at our own eyes in the mirror, we never see that the eyelids are closed.  

The definition of the body in terms of a set of objects $\mathcal{B}$ is just an attempt to be semantically clear. The essential physical quantity associated with the body is the reduced state $\check{S}_{i}$. If we break the body down to minimal objects like electrons and quarks, a description of the body in terms of objects would just read $X$ electrons, $Y$ quarks, and so on. Such a list provides no useful information. It is the state $\check{S}_{i}$ of these objects that is important, their relations, the conditional knowledge that defines the larger structures they form.

Having defined the body, and the state of the body, it possible to state the assumption of detailed materialism (Asssumption \ref{localmaterialism}) more precisely. In a materialistic world, each time a subject potentially perceives a distinct change of the world, there is also a distinct change of her body. By definition, at each temporal update $n\rightarrow n+1$ there is at least one subject that perceives a distinct change of the world.

\begin{assu}[\textbf{Detailed materialism 2}] At each time $n$ there is a subject $k$ such that $\check{S}_{i}^{k}(n+1)\cap \check{S}_{i}^{k}(n)=\varnothing$.
\label{localmaterialism2}
\end{assu}

\begin{figure}[tp]
\begin{center}
\includegraphics[width=80mm,clip=true]{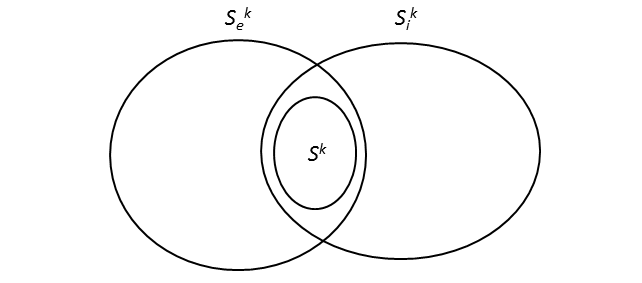}
\end{center}
\caption{The physical state corresponding to the knowledge of subject $k$ can be divided into two parts: the state $S_{e}^{k}$ to the knowledge of the external world, and the state $S_{i}^{k}$ corresponding to the knowledge or perception of her own body.}
\label{Fig41a1c}
\end{figure}

The fact that some of the perceived objects are part of the body makes it possible to divide this set of objects into two groups: the objects $O_{i}^{k}$ that belong to the body of subject $k$ and those objects $O_{e}^{k}$ that do not. We associate the states $S_{i}^{k}$ and $S_{e}^{k}$ to these two objects, respectively (Fig. \ref{Fig41a1c}). The physical state corresponding to the individual potential knowledge of $k$ fulfils

\begin{equation}
S^{k}\subseteq S_{e}^{k}\cap S_{i}^{k}.
\end{equation}

We concluded in section \ref{identifiability} that the entire world with state $S$ has to consist of several objects to be able to preserve its identity as time passes. In the same way we may conclude that the internal state $S_{i}^{k}$ of subject $k$ has to consist of several objects if she is to be able to preserve her subjective identity as time passes. At least one internal object has to stay the same when she perceives that another one changes. In the same way, $S_{e}^{k}$ must consist of several object if the external world should be able to preserve its identity in the eyes of subject $k$.

Note that the somewhat paradoxical situation might occur, where $k$ experiences a loss of individual identity, but the external world nevertheless preserves its identity in her eyes. This situation occurs if all internal objects undergo a distinct change at some time, whereas some external objects subjectively remain the same. Note also that identity may be preserved at the collective level even if it is lost at the individual level.

\begin{figure}[tp]
\begin{center}
\includegraphics[width=80mm,clip=true]{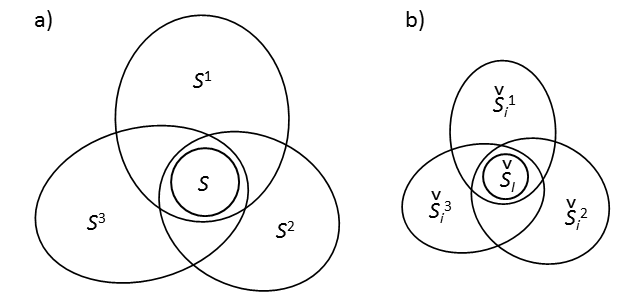}
\end{center}
\caption{The physical state $S$ belong to the common part of the physical states $S^{k}$ that correspond to individual potential knowledge. The reduced collective state of the bodies $\check{S}_{I}$ of a group of subjects belong to the common part of the individual states of these bodies.}
\label{Fig41a}
\end{figure}

Let us turn to this collective level. Let $\check{S}_{I}$ be the reduced state of the composite object consisting of all bodies in the world (Fig. \ref{Fig41a}). Epistemic consistency (Assumptions \ref{epconsistency}, \ref{epconsistency1} and \ref{epconsistency2}) requires

\begin{equation}\begin{array}{lll}
S(n) & \subseteq & \bigcap_{k} S^{k}(n)\\
\check{S}_{I}(n) & \subseteq & \bigcap_{k}\check{S}_{i}^{k}(n).
\label{bodystate1}
\end{array}\end{equation}

Assume that these relation were not true, and that there were an exact state $Z\in S(n)$ such that $Z\in S^{1}(n)$, but $Z\notin S^{2}(n)$ (Fig. \ref{Fig41a}). Then, at a later time $n'>n$ subject 2 could meet subject 1 and tell her that she excluded $Z$ as a possible state of the world already at time $n$. That could lead to a contradiction knowable to subject 1, since the evolution of her state $S^{1}(n)\rightarrow S^{1}(n')$ depends irreducibly on $S(n)$ (Statement \ref{irreduciblelaw}), just like the state of any object.   

The potential knowledge $PK(n)$ of the world consists of the perceptions of all subjects $k$, corresponding to individual internal states $\check{S}_{i}^{k}$ of their bodies, giving the collective internal state $\check{S}_{I}$. By definition, $PK(n)$ also corresponds to $S(n)$. We conclude that there is a one-to-one correspondence between $\check{S}_{I}(n)$ and $S(n)$:

\begin{equation}
S(n)\leftrightarrow \check{S}_{I}(n).
\label{bodystate2}
\end{equation}

This relation embodies intertwined dualism and detailed materialism. The correspondence is not an identity, but a functional relationship where physical law defines the function. Given $S(n)$, we can deduce $\check{S}_{I}(n)$ from physical law, and given $\check{S}_{I}(n)$ there is exactly one possible state of knowledge $PK(n)$ and a corresponding physical state $S(n)$.

\begin{figure}[tp]
\begin{center}
\includegraphics[width=80mm,clip=true]{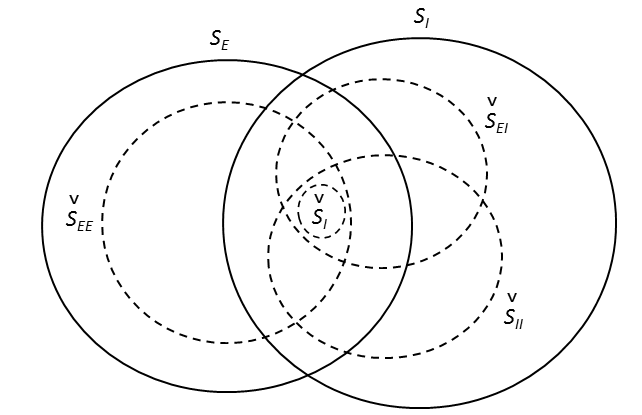}
\end{center}
\caption{Relation between various physical states derived from the distinction between subject and object, the body and the external world. All states are derived from the collective potential knowledge $PK$ of all individual subjects. The subscripts $I$ and $E$ stands for internal and external, respectively. See text for further explanation.}
\label{Fig41a3}
\end{figure}

We may define $S_{I}$ as the state of the composite object $O_{I}$ that consists of the parts of all the bodies that are directly perceived by somebody. $S_{E}$ is then the state of all the other parts $\Omega_{E}$ of the world, those that do not belong to the body of anyone (Fig. \ref{Fig41a3}). We have

\begin{equation}
S(n)\subseteq S_{E}(n)\cap S_{I}(n).
\label{bodystate3}
\end{equation}

Let $\check{S}_{II}$ be the reduced state of the quasiobjects $\hat{O}_{II}$ in the bodies that are necessary to account for the perception $O_{I}$ of the corresponding subjects. Likewise, let $\check{S}_{EI}$ be the reduced state of the quasiobjects $\hat{O}_{EI}$ in the bodies that are necessary to account for the perception of the external world $\Omega_{E}$. We may also define the reduced state $\check{S}_{EE}$ of those quasiobjects $\hat{O}_{EE}$ that are possible to deduce from the directly observed external objects in $\Omega_{E}$ and physical law. The relation between these sets are illustrated in Fig. \ref{Fig41a3}.

Clearly, we must have

\begin{equation}
\check{S}_{I}\subseteq \check{S}_{II}\cap \check{S}_{EI}\cap \check{S}_{EE},
\end{equation}
and
\begin{equation}
\check{S}_{I}\subseteq S.
\end{equation}
These relations should be compared with Eq. (\ref{bodystate1}), (\ref{bodystate2}), and (\ref{bodystate3}).

The (composite) quasiobjects $\hat{O}_{II}$, $\hat{O}_{EI}$, and $\hat{O}_{EE}$ that are defined above, can be used to make a more elaborate version of Fig. \ref{Fig8}. We let $PK_{I}\leftrightarrow S_{I}$ denote the potential knowledge of the directly perceived internal objects $O_{I}$, and we let $PK_{E}\leftrightarrow S_{E}$ denote the potential knowledge of the directly perceived external objects in $O_{E}$. Clearly $PK \supseteq PK_{I}\cup\ PK_{E}$. The length of the interval covered by the various composite objects indicate how many distinct objects are contained in the representation of the corresponding physical state. In short, we can deduce more (quasi)objects than we can see directly.

\begin{figure}[tp]
\begin{center}
\includegraphics[width=80mm,clip=true]{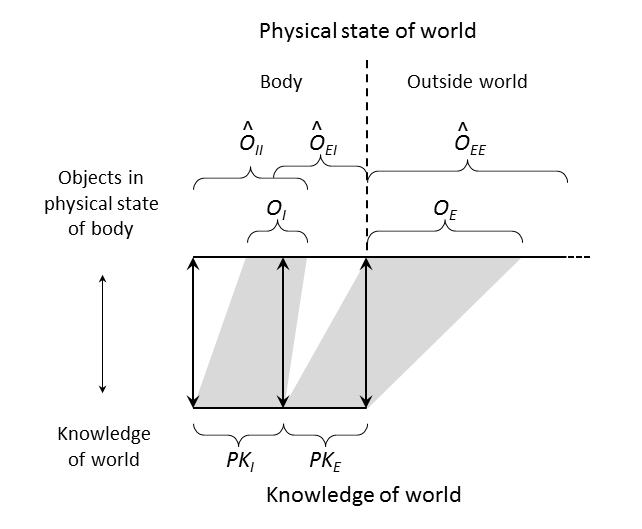}
\end{center}
\caption{An elaborate version of Fig. \ref{Fig8}. The relations between the corresponding states of these objects are shown in Fig. \ref{Fig41a3}. The discrete sequences of objects along the upper and lower lines shown in Fig. \ref{Fig8} are suppressed for clarity.}
\label{Fig41a2}
\end{figure}

Basically, the physical state of the internal and external world is an encoding of our knowledge about these two worlds. This knowledge is in turn encoded in the physical state of the internal world, so that the state of both the internal and the external world is encoded in the state of the internal world. As we discussed in section \ref{limits}, this implies that the knowledge of both worlds is incomplete.

\section{Options, intention and choice}
\label{aic}

We return to the concept of \emph{alternatives}, introduced in section \ref{law}, and further discussed in the following sections. In our terminology, an alternative is a physical state that corresponds to the present potential knowledge \emph{plus} a specific property of some object, the value of which is not known at present. The object may or may not turn out to have this property when it is observed at some later time. We let an \emph{option} be an alternative that is \emph{subjectively preconceived}.

To highlight the difference between an alternative and an option, suppose that a cat is hiding in the bushes. If someone walks along a street and approaches the bushes without thinking about the possibility that there may be a cat nearby, and it suddenly walks out before her, then it is an alternative that comes true. Her individual state changes so that it becomes a subset of the part of state space defined by the property "there is a cat nearby". If she walks along thinking that her neighbour's cat might come to greet her in the street any second, and it actually does, the alternative would also be an option that comes true.

\begin{figure}[tp]
\begin{center}
\includegraphics[width=80mm,clip=true]{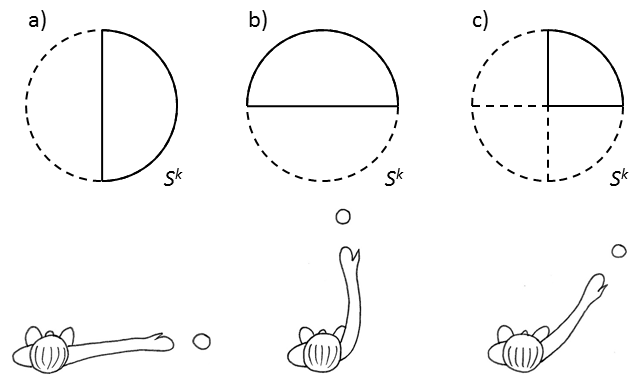}
\end{center}
\caption{States $S^{k}$ of a person standing with a ball in her hands. (a) The idea to throw the ball to the left or to the right comes into her mind, splitting the state into two future options. She decides to throw it to the right. The solid half circle becomes the new state. (b) The idea to throw the ball forward or backward comes into her mind, and she decides to throw it forward. (c) The composite idea to throw the ball to the right and forward comes into her mind, and she makes it happen.}
\label{Fig41a3b}
\end{figure}

Imagine next that someone is standing with a ball in her hand. Suddenly, the two options to throw it to the left or the right materialize in her mind. This event corresponds to a distinct change of the state $\check{S}_{i}^{k}$ of her body, according to detailed materialism. At this stage, suppose that there is nothing in her state of mind that favours one option above the other. Half of the exact states that conforms with the present state of subject $k$ must then correspond to an evolution where the ball moves to the left, and the other half to a future scenario where the ball moves to the right (Fig. \ref{Fig41a3b}).

In the same way, the options to throw the ball forward och backward may appear in her head [Fig.\ref{Fig41a3b}(b)]. In this case there is an inherent asymmetry of the options, but for the sake of illustration we may again assign the two options the same chance of realization.

The idea to throw the ball to the right is rather imprecise. Consequently, there are many exact states that are consistent with this option [the solid half-circle in Fig. \ref{Fig41a3b}(a)]. Options may be combined, of course. The idea may appear in her mind to throw the ball to the right and forward. This option is more precisely defined and thus correspond to a smaller area in state space - the upper left quarter of the circle [Fig.\ref{Fig41a3b}(c)].

In this case the combined option may be realized without any problem. But since knowledge is always incomplete, there is  a limit to how precisely an option can be defined, if we require that it should be possible to decide whether it comes true or not. Not every alternative is realizable (Definition \ref{realizablealt}). Suppose that the state depicted in Fig. \ref{Fig41a3b} corresponds to a single electron rather than a person with a ball in her hands. Suppose further that the options to throw the ball to the left or to the right are replaced by the options that the $x$-component $s_{x}$ of the spin of the electron turns out to be positive or negative, and that the ball flying forward or backward corresponds to the cases where the $y$-component $s_{y}$ of the spin is positive or negative, respectively. Then we know experimentally that it can never be decided whether the combined option  $s_{x}>0$ and $s_{y}>0$ is realized. It is epistemically meaningless.

The lesson is the usual one: in a graphical representation of a state $S$ as a set, where realizable alternatives $S_{j}$ divides $S$ into subsets, each such subset always contains more than one exact state $Z$.

\begin{figure}[tp]
\begin{center}
\includegraphics[width=80mm,clip=true]{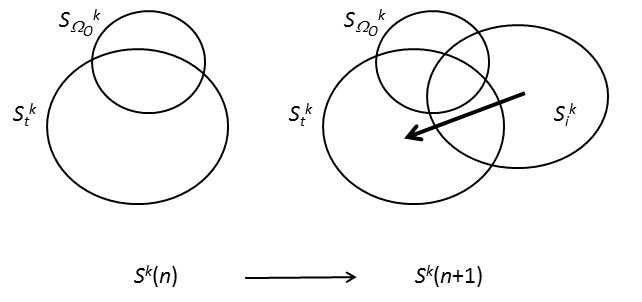}
\end{center}
\caption{That subject $k$ start to visualize an option for object $O_{t}$ with present state $S_{t}^{k}$ corresponds to a temporal update $n\rightarrow n+1$. $\Omega_{O}$ is the complement to $O_{t}$ and $O_{i}$, such that $S^{k}\subseteq S_{t}^{k}\cap S_{i}^{k}\cap S_{\Omega_{O}}^{k}$ when these states are represented in the full state space $\mathcal{S}$. The visualized object $O_{i}$ is an internal object in subject $k$ with state $S_{i}^{k}$ that appears at the temporal update, together with a directed relational attribute that associates $S_{i}^{k}$ with a possible future state of the target object $O_{t}$. This relation is represented by an arrow.}
\label{Fig41a4}
\end{figure}

Let us make the discussion more formal. We have a target object $O_{t}$, which subject $k$ perceives, with state $S_{t}^{k}$. In the example above it is the ball. Alternative future states of this target object takes the form of imagined, internal objects $O_{i}$, which are subjectively associated with $O_{t}$. The state of $O_{i}$ is $S_{i}^{k}$. The interpretation of $S_{i}^{k}$ as an option, as possible future state of $O_{t}^{k}$, can be described as a directed relational attribute, that points from the imagined object $O_{i}$ to the target object $O_{t}$.

\begin{defi}[\textbf{Image attribute} $I$] The directed relational attribute $I$ is such that $I_{it}^{k}(n)=1$ if subject $k$ interprets an internal object $O_{i}$ to be an image of another object $O_{t}$ that differs from the present state of $O_{t}$, i.e. $S_{i}^{k}(n)\cap S_{t}^{k}(n)=\varnothing$ when $S_{i}^{k}$ and $S_{t}^{k}$ are represented in object state space $\mathcal{S}_{O}$ (Definition \ref{objectstatespace}). Otherwise $I_{it}^{k}(n)=0$.
\label{imageattribute}
\end{defi}

Note that we do not specify whether the target object $O_{t}^{k}$ is external or internal. Often it is an external object we wish to manipulate, but it may also be an internal object. We may want to steer our fantasies in a given, preconceived direction. That is, the target object $O_{t}$ may be a pure fantasy, just like the imagined future state $O_{i}$ of this fantasy.

The imagined object $S_{i}$ that points to $O_{t}$ may be a composite object. This is the case for the combined option where your intention is to throw the ball to the left \emph{and} forward. It is an intention with two parts $O_{i1}$ and $O_{i2}$. However, the incompleteness of knowledge prevents the imgained object $O_{i}$ from being composed of too many objects - if the corresponding option should have any chance to come true. In the case of spin directions of an electron $O_{t}$, already two constituent objects $O_{i1}$ and $O_{i2}$ is one too many, as discussed above.

The formation of an option for the target object $O_{t}$ in someone's mind can be expressed as a temporal update $n\rightarrow n+1$ in which the state $S^{k}$ of subject $k$ undergoes a distinct change where an object $O_{i}$ is created such that $I_{it}^{k}(n+1)=1$ (Fig \ref{Fig41a4}). As usual, the complement $\Omega_{O}$ with state $S_{\Omega_{O}}^{k}$ is defined as the part of the world in the state of $k$ that is not $O_{t}$, nor $O_{i}$ (Definition \ref{complement}).

\begin{defi}[\textbf{Option}]An option is a subjectively imagined alternative. It is a future state $S_{i}^{k}$ of an object, preconceived by some subject $k$. Necessary and sufficient ingredients are the target object $O_{t}$ with state $S_{t}^{k}$, an internal object $O_{i}$ representing the imagined future state $S_{i}^{k}$ of $O_{t}$, and the attribute value $I_{it}^{k}=1$ according to definition \ref{imageattribute}.
\label{option}
\end{defi}

It is convenient to represent options as a partition of the state of the target object, just like we represented alternatives in the preceding sections. Let us establish the link between this representation and the formalism described above.

Suppose that we have a set $\{S_{ij}^{k}\}$ of distinct imagined future states of the target object. This set of states corresponds to a set of internal objects $\{O_{ij}\}$ in subject $k$. We have

\begin{equation}
S_{ij}^{k}\cap S_{ij'}^{k}\neq\varnothing
\end{equation}
since they correspond to \emph{different} objects that can be present \emph{at the same} time in the mind of $k$. However, each state $S_{ij}^{k}$ correspond to a distinct property value

\begin{equation}
p_{j}\leftrightarrow S_{ij}^{k}
\label{poe}
\end{equation}
of some property $P$ of the target object $O_{t}$, such that

\begin{equation}
\mathcal{P}_{j}\cap\mathcal{P}_{j'}=\varnothing.
\end{equation}

The property values $p_{j}$ of the target object may or may not become realized.

\begin{defi}[\textbf{Realizable option}]The option defined by the imagined state $S_{ij}^{k}$ is realizable if and only if it is equivalent to a realizable property value $p_{j}$ according to Eq. [\ref{poe}].
\label{realoption}
\end{defi}

We may now define present and future options in the same way as we defined present and future alternatives in Definitions \ref{presentalt} and \ref{futurealt}, and we can also define complete sets of such options just like in Definitions \ref{setpresentalt} and \ref{setfuturealt}. From now on, for the sake of illustration, we focus mainly on sets of future options $\{\tilde{S}_{j}\}$ that define a partition of the state $S_{t}$ of the target object $O_{t}$ such that $\tilde{S}_{j}\subseteq S_{t}$ and $\tilde{S}_{j}\cap\tilde{S}_{j'}=\varnothing$. However, we drop the tildes to avoid cluttered notation.

\begin{figure}[tp]
\begin{center}
\includegraphics[width=80mm,clip=true]{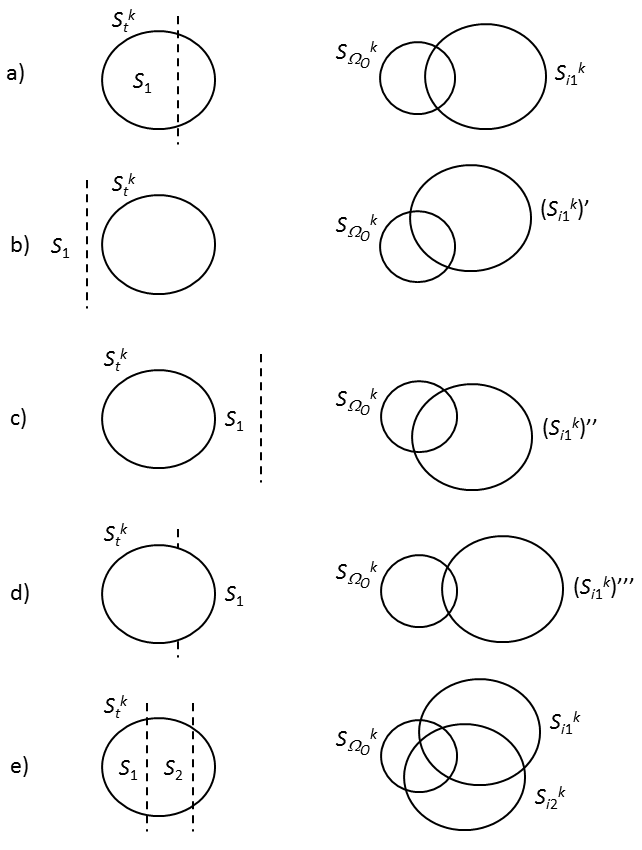}
\end{center}
\caption{The relation between the representation of options as intersecting states (Fig. \ref{Fig41a4}) and the representation as a partition of a present state (section \ref{law}). The target state $S_{t}^{k}$ is separated from the states $S_{i}^{k}$ and $S_{c}^{k}$ for clarity. (a) The option $S_{i1}^{k}$ can be realized. It corresponds to the subset $S_{1}$ in the partition. There is also an implicit complementary option $S_{2}$. (b) the option cannot be realized. (c) The option is realized with certainty at some future time. (d) It will never be decided whether the option is realized. (e) Two options $S_{i1}^{k}$ and $S_{i2}^{k}$ are imagined, corresponding to subsets $S_{1}$ and $S_{2}$ in the partition, respectively. There is also an implicit alternative $S_{3}$ that completes the set of options.}
\label{Fig41a5}
\end{figure}

Suppose that there is just one realizable option $S_{i1}^{k}$, and that the corresponding alternative $S_{1}$ is a proper subset of $S_{t}^{k}$ [Fig. \ref{Fig41a5}(a)]. Then a complementary alternative $S_{2}$ is automatically defined as $S_{2}=S_{t}^{k}/S_{1}$, so that $S_{t}^{k}=\bigcup_{j}S_{j}$ with $S_{j}\cap S_{j'}=\varnothing$ for $j\neq j'$, as described in section \ref{law}. For instance, if you get the idea to throw the ball to the right, the alternative that it nevertheless is thrown to the left is implicitly defined, even if you do not consider it as an option.

It may also be the case that the option $S_{i1}^{k}$ can never be realized, but the complementary option $S_{2}$ is automatically realized at some future time. We have $S_{t}\subseteq\mathcal{NP}_{1}$ and $S_{t}\subseteq\tilde{\mathcal{P}}_{2}$, according to Fig. \ref{Fig33b} and Eq. [\ref{pfp}]. This situation may be illustrated as in Fig. \ref{Fig41a5}(b). It may also be the case that the option is necessarily realized at some future time $n'>n$ [Fig. \ref{Fig41a5}(c)]. There is no complementary option $S_{2}$.

Note that in a deterministic world with complete knowledge, one of the cases in Fig. \ref{Fig41a5}(b) or \ref{Fig41a5}(c) always apply. The physical state is an exact state $Z$, a point in state space, and cannot be partitioned into two or more options. Then it is meaningless to introduce the concepts of alternatives and options as basic ingredients in the formulation of physical law. There is no chance and no `free will' (Definition \ref{freewill}).

In a non-deterministic world there is one possibility left. It may be the case that neither the option $S_{1}$ nor its complement $S_{2}$ can ever be realized [Fig. \ref{Fig41a5}(d)]. There are two possible reasons for this.

The first possibility is that physical law makes it impossible to gain enough knowledge. The option is too specific, like $s_{x}>0$ \emph{and} $s_{y}>0$ for the spin of an electron.

The second possibility is that the initial state $S_{t}^{k}(n)$ of the target object $O_{t}$ happens to be such that $S_{t}^{k}(n')$ for any $n'>n$ will have parts on both sides of the dashed line separating $S_{1}$ from $S_{2}$. One example is the double slit experiment arranged in such a way that it is impossible to ever tell which slit the particle passed. Another initial condition $S'_{t}(n)$ could make it possible for one option to come true, as in Fig. \ref{Fig41a5}(a). That is, the double slit experiment could be prepared with a detector at one of the slits.  

Options may materialize one by one in the mind of subject $k$. We get an extended set of imagined, future states $\{S_{ij}^{k}\}$ of the target object $O_{t}$. These may represent a set $\{O_{ij}\}$ of identifiable objects that exist `in the back of your head' for some time. In the morning, you may wonder what to cook for supper. During the day, more and more ideas may pop up in your head, options that are stored in your memory until you finally decide on your way to the supermarket. This situation is illustrated in Fig. \ref{Fig41a5}(e). The circular set $S_{t}^{k}$ may then represent the state of the round dinner table $O_{t}$. If the set of options is not exhaustive, it is always completed by the automatically generated complementary options `not any of the above', as discussed above and in section \ref{law}.

Let us turn from the characterization of options to the choice of one option. As usual, we are very allowing in the definitions. We are not concerned with the question whether the chosen option comes true, or how the choice comes about. We just want to combine the concepts and quantitities already introduced to define a sequence of events that any detailed description of the process of choice must follow.

For the first time, we will have to distinguish the current knowledge $K(n)$ at time $n$ from the potential knowledge $PK(n)$, where $K(n)\subseteq PK(n)$ (Fig. \ref{Fig3}). That is, we will contrast current awareness with potential awareness, or, using a popular phrase, we will distinguish the conscious from the subconscious. An object $O$ with state $S_{O}$ may belong to the current knowledge $K^{k}$ of subject $k$ or not. If so, we may say that $O$ is in the attention of $k$.

\begin{defi}[\textbf{Current awareness attribute} $Ca$]
The binary attribute $Ca$ is such that $Ca_{O}^{k}(n)=1$ if $O\in K^{k}(n)$, and $Ca_{O}^{k}(n)=0$ otherwise.
\label{currentawarenessdef}
\end{defi}

Just as the binary presentness attribute $Pr$, we let $Ca$ be an attribute that is defined for all objects. They are both seen as fundamental parts of the epistemic formalism.

\begin{figure}[tp]
\begin{center}
\includegraphics[width=80mm,clip=true]{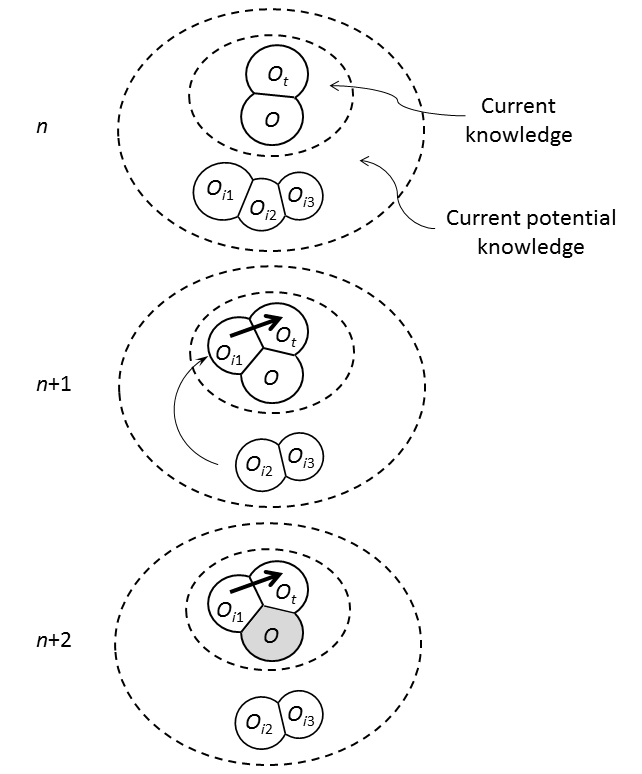}
\end{center}
\caption{One way in which a choice may come about. The knowledge space of one subject is shown. At time $n$ her attention is focused on the target object $O_{t}$ and she is ready for action, as represented by the awareness of the internal object $O$, the change of which defines the action. She has three already formed options $O_{i1}$, $O_{i2}$, and $O_{i3}$ in the back of her head. The next instant (time $n+1$) her intention is to choose option $O_{i1}$, and the instant after that (time $n+2$) she has taken action, represented by a change in the state of $O$, which is part of her body. The arrow represents the interpretation of $O_{i1}$ as an object that defines an option for $O_{t}$ ($I_{it}^{k}=1$).}
\label{Fig41a6}
\end{figure}

A choice by subject $k$ is a sequence of two subsequent events (Fig. \ref{Fig41a6}):

\begin{enumerate}
\item As $n\rightarrow n+1$ one option $S_{i}^{k}$ becomes part of current knowledge, $C_{i}^{k}(n)=0\rightarrow C_{i}^{k}(n+1)=1$.
\item As $n+1\rightarrow n+2$ the state $S_{O}^{k}$ of some identifiable object $O$ undergoes a distinct change $S_{O}^{k}(n+1)\cap S_{O}^{k}(n+2)=\varnothing$. $O$ is always an internal object of $k$, who is aware of the change: $C_{O}(n+1)=C_{O}(n+2)=1$. It may or may not be the target object $O_{t}$, but it is not $O_{i}$. The latter object stays in the current knowledge, so that $C_{i}(n+2)=1$, without any knowable changes: $S_{i}(n+1)\cap S_{i}(n+2)\neq\varnothing$.
\end{enumerate}

Simply put, a choice is an action with a future state of an object in mind. The two steps may be said to represent intention and action, respectively. The action is the aware change of the internal object $O$. If the target object $O_{t}$ is a piece of chocolate, $O$ may be a sudden tension in the muscles of the arm, followed by the vision of the arm stretching out to grab the target object. Just as we do not care if the action is the beginning of a sequence of events that leads to the realization of the option, we are not concerned with the question if the awareness of the option in some sense \emph{cause} the action, or it \emph{just happen}. Our treatment is purely descriptive.

The two-step process of choice is illustrated in Fig. \ref{Fig41a6}. It is not specified whether the option $S_{i}^{k}$ that comes into attention ($Ca=1$) is created in the update $n\rightarrow n+1$ ($Pr=1$), or it is fetched as a memory ($Pr=0$) from latent knowledge ($Ca=0$). The same goes for the target object. Of course, if the option is fetched from memory, there has to be a memory of the target object also.

Let us try to express the essence of choice in condensed form.

\begin{defi}[\textbf{Choice}] A choice is two-step process $n\rightarrow n+1\rightarrow n+2$. Either $Ca_{i}^{k}(n)=0$, or $O_{i}$ is not defined at time $n$. We have $Ca_{i}^{k}(n+1)=Ca_{i}^{k}(n+2)=1$. Further, $Ca_{t}(n+1)=Ca_{t}(n+2)=1$. There is also an internal object $O$ which is not $O_{i}$ such that $Ca_{O}(n+1)=Ca_{O}(n+2)=1$ and $S_{O}^{k}(n+1)\cap S_{O}^{k}(n+2)=\varnothing$.
\label{choice}
\end{defi}

We may have a sequence of choices with the target $O_{t}$ in mind. Then, $O_{i}$ stays in the current knowledge without any knowable changes for a sequence of temporal updates. We may call this a `sustained action', an action in several steps to make the option $S_{i}^{k}$ come true.

We may have an intention without making a choice. In the present formalism, this means that the conditions in Definition \ref{choice} for the update $n\rightarrow n+1$ are fulfilled, but an external object defines the next update $n+1\rightarrow n+2$ rather than an internal object $O$ as in Fig. \ref{Fig41a6}. We may have a future option in mind, but instead of taking action with our own body, we listen to a dripping faucet in the kitchen. Of course, we may make a choice and take action at a later time $n+m$, if we revive the intention in our head.

\begin{defi}[\textbf{Intention}]An intention is a temporal update $n\rightarrow n+1$ such that $Ca_{i}^{k}(n)=0$, or $O_{i}$ is not defined at time $n$, but $Ca_{i}^{k}(n+1)=Ca_{t}(n+1)=1$.
\label{intention}
\end{defi}

We have not specified whether the imagined future state $S_{i}^{k}$ is something the subject wants to achieve or to avoid. This is not needed in the above formal description, although it is an elementary aspect of making a choice. Sometimes the option that gets your attention is negative: if you stand in the street and see a car approaching, you imagine the situation where it hits you, and jump away. In other cases the option you imagine is positive: you see a coin in the street, imagine that it will become yours, and bend down to grab it.

No matter how fuzzy an intention may be, it can be captured by the above formalism. A fuzzy intention simply means that the state $S_{i}^{k}$ of the imagined object $O_{i}$ is large. This situation is sometimes combined with a fuzzy target object $O_{t}$ with a corresponding large state $S_{t}^{k}$. We may say that an intention $I=\{S_{i}^{k},S_{t}^{k}\}$ becomes more precise if it is replaced by $I'=\{(S_{i}^{k})',(S_{t}^{k})'\}$ with $(S_{i}^{k})'\subset S_{i}^{k}$ and $(S_{t}^{k})'\subset S_{t}^{k}$.

The name `intention' given to the event described in Definition \ref{intention} may seem too narrow. We may have a mental image of some target object in our mind even if we do not want to manipulate it. We neither want to achieve nor avoid the imagined state. We call it intention since we will only use the concept in conjunction with a subsequent action, forming a choice.

It is possible to define other kinds of associations between objects than that between the imagined object $O_{i}$ and the target object $O_{t}$, which makes $S_{i}$ an option. To do so we have to introduce other relational attributes than the image attribute $I$ (Definition \ref{imageattribute}). For example, we might use our formalism to define a symbol. We do not pursue these matters here, since they will not be needed in the following.

\section{Individual subjects}
\label{individualsubjects}

We can imagine several alternatives at the same time. The composite wish to throw a ball to the left \emph{and} forward is one example. However, we cannot have two \emph{mutually exclusive} alternatives $O_{i1}$ and $O_{i2}$ in our mind simultaneously, while taking action (Fig. \ref{Fig41a6}). You cannot have the mental images of throwing the ball to the right \emph{and} to the left in your mind at the same time as you make your choice, and takes action.

This is reasonable, but it is nevertheless possible to question the statement. To circumvent the problem, we may simply \emph{define} a subject in such a way that a single individual can never have two contradicting alternatives in mind while making a choice. If there are two such alternatives at work simultaneously, there are, by definition, two or more subjects present.

Such a definition of a subject is appealing since it makes it possible to express what we mean by an individual without having to introduce additional concepts and quantitites. Up to now we have repeatedly referred to individual subjects $k$ without defining what they are - the relation between the individual potential knowledge $PK^{k}$ and the collective potential knowledge $PK$ that corresponds to the physical state has never been clarified. (This matter was briefly discussed in section \ref{collective}.)

The prize to pay is that such a definition of an individual needs the concept of choice as a fundamental ingredient. Since the common potential knowledge $PK$ is constructed from the knowledge of all individuals, and since $PK$ is the basis of the physical state, as defined in this text, the concept of choice becomes fundamental in the physical world view.

\begin{figure}[tp]
\begin{center}
\includegraphics[width=80mm,clip=true]{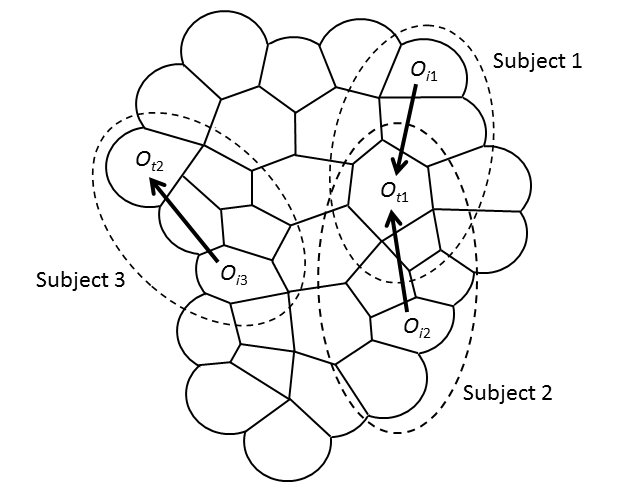}
\end{center}
\caption{The collective current knowledge $K(n)$, consisting of a group of objects in the foam picture (Fig. \ref{Fig5b}). There are three arrows of intention, so that at least three subjects exist at this time. Subjects 1 and 2 are currently aware of two common objects, as indicated by the overlapping dashed ovals, whereas subject 1 is isolated from the other two at this moment. Furthermore, subjects 1 and 2 have independent intentions for the same target object $O_{t1}$.}
\label{Fig41a7}
\end{figure}

Figure \ref{Fig41a7} shows the currently known objects at a given time $n$ in the collective state of potential knowledge, those objects that someone is currently aware of, those objects that have current awareness attribute $Ca_{O}=1$ (Definition \ref{currentawarenessdef}). The relation between the imagined alternatives $O_{i}$ and the target objects $O_{t}$ that interprets the former to be an image of the latter can be regarded as an association object $O_{it}$ that carries an image attribute $I_{it}=1$ according to Definition \ref{imageattribute}.

\begin{defi}[\textbf{Association object} $O_{it}$]
An association object $O_{it}$ is the interpretation of an imagined object $O_{i}$ as an image of another target object $O_{t}$. If $O_{it}$ is part of the potential knowledge $PK(n)$, it carries the internal attributes $Pr_{Oit}(n)=1$ and $I_{it}(n)=1$ (Definitions \ref{presentness} and \ref{currentawarenessdef}). Also, $O_{it}\subset K(N)\subseteq PK(n)$.
\label{assobject}
\end{defi}

These objects are indicated as arrows rather than sets (Fig. \ref{Fig41a4}). The use of arrows is justified since the association is a directed relation: the imagined object represent the target object, but not vice versa. At this stage, no superscripts $k$ labeling individual subjects are used, since we have not yet defined the meaning of an individual.

\begin{defi}[\textbf{Individual subjects}]If there are $M$ association objects $O_{it}^{k}$ in the state of current knowledge $K(n)$ at time $n$, then there are at least $M$ individual subjects at time $n$.
\label{individuals}
\end{defi}

Loosely speaking, individuals are subsets of current awareness capable of making independent choices. We allow the possibility that there are individuals that do not make choices at time $n$, that just `go with the flow'. They perceive but do not act deliberately. Such individuals are invisible in the snapshot $K(n)$ in Fig. \ref{Fig41a7}, they have no associated `arrows of intention'. They may eventually resume activity, while others become passive. Arrows may appear and disappear in the sequence $\{K(n),K(n+1),\ldots\}$ even if no subject is born or dies. The number of arrows is not conserved.

\begin{figure}[tp]
\begin{center}
\includegraphics[width=80mm,clip=true]{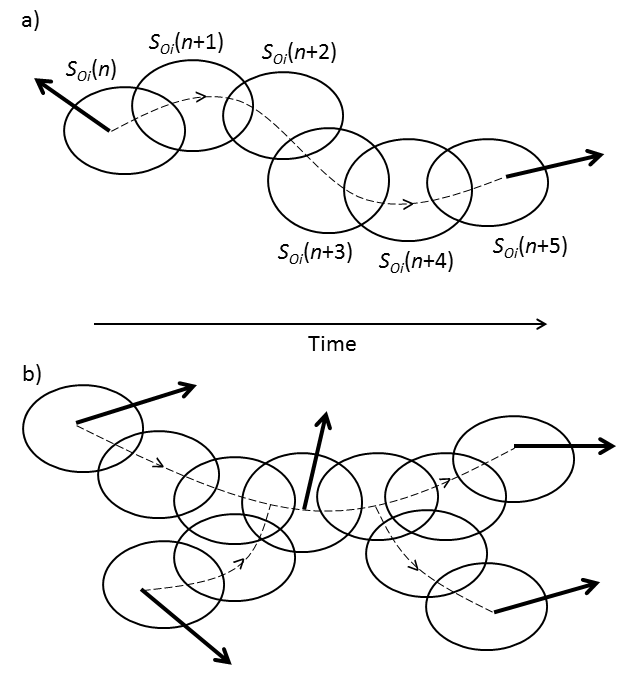}
\end{center}
\caption{(a) Two arrows of intention that can be linked by a chain of imagined objects belong to the same subject. (b) We exclude by assumption the possibility that chains that link different arrows of intention branch. Subjects do not merge or divide. The chains of objects in $PK(n)$ form worms, but no trees or loops. Subjective experiences are personal. It is hard to make sense of the statement that two subjects share the same perception of an object. (The branching object states are represented in object state space $\mathcal{S}_{O}$.}
\label{Fig41a7b}
\end{figure}

This picture calls for a way to associate arrows of intention present at different times to the same individual subject. Such an association requires a subjective link of identifiable objects between two times $n$ and $n''$ where an arrow of intention is present (Fig. \ref{Fig41a7b}). This link consists of a chain of objects with overlapping states, such that you can jump from object to object from intention $A$ at time $n$ to intention $B$ at time $n''$.

However, we cannot allow \emph{any} kind of object in the chain. By definition, the chain starts at time $n$ with the image object $O_{1}\equiv O_{i}$, the source of the arrow that defines the intention $A$. The next object $O_{2}$ in the chain must overlap $O_{i}$. Therefore, in a sense, $O_{2}$ is also an image object associated with an intention. However, this intention does not have to be `active' at time $n+1$. We do not need to require that there is an association object $O_{it}\subset K(n+1)$. Finally, the evolved image object surfaces again, becoming the source for a new intention $B$ at time $n''$, possibly targeting a new object $O_{t''}$.

If this kind of link is present, the two intentions $A$ and $B$ can be said to belong to the same individual. Admittedly, we have dived into deep waters in this discussion. We should think of it merely as a game, where we see how far we can go with the formal concepts that we have introduced. In any case, the picture that emerges resembles the psychological one, where an unbroken flow of imaginations, which may sometimes belong to the subconscious, sometimes surfaces to become sources of conscious intentions and choices that affect the world around us.

The (possibly) new idea here is that the existence of such a continuous flow of internal images \emph{defines} the individual. If the flow is interrupted at some time $n'$ between $n$ and $n''$, we do no longer say that the intentions $A$ and $B$ belong to the same person.

\begin{defi}[\textbf{Link between intentions}]Consider two intentions $A$ and $B$ at times $n$ and $n''$, respectively, with association objects $(O_{it})_{A}$ and $(O_{it})_{A}$ referring to the image objects $(O_{i})_{A}$ and $(O_{i})_{B}$, respectively. Let $(O_{i})_{A}\equiv O_{1}(n)$ and $(O_{i})_{B}\equiv O_{n''-n+1}(n'')$. $A$ and $B$ are linked if and only if there is a chain of objects $\{O_{1}(n),O_{2}(n+1),O_{3}(n+2),\ldots,O_{n''-n+1}(n'')\}$ such that $S_{n'-n+1}(n')\cap S_{n'-n+2}(n'+1)\neq\varnothing$ for each $n\leq n'<n''$.
\label{intentionlink}
\end{defi}

We have defined $O_{1}(n)$ and $O_{n''-n+1}(n'')$ to be such that they belong to the current knowledge, that is, $O_{1}(n)\in K(n)$ and $O_{n''-n+1}(n'')\in K(n'')$. However, for each $n<n'<n''$ it suffices that $O_{n'-n+1}(n')\in PK(n')$.

\begin{defi}[\textbf{Intentions and subjects}]
Two intentions $A$ and $B$ belong to the same subject if and only if there is a link between them according to Definition \ref{intentionlink}.
\label{intentsubject}
\end{defi}

Note that this definition allows for a subject $k$ to become unconsciuos for a while, and still preserve her identity as a subject when she wakes up. According to the discussion in Section \ref{time}, and in particular Fig. \ref{Fig22}, her state of potential knowledge $PK^{k}$ is resting in a `frozen' state if the events that define the temporal update happen to another subject $k'$. It is just carried along, meaning that $PK^{k}(n+1)=PK^{k}(n)$. That is, from the perspecive of subject $k$ there is no temporal lacuna.

\begin{assu}[\textbf{Separate subjects}]The links between intentions never branch. There is no chain of image objects according to Definition \ref{intentionlink} such that two or more arrows of intention at time $n$ can be linked to a single arrow of intention at time $n'>n$. There is no arrow of intention at time $n'$ such that it can be linked to two or more arrows of intention at time $n''>n'$.
\label{separatesubjects}
\end{assu}

This means that subjects can be born and die, but they cannot multiply by division, and they cannot merge. Note that subjects described in this way do not mirror objects, which can indeed both divide and merge, and still keep their identity (section \ref{divideconserve} and \ref{objectmerging}).

\begin{figure}[tp]
\begin{center}
\includegraphics[width=80mm,clip=true]{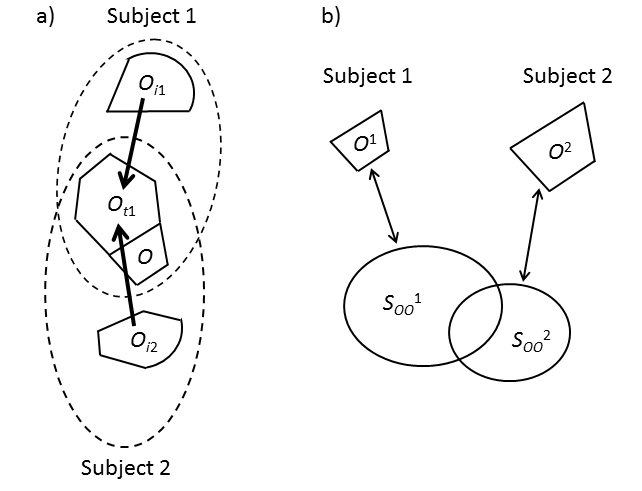}
\end{center}
\caption{(a) The two objects $O_{t1}$ and $O$ in the current knowledge shown in Fig. \ref{Fig41a7} that subjects 1 and 2 both perceive. (b) That they both perceive object $O$ means that the states $S_{OO}^{1}$ and $S_{OO}^{2}$ of their personal perceptions $O^{1}$ and $O^{2}$ of $O$ overlap. Since no stronger condition that is epistemically sound can be formulated, this necessary condition for identity is also treated as a sufficient condition.}
\label{Fig41a9}
\end{figure}

The assumption that subjects are separate and cannot be linked to each other raises the question what links them to the same world. Intuitively, different subjects in the same world have to be able to perceive \emph{the same} object. Let us return to Fig. \ref{Fig41a7}. The dashed ovals encircle the objects perceived by each of the three subjects. Some objects are encircled by more than one oval. The situation is depicted in Fig. \ref{Fig41a9}(a), where subjects 1 and 2 in Fig. \ref{Fig41a7} perceive the common objects $O_{t1}$ and $O$. The former object is the target for two independent intentions, whereas the latter is passively perceived by both.

The statement that two subjects 1 and 2 perceive the same object $O$ cannot mean that their perceptions are identical. If they were, it would not be possible to distinguish the two subjects. Therefore, we cannot write $O^{1}=O^{2}$, where $O^{k}$ denotes the object $O$ as perceived by subject $k$. Rather, we should demand that it cannot be excluded that the two perceived objects are the same [Fig. \ref{Fig41a9}(b)].

\begin{defi}[\textbf{Perception of the same object by two subjects at time} $n$]Subjects 1 and 2 perceive the same object $O$ at time $n$ if and only if $S_{OO}^{1}(n)\cap S_{OO}^{2}(n)\neq\varnothing$, where the states $S_{OO}$ are subsets of the object state space $\mathcal{S}_{O}$ (Definition \ref{objectstatespace}).
\label{sameobject}
\end{defi} 

Equivalently, we may say that the corresponding reduced states must overlap: $\check{S}_{O}^{1}(n)\cap \check{S}_{O}^{2}(n)\neq\varnothing$. The overlap of these states means that there are exact states $Z$ represented by minimal quasiobjects that are consistent with both perceptions $O^{1}$ and $O^{2}$. The notational exercise may seem obscure, but this is just the usual scientific view of a necessary condition for the statement that two people observe the same object: their perceptions of the object are consistent in terms of elementary particles, atoms, molecules, spectrum of light, temperature, pressure and so on. The perception of the same object by two subjects defines a mutual relation between these subjects, as illustrated in Fig. \ref{Fig7b}.

\begin{defi}[\textbf{Perception of the same identifiable object by two subjects during a time interval}]Subjects 1 and 2 perceive the same object $O$ during the time interval $[n,n+m]$ if and only if 1) they perceive the same object at each time $n\leq n'\leq n+m$ according to Definition \ref{sameobject}, and 2) objects $O^{1}$ and $O^{2}$ are identifiable throughout this time interval by subjects 1 and 2, respectively.
\label{sameobjectinterval}
\end{defi}

We may say that the states of objects $O^{1}$ and $O^{2}$ are locked onto each other during an extended period of time. Definitions \ref{sameobject} and \ref{sameobjectinterval} should be compared to the corresponding Definitions \ref{identifiable} and \ref{identifiableinterval} of identifiable objects. We defined identifiability so that two objects that cannot be told apart at two subsequent times are considered to be the same. Analogously, we say that if two subjects perceive objects with physical states that cannot be told apart, the subjects perceive the same object. The necessary condition for identity is used as the defining condition for identity.

\begin{figure}[tp]
\begin{center}
\includegraphics[width=80mm,clip=true]{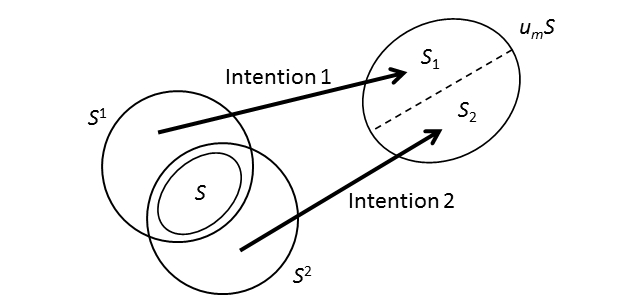}
\end{center}
\caption{Battle of wills. Two subjects can have mutually exclusive intentions $O_{it}^{1}\leftrightarrow\{O_{i}^{1},O_{t}^{1}\}$ and $O_{it}^{2}\leftrightarrow\{O_{i}^{2},O_{t}^{2}\}$ concerning the same target object $O_{t}$, and take action accordingly. In this situation, in the simplest case, the total physical state $S$ is divided into two distinct alternatives $S_{1}$ and $S_{2}$ that evolve into fulfilment of intention 1 and 2, respectively.}
\label{Fig41a8}
\end{figure}

We may apply the considerations about identity to the object that corrsponds to the total potential knowledge $PK^{k}$ of subject $k$. This is the entire world, as subject $k$ perceives it. It must be the same world as another subject $k'$ perceives. If we assume that two subjects that are born into the same world live in the same world until one of them dies, we conclude that the overlapping states shown in Fig. \ref{Fig41a} must overlap as long as all three subjects persist.

\begin{assu}[\textbf{Subjects are locked into the common world in which they live}]If $S^{1}(n)\cap S^{2}(n)\neq\varnothing$, then $S^{1}(n')\cap S^{2}(n')\neq\varnothing$ for all $n'>n$.
\label{commonworld}
\end{assu}

We do not explicitly have to account for the death of subject 1 or 2 in the above assumption if we make the following definition.

\begin{defi}[\textbf{The death of a subject}]Subject $k$ dies at time $n$ if and only if $S^{k}(n-1)\subset S^{k}(n)$ and $S^{k}(n)=\mathcal{S}$.
\label{death}
\end{defi}

In words, death means that the individual state expands to fill the entire state space. Individual potential knowledge becomes zero. Analogously,

\begin{defi}[\textbf{The birth of a subject}]Subject $k$ is born at time $n$ if and only if $S^{k}(n-1)=\mathcal{S}$ and $S^{k}(n)\subset S^{k}(n-1)$.
\label{birth}
\end{defi}

The death of a subject $k$ means that there is no future subject that can have direct access to the memories of $k$; the chain of objects that link the intentions of $k$ to a continuous, personal history according to Definition \ref{intentsubject} is broken at time $n$. In the same way, the birth of $k$ means that there is no objects that bridge her personal experiences to any objects present before her birth.

\begin{figure}[tp]
\begin{center}
\includegraphics[width=80mm,clip=true]{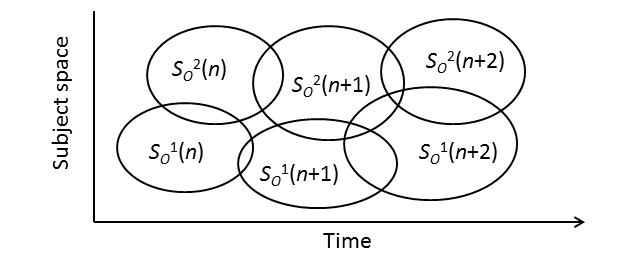}
\end{center}
\caption{To preserve the notion of one single world that several subjects perceive during an extended period of time, the states of some object must overlap, both along the temporal direction, and among the set of subjects that perceive it. Metaphorically speaking, the unity of the world corresponds to a nest of linked rings of quasiobjects, where no ring is completely loose.}
\label{Fig41a10}
\end{figure}

To conclude, the crucial identifier for an individual subject is the capability of independent intention. The ultimate manifestation of independence is the possibility that two subjects have contradictory intentions (Fig. \ref{Fig41a8}). The states of some personally preceived objects always overlap, to create a temporally connected world (Definition \ref{worldidentity}). Also, they overlap the states of some objects perceived by other persons, to create a single world in which we all live. In other words, the unity of the world means that some object states overlap both along the temporal axis $n,n+1,n+2,\ldots$, and along the axis defined by the set of subjects $k,k+1,k+2,\ldots$ (Fig. \ref{Fig41a10}).

\section{Individuality and time}
\label{indtime}

This means that there is an analogy between the degrees of freedom represented by time and individuality, respectively. The  difference is that time is directed, whereas the space of individuals is not - they are not inherently ordered. This analogy is visible in language. The relation between the pronouns `I' and `you' resembles the relation between the adverbs `now' and `then', except for the directionality of the latter pair. Just like everyone is `I' from her own perspective, every moment is `now' from its own perspective. The analogy can be extended, so that reference to third person, `She' or `he', corresponds to a temporally neutral `at five o'clock' or the physicist's `at time $t$'. The epistemic view explored in this text is that the latter neutral perspective is not enough to account for physical law. In fact, it exists only as a perspective a given subject may take at a given time. 

\begin{figure}[tp]
\begin{center}
\includegraphics[width=80mm,clip=true]{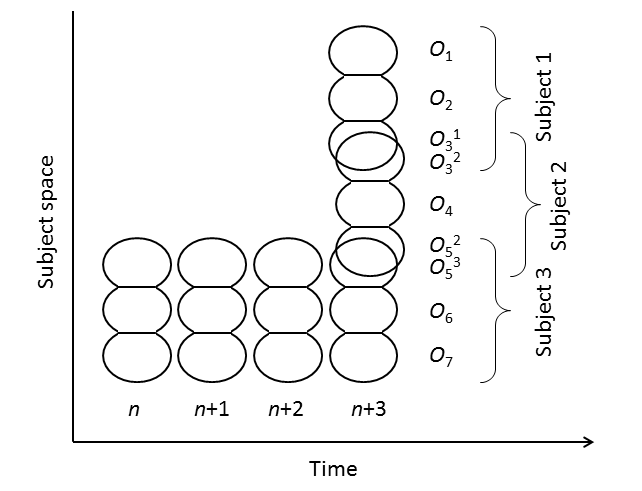}
\end{center}
\caption{The neutral perspective on the space spanned by the set of sequential times and individual subjects. Each group of three objects is perceived at a given time by a given subject. No single subject can have perfect neutral overview like this, transcending time and individuality. Instead, the knowledge contained in the perceptions at other times, or by other subjects, is partially represented in my own perceptions here and now, which may be identified with the three objects $O_{5-7}$ perceived by subject $3$ at time $n+3$.}
\label{Fig41a11}
\end{figure}

Consider the neutral perspective taken in Fig. \ref{Fig41a11}. Each subject is represented by three perceived objects. Subjects 1 and 2 both perceive object $O_{3}$, and subjects 2 and 3 both perceive object $O_{5}$ (compare Fig. \ref{Fig41a9}). However, as discussed above, the identity of the objects perceived by different subjects is indirect. In the same way, the identity of the objects perceived by subject 3 at different times is indirect.

Each group of three objects shown in Fig. \ref{Fig41a11} is separated from all the other groups in the sense that no one can jump to perceive directly a group of objects belonging the past, or a group of objects belonging to another subject. It is impossible to transcend `I' and `now'. This is a consequence of the concept of time that was introduced in section \ref{time}, and of the assumption of separate subjects (Assumption \ref{separatesubjects}).

\begin{state}[\textbf{No transcendence}]No subject can transcend her own perceptions to see through the eyes of another subject or see directly into the past or the future.
\label{notranscendence}
\end{state}

\begin{figure}[tp]
\begin{center}
\includegraphics[width=80mm,clip=true]{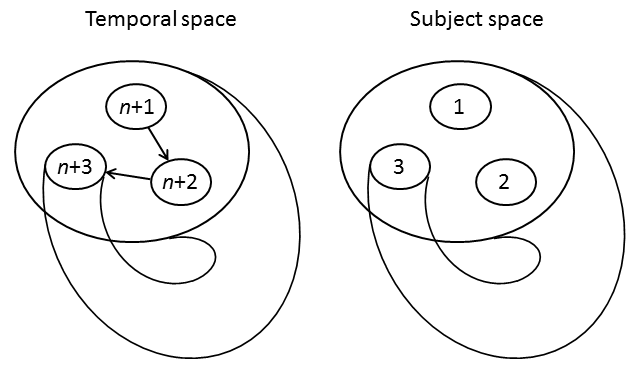}
\end{center}
\caption{(a) Representation of the perceptions at times $n+1$ and $n+2$ in the perception at time $n+3$.
(b) The perceptions of subjects $1$ and $2$ as represented in the perceptions of subject $3$. The analogy between the cases is only partial. Time is directed, so that time $n+2$ is represented in time $n+3$, but not vice versa. In contrast, each subject is represented in each of the other subjects. Compare Fig. \ref{Fig41a11}.}
\label{Fig41a12}
\end{figure}

Instead, the other groups of objects are represented inside the group of objects that `I' perceive `now' (Fig. \ref{Fig41a12}). In Fig. \ref{Fig41a11}, we let time $n+3$ represent `now' and subject 3 represent `I'. In the temporal degree of freedom, the representation of past objects take the form of memories. These are nevertheless perceived right now. As discussed in section \ref{time}, we have to associate an entire space-time to each sequential time $n+3$. All the past objects in this space-time, from times $n+2$, $n+1$, and so on, forms the representation of the past in the present, as illustrated in Fig. \ref{Fig41a12}(a). Since the memory may not be perfect, this representation may not be perfect. In the same way, the representation of perceptions of other subjects in our own perception may not be perfect [Fig. \ref{Fig41a12}(b)].

These relations, where the whole is represented in a part of the whole, resembles the relation between the body and the world expressed in Fig. \ref{Fig1}. There, the world (the body and the external world) is represented in a part of the world (the body). This led to the conclusion that knowledge is incomplete. By the same argument, it can be shown that the representation of other people's perceptions in my own perception must also be incomplete or imperfect.

It is not possible to conclude, using the same argument, that past times cannot be perfectly represented in the present. The reason is the directionality of time. In a perfect representation, the state at a given time must contain a representation of the states of all previous times, but not of any future states. The contradiction is avoided. However, a perfect representation of the past would mean that the amount of potential knowledge is a strictly increasing function of time. If the amount of new knowledge gained at each time instant is approximately constant, this leads to a situation where potential knowledge is increasing linearly with time, so that an exact state $Z$ is approached, contradicting the incompleteness of knowledge. Admittedly, this is just a plausibility argument.

\begin{defi}[\textbf{States of the present and of the past}]Let $S_{M}(n)$ be the physical state of the past, as represented in the present state $S(n)$ at time $n$. That is, $S_{M}(n)\leftrightarrow M[PK(n-1)]$ (Fig. \ref{Fig20} and Definition \ref{presentness}). Let $S_{N}(n)$ be the physical state of the present at time $n$, that is the state of all objects with presentness attribute $Pr=1$. We may write $S_{N}(n)\leftrightarrow PKN(n)$.
\label{presentpast}
\end{defi}

\begin{state}[\textbf{Incomplete representation of past times and of other subjects}]At all times $n$ we have $S(n-1)\subset S_{M}(n)$ and $S^{k}(n)\subset S(n)$, given that there is more than one subject.
\label{incompleterepresent}
\end{state}

Apart from these incomplete representations, there is another sense in which the states of past times and other subjects leave their fingerprints on the present personal state $S^{k}(n)$ - they govern its evolution. In general, we may write

\begin{equation}
S^{k}(n+1)\subseteq u_{1}^{k}S^{k}(n)=f[S(n),S^{k}(n)],
\label{externalcontrol}
\end{equation}
where $u_{1}^{k}$ is the `personal' evolution operator of subject $k$. The crucial fact about $u_{1}^{k}$ is that it depends not only on the personal state $S_{k}$, but also on the collective state $S$.

\begin{figure}[tp]
\begin{center}
\includegraphics[width=80mm,clip=true]{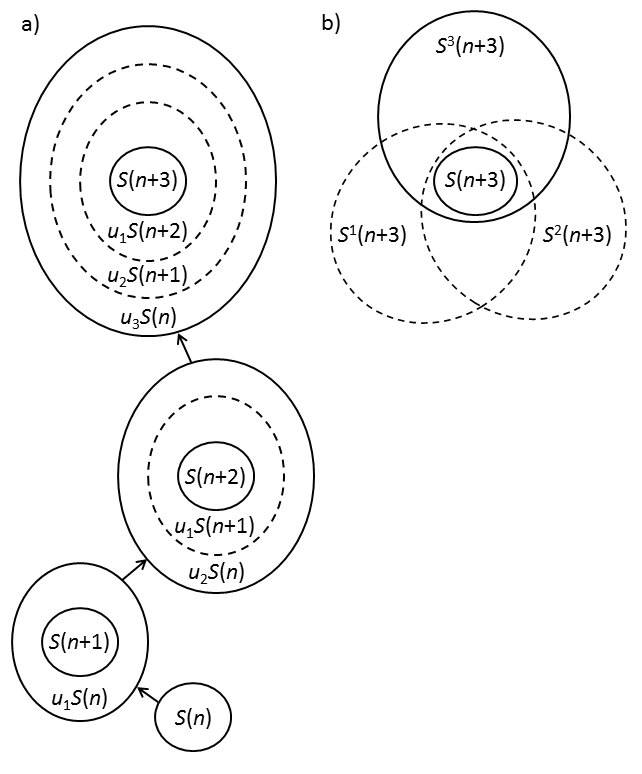}
\end{center}
\caption{(a) Indirect representation of past states $S(n+1)$ and $S(n+2)$ in the present state $S(n+3)$. The existence of these past states forces the present state to be contained inside the dashed ovals, thus constraining its evolution. (b) Indirect representation of the states of subjects 1 and 2 in the state of subject 3. The existence of these other subjects forces the total state $S$ to be contained inside the intersection of the three states, thus constraining the evolution of the state of subject $3$. Compare Fig. \ref{Fig41a11}.}
\label{Fig41a13}
\end{figure}

Consider Fig. \ref{Fig41a13}(a), showing the evolution of the total state $S$ from time $n$ to time $n+3$. The state $S(n+3)$ is constrained by the very fact that the intermediate states $S(n+1)$ and $S(n+2)$ exist. The constraining effect of these intermediate states are shown in the uppermost `russian doll' of states. The existence of the state $S(n+1)$ implies that $S(n+3)$ has to be contained inside outer dashed oval, and the existence of the state $S(n+2)$ implies that it has to be contained inside the inner dashed oval. The two intermediate states correspond to a sequence of two subjectively perceived changes, to two independent observations. The content of these observations, the states they resulted in, determines in part the evolution of my own personal state $S^{k}$ according to Eq. [\ref{externalcontrol}]. It does not matter who made the intermediate observations.

Consider now Fig. \ref{Fig41a13}(b). Quite analogously, the very fact that the other subject $1$ exists constrains $S(n+3)$ to be contained inside $S^{k}(n+3)$ under the top of the left dashed oval. The existence of subject $2$ further constrains $S(n+3)$ so that it also has to be contained under the top of the right dashed oval. The existence and the state of these other subjects determines in part the evolution of the personal state $S^{3}$ of subject 3 according to Eq. \ref{externalcontrol}.

As noted above, the analogy between the degrees of freedom defined by time and individuality is not perfect. The directionality of time gives rise to the russian doll structure of the constraints defined by sequential observations. In contrast, the roles of the three different subjects in Fig. \ref{Fig41a13} are symmetrical, as are the constraints they define for each other when it comes to the evolution of the personal state $S^{k}$.

As a final remark, note that $S(n)$ does not define which objects belong to the perception of which subjects; it does not define the number of subjects nor their individual states $S^{k}$. In this sense, $S(n)$ is an incomplete description of the state of the world. It is defined in the opposite direction: \emph{given} a number of subjects with individual states $S^{k}$, the physical state $S(n)$ corresponds to the union of their potential knowledge.

\begin{state}[\textbf{The physical state is incomplete}]Information about individuality is not contained in the physical state. We have $S=f(\{S^{k}\})$, but $\{S^{k}\}\neq f(S)$.
\label{incompletestate}
\end{state}

\section{The evolution of individual aware knowledge}
\label{knowledgeevolution}
We have defined physical law essentially as an operator $u_{1}$ that acts on the physical state $S$ (Definition \ref{physicallaw}). The physical state $S$ corresponds to the state of potential knowledge $PK$. In addition to $PK$ we have also introduced the state of aware knowledge $K\subseteq PK$ (Fig. \ref{Fig25}). Clearly, physical law does not refer to $K$. What role can then be given to $K$ in the evolution of the world? It should have a crucial role, since we have used $K$ to define such basic notions as intention and choice, as well as the meaning of an individual.

\begin{figure}[tp]
\begin{center}
\includegraphics[width=80mm,clip=true]{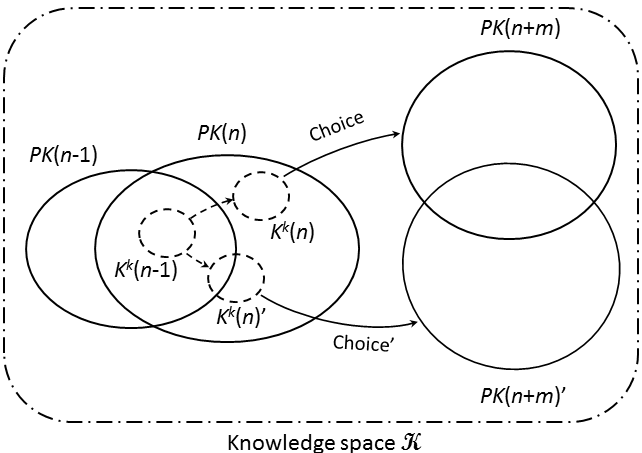}
\end{center}
\caption{The evolution of potential knowledge $PK$ is determined by physical law, but the evolution of the aware knowledge $K^{k}$ of subject $k$ is more free. There may appear two distinct states of aware knowledge $K^{k}(n)$ and $K^{k}(n)'$ from a single aware state $K^{k}(n-1)$. This is a formalisation of the subjective feeling that the target of ones own attention is volatile. The different possible states of aware knowledge enables different sets of choices, which give rise to different future states of potential knowledge $PK(n+m)$, and thus different future physical states $S(n+m)$. The volatile aware individual knowledge may therefore affect the evolution of the world without being fully determined by physical law.}
\label{Fig155}
\end{figure}

That physical law does not refer to the aware state of knowledge $K^{k}$ of a subject $k$ means that we cannot exclude that a single such state $K^{k}(n-1)\subseteq PK^{k}(n-1)\subseteq PK(n-1)$ may evolve into any of two distinct states $K^{k}(n)\cap K^{k}(n)'=\varnothing$ such that $K^{k}(n)\subseteq PK^{k}(n)\subseteq PK(n)$ and $K^{k}(n)'\subseteq PK^{k}(n)\subseteq PK(n)$. This is a formalisation of the subjectively perceived volatility of our attention (Fig. \ref{Fig155}). We may imagine a hunter waiting for prey. At a given moment $n$, her attention may be caught by a faint flapping of wings to the left, corresponding to $K^{k}(n)$, or by the rustle of some leaves in the bushes to the right, corresponding to $K^{k}(n)'$.

Each of these two states of aware knowledge may give rise to an intention and a choice, according to the discussion in Section \ref{aic}. These choices may affect the future physical state of the world in different ways, so that we get $S(n+m)\cap S(n+m)'=\varnothing$ for some $m$. In knowledge space, $PK(n+m)$ and $PK(n+m)'$ will be different (Fig. \ref{Fig155}). If the attention of the hunter is caught by the flapping of wings, the hunter may choose to shoot at the presumed bird. If the attention is caught by the rustle, she may choose to look at the bushes to see whether an animal is moving there. The state of the world will be different in the two cases. A shot is fired in the first case but not in the second, and a head is turned towards the bushes in the second case but not in the first.

This means that the volatility of attention can have real physical effects even if it is not restrained by physical law. This is an expression of the `white hole' in physical law discussed in Section \ref{twoways}. There seems to be an action $\mathcal{A}$ emanating from the subjetive aspect of the world that cannot be explained in terms of the observation $\mathcal{O}$ of a sequence of events governed by physical law (Fig. \ref{Fig24}).´

We may identify events "caused" by an action $\mathcal{A}$ with events or choices that cannot be predicted deterministically or probabilistically. The choice of measurement in quantum mechanics is an example of such an event. Given the choice to measure the spin of an electron along the $x$-axis, the probabilities for different outcomes are known, and given the choice to measure the spin along the $y$-axis, the corresponding probabilities are also known. But there are no probabilities associated with the initial choice of axis. This event may be traced back to a volatile shift of attention, followed by an intention and a choice, just as in the example of the hunter who takes different actions depending on which sound happens to catch her attention.

The necessity to consider events in the bodies of subjects without deterministic or probabilistic causes is discussed in Section \ref{probabilities}. The question how such events may go together with detailed materialism (Assumption \ref{localmaterialism}) is discussed in Section \ref{neurological}.

It would be satisfying if the picture painted here holds water, since that would be an expression of epistemic closure (Assumption \ref{closure}). The epistemic distinction between potential knowledge and aware knowledge would correpond to the physical distinction between events with and without cause, respectively.

\section{Object division and conservation laws}
\label{divideconserve}

In this section, we discuss how the concept of object division and the requirement of identifiability lead to conservation laws.

Object division means that the state of potential knowledge about an identifiable object increases in such a way that it is observed to be composite, to consist of two or more parts. It is not appropriate to say that these parts always are "smaller" than the original object. If the object is minimal, its parts belong to the same set of minimal objects as the original object. Since object division corresponds to a change of potential knowledge, it is always associated with an update $n\rightarrow n+1$ of sequential time.

\begin{figure}[tp]
\begin{center}
\includegraphics[width=80mm,clip=true]{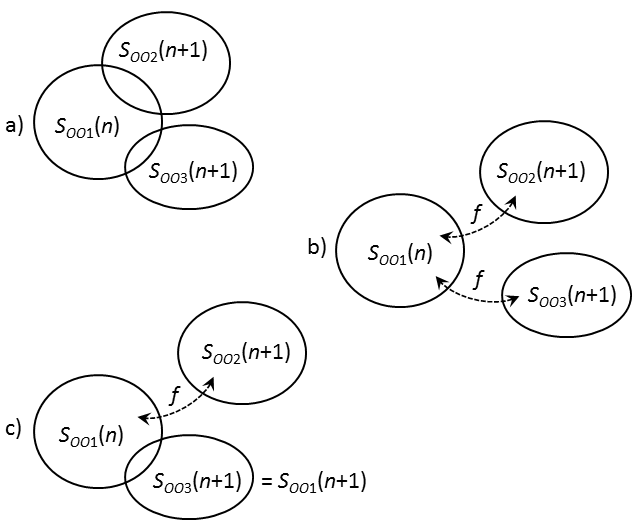}
\end{center}
\caption{When an object divides, corresponding to a temporal update $n\rightarrow n+1$, there are two ways to identify one of its components as a part of the original object. Either the state of the component object overlaps with the state of the original object, or the non-overlapping attribute values are functionally related to the corresponding values of the original object. When the component is actually possible to distinguish from the original object, such an exact attribute relationship is necessary to be sure they belong together. When an object divides into two, these possibilites can be combined into three cases a), b) and c). The object states are represented in object state space $\mathcal{S}_{O}$. Compare Fig. \ref{Fig40}.}
\label{Fig42}
\end{figure}

To be able to speak about object division, there must be a way to identify the parts with the object they are parts of. Figure \ref{Fig40} shows the temporal evolution of an identifiable object $O$ that does not divide. Suppose instead that a division into two parts takes place at time $n$. (An object may divide into three or more parts in the same temporal update, but for clarity we consider first division into two parts.) There are three possible ways to ensure that the two parts can be identified with the original object (Fig. \ref{Fig42}).

\begin{enumerate}
	\item The physical states of the objects $O_{2}$ and $O_{3}$ that correspond to the parts both overlap the state of the original object $O_{1}$, but their states do not overlap with each other, when represented in object state space $\mathcal{S}_{O}$. That is, $S_{OO1}(n)\cap S_{OO2}(n+1)\neq\varnothing$ and $S_{OO1}(n)\cap S_{OO3}(n+1)\neq\varnothing$, but $S_{OO2}(n+1)\cap S_{OO3}(n+1)=\varnothing$.
	
	\item There is no state overlap between any of these three objects, but their attribute values are functionally related, providing the link. We have to require that $O_{1}$, $O_{2}$ and $O_{3}$ share the same set of attributes $\mathcal{A}=\{A_{1},\ldots,A_{m}\}$. Then, for each value $\upsilon_{i1}\in\Upsilon_{i1}$ of any attribute $A_{i}\in\mathcal{A}$ of object $O_{1}$, there is a function $f_{i}$ such that we have the following conditional knowledge: $\upsilon_{i1}=f_{i}(\upsilon_{i2},\upsilon_{i3})$, where $\upsilon_{i2}$ and $\upsilon_{i3}$ are attibute values of objects $O_{2}$ and $O_{3}$, respectively.
	
	\item The state of one part $O_{3}$ overlaps the state of $O_{1}$ (case 1), whereas each attribute value of the other part $O_{2}$ is functionally related to the corresponding attribute value of $O_{1}$ (case 2). That is, $\upsilon_{i1}=\tilde{f}_{i}(\upsilon_{i2})$. Since $O_{3}$ and $O_{1}$ are indistinguishable, they must be seen as the same object $O_{1}$, emitting object $O_{2}$. 
\end{enumerate}

Case 1 corresponds to the situation where the only new knowledge we get at time $n+1$ is that there are now at least two objects. We cannot with certainty distinguish them from the original object, but we can tell them apart. We may listen to a tone, and suddenly realize that there is slight disharmony, so that there are at least two tones, but we we cannot decide with certainty that any of these is different from the single tone we heard to begin with.

In this situation, neither of the parts $O_{2}$ and $O_{3}$ can be said to be the same as the original object $O_{1}$. If we tried to identify one of them with $O_{1}$, we had to do the other component the same favour. Since they are distinct, this leads to a contradiction. In contrast, in case 3 it causes no problems to identify object $O_{3}$ with $O_{1}$, and we should therefore do so.

In cases 2 and 3, the functional relationship of attribute values between two non-overlapping object states can be seen either as a necessary requirement to be sure the objects can be associated, or as a mathematical definition of what mutual association means. A typical example is the relationship between internal attribute values of a decaying particle and its decay products.

\begin{figure}[tp]
\begin{center}
\includegraphics[width=80mm,clip=true]{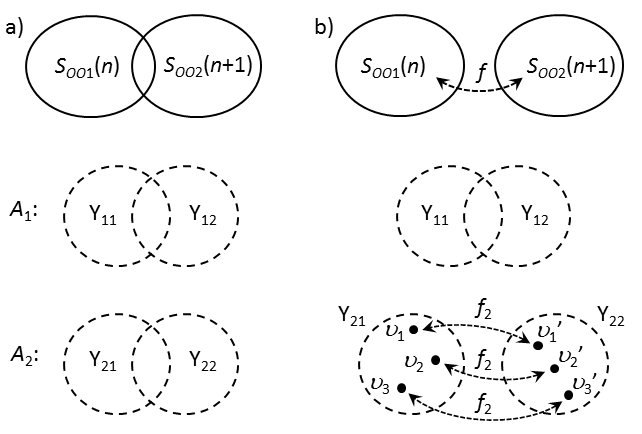}
\end{center}
\caption{(a) The state of an object $O_{2}$, that is a divided part of $O_{1}$, overlaps the state of $O_{1}$, as in Fig. \ref{Fig42}, case 1. The sets $\Upsilon_{i1}$ and $\Upsilon_{i2}$ of possible values of each of their common attributes $A_{i}$ must overlap. (b) The states of $O_{1}$ and $O_{2}$ do not overlap, as in Fig. \ref{Fig42}, case 2 or 3. There is at least one attribute such that $\Upsilon_{i1}$ and $\Upsilon_{i2}$ do not overlap. The possible values of this attribute must each be related according to Eqs. (\ref{frelation1}) or (\ref{frelation}).}
\label{Fig43}
\end{figure}

We talk here about attributes and attribute values that can be assigned to individual objects. Thus we talk in essence about internal attributes. However, in some circumstances relational attributes may be called `pseudo-internal'. If we study a set $\mathcal{O}$ of objects and choose another set $\mathcal{R}$ of objects as a reference frame, spatio-temporal attributes such as linear and and angular momentum can be assigned to each object in $\mathcal{O}$, so that these attributes become pseudo-internal. To exemplify, the total angular momentum of an object is an internal attribute, whereas the angular momentum in a given direction is a pseudo-internal attribute. Additional objects are required to define the direction.

The two possibilities that two objects $O_{1}$ and $O_{2}$ can be associated either by state overlap, or by functional relations of attribute values, can be brought down from the object level to the attribute level (Fig. \ref{Fig43}).

If the states of $O_{1}$ and $O_{2}$ overlap, then the sets of possible values $\Upsilon_{il}$ and $\Upsilon_{il'}$ of \emph{each} common attribute $A_{i}\in\mathcal{A}$ overlap. Just as $S_{OO1}(n)\cap S_{OO2}(n+1)\neq\varnothing$, we have $\Upsilon_{i1}\cap\Upsilon_{i2}\neq\varnothing$ for all $i$. In contrast, if the states of $O_{1}$ and $O_{2}$ do \emph{not} overlap, then we have $\Upsilon_{i1}\cap\Upsilon_{i2}=\varnothing$ for at least one $i$.

In the latter case, suppose that $\Upsilon_{i1}=\{\upsilon_{1},\ldots,\upsilon_{u}\}$, that $\Upsilon_{i2}=\{\upsilon_{1}',\ldots,\upsilon_{u}'\}$, and that $\Upsilon_{i3}=\{\upsilon_{1}'',\ldots,\upsilon_{u}''\}$. Then any indisputable association of the values of attribute $A_{i}$ must be expressed as

\begin{equation}
\upsilon_{j}=\tilde{f}_{i}(\upsilon_{j}')
\label{frelation1}
\end{equation}
for any $j\leq u$. In case 3 [Fig. \ref{Fig42}(c)] this relation holds for the values of at least one attribute $A_{i}$.

In case 2 [Fig. \ref{Fig42}(b)], it may or may not hold for some attribute values. Instead, we require that

\begin{equation}
\upsilon_{j}=f_{i}(\upsilon_{j}',\upsilon_{j}'').
\label{frelation}
\end{equation}
for the values of at least one attribute.

Expressed as conditional knowledge, this relation should be read: "If the values of attribute $A_{i}$ for the divided objects $O_{2}$ and $O_{3}$ are $\upsilon_{j}'$ and $\upsilon_{j}''$, the corresponding value for the original object $O_{1}$ is $\upsilon_{j}=f_{i}(\upsilon_{j}',\upsilon_{j}'')$, for some function $f_{i}$ that is known \emph{a priori}". That the function is known `\emph{a priori}' is another way to say that it should be considered part of physical law. Equation (\ref{frelation1}) should be read in a corresponding way. The implications cannot automatically be turned around. For example, if we know $\upsilon_{j}$ in Eq. (\ref{frelation1}), $\upsilon_{j}'$ is only fixed if $\tilde{f}_{i}$ happens to be invertible.

According to Statement \ref{irreduciblelaw}, physical law should not refer to exact states $Z$. More generally, according to epistemic minimalism (assumption \ref{explicitepmin}), it should not refer to any state that cannot correpond to actual, subjective potential knowledge. The conditions (\ref{frelation1}) or (\ref{frelation}) should therefore only be applied in situations where it can be checked that they hold exactly. This is only possible if 1) the values of $A_{i}$ are known to be discrete, or 2) if there is some conservation law, deduced by other means, that ensures that they are fulfilled.

Regarding condition 1), a continuous set of attribute values means that each of the three values $\upsilon_{j}$, $\upsilon_{j}'$ and $\upsilon_{j}''$ have to be measured with infinite precision to check relation (\ref{frelation}). An uncountable number of bits is required to encode such infinite precision, or, equivalently, an uncountable number of objects. According to Statement \ref{countobjects}, the number of objects is always countable. Thus infinite precision can never be achieved, and to be able to check relations ((\ref{frelation1}) and \ref{frelation}) we have to assume discrete attribute values, separated in such a way that they can be told apart by observation, at least in principle.

As an example of a situation where condition 2) applies, take linear momentum $\mathbf{p}$ (a pseudo-internal attribute). Conservation of momentum can be deduced from the translational symmetry of space. Thus we know, \emph{a priori}, that $\mathbf{p}=\mathbf{p}'+\mathbf{p}''$, without having to rely on the assumption that the momenta take discrete values.

In fact, established physical understanding suggests that \emph{all} conservation laws `deduced by other means' concern relational spatio-temporal attributes. It seems that all of these laws follow from the redundancy in the mathematical description that arises when relational attributes (such as distance) are described as pseudo-internal attributes (such as position). This matter is further discussed in Section \ref{symmetries}.

\begin{state}[\textbf{Conservation laws for relational and internal attributes}]
General conservation laws can be derived for relational attributes from redundancies in their mathematical representation. No such conservation laws can be derived for internal attributes.
\label{deriveconserve}
\end{state}

One may argue that it is not important to be sure that one object can be seen as a part of another. But we have assumed that physical law can be expressed in terms of identifiable minimal objects (Statement \ref{evred}), and we have argued that these must be allowed to divide into other minimal objects. Minimal objects have a precise definition in terms of its attributes and attribute values (Section \ref{minimal}). Therefore, to uphold the assumed identifiability, the rules that associate the divided minimimal objects from the original minimal object must be exact. Thus, for minimal objects, the above discussion is essential. If Statement \ref{allminimalidentity} is accepted, the following statement follows.

\begin{state}[\textbf{Minimal objects have discrete attribute values}]
For minimal objects, the values of all internal attributes are discrete.
\label{minidiscrete}
\end{state}


Consider a set of minimal object species $\mathcal{M}=\{M_{1},M_{2},\ldots,M_{m}\}$. Whenever an object $O_{Ml}$ of species $M_{l}$ is divided into another, distinct, object $O_{Ml'}$ of species $M_{l'}$ we have case 2 or 3 in Fig. \ref{Fig42}. Thus there is an attribute, such as $A_{2}$ in Fig. \ref{Fig43}(b), for which each allowed attribute value of $O_{Ml}$ can be paired with the corresponding allowed attribute value of $O_{Ml'}$. Consider one of these attribute value pairs, for instance $(\upsilon_{3},\upsilon_{3}')$.

Generalize such a pair to a multiplet of corresponding attribute values $(\upsilon_{(1)},\upsilon_{(2)},\ldots,\upsilon_{(m)})$ for all object species in $\mathcal{M}$. We drop the index referring to value number in the pool of allowed values of $A_{2}$, but keep track of which attribute $i$ the values belong to [in fig. \ref{Fig43}(b) we have $i=2$].

We may generalize Eq. (\ref{frelation}) to represent a division of one object $M_{l}\in\mathcal{M}$ to any number of objects in $\mathcal{M}$:

\begin{equation}
\upsilon_{(l)}=f_{i}
\left(\begin{array}{llll}
\upsilon_{(1)} & \upsilon_{(2)} & \ldots & \upsilon_{(m)}\\
q_{1} & q_{2} & \ldots & q_{m}
\end{array}\right).
\end{equation}
The integer $q_{l'}$ is the number of objects of species $M_{l'}$ that is produced in the division. Several occurences of the same attribute value $\upsilon_{(l')}$ among the arguments of $f_{i}$ may be represented by the single number $q_{l'}$ in this way, since there is no inherent ordering of objects. For the same reason, $f_{i}$ should be invariant under variable exchange
 
What form may the function $f_{i}$ take? Without loss of generality, we look for a function such that a division of $O_{Ml}$ to a set of minimal objects having species that form a subset of all species in $\mathcal{M}$ corresponds to setting the attribute values of all object species that do not occur to zero. For example, if $\mathcal{M}=\{M_{1},M_{2},M_{3},M_{4}\}$ and $O_{M1}$ divides to one object $O_{M1}$ belonging to species $M_{1}$ and two objects $O_{M3}$ and $O_{M3'}$ belonging to the same species $M_{3}$, we have

\begin{equation}
\upsilon_{(1)}=f_{i}
\left(\begin{array}{cccc}
\upsilon_{(1)} & 0 & \upsilon_{(3)} & 0\\
1 & 0 & 2 & 0
\end{array}\right).
\end{equation}

\begin{figure}[tp]
\begin{center}
\includegraphics[width=80mm,clip=true]{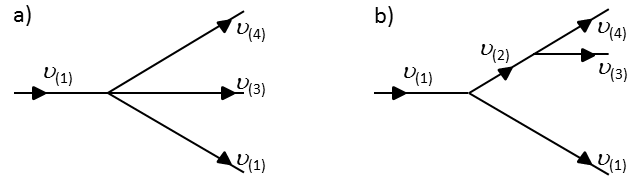}
\end{center}
\caption{Object division in a set of minimal object species with at least four members $M_{1}$, $M_{2}$, $M_{3}$ and $M_{4}$. $\upsilon_{(l)}$ is a value of attribute $A_{i}$ of $M_{l}$ that can be paired with the corresponding value $\upsilon_{(l')}$ of $M_{l'}$ (c.f. Fig. \ref{Fig43}). To maintain identifiability if the attribute values are different for the objects invloved in a division, these values must be functionally related. Since $\upsilon_{(l)}$ is an inherent property of object species $M_{l}$, it does not depend on the way in which a particular object $O_{Ml}$ of this species was created. (a) $O_{M1}$ divides into one object $O_{M1}$, one object $O_{M3}$, and one object $O_{M4}$. (b) $O_{M1}$ divides first into one object $O_{M1}$ and one object $O_{M2}$. At a second temporal update, one object $O_{M2}$ divides into one object $O_{M3}$ and one object $O_{M4}$. The end result is the same in both processes. If the values $\upsilon$ of attribute $A_{i}$ are not all the same, the functional relation must therefore be $\upsilon_{(1)}=\upsilon_{(1)}+ \upsilon_{(3)}+\upsilon_{(4)}$ in process (a). In process (b) we must have $\upsilon_{(1)}=\upsilon_{(1)}+\upsilon_{(2)}$ in the first event, and $\upsilon_{(2)}=\upsilon_{(3)}+\upsilon_{(4)}$ in the second. We have a conservation law for the values of internal attribute $A_{i}$.}
\label{Fig44}
\end{figure}

The value $\upsilon_{(l)}$ is an inherent property of object species $M_{l}$; it does not depend on the division processes that ended up producing the object $O_{Ml}$ that belong to this species. Consider, for instance, the two processes in Fig. \ref{Fig44}. Both have one object of species $M_{1}$, one of species $M_{3}$, and one of species $M_{3}$ as final products. The first process is characterized by the conditional knowledge
\begin{equation}
\upsilon_{(1)}=f_{i}
\left(\begin{array}{cccc}
\upsilon_{(1)} & 0 & \upsilon_{(3)} & \upsilon_{(4)}\\
1 & 0 & 1 & 1
\end{array}\right).
\label{cond1}\end{equation}
The second process is characterized by

\begin{equation}\begin{array}{lll}
\upsilon_{(1)} & = & f_{i}\left(\begin{array}{cccc}\upsilon_{(1)} & \upsilon_{(2)} & 0 & 0\\1 & 1 & 0 & 0\end{array}\right)\\
& &\\
& = & f_{i}\left(\begin{array}{cccc}\upsilon_{(1)} & f_{i}\left(\begin{array}{cccc}0 & 0 & \upsilon_{(3)} & \upsilon_{(4)}\\0 & 0 & 1 & 1\end{array}\right) & 0 & 0\\1 & 1 & 0 & 0
\end{array}\right).
\end{array}\label{cond2}\end{equation}
In principle, an endless number of relations like Eq. (\ref{cond1}) and Eq. (\ref{cond2}) can be produced, and they can all be equated. Physical law may, however, contain selection rules that limit the number of allowed combinations. Anyhow, the only two options that are \emph{guaranteed} to allow a solution to the set of all possible equations of this type are the following:

\begin{equation}\begin{array}{ll}
1) & f_{i}\left(\begin{array}{llll}\upsilon_{(1)} & \upsilon_{(2)} & \ldots & \upsilon_{(m)}\\q_{1} & q_{2} & \ldots & q_{m}
\end{array}\right)=\sum_{l'=1}^{m}q_{l'}\upsilon_{(l')}\\
& \\
2) & \upsilon_{(1)}=\upsilon_{(2)}=\ldots=\upsilon_{(m)}
\end{array}\label{additivity}\end{equation}
For some internal attributes $A_{i}$, option 1) may apply, and for others option 2) may apply.

Clearly, option 1) corresponds to a conservation law for the values $\upsilon$ of attribute $A_{i}$. Whenever an object that possesses attribute $A_{i}$ divides, the attribute values of the divided parts add upp to the original value. If the potential knowledge of the values is defocused, this additivity holds true for each possible value - a kind of detailed balance. Attributes with values that fulfil option 2) may be described as inherited identity markers: `objects of a certain type divide into objects of the same type'.

In Fig. \ref{Fig44}(b), the first process belongs to case 3 in Fig. \ref{Fig42}, and the second process belongs to class 2. If Eq. (\ref{additivity}) is fulfilled by all attributes $A_{i}$ in case 3, we must either have $\upsilon_{(l')}=0$ and $\upsilon_{(l)}\neq 0$ (option 1), or $\upsilon_{(l')}=\upsilon_{(l)}$ (option 2). Clearly, since $M_{Ol'}$ is distinct from $M_{Ol}$ by assumption, we must have $\upsilon_{(l')}=0$ and $\upsilon_{(l)}\neq 0$ for some attribute. For minimal objects, this condition is expressed in Eq. (\ref{distinctobjects}).

If some division processes are forbidden \emph{a priori}, Eq. (\ref{additivity}) does not necessarily apply, as noted above. Imagine, for instance, that we have the following selection rule:  "if a minimal object of species $M_{1}$ divides, it divides into two minimal objects, one of which belongs to species $M_{2}$". This condition excludes Eq. (\ref{cond1}). We may also have backward-referring selection rules, for instance: "minimal objects of species $M_{3}$ and $M_{4}$ always appear in pairs created when a minimal object of species $M_{2}$ divides". This condition again excludes Eq. (\ref{cond1}).

Either of these conditions make it possible to have $\upsilon_{(2)}\neq \upsilon_{(1)}$ as well as $\upsilon_{(2)}\neq 0$ for the process in Fig. \ref{Fig44}(b), in disagreement with Eq. (\ref{additivity}). Equations (\ref{cond1}) and (\ref{cond2}) cannot be used together to exclude the possibility. To be able to associate $O_{M2}$ with its parent $O_{M1}$ in such a case, there must be a function $\upsilon_{(1)}=\tilde{f}_{i2}(\upsilon_{(2)})$. (In the following discussion we omit the second row of arguments of $f_{i}$ to make the notation less cluttered.) 

A minimal object such as $O_{M2}$ with attribute values that causes disagreement with Eq. (\ref{additivity}) when it is created may be given a special mark. We may write $\tilde{O}_{M2}$ instead of $O_{M2}$. The root of such disagreement is always that for some attribute $A_{i}$, option 2) applies for all involved objects except $\tilde{O}_{M2}$ at the same time as option 1) is \emph{not} fulfilled. If Eq. (\ref{additivity}) is fulfilled for all other attributes (or $\tilde{M}_{2}$ causes trouble in the same way as we just described), all the other involved object can be left unmarked.

\begin{figure}[tp]
\begin{center}
\includegraphics[width=80mm,clip=true]{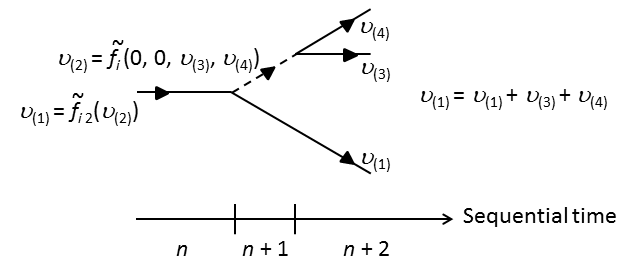}
\end{center}
\caption{Equation (\ref{additivity}) not necessarily applies if there are selection rules that forbid some division processes. In this example, the same objects as in Fig. \ref{Fig44} are involved. There is apriori knowledge that minimal objects of species $M_{3}$ and $M_{4}$ are always created pair-wise when a minimal object of species $\tilde{M}_{2}$ divides. If objects $O_{M1}$, $O_{M3}$ and $O_{M4}$ are observed at some time $n+2$, this conditional knowledge  excludes the process in Fig. \ref{Fig44}(a). At time $n+1$, object $\tilde{O}_{M2}$ breaks Eq. (\ref{additivity}) by having a non-zero attribute value $\upsilon_{(2)}\neq\upsilon_{(1)}$. This makes $\tilde{O}_{M2}$ an `exceptional' object. To be able to associate it with its parent $O_{M1}$ and its children $O_{M3}$ and $O_{M4}$, these objects must nevertheless be related by some functions $\tilde{f}_{i2}$ and $\tilde{f}_{i}$. $O_{M1}$ may be interpreted as an electron emitting a chargeless boson $\tilde{O}_{M2}$ ($\gamma$ or $Z^{0}$), that divides into a neutrino $O_{M3}$ and an anti-neutrino $O_{M4}$. $\tilde{O}_{M2}$ is exceptional in the sense that it is a boson, whereas all the other objects are fermions. The attribute $A_{i}$ is the total spin, with values $\upsilon_{(1)}=\upsilon_{(3)}=\upsilon_{(4)}=1/2$ and $\upsilon_{(2)}=1$.}
\label{Fig45}
\end{figure}

We may consider more complicated possibilities where we have to mark minimal objects in several different ways, but let us concentrate on the case with two object classes. In Fig. \ref{Fig45} the exceptional object $\tilde{O}_{M2}$ is shown as a dashed line. When $\tilde{O}_{M2}$ divides, we have two possibilities: either its components belong to the same marked class as $\tilde{O}_{M2}$, or to the unmarked class. If Eq. (\ref{additivity}) is fulfilled, they should belong to the same marked class as their parent $\tilde{O}_{M2}$. Equation (\ref{additivity}) makes it possible to identify the parent at time $n+1$ with the children at time $n+2$ class-wise as well as individually.

If, on the other hand, all objects $O_{M1}$, $O_{M3}$ and $O_{M4}$ that are present at time $n+2$ fulfil Eq. (\ref{additivity}) in relation to the original object $O_{M1}$ at time $n$, they can all be class-wise identified with $O_{M1}$ as unmarked objects represented by solid lines. Still, $O_{M3}$ and $O_{M4}$ have to be associated with their parent $\tilde{O}_{M2}$ on the individual level. As usual, this means conditional knowledge summarized as $\upsilon_{(2)}=\tilde{f}_{i}(0,0,\upsilon_{(3)},\upsilon_{(4)})$. Since $\tilde{f}_{i}$ refers to the exceptional object $\tilde{O}_{M2}$, it does not have to fulfil Eq. (\ref{additivity}).

It is tempting to interpret the two object classes introduced in this way as fermions (solid lines) and elementary bosons (dashed lines). Of course, it does not follow from our discussion that there has to be more than one object class, or, if so, that there should be two of them. Anyhow, the process shown in Fig. \ref{Fig45} could represent an electron emitting a chargeless boson, that divides into a neutrino and an anti-neutrino. $\tilde{O}_{M2}$ is the boson.

The displayed attribute values $\upsilon$ may then represent total spin. We have $\upsilon_{(1)}=\upsilon_{(3)}=\upsilon_{(4)}=1/2$ and $\upsilon_{(2)}=1$, so that Eq. (\ref{additivity}) is not fulfilled in the first division. Further, $\tilde{f}_{i2}(x)=x/2$ and $\tilde{f}_{i}(0,0,x,y)=x+y$. The latter function happens to conform with option 1) in Eq. (\ref{additivity}) even if this does not follow from the above discussion as a necessity. The total spins of the final set of objects at time $n+2$ are the same as the total spin of the initial object at time $n$. In other words, they fulfil option 2) in Eq. (\ref{additivity}). Thus, all internal attributes values of the objects at time $n+2$ are related to the corresponding attribute values of the object at time $n$ via Eq. (\ref{additivity}). Therefore they shall be associated class-wise as fermions and be represented as solid lines.

We have to check that the behaviour of the other attributes conforms with this conclusion. In the list (\ref{attributelist}), total spin is called $A_{1}$. Let us assign zero values of generation $A_{2}$, baryon number $A_{3}$, and lepton number $A_{4}$ to the boson $\tilde{O}_{M2}$. In the first division, option 1) in Eq. \ref{additivity}) is fulfilled for these three attributes. (For baryon number, option 2) is also fulfilled.) In the second division we have $\tilde{f}_{i}(0,0,x,y)\equiv 0$ for these three attributes, which is perfectly allowed. Regarding electric charge $A_{5}$, option 1) is fulfilled in the first division, and options 1) and 2) are fulfilled in the second division. For colour charge, options 1) and 2) are both fulfilled in both divisions.

Let us summarize some qualities that minimal objects have. Some of them have already been motivated, others are motivated briefly below.

\begin{state}[\textbf{All minimal objects have the same internal attributes}]
All minimal objects $O_{Ml}$ of any species $M_{l}$ in a set $\mathcal{M}$ share the same attributes $A_{i}$.
\label{sameattri}
\end{state}

This is necessary to maintain identifiability when they divide. The condition is easily met by assigning the value zero to an attribute that some object in $\mathcal{M}$ apparently lack.

\begin{state}[\textbf{The number of allowed values of a given internal attribute is the same for all minimal objects}]
The number of elements $\upsilon_{ij(l)}$ in the set of allowed values $\Upsilon_{i(l)}$ of attribute $A_{i}$ is the same for each minimal object species $M_{l}\in\mathcal{M}$.
\label{samevalues}
\end{state}

This is necessary in division processes, given the detailed balance of the functional relations between the each attribute value that is required (Fig. \ref{Fig43}).

\begin{state}[\textbf{The potential knowledge of the internal attribute values does not change in a division}]
The number of elements $\upsilon_{ij(l)}$ according to Statement \ref{samevalues} that are consistent with potential knowledge is the same before and after division.
\label{sameknow}
\end{state}

Potential knowledge of attribute $A_{i}$ is characterized by the set $\Upsilon_{il}$ of values that are not excluded by this knowledge. $\Upsilon_{il}$ is a subset of $\Upsilon_{i(l)}$. To avoid notational confusion, $\Upsilon_{il}$ are the values of $A_{i}$ that are allowed by potential knowledge of an \emph{individual} minimal object $O_{Ml}$. In contrast, $\Upsilon_{i(l)}$ are the values allowed in the definition of minimal object \emph{species} $M_{l}$.

To maintain the detailed balance described above, the number of elements in $\Upsilon_{il}$ must be the same for all component objects that emerge in a division as it was in the original object. This may be seen as a matter of definition. What we do is to distinguish the mere division from the observations that subsequently can be made on the objects that emerge from the division. For example, we may, as a matter of definition, collect all generations of fermions into four minimal objects: a up-like quark, a down-like quark, an electron-like lepton, and one neutrino. If the generations are superposed in the original object just before a division, they are superposed in the emerging component objects, until a measurement increases potential generation knowledge, deciding, for instance, that it was a muon that emerged from the partice reaction rather than an electron.

\begin{state}[\textbf{Conservation laws for internal attributes}]
When a minimal object divides, the only general way to maintain identifiability is that one of two options hold for each attribute value: 1) the values of the emerging objects add up to the value of the original object, or 2) each emerging object inherits the value of the original object.
\label{conservationtypes}
\end{state}

As an example of a law of type 1) we may take charge conservation, and as an example of a law of type 2) we may take the fact that all objects in the same class involved in a division have the same total spin $1/2$.

Among the six internal attributes for elementary fermions that we listed in section \ref{minimal}, there is one that is neither of type 1), nor type 2), namely the generation $A_{2}$. We assigned three possible values $1,2,3$ to this attribute, but these numbers are not conserved in flavor changing weak interactions. A quark of generation 3 (say a top quark) may decay to a quark of generation 2 (a strange quark) without the appearance of another decay product that carries the `lost' generation unit.

Thus the generation attribute has a special status. In section \ref{evconsequences} we will very superficially discuss the idea that this attribute is analogous to the quantum number $n$ in atomic physics, which lists the possible energy eigenvalues for the orbiting electron for given values of the quantum numbers $l$ and $m$ (corresponding to angular momentum and magnetic moment, respectivley). In the case of elementary fermions, the task corresponding to the calculation of energy eigenvalues is the calculation of possible rest mass eigenvalues for given values of baryon number, lepton number and charge. Just as there are several possible quantum numbers $n$ (energies $E$) for each pair $\{l,m\}$, there may be several possible generations (rest masses) for each triplet of quantum numbers that specify baryon number, lepton number and charge. And just as $n$ is not conserved in atomic transitions, the generation quantum number does not need to be conserved in transitions between minimal objects within the same set $\mathcal{M}$ of minimal fermions.

\begin{assu}[\textbf{Minimal objects cannot be copied}]
There is no division process in which option 2) above is fulfilled for all attributes.
\label{nocopy}
\end{assu}

This assumption can be seen as a way to prevent the amount of matter to increase without bound. It excludes case 1) in Fig. \ref{Fig42} for minimal objects.

\begin{state}[\textbf{Classes of minimal objects}]
If there is some attribute for which none of the options in Statement \ref{conservationtypes} are fulfilled, but identifiability can still be upheld, it is possible to divide the minimal objects in two or more classes, in such a way that within each class, one of the options is fulfilled for each attribute.
\label{classes}
\end{state}

This statement is vague. We refer to the discussion in relation to Fig. \ref{Fig45} for more details. To formulate the statement precisely would require quite a bit of notation and technical distinctions that are not very interesting. The reason the statement is made is that elementary fermions and bosons can be seen as two such classes of objects, according to the example discussed above (Fig. \ref{Fig45}).)

However, we will qualify this conclusion in sections \ref{eveqi} and \ref{fermbos}. We will argue that elementary bosons are not really objects at all, in the sense that they cannot be observed. Referring to Fig. \ref{Fig45}, this means that the first division does not correspond to a temporal update $n\rightarrow n+1$. All that is observed are the incoming object $O_{M1}$ at time $n$ and the outgoing object triplet $(O_{M1},O_{M3},O_{M4})$ at time $n+2$. Another event, not pertaining to the division process, may then define the update $n\rightarrow n+1$. If there is no such event, time $n+2$ should be renamed $n+1$. The intermediate step is merely a deduced association between $O_{M1}$ and the object pair $(O_{M3},O_{M4})$, as symbolized an abstract entity $\tilde{O}_{M2}$. This `degradation' of elementary bosons indeed justify the notion that they correspond to an entirely different class of minimal `objects' than the minimal fermions.

\begin{state}[\textbf{Minimal objects do not transform without division}]
If a minimal object $O$ is identified to belong to species $M_{l}$ at time $n$, it cannot be found to belong to another species $M_{l'}$ at a later time, unless $M_{l'}$ is the species of an object that emerged in a division that $O$ underwent at some intermediate time.
\label{notransform}
\end{state}

This is a direct consequence of the temporal invariance expressed in Eq. (\ref{constantobjects}), as part of the definition of a minimal object. 

Up to now we have considered identifiability of internal attributes in division processes. This ensures that object \emph{species} are preserved or functionally related. To make sure that \emph{individual} objects remain properly identifiable in division processes, we need to focus on relational (or pseudo-internal) attributes also.

\begin{figure}[tp]
\begin{center}
\includegraphics[width=80mm,clip=true]{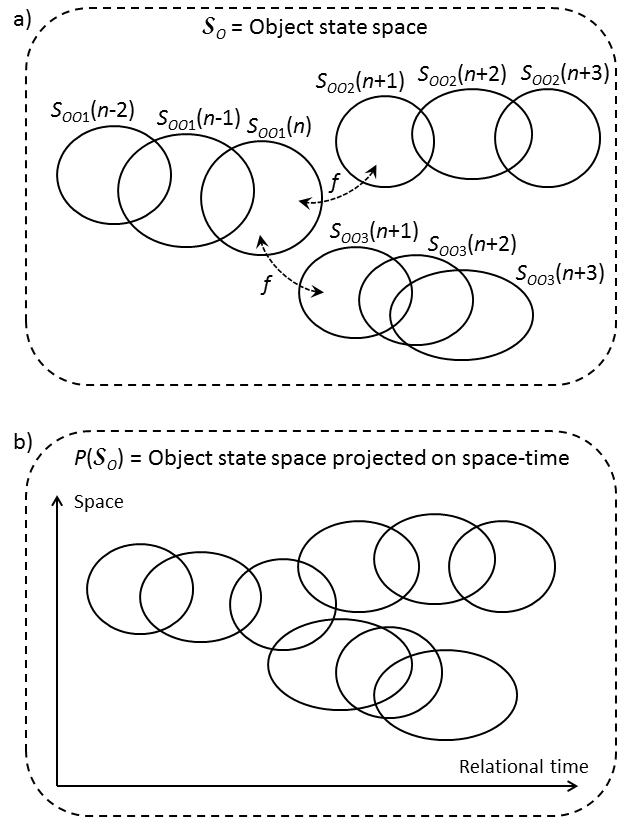}
\end{center}
\caption{(a) In a division, the states of the parts $O_{2}$ and $O_{3}$ at time $n+1$ do not always overlap with the state of the original object $O_{1}$ at time $n$ [\ref{Fig42}(b)]. Instead, the internal attributes of the three objects are linked by the function $f$. (b) However, if we project the object state space $\mathcal{S}_{O}$ onto space-time, the projections of these three states must overlap, to maintain spatio-temporal identifiability. For minimal objects, such an overlap is never present if wo do not make such a projection. That would correspond to copying, which is prohibited by Assumption \ref{nocopy}. Compare Fig. \ref{Fig40}.}
\label{Fig46}
\end{figure}

The conservation of linear and angular momentum provide perfect means to associate the original object with its divided components. For position there is no such conservation law. Let $\mathbf{r}_{4}=(\mathbf{r},t)$ be the pseudo-internal spatio-temporal attribute of an object that is constructed from a set of distances to a group of reference objects. The necessary condition for identifiability is then that the set of values $\Upsilon_{\mathbf{r}_{4}1}$ allowed by potential knowledge of $\mathbf{r}_{4}$ of the original object $O_{1}$ overlaps all the corresponding sets $\Upsilon_{\mathbf{r}_{4}2},\Upsilon_{\mathbf{r}_{4}3},\ldots$ of the objects $O_{2},O_{3},\ldots$ that emerge from the division. The situation is illustrated in Fig. \ref{Fig46}.

\begin{figure}[tp]
\begin{center}
\includegraphics[width=80mm,clip=true]{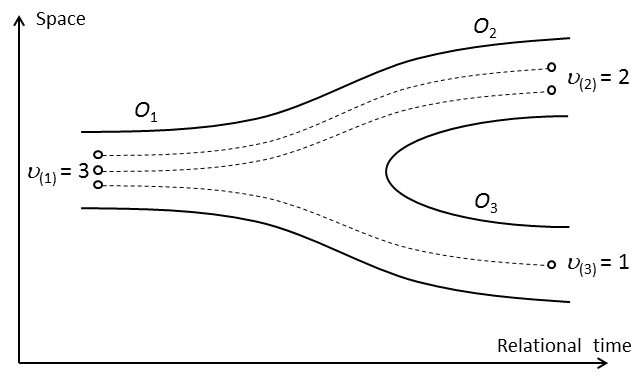}
\end{center}
\caption{The region of space-time consistent with the potential knowledge of a moving object forms a tube. When the object divides, the tube splits into two (compare Fig. \ref{Fig46}). We may imagine that attribute value substance flows freely in the tube. When a minimal (or any other) object $O_{1}$ divides, some substance chooses one direction, and the rest the other direction, leading to the additive rule in Eq. (\ref{additivity}).}
\label{Fig47}
\end{figure}

One may argue that minimal objects do not divide, but rather transform into other minimal objects. But this process can also be described as a division. The dividing worm of object states projected onto space-time (Fig. \ref{Fig46}) stays connected during the transformation or division. Imagine that we sketch the envelope of this worm as a smooth, dividing tube, as in Fig. \ref{Fig47}. According to Statement \ref{conservationtypes}, we have $\upsilon_{(1)}=\upsilon_{(2)}+\upsilon_{(3)}$ for the values of some internal attribute, at least if all three objects belong to the same class. if not, the relation holds anyway for linear and angular momentum. Such conserved attributes may be seen as `attribute substance' that flows inside the tube. When the tube splits, some amount of this substance chooses one road, and the rest chooses the other road. (The use of the smooth envelope of the sequence of spatio-temporal object states is justified in section \ref{evolutionparameter}, where the evolution parameter is introduced.)

This picture conforms with Kant's conclusion that `das Ding an sich' is inaccessible in principle. According to epistemic minimalism, physics should therefore not refer to such a thing. A minimal object species is not defined as something with given identity per se, but rather a prescribed set of attribute values that are confined to a specific region of space-time. These values are just a numerical encoding of the subjective observations of this region that can be made in principle. What is assumed to exit `an sich' is the physical law that limits what attribute values can be observed, determines how often and how the specific region in space-time divides into several such regions, and how the attribute values recombine when this happens.

\section{Object merging}
\label{objectmerging}

Objects sometimes must merge, to avoid that potential knowledge grows without bound. This is excluded, at least in spirit, by Statement \ref{incompleteknowledge}.

Object merging is the reverse of object division. Let us focus on the internal attributes of minimal objects that belong to the set of species $\mathcal{M}$. Let a set $\{O_{Ml}\}$ of minimal objects whose species all belong to the set $\mathcal{M}$ merge into a single minimal object $O_{Ml'}$ with species $M_{l'}\in\mathcal{M}$. Going through all possible cases in the preceding section in reverse, we see that for any such set $\{O_{Ml}\}$, each set $\Upsilon_{il'}$ of allowed attribute values of attribute $A_{i}$ of the final object $O_{Ml'}$ is uniquely determined by the corresponding sets $\{\Upsilon_{il}\}$  of $\{O_{Ml}\}$.

\begin{state}[\textbf{The result is unique when minimal objects merge}]
For any initial set of minimal objects that merge into one, the species of the final minimal object is uniquely determined by the initial set of minimal objects.
\label{uniquemerge}
\end{state}

Physical law may not allow all sets $\{O_{Ml}\}$ of minimal objects to merge into one. The distinct sets of attribute values used to define each initial object $O_{Ml}$, together with the deduced conservation laws for internal attributes, may exclude some combinations. If we allow merging into one object in several time steps, physical law may be such that more sets $\{O_{Ml}\}$ are allowed than if we require that they all merge into one object at once.  

\chapter{\normalfont{MATHEMATICAL REPRESENTATION}}
\label{mathematical}

\section{The evolution parameter}
\label{evolutionparameter}

The statement that any physical law can be encapsulated by an evolution operator $u_{1}$ that tells us what can be said about the temporal update $PK(n)\rightarrow PK(n+1)$ does not capture the apparent continuity of evolution. Subsequent states of knowledge $PK(n)$ and $PK(n+1)$ tend to be more similar than an arbitrary pair of states $PK(n)$ and $PK(n')$. In terms of physical state representations, the numerical values of the attributes that define $\bar{S}(n)$ and $\bar{S}(n+1)$ are close. This is what makes the concept of time meaningful, and makes it natural to use the temporal attribute $t$ in the formulation of physical law. The relational attribute $t$ can be seen as a measure of the degree of dissimilarity of two states $S(n)$ and $S(n')$. Further, the apparent smoothness of evolution makes it tempting to disregard sequential time $n$ altogether; if the degree of dissimilarity is roughly proportional to $n$ for small enough $n$, then we may use an approximate parametrization $n=n(t)$.

The $m$-step evolution operator $u_{m}$ was introduced in Definition \ref{determinism3} as

\begin{equation}
u_{m}S(n)=(u_{1})^{m}S(n).
\end{equation}
That is, $u_{m}S(n)$ is the set of allowed exact states $Z$ at time $n+m$ given the state $S(n)$ at time $n$. In other words, $u_{m}S(n)$ is the smallest set $C'$ such that we can be sure at time $n$ that $S(n+m)$ belongs to $C'$. Clearly,

\begin{equation}\begin{array}{llll}
S(n+m) & \subseteq & u_{1}S(n+m-1) & \subseteq\\
& \subseteq & u_{2}S(n+m-2) & \subseteq\\
& \vdots & &\\
& \subseteq & u_{m}S(n). &
\end{array}\end{equation}

\begin{figure}[tp]
\begin{center}
\includegraphics[width=80mm,clip=true]{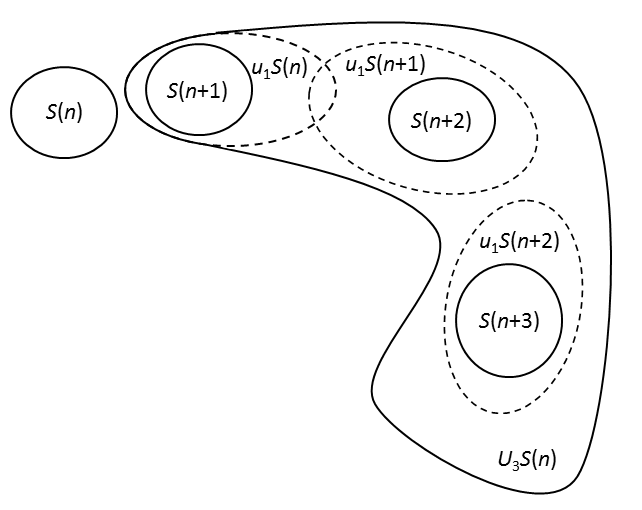}
\end{center}
\caption{The action of the stepwise evolution operator $u_{1}$, and its relation to the world tube defined by $U_{M}$. Since the states $S(n)$ and $S(n+1)$ are distinct by definition, $S(n)$ and $u_{1}S(n)$ does not overlap.}
\label{Fig49}
\end{figure}

We may define

\begin{equation}
U_{M}S(n)=\bigcup_{m=1}^{M}u_{m}S(n).
\label{intev}
\end{equation}
In words, given the state $S(n)$ at time $n$, $U_{M}S(n)$ is the smallest set $C''$ such that any exact state $Z$ outside $C''$ can be excluded as a description of the physical state within the time interval $[n,n+M]$. As we let $M$ grow, the `world tube' defined by $U_{M}S(n)$ gets longer and longer in state space $\mathcal{S}$. The relation between these operators and sets is illustrated in Fig. \ref{Fig49}.

Consider now the evolution of the state $S_{O}$ of object $O$ rather than the total state $S$ (Definition \ref{objectstate}). Let $\Omega_{O}$ be the complement to $O$ (Definition \ref{environmentstate}), representing `all the other objects', so that $S\subseteq S_{O}\cap S_{\Omega_{O}}$. Because of the assumed identifiability of all objects, $O$ and $\Omega_{O}$ can be separated at any time after the partition is defined, regardless the fact that there may be object divisions and mergings in $O$ or in $\Omega_{O}$. Thus it is possible to apply $u_{1}$ to $S_{O}$ as well as to $S_{\Omega_{O}}$ and write

\begin{equation}\begin{array}{lll}
S_{O}(n+1) & \subseteq & u_{1}S_{O}(n)\\
S_{\Omega_{O}}(n+1) & \subseteq & u_{1}S_{\Omega_{O}}(n),
\end{array}\end{equation}
and

\begin{equation}
S(n+1)\subseteq u_{1}S(n) \subseteq (u_{1}S_{O}(n)\cap u_{1}S_{\Omega_{O}}(n)).
\end{equation}
(Compare Fig. \ref{Fig39}.) Remember that epistemic invariance makes $u_{1}$ independent of the state we apply it to - it gives the minimal future state $C$ of $S$, as well as of $S_{O}$ and $S_{\Omega_{O}}$. However, there are two differences to keep in mind when we apply $u_{1}$ to $S$ and $S_{O}$, respectively.

\begin{itemize}
\item We have to write $u_{1}(S)S_{O}$, since the evolution of object $O$ depends in its environment. In contrast, $u_{1}$ has no argument when it is applied to the total state $S$.

\item We always have $S(n)\cap u_{1}S(n)=\varnothing$, whereas we may have $S_{O}(n)\cap u_{1}S_{O}(n)\neq\varnothing$. The latter situtation occurs when a distinct change of $O$ does not define the temporal update $n\rightarrow n+1$.
\end{itemize}

\begin{figure}[tp]
\begin{center}
\includegraphics[width=80mm,clip=true]{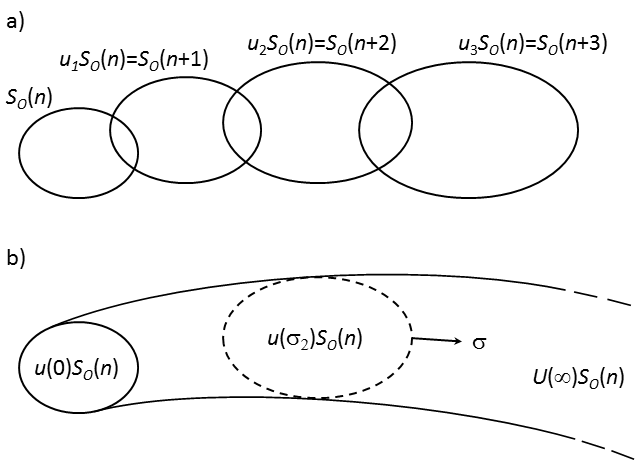}
\end{center}
\caption{(a) If object $O$ is not directly observed for a while after time $n$, its subsequent states $S_{O}(n+1)$, $S_{O}(n+2)$, and so on, are determined by physical law alone: $S_{O}(n+1)=u_{1}S_{O}(n)$, $S_{O}(n+2)=u_{1}S_{O}(n+1)=u_{2}S_{O}(n)$, and so on. (b) The apparent fact that physical evolution is gradual can be expressed as the possibility to find a representation $u(\sigma)$ of the stepwise evolution operator $u_{m}$ that depends continuously on an evolution parameter $\sigma$. Given the state $S_{O}(n)$ of object $O$ at time $n$, the world tube $U(\infty)S_{O}(n)$ expresses all future states of $O$ allowed by physical law and the present total state $S(n)$.}
\label{Fig50}
\end{figure}

In that latter case there is no complete set of present alternatives $\{S_{j}\}$ (Definition \ref{setpresentalt}) such that $S_{O}(n+1)=u_{1}S_{j}$ for some $j$. No reduction of the object state $S_{O}$ occurs at time $n+1$. We may write

\begin{equation}
S_{O}(n)\cap u_{1}S_{O}(n)\neq\varnothing\Rightarrow S_{O}(n+1)=u_{1}S_{O}(n).
\end{equation}
A sequence of such states is shown in Fig. \ref{Fig50}(a). In the same way as for the total state $S$, we can define a world tube $U_{M}S_{O}$ that defines the `trace' of $S_{O}$ in state space during the time interval $[n,n+M]$. 

The apparent fact that evolution is continuous can be expressed as the assumption that it is possible to parameterize the evolution of an object $O$ with an evolution parameter $\sigma\in\mathbf{\mathbb{R}}$ in such a way that

\begin{equation}\begin{array}{lll}
u(\sigma_{1})S_{O}(n) & \leftrightarrow & u_{1}S_{O}(n)\\
u(\sigma_{2})S_{O}(n) & \leftrightarrow & u_{2}S_{O}(n)\\
u(\sigma_{3})S_{O}(n) & \leftrightarrow & u_{3}S_{O}(n)\\
\vdots & &
\end{array}
\label{sigmadef}
\end{equation}
where $\sigma_{1}<\sigma_{2}<\sigma_{3}$, and $u(\sigma)S_{O}$ depends continuously on $\sigma$ [Fig. \ref{Fig50}(b)]. The $S$-dependence of $u_{m}$ is suppressed for clarity. It is natural to choose the parameterization so that $u(0)$ is the unit operator:

\begin{equation}
u(0)S_{O}(n)=S_{O}(n).
\end{equation}

Note that it is not very meaningful to apply the parameterized evolution operator $u(\sigma)$ to the total state and write $u(\sigma)S$, since all that can be said about $S(n+1)$ is a function of $S(n)$. Thus the stepwise evolution $u_{1}$ suffices. There is no freedom to vary the instant at which the entire world is observed, corresponding to a distinct change of $S$. By definition there is no observer outside the world with a clock that can choose to study the changes of the world at certain times.

In contrast, according to the discussion in section \ref{graphical}, there may or may not be lower and upper time limits $\hat{n}$ and $\check{n}$ such that a distinct change of object $O$ occurs with certainty between times $n+\hat{n}$ and $n+\check{n}$:

\begin{equation}\begin{array}{lll}
S_{O}(n+m)\cap S_{O}(n)=\varnothing & \Rightarrow & m\geq\hat{n}\\
S_{O}(n+m)\cap S_{O}(n)\neq\varnothing & \Rightarrow & m\leq\check{n}.
\end{array}\end{equation}

The existence of these time limits is a function of the total state $S(n)$, and so are their values - there is nothing else they can be a function of. However, when we would like to study the evolution of some system $O$, we most often separate $O$ from its complement $\Omega_{O}$ in such a way that this functional dependence resides in $\Omega_{O}$. In theoretical analyses, we often want to evolve a system for an arbitrary amount of time, and then ask the question: what can we say about the state of the system if we observe it at that time? This is, of course, what predictions of the outcome of experiments is all about. In our formalism, `time' in the above sense corresponds to the parameter $\sigma$. We seek

\begin{equation}
S_{O}(\sigma)=u(\sigma)S_{O}(n),
\end{equation}
and ask what present alternatives $S_{j}$ are possible as the outcome of an observation at a time $n+m+1$, given that $S_{O}(n+m)=S_{O}(\sigma)$. In analogy with Eq. (\ref{intev}), we may define the `dressed world tube'

\begin{equation}
U(\sigma)S_{O}(n)
\end{equation}
as the union of all states $u(\sigma')S_{O}(n)$ for $0\leq\sigma'\leq\sigma$ [Fig. \ref{Fig50}(b)].

We can use the evolution parameter $\sigma$ to define what we mean by the statement that an object that we observe at two different times is \emph{the same}, even if it has undergone a distinct change in the meantime. This will be an alternative formulation of the condition of \emph{quasi-identifiability}, as introduced in Defintion \ref{quasiidentifiable}.

\begin{defi}[\textbf{Quasi-identifiable object at times} $n$ \textbf{and} $n+m$]Assume that $S_{O}(n)\cap S_{O}(n+m)=\varnothing$. The object $O$ is quasi-identifiable at times $n$ and $n+m$ if and only if there is a $\sigma>0$ such that $S_{O}(n+m)\subseteq u(\sigma)S_{O}(n)$.
\label{quasiidentifiable2}
\end{defi}

In other words, the object $O$ observed at two different times is considered to be the same if and only if this interpretation is consistent with physical law.

\begin{figure}[tp]
\begin{center}
\includegraphics[width=80mm,clip=true]{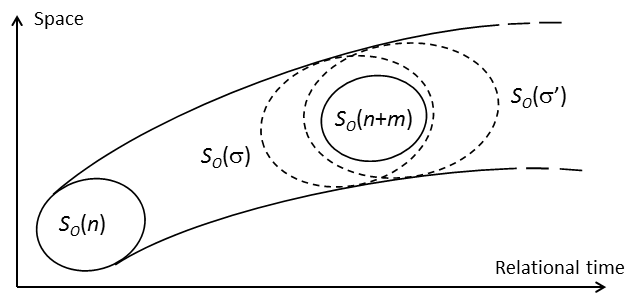}
\end{center}
\caption{Projection of the world tube on space-time (compare Fig. \ref{Fig50}). The projection operator $P$ acting on the states is suppressed for clarity. It is not possible to deduce the value of $\sigma$ uniquely, given the two states $S(n)$ and $S(n+m)$. The evolution parameter $\sigma$ is not an exact measure of the amount of time that has passed between two observations of object $O$, or, speaking relativistically, between two events.}
\label{Fig51}
\end{figure}

The evolution parameter $\sigma$ is no attribute; it is not an observable quantity. Consider the world tube $U(\sigma)S_{O}$ in state space projected on space-time (Fig. \ref{Fig51}). Let $PU(\sigma)S_{O}$ represent this projection.

Suppose that after time $n$, the object $O$ is not observed to undergo a distinct change until time $n+m$. That is, the observed distinct change of $O$ defines the temporal update $n+m-1\rightarrow n+m$. If this observation corresponds to a state reduction (Definition \ref{statereduction}), we have

\begin{equation}\begin{array}{ll}
\forall\sigma>0: & PS_{O}(n+m)\neq PS_{O}(\sigma)\\
\exists\sigma>0: & PS_{O}(n+m)\subset PS_{O}(\sigma).
\end{array}\end{equation}
Since $PS_{O}(\sigma)$ evolves continuously with $\sigma$, it follows from these two statements and the definition of $PU(\sigma)S_{O}$ that there are (at least) two evolution parameter values $\sigma\neq\sigma'$ such that $PS_{O}(n+m)\subset PS_{O}(\sigma)$ and $PS_{O}(n+m)\subset PS_{O}(\sigma')$. Therefore it is not possible to deduce the value of $\sigma$ from the observation of object $O$ at time $n+m$ (Fig. \ref{Fig51}). In other words, $\sigma$ is not an observable measure of the exact amount of time passed between sequential time instants $n$ and $n+m$.

We have not assumed \emph{\emph{a priori}}, however, that a state reduction always takes place when an object is observed to undergo a distinct change. If $m$ is large enough, we expect that $u_{m}S_{O}\cap S_{O}=\varnothing$ (compare Fig. \ref{Fig40}), so that we may have $S_{O}(n+m)\cap S_{O}(n)=\varnothing$, but still $S_{O}(n+m)=u_{m}S_{O}(n)$. Does this mean that we can deduce the values of $\sigma$? If the object state $S_{O}$ were exactly knowable, that would indeed be the case. Given a known $S_{O}(n)$ and a known $S_{O}(n+m)=u_{m}S_{O}(n)$, the value $\sigma=\sigma_{m}$ is determined according its definition in Eq. [\ref{sigmadef}]. But we have argued in section \ref{knowstate} that the boundary of a state cannot be known exactly unless the state is an exact state $Z$. This is never the case since knowledge is incomplete (Statement \ref{incompleteknowledge}). Therefore $\sigma$ cannot be exactly known, either. This remains true also if the state of the object has not knowably changed between the two observations, and for the same reason.

\begin{state}[\textbf{The exact value of the evolution parameter is unknowable}]Suppose that a given object $O$ is observed at times $n$ and $n+m$. Suppose further that we fix a parametrization such that $\sigma(n)=\sigma_{n}$, and such that $\sigma(n-\mu)$ is assigned a definite value $\sigma_{m-\mu}$ for each previous time $n-\mu$ at which $O$ was observed. Even so, the knowledge of the states $S_{O}(n+m)$ and $S_{O}(n)$ is never sufficient to deduce the value $\sigma_{n+m}$ exactly.
\label{unknownsigma}
\end{state}

\begin{figure}[tp]
\begin{center}
\includegraphics[width=80mm,clip=true]{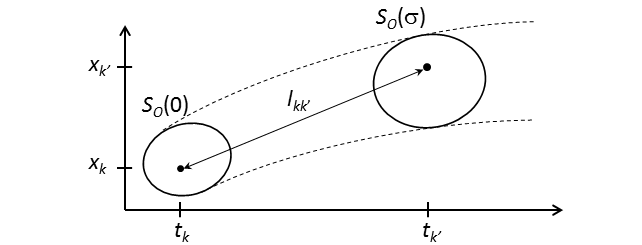}
\end{center}
\caption{(a) The knowledge of the position in space-time of object $O$ at time $n$ is represented as a set $PS_{O}$ of allowed pairs of coordinates $(x_{k},t_{k})$. The evolved states are characterized by corresponding sets of allowed coordinates $(x_{k'},t_{k'})$. The change of position related to $\sigma$ is thus represented by the set of Lorentz distances $l_{kk'}$. The projection $P$ of the states to space-time is suppressed for clarity. Compare Fig. \ref{Fig53}.}
\label{Fig52}
\end{figure}

Since $\sigma$ is not an observable attribute, it is not constrained by the incompleteness of knowledge of attributes: we can simply assign a definite value to it. The evolution parameter is a way to express physical law, it parameterizes the work of the bird that winds up the world; it acts behind the backdrops. Nevertheless, it is closely related to the passage of time. Naively, we would like to have

\begin{equation}
\frac{d\left\langle t\right\rangle}{d\sigma}=constant,
\end{equation}
where we take the appropriate average over the fuzziness of our knowledge of the relational temporal attribute $t$. (That the derivative is constant means that it does not depend on $\sigma$.) However, we have to respect Lorentz invariance. A natural choice of parameterization then fulfils

\begin{equation}
\frac{d\left\langle l\right\rangle}{d\sigma}=constant,
\end{equation}
where $l=\sqrt{c^{2}t^{2}-x^{2}}$ (Fig. \ref{Fig52}). The average should be taken over all allowed pairs of points $(k,k')$. The terms in the sum should not be weighted, since no measure is defined on the exact states $Z\in S_{O}$. Naturally, we exclude all time-like pairs $(k,k')$, for which $s$ becomes imaginary. In Section \ref{eveq} we will derive a simple differential evolution equation bases on these ideas in the particular situation in which we have a well-defined experimental context where a wave function can be defined.

Let us relate the theoretical family of states $S_{O}(\sigma)$ to the actual states of object $O$ in the situation where $O$ is observed at time $n$ and then again at time $n+m$. As $n\rightarrow n+1$, the state of $O$ at time $n$ becomes a memory; its presentness attribute $Pr$ changes from $1$ to $0$. Let $S_{O}(n;n')$ denote the memory at time $n'\geq n$ of the state of $O$ at time $n$, with $S_{O}(n;n)=S_{O}(n)$. Since memory of the past cannot improve with time, or, more formally, because of epistemic consistency (Assumptions \ref{epconsistency}, \ref{epconsistency1} and \ref{epconsistency2}):

\begin{equation}
S_{O}(n)=S_{O}(n;n)\subseteq S_{O}(n;n+1) \subseteq S_{O}(n;n+2)\ldots.
\end{equation}

Before time $n+m$ we get no new information about object $O$. We have $S_{O}(n+1)=u_{1}S_{O}(n)$, $S_{O}(n+2)=u_{2}S_{O}(n)$, and so on. Finally, at time $m$, new potential knowledge is gained: $S_{O}(n+m)\subset u_{1}S_{O}(n+m-1)$. Thus, the potential knowledge at time $n+m$ of the recent evolution of object $O$ corresponds to:

\begin{equation}\begin{array}{lll}
S_{O}(n) & \subseteq & S_{O}(n,n+m)\\
S_{O}(n+m) & \subset & u_{m}S_{O}(n).
\end{array}\end{equation}

\begin{figure}[tp]
\begin{center}
\includegraphics[width=80mm,clip=true]{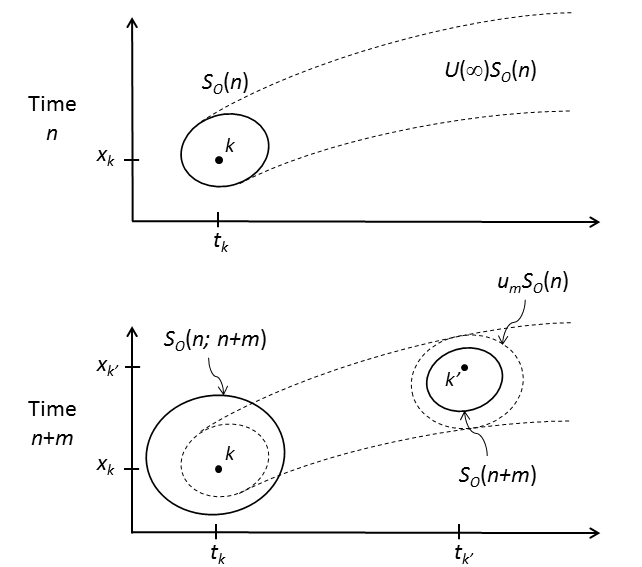}
\end{center}
\caption{(a) At time $n$ the spatio-temporal knowledge of object $O$ is represented by a set of points $PS_{O}(n)$ in a coordinate system. The dashed world tube represents all future states allowed by physical law. (b) The next observation of object $O$ takes place at time $n+m$. Its state turns out to be $S_{O}(n+m)$. The potential knowledge at time $n+m$ of the state of $O$ at the past time $n$ is represented as $S_{O}(n;n+m)$. This state will have greater volume in space-time than $S_{O}(n)$ if memory is imperfect. The projection $P$ of the states to space-time is suppressed for clarity. Compare Fig. \ref{Fig52}.}
\label{Fig53}
\end{figure}

Figure \ref{Fig53} illustrates the relation between these `actual' states and the family of evolved `theoretical` states $S_{O}(\sigma)$ shown in Fig. \ref{Fig52}. Again, we project the states onto space-time. The coordinate system has to be regarded as an object (or rather a set of objects) that stays the same during the time interval $[n,n+m]$. That is, it does not undergo any distinct changes during this time. Thus it makes it possible to define $x$ and $t$ as pseudo-internal attributes of object $O$ valid both at time $n$ and at time $n+m$, and also to define the Lorentz distance $l$ that object $O$ has travelled during this time.

We see that the actual potential knowledge of $l$ at time $n+m$, corresponding to the set of distances between the set of allowed pairs of points $(k,k')$, is not the same as the `theoretical' knowledge, since the states marked with solid lines are different from the states marked with dashed lines. The potential knowledge of $l$ tends to become more defocused because of imperfect memory.

It becomes clear in Fig. \ref{Fig53} that there is one space-time defined for each sequential time $n$. In general, each such space-time consists of objects (or events) belonging to the present as well as objects belonging to the past. In Fig. \ref{Fig53}(b), the state $S_{O}(n+m)$ corresponds to a present object, and the state $S_{O}(n;n+m)$ corresponds to an object of the past.

Note that the states of two past objects may overlap along the temporal axis, meaning that memory of which event occurred first has been lost. More formally, consider two states $S_{O}$ and $S_{O'}$ projected onto space-time. Let $O$ be a past object with presentness attribute $Pr=0$ and characterized by a set of possible positions $k$. If $O'$ is a present object with $Pr'=1$ and possible positions $k'$, then we have the following conditional knowledge, dictated by relativity:

\begin{equation}\begin{array}{ll}
Pr'=1: & c t_{kk'}\geq \left|x_{kk'}\right|,
\end{array}\end{equation}
where $t_{kk'}=t_{k'}-t_{k}$ and $x_{kk'}=x_{k'}-x_{k}$. If, instead, $O'$ is a past object just like $O$, we should write

\begin{equation}\begin{array}{ll}
P'=0: & c\left|t_{kk'}\right|\geq \left|x_{kk'}\right|,
\end{array}\end{equation}
expressing the fact that we cannot be sure \emph{a priori} about the sign of $t_{kk'}$. Finally, if $P=P'=1$ the condition simply reads $t_{kk'}=0$.

It is important to note that the Lorentz distance $l$ is an attribute that relates two objects belonging to \emph{the same} space-time, defined for a given sequential time instant. In contrast, the evolution parameter $\sigma$ makes it possible to jump from one space-time to the next; it makes it possible to foresee, to some degree, the content of future space-times.

\begin{figure}[tp]
\begin{center}
\includegraphics[width=80mm,clip=true]{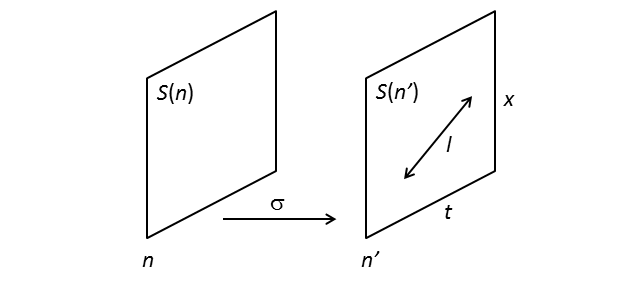}
\end{center}
\caption{Schematic illustration of the relation between the three temporal entities $n$, $t$ and $\sigma$.}
\label{Fig54}
\end{figure}

We have made the concept of time more involved: instead of a single variable $t$, we argue that we need a discrete, sequential time $n$, a relational attribute $t$ (the knowledge of which may be incomplete), and also an evolution parameter $\sigma$ in order to express physical law conveniently. The relation between these three entities is illustrated schematically in Fig. \ref{Fig54}.

Why do we allow the presence of a mysterious parameter $\sigma$ that can never be measured, at the same time as we try to adopt a strictly epistemic perspective on physics? As we discussed at the beginning of this text, we must assume that physical law is absolute, that it is independent of our perception of it. It transcends epistemology. The evolution parameter is simply a convenient way to express such a physical law, given the apparent continuity of evolution it gives rise to.

Finally, let me modify a statement I made at the beginning of this section. It is not $t$ that measures the degree of dissimilarity between two states $S(n)$ and $S(n')$, but rather $\sigma$. (We have also seen that we should focus on the state $S_{O}$ of a group of objects rather than the total state $S$.) What $t$ actually does is to measure the degree of dissimilarity between the memories of two states $S_{O}(n)$ and $S_{O}(n')$ at a later time $n''$. These memories are two parts of the single state $S(n'')$.

\section{Probability}
\label{probabilities}
To make the assignment of a probability to an alternative $S_{j}$ epistemically meaningful, the following conditions should be fulfilled.

\begin{enumerate}
\item Probabilities should only be assigned to options $S_{j}$ that are predefined in the mind of some subject $k$, as described in section \ref{aic}. Otherwise it will never be decided by anyone whether the alternative come true or not, and the concept becomes meaningless.

\item Probabilities should only be assigned to realizable options (Definition \ref{realoption}). It is meaningless to speak of a probability associated with an exact state $Z$, or another state that is too specific to represent actual potential knowledge. The electron with a given spin in both the $x$- and $y$-directions is one example.

\item Probabilities should only be assigned to future alternatives or options $\tilde{S}_{j}$ since probabilities presuppose a predefined test situation. Some time passes between the formation of the options $\tilde{S}_{j}$ at some time $n$ and the realization of one of these alternatives at a later time $n+m$. At least one intermediate time instant $n+1$ has to be allowed to deduce the probabilities from the options (Eqs. [\ref{deduction1}] and [\ref{deduction2}]), before the alternative is realized, so that $m\geq 2$.

\item Probabilites should only be assigned to a complete set of such future alternatives (Definition \ref{setfuturealt}) at knowability level 3 (Table \ref{levels}). We must be sure that one alternative in the complete set is indeed realized within a time limit $\check{n}$ known \emph{a priori}. There has to be a predefined moment of decision. If it may happen that no actual event that corresponds to the realization of a property value $p_{j}$ occurs, the complete set has to contain an additional alternative `nothing happens before the upper time limit $n+M$'. For this alternative we have $\hat{n}=\check{n}=n+M$

\item Probabilities must be knowable \emph{a priori}, before the trial is carried out.

\label{probrequire}
\end{enumerate}

The last requirement can only be fulfilled if the set of options is repeatable, or if it is possible to deduce the probabilities beforehand using symmetries of the observed system $O$ to which the probabilities apply. If the set of options is repeatable, then the probabilities can be calculated before a given trial by the observation of the relative frequency of the realization of the different options in a large collection of identical previous trials. Such repeatability requires that $O$ is sufficiently isolated from the environment, to avoid that the necessary changes in the subsequent states of the environment $\Omega_{O}$ make different repetitions inquivalent.

\begin{figure}[tp]
\begin{center}
\includegraphics[width=80mm,clip=true]{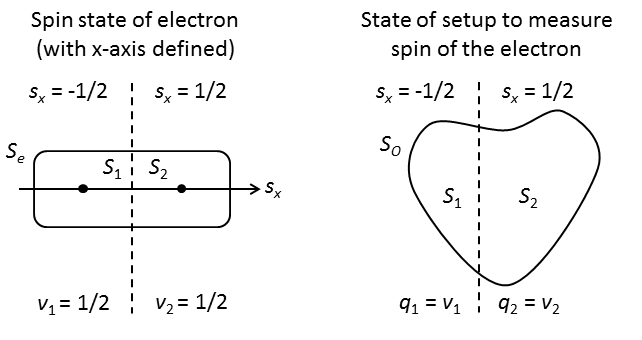}
\end{center}
\caption{Probability as a macroscopic quantity. a) The projection of state space onto the spin in the $x$-direction of a single electron. The projected state $S_{e}$ contains one or both of the (projected) exact states $Z_{1}:\;s_{x}=-1/2$ and $Z_{2}:\;s_{x}=-1/2$. We have $V[S_{e}]=1$ if the spin direction is known, $V[S_{e}]=2$ otherwise. No probability can be assigned to each alternative in the latter case; the relative volumes $v_{1}=v_{2}=1/2$ provide no such information (Definition \ref{relvol}). b) The state $S_{O}$ of an experimental setup aimed at measuring the spin of the electron in the $x$-direction. The futue alternatives $\tilde{S}_{1}$ and $\tilde{S}_{2}$ with probabilities $q_{1}=v_{1}$ and $q_{2}=v_{2}$ correspond to the perception of a detector showing one of two possible results (Statement \ref{probrel}). Once this happens, the corresponding spin of the electron quasiobject can be deduced from the conditional knowledge that relates detector state with spin state. This conditional knowledge is encoded in the state $S_{O}$.}
\label{Fig57}
\end{figure}

It is immediately clear that the five conditions are not fulfilled in all state reductions $S(n)\rightarrow S(n+1)\subset u_{1}S(n)$. Therefore probability is not a fundamental concept in the present approach to physics; it cannot always be used to quantify the chance to observe those things that are not dictated by a deterministic law. While the evolution of any state can be expressed in terms of minimal objects (Statement \ref{evred}), probability should therefore not always enter in this expression. It may be lost somewhere along this reductionist path since it is a measure on an subjectively perceived option $\tilde{S}_{j}$, which is a macroscopic state by definition. Consider the alternative $S_{j}$: \emph{The spin in the $x$-direction of an electron is} $+1/2$. If this statement should have any chance to come true in a knowable sense, the proper future alternative $\tilde{S}_{j}$ must involve a detector and an observer. Therefore, from the present perspective, the probability we assign to the spin direction of an electron is a measure not on the state of the electron, but on the state of the electron together with the states of the detector and the observers (Fig. \ref{Fig57}).

The only measure that can be defined for any set $\Sigma$ is state space volume (Definition \ref{voldef}). For any subset $\Sigma_{j}\subseteq \Sigma$ we may use the state space volume to define the relative volume of $\Sigma_{j}$.

\begin{defi}[\textbf{Relative volume} ]
For any partition $\Sigma=\bigcup \Sigma_{j}$ such that $\Sigma_{j}\cap \Sigma_{j'}=\varnothing$ whenever $j\neq j'$, we let $v[\Sigma_{j},\Sigma]\equiv V[\Sigma_{j}]/V[\Sigma]$.
\label{relvol}
\end{defi}

Whenever the probability associated with $\Sigma_{j}$ exists, we will see that it is given by this relative state space volume. No other determinants are needed.

The only entities there are (in the present approach to physics) on which the probability $q_{j}$ to see the propery value $p_{j}$ may depend are the future alternative $\tilde{S}_{j}$ that corresponds to $p_{j}$, the other alternatives $\{S_{j'}\}$ in the complete set, the system state $S_{O}$, and the total state $S$. We may write

\begin{equation}
q_{j}=f[\tilde{S}_{j},\{S_{j'}\}, S_{O}, S].
\end{equation}

The total state $S$ enters the picture if the system $O$ is not isolated from the environment. As discussed above, such isolation is necessary if $q_{j}$ cannot be deduced from the symmetries of $S_{O}$. Even if $q_{j}$ can be deduced beforehand in this way, these symmetries must include all aspects of the environment that may affect the outcome. That is, we should enlarge $O$ to include all parts of the environment that are relevant to the trial. This means that we can drop the dependence of $q_{j}$ on $S$.

In this way we also exclude hypothetical 'mental influences' on the outcome. Such influences have to be attributed to states of the bodies of observers that does not belong to the system $O$ under study and cannot affect it by means of ordinary physical law. The exclusion of mental influences also makes it possible to drop the dependence of $q_{j}$ on the \emph{other} future alternatives $\{\tilde{S}_{j'}\}$ in the complete set. You cannot influence the probability that something will happen by imagining other alternatives.

\begin{assu}[\textbf{No mental influence on probability}]
For any system $O$ in the state $S_{O}$, whenever the probability $q_{j}$ can be assigned to a future alternative $\tilde{S}_{j}\subset S_{O}$ we have $q_{j}=f[\tilde{S}_{j},S_{O}]$ for some function $f$.
\label{nomental}
\end{assu}

Sets of realizable alternatives $\{S_{j}\}$ (Definition \ref{realizablealt}) correspond to mutually exclusive events. Given Assumption \ref{nomental}, the axioms of probability can therefore be expressed as  

\begin{equation}\begin{array}{rcl}
f[\tilde{S}_{j},S_{O}] & \geq & 0\\
f[S_{O},S_{O}] & = & 1\\
f[\tilde{S}_{j}\cup \tilde{S}_{j'},S_{O}] & = & f[\tilde{S}_{j},S_{O}]+f[\tilde{S}_{j'},S_{O}].
\end{array}
\label{probcond}
\end{equation}

These relations are fulfilled for all complete sets of future alternatives (Definition \ref{setfuturealt}) for which probabilities can be assigned if and only if we identify probability $q_{j}$ with relative volume $v_{j}\equiv v[\tilde{S}_{j},S_{O}]$, meaning that

\begin{equation}
q_{j}=f[\tilde{S}_{j},S_{O}]=V[\tilde{S}_{j}]/V[S_{O}].
\end{equation}

This follows from a direct comparison of Eq.[\ref{probcond}] with the definitions of volume and relative volume (Definitions \ref{voldef} and \ref{relvol}).

\begin{state}[\textbf{Probability is relative volume}]If a probability $q_{j}$ can be assigned to all future alternatives $\tilde{S}_{j}$ in a complete set $\{\tilde{S}_{j}\}$, then $q_{j}=v_{j}\equiv v[\tilde{S}_{j},S_{O}]$.
\label{probrel}
\end{state}

This means that the probability of an alternative does not depend on the shape of the boundary $\partial S_{j}$, but only on the `size' of the region in state space that $\tilde{S}_{j}$ encloses. However, state space volumes and relative volumes are primary in the description of physical law whereas probabilities are secondary.

We might argue that even if a probability cannot be known it still exists in principle. The potential knowledge about any object determine its state $S_{O}$ in principle, so that $V[S_{O}]$ is defined, and the imagination of an option determine $\tilde{S}_{j}$ in principle, so that $V[\tilde{S}_{j}]$ is defined. Therefore $v[\tilde{S}_{j},S_{O}]$ is always defined. Further, $u_{1}S_{O}$ is always defined, so that we can speak about the evolution $u_{1}v[\tilde{S}_{j},S_{O}]$ of the relative volume. However, physical states are unknowable in their details according to Section \ref{knowstate}, so that $v[\tilde{S}_{j},S_{O}]$ is not knowable in general. Therefore we should not use these numbers in general representations $\bar{S}$ of $\bar{S}_{O}$ of physical states and physical law $\bar{u}_{1}$ according to the assumption of epistemic minimalism. We may say that if $v[\tilde{S}_{j},S_{O}]$ is knowable (and the other requirements in the list \ref{probrequire} are fulfilled), then the probability of the alternative $\tilde{S}_{j}$ exists, otherwise not.

\begin{state}[\textbf{The relative volume corresponds to a probability in the frequentist sense}]
Consider a system or object $O$ that to arbitrary precision can be prepared in a given state $S_{O}$ an arbitrarily large number $N$ of times, and can be arbitrarily well isolated from the environment so that $u_{1}(S)S_{O}=u_{1}(S_{O})S_{O}$. Suppose that we can associate to $S_{O}$ a complete set of future alternatives $\{\tilde{S}_{j}\}$ at knowability level 3 (Table \ref{levels}) that corresponds to the set of values $\{p_{j}\}$ of some property $P$. Let $O_{N}$ be a system in which $P$ is repeatedly observed $N$ number of times according to the above, with initial state $S_{ON}$. Let $n_{-}(N)$ be the largest integer $n\geq 0$ such that $n_{-}(N)/N\leq v[\tilde{S}_{j},S_{O}]-\epsilon$, and let $n_{+}(N)$ be the smallest integer $n\geq 0$ such that $v[\tilde{S}_{j},S_{O}]+\epsilon\leq n_{+}(N)/N$. Let $\tilde{S}_{j,N,\epsilon}$ be the future alternative defined for the state $S_{ON}$ that corresponds to the fulfilment of the condition that the number of times $R$ that $p_{j}$ is observed obeys $n_{-}(N)\leq R\leq n_{+}(N)$. Then $\lim_{N\rightarrow\infty}v[\tilde{S}_{j,N,\epsilon},S_{ON}]=1$ for any $\epsilon>0$ such that $0\leq v[\tilde{S}_{j},S_{O}]\pm\epsilon\leq 1$.
\label{volprob}
\end{state}

\begin{figure}[tp]
\begin{center}
\includegraphics[width=80mm,clip=true]{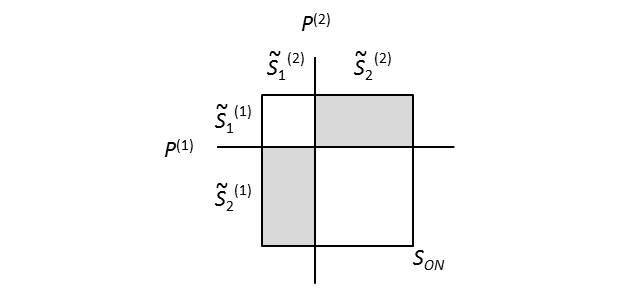}
\end{center}
\caption{A system $O_{N}$, with initial state $S_{ON}$, in which the same property $P$ is repeatedly observed $N$ times with the same set of alternative values $\{p_{j}\}$. In this example we let $P^{(k)}$ represent the $k$:th observation of $P$. A complete set $\{\tilde{S}_{1},\tilde{S}_{2}\}$ of future alternatives is defined by the binary outcome `$p_{1}$' and `not $p_{1}$' for $N=2$. For the system $O_{N}$ we my define a meta-alternative $\tilde{S}_{1,N,m}$ as `$p_{1}$ is observed $m$ times in $N$ trials' (corresponding to the shaded region for $m=1$). Letting $v_{1}\equiv v[\tilde{S}_{1},S_{O}]$, we see that $v[\tilde{S}_{1,2,1},S_{ON}]=2v_{1}(1-v_{1})$. The relative volumes behave just like probabilities, in the axiomatic sense. Temporal invariance of probability means that the volumes of the four compartments are invariant under the interchange $P^{(1)}\leftrightarrow P^{(2)}$. Compare Fig. \ref{Fig34}.}
\label{Fig144}
\end{figure}

In plain language, the relative frequency $F_{j}$ with which an alternative is fulfilled approaches the relative volume $v[\tilde{S}_{j},S_{O}]$ with probability one as the number of trials increases without bound. Put differently, in this limit the future alternative for which property value $p_{j}$ is observed with relative frequency $v_{j}$ becomes identical with $S_{ON}$, meaning that there is no room in the initial state $S_{ON}$ for other alternatives for which the relative frequency turns out to be different than $v_{j}$. This statement can be seen as a variant of Borel's law of large numbers \cite{borel}, expressed in the vocabulary used in this text. Figure \ref{Fig144} provides a rudimentary illustration.

Statement \ref{timeindependence} expresses the temporal invariance of the evolution $u_{1}$. This is the deterministic aspect of physical law. Statement \ref{volprob} can be seen as a corresponding expression of the temporal invariance of the probabilistic aspect of physical law, in those cases such an aspect exists. The probabilistic temporal invariance follows from the deterministic counterpart. The relative frequency $F_{j}$ is potentially a function of the time $n$ at which the data collection from system $O$ begins, the environment $\Omega_{O}$, the number $N$ of repetitions, and the detailed sequence of outcomes $p_{1},p_{2},\ldots,p_{N}$. Temporal invariance means that the outcome $p_{j}$ of the $j$:th observation depends neither on $n$, nor on the preceding sequence of outcomes $p_{1},\ldots,p_{j-1}$ (Fig. \ref{Fig144}). We have assumed that the evolution of $O$, as an idealisation, is independent of the environment. This means that each outcome $p_{j}$ in the sequence must be seen as an independent event, in a temporal as well as a spatial sense. Statement \ref{volprob} is a consequence of that fact.

Sometimes a large number of repetitions can be performed at the same time. This is the case if we are dealing with a system $O$ that can be divided in a large number $N$ of subsystems $O_{l}$ that evolve independently, and where each subsystem is identically prepared. That is, $S_{Ol}=S_{Ol'}$ for all pairs $(l,l')$. In that case we may write $S_{O}=(S_{Ol},N)$ meaning that the state of the system $O$ is the same as the state of any of its parts $O_{l}$, apart from the fact that there is also a number $N$ indicating how many parts it has. The diffraction of light is one example of this situation. Statement \ref{volprob} implies that the same diffraction pattern appears each time an experiment is performed, and that it can be used to decide the probability that a photon hits a given position on the detector screen.

\vspace{5mm}
\begin{center}
$\maltese$
\end{center}
\paragraph{}

\begin{state}[\textbf{No universal probabilities}]It is not possible to define probabilites for future alternatives $\tilde{S}_{j}$ for the total state $S$, only for states $S_{O}\supset S$ of objects $O$.
\label{nouniversalprob}
\end{state} 

To define a probability, there has to be a subject who observes an object and formulate alternatives about its future, who defines options. By the definition of a universe, there cannot be any outside subject that formulate alternatives about it. While this provides an intuitive explanation, I think the statement is most easily motivated in knowledge space, where each object is represented by a `bubble' (Fig. \ref{Fig5b}). In the normal situation, the system $O$ about which someone formulates alternatives consists of a set of bubble objects. There is also a set of objects $O_{A}$ belonging to the body of the observing subject that correspond to the alternatives that she formulate. These must be separate from all objects of the observed system. If we want to include them, we have a new system $O'$, about which another set of alternative-defining objects $O_{A'}$ must be formulated, objects that are separate from $O'$. If $O'$ is the entire universe, this procedure is not possible.

\begin{state}[\textbf{No frequentist probabilities for bodies of subjects}]No probabilities can be assigned by repetition for future alternatives $\tilde{S}_{j}$ that concern the state $S_{O}=\check{S}^{k}_{i}$ of the system defined by the body $\mathcal{B}^{k}$ of a subject $k$ that have memory and finite life time.
\label{nobodyprob}
\end{state} 

The reason is that the condition of repeatability cannot be fulfilled. If a subject $k$ remembers the outcome of the previous repetitions made to verify an assigned probability, she is not in her original state $S_{O}$ just before the next repetition. If she remembers all previous attempts to see which alternative comes true, she is put into a new state at each new repetition. Repeatability can only be upheld in this case if the subject has no memory of previous repetitions, or in the far-fetched case where the life time of $k$ is infinite and her memory only streteched a finite time $\Delta n$ backwards. Then the required limit $N\rightarrow\infty$ can be taken if the time span between each repetition is greater than $\Delta n$.

Apart from the ability to prepare the system in the original state over and over again, another necessary condition for repeatability is that the system can be arbitrarily well isolated from the environment, so that it evolves in the same way according to $u_{1}$ after each preparation. No part of the body of a given subject fulfils this condition. Therefore we cannot even assign probabilities to alternatives that concern body parts whose state are not immediatley associated with memories. These will nevertheless be indirectly affected when memories are formed.

\begin{state}[\textbf{No frequentist probabilities for the union of all bodies}]No probabilities can be assigned by repetition for future alternatives $\tilde{S}_{j}$ that concern the state $S_{O}=\check{S}_{I}$ of the system defined by the union of the bodies $\mathcal{B}^{k}$ of all subjects $k$. The alternative $\tilde{S}_{j}$ should not be possible to reduce to an alternative concerning the states of a proper subgroup of subjects only.
\label{nocbodyprob}
\end{state}

In this collective case, we can drop the memory condition, since if no subject has any memory, even if the system or object that consists of all bodies can, in principle, be repeatedly put in the same original state, there is no one that can remember the outcome of all trials and verify that an assigned probability that alterntive $\tilde{S}_{j}$ comes true is correct.

We have left the possibility open that a probability can be assigned \emph{a priori} by deduction rather than `experimentally' by repetition. By a simple self-reference argument it will be shown below, however, that there are some alternatives and states for which this is not possible.

\begin{state}[\textbf{There are states of some objects for which no probabilities exist}]There are future alternatives $\tilde{S}_{j}$ that concern the state $S_{O}=\check{S}^{k}_{i}$ of an object $O$ that contains the body of a subject that have memory and finite life time, for which no probabilites exist.
\label{noprob}
\end{state}

To verify this statement, consider an experiment where a subject is asked to press one of two buttons a given number $t$ of seconds after the ringing of a bell. The pressing of button 1 corresponds to the future alternative $\tilde{S}_{1}$ and the pressing of button 2 corresponds to the future alternative $\tilde{S}_{2}$. The state $S_{O}$ of the entire experimental setup - including the body of the subject - just after the ringing is such that both alternatives $\tilde{S}_{1}$ and $\tilde{S}_{2}$ are realizable, so that $S_{O}=\tilde{S}_{1}\cup \tilde{S}_{2}$ with $V[\tilde{S}_{1}]>1$ and $V[\tilde{S}_{2}]>1$.

Before moving on, let us problematize this setup. First, in a deterministic world it cannot be realized. Then we have $V[S_{O}]=1$ so that $S_{O}$ cannot be divided into two realizable alternatives $\tilde{S}_{1}$ and $\tilde{S}_{2}$, each having a non-zero volume. However, in an indeterministic world such as ours where potential knowledge is incomplete, such alternatives can be considered in principle. But can they occur in practice? Let us argue by contradiction that they can.

\begin{figure}[tp]
\begin{center}
\includegraphics[width=80mm,clip=true]{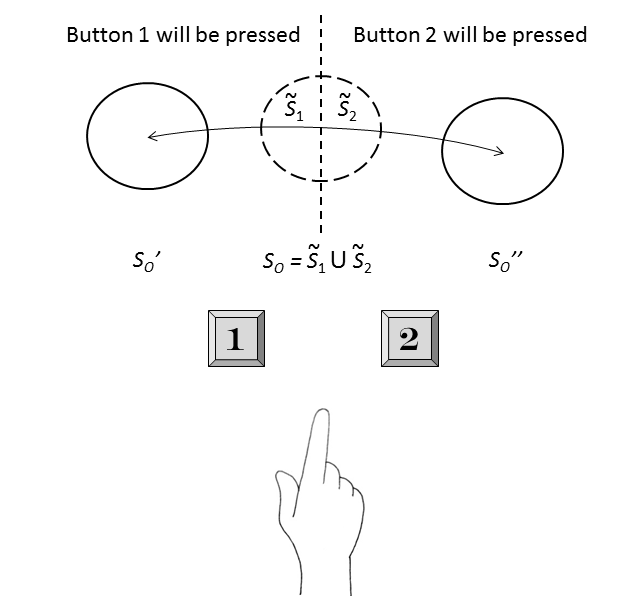}
\end{center}
\caption{A system $O$ that contains a subject that is asked to press one of two buttons. If there are states $S_{O}$ for which it is knowable in principle which button will be pressed, then, by interpolation, there should be a realizable state that is divided by the line in property space that separates the two alternatives. In other words, there are realizable states for which the outcome of such a choice is unknowable in principle.}
\label{Fig58}
\end{figure}

Assume therefore that there cannot be two realizable alternatives $\tilde{S}_{1}$ and $\tilde{S}_{2}$ even if the world is indeterministic. Then either $S_{O}\subset\mathcal{P}_{1}$ or $S_{O}\subset\mathcal{P}_{2}$, where $\mathcal{P}_{j}$ is the region in state space in which button $j$ will be pressed (Eq. [\ref{pfp}]). We may make the experimental setup symmetrical so that there is nothing special about button 1 or button 2. This means that if we may have $S_{O}\subset\mathcal{P}_{1}$, then we may also have $S_{O}\subset\mathcal{P}_{2}$. Simply put, is is possible to press both buttons. Choose two such states $S_{O}'\subset\mathcal{P}_{1}$ and $S_{O}''\subset\mathcal{P}_{2}$ (Fig. \ref{Fig58}). In an indeterministic world we always have $S_{O}\subset S$ so that $V[S_{O}]>V[S]>1$. Therefore $V[S_{O}']>1$ and $V[S_{O}'']>1$. Imagine that we move $S_{O}$ gradually from $S_{O}'$ to $S_{O}''$. We may define a family of states $S_{O}(\alpha)$ parameterized by $\alpha\in [0,1]$ such that $S_{O}(0)=S_{O}'$ and $S_{O}(1)=S_{O}''$. There is no circumstance \emph{\emph{a priori}} that excludes any state $S_{O}(\alpha)$ in such an interpolation. Since $V[S_{O}(\alpha)]>1$ for all $\alpha$, there will be at least one realizable state $S_{O}\in \{S_{O}(\alpha)\}$ such that $V[S_{O}\cap\mathcal{P}_{1}]>1$ and  $V[S_{O}\cap\mathcal{P}_{2}]>1$. These two intersections ought to correspond to two realizable alternatives $\tilde{S}_{1}$ and $\tilde{S}_{2}$. If so, we have contradicted our assumption, and the outcome of the choice between the buttons may very well be unknowable.

Having escaped the problematization of $\tilde{S}_{1}$ and $\tilde{S}_{2}$, let us move on. The task is to construct an example involving these alternatives that confirms Statement \ref{noprob}. Suppose that there is some procedure by which the probabilities $q_{1}=v[\tilde{S}_{1},S_{O}]$ and $q_{2}=v[\tilde{S}_{2},S_{O}]$ can be determined after the ringing of the bell, and that the result is ready to be announced within $t$ seconds, before the subject presses one of the buttons. Suppose also that this procedure is part of the experimental setup, as well as a mechanism with which the result is announced to the subject. Further, the setup is assumed to be such that the subject is instructed to press button 1 if $q_{1}<q_{2}$, and to press button 2 if $q_{1}\geq q_{2}$. Clearly, regardless the procedure used to deduce the probablities, the result will be wrong. This means that we have an example of a state $S_{O}$ with two future alternatives involving the state of a subject for which no probabilities can be deduced before one of the alternatives is realized. Therefore Statement \ref{noprob} holds.

This gedankenexperiment can be related to those experiments \cite{freew,soon} in which analysis of brain activity is used to predict a decision before the subject is aware of it herself. Benjamin Libet let subjects decide at a time of their own choosing to move a hand. He analysed so called readiness potentials to conclude that the decision was made 500 ms before the subject reported that she consciously decided to take action. Soon et al. went a step further and deduced in advance not only \emph{that} a decision had been made but also \emph{which}. The subjects were asked to press one of two available buttons. By the use of functional magnetic resonance imaging (fMRI) the researchers were able to predict, with significant accuracy, whether a subject was going to press the left or the right button, several seconds before the subject herself reported that she made up her mind.

It is hard to define free will. A negative definition, based on what it is not, is the following.

\begin{defi}[\textbf{Free will}]
The possibility to take one of several actions the choice of which is neither dictated by necessity nor governed by probabilities.
\label{freewill}
\end{defi}

The reported experiments demonstrate that some choices that are subjectively perceived as free are indeed goverened by probabilities that can be deduced beforehand, so that there is no free will in the above sense.

These experiments do not rule out free will for all types of choices, however. Sometimes alternatives unknown beforehand materialize within a second before a conscious decision is made. A hunter may suddenly spot something that moves in the bushes, and must decide immediately whether to shoot or not. In such situations, probabilistic predictions of the outcome several seconds in advance are, of course, impossible, suggesting that the kind of brain activity Soon et al. analyzed to predict the outcome cannot be used for all kinds of conscious choices. In a controlled experiment, a pair of random objects may appear simultaneously on a computer screen, and the subject may be asked to click on one of them as fast as she can.

Libet himself introduced the idea of `free won't', the possibility to veto, until the last moment, a choice that has been building up consciuously or unconsciously in the brain for an extended period of time. The self-referential gedankenexperiment presented above may be described in this way: the subject negates any attempt to predict her behaviour by doing the opposite to what is predicted. Soon et al. could have implemented this setup by telling the subject the prediction of which button she was going to press during those seconds that were available after the prediction, but before she actually did press one of the buttons.

The conclusion expressed in Statement \ref{noprob} that there are realizable alternatives, involving states of subjects, with no associated probabilities, means that in some cases will is indeed free according to Definition \ref{freewill}. This conforms with the primary role given to intention in section \ref{aic}. In Definition \ref{intention}, intention is described as the appearance of an option in the active mind of a subject, an internal object that is an image of a possible future state of another object. If will is not free, intention should be regarded a secondary, deduced concept. It should not be given a crucial role in a fyndamental formalism that aims to describe physics.

It can be argued that the appearance of a set of options cannot be the result of a probabilistic or deterministic process. Just as in the motivation why there cannot be any universal probabilities (Statement \ref{nouniversalprob}), we end up in infinite regress if we try.

To calculate the probability for the appearance of a set of options $\mathcal{A}=\{S_{j}\}$ in the mind of some subject $k$ in state $S_{O}$, this set of alternatives must already have appeared, in the following sense. $\mathcal{A}$ can be seen as a meta-option that is chosen by $k$ rather than any other set of alternatives $\mathcal{A}'$. To deduce the probability for this event, the relative state space volume $V[S_{\mathcal{A}}]/V[S_{O}]$ must be calculated by some subject $k'$, possibly with the help of some apparatus. In any case, $\mathcal{A}=\{S_{j}\}$ must already have appeared in $k'$. To deduce the probability for the appearance of $\mathcal{A}=\{S_{j}\}$ in $k'$ we must then invoke a third subject $k''$, and so on. In conclusion, either the set of options $\mathcal{A}$ has always been present in the collective state of potential knowledge $PK$, or it appeared in some subject $k$ at some time as an act of free will.

\begin{state}[\textbf{Appearance of intention is the result of free will}]There is no probability associated with the subjective appearance of one complete set of options rather than another.
\label{freeintention}
\end{state}

From a symmetry point of view, the primary status of intention goes together well with the assumption of intertwined dualism (Fig. \ref{Fig24}). It is the channel through which the subjective aspect of reality affects the objective aspect, just as physical law is the channel through which the objective aspect affects the subjective aspect. None of them can be reduced to the other, they are both primary.

It should be noted that the self-reference argument used to motivate Statement \ref{noprob} does not rely on the presence of a subject. She could be replaced by an apparatus that is programmed to act against the prediction. For example, in a double slit experiment, the task may be to deduce the probability of the passage of an electron through each slit given the presence of a probability calculating device that is coupled to a generator that creates an electric field that steers the electron towards the slit with the lowest calculated probability.

In this case, however, true probabilities do nevertheless exist. For a given probability calculating device, the setup can be isolated and the experiment repeated an arbitrary number of times so that probabilities can be estimated in the frequentist sense. That the probabilities deduced \emph{\emph{a priori}} are all wrong does not matter.

At the beginning of this section we stated boldly that it is immediately clear that there are some state reductions $S(n)\rightarrow S(n+1)\subset u_{1}S(n)$ to which no probability can be associated, taking the four requirements into account. After that we argued that this is, in fact, always the case when we consider the total state $S(n)$ (Statement \ref{nouniversalprob}). It is therefore sufficient to contemplate the reduction of object states $S_{O}(n)\rightarrow S_{O}(n+1)\subset u_{1}S_{O}(n)$. In so doing, we gave examples of such state reductions without probabilities, namely those where the observed obect $O$ contains bodies of subjects (Statement \ref{noprob}).

There are other examples. Suppose that we see a branch falling off a tree while we pass it in a park. We may imagine a probability that tells us the odds for the event that it will fall in the time span $T$ during which we are able to see it as we walk by. However, in that case there are no predefined options, it is not possible to recreate the initial state of the tree perfectly an arbitrary number of times in order to determine the frequentist probability, and there is no apparent symmetry that makes it possible to deduce it. Nevertheless, if we inspect the fallen branch afterwards, we might notice that its base is rotten. This could make it possible to \emph{estimate} the probability \emph{a posteriori}. Two questions arise. How should treat `fuzzy' probabilities which cannot be determined exactly, and how should we treat options and probabilites that are defined and determined after the event has taken place?

In a sense, all probabilities are fuzzy, since the condition of perfect repeatability is an idealization. However, we will treat all probabilities that are introduced in our formalism \emph{as if} they were perfectly known. One may ask if the formalism that we are going to discuss should be generalized to allow fuzzy probabilities, just like we argued in Section \ref{knowstate} that the physical state $S$ itself is fuzzy. We will not address this question here.

Regarding probabilities determined \emph{a posteriori}, the crucial matter is whether they are \emph{potentially} knowable beforehand or not. If such potential knowledge about symmetric alternatives is present before the event, making it possible to deduce probabilities, then these probabilities are proper elements of a representation of a reduced state  $\check{S}_{O}(n)$, otherwise not. It does not matter whether these probabilities are actually known to someone before the event $S_{O}(n)\rightarrow S_{O}(n+1)\subset u_{1}S_{O}(n)$. This is just the basic assumption in our approach to physics that it is the potential knowledge $PK(n)$ that corresponds to the physical state rather than the aware knowledge $K(n)$. Regarding the question if knowledge deduced afterwards, say at time $n+m$, belongs to $PK(n)$ or to $PK(n+m)$, we refer to the discussion in Section \ref{knowledge}.

\vspace{5mm}
\begin{center}
$\maltese$
\end{center}
\paragraph{}

Let us approach the more concrete matter of probability calculation. Consider a complete set of future alternatives $\{\tilde{S}_{j}\}$ (Definition \ref{setfuturealt}) with relative volumes $\{v_{j}\}\equiv \{v[\tilde{S}_{j},S_{O}]\}$. Since actual probabilities are only defined for object states $S_{O}$ (Statement \ref{nouniversalprob}), we focus on this case. For any such state and any such set of alternatives we may write

\begin{equation}
\bar{S}_{O}\equiv\left[\begin{array}{cccc}
\tilde{S}_{1} & \tilde{S}_{2} & \ldots & \tilde{S}_{m}\\
v_{1} & v_{2} & \ldots & v_{m}
\end{array}\right]\hookrightarrow S_{O}.
\label{rep}
\end{equation}
The interpretation of this schema should be obvious. The schema is a representation $\bar{S}_{O}$ of $S_{O}$. There may, of course be several complete sets of alternatives, corresponding to different properties $P$ and $P'$. We may then also write

\begin{equation}
\bar{S}_{O}'\equiv\left[\begin{array}{cccc}
\tilde{S}_{1}' & \tilde{S}_{2}' & \ldots & \tilde{S}_{m}'\\
v_{1}' & v_{2}' & \ldots & v_{m}'
\end{array}\right]\hookrightarrow S_{O}.
\label{rep2}
\end{equation}

\begin{figure}[tp]
\begin{center}
\includegraphics[width=80mm,clip=true]{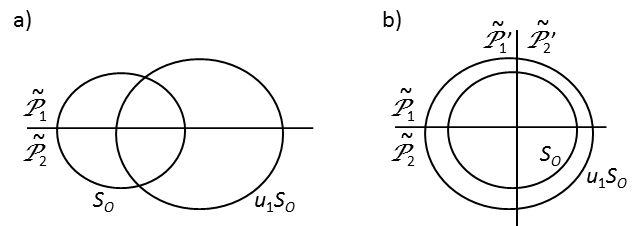}
\end{center}
\caption{The relative volumes of future alternatives are invariant under evolution $u_{1}$ (Statement \ref{invariantvol}). a) If only one property $P$ is defined, the proportions of the state volume $V[S_{O}]$ that belong to $\tilde{\mathcal{P}}_{1}$ and $\tilde{\mathcal{P}}_{2}$ stay the same. b) If two properties $P$ and $P'$ are defined, the relative volume of each quadrant stays the same. The position of the state is `nailed' by the two coordinate axes in this two-dimensional projection of state space.}
\label{Fig57b}
\end{figure}

It follows from Statement \ref{invariantregions} that for any state $S\subseteq\tilde{\mathcal{P}}_{j}$ we have $u_{1}S\subseteq\tilde{\mathcal{P}}_{j}$. The definition of state space volume and relative volume (Definitions \ref{voldef} and \ref{relvol}) then imply that the relative volume of any future alternative $\tilde{S}_{j}$ is invariant under the action of $u_{1}$ (Fig. \ref{Fig57b}).

\begin{state}[\textbf{Relative volumes of future alternatives are invariant under evolution $u_{1}$}]Let $v_{j}\equiv v[\tilde{S}_{j},S_{O}]$ and $u_{1}v_{j}\equiv v[u_{1}\tilde{S}_{j},u_{1}S_{O}]$. Then we have $u_{1}v_{j}=v_{j}$.
\label{invariantvol}
\end{state}

\begin{figure}[tp]
\begin{center}
\includegraphics[width=80mm,clip=true]{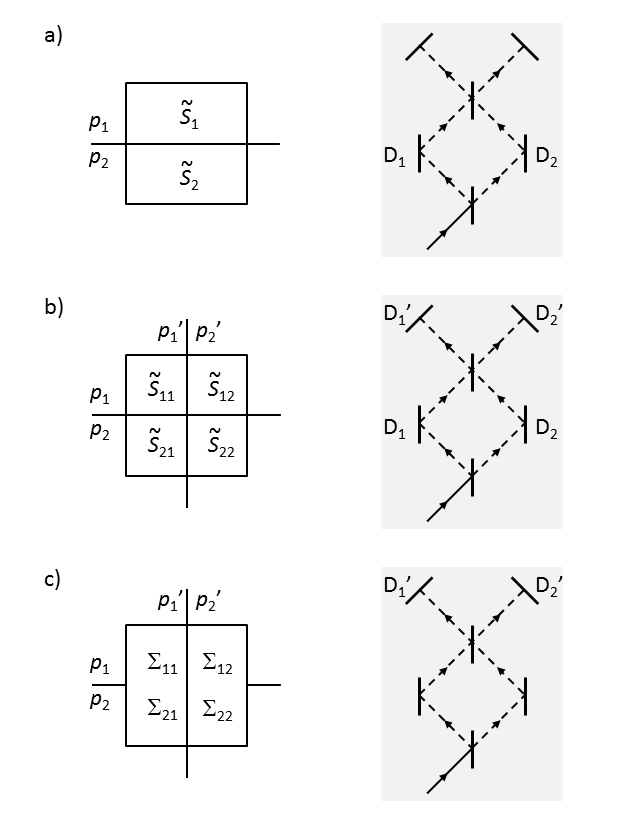}
\end{center}
\caption{The basic cases in which probabilities can be defined. The state $S_{O}$ is schematically represented as in section \ref{graphical}. The solid lines divide future alternatives corresponding to a property $P$ for which the value $p_{j}$ will be decided with certainty. (a) A single such property $P$ is defined. (b) Two such properties $P$ and $P'$ are defined, where the value of $P$ is decided first, then the value of $P'$. (c) It is known that the value of $P$ cannot be decided (knowability level 1). The symbols $\Sigma_{ij}$ are labels on the quadrants of $S_{O}$ that not correspond to realizable alternatives $\tilde{S}_{ij}$. Each of the cases can be implemented by an adjustable Mach-Zehnder interferometer. Property $P$ corresponds to the passage of the left or right mirror, and $P'$ corresponds to the final absorption to the left or right. The prescence of a detector to decide the value of the property is marked by the letter $\mathrm{D}$.}
\label{Fig59}
\end{figure}

Suppose that we are dealing with two sets of complete future alternatives $\{\tilde{S}_{1},\tilde{S}_{2}\}$ and $\{\tilde{S}_{1}',\tilde{S}_{2}'\}$, which correspond to two properties $P$ and $P'$. Define the quadrants $\Sigma_{11}\cup\Sigma_{12}\cup\Sigma_{21}\cup\Sigma_{22}=S_{O}$ like in Fig. \ref{Fig59}. The relative volumes $v[\Sigma_{ij}]$ are then invariant under evolution $u_{1}$ [Fig. \ref{Fig57b}(b)].  

As long as no state reduction occurs, the total state $S$ such that the state $S_{O}$ of the studied system is affected, the following holds true.

\begin{state}[\textbf{Relative volumes of future alternatives are constant in time}]As long as the state $S$ is not reduced in a way that affects $S_{O}$, the relative volumes $v_{j}$ stay the same.
\label{samevol}
\end{state}

The total state $S$ may be reduced in two ways that affect $S_{O}$. In the first case, the state reduction of $S$ simply means that $S_{O}$ is also reduced, that is, an alternative $\tilde{S}_{j}$ for the system $S_{O}$ is realized. In the second case, the reduction of $S$ means that the conditional knowledge that may relate $O$ with its environment changes. Recall that the object state $S_{O}$ is defined in such a way that this conditional knowledge is neglected; the object $O$ is treated as if it was isolated (Definition \ref{objectstate}).   

For future alternatives we have, from Statement \ref{invariantvol}:

\begin{equation}
\bar{u}_{1}\bar{S}_{O}\equiv\left[\begin{array}{cccc}
u_{1}\tilde{S}_{1} & u_{1}\tilde{S}_{2} & \ldots & u_{1}\tilde{S}_{m}\\
v_{1} & v_{2} & \ldots & v_{m}
\end{array}\right]\hookrightarrow u_{1}S_{O}.
\label{uprep}
\end{equation}

The same statement would not necessarily hold true if we considered present alternatives. For such alternatives we would have to write

\begin{equation}
\bar{u}_{1}\bar{S}_{O}\equiv\left[\begin{array}{cccc}
u_{1}S_{1} & u_{1}S_{2} & \ldots & u_{1}S_{m}\\
u_{1}v_{1} & u_{1}v_{2} & \ldots & u_{1}v_{m}
\end{array}\right]\hookrightarrow u_{1}S_{O}.
\label{uprep2}
\end{equation}

The set of property values $\{p_{j}\}$ to which the alternatives $\{\tilde{S}_{j}\}$ (or $\{S_{j}\}$) correspond can be used to define a partial representation of the state $S_{O}$:

\begin{equation}
\bar {S}_{Op}\equiv\left[\begin{array}{cccc}
p_{1} & p_{2} & \ldots & p_{m}\\
v_{1} & v_{2} & \ldots & v_{m}
\end{array}\right]\rightharpoonup S_{O}.
\label{prep}
\end{equation}
The representation is partial since the shape of boundary $\partial S_{j}$ of each alternative $\tilde{S}_{j}$ is not represented, only its relative volume. The set of realizable property values $\{p_{j}\}$ may be said to define a coordinate system, and we may interpret the relative volumes as coordinates of a vector $\bar{S}_{Op}$ that represents the state $S_{O}$ (partially). This algebraic picture is developed further in Section \ref{qmpostulates}.

\begin{state}[\textbf{Stationary state}]The state $S_{O}$ of an object $O$ equipped with a complete set $\{\tilde{S}_{j}\}$ of future alternatives is stationary with respect to property $P$ during the time interval $[n,n+\hat{n}]$ if and only if $\bar{S}_{Op}(n)\rightharpoonup u_{m}S_{O}(n)$ for each $m$ such that $1\leq m\leq\hat{n}$.
\label{stationarystate}
\end{state}

We may say that the state $S_{O}$, as represented by a set of future alternatives $\{\tilde{S}_{j}\}$, is stationary with respect to the corresponding coordinate system $\{p_{j}\}$ as long as none of these alternative is realized, and the object $O$ is not affected by the environment, as discussed in relation to Statement \ref{samevol}.

The above list of five conditions that have to be fulfilled in order to assign probabilities to alterntives leaves us with a few basic cases for which we need rules to calculate probabilities (Fig. \ref{Fig59}). The simplest case is when we have a single property $P$ for which we can assign probabilities to its values. The more complex cases arise when we have two properties $P$ and $P'$ whose values $p_{j}$ and $p_{j}'$ are realized in succession. Let $P$ be the property whose value $p_{j}$ is realized first. Then we need to consider two possibilities: either both values are realized with certainty at some future time (knowability level 3), or $p_{j}$ can never be known (knowability level 1). 

For the sake of clarity we will consider the case where $P$ and $P'$ have only two values each, corresponding to complete sets of future alternatives $\{\tilde{S}_{1},\tilde{S}_{2}\}$ and $\{\tilde{S}_{1}',\tilde{S}_{2}'\}$, respectively. The following discussion can be repeated straightforwardly when there are more than two alternatives for each property value. The results that we arrive at will be analogous. In a similar fashion, the cases when there are more than two properties $P,\;P',\;P''$ whose values are determined in succession can be straightforwardly reduced to the case when there is a series of two properties whose values are determined in succession.

The subsets of $S_{O}$ shown in Fig. \ref{Fig59} that are called $\Sigma_{ij}$ rather than $S_{ij}$ do not correspond to realizable alternatives for which evolution $u_{1}$ can be defined. They are just parts of the state that are created by a theoretical division. Consider the state in Fig. \ref{Fig59}(c). We know by construction that the actual value of property $P$ will never be known, so that the state $S_{O}$ of the system can never be found entirely on the upper half plane. 

Figure \ref{Fig59} also shows a Mach-Zehnder interferometer that can be adjusted to correspond to each of the basic cases described above. There is a monocromatic light source whose intensity can be reduced so much that individual photons are sent out one by one. A beam splitter transmits some fraction of the light, and reflects the rest. A phase difference with a fixed magnitude between the transmitted and reflected beam arises. The two beams are reflected at two mirrors. Depending on which case we are considering, there may be a pair of detectors $\mathrm{D}_{1}$ and $\mathrm{D}_{2}$ at these mirrors, which registers the reflection of the photon without disturbning its direction of motion. The two beams are recombined at a second beam splitter. The two outgoing beams are absorbed at detectors $\mathrm{D}_{1}'$ and $\mathrm{D}_{2}'$, which may or may not be switched on, depending on the case.

In the simplest case with a single measurement by detector $\mathrm{D}_{1}$ or $\mathrm{D}_{2}$ [Fig. \ref{Fig59}(a)], the probabilities can simply be expressed as

\begin{equation}
q_{j}=q(p_{j}).
\label{simplestcase}
\end{equation}
We may also write

\begin{equation}
q_{j}=v[\tilde{S}_{j}],
\end{equation}
since the relative volumes stay constant until the measurement or state reduction takes place (Statement \ref{samevol}), even if we may not know exactly when this happens. (We assume that the surroundings do not affect the experimental setup $O$ during this time.)

In the case where we make measurements in succession (Fig. \ref{Fig59}b), we get analogous probabilities $q_{j}=q(p_{j})$ for the outcome of the first measurement, and then

\begin{equation}
q_{j}'=q(p_{j}'|p_{1})q(p_{1})+q(p_{j}'|p_{2})q(p_{2})
\label{condprob}
\end{equation}
for the outcome of the second measurement, where $q(p_{j}'|p_{i})$ is the conditional probability that property $P'$ will have value $p_{j}'$ given that $P$ turned out to have value $p_{i}$. Suppose that this was decided at time $n$ and that the value of $P'$ will be decided at time $n'>n$. Let us write $S_{O}(m)=\tilde{S}_{1}'(m)\cup \tilde{S}_{2}'(m)$ for $n\leq m<n'$, where $\tilde{S}_{1}'(m)$ and $\tilde{S}_{2}'(m)$ are the two future alternatives for property $P'$. Then we may express

\begin{equation}
q(p_{j}'|p_{i})=v[\tilde{S}_{j}'(m)]
\end{equation}
at any time $m$ between the two state reductions. Remember again that the relative volumes stay constant if nothing affects the experimental setup between the two measurements.

In the case where it is assumed to be impossible in principle to determine whether the particle passed mirror $1$ or $2$, the probabilities $q_{j}=q(p_{j})$ are not defined, nor are the conditional probabilities $q(p_{j}'|p_{i})$. Just like in the first case (Eq. \ref{simplestcase}), we must simply write

\begin{equation}
q_{j}'=q(p_{j}').
\label{nocondprob}
\end{equation}

\begin{state}[\textbf{On conditional probabilities}]The probability $q(p_{j}'|p_{i})$ can be defined if and only if $P$ is a realizable property at knowability level 3, with a value $p_{i}$ that will be potentially known before the value $p_{j}'$ of property $P'$ becomes potentially known.
\label{oncondprob}
\end{state}

\begin{figure}[tp]
\begin{center}
\includegraphics[width=80mm,clip=true]{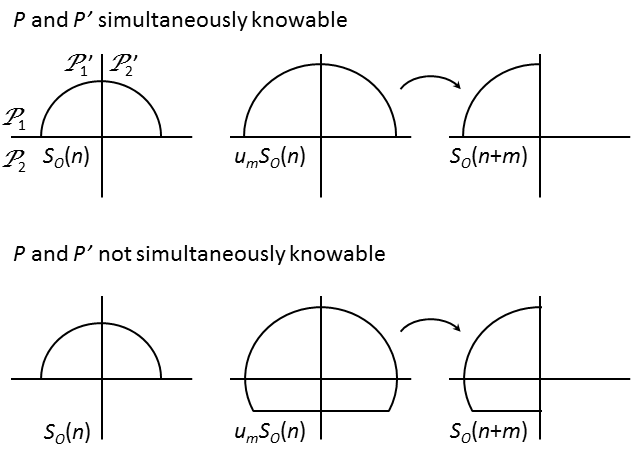}
\end{center}
\caption{The state $S_{O}$ just after the value of property $P$ is decided to have value $p_{1}$ in case b in Fig. \ref{Fig59}. (a) If $P$ and $P'$ are simultaneously knowable, the state may be contained in a single quadrant just after the value of $P'$ has been decided. (b) This may never happen if the properties are not simultaneously knowable.} 
\label{Fig60}
\end{figure}

It is also of interest to consider what can happen \emph{after} one of the alternatives in Fig. \ref{Fig59} has been realized. When we have two properties $P$ and $P'$ with corresponding alternatives that are realized in succession (panel b), we should distinguish between the cases where $P$ and $P'$ are simultaneously knowable or not. The fact that there are property pairs whose values cannot be simultaneously known follows from Statement \ref{simknowprop}.

For the specific implementation of these cases provided by the Mach-Zehnder interferometer, $P$ and $P'$ are indeed simultaneously knowable. We may know that the photon first passed the left mirror, and then arrived at the right detector. To discuss the general case, we therefore forget about this particular setup. 

Consider Fig. \ref{Fig60}. The observation of the value of property $P$ defines a temporal update $n-1\rightarrow n$. Thus the state $S_{O}(n)$ is contained in either the upper or lower half-plane that correspond to the regions $\mathcal{P}_{1}$ and $\mathcal{P}_{2}$. If $P$ and $P'$ are simultaneously knowable, it is possible that the state stays in the (say) upper half plane until the time $n+\hat{n}$ when observation of the value of $P'$ becomes possible (according to the definition of the knowability levels). Then $S_{O}(n+m)$ is contained in one of the quadrants for some $m\geq \hat{n}$.

Obviously, this scenario is forbidden when $P$ and $P'$ are not simultaneously knowable. Then the state must evolve from time $n$ to time $n+m$ to cover part of the lower half-plane. When the value of property $P'$ is observed, the state have parts that belong to $\mathcal{P}_{1}$ as well as $\mathcal{P}_{2}$, so that the outcome of a repeated observation of $P$ is no longer certain. 

We conclude that the state $S_{O}$ can never be contained in a single quadrant $\mathcal{P}_{i}\cap \mathcal{P}_{j}$ when $P$ and $P'$ are not simultaneously knowable. Thus evolution $u_{1}$ cannot be defined for such a hypothetical state (Statement \ref{evolutiondomain}). The classical example is the spin of an electron. Its component along two spatial directions cannot be known at the same time. There is no evolution equation for a spin vector where all components are specified.

\begin{figure}[tp]
\begin{center}
\includegraphics[width=80mm,clip=true]{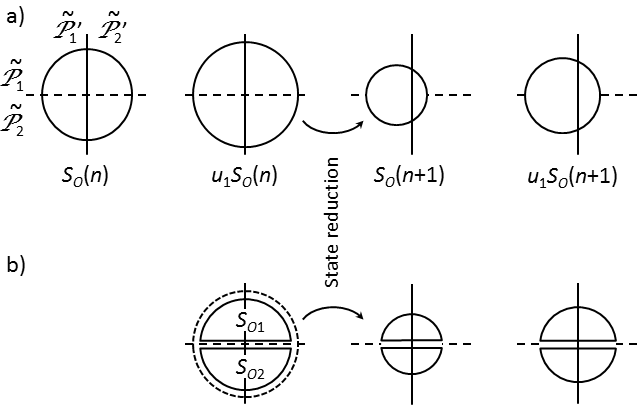}
\end{center}
\caption{The state $S_{O}$ before it is decided whether the value of property $P$ will ever be known or not. At the moment of decision, we end up in cases b or c in Fig. \ref{Fig59}. (a) This decision corresponds to a state reduction. The reduced state may not be left-right symmetric even if the original state is symmetric. This means that the probabilities to see values $p_{1}'$ or $p_{2}'$ may be unequal. (b) If we somehow know the value of $P$ beforehand, we start out with states $S_{1}$ or $S_{2}$. Then there is no symmetry breaking. Two subsets of a state $S_{O}$ need not stay subsets forever. This is an example of the irreducibility of physical law.} 
\label{Fig61}
\end{figure}

This irreducibility of physical law (Assumption \ref{irreduciblelaw}) is also helpful to understand the fact that a left-right symmetric experimental arrangement, like the interferometer in Fig. \ref{Fig59} with transmission ratio $1/2$ at each beam splitter, can produce unequal probabilities for the arrival of photons in the left and right detectors ($\mathrm{D}_{1}'$ and $\mathrm{D}_{2}'$). We know that this is the case due to interference effects that may arise because of the phase delay of the reflected beam at the beam splitters.

Strictly speaking, the experimental arrangement is not left-right symmetric, since the beam that emanates from the light source moves to the right (Fig. \ref{Fig59}). It is symmetric only in the sense that if we imagine an exact state $Z$ that corresponds to a photon taking a well-defined route through our `pinball machine', then each possible route is equally probable. Thus, if we want to illustrate the initial state $S_{O}$ of the setup as in the right panel of Fig. \ref{Fig57}, then the volumes of all quadrants $\Sigma_{ij}$ should have equal volumes (leftmost panel in Fig. \ref{Fig61}). At least, this must be true when it is possible in principle to decide whether the photon arrives at mirror 1 or 2 (the values of property $P$), and at detector $D_{1}'$ or $D_{2}'$ (the values of property $P'$). In other words, the volumes of all quadrants $\Sigma_{ij}$ must be the same if both properties $P$ and $P'$ have knowability level 2 or 3.

Assume that this is initially the case. Then, at some moment, it becomes certain that it will forever be unknowable whether the photon passed mirror 1 or 2. The detectors at these mirrors are switched off. The knowability level of property $P$ changes from 2 to 1. Such an event necessarily corresponds to a state reduction (Definition \ref{statereduction}), since knowability level 2 means that it is \emph{not} part of potential knowledge whether the value of property $P$ will ever be known. The physical state corresponding to this state of knowledge cannot evolve via $u_{1}$ alone to a state where it \emph{is} part of potential knowledge whether property $P$ will be known or not (levels 3 or 1). That would be a contradiction, since anything that can be deduced about the \emph{future} via physical law is part of the extended \emph{present} potential knowledge (Eq. \ref{extendedknowledge}), which is equivalent to the \emph{present} potential knowledge (Definition \ref{potknowstate}).

This state reduction can be expressed as $u_{1}S_{O}(n)\rightarrow S_{O}(n+1)\subset u_{1}S_{O}(n)$. Since we have assumed no physical law that dictates that the `center of mass' of the reduced state is the same as that of the state it is reduced from, the equality of the volume of the quadrants may very well be lost at this moment. In conclusion, such a symmetry breaking may occur at the same time as it is established that property $P$ will be forever unknowable. This is indicated in Fig. \ref{Fig61}(a). The unequal volumes $V[S_{O}(n+1)\cap\mathcal{P}_{1}]$ and $V[S_{O}(n+1)\cap\mathcal{P}_{2}]$ that follows in such a situation means that the probabilites to see the values $p_{1}'$ and $p_{2}'$ will be different.

Imagine another situation, in which hidden detectors are found along the beams between the first beam splitter and the two mirrors. They make it possible to decide the value of property $P$ before the photon hits one of the mirrors. It is still not known whether the detectors at the mirrors will be switched on or off at the time of the photon's passing, just like in the initial state of the previously studied case.

Just after the passing of the hidden detectors, the state $S_{O}$ is a subset of either $\tilde{\mathcal{P}}_{1}$ or $\tilde{\mathcal{P}}_{2}$, as shown in Fig. \ref{Fig61}(b). Here, both possible states $S_{O1}$ and $S_{O2}$ after the passing are shown. The first snapshot in bottom row is shown under the state $u_{1}S_{O}(n)$ in the top row, just to indicate that the splitting of the state by the hidden detectors takes place some time after the experiment starts. Since the presence of the hidden detectors amounts to more information about the setup - increased potential knowledge - the union of the splitted states is a proper subset of the corresponding state with the hidden detectors blanked out from perception. We may write $S_{O1}\cup S_{O2}\subset u_{1}S_{O}$. The latter state is marked as the dashed circle in the first panel of the bottom row.

In this case, the event where the detectors at the mirrors are switched off can have no effect on the relative volumes of the left and right halves of the splitted states $S_{1}$ and $S_{2}$, since we already know which of the switched off detectors will be passed. No symmetry can be lost at this moment. The result may be that $S_{O1}(n+1)\cup S_{O2}(n+1)\not\subseteq S_{O}(n+1)$; the splitted states have partially jumped out of the `mother state' $S_{O}$. This is no violation of epistemic invariance (Assumptions \ref{epistemicinvariance} and \ref{epistemicinvariance2}) since this principle only applies for ordinary evolution $u_{1}$, and may be broken when state reductions are involved. This is an example of the irreducibility of physical law - the sequence of states $S_{O}(n),S_{O}(n+1),S_{O}(n+2),\ldots$ cannot be modelled by rules that determine sequences of exact states $Z_{j}(n),Z_{j}(n+1),Z_{j}(n+2),\ldots$, where $Z_{j}(n)\in S_{O}(n)$.

\section{The postulate of \emph{a priori} equal probabilities}
\label{aprioriequal}

Statistical mechanics relies on the assumption that all microscopic states compatible with a set of macroscopic constraints should be assigned equal probabilities. The perspective adopted here is that it is only macroscopically observable alternatives that should be assigned probabilities. These probabilites are the relative volumes of the alternatives, according to Statement \ref{probrel}. The postulate of \emph{a priori} equal probabilities is then nothing more than a repetition of this statement, where we identify `macroscopic constraints' with the state $S_{O}$, which corresponds to our subjective potential knowledge of the system of interest, and the `microscopic states' with the exact states $Z$ that make up $S_{O}$.

The possibility that `microscopic states' have different `probabilities' is excluded in the present description, since there is no measure defined \emph{\emph{a priori}} on the elements $Z$ of $S_{O}$. The physical state is the set of exact states that are not excluded by potential knowledge, nothing more. Of course, measures that are functionals of $S_{O}$ as a whole can be defined, depending, for example, on its volume or the shape of its boundary. But that is a different matter.

We need to know the possible attribute values of exact states to be able to compute the relative volume of alternatives $S_{j}$. We return to this question below. At this stage we just note the simple fact that we may often group small sets of exact states together and use these groups as the `microscopic states' in statistical mechanical applications. For example, if we have a gas of particles that do not interact, and if there is no external magnetic fields, then we may disregard the spin, and group the possible spin states $(s_{x},s_{y},s_{z})$ together into one unit when we sum over all possible states in the partition function. Each such spin-unit $\Sigma_{s}$ will have the same volume $V[\Sigma_{s}]$ for any given array of values of the other attributes, such as position and momentum. They can therefore be given the same unit statistical weight. Such `microscopic units' may correspond to states that are observable in principle, in contrast to the exact states $Z$.

\section{The postulates of quantum mechanics}
\label{qmpostulates}

We seek an algebraic representation of the object state $S_{O}$ that uses schemas such as [\ref{rep}] as a starting point. The aim is to find algebraic rules operating on such a representation that extract everything that can be said about probabilities.

Formally, we may rearrange the schema [\ref{rep}] as follows:

\begin{equation}
\bar{S}_{O}=v_{1}\bar{S}_{1}+v_{2}\bar{S}_{2}+\ldots v_{m}\bar{S}_{m}.
\label{arep}
\end{equation}
To make such a representation meaningful, we must check that the algebraic rules for addition and multiplication hold. We put bars over the future alternatives $\tilde{S}_{j}$ to emphasize that they are now part of an (algebraic) representation rather than sets. On the other hand, we skip the tildes when we throw in the bars, to make the notation less cluttered.

\begin{figure}[tp]
\begin{center}
\includegraphics[width=80mm,clip=true]{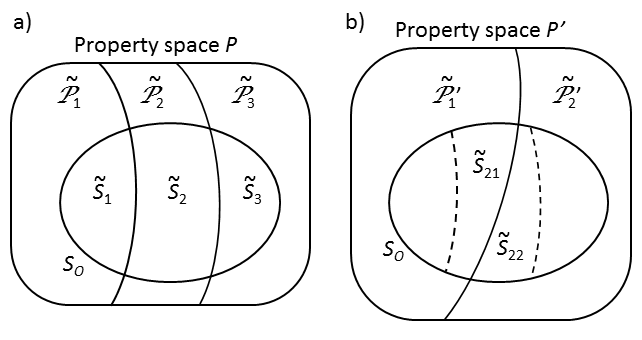}
\end{center}
\caption{The introduction of two simultaneously knowable properties $P$ and $P'$ at knowability level 3 can be used to motivate why the distributive law (Eq. [\ref{divrep2}]) holds for the relative volumes $v_{j}$. This is an indication that the set-theoretic description of physical states can be translated into an algebraic description. The volume measure $V$ is the bridge to the algebraic world.} 
\label{Fig61b}
\end{figure}

To each alternative $\tilde{S}_{j}$ is associated a value $p_{j}$ of property $P$. Suppose that these alternatives have knowability level 3 (Table \ref{levels}). That is, we know for sure that at some future time we will get to know which value $p_{j}$ applies. Consider now another property $P'$, which is simultaneously knowable (Fig. \ref{Fig60}), for which another set of future states $\{S_{j}'\}$ having knowability level 3 is defined. Suppose that these alternatives are related according to Fig. \ref{Fig61b}(b). We may then define a combined property $P''$ with possible values $(p_{1},p_{1}')$, $(p_{2},p_{1}')$, $(p_{2},p_{2}')$ and $(p_{3},p_{2}')$, and with corresponding complete set of future alternatives $\{\tilde{S}_{1},\tilde{S}_{21},\tilde{S}_{22},\tilde{S}_{3}\}$. We may also write

\begin{equation}
\bar{S}_{2}=\left[\begin{array}{cc}
\tilde{S}_{21} & \tilde{S}_{22} \\
v_{21} & v_{22}
\end{array}\right],
\label{altrep}
\end{equation}
with $v_{21}\equiv v[\tilde{S}_{21},\tilde{S}_{2}]$ and $v_{22}\equiv v[\tilde{S}_{22},\tilde{S}_{2}]$, just like in Eq. [\ref{rep}]. Using the notation of Eq. [\ref{arep}] this expression transforms to

\begin{equation}
\bar{S}_{2}=v_{21}\bar{S}_{21}+v_{22}\bar{S}_{22}
\label{arep2}
\end{equation}
From the definition of each involved relative volume it follows that

\begin{equation}
\bar{S}_{O}=\left[\begin{array}{ccc}
S_{1} & S_{2} & S_{3}\\
v_{1} & v_{2} & v_{3}
\end{array}\right]=
\left[\begin{array}{cccc}
S_{1} & S_{21} & S_{22} & S_{3}\\
v_{1} & v_{2}v_{21} & v_{2}v_{22} & v_{3}
\end{array}\right],
\label{divrep}
\end{equation}
where the equality sign means `represent the same state as'. We may therefore use Eqs. [\ref{arep2}] and [\ref{divrep}] to write

\begin{equation}\begin{array}{lll}
\bar{S}_{O} & = & v_{1}\bar{S}_{1}+v_{2}\bar{S}_{2}+v_{3}\bar{S}_{3}\\
& = & v_{1}\bar{S}_{1}+v_{2}(v_{21}\bar{S}_{21}+v_{22}\bar{S}_{22})+v_{3}\bar{S}_{3}\\
& = & v_{1}\bar{S}_{1}+v_{2}v_{21}\bar{S}_{21}+v_{2}v_{22}\bar{S}_{22}+v_{3}\bar{S}_{3}.
\label{divrep2}
\end{array}\end{equation}
This is reassuring, since these expressions conform with the distributive law.

Eq. [\ref{uprep}] transforms to

\begin{equation}
\bar{u}_{1}\bar{S}_{O}=v_{1}\bar{u}_{1}\bar{S}_{1}+v_{2}\bar{u}_{1}\bar{S}_{2}+\ldots v_{m}\bar{u}_{1}\bar{S}_{m},
\label{aurep}
\end{equation}
meaning that we, tentatively, should interpret the evolution operator as a linear operator in this (proto-algebraic) representation.

\subsection{Contextual states}
\label{contextualstates}

To proceed further, we need to take a step back and introduce a couple of new concepts. When we imagine a set of options, and when we subsequently identify which option $S_{j}$ comes true, we are most often interested in a particular aspect of the realized alternative, an aspect that can be coded as a value $p_{j}$ of property $P$. Other information that is part of the potential knowledge that corresponds to $S_{j}$ is considered irrelevant.

For example, in a scientific experiment the state $S_{O}$ that we use to define alternatives $S_{j}$ is the state of the object $O$ that consists of the experimental apparatus together with the specimen $OS$ to be examined. If the specimen is a single electron, the property of interest may be its spin direction or its position when it hits a detector. As we perceive the outcome of the experiment, it is irrelevant if we, at the same moment, perceive a new scratch on some metal part of the detector.

\begin{figure}[tp]
\begin{center}
\includegraphics[width=80mm,clip=true]{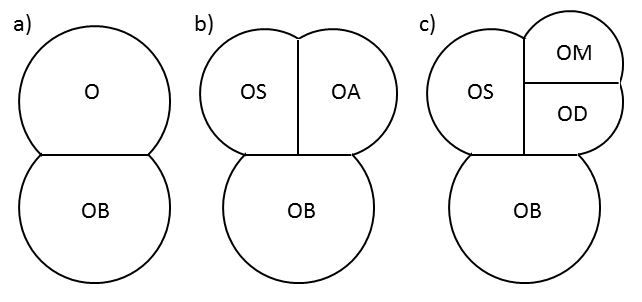}
\end{center}
\caption{Objects that have to be or may be parts of an observational setup. a) The observed object $O$ and the body of an observer $OB$ are necessary parts. b) The observed object may sometimes be divided into a specimen $OS$ and an apparatus $OA$ with which we decide a property of the specimen. We may then say that the observation is the outcome of an experiment. c) When the specimen is a deduced quasiobject, the apparatus can be divided into a machine $OM$ and a detector $OD$, where the subjective change in the state of the detector defines the outcome of the experiment.} 
\label{Fig61c}
\end{figure}

Let us schematically discuss the role of the specimen in the observational setup (Fig. \ref{Fig61c}). Such a setup necessarily consists of at least two subjectively distinct objects: the object $O$ whose state (or evolution) we observe, and the body $OB$ of an observer. In a controlled, scientific setting, $O$ is divided into at least two parts: the specimen $OS$ and the apparatus $OA$, with which we study the specimen, decide one of its properties. This is the crucial distinction introduced in this section.

Our aim is to formulate physical law in terms of the behavior of specimens. What kind of specimens can be scientifically investigated? First, the specimen must be `real' rather than imagined. We cannot expect physical law to hold for fantasies. But what does `reality' mean in this context?

Recall that we assumed the basic ability of a subject to make proper interpretations of perceptions in terms of internal or external, present or past, and so on (Fig. \ref{Fig2}). In fact, we defined knowledge as perceptions with proper interpretation (Definition \ref{knowledgedef}). Further, we associated physical states with states of knowledge. Therefore we may regard the interpretation of a perceived object as a set of internal physical attributes that describe the state of this object. We have already introduced one such attribute, namely the presentness attribute $Pr$, with possible values zero and one. Let us introduce a similar binary reality attribute $Re$ (Table \ref{objectinterpretation}) and let it rely on the following definition of reality.

\begin{defi}[\textbf{Real objects}]
Suppose that the object $O$ is identifiable during the time interval $[n,n+m]$ (Definition \ref{identifiableinterval}). The sequence of states $\{S_{O}\}=\{S_{O}(n),S_{O}(n+1),\ldots,S_{O}(n+m)\}$ will be consistent with at least one sequence of states $\{\check{S}_{OM}\}=\{\check{S}_{OM}(n),\check{S}_{OM}(n+1),\ldots,\check{S}_{OM}(n+m)\}$ describing the evolution according to physical law of a set of identifiable minimal objects (Definition \ref{minimalidentity}). Consider a relational attribute $A_{r}$ that relates $O$ with its complement $\Omega_{O}$ (Definition \ref{complement}). Then $O$ is real if and only if 1) there is a sequence $\{\check{S}_{OM}\}$ for which the expected value of any such $A_{r}$ for the set of minimal objects $\{O_{M}\}$ is the same at each time $n,n+1,\ldots,n+m$ as the expected value of $A_{r}$ for $O$ itself, and 2) for all $n$, the physical state $S(n)$ is consistent with the description of $S_{O}$ in terms of the sequence $\{\check{S}_{OM}\}$.
\label{realobjects}
\end{defi}

Basically, the definition says that it should be possible to model a real object with minimal objects that move along with the object in question, and that this model should be valid at all times, even after the object itself has disintegrated.

Consider a cloud, for example. It is possible to model this object as a set of water droplets, which in turn consists of elementary particles arranged in atoms and molecules. These water droplets move across the sky together with the cloud itself. This model is valid even after the cloud dissolves; it does not contradict any later observations. The water in the droplets simply evaporates. In contrast, if we watch a film of a cloud on a television screen, there are no water molecules that move across the screen together with the image of the cloud. If we watched the cloud dissolve on the screen and attempted such a model, we would not be able to account for the evaporated water by measuring the humidity in the living room where the TV set is located.

In the above example, velocity is the relevant relative attribute $A_{r}$. Let us discuss an example in which position is the relevant relative attribute. Suppose that you feel pain in your leg. This interpretation of the perception may be proper (Fig. \ref{Fig2}) in the sense that it can actually be modeled by a set of minimal objects in the leg engaged in a process that causes the pain, so that no future perception can contradict this model. Then the `pain in the leg' is a real object. If, on the other hand, your leg has been amputated, such a model would be contradicted once you recall your condition and look at your body. The `pain in the leg' would not be a real object.

However, the `pain' is still a real object - it is just the location `in the leg' that is a wrong interpretation. In the same way, the `cloud displayed on the TV-screen' is a real object - it can be modeled by minimal objects arranged as a TV set, with the screen emitting light according to the recipe provided by the input signal. It is just the interpretation `the cloud is moving across the sky' that is wrong. That every object has a proper interpretation that makes it `real' can be seen as an expression of detailed materialism (Definition \ref{localmaterialism}). The distinction between `real' and `unreal' objects can be seen as a simple reflection of the basic assumption that it is possible in principle to distinguish proper interpretations from improper ones.

\begin{table}
	\centering
		\begin{tabular}{|l||c|c|}
		\hline
		Attribute & Values & Definition \\
		\hline
		Reality & $Re=0,1$ & \ref{realobjects} \\
		Presentness & $Pr=0,1$ & \ref{presentness}\\
		Direct perception & $Dp=0,1$ & \ref{quasiobjectdefi}\\
		\hline
		\end{tabular}
	\caption{Binary internal attributes that apply to all object $O$ and concerns its basic interpretation. A quasiobject is an object $O$ with $Dp(O)=0$. A directly perceived specimen must be real: $Re(OS)=1$ if $Dp(OS)=1$.}
	\label{objectinterpretation}
\end{table}



This means that we treat the ability to distinguish between `real' and `unreal' objects as fundamental, just like the ability to distinguish the present from the past, and perceived objects from deduced ones. Therefore, from our epistemic perspective, the attributes listed in Table \ref{objectinterpretation} should enter the physical description at a fundamental level, just like the directed nature of time.

We may argue that a memory of a cloud that passes across the sky is not real in the same way as we argued that an image of a cloud on a TV screen is not real. The memory of the cloud cannot be modelled in terms of minimal objects that travels here and now across your field of vision. Instead, it must be modelled in terms of internal minimal objects in the brain that are quite stationary in relation to yourself.

\begin{state}[\textbf{Remembered objects are not real}]
Objects $O$ with $Pr[O]=0$ and $Dp[O]=1$ are not real in the sense of Definition \ref{realobjects}. However, all remembered objects become real if we let the proper interpretation `this is a past object' be part of the object.
\label{norealmemories}
\end{state}

The reverse of this statement is not true. We may have present objects that are not real. Two examples are the cloud on the TV screen and the phantom pain.

\begin{state}[\textbf{There are both real and unreal present objects}]
We may have both $Re[O]=1$ and $Re[O]=0$ for objects $O$ with $Pr[O]=1$ However, all present objects become real if we let the proper interpretation of its context be part of the object.
\label{somerealpresentobjects}
\end{state}


We argued above that the specimen must be `real'. Does this mean that it must be a present object, that we must have $Pr[OS]=1$? To answer this question we discuss first the possible role of quasiobjects in the experimental setup (Fig.\ref{Fig61c}).

The division of $O$ into a specimen $OS$ and an apparatus $OA$ opens up the possibility that $OS$ is a quasiobject. That is, we allow both $Dp[OS]=1$ and $Dp[OS]=0$. However, $OA$ cannot be a quasiobject, since something has to be actually observed. If $OS$ is a quasiobject the setup is such that there is conditional knowledge that relates the states of $OS$ and $OA$, so that new knowledge about the state of $OS$ is gained by deduction (using physical law) when new knowledge of $OA$ is gained by observation. Such a `quasiobservation' of $OS$ means that the extended potential knowledge grows, and the reduced physical state shrinks (Figs. \ref{Fig24c} and \ref{Fig28c}).

It does not make sense to talk about 'unreal' quasiobjects. Their role is to account for the evolution of real, directly perceived objects in a reductionistic, scientific way. In fact, minimal quasiobjects are used to \emph{define} the reality of directly perceived objects according to Statement \ref{alwaysquasi} and Definition \ref{realobjects}. Formally, $Dp[O]=0\Rightarrow Re[O]=1$.

These considerations open up the possibitily that we can use an apparatus $OA$ to deduce the past state of a specimen that is a quasiobject and still keep the requirement that specimens should be real. In that case we have $Re[OS]=1$, $Dp[OS]=0$, and $Pr[OS]=0$. For instance, we may use a telescope to observe a distant galaxy. The perceived luminous blob is a present, directly perceived object, but the galaxy as deduced to be located in a distant part of space in the distant past must be considered a past quasiobject.

\begin{defi}[\textbf{The specimen} $OS$ \textbf{and its state} $S_{OS}$]
Assume that $O$ is a composite object, and let $S_{OS}$ be the state of an object $OS$ that is part of $O$, and whose possible property values are used to define a complete set of future alternatives for $O$. If $OS$ is real (Defintion \ref{realobjects}), then $OS$ is a specimen with specimen state $S_{OS}$.
\label{specstatedef}
\end{defi}

From this definition, we immediately conclude that

\begin{equation}
S_{OS}\supset S_{O}.
\end{equation}

When the specimen $OS$ is a quasiobject, it is possible to divide the apparatus $OA$ into one detector $OD$ and one machine $OM$ (Fig. \ref{Fig61c}). The term `machine' may not be very illuminating, and we define it negatively as those parts of the apparatus $OA$ that is not a detector.

\begin{defi}[\textbf{The detector} $OD$]
Suppose that an experiment starts at time $n$, that the observation that defines its outcome occurs at time $n+m$, and that $OS$ is not directly perceived during the course of the experiment. Then $OD$ are those objects that are part of the apparatus $OA$, and are such that $S_{OA}(n)\cap S_{OA}(n')\neq\varnothing$ whenever $n\leq n'<n+m$ and $S_{OA}(n+m-1)\cap S_{OA}(n+m)=\varnothing$.
\label{detectordef}
\end{defi}

The distinct change of the detector state $S_{OD}$ thus defines the outcome of the experiment, and also defines the temporal update $n+m-1\rightarrow n+m$. The state of the machine $S_{OM}$, on the other hand, may undergo distinct changes during the course of the experiment, but may not change subjectively at time $n+m$.

\begin{defi}[\textbf{Property value state} $S_{Pj}$ \textbf{of a specimen}]
Consider a set of properties $\{P_{OS}\}$ that specify the nature of the specimen $OS$, with fixed, limited value ranges $\{\Upsilon_{POS}\}$ that are considered known \emph{a priori}. Consider also another property $P$ that can be defined for $OS$, but whose value may vary. Let $S_ {Pj}$ be the state of $OS$ that corresponds to the knowledge that the value of $P$ is $p_{j}$, in addition to knowledge about $\{\Upsilon_{POS}\}$.
\label{propertyvaluestate}
\end{defi}

\begin{figure}[tp]
\begin{center}
\includegraphics[width=80mm,clip=true]{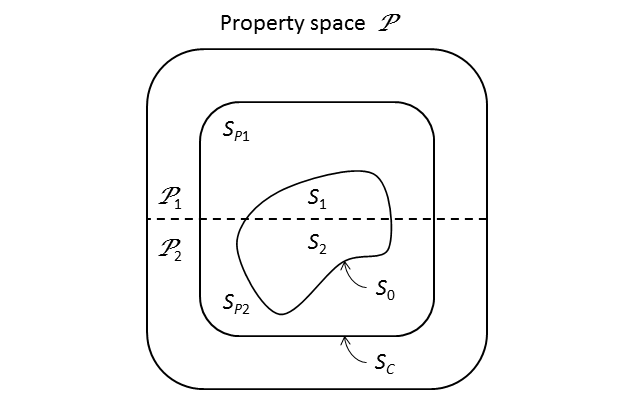}
\end{center}
\caption{The property value states $S_{Pj}$ and the contextual state $S_{C}(n)$ of the specimen $OS$ we investigate. If we forget all knowledge about the composite object $O$ except the nature of a specimen $OS$ that is part of it, and also know the present value $p_{j}$ of a property $P$ that applies to $OS$, we get the state $S_{Pj}$. If this knowledge is gained at time $n$, then $S_{C}(n)=S_{Pj}$. If the is no momentary knowledge of the value of $P$ (and no knowledge of any other property $P'$), then $S_{C}=\bigcup_{j}S_{Pj}$. If forget the very nature of $OS$ we get a state that is the entire property space $\mathcal{P}$. We assume here that $P$ is an independent attribute that has two allowed values (for instance the two possible spin directions of an electron). Then we have $V[S_{P1}]=V[S_{P2}]$, wheras we may have $V[S_{1}]\neq V[S_{2}]$ for the corresponding (present) alternatives that apply to the entire object $O$.} 
\label{Fig63}
\end{figure}

We assume in Definition \ref{detectordef} that the observation of $P$ occurs at time $n+m$. This means that any future alternative $\tilde{S}_{j}$ becomes a present alternative $S_{j}$ at time $n+m-1$ (Definitions \ref{presentalt} and \ref{futurealt}). The fact that $S_{Pj}$ corresponds to knowledge about nothing more than that the value of $P$ is $p_{j}$ means that

\begin{equation}
S_{j}\subset S_{Pj}
\end{equation}
for all such present alternatives $S_{j}$ just before the observation of $P$ is made (Fig. \ref{Fig63}). The union $\bigcup_{j}S_{Pj}$ is a state that corresponds to knowledge about the fixed ranges $\{\Upsilon_{POS}\}$ of the values of $\{P_{OS}\}$, that is, to knowledge of the nature of the specimen. It consists of all exact states $Z$ that are not excluded by the existence of a specimen of the given nature.

\begin{defi}[\textbf{An observational context} $C$]
The context $C$ is the potential knowledge contained in the state $S_{O}$, together with a sequence of complete sets of future alternatives $\{\tilde{S}_{j}\}, \{\tilde{S}_{j}'\},\ldots,\{\tilde{S}_{j}^{(F)}\}$ that correspond to values of properties $P, P',\ldots, P^{(F)}$, that are observed in sequence. These properties are defined for a specimen $OS$ that is a part of $O$, but do not belong to $\{P_{OS}\}$. Further, from a given time $n$, the knowability level associated with the values of each property should be 1 or 3, and the knowability level associated with $P^{(F)}$ should be 3.
\label{observationalcontext}
\end{defi}

We may say that the context $C$ is \emph{initiated} at time $n$. This is the time of no return, after that the properties will be observed in sequence, whether we like it or not. There may, however, be intermediate unobservable properties in the sequence, like $P$ in Fig. \ref{Fig59}(c). The important thing is that there are no propertes in the sequence that may or may not be observed; the observational context should correspond to a well-defined experiment. 

\begin{defi}[\textbf{Contextual state} $S_{C}$ \textbf{of a specimen}]
Consider a context $C$ in which $P, P',\ldots, P^{(F)}$ are observed in sequence at times $n+m,n+m',\ldots,n+m^{(F)}$. Then $S_{C}(n')$ is defined for $n\leq n'\leq n+m^{(F)}$ and corresponds to the potential knowledge of these properties at time $n'$, in addition to knowledge about the values of $\{P_{OS}\}$.
\label{contextualstate}
\end{defi}

\begin{figure}[tp]
\begin{center}
\includegraphics[width=80mm,clip=true]{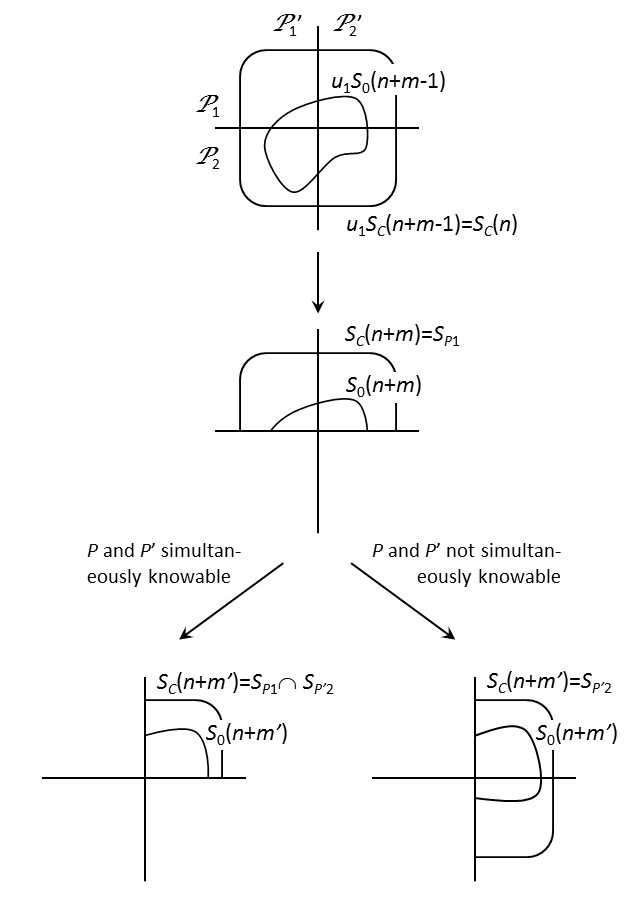}
\end{center}
\caption{Sequences of state reductions of the contextual state $S_{C}$ when properties $P$ and $P'$ are observed at times $n+m$ and $n+m'$, respectively. The final state $S_{C}(n+m')$ depends on whether $P$ and $P'$ are simultaneously knowable or not. Compare Figs. \ref{Fig60} and \ref{Fig63}.} 
\label{Fig145}
\end{figure}

At time $n$, before the first property $P$ is observed, we have

\begin{equation}
S_{C}(n)=\bigcup_{j}S_{Pj}=\bigcup_{j}S_{P'j}=\ldots=\bigcup_{j}S_{P^{(F)}j}.
\end{equation}
When the value of $P$ is observed to be $p_{j}$ at time $n+m$, the contextual state reduces to

\begin{equation}
S_{C}(n+m)=S_{Pj}.
\end{equation}
If $P$ and $P'$ are simultaneously knowable (Fig. \ref{Fig60}), the contextual state reduces further to

\begin{equation}
S_{C}(n+m')=S_{Pj}\cap S_{P'j'}
\end{equation}
when the value of $P'$ is observed to be $p_{j'}'$ at time $n+m'>n+m$. In contrast, if $P$ and $P'$ are not simultaneously knowable, then we may loose all knowledge of the value of $P$ at time $n+m'$, so that

\begin{equation}
S_{C}(n+m')=S_{P'j'}.
\end{equation}
We may, for example, know and remember that the specimen is an electron, and let $P$ be the spin in the $z$-direction and $P'$ be the spin in another direction that will be measured subsequently. These state reductions are illustrated in Fig. \ref{Fig145}. In general, we see that

\begin{equation}
S_{C}\supseteq S_{OS}\supset S_{O}.
\end{equation}

Assume that the specimen $OS$ indeed is an electron, for which we have just measured the spin in the direction defined by the unit vector $\bar{e}_{z}$. Let the complete set of alternatives $\{S_{1}',S_{2}'\}$ be defined by the possible values $-1/2$ and $+1/2$ in a subsequent measurement of the spin in a direction $\bar{e}_{n}$ that is deflected $15^{\circ}$ from $\bar{e}_{z}$. If object $O$ is taken to be the combined apparatus that perform both spin measurements, the knowledge of the outcome of the first measurement is encoded in the state of the apparatus before the second measurement takes place. This encoded knowledge makes it possible to predict which outcome of the second measurement is more likely. No such knowledge is encoded in the electron itself, however. This is the distinction expressed in Fig. \ref{Fig57}. In this case $S_{C}=S_{OS}$.

In contrast, if the specimen $OS$ is macroscopic, it is possible to encode its history in the specimen itself, to mark it in a way that makes it possible to predict its future property values. Then $S_{C}\supset S_{OS}$.

However, we want to avoid the terms `microscopic' and `macroscopic' since they are vague. The relevant distinction is that between an object and a quasiobject. The property values of a quasiobject (like an electron) can never be directly perceived, and therefore their history cannot be encoded in the object itself, but only in the state of its environment from which we deduce the very existence and the attributes of the quasiobject. This encoding may take place in the experimental setup, in the object $O$, or in the memory of a subject, in the body $OB$ of a experimenter (Fig. \ref{Fig61c}).

\begin{state}[\textbf{No information can be encoded in a permanent quasiobject}]
If the specimen $OS$ in an observational context $C$ is a quasiobject that is never directly perceived in the time period $[n,n+m^{(F)}]$ during which the set of observations in $C$ is performed (Definition \ref{contextualstate}), then $S_{OS}=S_{C}$.
\label{nocodedinfo}
\end{state}

We add the condition of permanence since the quality that an object is a quasiobject is not absolute; a given identifiable object may be a quasiobject at some times and be directly perceived at other times (Statement \ref{temporalquasi}).

Even if the specimen $OS$ is a directly perceived object, the values of the property $P$ of interest may not be directly perceivable. We may speak about `quasiproperties'. As an example we may take the blood sugar level of a human being. The physician cannot decide it by just looking at the patient, but must extract a blood sample an analyze it in an apparatus. The case of blood pressure is slightly different. To some extent it is possible to deduce it by looking at the skin color or the veins on the forehead, neck or hands. In other words, there is conditional knowledge that relates perceivable attributes to the `hidden' property.

\begin{defi}[\textbf{Quasiproperty}]
A property of an object whose values are not directly perceivable is a quasiproperty if and only if these values are not related by conditional knowledge to the values of any other directly perceivable property of the same object.
\label{quasiproperty}
\end{defi}

\begin{state}[\textbf{No information can be encoded in a specimen if only quasiproperties are observed}]
If all properties $P,P',\ldots,P^{(F)}$ of a specimen $OS$ that is observed in a context $C$ are quasiproperties, then $S_{OS}=S_{C}$ regardless whether $OS$ is a quasiobject or not.
\label{nocodeinfo2}
\end{state}

After this conceptual digression, we return to the task to develop an algebraic representation of physical states. The state that we will try to represent is the contextual state $S_{C}$. Assume that $P$ is an independent attribute (Definition \ref{indattributes}), and that the complete set of future alternatives $\{\tilde{S}_{j}\}$ corresponds to the set of all values of $P$ that are allowed by physical law. Then there is exactly one exact state $Z$ for which the value of $P$ is $p_{j}$ for each exact state for which the value is $p_{i}$. This is true for each $i$ and $j$. It follows that $V[S_{Pi}]=V[S_{Pj}]$ for all $i,j$. If the number of possible property or attribute values is finite and equals $m$ we may write $v[S_{Pj},S_{C}]=1/m$.

\begin{defi}[\textbf{Fundamental property}]
A property that corresponds to an individual independent attribute according to Definition \ref{indattributes}.
\label{fproperty}
\end{defi}

\begin{defi}[\textbf{Fundamental set of future alternatives}]
A complete set $\{\tilde{S}_{j}\}$ of future alternatives defined by the values $p_{j}$ of some property $P$, such that there is one alternative $\tilde{S}_{j}$ for each property value allowed by physical law.
\label{fundalt}
\end{defi}

\begin{defi}[\textbf{Fundamental context}]
An observational context $C$ in which all involved properties are fundamental, and all sets of future alternatives are fundamental.
\label{fundamentalcontext}
\end{defi}

It may not be possible to create an observatinal context $C$ such that the set of future alternatives becomes fundamental. This occurs for example, if the allowed property values of $P$ form a continuous set. The set of possible outcomes of an actual observation must form a discrete set because each outcome must be subjectively distinguishable from all the alternatives that were not realized.

Figure \ref{Fig63} illustrates the situation where we have a fundamental set of (discrete) future alternatives for a fundamental property. For such contextual states it is meaningless to represent $S_{Pj}$ as a schema like that in Eq. [\ref{rep}]; all the relative volumes are equal and carry no information. We get, with $v_{Pj}\equiv v[S_{Pj},S_{C}]$,

\begin{equation}\begin{array}{llll}
\bar{S}_{C} & \equiv & \left[\begin{array}{cccc}
S_{P1} & S_{P2} & \ldots & S_{Pm}\\
v_{P1} & v_{P2} & \ldots & v_{Pm}
\end{array}\right] & =\\
&&&\\
& & \left[\begin{array}{cccc}
S_{P1} & S_{P2} & \ldots & S_{Pm}\\
m^{-1} & m^{-1} & \ldots & m^{-1}
\end{array}\right] & \hookrightarrow S_{C}.
\end{array}
\label{vpjrep}
\end{equation}

If we consider a property that is not fundamental, or a set of future alternatives that is not fundamental, the relative volumes $v_{Pj}$ need not be the same. These numbers are nevertheless of no primary interest, since it is the relative volumes $v[S_{j},S_{O}]$ that can be identified with a probability for a perceivable outcome of an observation. The crucial point is that probability is a measure that refers to the entire observational context $C$, not just the specimen $OS$.

Assume that a mouse is known to be somewhere in a house, and let property $P$ be its location. Let the value of $P$ be $p_{1}$ when the mouse is in the bathroom, and let it be $p_{2}$ when it is somewhere else in the house. We have $V[S_{P1}]<<V[S_{P2}]$, since the bathroom is just a small part of the house. In general this means that it is more likely to find property value $p_{2}$ than $p_{1}$.

The mouse may be very shy however, so that is is very hard to get a glimpse of it unless you trap it. Assume that you place one mouse trap in the bathroom and one trap in the cloakroom. Then the probabilities of the outcomes $p_{1}$ and $p_{2}$ are nevertheless roughly the same. In this case the state $S_{O}$ is the state that corresponds to the knowledge of the house, the mouse and the traps. $S_{1}$ is the future alternative that corresponds to finding the mouse in the trap in the bathroom, and $S_{2}$ is the alternative in which it is caught in the cloakroom. The mouse is the specimen, with state $S_{OS}$. Clearly, the relevant relative volumes are $v[S_{1},S_{O}]\approx v[S_{2},S_{O}]\approx 1/2$, since these are the ones that relates to the actual observation.

\subsection{Born's rule}
\label{bornrules}

Instead of Eq. [\ref{vpjrep}] we attempt the following, more meaningful representation

\begin{equation}
\bar{S}_{C}=
\left[\begin{array}{cccc}
S_{P1} & S_{P2} & \ldots & S_{Pm}\\
a_{1} & a_{2} & \ldots & a_{m}
\end{array}\right],
\label{mrep}
\end{equation}
where the numbers $a_{j}$ are related in some as yet undetermined way to the relative volumes $v_{j}$ of the future alternatives $\tilde{S}_{j}$, and thus to the probability to find property value $p_{j}$ in an actual observation within the context $C$. In other words, the numbers $a_{j}$ are contextual, referring not primarily to the contextual `naked' state $S_{C}$ of the specimen $OS$ that we investigate, but to the state of the entire experimental setup $O$, of which the specimen is just a small part. The state of this setup may carry information about the past and present state of the specimen that is not carried by the specimen itself, as discussed above. And this information may be crucial to determine the probability to find property value $p_{j}$.

\begin{figure}[tp]
\begin{center}
\includegraphics[width=80mm,clip=true]{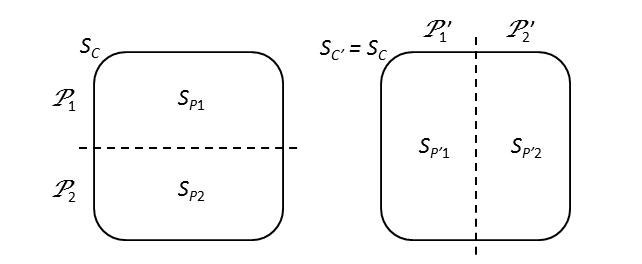}
\end{center}
\caption{The contextual states $S_{C}$ and $S_{C'}$ in two observational contexts aimed to measure the values of properties $P$ and $P'$, respectively, together with the corresponding property values states $S_{Pj}$ and $S_{P'j}$. We have $S_{C}=S_{C'}$, but the corresponding states $\bar{S}_{C}$ and $\bar{S}_{C'}$ defined within context may differ.} 
\label{Fig63b}
\end{figure}

The purpose of the state representation $\bar{S}_{C}$ is to strip the state $S_{O}$ of everything irrelevant. $\bar{S}_{C}$ represents, via the states $S_{Pj}$, just the possible property values that are about to be observed, and also aims to make it possible to calculate the probability to see these property values via the contextual numbers $a_{j}$. Therefore, $\bar{S}_{C}$ is just a partial representation of $S_{O}$:

\begin{equation}
\bar {S}_{C}\rightharpoonup S_{O}.
\label{cpartialrep}
\end{equation}
This means that different contexts $C$ and $C'$ with different initial states $S_{O}$ and $S_{O'}$ may have the same representation of its contextual state: $\bar {S}_{C}\rightharpoonup S_{O}$ and $\bar {S}_{C}\rightharpoonup S_{O'}$, or $\bar {S}_{C}=\bar{S}_{C'}$. In contrast, $\bar {S}_{C}$ is a complete representation of $S_{C}$, that is $\bar{S}_{C}\hookrightarrow S_{C}$. In fact, the representation is over-determined. As discussed above, it represents knowledge about the state of macroscopic experimental apparatus $OA$ apart from knowledge about the specimen $OS$. This means that the same state $S_{C}$ may have different representations $\bar{S}_{C}\hookrightarrow S_{C}$ and $\bar{S}_{C}'\hookrightarrow S_{C}$, depending on its experimental environment. For example, in the two cases where we aim to observe the value of properties $P$ and $P'$, respectively, we may write

\begin{equation}
\bar{S}_{C}=\left[\begin{array}{cc}
S_{P1} & S_{P2}\\
a_{1} & a_{2}
\end{array}\right],\;\;
\bar{S}_{C'}=\left[\begin{array}{cc}
S_{P'1} & S_{P'2}\\
a_{1}' & a_{2}'
\end{array}\right],
\end{equation}
as illustrated in Fig. \ref{Fig63b}. Even if $S_{C}=S_{C'}$ since these sets cover the same subset of state space, in general we have $\bar{S}_{C}\neq\bar{S}_{C'}$, since the sets of numbers $\{a_{j}\}$ and $\{a_{j'}\}$ may be different.

Let us leave the conceptual and notational subtleties aside. To fulfil the task of $\bar{S}_{C}$ to make it possible to calculate probabilities, a given set of numbers $\{a_{j}\}$ must correspond to a single set of relative volumes $\{v_{j}\}$, and thus to a given set of probabilities $\{q_{j}\}$. That is, we require

\begin{equation}
\{v_{j}\}=f(\{a_{j}\}).
\end{equation}
Of course, $S_{C}$ can still be seen as a legitimate state of an object (the specimen), and can therefore be represented by the schema [\ref{rep}] using relative volumes $v_{j}$ rather than by the schema [\ref{mrep}] using numbers $a_{j}$. This situation occurs when all knowledge of the context dissipates, so that $S_{O}$ grows to fill the entire state $S_{C}$ (Fig. \ref{Fig63}). To make these two representations consistent we require

\begin{equation}
v_{j}=f(a_{j}).
\label{vfa}
\end{equation}
There is no need to require \emph{a priori} that the function $f$ is invertible, given the purpose of the representation [\ref{mrep}] that we seek, which is to keep track of property values and probabilities.

To do so, we want to be free to do algebraic manipulations in the corresponding representation

\begin{equation}
\bar{S}_{C}=a_{1}\bar{S}_{P1}+a_{2}\bar{S}_{P2}+\ldots a_{m}\bar{S}_{Pm}.
\label{amrep}
\end{equation}

We concluded above that the distributive law [\ref{divrep2}] holds in the analogous representation [\ref{arep}] of future alternatives and relative volumes. We seek a function $f(a_{j})$ that makes it possible to uphold the same distributive law for the numbers $a_{j}$ in the representation [\ref{amrep}], as well as a distributive law for the property value states:

\begin{equation}\begin{array}{rcl}
a_{x}(a_{y}+a_{z})\bar{S}_{Pj} & = & (a_{x}a_{y}+a_{x}a_{z})\bar{S}_{Pj},\\
(a_{x}+a_{y})\bar{S}_{Pj} & = & a_{x}\bar{S}_{Pj}+a_{y}\bar{S}_{Pj}.
\label{distlaw}
\end{array}
\end{equation}

The bars put above the states $S_{Pj}$ are introduced just to close the notation, so that we get objects of the same type at both sides of the equality in equations like Eq. [\ref{amrep}], and thus are free to perform the desired algebraic manipulations.

We also want to define a contextual evolution operator $u_{C}$ whose representation is linear, just like the ordinary evolution operator $\bar{u}_{1}$ becomes linear in the analogous representation [\ref{arep}] according to Eq. [\ref{aurep}]:

\begin{equation}
\bar{u}_{C}\bar{S}_{C}=a_{1}\bar{u}_{C}\bar{S}_{P1}+a_{2}\bar{u}_{C}\bar{S}_{P2}+\ldots a_{m}\bar{u}_{C}\bar{S}_{Pm}.
\label{linearuc}
\end{equation}

The evolution of $\bar{S}_{C}$ depends on the entire experimental setup $O$ and its evolution. We may define a contextual evolution operator such that $\bar{u}_{C}\bar{S}_{C}(n)$ is a proper representation of the contextual state just before the observation of property $P$ at time $n+m$. That is, the temporal update $n+m-1\rightarrow n+m$ corresponds to a state reduction

\begin{equation}
u_{C}S_{C}(n)\rightarrow S_{C}(n+m)\subset u_{C}S_{C}(n),
\end{equation}
where $u_{C}=u_{C}[S_{O}(n),S_{OB}]$. Similarly, $\bar{u}_{C}\bar{S}_{C}(n+m)$ is a proper representation of the contextual state just before the observation of property $P'$ at time $n+m'$, so that the temporal update $n+m'-1\rightarrow n+m'$ corresponds to a state reduction

\begin{equation}
u_{C}S_{C}(n+m)\rightarrow S_{C}(n+m')\subset u_{C}S_{C}(n+m).
\end{equation}

Equations [\ref{vfa}], [\ref{distlaw}], and [\ref{linearuc}] express three desiderata that the formal representation [\ref{amrep}] should fulfill. To make such a representation useful, it should also be generally applicable. It should be possible to use same function $f(a_{j})$ regardless the details of the context. It should not matter how many properties we observe in succession, or their knowability level. These four conditions taken together make it possible to motivate a specific form of the function $f(a_{j})$, as we will now see.

\begin{figure}[tp]
\begin{center}
\includegraphics[width=80mm,clip=true]{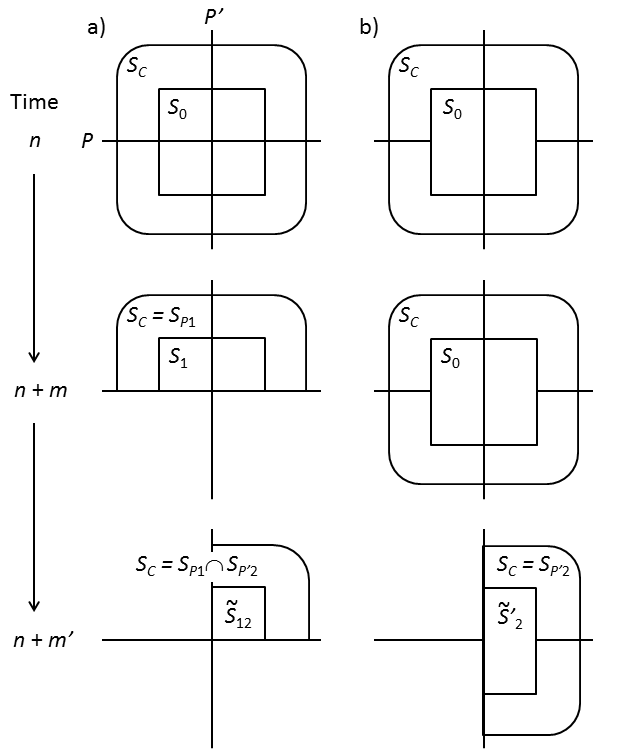}
\end{center}
\caption{The evolution in the setup depicted in Fig. \ref{Fig59} in terms of the contextual state $S_{C}$ and the property value states $S_{Pj}$. Panel a) shows the case in which properties $P$ and $P'$ both have knowability level 3, and the values $p_{1}$ and $p_{1}'$ are realized in succession. Panel b) shows the case in which property $P$ has knowability level 1. Again, property $P'$ has knowability level 3, and the value $p_{1}'$ is realized.} 
\label{Fig64}
\end{figure}

Consider Fig. \ref{Fig64}. As usual, the observational context is assumed to be initiated at time $n$, and properties $P$ and $P'$ attain their values at times $n+m$ and $n+m'$, respectively (Definitions \ref{observationalcontext} and \ref{contextualstate}). We use the vague phrase `attain their values' rather than `are observed', since we allow for the case where $P$ has knowability level 1 (Table \ref{levels}).

Consider first the case where property $P$ has knowability level 3 [Fig. \ref{Fig64}(a)]. A state reduction takes place at time $n+m$, at which value $p_{1}$ or $p_{2}$ is revealed:

\begin{equation}
\bar{S}_{C}(n+m)=\bar{S}_{Pj}
\end{equation} 
for $j=1$ or $j=2$. If the probabilities $q_{1}$ and $q_{2}$ of these alternatives are defined, we have $q_{j}=v_{j}=f(a_{j})$. At time $n+m'$ a second state reduction takes place, at which the value of $P'$ is revealed. The state that reduces is

\begin{equation}
\bar{u}_{C}\bar{S}_{C}(n+m)=
\left[\begin{array}{cc}
S_{PP'j1} & S_{PP'j2}\\
a_{j1} & a_{j2}
\end{array}\right]
\label{finalreduction}
\end{equation}
for $j=1$ or $j=2$, where

\begin{equation}
S_{PP'jk} \equiv S_{Pj}\cap S_{P'k}.
\end{equation}

If the probabilities $q_{1}'$ and $q_{2}'$ in this second set of alternatives are also defined, we get

\begin{equation}
q_{k}' = f(a_{1})f(a_{1k})+f(a_{2})f(a_{2k})
\label{classicprob}
\end{equation}
according to the classical axioms of probability.

Consider now the case where property $P$ has knowability level 1 [Fig. \ref{Fig64}(b)]. At time $n+m$ we know (contextually) that $P$ has attained value $p_{1}$ or $p_{2}$ but it is forever outside potential knowledge which of these values apply. No state reduction occurs, but epistemic completeness (assumption \ref{epcompleteness}) requires that we somehow account for the knowledge that `one of the two alternatives has occurred' in the representation of the contextual state. Most natural is to do it by writing

\begin{equation}
\bar{S}_{C}(n+m)=\bar{S}_{C}(n)=\left[\begin{array}{cc}
S_{C1} & S_{C2}\\
a_{1} & a_{2}
\end{array}\right],
\end{equation}
where $S_{C1}$ and $S_{C2}$ are two hypothetical contextual states that would have applied if the corresponding alternative had occurred. At time $n+m'$ a state reduction does finally take place. The state that reduces is

\begin{equation}
\bar{u}_{C}\bar{S}_{C}(n+m)=\left[\begin{array}{cc}
u_{C}S_{C1} & u_{C}S_{C2}\\
a_{1} & a_{2}
\end{array}\right],
\label{umrep}
\end{equation}
according to the linearity of $u_{C}$ in the formal algebraic representation (Eq. [\ref{linearuc}]. We should write

\begin{equation}
\bar{u}_{C}\bar{S}_{Cj}=\left[\begin{array}{cc}
S_{P'1} & S_{P'2}\\
a_{j1} & a_{j2}
\end{array}\right].
\label{umrep2}
\end{equation}
These expressions are different from Eq. [\ref{finalreduction}], since in the present case we have no knowledge about the value of $P$ after the final state reduction within context.

Let us express these relations in the formal algebraical representation, and manipulate them by making use of the algebraic rules [\ref{vfa}], [\ref{distlaw}], and [\ref{linearuc}] that this representation is supposed to fulfill. Consider first the case where both properties $P$ and $P'$ have knowability level 3. In the sequence of events shown in Fig. \ref{Fig64}(a):

\begin{equation}\begin{array}{rcl}
\bar{S}_{C}(n) & = & a_{1}\bar{S}_{P1}+a_{2}\bar{S}_{P2}\\
\bar{u}_{C}\bar{S}_{C}(n) & =  & a_{1}\bar{S}_{P1}+a_{2}\bar{S}_{P2}\\
\bar{S}_{C}(n+m) & = & \bar{S}_{Pj}\\
\bar{u}_{C}\bar{S}_{C}(n+m) & = & a_{j1}\bar{S}_{PP'j1}+a_{j2}\bar{S}_{PP'j2}\\
\bar{S}_{C}(n+m') & = & \bar{S}_{PP'jk}.
\end{array}
\label{classiccase}
\end{equation} 

Consider next the case where property $P$ haw knowability level 1 [Fig. \ref{Fig64}(b)]. Just like in the previous case we get

\begin{equation}\begin{array}{rcl}
\bar{S}_{C}(n) & = & a_{1}\bar{S}_{P1}+a_{2}\bar{S}_{P2}\\
\bar{u}_{C}\bar{S}_{C}(n) & =  & a_{1}\bar{S}_{P1}+a_{2}\bar{S}_{P2}.\\
\end{array}\end{equation}
However,

\begin{equation}\begin{array}{rcl}
\bar{S}_{C}(n+m) & = & a_{1}\bar{S}_{C1}(n+m)+a_{2}\bar{S}_{C2}(n+m),
\end{array}\end{equation}
and

\begin{equation}\begin{array}{rcl}
\bar{u}_{C}\bar{S}_{C}(n+m) & = & a_{1}\bar{u}_{C}\bar{S}_{C1}(n+m)+a_{2}\bar{u}_{C}\bar{S}_{C2}(n+m)\\
& = & a_{1}(a_{11}\bar{S}_{P'1}+a_{12}\bar{S}_{P'2})+a_{2}(a_{21}\bar{S}_{P'1}+a_{22}\bar{S}_{P'2})\\
& = & (a_{1}a_{11}+a_{2}a_{21})\bar{S}_{P'1}+(a_{1}a_{12}+a_{2}a_{22})\bar{S}_{P'2},
\end{array}
\label{algu}
\end{equation} 
where we have used Eq. [\ref{distlaw}] in the final equality. Finally,

\begin{equation}
\bar{S}_{C}(n+m')=\bar{S}_{P'k}.
\end{equation} 

Equation [\ref{algu}] means that the probability that the evolved contextual state $u_{C}S_{C}(n+m)$ will reduce to $S_{P'1}$ is $f(a_{1}a_{11}+a_{2}a_{21})$, and the probability that it will reduce to $S_{P'2}$ is $f(a_{1}a_{12}+a_{2}a_{22})$, provided that these probabilities exist. In short,

\begin{equation}
q_{k}'=f(a_{1}a_{1k}+a_{2}a_{2k}).
\label{quantprob2}
\end{equation}
We argued in Section \ref{closure} that we should not treat the case where the value of property $P$ is unknowable as if we can actually know it. According to the principle of explicit epistemic minimalism (Assumption \ref{explicitepmin}) we get the wrong answers if we do. In the present sitution this means that the probability [\ref{quantprob2}] must be different than in the case [\ref{classicprob}] where the value of $P$ becomes known, and can sometimes be associated with a probability $q_{j}$. That is,

\begin{equation}
f(a_{1}a_{1k}+a_{2}a_{2k})\neq f(a_{1})f(a_{1k})+f(a_{2})f(a_{2k}). 
\end{equation}
We may put the reason why different equations must hold in the two cases another way, as a consequence of epistemic completeness (Assumption \ref{epcompleteness}). The fundamental epistemic distinction between knowability levels 1 and 3 must correspond to a distinction in physical law. Such a distinction can be expressed only if different equations hold for the probabilities of the values of $P'$. To fulfill this condition we must require that

\begin{equation}
f(a)\neq a.
\label{fanota} 
\end{equation}

Given this fact, let us discuss which other functions $f$ are possible in Eq. [\ref{vfa}]. All relative volumes of a complete set of future alternatives add to one. We get the conditions

\begin{equation}\begin{array}{lll}
1 & = & f(a_{1})+f(a_{2})\\
1 & = & f(a_{11})+f(a_{12})\\
1 & = & f(a_{21})+f(a_{22})\\
1 & = & f(a_{1}a_{11}+a_{2}a_{21})+f(a_{1}a_{12}+a_{2}a_{22})
\end{array}
\label{acond}
\end{equation}

Since $f(a)$ corresponds to a relative volume, we must also require

\begin{equation}
0\leq f(a)\leq 1
\label{positivef}
\end{equation}
for all $a$ in the domain of $f$. To determine $f$ from these conditions we make use of the assumption that the parts of the observational context $O$ that correspond to properties $P$ and $P'$ can be arranged independently from each other.

One way to express this fact is to say that the parameters describing the experimental setup that determine the mode of observation of property $P'$ are independent from the corresponding parameters that determine the mode of observation of property $P$. If both $P$ and $P'$ have knowability level 3 (Table \ref{levels}), then this vague statement can be translated to a statement about relative volumes.

\begin{state}[\textbf{Relative volume independence}]Consider the set $\{C\}$ of all observational contexts $C$ with a given sequence $\ldots, P, P', \ldots $ of observed properties where $P$ and $P'$ have knowability level 3, and with given sets of possible values $\ldots,\{p_{j}\}, \{p_{k}'\}, \ldots $. There are enough elements $C$ in in $\{C\}$ so that the relative volumes $\{v_{jk}\}$ that describe the measurement of property $P'$ can be chosen independently from the relative volumes $\{v_{j}\}$ that describe a preceding measurement of $P$.
\label{volind}
\end{state}
This statement is quite trivial and follows from the fact that the only condition that the relative volumes has to fulfil \emph{a priori} is that they add to one: $1=\sum_{j}v_{j}$ and $1=\sum_{k}v_{jk}$. These relations do not mix relative volumes belonging to $\{v_{j}\}$ with those belonging to $\{v_{jk}\}$. Therefore relative volumes associated with a property $P$ are independent from those associated with another property $P'$. 

If the values of property $P$ are unknowable, then we must generalize the above statement to account for the fact that the relative volumes $\{v_{jk}\}$ may not be knowable either, since they do not correspond to a probability $q(p_{k}'|p_{j})$ that is possible to determine by repeating the experiment a large number of times. Nevertheless, $\{v_{jk}\}$ can still be formally defined in those cases, as

\begin{equation}
v_{jk}=f(a_{jk})=\frac{V[u_{1}S_{O}(n+m'-1)\cap\tilde{\mathcal{P}}_{j}\cap\tilde{\mathcal{P}}_{k}']}{V[u_{1}S_{O}(n+m'-1)\cap\tilde{\mathcal{P}}_{j}]}.
\label{vjk}
\end{equation}
However, in a strict epistemic approach we should not refer to potentially unknowable quantities in a physical statement about the independence of the properties $P$ and $P'$, just as we do not refer directly to the exact states $Z$ when we make statements about the evolution $u_{1}$ of the physical state $S$. A more general version of Statement \ref{volind} is then the following, which must now be seen as an assumption, since it cannot be motivated in the same straightforward way.

\begin{assu}[\textbf{Property independence}]Consider the set $\{C\}$ of all possible observational contexts $C$ with a given sequence $\ldots, P, P', \ldots$ of observed properties where $P'$ has knowability level 3, and with given sets of possible values $\ldots,\{p_{j}\}, \{p_{k}'\}, \ldots $. Let $\{\alpha_{j}\}$ be a set of knowable property values that describe the part of the observational setup that is related to the observation of $P$, and let $\{\alpha_{k}'\}$ be a corresponding set relating to $P'$. Suppose that these parameter sets are minimal in the sense that they determine the probability $q_{k}'$ to get the value $p_{k}'$ for each $k$, so that we may write $\{q_{k}'\}=f(\{\alpha_{j}\},\{\alpha_{k}'\})$, but if we take away one parameter $\alpha_{j}$ or $\alpha_{k}'$ this is no longer true. Then there are enough elements $C$ in $\{C\}$ so that $\{\alpha_{j}\}$ can be chosen independently from $\{\alpha_{k}'\}$.
\label{propind}
\end{assu}

If property $P$ has knowability level 3, we can choose $\{\alpha_{j}\}=\{v_{j}\}$ and $\{\alpha_{k}\}=\{v_{jk}\}$, and we regain Statement \ref{volind}. These sets are minimal since $q_{k}'=\sum_{j}v_{j}v_{jk}$ for each $k$, but we cannot take away any element from these sets and still determine all probabilities $q_{k}'$.

We may ask how many elements are contained in the two minimal sets of independent parameters $\{\alpha_{j}\}$ and $\{\alpha_{k}'\}$ that pertain to properties $P$ and $P'$, respectively. Suppose that there are $M$ and $N$ possible values of these properties in the context $C$. There is then $M-1$ independent values of $v_{j}$ and $M(N-1)$ independent values of $v_{jk}$, taking into account the relations $1=\sum_{j}v_{j}$ and $1=\sum_{k}v_{jk}$. These numbers give the requested number of elements in $\{\alpha_{j}\}$ and $\{\alpha_{k}'\}$ if both $P$ and $P'$ have knowability level 3, since then we can choose $\{\alpha_{j}\}=\{v_{j}\}$ and $\{\alpha_{k}\}=\{v_{jk}\}$. We may argue that the amount of freedom to choose the experimental setup should never be less than in this case. 

\begin{assu}[\textbf{Experimental freedom}]Consider the set $\{C\}$ of all possible observational contexts $C$ with a given sequence $\ldots, P, P', \ldots$ of observed properties where $P'$ has knowability level 3, and with given sets of possible values $\ldots,\{p_{j}\}, \{p_{k}'\}, \ldots $. Suppose that $\{p_{j}\}$ and $\{p_{k}'\}$ contain $M$ and $N$ values, respectively. Then the sets of independent parameters $\{\alpha_{j}\}$ and $\{\alpha_{k}'\}$, as defined in Assumption \ref{propind}, contain at least $M-1$ and $M(N-1)$ values, respectively.
\label{expfree}
\end{assu}
If we add the numbers $M-1$ and $M(N-1)$ we conclude that there is always at least $MN-1$ free parameters to describe the observations of $P$ and $P'$.

We look for a function $f(a)$ such that the numbers $a_{j}$ and $a_{jk}$ can always be used to parametrize the necessary level of experimental freedom, just as $v_{j}$ and $v_{jk}$ can in the cases where both $P$ and $P'$ have knowability level 3. Otherwise the algebraic representation $\bar{S}_{C}$ of the contextual state becomes useless, since it cannot be used to calculate probabilities for the possible outcomes in all kinds of experiments. The whole point of the search for $f(a)$ is that we should find a function that makes the form of the representation [\ref{amrep}] generally applicable, regardless the number of observed properties, their knowability level, and their sets of possible values. This means that we should be able to write

\begin{equation}\begin{array}{lll}
\{\alpha_{j}\} & = & F(\{a_{j}\})\\
\{\alpha_{k}'\} & = & F'(\{a_{jk}\}).
\end{array}
\label{bigf}
\end{equation}
It then follows from property independence that $f(a)$ is also such that the elements in $\{a_{j}\}$ and $\{a_{jk}\}$ can be chosen independently. 

If both $P$ and $P'$ have knowability level 3 [Fig. \ref{Fig64}(a)] we could identify $\{\alpha_{j}\}=\{v_{j}\}$ and $\{\alpha_{k}'\}=\{v_{jk}\}$ with the independent parameter sets introduced in Assumption \ref{propind}. We cannot in general do the corresponding identifications $\{\alpha_{j}\}=\{a_{j}\}$ and $\{\alpha_{k}'\}=\{a_{jk}\}$. The parameters $\alpha_{j}$ and $\alpha_{k}$ are defined in Assumption \ref{propind} as values of properties that define the design of the observational context $C$, which must be assumed to be known \emph{a priori}. This means that they should not only be knowable, but already known at the start of the experiment at time $n$. In contrast, the relative volumes $v_{j}$ and $v_{jk}$, as well as the numbers $a_{j}$ and $a_{jk}$ may be unknowable in principle when it is unknowable which value of $P$ is attained, when this property has knowability level 1. In that case the functions $F$ and $F'$ in Eq. [\ref{bigf}] are unknowable, even though it is known that they exist.

The assumptions \ref{propind} and \ref{expfree} make it possible to pinpoint a single acceptable function $f(a)$, given the other requirements discussed above.

We note first, however, that if properties $P$ and $P'$ both have knowability 3 within context, then any function $f(a)$ will do. In that case property independence and experimental freedom is automatically fulfilled. The fourth condition in Eq. \ref{acond} is replaced by $1=q_{1}'+q_{2}'= f(a_{1})f(a_{11})+f(a_{2})f(a_{21})+f(a_{1})f(a_{12})+f(a_{2})f(a_{22})$ according to the classical axioms of probability (see Eq. [\ref{classicprob}]). This condition follows from the first three conditions in Eq. [\ref{acond}], and is therefore not independent. This circumstance accounts for the fact that no restriction on $f(a)$ can be derived. 

In what follows, we therefore focus on the case where $P$ has knowablity level 1 and $P'$ has knowability level 3. We will see that the existence of the fourth condition in Eq. [\ref{acond}] is crucial, together with the requirement that the representation allows experimental freedom. Let us first make a general observation.

\begin{state}[\textbf{We must allow complex} $a$]No function $f:\mathbb{R}\rightarrow\mathbb{R}$ such that $f(a)\neq a$ fulfils conditions [\ref{acond}], and can also be used to parametrize the necessary level of experimental freedom (Assumption \ref{expfree}).
\label{complexa}
\end{state}

If all $a_{x}$ are restricted to be real and $f(a)\neq a$, we have four conditions in Eq. [\ref{acond}] that relate six real numbers $a_{1},a_{2},a_{11},a_{12},a_{21}$ and $a_{22}$. In that case two independent real parameters are sufficient determine the probabilities $q_{k}'$, whereas experimental freedom (Assumption \ref{expfree}) requires that at least $MN-1=3$ are necessary, since $M=N=2$.

\begin{figure}[tp]
\begin{center}
\includegraphics[width=80mm,clip=true]{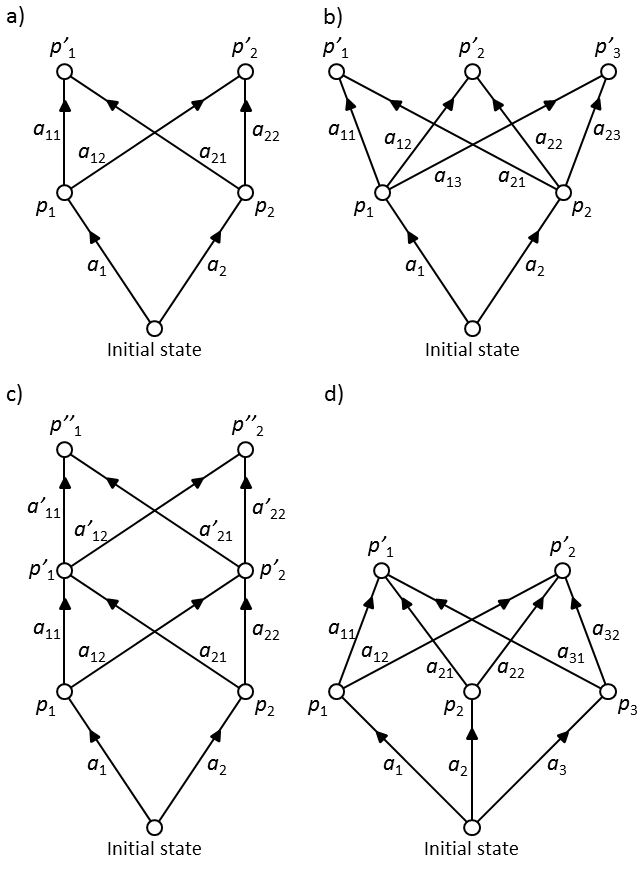}
\end{center}
\caption{Observational contexts depicted as `networks of alternatives'. The nodes at each row correspond to the values of a given property that are possible in the given context. The numbers $a_{x}$ are seen as relations between the possible values of sequentially realized properties. Time flows upwards, as indicated by the directed edges. The context in panel a) is the one shown in \ref{Fig59}. In more complex contexts, the visual representation introduced here is easier to interpret. We do not distinguish graphically between different knowability levels in this figure, but this can, of course, be accomplished by marking the nodes or the edges in different ways.} 
\label{Fig65}
\end{figure}

This consideration applies to the simple observational context with only two possible values for $P$ as well as for $P'$. Figure \ref{Fig65} shows how one can visualize more complex situations with $M$ possible values of $P$ and $N$ possible values of $P'$. Since we seek a function $f$ and numbers $a$ that apply generally, the fact that a restriction to real numers is impossible in the simplest context means that we must allow complex numbers.

One could try to identify the quantities $a_{x}$ with members of another collection (ring) of mathematical objects than the complex numbers (requiring that addition and multiplication is defined and yield another member of the same collection). We might, for example, consider vectors of three real numbers, or matrices. The discussion below will show, however, that nothing is gained in terms of property independence (Assumption \ref{propind}) if we use more complicated mathematical objects than complex numbers, objects defined by more than two real numbers.

Thus, we assume that $f: \mathbb{C}\rightarrow\mathbb{R}$. We may then write $v=f(a)=g(x,y)$, where $a=x+iy$ and $g: \mathbb{R}^{2}\rightarrow\mathbb{R}$. We then ask which functions $g(x,y)$ fulfil the requirements expressed in Eqs. [\ref{fanota}], [\ref{acond}], and [\ref{positivef}], as well as property independence and experimental freedom (Assumptions \ref{propind} and \ref{expfree}). We argue without proof that the only function that does the job is $g(x,y)=x^{2}+y^{2}$, that is, $f(a)=|a|^{2}$. This can be seen by inserting the Taylor expansion $g(x,y)=\sum_{m,n=0}^{\infty}d_{mn}x^{m}y^{n}$ into Eq. [\ref{acond}] and check in what cases property independence and experimental freedom can be upheld. The general expressions become messy, but the lesson is that more terms and higher exponents make it impossible to comply with these assumptions if we insist on fulfilling the other requirements. We take a shortcut through this mess by arguing that $f(a)$ should fulfil one additional condition, which makes the argument why $f(a)=|a|^{2}$ is the only possible choice much easier.

If we write the evolution of the contextual state representation $\bar{S}_{C}$ sequentially in the case where both $P$ and $P'$ have knowability level 3 (Fig. \ref{Fig64}(a)) we get Eq. [\ref{classiccase}]. The probability $q_{jk}$ to see the property values $p_{j}$ and $p_{k}'$ becomes $q_{jk}=f(a_{j})f(a_{jk})$. However, we may also regard $(P,P')$ as one single property with values $(p_{j},p_{k}')$. This value is decided at time $n+m'$. We may therefore write

\begin{equation}
\bar{u}_{C}\bar{S}_{C}(n+m)=a_{1}a_{11}\bar{S}_{PP'11}+a_{1}a_{12}\bar{S}_{PP'12}+a_{2}a_{21}\bar{S}_{PP'21}+a_{2}a_{22}\bar{S}_{PP'22}.
\end{equation}
In this way we see that $q_{jk}=f(a_{j}a_{jk})$. Thus the function $f(a)$ must also fulfil

\begin{equation}
f(a_{x}a_{y})=f(a_{x})f(a_{y}).
\end{equation}
There are only two operations on a pair of complex numbers that have this property, namely complex conjugation and exponentiation. That is, we must have $f(a)=(a^{*})^{m}a^{n}$. Since $f(a)$ is real we hve to require $m=n$, so that

\begin{equation}
f(a)=|a|^{2n},
\label{exponentialf}
\end{equation}
where $n$ is a positive integer.

Let us check first that the choice $f(a)=|a|^{2}$ is acceptable, as claimed. We start by considering the simple case where property $P$ has knowability level 1, and $P'$ has knowability level 3, and there are two possible values for each of these properties (Figs. \ref{Fig64}(a) and \ref{Fig65}(a)). Inserting the ansatz in Eq. [\ref{acond}] we get

\begin{equation}\begin{array}{lll}
1 & = & |a_{1}|^{2}+|a_{2}|^{2}\\
1 & = & |a_{11}|^{2}+|a_{12}|^{2}\\
1 & = & |a_{21}|^{2}+|a_{22}|^{2}\\
0 & = & a_{1}a_{2}^{*}(a_{11}a_{21}^{*}+a_{12}a_{22}^{*})+\\
& & a_{1}^{*}a_{2}(a_{11}^{*}a_{21}+a_{12}^{*}a_{22}).
\end{array}
\label{a2cond}
\end{equation}
Suppose that a given choice $\{a_{j}\}$ and $\{a_{jk}\}$ satisfies the last equation. Property independence (Assumption \ref{propind}) then means that we should be allowed to vary $\{a_{jk}\}$ freely for the given choice of $\{a_{j}\}$, or vice versa, and the equality would still hold. To make this possible we must require that the following relation is always fulfilled.

\begin{equation}
0=a_{11}a_{21}^{*}+a_{12}a_{22}^{*}.
\label{protoinnerprod}
\end{equation}

This relation corresponds to two equations that relate the eight real parameters in the set $\{x_{jk},y_{jk}\}$. The second and third lines in Eq. [\ref{a2cond}] give two more conditions, so that we have four free parameters that are related to the setup to measure the value of $P'$. The necessary minimum number of free parameters is two according to Assumption \ref{expfree}. All in all, we get five equations relating ten real parameters $\{x_{j},y_{j}\}$ and $\{x_{jk},y_{jk}\}$. This means that five free parameters are left to specify the probabilities, whereas the minimum possible number is three in order to respect experimental freedom. We conclude that the choice $f(a)=|a|^{2}$ for complex $a$ is acceptable in experimental contexts involving two properties $P$ and $P'$ with two possible values each. 

Let us next try $f(a)=|a|^{4}$. If we require property independence, we get the following conditions that relate the numbers $a_{jk}$, describing the setup to measure property $P'$:

\begin{equation}\begin{array}{lll}
0 & = & a_{11}a_{21}a_{21}^{*}a_{21}^{*}+a_{12}a_{22}a_{22}^{*}a_{22}^{*}\\
0 & = & a_{11}a_{11}a_{21}^{*}a_{21}^{*}+a_{12}a_{12}a_{22}^{*}a_{22}^{*}\\
0 & = & a_{11}a_{11}^{*}a_{21}a_{21}^{*}+a_{12}a_{12}^{*}a_{22}a_{22}^{*}\\
0 & = & a_{11}a_{11}a_{11}^{*}a_{21}^{*}+a_{12}a_{12}a_{12}^{*}a_{22}^{*}\\
\end{array}\label{messycond}
\end{equation}
These relations should be compared to the corresponding Eq. [\ref{protoinnerprod}] that holds for $f(a)=|a|^{2}$. They correspond to eight conditions that relate the eight real parameters in the set $\{x_{ij},y_{ij}\}$. The second and third lines in Eq. [\ref{a2cond}] give two more conditions, so that we have ten independent conditions, preventing any solution at all with property independence. Thus the choice $f(a)=|a|^{4}$ for complex $a$ is not acceptable.

It is easy too see that the higher exponent $n$ is used in Eq. [\ref{exponentialf}], the more independent conditions like those in Eq. [\ref{messycond}] appear when we demand property independence. This fact spoils all chances to get any solution at all for $n>1$. Having concluded in this way that $f(a)=|a|^{2}$ is therefore the only acceptable choice in the simple situation with two properties $P$ and $P'$ with two property values each, we must also check that it is acceptable in more complex situations with more than two properties that can take more than two values.

Figure \ref{Fig65} shows some of the simplest of these more complex cases. Sequential time $n$ flows upwards in these diagrams, so that the order in which the values of properties $P$, $P'$ and $P''$ occur is assumed to be known \emph{\emph{a priori}}. Each circle represents an `event' defined by the occurrence of property value $p_{x}$. The contextual numbers $a_{x}$ encapsulate the relative volume of the event at the end point of the associated arrow given the event at the starting point. If this latter event is knowable (if the corresponding property has knowability level 3), this relative volume may correspond to a probability for the event at the end point of the arrow given the event at the starting point.

The context depicted in Fig. \ref{Fig65}(b) gives rise to the following relation for the numbers $a_{jk}$ when $P$ has knowability level 1, and we require property independence (compare Eq. [\ref{protoinnerprod}]):

\begin{equation}
0=a_{11}a_{21}^{*}+a_{12}a_{22}^{*}+a_{13}a_{23}^{*}.
\label{protoinnerprod2}
\end{equation}
This relation corresponds to two conditions that relate twelve real parameters. The equations that correspond to the requirement $1=\sum_{j}v_{jk}$ give three more conditions, so that we have left seven independent real parameters describing the setup to measure property $P'$. This is more than enough to fulfil experimental freedom.

If we have more than two properties with possible property values each that are observed in succession, we get a new set of relation of the type in Eqs. [\ref{protoinnerprod}] and [\ref{protoinnerprod2}] for each such property at knowability level 1 when we demand property independence. For the context in Fig. \ref{Fig65}(c) we get

\begin{equation}\begin{array}{lll}
0 & = & a_{11}a_{21}^{*}+a_{12}a_{22}^{*}\\
0 & = & a_{11}'(a_{21}')^{*}+a_{12}'(a_{22}')^{*}.
\end{array}\label{protoinnerprod3}
\end{equation}
It is easily understood that experimental freedom is respected in cases such as that as well. If we have more than two possible values of $P$, we also get more than one relation of the same type when property independence is required. For the context in Fig. \ref{Fig65}(d) we get

\begin{equation}\begin{array}{lll}
0 & = & a_{11}a_{21}^{*}+a_{12}a_{22}^{*}\\
0 & = & a_{11}a_{31}^{*}+a_{12}a_{32}^{*}\\
0 & = & a_{21}a_{31}^{*}+a_{22}a_{32}^{*}.
\end{array}\label{protoinnerprod4}
\end{equation}
Here we have six conditions relating twelve real parameters. The requirement $1=\sum_{j}v_{jk}$ gives three conditions more. Thus we are left with three real parameters which equals the minimum number $M(N-1)$ needed to respect experimental freedom (Assumption \ref{expfree}). Here we are approaching a problem. If we increase the number of possible values $M$ of $P$ from three to four, and still have $N=2$ possible values of $P'$, we get six relations of the type in Eq. [\ref{protoinnerprod4}]. This means twelve conditions relating sixteen real parameters. The requirement $1=\sum_{j}v_{jk}$ gives four conditiions more, so that we have at most a unique solution $\{a_{jk}\}$ when we demand property independence. Experimental freedom is obviously not respected. This problem arises when $M>N$, and gets worse when $M-N$ gets bigger. 

Does this mean that there are contexts in which the choice $f(a)=|a|^{2}$ is not acceptable, so that we fail in the search for a function $f(a)$ that makes the representation [\ref{amrep}] generally applicable? No, we are saved by a trick, by a bit of sophistry, by the elusive relation between the numbers $a_{jk}$ and the tangible physical properties that describe the experiment. We may simply regard the set of contexts in which $M>N$ as contexts in which $M=N$ where the set of probabilities $\{q_{N+1}', q_{N+2}', \ldots, q_{M}'\}$ to see the last $M-N$ values of property $P'$ are set to zero. In such contexts we regain the necessary experimental freedom (Assumption \ref{expfree}) that should be reflected in the representation. This is so since we add $M(M-N)$ numbers $a_{jk}$, corresponding to $2M(M-N)$ new real parameters, while we only add $M-N$ conditions that relate all real parameters that occur in the representation, corresponding to  $q_{N+1}'=q_{N+2}'= \ldots= q_{M}'=0$.

We may argue that we should also set all corresponding relative volumes $v_{j(N+1)},v_{j(N+2)},\ldots,v_{jM}$ and numbers $a_{j(N+1)},a_{j(N+2)},\ldots,a_{jM}$ to zero, in order to really erase the ghostly presence of the imagined extra $M-N$ values of property $P'$. If we do this, the point of the trick is lost, since then we introduce as many new real parameters as we introdude conditions relating them. The degree of experimental freedom in the representation does not increase.

However, the procedure to add hypothetical values to $P$ is needed only when the knowability level of $P$ is 1, when the value this property attains is forever unknowable. In that case $a_{jk}$ and $v_{jk}$ are also unknowable in principle. (To determine $v_{jk}$ would mean to repeat the experiment many times until the conditional probability $q(p_{k}'|p_{j})$ is determined. But this would require that we know in which repetitions the value $p_{j}$ of $P$ was attained.) Therefore we should not refer to $a_{jk}$ or $v_{jk}$ explicitly when we demand that the context is not physically arranged so that imagined new values of $P'$ are observable. That would go against explicit epistemic minimalism (Assumption \ref{explicitepmin}). We should only refer to the knowable final probabilities $q(p_{k}')=q_{k}'$.

If, on the other hand, $P$ has knowability level 3, just as $P'$, then it is equivalent to demand that $q_{k}'=0$ and to demand that $q(p_{k}'|p_{j})=v_{jk}=f(a_{jk})=0$ for all $j$. We may introduce imagined values of $P$ and erase them again by setting the relevant probablities or relative volumes to zero, but there is absolutely no point to it - nothing changes in the representation.

\begin{state}[\textbf{Born's rule}]
Consider the formal algebraic representation [\ref{amrep}] of the contextual state $S_{C}$, where the relative volume $v_{j}$ of the corresponding future alternative $\tilde{S}_{j}$ is given by $v_{j}=f(a_{j})$. A generally applicable choice of function $f(a)$ is such that it makes this representation fulfil the distributive laws [\ref{distlaw}] in all observational contexts $C$ (Definition \ref{observationalcontext}), as well as the linearity of evolution [\ref{linearuc}]. Such a choice should also make sure that the representation respects explicit epistemic minimalism (Assumption \ref{explicitepmin}) in all contexts, and also expresses property independence (Assumption \ref{propind}) as well as the necessary degree of experimental freedom (Assumption \ref{expfree}). The choice $f(a)=|a|^{2}$, where $f: \mathbb{C}\rightarrow\mathbb{R}$, is generally applicable in this sense, and it is the only generally applicable choice.
\label{acceptablef}
\end{state}

\subsection{Contexts and Hilbert spaces}
\label{contextrep}

The form of relations [\ref{protoinnerprod}], [\ref{protoinnerprod2}], [\ref{protoinnerprod3}], and [\ref{protoinnerprod4}] resembles that of inner products in a vector space. We may treat them as actual Hermitian inner products in a complex vector space if we formally define the orthonormality relation

\begin{equation}
\delta_{ij}=\langle \bar{S}_{Pi},\bar{S}_{Pj}\rangle,
\label{ort1}
\end{equation}
where $\delta_{ij}$ is the Kronecker delta. Then the condition [\ref{protoinnerprod}] translates to

\begin{equation}\begin{array}{lll}
0 & = & a_{11}a_{21}^{*}+a_{12}a_{22}^{*}\\
& = & \langle a_{11}\bar{S}_{P'1}+a_{12}\bar{S}_{P'2},a_{21}\bar{S}_{P'1}+a_{22}\bar{S}_{P'2}\rangle\\
& = & \langle \bar{u}_{C}\bar{S}_{C1},\bar{u}_{C}\bar{S}_{C2}\rangle
\end{array}
\end{equation}

Consider the general case where a property $P$ at knowability level 1, with an arbitrary number of possible values, is observed before another property $P'$, which also has an arbitrary number of possible values. Then the conditions corresponding corresponding to Eq. [\ref{protoinnerprod}] can be collapsed to the relation

\begin{equation}
\forall ij: \,\,\,\delta_{ij}=\langle \bar{u}_{C}\bar{S}_{Ci},\bar{u}_{C}\bar{S}_{Cj}\rangle.
\label{ort2}
\end{equation}

The conditions of the form [\ref{protoinnerprod}] arose because we required that the representation [\ref{amrep}] should be able to express property independence. Therefore the orthonormality relation [\ref{ort2}] can be seen as a consequence of this requirement. In words, the hypothetical contextual state representations $\bar{S}_{Cj}$ that would have applied if we knew the value of $P$ are orthogonal to each other, and stay orthogonal as they evolve according to $\bar{u}_{C}$.

\begin{figure}[tp]
\begin{center}
\includegraphics[width=80mm,clip=true]{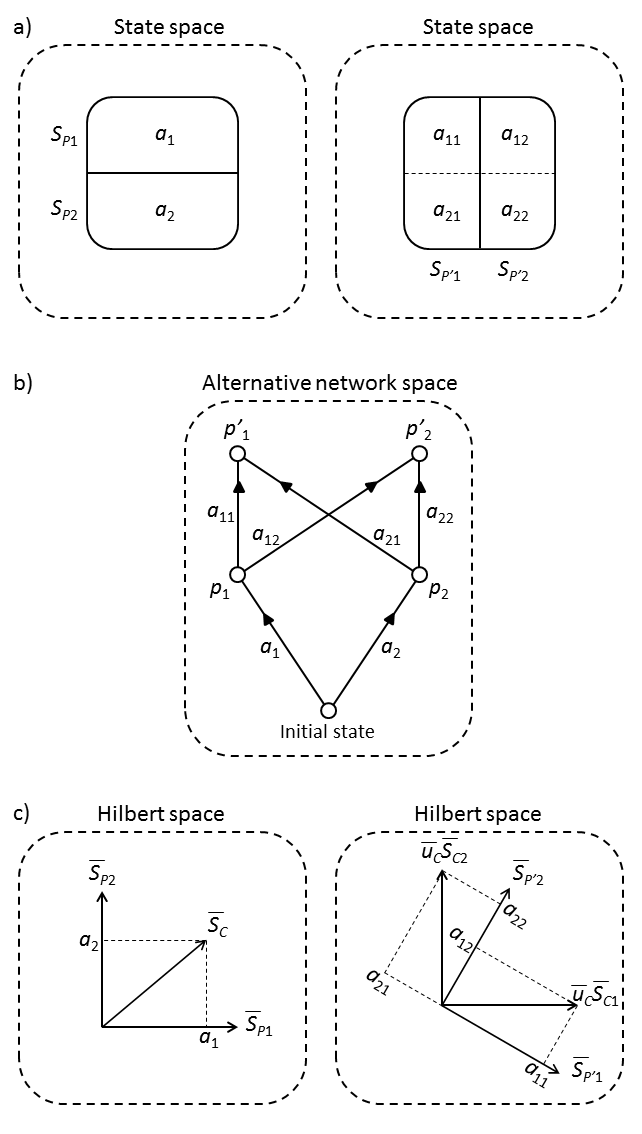}
\end{center}
\caption{Three representations of an observational context in three spaces. Compare Figs. \ref{Fig59} and \ref{Fig65}.} 
\label{Fig66}
\end{figure}

Let us focus again on the basic context in Fig. \ref{Fig65}(a). We see that we have now developed three ways to describe the same situation, as shown in Fig. \ref{Fig66}. We have the familiar state space description, the description as a network of alternatives, and now we also have a description in terms of a complex vector space, which we may call $\mathcal{H}_{C}$. It is sufficient to choose a two-dimensional vector space, in which case the orthonormality relations [\ref{ort1}] and [\ref{ort2}] implies that the contextual states $S_{C}$ and property value states $S_{Pj}$ become elements or normalized vectors $\bar{S}_{C}$ and $\bar{S}_{Pj}$ in this space [Fig. \ref{Fig66}(c)].

Since the numbers $a_{x}$ are contextual, so is the entire vector space $\mathcal{H}_{C}$. More precisely, it is defined only within an observational context $C$, as introduced in Defintion \ref{observationalcontext}.

In a general context where there are more than two possible values of some property, or there are more than two properties that will be realized in succession, the smallest possible dimension $D_{H}$ of $\mathcal{H}_{C}$ will be larger than two, and the property value states $S_{Pj}$ do not necessarily correspond to elements of $\mathcal{H}_{C}$, but should be interpreted more generally as orthogonal subspaces. Suppose that we are dealing with two properties $P$ and $P'$ with $M$ and $N$ possible values, respectively. If $P$ has knowability level 1 and $P'$ has level 3, we need the dimension $D_{H}=\max\{M,N\}$ to describe the context. If both $P$ and $P'$ has knowability level 3 and are simultaneously knowable, we need the dimension $D_{H}=M\times N$.

Let us analyze and generalize these statements. We start with the case where there are one or more property at knowability level 1, before the final observation of $P^{(F)}$ is made. Since Eq. [\ref{ort1}] should hold for all pairs of property value states $S_{Pi}$ and $S_{Pj}$ associated with any property $P$ that is part of the context, we need to have $D_{H}\geq M_{\max}\equiv\max\{M,M',\ldots, M^{(F)}\}$, where $M$ is the number of possible values of property $P$. This is necessary since we want to embed \emph{all} sets of orthogonal subspaces $\{\bar{S}_{Pj}\},\{\bar{S}_{Pj}'\},\ldots,\{\bar{S}_{Pj}^{(F)}\}$ in the \emph{same} vector space $\mathcal{H}_{C}$ that is supposed to describe the entire context, not just a particular property. Let $P$ be a property that is part of the context and has $M_{\max}$ possible values. Without loss of generality, we can choose the dimension of the associated subspaces $\bar{S}_{Pj}$ as small as possible, that is, let them be normalized vectors. In other words we may choose $D_{H}=M_{\max}$.

We may define the property space $\bar{S}_{P}$ as follows.

\begin{equation}
\bar{S}_{P} = \oplus_{j}\bar{S}_{Pj}
\label{propertyspaces}\end{equation}

According to the above discussion the dimension of such a property spaces is equal to or less than the dimension of the vector space $\mathcal{H}_{C}$ itself in the case where there are one or more property at knowability level 1. That is, $D_{S_{P}}\leq D_{H}$. The situation is different when all properties have knowability level 3. Then we must require $D_{\bar{S}_{P}}=D_{\bar{S}_{P'}}=D_{\bar{S}_{P}^{(F)}}= D_{H}$ for all properties $P, P',\ldots,P^{(F)}$ that are observed within context, as we will discuss below.

Generally speaking, if we try to identify the property value state $S_{Pj}$ with a subspace of a complex vector space, the dimension of $\bar{S}_{Pj}$ must be at least as large as the number of distinct outcomes there are from the context in which the value of $P$ is known to be $p_{j}$. Referring to the principle of epistemic minimalism, the dimension should not be greater than the number of such distinct outcomes either. That would introduce a redundancy of description, since such formal degrees of freedom would have no epistemic counterpart.

\begin{figure}[tp]
\begin{center}
\includegraphics[width=80mm,clip=true]{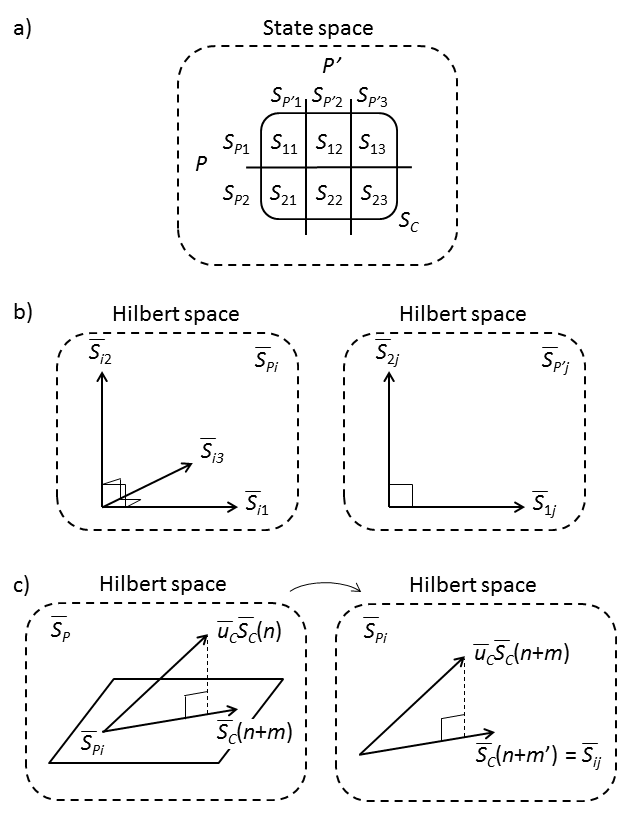}
\end{center}
\caption{In a context with two simultaneously knowable properties $P$ and $P'$ at knowability level 3, we must choose a vector space with dimension $D_{H}=M\times M'$. In this example we get $D_{H}=2\times 3=6$. The property value spaces $\bar{S}_{Pi}$ and $\bar{S}_{P'j}$ become subspaces rather than vectors. In this case we get $D[\bar{S}_{Pi}]=3$ and $D[\bar{S}_{P'j}]=2$ (panel b). The observation of the value of $P$ corresponds to a projection of the contextual state representation $\bar{S}_{C}$ down to $\bar{S}_{Pi}$ (panel c). The subsequent observation of the value of $P'$ corresponds to a further projection down to a vector $\bar{S}_{ij}$, which corresponds to one of the six compartments in state space shown in panel a).} 
\label{Fig68}
\end{figure}

The reasonable description is then to identify each possible state of final potential knowledge with a vector in $\mathcal{H}_{C}$ that is perpendicular to all other such vectors, which correspond to the other possible states of final potential knowledge. The situation is illustrated in Fig. \ref{Fig68}(a), where we have two properties $P$ and $P'$ at knowability level 3 with $M=2$ and $M'=3$, giving six distinct outcomes from the context as a whole, represented by the sets $S_{ij}$. We then let these sets correspond to perpendicular vectors, that is, $\langle \bar{S}_{ij},\bar{S}_{kl}\rangle=\delta_{(ij),(kl)}$.

The dimension of $\bar{S}_{P1}$ and $\bar{S}_{P2}$ is three, since there are three possible final states for each value of $P$, and the dimension of $\bar{S}_{P'1}$, $\bar{S}_{P'2}$ and $\bar{S}_{P'3}$ is two, since there are two possible final states for each value of $P'$. We may write $\bar{S}_{Pi}=\oplus_{j=1}^{D_{H}/M}\bar{S}_{ij}$ and $\bar{S}_{P'j}=\oplus_{i=1}^{D_{H}/M'}\bar{S}_{ij}$. This is illustrated in Fig. \ref{Fig68}(b). Equation [\ref{propertyspaces}] holds also in the present case where all involved properties have knowability level 3, so that both property spaces $\bar{S}_{P}$ and $\bar{S}_{P'}$ span the entire vector space $\mathcal{H}_{C}$, implying $D_{H}=D_{\bar{S}_{P}}=D_{\bar{S}_{P'}}$.

In this vector space picture, the observation of the value $p_{j}$ of property $P$ corresponds to a projection of the contextual state vector onto the subspace $\bar{S}_{Pi}$, and the subsequent observation of value $p_{j}'$ of property $P'$ corresponds to a further projection down onto the line defined by the vector $\bar{S}_{ij}$. This process is illustrated in Fig. \ref{Fig68}(c).

The above description is straightforwardly generalized to contexts with an arbitrary number of properties at knowability level 3, each having an arbitrary number of possible values. We get $D_{H}=M\times M'\times\ldots\times M^{(F)}$, and each subseqent observation of property values correspond to successive projections onto smaller and smaller subspaces of $\mathcal{H}_{C}$, until the state is described by one of the normalized vectors that correspond to one of the distinct final states $S_{ii'\ldots i^{(F)}}$. Each property space $\bar{S}_{P}$ spans the entire context space $\mathcal{H}_{C}$, implying $D_{H}=D_{\bar{S}_{P}}=D_{\bar{S}_{P'}}=\ldots D_{\bar{S}_{P^{(F)}}}$. Further, each property value space $\bar{S}_{Pj}$ is perpendicular to all other property value spaces associated with the same property, meaning that $\langle \bar{v}_{i},\bar{v}_{j}\rangle=0$ whenever $\bar{v}_{i}\in \bar{S}_{Pi}$, $\bar{v}_{j}\in \bar{S}_{Pj}$, and $i\neq j$.

A guiding principle for choosing the dimension $D_{H}$ of the vector space $\mathcal{H}_{C}$ has been that two states that are subjectively distinguishable, and both may occur at a given future time, should be represented by two perpendicular vectors in $\mathcal{H}_{C}$ whenever the context $C$ describes the possible states at this future time. To avoid formal redundancy with epistemic basis, $D_{H}$ is chosen as the smallest dimension for which this principle is fulfilled.

\vspace{5mm}
\begin{center}
$\maltese$
\end{center}
\paragraph{}

\begin{figure}[tp]
\begin{center}
\includegraphics[width=80mm,clip=true]{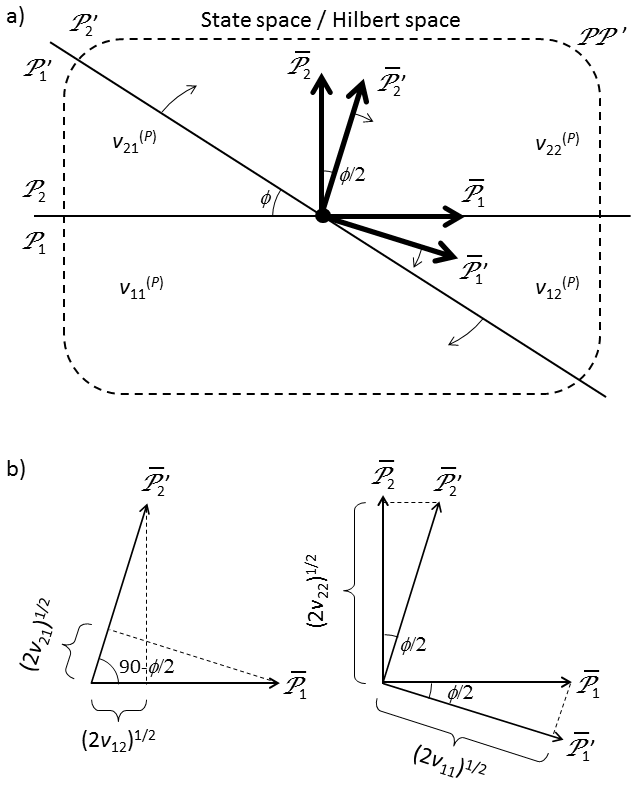}
\end{center}
\caption{Relation between the state space and vector space representation of two properties $P$ and $P'$ that are not simultaneously knowable. $P$ and $P'$ are assumed to have two possible values each. (a) The region of state space enclosed by the dashed curve is the set in which both $P$ and $P'$ are defined. When the tilted line associated with $P'$ is rotated to the left, so that $P$ and $P'$ become `more different', the corresponding basis in the vector space is also rotated, but only half the angle. We assume that the properties $P$ and $P'$ are fundamental (Definition \ref{fproperty}) so that the state space volumes at each side of the horizontal line are equal, as well as the volumes at each side of the tilted line. (b) This fact implies that $v_{ij}^{(P)}=v_{ji}^{(P)}$ and $v_{ii}^{(P)}=v_{jj}^{(P)}$, making it possible to represent the property value spaces associated with $P$ and $P'$ as two orthonormal bases in a vector space.} 
\label{Fig69}
\end{figure}

These considerations may also be applied to contexts in which we have two properties $P$ and $P'$ that both have knowability level 3, but are not simultaneously knowable. Say that there are two possible values of each property, like in Fig. \ref{Fig65}(a). Since they both have knowability level 3, one may argue that we should choose $D_{H}=M\times M'=4$. However, since there is no final outcome in which the state is one of the four possible property value combinations, it sufficient to let $D_{H}=\max\{M, M'\}=2$. This is the maximum number of degrees of freedom in terms of alternatives at any given time $n$.

Let us see if we can fit contexts with a sequence of such observations into the vector space formalism. However, before we do, let us consider the combined property space $\mathcal{P}\mathcal{P}'$ of two such properties. Having set this abstract stage, it will be easier to see how to handle actual observations.

For the sake of illustration, we will focus mainly on the case $M=M'=2$, that is, we consider one property $P$ with a set of possible values $\{p_{1},p_{2}\}$, and one property $P'$ with a set of possible values $\{p_{1}',p_{2}'\}$. Unfortunately, the reasoning does not generalize straightaway to situations in which $M=M'>2$. In due course we will return to these cases. In the meantime, the reader should take note that some statements only apply when $M=M'=2$.

\begin{state}[\textbf{Equipartition of property space}]Let $v[\mathcal{P}_{j}]=V[\mathcal{P}_{j}]/V[\mathcal{P}]$. If $P$ is fundamental according to Definition \ref{fproperty}, then $v[\mathcal{P}_{i}]=v[\mathcal{P}_{j}]$ for any pair $(p_{i},p_{j})$ of property values allowed by physical law.
\label{equiprop}
\end{state}

The reason why the relative volumes are equal is simply that for each exact state $Z$ for which the value of $P$ turns out to be $p_{i}$ there exists exactly one exact state $Z'$ for which the value is $p_{j}$, but all other attributes that defines the exact state are the same. This follows from the assumed independence of the attribute that defines the fundamental property $P$. Actually, Statement \ref{equiprop} is just a reformulation of a part of the definition of state space volume (Definition \ref{voldef}) in the vocabulary used here. Namely, a fundamental property is nothing but an independent attribute $A$, which means that we can write $\mathcal{P}_{i}=S(A,\upsilon_{i})$ according to Definition \ref{valuespacedef}. If we refer directly to the definition of $V[S]$ we do not have to bother about whether the possible values of $P$ are discrete or continuous. (The notation used here with subscripts implicitly assumes a discrete underlying set of property values, but this is not essential.) 

Consider Fig. \ref{Fig69}(a). Let us define the space $\mathcal{PP}'\subseteq\mathcal{S}$ as the set of exact states $Z$ for which there is at least one object for which both $P$ and $P'$ are defined. Let $v_{ij}^{(P)}=v[\Sigma_{ij}]$, where $\Sigma_{ij}$ is the region in $\mathcal{PP}'$ inside which the value of $P$ is $p_{i}$ and the value of $P'$ is $p_{j}'$. Explicitly,

\begin{equation}
v_{ij}^{(P)}\equiv\frac{V[\mathcal{P}_{i}\cap\mathcal{P}_{j}'\cap\mathcal{PP}']}{V[\mathcal{PP}']}.
\label{vjjp}
\end{equation}

Note that if $P$ and $P'$ is not simultaneously knowable, this region does not correspond to any alternative; there is no object $O$ that can have a state $S_{O}$ such that $S_{O}\subseteq\Sigma_{ij}$.

In any case, we see in Fig. \ref{Fig69}(a) that the equipartition of property value spaces (Statement \ref{equiprop}) implies that

\begin{equation}\begin{array}{rcll}
v_{ij}^{(P)} & = & v_{ji}^{(P)} & \forall(i,j)\\
v_{ii}^{(P)} & = & v_{jj}^{(P)} & \forall(i,j)
\end{array}
\label{vectorcond}
\end{equation}
in the case $M=M'=2$. This follows algebraically from the equipartition criteria $v_{11}^{(P)}+v_{12}^{(P)}=v_{21}^{(P)}+v_{22}^{(P)}=v_{11}^{(P)}+v_{12}^{(P)}=v_{12}^{(P)}+v_{22}^{(P)}=1/2$. (The corresponding criteria do not necessarily imply Eq. [\ref{vectorcond}] when $M=M'>2$.) 

Equation [\ref{vectorcond}] makes it possible to represent the two sets of property value spaces $\{\mathcal{P}_{1},\mathcal{P}_{2}\}$ and $\{\mathcal{P}_{1}',\mathcal{P}_{2}'\}$ as two orthonormal bases in one vector space, as illustrated in Fig. \ref{Fig69}(b). These bases are specified by the following conditions.

\begin{equation}\begin{array}{rcl}
\langle \bar{\mathcal{P}}_{i}, \bar{\mathcal{P}}_{j}\rangle & = & \delta_{ij}\\
\langle \bar{\mathcal{P}}_{i}', \bar{\mathcal{P}}_{j}'\rangle & = & \delta_{ij}\\
\langle \bar{\mathcal{P}}_{i}, \bar{\mathcal{P}}_{j}'\rangle & = & \sqrt{2v_{ij}^{(P)}}\exp(i\theta_{ij})
\label{vectorrepp}
\end{array}
\end{equation}
The relation between the inner product and the relative volume in the bottom row is chosen to conform with Born's rule in the following general sense.

\begin{state}[\textbf{Generalized Born's rule}]If the vectors $\bar{v}$ and $\bar{w}$ are associated with two regions $\Sigma_{v}\subset\mathcal{S}$ and $\Sigma_{w}\subset\mathcal{S}$ such that

\begin{equation}
V[\Sigma_{v}]=V[\Sigma_{w}],
\label{ncondv}
\end{equation}
then

\begin{equation}
|\langle \bar{v},\bar{w}\rangle|^{2}=\frac{V[\Sigma_{v}\cap\Sigma_{w}]}{V[\Sigma_{v}]}.
\end{equation}
\label{generalborn}
\end{state}

The requirement [\ref{ncondv}] is necessary to make sense of the representation of $\Sigma_{v}$ and $\Sigma_{w}$ as vectors or subspaces in a vector space.

By conforming to this rule in vector representations of all kinds of contexts, we can use the same formalism throughout. Recall that, according to Statement \ref{acceptablef}, Born's rule is forced upon us in contexts where two properties $P$ and $P'$ are observed, and where the alternatives of $P$ have knowability 1, as illustrated in Fig. \ref{Fig66}. In other kinds of contexts we may just choose to conform to it for simplicity.

Clearly, the conditions [\ref{vectorrepp}] just specify the relation between the bases completely up to a set of arbitrary, independent phases $\{\theta_{ij}\}$. In other words, if the property value spaces $(\{\mathcal{P}_{i}\},\{\mathcal{P}_{j}'\})$ of two properties $P$ and $P'$ are represented by the two bases $(\{\bar{P}_{i}\},\{\bar{P}_{j}'\})$ in a vector space, then this representation is preserved under the transformation

\begin{equation}
(\{\bar{P}_{i}\},\{\bar{P}_{j}'\})\rightarrow (\{\exp(i\theta_{i})\bar{P}_{i}\},\{\exp(i\theta_{j})\bar{P}_{j}'\}),
\label{phasered1}
\end{equation}
a symmetry that just reflects the redundancy of the representation. The need to consider complex vector spaces follows from the fact that we must allow complex contextual numbers $a_{j}$ (Statement \ref{complexa}).    

We may look at the boundary between $\mathcal{P}_{1}'$ and $\mathcal{P}_{2}'$ in Fig. \ref{Fig69} as a line that is fixed at the centre of $\mathcal{PP}'$ and may be rotated to alter the relative volumes $v_{ij}^{(P)}$. If the line is vertical we have $v_{11}^{(P)}=v_{12}^{(P)}=v_{21}^{(P)}=v_{22}^{(P)}$. The corresponding basis $(\bar{\mathcal{P}}_{1}',\bar{\mathcal{P}}_{2}')$ is tilted $45^{\circ}$ in relation to the basis $(\bar{\mathcal{P}}_{1},\bar{\mathcal{P}}_{2})$. If the boundary line is horizontal the two properties $P$ and $P'$ are not independent and can be seen as two manifestations of the same property. We have $v_{11}^{(P)}=v_{22}^{(P)}=1/2$ and $v_{12}^{(P)}=v_{21}^{(P)}=0$. The corresponding bases in vector space coincide. In general, if we rotate the boundary line in state space between $\mathcal{P}_{1}'$ and $\mathcal{P}_{2}'$ the angle $\phi$, the corresponding basis in vector space is rotated the angle $\phi/2$.

Let us now descend from the abstract realm to actual contexts in which $P$ and $P'$ are observed. From the definition of the contextual state $S_{C}$ and the property value states $S_{Pi}$ we see that we can formulate another equipartition principle as follows, analogous to that in Statement \ref{equiprop}.

\begin{state}[\textbf{Equipartition of the contextual state}]Consider the contextual state $u_{C}S_{C}(n)$ before the observation of the first property $P$ is made at time $n+m$. Let $v[S_{Pi}]=V[S_{Pi}]/V[u_{C}S_{C}(n)]$. If $C$ is fundamental according to Definition \ref{fundamentalcontext}, then $v[S_{Pi}]=v[S_{Pj}]$ for any pair $(p_{i},p_{j})$ of property values allowed by physical law.
\label{equicstate}
\end{state}

Figure \ref{Fig63} illustrates the close relation between $S_{C}$ before any observation is made and $\mathcal{P}$, and also between $S_{Pj}$ and $\mathcal{P}_{j}$.

Consider a fundamental context in which the $P$ and $P'$ are observed, and the alternatives correspond to the two sets $(p_{1},p_{2})$ and $(p_{1}',p_{2}')$ of values allowed by physical law. Then we can clearly represent $(S_{P1},S_{P2})$ and $(S_{P'1},S_{P'2})$ as two bases $(\bar{S}_{P1},\bar{S}_{P2})$ and $(\bar{S}_{P'1},\bar{S}_{P'2})$ in a vector space $\mathcal{H_{C}}$ in the same way as we did for the property value spaces in Fig. \ref{Fig69}.

However, we have to be more careful when we interpret the representation of the context than that of the property value spaces, since now we are representing a sequence of events rather than an abstract relation between properties. Clearly, in a given context there is a predefined order in which two properties are observed. This introduces a hierarchy among the corresponding bases used in the vector space representation; we have the basis associated with the first observation and that associated with the second observation. To obtain a more symmetric picture we need to consider contexts in which $P'$ is observed after $P$ in conjunction with those contexts in which $P'$ is observed first, and then $P$.

\begin{defi}[\textbf{Reverse context} $\breve{C}$]
Let $C$ be a context in which two properties $P$ and $P'$ are observed in succession. Let the sets of future alternatives associated with the two observations correspond to sets of possible property values $\{p_{i}\}$ and $\{p_{j}'\}$, respectively, where the number of members in each set is the same $(M=M')$. A context $\breve{C}$ is the reverse of $C$ if and only if the same properties $P$ and $P'$ are observed, but in reverse order, and the sets of possible property values are the same as in $C$.
\label{reversecontext}
\end{defi}

\begin{figure}[tp]
\begin{center}
\includegraphics[width=80mm,clip=true]{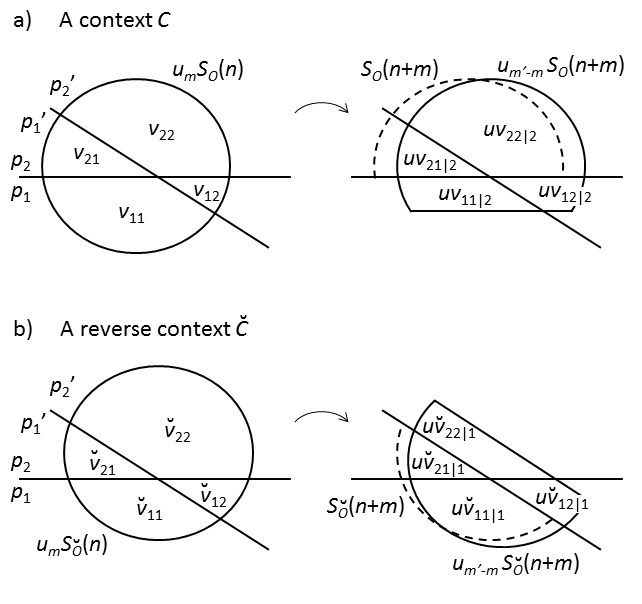}
\end{center}
\caption{a) A context $C$ in which $P$ is observed first, then $P'$. b) A reverse context $\breve{C}$ in which $P'$ is observed first, then $P$. The observation of the first property takes place at time $n+m$, at which time the states $S_{O}$ and $S_{\breve{O}}$ reduce (left panels). After the first observation in $C$, the state may evolve so that the relative volume $v_{ij}$ may change to $uv_{ij}$ just before the second observation. In this example, value $p_{2}$ of $P$ is observed. The observation of $P'$ takes place a time $n+m'$. The probability to see $p_{j}'$ given the observed value $p_{2}$ is $q(p_{j}'|p_{2})=uv_{1j}+uv_{2j}$. If $P$ and $P'$ had been simultaneously knowable, we would have had $uv_{11}=uv_{12}=0$. Compare Fig. \ref{Fig60}.} 
\label{Fig69b}
\end{figure}

A context $C$ and its reverse $\breve{C}$ are shown in Fig. \ref{Fig69b}. The properties $P$ and $P'$ and their relation in state space are assumed to be the same as those in Fig. \ref{Fig69}. We obviously have $S_{O}\subset\mathcal{P}\mathcal{P}'$. The left part of panel a) shows the state $S_{O}$ just before the observation of $P$, an event which is assumed to define $n+m-1\rightarrow n+m$. The state $S_{O}$ reduces when $P$ is observed. In Fig. \ref{Fig69b}(a) it reduces to a subset of the property value space $\mathcal{P}_{2}$ as the value of $P$ is found to be $p_{2}$. There will always pass some time after this event before $P'$ is observed. This second event defines $n+m'-1\rightarrow n+m'$. Between these two events, the object state $S_{O}$ have had time to move, governed by the evolution $u_{1}$, as indicated by the dashed and solid half-circles in the right part of panel a). If we wish, we may describe this motion with the continuous evolution parameter $\sigma$, as discussed in section \ref{evolutionparameter}. We let

\begin{equation}
u\equiv u_{m'-m},
\end{equation}

and define

\begin{equation}\begin{array}{lll}
v_{ij} & \equiv & \frac{V[\mathcal{P}_{i}\cap\mathcal{P}_{j}'\cap u_{m}S_{O}(n)]}{V[u_{m}S_{O}(n)]}\\
&&\\
uv_{ij|k} & \equiv & \frac{V[\mathcal{P}_{i}\cap\mathcal{P}_{j}'\cap uS_{O}(n+m)]}{V[uS_{O}(n+m)]},
\end{array}
\label{relvolreg}
\end{equation}
and correspondingly for the reverse context $\breve{C}$. In the second row, the evolved volumes $uv_{ij|k}$ depends on which value $p_{k}$ was found for property $P$. Recall that the regions with volumes $v_{ij}$, $uv_{ij|k}$, $\breve{v}_{ij}$ or $u\breve{v}_{ij|k}$ do not correspond to any realizable alternatives if $P$ and $P'$ are not simultaneously knowable. The object state can never be contained inside any single such region, and the evolution of such a region is not defined. Referring back to Fig. \ref{Fig65} we have

\begin{equation}\begin{array}{lll}
|a_{i}|^{2} & = & v_{i1}+v_{i2}\\
|a_{kj}|^{2} & = & uv_{1j|k}+uv_{2j|k}.
\end{array}
\end{equation}

For the reverse context the corresponding relations hold.

\begin{equation}\begin{array}{lll}
|\breve{a}_{i}|^{2} & = & \breve{v}_{i1}+\breve{v}_{i2}\\
|\breve{a}_{kj}|^{2} & = & u\breve{v}_{1j|k}+u\breve{v}_{2j|k}.
\end{array}
\end{equation}

\begin{figure}[tp]
\begin{center}
\includegraphics[width=80mm,clip=true]{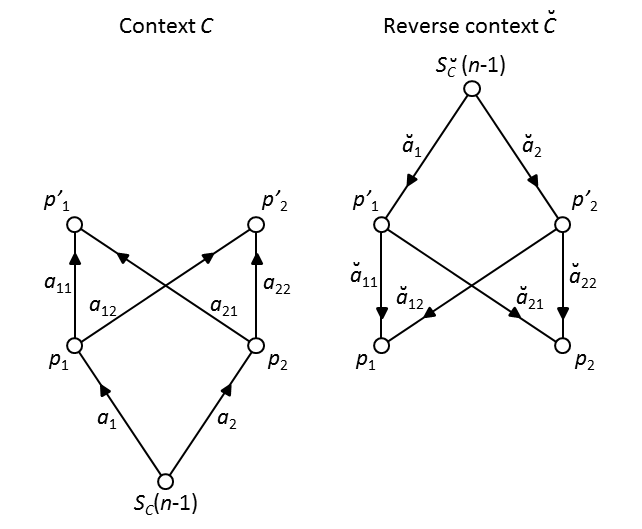}
\end{center}
\caption{A context $C$ in which two properties $P$ and $P'$ with two alternative values is observed, together with a reverse context $\breve{C}$ in which the order of observation is reversed. The contextual numbers $a_{x}$ and the reverse numbers $\breve{a}_{x}$ are not necessarily related in any particular way.} 
\label{Fig69a}
\end{figure}

The relation between $C$ and $\breve{C}$ in terms of these contextual numbers is illustrated in Fig. \ref{Fig69a}.

To be able to express contexts and reverse contexts in \emph{the same} vector space representation, corresponding to that in Fig. \ref{Fig69}, it should be possible to choose bases $(\bar{S}_{P1},\bar{S}_{P2})$ and $(\bar{S}_{P'1},\bar{S}_{P'2})$ so that these have the same mutual relation as $(\bar{\mathcal{P}}_{1},\bar{\mathcal{P}}_{2})$ and $(\bar{\mathcal{P}}_{1}',\bar{\mathcal{P}}_{2}')$.  This relation should not depend on the details of the context, just the inherent relation between the property values, as illustrated in Fig. \ref{Fig69}. Put differently, given a pair of bases $(\bar{S}_{P1},\bar{S}_{P2})$ and $(\bar{S}_{P'1},\bar{S}_{P'2})$, it should be possible to represent variations in the contextual details by a transformation of the contextual state vector $\bar{S}_{C}$ rather than a transformation of the relation beteen $(\bar{S}_{P1},\bar{S}_{P2})$ and $(\bar{S}_{P'1},\bar{S}_{P'2})$.

\begin{defi}[\textbf{Neutral context}]
Consider a context $C$ in which two not simultaneously knowable properties $P$ and $P'$ are observed, and suppose that we observe value $p_{i}$ of $P$. Then $C$ is neutral if and only if it is possible to assign labels so that $uv_{1j|k}+uv_{2j|k}+\ldots+uv_{Mj|k}=Mv_{kj}^{(P)}$ for all $(k,j)$, or, equivalently, $|a_{kj}|^{2}=|\langle\bar{\mathcal{P}}_{k},\bar{\mathcal{P}'}_{j}\rangle|^{2}$.
\label{neutralcontext}
\end{defi}

\begin{figure}[tp]
\begin{center}
\includegraphics[width=80mm,clip=true]{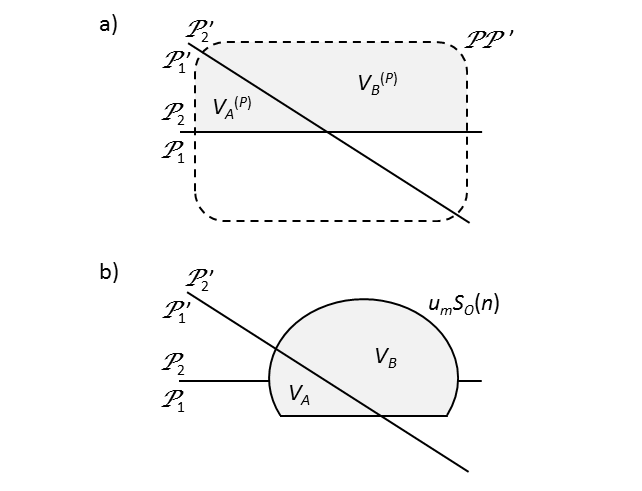}
\end{center}
\caption{Illustration of a neutral context. a) The relation between the volumes $V_{A}^{(P)}$ and $V_{B}^{(P)}$ in property space determines the angle between $\bar{\mathcal{P}}_{1}$ and $\bar{\mathcal{P}}_{1}'$, as well as between $\bar{\mathcal{P}}_{2}$ and $\bar{\mathcal{P}}_{2}'$ (Fig. \ref{Fig69}). b) A neutral context in which value $p_{2}$ of $P$ has been observed, and in which we are about to observe $P'$. The relation between $V_{A}$ and $V_{B}$ is the same as that between $V_{A}^{(P)}$ and $V_{B}^{(P)}$. More precisely, $V_{A}/V_{B}=V_{A}^{(P)}/V_{B}^{(P)}$. We get $q(p_{1}'|p_{2})=V_{A}/(V_{A}+V_{B})$ and $q(p_{2}'|p_{2})=V_{B}/(V_{A}+V_{B})$, where $q(p_{j}'|p_{2})$ is the probability to observe value $p_{j}'$ given value $p_{2}$. Analogous relations would hold if $p_{1}$ would have been observed. We can therefore use the property bases $(\bar{\mathcal{P}}_{1},\bar{\mathcal{P}}_{2})$ and $(\bar{\mathcal{P}}_{1}',\bar{\mathcal{P}}_{2}')$ to represent the context in a vector space.} 
\label{Fig69a2}
\end{figure}

Expressed in words, the relative volumes in the state $S_{O}$ that determines the contextual relation between $\bar{S}_{Pk}$ and $\bar{S}_{P'j}$ should be the same as the relative volumes in the entire property space $\mathcal{P}\mathcal{P}'$ that determine the relation between $\bar{\mathcal{P}}_{k}$ and $\bar{\mathcal{P}}_{j}'$. Figuratively speaking, a neutral context is transparent; it does not color the inherent relation between the properties we observe. An example of a neutral context is shown in Fig. \ref{Fig69a2}.

We can always formulate the formal algebraic representation

\begin{equation}\begin{array}{rcl}
\bar{u}_{C}\bar{S}_{C}(n) & = & a_{1}\bar{S}_{P1}+a_{2}\bar{S}_{P2}\\
\bar{S}_{C}(n+m) & = & \bar{S}_{Pj}\\
\bar{u}_{C}\bar{S}_{C}(n+m) & = & a_{k1}\bar{S}_{P'1}+a_{k2}\bar{S}_{P'2},
\end{array}
\label{typecevol}
\end{equation}
where we have assumed in the second and third row that the value $p_{k}$ of property $P$ was observed at time $n+m$. We may alternatively say that the context is neutral if and only if

\begin{equation}
|a_{kj}|^{2}=v_{kj}^{(P)}
\label{transparenta}
\end{equation}
for each pair of indices $(k,j)$, where $v_{kj}^{(P)}$ is defined in Eq. [\ref{vjjp}].

Let us define $q(p_{j}'|p_{k})$ as the conditional probability to find value $p_{j}'$ given that we have found value $p_{k}$. Analogously, let $q(p_{k}|p_{j}')$ be be conditional probability to find $p_{k}$ given $p_{j}'$ in a reverse context.

\begin{state}[\textbf{A principle of detailed balance}]
Consider a neutral context $C$ in which two properties $P$ and $P'$ are observed in succession, and a neutral reverse context $\breve{C}$. Then we have $q(p_{j}'|p_{k})= q(p_{k}|p_{j}')$ for all $(k,j)$.
\label{detailedbalance}
\end{state}

This statement follows directly from the fact that the same basis pairs $(\bar{\mathcal{P}}_{1},\bar{\mathcal{P}}_{2})$ and $(\bar{\mathcal{P}}_{1}',\bar{\mathcal{P}}_{2}')$ can be used to describe both $C$ and $\breve{C}$ if they are both neutral.

With the notion of neutral contexts at hand, let us clarify the issue when and how contexts dealing with not simultaneously knowable properties can be represented in a vector space. Whenever this is possible, we identify this vector space with that which is defined for the involved properties themselves, as indicated above. 

\begin{state}[\textbf{Existence of the property vector space} $\mathcal{H}_{PP'}$]
Consider two fundamental properties $P$ and $P'$ with the same number of possible values $(M=M')$, which fulfil $v_{ij}^{(P)}=v_{ji}^{(P)}$ for all $(i,j)$. Such a pair $(P,P')$ can be represented in a $M$-dimensional vector space $\mathcal{H}_{PP'}$ spanned by two orthonormal bases $(\bar{\mathcal{P}}_{1},\bar{\mathcal{P}}_{2})$ and $(\bar{\mathcal{P}}_{1}',\bar{\mathcal{P}}_{2}')$ that are related accordning to Eq. [\ref{vectorrepp}].
\label{existhpp}
\end{state}

The requirement that $P$ and $P'$ are fundamental (Definition \ref{fproperty}) is needed to fulfil the necessary condition [\ref{ncondv}] for a vector space representation. In the present case this condition requires the equipartition of the combined property space $\mathcal{P}\mathcal{P}'$, and such an equipartition is only guaranteed for fundamental properties (Statement \ref{equiprop}). This does not mean that it is impossible to construct vector space representations for other kinds of properties. We just have to make sure `by hand' that the property space is equipartitioned (and also that the conditions $v_{ij}^{(P)}=v_{ji}^{(P)}$ and $M=M'$ are fulfilled).

\begin{state}[\textbf{Representation of neutral contexts in a vector space} $\mathcal{H}_{C}$]
Consider a pair of properties $P$ and $P'$ such that the vector space $\mathcal{H}_{PP'}$ exists. Suppose that the same two properties are observed in a neutral context $C$, that they are not simultaneously knowable, and that the sets of future alternatives correspond to the sets of values $\{p_{i}\}$ and $\{p_{j}'\}$ that are allowed by physical law. Then $C$ can be represented in a vector space $\mathcal{H}_{C}$ which we may formally identify with $\mathcal{H}_{PP'}$ so that $\bar{S}_{Pi}=\bar{\mathcal{P}}_{i}$ and $\bar{S}_{P'j}=\bar{\mathcal{P}}_{j}'$ for all $1\leq i,j\leq M$.
\label{existhc}
\end{state}

Note that if $P$ and $P'$ were simultaneously knowable, it would not be possible to identify $\mathcal{H}_{C}$ with $\mathcal{H}_{PP'}$, since then we have to use a contextual vector space $\mathcal{H}_{C}$ with dimension $M\times M'=M^{2}$, whereas the dimension of $\mathcal{H}_{PP'}$ is $M$ (Fig. \ref{Fig68}).

\begin{figure}[tp]
\begin{center}
\includegraphics[width=80mm,clip=true]{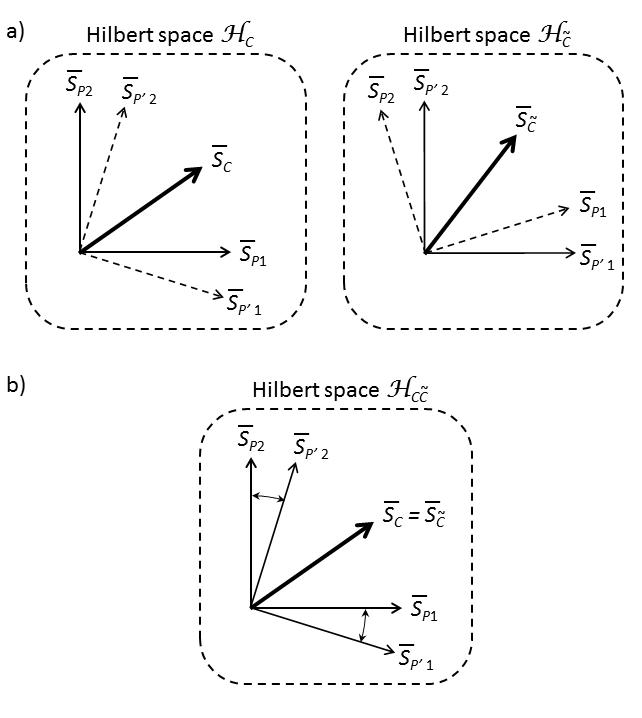}
\end{center}
\caption{Vector space representation of a context $C$ with a reciprocal $\tilde{C}$ where two not simultaneously knowable properties $P$ and $P'$ are observed. a) In $C$, property $P$ is observed first, and $\{\bar{S}_{P1},\bar{S}_{P2}\}$ is therefore the primary basis. In $\tilde{C}$ property $P'$ is observed first, and $\{\bar{S}_{P'1},\bar{S}_{P'2}\}$ beomes the primary basis. b) If we consider $C$ and $\tilde{C}$ together, we may identify $\bar{S}_{C}=\bar{S}_{\tilde{C}}$ and consider the two bases to be interchangeable in the common vector space.}
\label{Fig69d}
\end{figure}

We should keep in mind that for a given context $C$, the two bases $\{\bar{S}_{P1},\bar{S}_{P2}\}$ and $\{\bar{S}_{P'1},\bar{S}_{P'2}\}$ are not on equal footing, since the contextual state vector $\bar{S}_{C}$ is first projected onto $\bar{S}_{P1}$ or $\bar{S}_{P2}$ when $P$ is observed, then onto $\bar{S}_{P'1}$ or $\bar{S}_{P'2}$ when $P'$ is observed - but not the other way around. The order in which the projections occur is reversed in a reverse neutral context $\breve{C}$, of course. However, in order to treat the bases as equivalent in the sense that we change bases in the manner we are used to in quantum mechanics, we need to consider a reverse context $\breve{C}$ that is `the same' as $C$, apart from the reversion (Fig. \ref{Fig69d}).

\begin{defi}[\textbf{The reciprocal context} $\tilde{C}$]
Consider a neutral context $C$ that has a vector space representation $\mathcal{H}_{C}$. A reciprocal context $\tilde{C}$ is a reverse context $\breve{C}$ to $C$ for which $\breve{a}_{j}=a_{1}a_{1j}+a_{2}a_{2j}$ for all $j$, and $\breve{A}=A^{-1}$, where

\begin{equation}\begin{array}{rccccl}
\breve{A} & \equiv & \left(\begin{array}{cc}
\breve{a}_{11} & \breve{a}_{12}\\
\breve{a}_{21} & \breve{a}_{22}\end{array}\right), &
A & \equiv & \left(\begin{array}{cc}
a_{11} & a_{12}\\
a_{21} & a_{22}\end{array}\right) 
\end{array}
\end{equation}
in the case $M=M'=2$.
\label{reciprocalcontext}
\end{defi}

Such a choice of reciprocal context $\tilde{C}$ is chosen since it gives rise to an evolution of $\bar{S}_{\tilde{C}}(n)$ as if we set $\bar{S}_{\tilde{C}}(n)=\bar{S}_{C}(n)$ in Eq. [\ref{typecevol}], and make an algebraic change of basis from $\{\bar{S}_{P1},\bar{S}_{P2}\}$ to $\{\bar{S}_{P'1},\bar{S}_{P'2}\}$, using the fact that contextually we have $\bar{S}_{Pj}=a_{j1}\bar{S}_{P'1}+a_{j2}\bar{S}_{P'2}$ according to Eq. [\ref{typecevol}].

Clearly, $C$ is a reciprocal context to $\tilde{C}$ whenever $\tilde{C}$ is a reciprocal context to $C$.

\begin{defi}[\textbf{A pair of reciprocal contexts} $(C,\tilde{C})$]
The members of such a context pair are reciprocals to each other.
\label{reciprocalpair}
\end{defi}

The concept of pairs of reciprocal context makes it possible to motivate from the present vantage point the cavalier manner in which we change bases in quantum mechanical Hilbert spaces. We see that we may put the two bases on completely equal footing if we consider the context pair $(C,\tilde{C})$ together, and if we make a formal identification $u_{1}\bar{S}_{C}(n+m-1)=u_{1}\bar{S}_{\tilde{C}}(n+m-1)$ [Fig. \ref{Fig69d}(b)]. We include the temporal arguments to emphasize that the two state vectors are treated as equal before any observation is made in either context.

\begin{state}[\textbf{Physical interpretation of a change of basis}]
When we change the basis in a quantum mechanical Hilbert space to express the same state, what we are actually doing is to switch focus from a context $C$ to a reciprocal context $\tilde{C}$, and the vector space in which we change the bases is the combined vector space $\mathcal{H}_{C\tilde{C}}$ in which we consider the reciprocal context pair $(C,\tilde{C})$ together.
\label{physicalint}
\end{state}

The most natural illustration of these considerations is the subsequent measurements of the spin of an electron along two axes that are tilted the angle $\phi$ in relation to each other. Property $P$ is then the spin direction along the first axis, and $P'$ is the spin direction along the second axis.

There are, of course, angular momentum properties with more than two possible values. Let us therefore address the more general situation where we have two not simultaneously knowable properties for which $M=M'>2$. Then the equipartition conditions

\begin{equation}\begin{array}{rcll}
\sum_{j}v_{ij}^{(P)} & = & \sum_{j}v_{kj}^{(P)} & \forall(i,k)\\
\sum_{i}v_{ij}^{(P)} & = & \sum_{i}v_{il}^{(P)} & \forall(j,l)
\end{array}
\label{genequicon}
\end{equation}
for a fundamental property pair $(P,P')$ do not automatically imply $v_{ij}^{(P)}=v_{ji}^{(P)}$, which is a necessary condition for a vector space representation of $C$. Which property pairs $(P,P')$ that have more than two possible values each do actually fulfil the necessary additional condition $v_{ij}=v_{ji}$? Consider again the angular momentum $J$, observed along an arbitrary $z$-axis. Let us call this property $J_{z}$. The angular momentum component along another axis may be called $J_{z'}$. The relation between these properties are defined by the angle $\phi_{zz'}$ between the two axes. When we speak of the relation between two properties, we speak of the way the property values spaces $\mathcal{P}_{i}$ and $\mathcal{P}_{j}'$ relate to each other in the combined property space $\mathcal{P}\mathcal{P}'$ (Fig. \ref{Fig69}). We are therefore not allowed to refer to any particular context containing objects that can be used to define an external coordinate system in which we can assign coordinates to the $z$- and $z'$-axes. The only relation between $J_{z}$ and $J_{z'}$ is thus given by the angle $\phi_{zz'}$. An angle has no direction:

\begin{equation}
\phi_{zz'}=\phi_{z'z}.
\label{angleredundancy}
\end{equation}
This relation should be understood in the following sense. The permutation of the indices $z$ and $z'$ has no epistemic meaning if we consider the properties stripped from any context. Therefore, in a symbolic representation where the permutation nevertheless makes a graphical difference, like the one in Eq. [\ref{angleredundancy}], we must compensate for this redundancy in the representation by invoking a symmetry relation in this representation. This reasoning resembles the discussion in section \ref{minimalism}, concerning the ability to distinguish left from right (Fig. \ref{Fig17}). We conclude immediately that for the property pair $(J_{z},J_{z'})$ we must have $v_{ij}=v_{ji}$ because of such a redundancy of representation.

\begin{defi}[\textbf{Mutually defined property pairs}]A pair of properties $(P,P')$ is mutually defined if and only if the difference between them is completely specified by one or several relational quantities that lack direction, such as angles or spatial distances.
\label{mutualprop}
\end{defi}

It follows that mutually defined property pairs must have the same number of possible values, since a difference in number has a direction. A pair of integers can be ordered according to size. We also see that two properties related according to a temporal difference are not mutually defined, since time is directed.

\begin{state}[\textbf{Mutual properties have symmetrical relative volumes}]For any mutually defined fundamental property pair we have $v_{ij}^{(P)}=v_{ji}^{(P)}$ for all index pairs $(i,j)$.
\label{mutualequal}
\end{state}

If we let $P$ be the position and $P'$ be the momentum of the same object, there is no inherent relation between them at all, directed or not, that allows the observation of a given position $x_{i}$ to make it more probable to observe some momentum $p_{j}$ rather than $p_{l}$. This may be so in a particular experimental setup, but not in the entire property space $\mathcal{PP}'$. This becomes even more clear if we consider the fact that the numerical value of $x_{i}$ is just a matter of choice of a coordinate system, which requires external objects to define. In the abstract property space without any specific context, the numerical value assigned to $x_{i}$ is completely arbitrary. The same line of reasoning applies to $p_{j}$, since velocity is also defined in relation to an external coordinate system. We must set $v_{ij}=v_{kl}$ because we cannot define any quantity that relates $x_{i}$ and $p_{j}$ on which any difference could depend. 

\begin{defi}[\textbf{Independent property pairs}]A property pair $(P,P')$ is independent if and only if there is no set of relational attributes that can be used to specify the difference between them.
\label{indprop}
\end{defi}

\begin{state}[\textbf{Independent fundamental properties have equal relative volumes}]For any independent and fundamental pair of properties $(P,P')$ we have $v_{ij}^{(P)}=v_{kl}^{(P)}$ for all index pairs $(i,j)$ and $(k,l)$.
\label{indpropequal}
\end{state}

In particular, it follows that independent, fundamental properties have symmetrical relative volumes $v_{ij}^{(P)}=v_{ji}^{(P)}$, so that a vector space representation exists. Referring to Eqs. [\ref{vectorrepp}] and [\ref{phasered1}], for independent property pairs we must choose a pair of bases $(\{\bar{\mathcal{P}}_{i}\},\{\bar{\mathcal{P}}_{j}'\})$ in a common vector space such that

\begin{equation}
|\langle\bar{\mathcal{P}}_{i},\bar{\mathcal{P}}_{j}'\rangle|=|\langle\bar{\mathcal{P}}_{k},\bar{\mathcal{P}}_{l}'\rangle|,\ \forall(i,j,k,l).
\end{equation}

The reason why we include the technical condition in Statements \ref{mutualequal} and \ref{indpropequal} that the properties should be fundamental is that otherwise we could define one property value to consist of two possible property values of the corresponding fundamental property. Then the symmetry between a property value pair $(p_{i},p_{j}')$ could be broken, and we could motivate an assignment $v_{ij}^{(P)}\neq v_{kl}^{(P)}$.

\begin{figure}[tp]
\begin{center}
\includegraphics[width=80mm,clip=true]{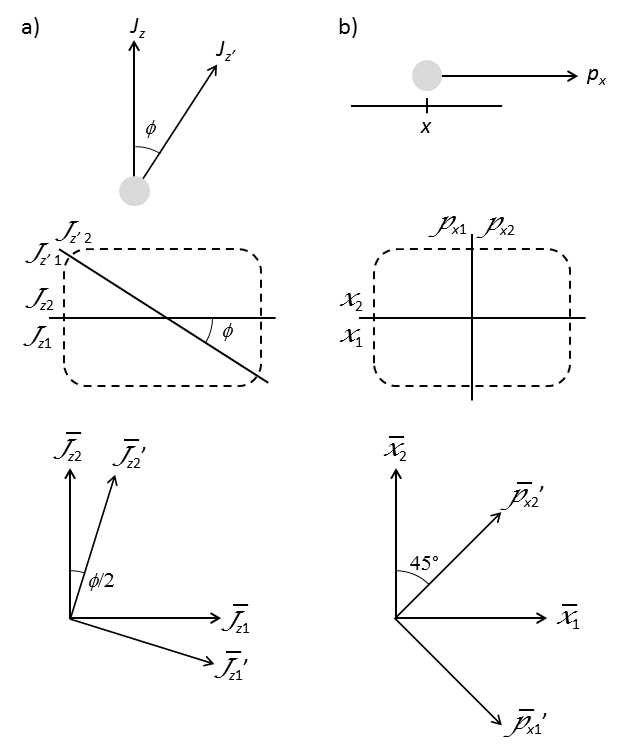}
\end{center}
\caption{Representation of pairs of not simultaneously knowable properties $P$ and $P'$. a) Representation of a pair of mutually defined such properties, exemplified by the angular momentum $J$ of a given object (grey) along two axes $\bar{e}_{z}$ and $\bar{e}_{z'}$. In the state space representation (middle panel), we have $v_{ij}=v_{ji}$ because of the symmetry in the mutual definition of the property pair. In the Hilbert space representation (bottom panel), the two bases that are associated with $P$ and $P'$ are tilted in relation to one another in a way that depends on the relational attribute $\phi$ that defines the property pair. b) Representation of a pair of independent such properties, exemplified by the position $x$ an momentum $p_{x}$ of a given object. In the state space representation all relative volumes $v_{ij}$ are the same. This holds true even though the possible values form a continuous set. In the Hilbert space representation, the two associated bases are tilted $45^{\circ}$ in relation to one another, given that there are only two possible values of each property. Compare Fig. \ref{Fig69}.} 
\label{Fig69e}
\end{figure}

\begin{state}[\textbf{Representable fundamental property pairs}]Consider a pair $(P,P')$ of not simultaneously knowable, fundamental properties, for which $M=M'$. If $(P,P')$ are mutually defined (Definition \ref{mutualprop}), or independent (Definition \ref{indprop}), they can be represented as two orthonormal bases in a vector space $\mathcal{H}_{PP'}$ with dimension $M$.
\label{representableprop}
\end{state}
This statement is illustrated in Fig. \ref{Fig69e}.

As an example of an independent property pair, we discussed the position and momentum of the same object. If we consider the pair of properties defined by the momenta of two different objects, the distance between them specifies the difference between the properties. They are therefore not independent according to Definition \ref{indprop}. Does this mean that a vector space representation does not exist? No, since the two properties in this case are simultaneously knowable, they can be represented in a common orthonormal basis of higher dimension ($M\times M'$). This is discussed in relation to Fig. \ref{Fig68} for specific contexts $C$. We can, of course, make the same kind of representation for the abstract properties themselves.

Let us finally discuss the case where we have a continuous set of property values allowed by physical law. In the case of $x$ or $p_{x}$, for example, we cannot exclude \emph{a priori} any real value of these quantities. Since the number $M$ of alternatives in an actual context $C$ where we observe such properties is always finite, $C$ is never fundamental (Definition \ref{fundamentalcontext}). We therefore cannot make the identification $\bar{S}_{Pi}=\bar{\mathcal{P}}_{i}$ straightaway. Instead, we must try to represent a non-fundamental version $P^{M}$ of such a property $P$, where we group together the possible values of the fundamental $P$ into $M$ bins that match the resolution of the observation.

Is it possible to perform such a discretization and still fulfill the necessary equipartition $v(\mathcal{P}_{i}^{(M)})=v(\mathcal{P}_{k}^{(M)})$ for all $(i,k)$? (If so, the equally necessary symmetry $v_{ij}^{(P)}=v_{ji}^{(P)}$ follows for a pair $[P^{M},(P^{M})']$ of such properties if they are mutual or independent.) Yes, any chopping up in finite pieces of the axis in state space defined by $P$ will do the job. The reason is that the definition of the volume $V[S]$ (Definition \ref{voldef}) does not care about particular coordinate systems applied to the attribute axes in state space. (Recall that we identify a fundamental property with an independent attribute.) The coordinate system is not fundamental to state space, as discussed in section \ref{statespaces}.

In that section, we introduced the attribute value space $S(A,\upsilon)$ (Definition \ref{valuespacedef}) and the attribute interval space (Definition \ref{valueintspace}). The latter concerns an entire interval of values $\Delta\upsilon$ rather than just one value. We used it to conclude that the volumes of any pair of such intervals of values of a given attribute are equal (Statement \ref{equivalue}). If we consider two independent attributes $A$ and $B$, we can make the analogous statement for the `areas' $\Delta\upsilon_{A}\Delta\upsilon_{B}$.

\begin{state}[\textbf{Equipartition of attribute value areas}]
Let $A$ and $B$ be two independent attributes. Also, let $(\Delta\upsilon_{A},\Delta\upsilon_{A}')$ be a pair of sets according to Definition \ref{valueintspace} that belong to $A$, and let $(\Delta\upsilon_{B},\Delta\upsilon_{B}')$ be a pair of corresponding sets that belong to $B$. Further, let $S(A,\Delta\upsilon_{A};B,\Delta\upsilon_{B})$ be the set of exact states $Z$ for which there is at least one object for which $A$ and $B$ are defined, and for which the value of $A$ belongs to $\Delta\upsilon_{A}$ and the value of $B$ belongs to $\Delta\upsilon_{B}$. Then $V[S(A,\Delta\upsilon_{A};B,\Delta\upsilon_{B})]=V[S(A,\Delta\upsilon_{A}';B,\Delta\upsilon_{B}')]$.
\label{equiarea}
\end{state}

\begin{figure}[tp]
\begin{center}
\includegraphics[width=80mm,clip=true]{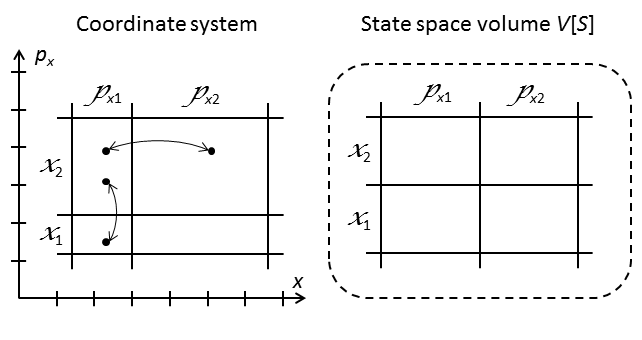}
\end{center}
\caption{A partition of the values of the independent, continuous properties $x$ and $p_{x}$ into bins with widths $\Delta x_{i}$ and $\Delta p_{xj}$, where each such interval corresponds to a given value $x_{i}$ and $p_{xj}$ of the discretized properties $x^{M}$ and $p_{x}^{M}$. Here $M$ is the number of bins used to partition the domain of each continuous property. a) The bins shown in a traditional phase space $(x,p_{x})$ with a coordinate system with which we can calculate the area of the elements $\Delta x\Delta p_{x}$. b) The bins shown in state space, where the only measure with which we can compare bin sizes is $V[S]$. The one-to-one correspondence between points in two bins in the vertical direction, as well as in the horizontal direction, implies that the volume of all bins is the same.}
\label{Fig69f}
\end{figure}

Let us define the values of a discretized property $P^{M}$ such that $p_{i}^{M}=\{p: p\in\Delta p_{i}\}$, where the interval $\Delta p_{i}$ contains a continuous infinity of property values $p$ of $P$. If we discretize a second property $P'$ in the same way, we see that we fulfill the necessary equipartition and symmetry $v_{ij}^{(P^{M})}=v_{ji}^{(P^{M})}$ for a vector space representation.

\begin{state}[\textbf{Representable non-fundamental property pairs}]
Consider a pair $(P^{M},(P^{M})')$ of not simultaneously knowable, non-fundamental properties, for which $M=M'$. Suppose that they are defined by the grouping together of continuous property values of the fundamental properties $P$ and $P'$, respectively. If both $P$ and $P'$ are mutually defined (Definition \ref{mutualprop}), or independent (Definition \ref{indprop}), they can be represented as two orthonormal bases in a vector space $\mathcal{H}_{P^{M}(P^{M})'}$ with dimension $M$.
\label{repnofund}
\end{state}

To exemplify the procedure, imagine that we have a double slit, towards which we shoot a specimen. There is a detector at each slit. There are three alternative outcomes in this context. The specimen may be detected at detector 1 or 2, or it may not be detected at all. We may then discretize the positions on the plane that is defined by the double slit screen into three compartments: those points that define slit 1 may be called $\Delta x_{1}$, those that define slit 2 may be called $\Delta x_{2}$, and the set of all other points are placed in the trashbin $\Delta x_{3}$. Then we may make a three-dimensional vector space representation. We may also combine the position measurement with a measurement of the momentum, which we should then discretize into three bins to enable a common algebraic representation.

In fact, the coordinate system independence of the possibility to make vector space representations of discretized continuous properties is crucial. Since the choice of coordinate system is arbitrary, basic physical distinctions, such as that between contexts that can be algebraically represented or not, should not depend on such arbitrary choices. Furthermore, when we take relativity into account, different observers may assign different coordinate values to parts of \emph{the same} context. We have chosen to build the algebraic representation upon the measure $V[S]$ on state space. We conclude that the coordinate independence of this measure is essential.

It may seem counter-intuitive that the choice of bin widths is irrelevant when we construct the vector space representation. Consider, for example, a context in which we observe position and momentum in succession, and in which there are just two alternative values for each property. This means that the bins may be `very big'. Even more, we can make one of the position bins, say $x_{2}$, very wide compared to the other [Fig. \ref{Fig69f}(a)]. Say that we indeed find value $x_{2}$ of the discretized position property. Even so, the outcome of the momentum measurement is completely undetermined, since the position and momentum bases are tilted $45^{\circ}$ in relation to one another regardless the details of the discretization. One might think that the finding of $x_{2}$ correspond to such a wide interval $\Delta x_{2}$ that Heisenberg would have allowed us to determine momentum well enough to be sure whether we would find $p_{x1}$ or $p_{x2}$, corresponding to the intervals $\Delta p_{x1}$ and $\Delta p_{x2}$, which may also be wide.

Recall, however, that the vector space representation is only possible for neutral contexts (Definition \ref{neutralcontext} and Fig. \ref{Fig69a2}), which are constructed to preserve the indeterminacy inherent in the abstract relation between $x$ or $p_{x}$. It may be harder to obtain such neutrality if we use a context with large bins. Recall also that the state that we represent in vector space is the memoryless contextual state $S_{C}$ of the specimen $OS$ (Definition \ref{contextualstate}). Whenever $OS$ is not a permanent quasiobject (like an electron), we may have more knowledge about its state than is encoded in $S_{C}$ (Statement \ref{nocodedinfo}). In other words, we may have $S_{OS}\subset S_{C}$. That is, the vector space description does not always represent all knowledge about a specimen. In particular, if the specimen is directly perceived object like a ball, we obviously can keep track of position $x_{i}$ and momentum $p_{xj}$ simultaneously if we choose large enough bins $\Delta x_{i}$ and $\Delta p_{xj}$.

\vspace{5mm}
\begin{center}
$\maltese$
\end{center}
\paragraph{}

We have been speaking of the complex vector space $\mathcal{H}_{C}$, and we have also introduced the combined vector space $\mathcal{H}_{C\tilde{C}}$ in Statement \ref{physicalint}. Let us be a little more precise about what we mean by these spaces.

\begin{defi}[\textbf{The complex vector space} $\mathcal{H}_{C}$]
Consider the set $SC$ of all observational contexts $C$ allowed by physical law for which the specimen $OS$ is the same, meaning that the specimens in all contexts are described by the same set of properties $P_{OS}$ with the same set the fixed value ranges $\Upsilon_{POS}$ (Definition \ref{observationalcontext}). Suppose also that the same sequence of properties $P,P',\ldots,P^{(F)}$ with the same sequence of knowability levels, and the same sets of realizable values $\{p_{j}\},\{p_{j}'\},\ldots,\{p_{j}^{(F)}\}$ are observed within each context $C\in SC$. To each such set $SC$ we associate one vector space $\mathcal{H}_{C}=\{\bar{S}_{C}(n)\}$, where $\{\bar{S}_{C}(n)\}$ is the set of all initial contextual states representations that corresponds to contexts in $SC$.
\label{hcdef}
\end{defi}

In plain language, we may say that each context that makes it possible to observe a given sequence of properties of a given specimen corresponds to a unique point in $\mathcal{H}_{C}$. At the next level, to each type of specimen and to each sequence of observed properties of this specimen corresponds a unique vector space $\mathcal{H}_{C}$. The dimension $D_{H}$ of $\mathcal{H}_{C}$ depends on the sequence of knowability levels of the observed properties, as well as on the fact whether these properties are simultaneously knowable or not. These matters are discussed above.

Definition \ref{hcdef} formally specifies the elements of $\mathcal{H}_{C}$, these being vectors of unit length with complex coordinates. It does not define any operations that we may perform on these vectors, and it does not tell us when $\mathcal{H}_{C}$ can be used in practice. These matters are summarized in Statement \ref{hilbertrep} below.

Operations on the elements of $\mathcal{H}_{C}$ have physical meaning only in certain circumstances. Well-defined projections occur at given instants during the course of the context. A change of basis is only justified if we consider a pair of reciprocal contexts $C$ and $\tilde{C}$, where the basis change corresponds to a change of perspective from one of these contexts to the other (Statement \ref{physicalint}). This viewpoint is necessary to make epistemic sense of the statement that \emph{one and the same} vector is expressed in two different bases (Fig. \ref{Fig69d}). All these qualifications express the same moral: $\mathcal{H}_{C}$ is not a universally defined state space in which we can follow the evolution of the state vector from a place for spectators, and look at it from different angles. Rather, we construct it ourselves when we build experimental setups, and it is defined during the course of the experiment only. When we build another experimental setup, we define another vector space $\mathcal{H}_{C'}$, independent from the first.

We have defined the inner product $\langle \bar{S}_{Pi},\bar{S}_{Pj}\rangle$ between basis vectors (Eq. [\ref{ort1}]), and we know that we have to deal with complex coordinates (Statement \ref{complexa}). To be able to call $\mathcal{H}_{C}$ a complex Hilbert space we must also argue that it is complete. If it is not complete, then there must be coordinates $a_{j}$ which can take a dense set of values in $H_{C}$, where this dense set contains holes in which a Cauchy sequence can end up.

We may write $a_{j}=\sqrt{v_{j}}\exp(i\phi_{j})$, where $v_{j}=V[\tilde{S}_{j}]/V[S_{O}]$. In our argument why $a_{j}$ has to be complex, we could not exclude any phase $\phi$. It must be allowed to take any value in $[0,2\pi)$. This means that there cannot be any holes in the set of allowed phases. We must look for them in the set of possible relative volumes $v_{j}$ instead. In the set $SC$ of contexts that contribute elements to $\mathcal{H}_{C}$ according to Definition \ref{hcdef}, there are experimental setups $O$ of all possible sizes, shapes and compositions. This means that $S_{O}$ can have a very large number of sizes, shapes and positions in state space, and consequently that the future alternative $\tilde{S}_{j}=S_{O}\cap\tilde{P}_{j}$ can have any conceivable size and shape (Fig. \ref{Fig63}). This means in turn that virtually any value of $v_{j}$ is possible, so that they may very well form a dense set. Can there be holes in this set? The existence of such a hole would mean that we can exclude a given pair of volumes $(V[\tilde{S}_{j}],V[S_{O}])$ that applies to context $C$, but not another pair $(V[\tilde{S}_{j}'],V[S_{O}'])$, which is arbitrarily close, that applies to a very similar context $C'$. Such a distinction would correspond to arbitrarily exact knowledge about the boundaries $\partial\tilde{S}_{j}$ and $\partial S_{O}$. That kind of knowledge cannot be attained according to the discussion in section \ref{knowstate}. Therefore there cannot be any holes between the values of $a_{j}$ even if they form a dense set, and therefore $\mathcal{H}_{C}$ is complete. 

\begin{state}[\textbf{The complex vector space} $\mathcal{H}_{C}$ \textbf{is a Hilbert space}]We have introduced an inner product in the construction of $\mathcal{H}_{C}$ and we have argued heuristically that it is complete.
\label{hchilbert}
\end{state}

\vspace{5mm}
\begin{center}
$\maltese$
\end{center}
\paragraph{}

Let us summarize our discussion about the complex Hilbert space representation of a sequence of observations within a given context. We have considered three kinds of contexts in which two properties with a given set of alternative values are known from the outset:

\begin{enumerate}
\item those where only the second property is observed, whereas the attained value of the first is outside potential knowledge,
\item those where both properties are observed in succession, and they are simultaneously knowable,
\item those where both properties are observed in succession, but they are not simultaneously knowable.
\end{enumerate}

These three kinds of situations play different roles in the translation from a set-theoretic state space description to the algebraic representation. Contexts of the first kind \emph{forces us} to choose a complex Hilbert space representation among all possible algebraic representations, whereas contexts of the second and third kind \emph{can} also be represented in this way. Observational contexts $C$ in which more than two properties are observed can be described as a combination of contexts of the above three kinds. There may, for example, be more than one unobserved property in contexts of the first kind (having knowability level 1) before any actual observation takes place. We have not mentioned the simplest kind of context of them all: those in which just one property is observed, and nothing more happpens. Such contexts can trivially be represented in complex vector spaces. We conclude that almost all context of interest can be represented in a complex Hilbert space. We refer to the preceding discussion for exceptions and limitations. The condition of neutrality (Definition \ref{neutralcontext}) in contexts of the third kind is maybe the most severe limitation.

\begin{state}[\textbf{Hilbert space context representation}]
Suppose that an observational context $C$ is such that the relative volumes of $S_{O}$ associated with the alternatives are known, so that the corresponding probabilities are defined. Then almost all contexts $C$ of practical interest can be represented in a complex Hilbert space $\mathcal{H}_{C}$ according to Definition \ref{hcdef}. When an observation corresponding to a state reduction occurs at time $n+m$, the evolved state vector that reduces can be written $\bar{u}_{C}\bar{S}_{C}(n)=\sum_{j}a_{j}\bar{S}_{Pj}$, where $\{\bar{S}_{Pj}\}$ is a set of orthogonal subspaces of $\mathcal{H}_{C}$. The reduced state vector $\bar{S}_{C}(n+m)$ is a projection of $\bar{u}_{C}\bar{S}_{C}(n)$ to one of the subspaces $\bar{S}_{Pj}$. The probability associated this projection is $|a_{j}|^{2}=|\langle\bar{u}_{C}\bar{S}_{C}(n),\bar{S}_{Pj}\rangle|^{2}$.
\label{hilbertrep}
\end{state}

Note that the Hilbert space representation is only partial; the exact shape and position of the sets $S_{O}$, $S_{OS}$ and $\tilde{S}_{j}$ in state space are not represented, just the associated relative volumes $v_{j}$. This fact is interesting just as a matter of principle, since the relevant quantities in an observational situation are the probabilities we identify with $v_{j}$. These probabilites are not always defined, as discussed in section \ref{probabilities}.

Note also that the Hilbert space representation is only meaningful as a way to calculate the outcome of an experiment if the specimen $OS$ is a permanent quasiobject (Definition \ref{nocodedinfo}), where `permanent' in this setting means that is never directly perceived during the course of the observational context $C$. 

\begin{state}[\textbf{Unique algebraic context representation}]Assume that the function $f(a)$ defined in Eq. [\ref{vfa}] is infinitely differentiable, and that we require that the distributive law expressed in Eq. [\ref{distlaw}] should hold. Then the Hilbert space description expressed in Statement \ref{hilbertrep} is the only possible algebraic representation that applies to almost all kinds of contexts $C$.
\label{uniquerep}
\end{state}

\subsection{Properties and operators}
\label{propop}

Having translated the alternatives and their probabilites or relative volumes to Hilbert space language, we may ask how the properties and the property values that the alternatives correspond to can be represented in the same language. We have discussed the subject to some extent in relation to the observation of properties that are not simultaneously knowable. Here we continue that discussion.

A property transcends the object for which it is defined, as well as the context in which it can be observed or measured. Therefore we should consider the entire property space $\mathcal{P}$, which contains the states of all such objects and contexts. Also, we should consider the set $\{p_{j}\}$ of all possible values of the property allowed by physical law, not just the set of possible values that are possible to see in a particular object.

To each value $p_{j}$ is associated a set $\mathcal{P}_{j}\subseteq\mathcal{P}$ such that $\mathcal{P}_{i}\cap\mathcal{P}_{j}=\varnothing$ for $i\neq j$ and $\mathcal{P}=\bigcup_{j}\mathcal{P}_{j}$. As discussed in Section \ref{propspaces}, it may not be possible to order property values subjectively, in contrast to attribute values. We may nevertheless label the property value spaces and the associated values with an index $j$ according to the above. We just have to remember that the labelling is sometimes arbitrary. If there is no inherent ordering among the property values, we can nevertheless assign numerical values to them, for instance according to $p_{j}=j$.

\begin{figure}[tp]
\begin{center}
\includegraphics[width=80mm,clip=true]{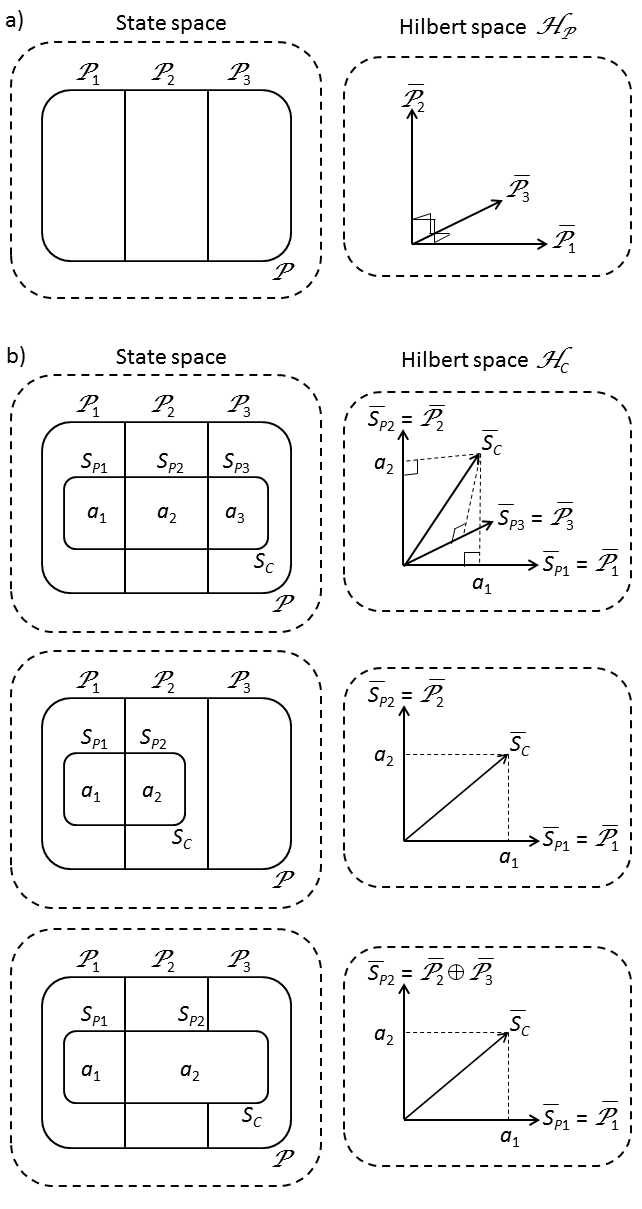}
\end{center}
\caption{a) The property space $\mathcal{P}$ as a Hilbert space $\mathcal{H}_{\mathcal{P}}$ spanned by the eigenvectors $\mathcal{P}_{j}$ of property operator $\bar{P}$. b) Relation between the property space $\mathcal{P}$ and the contextual state $S_{C}$ and the corresponding relation between the Hilbert spaces $\mathcal{H}_{\mathcal{P}}$ and $\mathcal{H}_{C}$. Three cases are shown. Top panel: all possible property values can be realized within context. Middle panel: Some property values cannot be observed. Bottom panel: some alternatives ($\tilde{S}_{2}$ in this example) correspond to several property values - the observation is not always precise.} 
\label{Fig70}
\end{figure}

We can formally construct a Hilbert space $\mathcal{H}_{\mathcal{P}}$ such that each property value space $\mathcal{P}_{j}$ corresponds to a vector $\bar{\mathcal{P}}_{j}$ such that $\langle\bar{\mathcal{P}}_{i},\bar{\mathcal{P}}_{j}\rangle=\delta_{ij}$ [Fig. \ref{Fig70}(a)]. This Hilbert space is analogous to $\mathcal{H}_{C}$ except for the fact that the coordinates $a_{j}$ are not determined by any context in the case of $\mathcal{H}_{\mathcal{P}}$. In the Hilbert space language, each property $P$ is uniquely specified by the complete basis $\{\bar{\mathcal{P}}_{j}\}$ with associated real numbers $\{p_{j}\}$. Conversely, each such pair of sets $(\{\bar{\mathcal{P}}_{j}\},\{p_{j}\})$ in $\mathcal{H}_{\mathcal{P}}$ defines at least one property $P$.

\begin{equation}
P\leftrightarrow\left\{\begin{array}{ll}
\mathcal{H}_{\mathcal{P}}\\
\{\bar{\mathcal{P}}_{j}\}, & \mathrm{a\:complete\:basis\:for}\:\mathcal{H}_{\mathcal{P}}\\
\{p_{j}\}, & p_{j}\in\mathbb{R}\:\mathrm{for\:all}\:j
\end{array}\right.
\end{equation}
A linear operator is uniquely defined by its eigenvectors and eigenvalues if the basis of eigenvectors is complete. Therefore we can associate $P$ with exactly one linear operator $\bar{P}$ with domain $\mathcal{H}_{\mathcal{P}}$, with a complete basis of eigenvectors $\bar{\mathcal{P}}_{j}$, and with real eigenvalues $p_{j}$. Any such operator $\bar{P}$ is necessarily self-adjoint: $\langle \bar{P}\bar{v},\bar{w}\rangle=\langle\bar{v},\bar{P}\bar{w}\rangle$ whenever $\bar{v},\bar{w}\in\mathcal{H}_{\mathcal{P}}$.

For any observational context $C$ in which property $P$ is observed, we have $S_{C}\subseteq\mathcal{P}$. It would therefore be natural to define $\mathcal{H}_{\mathcal{P}}$ so that $\mathcal{H}_{C}\subseteq\mathcal{H}_{\mathcal{P}}$. To relate the two Hilbert spaces, we define the action of $\bar{P}$ on a vector $\bar{S}_{C}\in\mathcal{H}_{C}$ so that

\begin{equation}
S_{C}\subseteq\mathcal{P}_{j}\Rightarrow\bar{P}\bar{S}_{C}\equiv p_{j}\bar{S}_{C}.
\label{eigenstate}
\end{equation}
If the property value state fulfils $S_{Pk}\subseteq\mathcal{P}_{j}$, then we may identify $\bar{S}_{Pk}=\bar{\mathcal{P}}_{j}$ since both quantities are defined as an eigenspace of the same operator with the same eigenvalue. If we define $\mathcal{H}_{\mathcal{P}}$ according to $\mathcal{H}_{\mathcal{P}}=\oplus_{j}\bar{\mathcal{P}}_{j}$, then we indeed get $\mathcal{H}_{C}\subseteq\mathcal{H}_{\mathcal{P}}$, as desired. In this picture we must regard $\bar{\mathcal{P}}_{j}$ as an eigenspace rather than an eigenvector, since $\bar{S}_{Pk}$ in some contexts must be seen as subspaces of $\mathcal{H}_{C}$ with dimension two or higher. This may happen when the context $C$ contains several properties at knowability level 3, as illustrated in Fig. \ref{Fig68}.

Assume that we have a fundamental set of future alternatives (Definition \ref{fundalt}). Then we have $S_{Pj}\subseteq\mathcal{P}_{j}$ for all $j$, so that we may write $\bar{S}_{Pj}=\bar{\mathcal{P}}_{j}$ for each $j$ [Fig. \ref{Fig70}(b), top panel]. The set $\{\bar{\mathcal{P}}_{j}\}$ becomes a complete set of orthogonal subspaces in $\mathcal{H}_{C}$ in the sense that $\mathcal{H}_{C}=\oplus_{j}\bar{\mathcal{P}}_{j}$, and $\langle \bar{v}_{i},\bar{v}_{j}\rangle=0$ whenever $\bar{v}_{i}\in\bar{\mathcal{P}}_{i}$ and $\bar{v}_{j}\in\bar{\mathcal{P}}_{j}$, where $i\neq j$. We have $\mathcal{H}_{C}=\mathcal{H}_{\mathcal{P}}$. The outcome of an observation within such a context is a property value that corresponds to exactly one eigenvalue $p_{j}$ of $\bar{P}$.

Assume now that we still have $S_{Pj}\subseteq\bar{\mathcal{P}_{j}}$ for all $j$, but that the complete set of alternatives is not fundamental. This simply means that there is one or more property values allowed by physical law that cannot be observed in the given context. The situation is illustrated in the middle panel of Fig. \ref{Fig70}(b), where $p_{3}$ cannot be realized. The angular momentum along a given axis is a typical example. Physical law allows a discrete sequence of values in the range from $-\infty$ to $\infty$, but the total angular momentum of the objects in the context truncates this sequence so that the absolute value of the angular momentum along the chosen axis never becomes greater than the total angular momentum. In this situation, we can still identify $\bar{S}_{Pj}=\bar{\mathcal{P}}_{j}$ for each set $S_{Pj}$ that occur in the complete set of alternatives, but the dimension of $\mathcal{H}_{C}$ is less than that of $\mathcal{H}_{\mathcal{P}}$. Such a statement is meaningful only if the dimension is finite. In general we may write $\mathcal{H}_{C}\subset \mathcal{H}_{\mathcal{P}}$, where $\mathcal{H}_{C}$ is spanned by a set of subspaces that is a proper subset of the complete set of eigenspaces $\{\bar{\mathcal{P}}_{j}\}$. Again, the outcome of an observation within this kind of context is a property value that corresponds to exactly one eigenvalue $p_{j}$ of $\bar{P}$, just like in the case of a fundamental set of alternatives.

Assume finally that the resolution of the observation of a property value is imperfect [Fig. \ref{Fig70}(b), bottom panel]. In other words, the property value that defines the future alternative $\tilde{S}_{j}$ corresponds to more than one of the property values allowed by physical law. Technically, there is a $k$ such that for each $j$ we have $S_{Pk}\not\subseteq\mathcal{P}_{j}$. Then we have to make a formal distinction between the imperfectly resolved `contextual property' $P_{C}$ and the `proper property' $P$. This situation always occurs for properties that are continuous, such as distance. This case was discussed in section \ref{contextrep} in relation to Fig. \ref{Fig69f}. (We may identify the contextual property with the discretized property $P^{M}$.) Of course, we may also have imperfect resolution among the alternatives that correspond to discrete-valued properties. If we allow ourselves to use the same index notation in both the continuous and the discrete case, we may say the following: sometimes we have to identify $\bar{S}_{Pk}$ with an entire set of eigenspaces: $\bar{S}_{Pk}=\oplus_{j=j0}^{i1}\bar{\mathcal{P}}_{j}$. Correspondingly, the outcome of the observation is a set of eigenvalues $p_{Ck}=\{p_{j0},p_{j0+1},\ldots,p_{j1}\}$, representing the property values that cannot be excluded by the acquired knowledge.

In the above discussion about the representation of properties as linear self-adjoint operators, we did not need to assume that the property we consider is fundamental, that it is a single independent attribute (Definition \ref{fproperty}). If we add this assumption to the case shown in the top panel of Fig. \ref{Fig70}(b), then we are dealing with a fundamental context. It is then possible to formulate the following clear-cut statement about the relationship between observations, eigenspaces and eigenvalues.

\begin{state}[\textbf{Properties as self-adjoint operators in} $\mathcal{H}_{C}$ \textbf{in fundamental contexts} $C$]
To each property $P$ that is observed in a fundamental context $C$, there corresponds exactly one self-adjoint linear operator $\bar{P}$ with a complete set of eigenvectors $\{\bar{\mathcal{P}}_{j}\}$ that span $\mathcal{H}_{\mathcal{P}}$. The set of property values possible to observe in $C$ is the set of eigenvalues $\{p_{j}\}$. We may identify $\mathcal{H}_{C}=\mathcal{H}_{\mathcal{P}}$ and $\bar{S}_{j}=\bar{\mathcal{P}}_{j}$. We have $\bar{P}\bar{S}_{Pj}=p_{j}\bar{S}_{Pj}$. If value $p_{j}$ is observed at time $n+m$, then $S_{C}(n+m)\subseteq\mathcal{P}_{j}$ and $\bar{S}_{C}(n+m)=\bar{S}_{Pj}$.
\label{propopfun}
\end{state}

Few physical contexts are truly fundamental. In realistic cases the above statement should therefore be weakened appropriately, according to the discussion in conntection with Fig. \ref{Fig70}. We do not provide formal statements about the relationship between the property operator and the outcome of the observation in each possible case. We do however provide the weakest possible statement, that applies even if we have a poorly resolved obeservation of a non-fundamental property.

\begin{state}[\textbf{Properties as self-adjoint operators in} $\mathcal{H}_{C}$ \textbf{in any context} $C$]
To each property $P$ that is observed in a context $C$, there corresponds exactly one self-adjoint linear operator $\bar{P}$ with a complete set of eigenspaces $\{\bar{\mathcal{P}}_{j}\}$ that span $\mathcal{H}_{\mathcal{P}}$. Each property value $p_{k}$ possible to observe in $C$ corresponds to a proper subset $p_{k}=\{p_{j0},p_{j0+1},\ldots,p_{j1}\}$ of the set of eigenvalues $\{p_{j}\}$. We may write $\mathcal{H}_{C}\subseteq\mathcal{H}_{\mathcal{P}}$ and $\bar{S}_{Pk}=\oplus_{j=i0}^{j1}\bar{\mathcal{P}}_{j}$. We have $\bar{P}_{C}\bar{S}_{Pk}=p_{k}\bar{S}_{Pk}$, where $\bar{P}_{C}$ is defined below. If value $p_{k}$ is observed at time $n+m$, then $S_{C}(n+m)\subseteq\mathcal{P}_{j0}\cup\mathcal{P}_{j0+1}\cup\ldots\cup\mathcal{P}_{j1}$ and $\bar{S}_{C}(n+m)=\oplus_{j=j0}^{j1}\bar{\mathcal{P}}_{j}$.
\label{propopnofun}
\end{state}

\begin{defi}[\textbf{The contextual property} $P_{C}$]
Consider Statement \ref{propopnofun}. If $\mathcal{H}_{C}\subset\mathcal{H}_{\mathcal{P}}$, then the property observed in context $C$ is not fundamental. $P_{C}$ is the property associated with $P$ with $M$ possible values that is actually observed in $C$. The corresponding self-adjoint linear operator is $\bar{P}_{C}$, with a complete set of eigenspaces $\{\bar{S}_{Pk}\}$ that span $\mathcal{H}_{C}$, and a corresponding set of eigenvalues $\{p_{k}\}$.
\label{contprop}
\end{defi}

Let us check that the operators $\bar{P}$ fulfil the familiar commutation rules. Consider a context in which two properties $P$ and $P'$ are observed in succession. If they are simultaneously knowable we have $S_{C}(n+m')=S_{Pi}\cap S_{P'j}$ at the time $n+m'$ when we have just observed the value of $P'$. We have $S_{Pi}\subseteq \mathcal{P}_{i}$ and $S_{P'j}\subseteq \mathcal{P}_{j}'$. According to Eq. [\ref{eigenstate}] this means that $\bar{P}S_{C}(n+m')=p_{i}S_{C}(n+m')$ and $\bar{P}'S_{C}(n+m')=p_{j}'S_{C}(n+m')$, so that $\bar{P}\bar{P}'\bar{S}_{C}(n+m')=\bar{P}'\bar{P}\bar{S}_{C}(n+m')=p_{i}p_{j}'\bar{S}_{C}(n+m')$. This holds true for any final contextual state. These final states are the same as those called $S_{ij}$ in Fig. \ref{Fig68}, and the corresponding subspaces $\bar{S}_{ij}$ can be identified with a complete basis for $\mathcal{H}_{C}$. Thus, for any vector $\bar{v}\in\mathcal{H}_{C}$ we get

\begin{equation}
\bar{P}\bar{P}'\bar{v}=\bar{P}'\bar{P}\bar{v}.
\label{simcom}
\end{equation}

The situation is different if $P$ and $P'$ are not simultaneously knowable. Then $S_{C}(n+m')=S_{P'j}$. We have $\bar{P}\bar{P}'\bar{S}_{C}(n+m')=p_{j}'\bar{P}\bar{S}_{C}(n+m')$, but $\bar{P}\bar{S}_{C}(n+m')\neq p_{i}\bar{S}_{C}(n+m')$ for all $i$, as expressed in Fig. \ref{Fig60}. Therefore we cannot arrive at the conclusion that $\bar{P}\bar{P}'\bar{S}_{C}(n+m')=\bar{P}'\bar{P}\bar{S}_{C}(n+m')$ in the same way as before.

We could nevertheless try to use the same kind of basis $\{\bar{S}_{ij}\}$ for $\mathcal{H}_{C}$ as in the case when $P$ are simultaneously knowable. Then we could try to apply $\bar{P}$ to $\bar{S}_{C}(n+m')$ written as a linear combination of these basis vectors. We would get the same commutation relation [\ref{simcom}] as for simultaneously knowable properties. Why is this approach forbidden? The reason is that the corresponding sets $\Sigma_{ij}$ in state space do not correspond to realizable alternatives (Definition \ref{realizablealt} and Statement \ref{realizabledomain}). Referring again to Fig. \ref{Fig60}, we understand that $S_{C}$ can never be squeezed into any of the individual compartments $S_{ij}$ in Fig. \ref{Fig68}(a). The property operators simply cannot be applied to vectors that correspond to such hypothetical states. Therefore it does not help to write $\bar{S}_{C}(n+m')$ as a linear combination of vectors $\bar{\Sigma}_{ij}$, since we cannot apply $\bar{P}$ to the outcome.

\begin{state}[\textbf{Property operators act only on state vectors that can occur within context}]The expression $\bar{P}\bar{v}$ is defined only for vectors $\bar{v}$ in Hilbert spaces $\mathcal{H}_{C}$ defined so that all $\bar{v}\in\mathcal{H}_{C}$ correspond to states $S_{C}$ that can be realized within the context $C$.
\label{propopdomain}
\end{state}

This is analogous to the fact that the evolution $u_{1}$ is defined for physical states and realizable alternatives only (Statement \ref{realizabledomain}), not for exact states $Z$ or other physical states that can never be observed.

These considerations block the road for any attempt to derive the result that $\bar{P}$ and $\bar{P}'$ commute when $P$ and $P'$ are not simultaneously knowable. As we discussed above, in this case we should choose a Hilbert space $\mathcal{H}_{C}$ with dimension $\max\{M,M'\}$, rather than $M\times M'$. This reduction of dimensionality means that $\mathcal{H}_{C}$ will be spanned by two `competing' orthogonal bases, one associated with $P$ and one with $P'$, as explained in connection with Fig. \ref{Fig69}, rather than a combined basis that spans a larger space. Since the two competing bases are not orthogonal to each other, we immediately see that the corresponding two operators cannot commute.

\begin{state}[\textbf{Commutation rules for property operators} $\bar{P}$]Assume that we have a context $C$ in which two properties $P$ and $P'$ are observed in succession. Then the pair of operators $\bar{P}$ and $\bar{P}'$ defined according to Statement \ref{propopfun} or \ref{propopnofun} fulfil the following commutation rules: $[\bar{P},\bar{P}']\equiv\bar{0}$ if $P$ and $P'$ are simultaneously knowable, and $[\bar{P},\bar{P}']\not\equiv\bar{0}$ if they are not.
\label{communationrules}
\end{state}

Here $\bar{0}$ is the zero operator that maps any vector to the origin.

The direction of the reasoning behind Statements \ref{propopfun} and \ref{propopnofun} can be reversed - just like a property corresponds to a self-adjoint operator, a self-adjoint operator corresponds to a property.

\begin{state}[\textbf{Self-adjoint operators in} $\mathcal{H}_{C}$ \textbf{as properties}]
Consider a context $C$ in which $P$ is the last property to be observed. To each linear, self-adjoint operator $\bar{P}'$ that acts upon any vector $\bar{v}\in\mathcal{H}_{C}$ and has a complete set of $D_{H}$ orthonormal eigenvectors $\bar{\mathcal{P}}_{j}'$ that span $\mathcal{H}_{C}$, there corresponds at least one property $P'$ with $D_{H}$ possible values in the following sense: there exists another context $C'$ that is the same as $C$ except that another property $P'$ is observed after $P$, and this property corresponds to the operator $\bar{P}'$, in the sense expressed in Statement \ref{propopfun} or \ref{propopnofun}. $P'$ and $P$ are not simultaneously knowable whenever $\bar{P}'\neq\bar{P}$.
\label{opisprop}
\end{state}

\begin{figure}[tp]
\begin{center}
\includegraphics[width=80mm,clip=true]{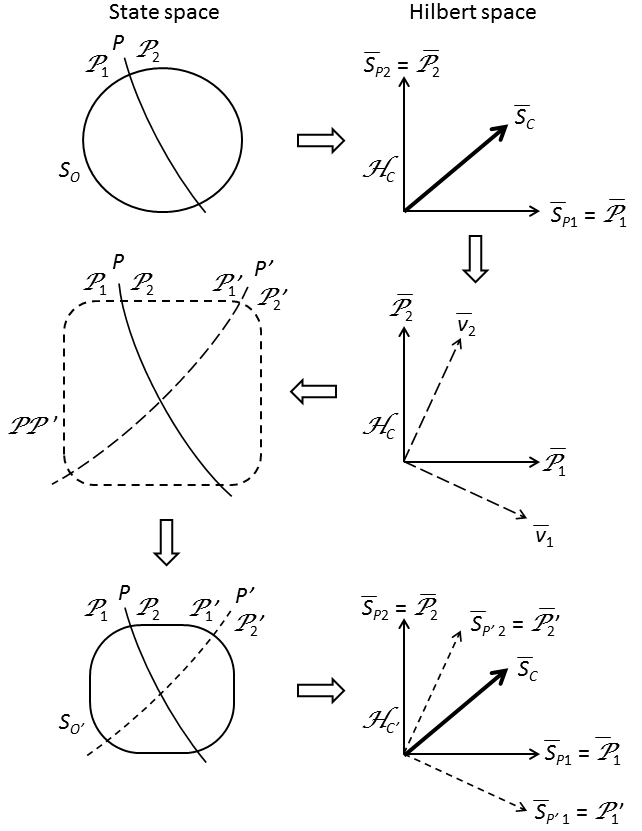}
\end{center}
\caption{In a context $C$, described by the state $S_{O}$, in which property $P$ is observed with possible values $\{p_{i}\}$, we may define a Hilbert space $\mathcal{H}_{C}$ with basis $\{\bar{S}_{Pi}\}$. We may define another arbitrary basis $\{\bar{v}_{j}\}$ in $\mathcal{H}_{C}$. Such a basis can always be associated with another property $P'$, where we can identify $\{\bar{v}_{j}\}=\bar{S}_{P'j}$ in another context $C'$ in which $P'$ is observed after $P$. This context $C'$ is described by the state $S_{O'}$, and must be such that a reciprocal $\tilde{C}'$ exists. Compare Figs. \ref{Fig69} and \ref{Fig69d}.}
\label{Fig70b}
\end{figure}

To see why this statement is reasonable, consider Fig. \ref{Fig70b}. In the top panel we show the state space and the vector space representation in $\mathcal{H}_{C}$ of the original context $C$. Choose an arbitrary self-adjoint operator $\bar{P}'$ that acts in $\mathcal{H}_{C}$. It will have a set of orthogonal eigenvectors $\{\bar{v}_{j}\}$ that spans $\mathcal{H}_{C}$ (middle panel, right). To define the operator uniquely, we also have to fix a set of eigenvalues $\{\epsilon_{j}\}$. We may write

\begin{equation}
\bar{P}'\leftrightarrow\left\{\begin{array}{l}
\{\bar{v}_{j}\}\\
\{\epsilon_{j}\}
\end{array}\right.
\end{equation}

To interpret the basis $\{\bar{v}_{j}\}$ as the vector space representation of the property value spaces $\mathcal{P}_{j}'$ of another property $P'$, we have to make a partition of the presumed property space

\begin{equation}\begin{array}{rcl}
\mathcal{P}\mathcal{P'} & = & \mathcal{P} \cap \left(\bigcup\mathcal{P}_{j}'\right)\\
\mathcal{P}_{j}'\cap\mathcal{P}_{l}' & = & \varnothing,\ \ \forall(j,l)
\end{array}\end{equation}
that conforms with such an interpretation. To start with, we construct the spaces $\mathcal{P}_{j}$ so that the property space is equipartitioned with respect to $P'$ (Statement \ref{equiprop}), since this is a necessary condition for the vector space representation (Statement \ref{vectorrepp}). Then we let the relative volume $v_{ij}^{(P)}$ of the intersections between $\mathcal{P}_{i}$ and $\mathcal{P}_{j}$ be given by the condition in the third row of Eq. [\ref{vectorrepp}]. That is, the spaces $\mathcal{P}_{j}'$ may be chosen so that

\begin{equation}\begin{array}{rcl}
v[\mathcal{P}_{j}'] & = & v[\mathcal{P}_{l}'],\ \ \forall(j,l)\\
v_{ij}^{(P)} & = &  |\langle \bar{\mathcal{P}}_{i}, \bar{v}_{j}\rangle|^{2}/2.
\end{array}
\end{equation}

We see that sets of the desired kind always exist, but they may not be uniquely determined by the operator $\bar{P}'$ and its eigenvectors $\{\bar{v}_{j}\}$. First, we ignore the phase factor in the third row of Eq. [\ref{vectorrepp}], so that several sets $\{\bar{v}_{j}\}$ of eigenvectors to $\bar{P}'$ may be assigned the same set of property value spaces $\{\mathcal{P}_{j}'\}$. Second, the set of relative volumes $\{v_{ij}^{(P)}\}$ does not uniquely determine $\{\mathcal{P}_{j}'\}$. The boundaries $\partial\mathcal{P}_{j}'$ may wiggle around some mean position without changing the relative volumes.

As a last step, we have to assign values $p_{j}'$ to the new property $P'$. Since the property value spaces $\mathcal{P}_{j}'$ are distinct by construction, they correspond to states of the world that are subjectively distinct. As such, they can be encoded by a set of values $\{p_{j}'\}$. We may set $p_{j}'=\epsilon_{j}$ without loss of generality, since this is just a matter of choosing a coordinate system.

\subsection{Composite specimens}
\label{compspec}

Until now we have treated the specimen $OS$ as a single object. Here we discuss the cases when it is known to be composite, and when we investigate how many objects it actually contains. The presentation will be less detailed than in the previous sections; we just want to indicate that the concepts and the formalism developed so far can handle these cases in a natural way.

When the possibility that the specimen is composite is allowed for, the possible outcome of an observation is affected in three ways. First, the possible values of a property of an individual object within the specimen may be affected by the presence of other objects. Second, we may choose to investigate how many objects the specimen is made of; we can introduce a `number property' $P_{N}$. Third, we may choose to observe collective properties, that is, properties $P_{c}$ whose values $p_{c}$ are a function of property values $p_{l},p_{l'},\ldots$ of several different objects $O_{l},O_{l'},\ldots$:

\begin{equation}
p_{c}=f(p_{l},p_{l'},\ldots).
\label{cprop}
\end{equation} 
A familiar example is the total energy $E_{c}$ of a composite system, where $E_{c}=E_{l}+E_{l'}+\ldots$.

Let $P_{c}$ be the collective property, and $P_{l}$ the property that refers to the state of the indivdual object $O_{l}$. We may then distinguish two kinds of contexts where collective properties are observed. In the first kind, one or more properties $P_{c}$ are observed in succession, together with one or more properties $P_{l}$. In the second, only collective properties $P_{c}$ are observed.

In the first kind of context, the relation [\ref{cprop}] may define conditional knowledge - the observation of the sequence of values $\{p_{l},p_{l'},\ldots\}$ may limit the set of possible values $p_{c}$ of $P_{c}$, or vice versa. This situation occurs in experiments of EPR type. Suppose that we measure the total angular momentum $L$ of a specimen that is known to consist of two objects $O_{1}$ and $O_{2}$. After that we measure the angular momentum $L_{z1}$ in the $z$-direction of $O_{1}$. If we find that $L=0$ and $L_{z1}=1/2$, then we can exclude all possible values of $L_{z2}$ except $L_{z2}=-1/2$.

In the second kind of context we do not actually make use of the fact that the specimen is composite. It determines the possible values of the collective property, but the algebraic formalism will be the same as in the case when the specimen is a single object. Two collective properties $P_{c}$ and $P_{c}'$ may or may not be simultaneoulsy knowable, depending on the functions $f$ and $f'$ that define them according to Eq. [\ref{cprop}]. The same holds for two individual properties $P_{l}$ and $P_{l}'$, defined for the same object $O_{l}$. We may say that the formalism is blind to the possibility that a specimen is composite, as long as its parts are not explicitly investigated.

Contexts in which we do not observe any collective properties $P_{c}$ at all, just a succession of individual properties $P_{l}$ of one or several specimen parts, do not introduce much new from a conceptual point of view, either. There is little difference between the observation of a sequence of properties $P,P',\ldots$ that refer to a single object, and the observation of $P_{l},P_{l'},\ldots$ that refer to different objects $O_{l},O_{l'},\ldots$. The only thing to keep in mind is that properties that refer to different objects are always simultaneously knowable. This is true for independent objects, at least. We assume that we are always dealing with such objects, since two dependent objects cannot be considered truly distinct. 

\begin{defi}[\textbf{Independent objects and simultaneous knowability}]
Two independent objects $O_{1}$ and $O_{2}$ are such that any pair of independent attributes $(A_{1},A_{2}')$ are simultaneously knowable, where $A_{1}$ is an internal attribute of $O_{1}$ and $A_{2}'$ is an internal attribute of $O_{2}$.
\label{multipleknowable}
\end{defi}

This condition sharpens the previous definitions of independent objects and independent attributes (Definitions \ref{indobjects} and \ref{indattributes}). Not only shall any value of the independent attribute $A_{2}'$ be \emph{possible} given the value of $A_{1}$, the value of $A_{2}'$ shall also be possible to \emph{determine} at the same time as that of $A_{1}$, so that the independence of $A_{2}'$ becomes manifest. This condition guarantees that the arguments in Eq. [\ref{cprop}] can be known at the same time, which is necessary in order to make the collective property value $p_{c}$ knowable, which in turn is necessary to make the collective property $P_{c}$ well defined. Note also that this is a statement of principle, a statement about what is allowed by physical law. A given observational context $C$ may be such that it is not possible to verify the independence of the attributes of two independent objects within this particular context. 

To conclude, the fact that we allow the specimen to be composite introduces two new possibilites, from a conceptual point of view. First, conditional knowledge may arise from the interplay between the observation of collective and individual properties. Second, we may investigate how many objects the specimen is made of. In other words, we may introduce a `number property' $P_{N}$ with values $p_{N}=1,2,\ldots$.

After this general orientation, let us look at the appropriate formalism. Assume that $OS$ is known to consist of two objects $O_{1}$ and $O_{2}$ with states $S_{O1}$ and $S_{O2}$, respectively. If there is no conditional knowledge relating the two objects, then $S_{OS}=S_{O1}\cap S_{O2}$. The existence of conditional knowledge means that some exact states $Z\in S_{O1}\cap S_{O2}$ can be excluded, so that $S_{OS}\subset S_{O1}\cap S_{O2}$. Let us try to define the algebraic representation of the contextual state $S_{C}\supseteq S_{OS}$ of the composite specimen appropriately, with the possibility of conditional knowledge in mind.

Any conditional knowledge that relates the attributes of objects $O_{1}$ and $O_{2}$ must arise either from some previous observation of a collective property, or from the ability to exclude a set of possible attribute values of $O_{1}$ or $O_{2}$ from the very fact that they belong to the same specimen, according to the observation above that the set of allowed individual property values may change due to the interaction between the objects within the specimen. However, the latter case can be seen as an example of the former, where the knowledge that the objects are parts of a specimen with a given nature can be regarded as the result of `an observation of a collective property'.

Assume first that there is no conditional knowledge that relates the attributes of objects $O_{1}$ and $O_{2}$. Then the same holds true for any property $P$ that is a function of the attributes of one of the objects only. Say that we are about to observe such a property $P$ of object $O_{1}$. No observation is made of the other object $O_{2}$; we assume that it is just a passive passenger within the specimen. Focusing on the relevant object, we may write, as before,

\begin{equation}
\bar{S}_{C1}(n)=\sum_{j}a_{j}\bar{S}_{Pj1}.
\label{compres}
\end{equation}
To indicate the knowledge of the existence of $O_{2}$ we may simply attach its state to this expression, and represent the contextual state of the enitre specimen as follows:

\begin{equation}\begin{array}{lll}
\bar{S}_{C}(n) & = & \bar{S}_{C2}(n)\bar{S}_{C1}(n)\\
& = & \bar{S}_{C2}(n)\sum_{j}a_{j}\bar{S}_{Pj1}.
\end{array}\label{pexp}
\end{equation}
Note that we have not yet assigned any algebraic meaning to this side-by-side notation. As usual, we suppose that the observation of $P$ takes place at time $n+m$. The state of $O_{2}$ does not change in the temporal update $n+m-1\rightarrow n+m$, since, by assumption, no observation is made of $O_{2}$. We should therefore write

\begin{equation}\begin{array}{lll}
\bar{S}_{C}(n+m) & = & \bar{S}_{C2}(n)\bar{S}_{Pj1}\\
& = & \bar{S}_{C2}(n+m)\bar{S}_{C1}(n+m),
\end{array}
\label{pres}
\end{equation}
for one of the $j$:s, to keep the notation consistent with Eq. [\ref{pexp}]. The purpose of such an expansion is to express the possible outcomes of a complete set of future alternatives $\tilde{S}_{j}$. Since the possible contextual states after observation are given by the top row of Eq. [\ref{pres}], we may replace Eq. [\ref{pexp}] with

\begin{equation}
\bar{S}_{C}(n)=\sum_{j}a_{j}\bar{S}_{C2}(n)\bar{S}_{Pj1}.
\label{pexp2}
\end{equation}

We have identified the distributive law

\begin{equation}
\bar{S}_{C\alpha}(a_{2}\bar{S}_{C\beta}+a_{3}\bar{S}_{C\gamma})=a_{2}\bar{S}_{C\alpha}\bar{S}_{C\beta}+a_{3}\bar{S}_{C\alpha}\bar{S}_{C\gamma}.
\end{equation}
The indexation is generalized to make the algebraic rule stand out clearly, but this is allowed since the relation nevertheless don't make any operational sense except in contextual expressions such as Eqs. [\ref{pexp}] and [\ref{pexp2}], where the proper indexation is given by the circumstances. The ordering between the states in the side-by-side notation carries no meaning, since it is just introduced to indicate the existence of two objects. Neither does the ordering between the complex coefficient $a_{j}$ and a state $\bar{S}_{C}$, since $a_{j}$ just indicates the contextual relative volume associated with the combined states of the two objects. Therefore the following commutative laws also hold:

\begin{equation}\begin{array}{lll}
\bar{S}_{C\alpha}\bar{S}_{C\beta} & = & \bar{S}_{C\beta}\bar{S}_{C\alpha}\\
\bar{S}_{C}a_{j} & = & a_{j}\bar{S}_{C}.
\end{array}\end{equation}
We can therefore identify the side-by-side notation with ordinary multiplication:

\begin{equation}
\bar{S}_{C\alpha}\bar{S}_{C\beta}\leftrightarrow \bar{S}_{C\alpha}\times\bar{S}_{C\beta}.
\end{equation}
This fact completes the dictionary displayed in Table \ref{dictionary}, showing how relations between object states are expressed in knowledge space, in state space and in the (partial) algebraic representation that we have developed most recently.

\begin{table}
	\centering
		\begin{tabular}{|l||c|c|}
		\hline
		& AND & OR \\
		Knowledge space & $\cup$ & $\cap$ \\
		State space & $\cap$ & $\cup$ \\
		Algebraic space & $\times$ & $+$ \\
		\hline
		\end{tabular}
	\caption{Symbols expressing corresponding relations between two objects in knowledge space, in state space and in algebraic (Hilbert) space.}
	\label{dictionary}
\end{table}

Assume next that we still consider a context $C$ in which we observe one property $P$ of object $O_{1}$, and no property of $O_{2}$, but that there is now conditional knowledge that relates the value of $P$ with the values of some property $P'$ of object $O_{2}$. As discussed in section \ref{state} this means that there is a value $p_{j}$ of $P$ such that there is at least one value $p_{j'}'$ of $P'$ that can be excluded in the final state of potential knowledge given that $p_{j}$ is observed. To express this fact we must expand the contextual state $S_{C2}$ in the expression of $S_{C}$, even if no actual observation of object $O_{2}$ is made within the context.

\begin{equation}
\bar{S}_{C2}(n)=\sum_{j'}\bar{S}_{P'j'2}.
\label{noaexp}
\end{equation}
Note that since no future alternatives are defined for the state $S_{O2}$ of object $O_{2}$, there are no relative volumes associated with these alternatives, and consequently we cannot define any contextual numbers $a_{j'}$. Equation [\ref{noaexp}] is therefore an algebraic expression that cannot be seen as an element in a Hilbert space. We express

\begin{equation}
\bar{S}_{C}(n)=\sum_{jj'}a_{j}\partial_{jj'}\bar{S}_{Pj'2}\bar{S}_{Pj1},
\label{pexp3}
\end{equation}
where $\partial_{jj'}=0$ if the property value pair $(p_{j},p_{j'}')$ is excluded by conditional knowledge and $\partial_{jj'}=1$ otherwise. When observation of property $P$ takes place at time $n+m$ and value $p_{j}$ is found, we have

\begin{equation}\begin{array}{lll}
S_{C}(n+m) & = & \left(\sum_{j'}\partial_{jj'}\bar{S}_{Pj'2}\right)\bar{S}_{Pj1}\\
& = & S_{C2}(n+m)S_{C1}(n+m)
\end{array}\end{equation}

Note however that we must write

\begin{equation}
\bar{S}_{C}(n)\neq\bar{S}_{C2}(n)\bar{S}_{C1}(n)
\end{equation}
since $S_{C2}(n)$ as given by Eq. [\ref{noaexp}] multiplied by $S_{C1}(n)=\sum_{j}a_{j}\bar{S}_{Pj}$ does not yield Eq. [\ref{pexp3}]. The representation of the combined contextual state of two parts of the specimen does not factorize if there is conditional knowledge that relates these parts. This `frustration' of the state is released when an actual observation is made; then the conditional knowledge disappears. To express it more simply: if we have the conditional knowledge $A\Rightarrow B$, and observe whether $A$ holds true, we no longer have any conditional knowledge. We have instead the knowledge $A\;\mathrm{AND}\;B$ or $\mathrm{NOT}\;A$.

We may imagine that the values of $P$ are related by conditional knowledge to more than one property of object $O_{2}$. Such cases can be reduced to the case discussed above if we consider such a set of properties $\{P_{\alpha},P_{\beta},\ldots\}$ as a single property $P'$ with vectorial values $p'=(p_{\alpha}, p_{\beta},\ldots)$.

\begin{figure}[tp]
\begin{center}
\includegraphics[width=80mm,clip=true]{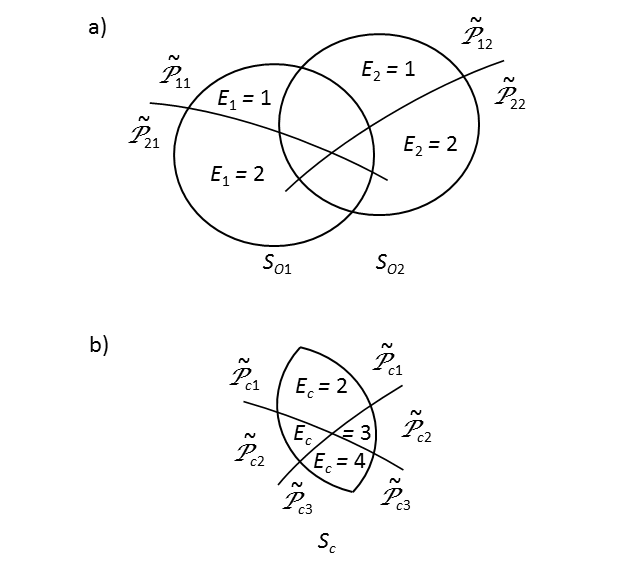}
\end{center}
\caption{A specimen that consists of two objects $O_{1}$ and $O_{2}$ may be treated as a single object $O_{c}$ if we do not make any observation of its parts, and they play no other individual role within the context. In this example, the objects are assumed to be non-interacting, and the energy $E_{c}=E_{1}+E_{2}$ of the entire specimen is measured. The individual energies can take the values $1$ or $2$. a) The specimen is viewed as composite, where the two parts have individual states $S_{O1}$ and $S_{O2}$. b) The specimen is viewed as a single object, with state $S_{c}=S_{O1}\cap S_{O2}$. The energy of the entire specimen can take the values $2$, $3$ or $4$.} 
\label{Fig71}
\end{figure}

We argued above that a context in which we observe a collective property of a composite specimen is formally equivalent to a context in which we observe an individual property of a specimen that consists of a single object. Let us make that statement a bit more precise. Consider the example illustrated in Fig. \ref{Fig71}. We assume that the individual objects do not interact, and we define a collective energy $E_{c}=E_{1}+E_{2}$. The energy values consistent with our knowledge of objects $O_{1}$ and $O_{2}$ are $E_{1}\in\{1,2\}$ and $E_{2}\in\{1,2\}$. Since we have not prepared a context to observe the individual energies, we cannot introduce any coefficients $a_{j}$ when we expand the individual contextual states, just like in the motivation of Eq. [\ref{noaexp}]. We get instead

\begin{equation}\begin{array}{lll}
\bar{S}_{C1}(n) & = & \bar{S}_{P11}+\bar{S}_{P21}\\
\bar{S}_{C2}(n) & = & \bar{S}_{P12}+\bar{S}_{P22},
\label{noaexp2}
\end{array}\end{equation}
and

\begin{equation}\begin{array}{lll}
\bar{S}_{C}(n) & = & a_{1}\bar{S}_{P11}\bar{S}_{P12}+a_{2}(\bar{S}_{P11}\bar{S}_{P22}+\bar{S}_{P21}\bar{S}_{P12})+\\
& & +a_{3}\bar{S}_{P21}\bar{S}_{P22},
\end{array}
\end{equation}
which reduces to

\begin{equation}\begin{array}{llll}
\bar{S}_{C}(n+m) & = & \bar{S}_{P11}\bar{S}_{P12} & \mathrm{OR}\\
\bar{S}_{C}(n+m) & = & \bar{S}_{P11}\bar{S}_{P22}+\bar{S}_{P21}\bar{S}_{P12} & \mathrm{OR}\\
\bar{S}_{C}(n+m) & = & \bar{S}_{P21}\bar{S}_{P22} &
\end{array}
\end{equation}
when the collective energy is measured at time $n+m$.

Let us define the collective object $O_{c}$ with state $S_{c}=S_{O1}\cap S_{O2}$, and future alternatives

\begin{equation}\begin{array}{lll}
\tilde{S}_{c1} & = & \Sigma_{11}\cap \Sigma_{12}\\
\tilde{S}_{c2} & = & (\Sigma_{11}\cap \Sigma_{22})\cup(\Sigma_{21}\cap \Sigma_{12})\\
\tilde{S}_{c3} & = & \Sigma_{21}\cap \Sigma_{22},
\end{array}\end{equation}
corresponding to the collective property values $E_{c}=2$, $E_{c}=3$, or $E_{c}=4$, respectively, with $\Sigma_{ij}=S_{Oj}\cap \tilde{\mathcal{P}}_{ij}$ according to Fig. \ref{Fig71}. (As usual, we denote subsets of states that do not correspond to realizable alternatives with the Greek letter $\Sigma$ rather than the Latin letter $S$.) Then we may equally well describe the context collectively:

\begin{equation}
\bar{S}_{cC}(n)=a_{1}\bar{S}_{cP1}+a_{2}\bar{S}_{cP2}+a_{3}\bar{S}_{cP3}
\label{collectivecontext}
\end{equation}
with collective property value states

\begin{equation}\begin{array}{lll}
\bar{S}_{cP1}& = & \bar{S}_{P11}\bar{S}_{P12}\\
\bar{S}_{cP2} & = & \bar{S}_{P11}\bar{S}_{P22}+\bar{S}_{P21}\bar{S}_{P12}\\
\bar{S}_{cP3}& = & \bar{S}_{P21}\bar{S}_{P22}.
\end{array}
\end{equation}
When the collective energy $E_{c}$ is measured, the state of the collective object reduces to

\begin{equation}
\bar{S}_{cC}(n+m)=\bar{S}_{cPj}
\end{equation}
for $j=1,2$ or $3$ with probability $|a_{j}|^2$.

We have discussed at length something that is quite simple to understand. The we have done so since the conclusion important as a matter of principle. The epistemic perspective dictates that if the assumption that an object $O_{c}$ consists of two objects $O_{1}$ and $O_{2}$ plays no role in the actual investigation or manipulation of $O_{c}$, then it should play no role in the formalism, and should not affect the outcome of calculations in any way. In this respect, the formalism described here differs from conventional quantum mechanics. There, coefficients $a_{j1}$ and $a_{j'2}$ would have been assigned to the property value states $\bar{S}_{Pj1}$ and $\bar{S}_{Pj'2}$ in Eq. [\ref{noaexp2}]. Here, we only consider realizable alternatives as elements in a Hilbert space, or alternatives that are known to play a role within the context even if it is outside potential knowledge which alternative is actually realized (knowability level 3). The passage of one of the two slits in the double slit experiment is the classical example.

Let us turn our attention to contexts in which the composition of the specimen certainly matters, namely those in which we count the number of objects. Sticking to the example shown in Fig. \ref{Fig71}, suppose that we have measured the total energy to be $E_{c}=3$. We are no longer sure that the specimen consists of two objects, but we still know that all objects $O_{l}$ that we may find within it have possible individual energies $E_{l}\in\{1,2\}$. Thus we are dealing with at most three objects; let us call them $O_{1},O_{2}$ and $O_{3}$. Obeying the formalism introduced above, we write

\begin{equation}\begin{array}{lll}
\bar{S}_{C}(n) & = & a_{1}\bar{S}_{P11}\bar{S}_{P12}\bar{S}_{P13}+\\
& & +a_{2}(\bar{S}_{P11}\bar{S}_{P22}+\bar{S}_{P11}\bar{S}_{P23}+\\
& & +\bar{S}_{P21}\bar{S}_{P12}+\bar{S}_{P21}\bar{S}_{P13}+\\
& & +\bar{S}_{P12}\bar{S}_{P23}+\bar{S}_{P22}\bar{S}_{P13}).
\end{array}
\label{permutable}
\end{equation}

We will see two objects with probability $|a_{1}|^2$ and three objects with probability $|a_{2}|^2$. By assumption, we investigate only the number of objects in the specimen, not their identity. Therefore we get

\begin{equation}\begin{array}{lll}
\bar{S}_{C}(n+m) & = & \bar{S}_{P11}\bar{S}_{P22}+\bar{S}_{P11}\bar{S}_{P23}+\bar{S}_{P21}\bar{S}_{P12}+\\
& & +\bar{S}_{P21}\bar{S}_{P13}+\bar{S}_{P12}\bar{S}_{P23}+\bar{S}_{P22}\bar{S}_{P13}
\end{array}
\end{equation}
if we observe two objects at time $n+m$. We may dicriminate between the six combinations of two objects in a subsequent observation of their internal attributes, and thereby reduce the state further. However, it is impossible to make such a discrimination if the objects $O_{1},O_{2}$ and $O_{3}$ are identical. Two objects are identical whenever the knowledge about the internal attributes is the same for both. This may be the case for minimal objects, for example. To make it meaningful to talk about \emph{different} objects in such a situation, it is necessary to assume that the values of some relational attributes can be distinguished, such as their position. To acknowledge that we are dealing with identical objects, we write $O_{l}=O_{1}=O_{2}=O_{3}$, but to indicate that disjunct sets of possible values are associated with their relational attibutes, we represent their contextual states as $\bar{S}_{Cl}, \bar{S}_{Cl}'$, and, if three objects appear, $\bar{S}_{Cl}''$. Equation [\ref{permutable}] transforms to

\begin{equation}
\bar{S}_{C}(n)=a_{1}\bar{S}_{P1l}\bar{S}_{P1l}'\bar{S}_{P1l}''+a_{2}\bar{S}_{P1l}\bar{S}_{P2l}'.
\label{idento}
\end{equation}
To arrive at this expression, remember that the symbol `+' in this representation does not mean that we should add things up, but that it corresponds to `OR', according to Table \ref{dictionary}. If the objects are identical, there are no longer six distinct alternative arrangements of two objects with energies $1$ and $2$, but just one. The three terms contract to one.

In cases where it is impossible to distinguish both the internal and relational attributes of two objects from each other, it does not make epistemic sense to talk about two objects at all. This matter will be discussed further in section \ref{fermbos}. If we nevertheless want to represent such a situation with the present formalism, we should drop the apostrophes, write $\bar{S}_{Cl}=\bar{S}_{C1}=\bar{S}_{C2}=\bar{S}_{C3}$ and

\begin{equation}
\bar{S}_{C}(n)=a_{1}\bar{S}_{P1l}\bar{S}_{P1l}\bar{S}_{P1l}+a_{2}\bar{S}_{P1l}\bar{S}_{P2l}.
\end{equation}
It is redundant to write the same state three times in the first term above. All states in expressions like these are identical, except possibly for the value of a well defined property (energy in our example). We may therefore use instead a variable $N$ indicating how many `objects' there are in the property value state that is placed after it:

\begin{equation}
\bar{S}_{C}(n)=a_{1}N_{11}\bar{S}_{P1l}+a_{2}N_{21}\bar{S}_{P1l}N_{22}\bar{S}_{P2l},
\label{identon}
\end{equation}
with $N_{11}=3$ and $N_{21}=N_{22}=1$.

Let us finally say a few words about the operator representation of collective properties specified as in Eq. [\ref{cprop}]. When we observe an individual property of an object $O_{l}$ in a specimen that contains other objects $O_{l'}$ as well, we need to specify which object the corresponding property operator acts upon. Let $\bar{P}$ be an operator that represents the property $P$, and let $\bar{P}_{l}$ be the operator defined in $\mathcal{H}_{C}$ that refers to property $P$ of object $O_{l}$.

We may then define there operators so that

\begin{equation}
\bar{P}_{l}\bar{S}_{Cl'}=\bar{S}_{Cl'}
\end{equation}
whenever $l\neq l'$. That is, the property operator for a given object in the specimen does not affect the contextual state of another object within the specimen. It follows that any two property operators that act on different independent objects commute: $\bar{P}_{l}\bar{P}_{l'}\bar{S}_{C}=\bar{P}_{l'}\bar{P}_{l}\bar{S}_{C}$ for any contextual state representation $\bar{S}_{C}$, or

\begin{equation}
[\bar{P}_{l},\bar{P}_{l'}]=0
\end{equation}
whenever $l\neq l'$. Therefore, if the function $f$ in Eq. [\ref{cprop}] is infinitely differentiable, we can use its Taylor expansion to define

\begin{equation}
\bar{P}_{c}=f(\bar{P}_{l},\bar{P}_{l'},\ldots).
\label{copprop}
\end{equation}

\subsection{Knowability of physical law}
\label{knowlaw}

Until now we have considered predefined contexts, fixed experimental setups for which the relevant relative volumes $v_{j}$ and corresponding amplitudes $a_{j}$ are constants. But to understand physical law means to be able to specify how the outcome of an observation depends on the parameters that specify the context. In other words, we need to let these parameters vary.

There are two ways to vary such parameters. We may vary the values of those attributes that specify the initial state $S_{O}(n)$ of the experimental setup that defines the context (Fig. \ref{Fig61c}). We may also vary the time at which we observe the outcome. However, these two kinds of variations are linked. The initial preparation of the specimen $OS$, and the final observation of $OS$ that define the `outcome' define two distinct object states, or two events. Clearly, the relational time $t$ passed between these two events is, loosely speaking, a function of the attributes that define the observational setup $O$.

Here, we do not consider variations in the initial state $S_{OB}(n)$ of the observer (Fig. \ref{Fig61c}). Such a change might in itself influence the time $t$ passed until the outcome is observed. However, we want to focus on physical law that governs the evolution of the observed specimen $OS$ rather than the evolution of the observer. Further, we do not consider variations in the initial state $S_{\Omega_{O}}(n)$ of the environment $\Omega_{O}$ to the experimental setup $O$ (Definition \ref{complement}). We assume that the interaction between $\Omega_{O}$ and $O$ can be made arbitrarily small, so that the physical law that governs the specimen $OS$ itself can be investigated to arbitrary precision. Implicit in such a statement is also the assumption that the influence on $OS$ from the environment decreases smoothly with the strength of the interaction. 

In terms of the evolution parameter $\sigma$, the fact that the state of the detector does not change appreciably before the observation of property $P$ at time $n+m$ (Definition \ref{detectordef}) can be expressed as

\begin{equation}
S_{OD}(\sigma_{0})\cap S_{OD}(\sigma)\neq\varnothing,
\end{equation}
where $S_{OD}(\sigma_{0})=S_{OD}(n)$, and we assume that $\sigma\geq\sigma_{0}$. (We have to assume that the state $S_{OD}$ is specified in terms of relativistically invariant attributes. If not, the attributes of the apparatus would change if it were not in the rest frame of the observer.) In contrast, nothing prevents the state of the specimen to vary during the course of the experiment, so that

\begin{equation}
S_{OS}(\sigma_{0})\cap S_{OS}(\sigma)=\varnothing.
\end{equation}

\begin{figure}[tp]
\begin{center}
\includegraphics[width=80mm,clip=true]{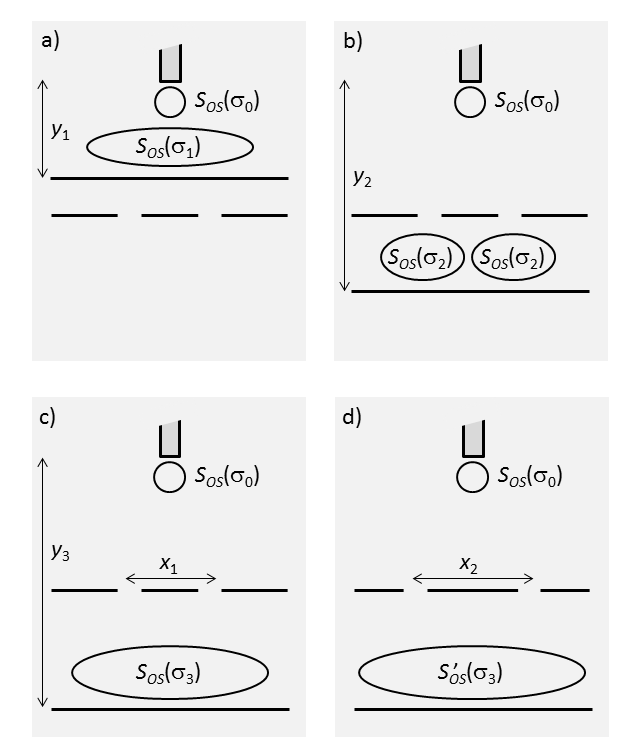}
\end{center}
\caption{An experimental setup whose initial state $S_{O}(n)$ depends on the attributes $x$ and $y$, which specify the state $S_{OA}(n)$ of the apparatus $OA$. We let the value of the (unknowable) evolution parameter just before the specimen $OS$ arrives at the detector screen $OD$ be $\sigma_{k}$ when the distance between the particle gun and the detector is $y_{k}$.}
\label{Fig72}
\end{figure}

Since the state of the specimen $OS$ is allowed to move arbitrarily far away from $S_{OS}(n)$ under the action of $u(\sigma)$ during the course of the experiment, so is the entire experimental setup $O$, consisting of apparatus and specimen. That is, we may have

\begin{equation}
S_{O}(\sigma_{0})\cap S_{O}(\sigma)=\varnothing.
\end{equation}
This is possible regardless the behavior of the machine $OM$, which may or may not undergo a sequence of distinct, directly perceived changes between times $n$ and  $n+m$. 

Figure \ref{Fig72} is intended to illustrate the abstract discussion above. This setup $O$, designed to execute a double-slit experiment, consists of particle gun, an ejected particle, a screen with slits, and a detector screen $OD$. The ejected particle is the specimen $OS$. The machine $OM$ is the particle gun and the screen with slits.

The distance between the particle gun and the detector screen is variable, as well as the distance between the slits. These two attributes belong to the apparatus $OA$, and their values $y$ and $x$ does not change significantly during the course of the experiment. The apparatus is fixed. That is, $S_{OD}(\sigma_{0})\cap S_{OD}(\sigma)\neq\varnothing$ and $S_{OM}(\sigma_{0})\cap S_{OM}(\sigma)\neq\varnothing$ during the entire experiment. The specimen moves through the slits towards the detector. Thus its state cannot be considered constant between the two events that mark the beginning and the end of the experiment. We must have $S_{OS}(\sigma_{0})\cap S_{OS}(\sigma)=\varnothing$ for some $\sigma$ during the course of the experiment. 

The final observation of the specimen at the detector screen at time $n+m$ corresponds to a state reduction, since the state of the specimen covers a vast area in the $(x,y)$-plane just before detection, whereas its spatial location is much more accurately known afterwards, provided the spatial resolution of the detector is reasonable. This fact implies that

\begin{equation}\begin{array}{rcl}
S_{OS}(n+m) & \subset & u_{1}S_{OS}(n+m-1)\\
S_{OD}(n+m) & \subset & u_{1}S_{OD}(n+m-1)\\
S_{O}(n+m) & \subset & u_{1}S_{O}(n+m-1).
\end{array}
\end{equation}
In contrast, the machine does not necessarily have to undergo a perceived change. We may have

\begin{equation}
S_{OM}(n+m)=u_{1}S_{OM}(n+m-1).
\end{equation}

As we seek to express the outcome of an observation as a function of the parameters that specify the initial state, it is wise to keep the complete set $\{\tilde{S}_{j}\}$ of future alternatives that define the possible outcomes constant as we vary these parameters. We have previously defined these in terms of a set of values $\{p_{j}\}$ of a property $P$, so that $\tilde{S}_{j}\equiv S_{O}\cap\tilde{P}_{j}$ (Definitions \ref{futurealt} and \ref{setfuturealt}). Here, we make the analogous definition, referring to the specimen $OS$ rather than the entire setup $O$. That is, $\tilde{S}_{j}\equiv S_{OS}\cap\tilde{P}_{j}$.

Under the condition that we keep $\{\tilde{S}_{j}\}$ constant, we look for the state $S_{OS}(n+m-1)$ just before the observation is made at time $n+m$, so that $S_{OS}(n+m)=S_{j}$ for some $j$. We let the initial state of the specimen be $S_{OS}(n)$. To make the notation simple, we introduce

\begin{equation}\begin{array}{lll}
S^{0} & = & S_{OS}(n)\\
S^{1} & = & S_{OS}(n+m-1)\\
S^{2} & = & S_{OS}(n+m).
\end{array}\end{equation}

Trying to use the conceptual framework presented above in a straightforward manner, understanding physical law could mean to know the function $f$ in the expression 

\begin{equation}
S^{1}=f(S^{0},\{\upsilon_{i}\})
\label{naivelaw}
\end{equation}
where $\{\upsilon_{i}\}$ is a set of attribute values that specify at time $n$ the apparatus $OA$ that is part of the observational context $C$. In the setup shown in Fig. \ref{Fig72} we may set $\upsilon_{1}\equiv x$ and $\upsilon_{2}\equiv y$.

There are problems with this expression, however. We cannot assume that knowledge is complete, neither about the specimen $OS$, nor about the apparatus $OA$. The irreducibility of physical law (Statement \ref{irreduciblelaw}) then makes Eq. [\ref{naivelaw}] invalid. The exact values $\{\upsilon_{i}\}$ should be replaced by intervals of values $\{\upsilon_{i}^{\min},\upsilon_{i}^{\max}\}$ that are not excluded by potential knowledge. We may try to write

\begin{equation}
S^{1}=f(S^{0},\{\upsilon_{i}^{\min},\upsilon_{i}^{\max}\}).
\label{betterlaw}
\end{equation}
However, this expression is valid only for states $S_{O}(n)$ specified by a number of independent intervals $\{\upsilon_{i}^{\min},\upsilon_{i}^{\max}\}$, meaning that the state takes the form of a rectangle in a state space where the values of each attribute are repesented along an axis perpendicular to all the others. This may not be the case if there is conditional knowledge that relate values of different attributes $\upsilon_{i}$ and $\upsilon_{i'}$. Furthermore, we cannot take for granted that the boundary values $\upsilon_{i}^{\min}$ and $\upsilon_{i}^{\max}$ are exactly knowable, according to the discussion in Section \ref{knowstate}.

This means that it is not possible in general to formulate physical law in terms of a well-defined response $\delta S^{1}$ in the evolved state $S^{1}$ to a change of an attribute value, or of a value interval:

\begin{equation}\begin{array}{lll}
\delta S^{1} & \neq & f'(d\upsilon_{i})\\
\delta S^{1} & \neq & f''(d\upsilon_{i}^{\min})\\
\delta S^{1} & \neq & f'''(d\upsilon_{i}^{\max}).
\end{array}
\end{equation}

This conclusion can be applied to relational time $t$, just like to any other attribute. It is therefore not possible to express physical law in a precise manner as

\begin{equation}
S^{1}\neq f(S^{0},t).
\end{equation}

It is most often possible \emph{in practice}, however. It goes without saying that this approach has been an outstanding success for centuries. In a similar way it is, of course, possible to express the response to changes in other attributes to a high degree of precision. For instance, we can calculate the response in the interference pattern when the distance $x$ between the slits is changed. The inherent uncertainty in the value of $x$, and conditional knowledge that relates different parts of the apparatus, most often play very little role.

Fair enough, but we are interested in those precise statements about physical law that we can make \emph{in principle}, statements based on what can be actually known about the physical state. Since we have excluded all attributes as numerical arguments in a functional expression of physical law, we are left with the evolution parameter $\sigma$ (section \ref{evolutionparameter}). With $S_{OA}(\sigma_{0})=S_{OA}(n)$, and choosing a parametrization such that $\sigma_{0}=0$ we may write

\begin{equation}
S^{1}=f(S^{0},S_{OA}(0),\sigma_{f}),
\label{plaw}
\end{equation}
where $\sigma_{f}$ is the `final' value of the evolution parameter at time $n+m-1$, just before the observation of the relevant property $P$. We assume, as an idealization, that the specimen and apparatus is isolated from the environment. Expression [\ref{plaw}] should be compared with the previous attempts [\ref{naivelaw}] and [\ref{betterlaw}].

The prize to pay for the functional precision is that $\sigma_{f}$ is not an observable quantity. With a proper parametrization we can, however, relate it closely to the relational time $t$ passed between sequential time instants $n$ and $n+m$, or the Lorentz distance $l^{2}$ in space-time travesed by the specimen. We may, for example, choose the parametrization so that $\sigma_{f}=\langle t\rangle$, or $\sigma_{f}=\langle l^{2}\rangle$, where $\langle\ldots\rangle$ denotes the expected value, defined in some convenient way.
  
\begin{figure}[tp]
\begin{center}
\includegraphics[width=80mm,clip=true]{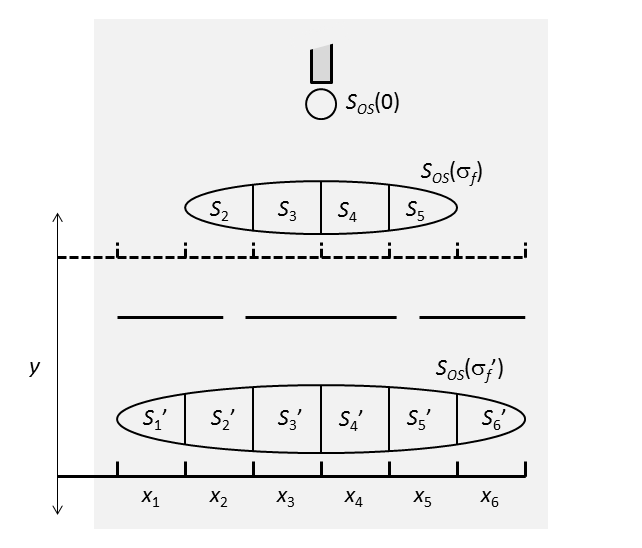}
\end{center}
\caption{Two members of a family $C(\sigma_{f})$ of observational contexts, where a change in the vertical position $y$ of the detector screen is simulated by a change in the value of $\sigma$ at which the observation takes place. The property $P$ that is observed is the position $x$ along the detector screen. The possible values are $p_{j}=x_{j}$. This set of values is the same in all contexts that belong to $C(\sigma_{f})$.}
\label{Fig73}
\end{figure}

The ideas are illustrated in Fig. \ref{Fig73}. We `simulate' the actual movement of the detector screen by a change of $\sigma_{f}$, roughly interpreted as the value of the evolution parameter when the detection takes place at time $n+m$. Two contexts $C$ and $C'$ are indicated, one in which the detector is placed before the passage of the slits, and one in which it is placed after the passage. In the first case (dashed detector), we have $S^{1}=S_{OS}(\sigma_{f})$, and in the second case (solid detector), we have $S^{1}=S_{OS}(\sigma_{f}')$, where $\sigma_{f}'>\sigma_{f}$. Contexts $C$ and $C'$ may be seen as members of a family of contexts $C(\sigma_{f})$. If we restict the possible initial settings $S_{OA}(0)$ of the apparatus to this one-parameter family, we may remove this argument in Eq. [\ref{plaw}] and write

\begin{equation}
S^{1}=f(S^{0},\sigma_{f}),
\label{law2}
\end{equation}

The property $P$ that is observed is the position along the $x$-axis at the detector screen. As noted above, we keep the alternatives defined by the possible values $\{p_{j}\}$ constant as we vary the context within the family $C(\sigma_{f})$. We have $\{p_{j}\}=\{x_{1},x_{2},\ldots,x_{6},p_{7}\}$, where property value $p_{7}$ is introduced to make the set of future alternatives complete. It corresponds to the event that the specimen never hits the detector in one of the six compartments of the detector. This event occurs if the specimen passes beside the detector, or if it never passes one of the slits since it hits the walls that define them.

The strategy to keep the alternatives fixed means that the relative volumes $v_{j}\equiv v[\tilde{S}_{j}]$ may depend on $\sigma_{f}$. In the setup shown in Fig. \ref{Fig73} we have, for example, $v[\tilde{S}_{1}]\equiv v[\tilde{S}_{1}(\sigma_{f})]=0$, whereas $v[\tilde{S}_{1}']\equiv v[S_{1}(\sigma_{f}')]\neq 0$. The same is true for the corresponding contextual numbers $a_{j}$, so that we may write

\begin{equation}\begin{array}{lll}
v_{j}& = & f_{j}(S^{0},\{\tilde{S}_{j}\},\sigma_{f})\\
a_{j}& = & g_{j}(S^{0},\{\tilde{S}_{j}\},\sigma_{f}).
\end{array}
\label{law3}
\end{equation}

These statements may seem confusing, since we have concluded that relative volumes of future alternatives are invariant under the application of the evolution operator $u_{1}$ (Statement \ref{invariantregions}), which means that they are independent of $\sigma$. However, we need to distinguish the `final' value $\sigma_{f}$ of the evolution parameter just before we observe the specimen $OS$ from the evolution parameter $\sigma$ itself, which interpolates between the initiation of the context at time $n$, and the observation at time $n+m$, according to the discussion in Section \ref{evolutionparameter}. For a given context in the family $C(\sigma_{f})$, the relative volumes are independent of $\sigma$, which may vary in the range $[0,\sigma_{f}]$.

Equation [\ref{law3}] summarizes what can be said in precise functional form about physical law. To be able to actually find the functions $f_{j}$ and $g_{j}$, we must be able to `execute' the same context over and over again, so that the relative volumes can be determined. In that case they can be interpreted as probabilities, as discussed in section \ref{probabilities}. This presupposes the ability to isolate the experimental setup $O$ from the environment to arbitrarily high degree. Further, this must be possible for all $\sigma_{f}$ in some interval of interest, corresponding to a one-parameter family of contexts $C(\sigma_{f})$. However, since $\sigma_{f}$ is not an observable that can be determined to arbitrarily high precision, it is not possible, even in principle, to repeat the experiment for fixed $\sigma_{f}$. Therefore the actual form of physical law, as expressed in Eq. [\ref{law3}], is not possible to determine from experiment to arbitrarily high precision. We may say that physical law in this form is formally well-defined, but unknowable in its details, just like the physical state (section \ref{knowstate}).

\begin{state}[\textbf{Physical law}]
Those aspects of physical law that in principle can be expressed in functional form with numerical arguments are captured by the functions $f_{j}$ and $g_{j}$ in Eq. [\ref{law3}].
\label{physicallawb}
\end{state}

\begin{state}[\textbf{Physical law is not exactly knowable}]
It is not possible to determine the functions $f_{j}$ and $g_{j}$ in Eq. [\ref{law3}] to arbitrary precision with the help of repeated observations. This holds true for any specimen in any initial state $S^{0}$, and for any family of contexts $C(\sigma_{f})$.
\label{unknowablelaw}
\end{state}

Figuratively speaking, the `detector comb' in Fig. \ref{Fig73} is our tool to dissect the state $S_{OS}$ and learn about it. By varying its position and orientation we can learn what it possible to know about the physical law that governs its behaviour. In essence, the unknowability of physical law stems from the facts that we cannot know the state $S_{OS}$ exactly, and we cannot know the exact position and orientation of the detector comb. This does not prevent us from hypothesizing exact forms of $f_{j}$ and $g_{j}$ via theoretic reasoning, of course. The point is that such hypotheses cannot be checked experimentally to arbitrary precision.

Let us nevertheless discuss an idealized situation. We assume that $f_{j}$ and $g_{j}$ are indeed exactly known, even if this contradicts Statement \ref{unknowablelaw}. Say that property $P$ is observed in a context $C$. We assume that $C$ is fundamental (Definition \ref{fundamentalcontext}), so that each property value $p_{j}$ that corresponds to the future alternative $\tilde{S}_{j}$ corresponds to a \emph{single} numerical value, not a set of values allowed by physical law. The expected value of the property $P$ just before we observe it at time $n+m$ will be $\langle p\rangle=\sum_{j}v_{j}p_{j}/\sum_{j}v_{j}$. (If physical law cannot exclude any real value of $P$, the sum should be replaced by an integral.) We may then write down the implicit relations

\begin{equation}\begin{array}{lll}
v_{j}& = & f_{j}(S^{0},\langle p\rangle)\\
a_{j}& = & g_{j}(S^{0},\langle p\rangle).
\end{array}
\label{law4}
\end{equation}
Note that we omit the complete set of future alternatives $\{\tilde{S}_{j}\}$ as an argument in these functions. Since $C$ is assumed to be fundamental, $\{\tilde{S}_{j}\}$ is predetermined to correspond to the entire set of values $\{p_{j}\}$ allowed by physical law.

If the property that we observe within context is the relational time $t$ passed since the initiation of the context at sequential time $n$, then we have

\begin{equation}\begin{array}{lll}
v_{j}& = & f_{j}(S^{0},\langle t\rangle)\\
a_{j}& = & g_{j}(S^{0},\langle t\rangle).
\end{array}
\label{law5}
\end{equation}
We have regained the familiar form of physical law, reinterpreting the conventional temporal parameter $t$ as the expected value $\langle t\rangle$ of the relational attribute $t$. Note, however, that these relations hold only in fundamental contexts $C$ in which $t$ is \emph{actually measured}. In that case we may speak about families of contexts $C(\langle t\rangle)$.

We will discuss in section \ref{eveq} the so called natural parametrization $d\langle t\rangle/d\sigma=1$ in fundamental contexts in which four-position $\mathbf{r}_{4}=(x,y,z,ict)$ is observed. Since $t$ is one of the observed properties we may express physical law either in the form [\ref{law3}] or in the form [\ref{law5}]. Simply put, since we may set $\sigma=\langle t\rangle$, we may identify the families of contexts $C(\sigma)$ and $C(\langle t\rangle)$.

We may choose to observe the energy $E$ of the specimen rather than $t$. Then we may use Eq. [\ref{law4}] to define a family of fundamental contexts $C(\langle E\rangle)$. This family has no trivial relation to $C(\langle t\rangle)$. The contexts in the two families are fundamentally different, since different properties are observed. We will discuss such contexts in section \ref{reciprocaleq}.

\begin{state}[\textbf{Idealized physical law}]
Suppose that property $P$ is observed in the fundamental context $C$, and that the functions $f_{j}$ and $g_{j}$ in Eq. [\ref{law3}] are exactly known. Then we may define a family of contexts $C(\langle p\rangle)$ where the `evolution' is given by Eq. [\ref{law4}].
\label{physicallaw2}
\end{state}

We put the world `evolution' within citation marks since Eq. [\ref{law4}] describes continuous temporal evolution only if $t$ is observed. If another property is observed, such as energy $E$, we have to make a more involved description of what the equations say. Generally, they specify the probabilities to see different values of $P$ at time $n+m$ as a function of the expected value of $P$, given the initial state of the specimen at time $n$.

\subsection{The wave function}
\label{wavef}

Suppose that all members in a family of contexts $C(\sigma_{f})$ are possible to express in a Hilbert space representation according to Statement \ref{hilbertrep}. Then the state of the specimen just before observation of property $P$ can be represented as

\begin{equation}
\bar{S}_{C}(\sigma_{f})=\sum_{j}a_{j}(\sigma_{f})\bar{S}_{Pj}
\label{sumrepf}
\end{equation}

The fact that we do not change the set of future alternatives $\{\tilde{S}_{j}\}$ as we let the context vary within the family $C(\sigma_{f})$ is reflected in the fact that the set of property value states $\{\bar{S}_{Pj}\}$ does not depend on $\sigma_{f}$ in the above equation. In order to simplify the notation, we will drop the subscript $f$ on the `final' evolution parameter $\sigma_{f}$. That is, we write

\begin{equation}
\bar{S}_{C}(\sigma)=\sum_{j}a_{j}(\sigma)\bar{S}_{Pj}.
\label{sumrep}
\end{equation}

We refer to the discussion in the preceding section to avoid confusion on the meaning of the symbol $\sigma$ in the following sections.

\begin{defi}[\textbf{The wave function} $a_{P}(p_{j},\sigma)$]
Suppose that we use variations in the evolution parameter $\sigma$ as a proxy for variations of a property that determines the physical setup of an observational context $C$. Then $a_{P}(p_{j},\sigma)\equiv a_{j}(\sigma)$, where $a_{j}(\sigma)$ is given by Eq. [\ref{sumrep}]. The wave function specifies the contextual state $S_{C}$ just before the observation of the contextual property $P_{C}$, which may have fewer possible values than a corresponding fundamental property $P$. The domain of $a_{P}(p_{j},\sigma)$ is $(\{p_{1},p_{2},\ldots,p_{M}\},[0,\sigma_{\max}])$ for some arbitrary $\sigma_{\max}$ that depends on the parametrization, and the details of the family of contexts $C(\sigma)$.
\label{wavedef}
\end{defi}
Note that we have to add the index $P$ to the wave function since it is only defined together with the property $P$ that we are about to observe. The function $a(x,y)$ has no physical meaning in itself.

By definition of relative volume, we have, for all $\sigma$ and for any complete set of future alternatives $\{\tilde{S}_{j}\}$ corresponding to the values $\{p_{j}\}$ of property $P$:

\begin{equation}\begin{array}{lll}
1 & = & \sum_{j}|a_{j}(\sigma)|^{2}\\
& = & \sum_{j}\sum_{j'}a_{j}^{*}(\sigma)a_{j'}(\sigma)\langle\bar{S}_{Pj},\bar{S}_{Pj'}\rangle\\
& = & \langle\bar{S}_{C}(\sigma),\bar{S}_{C}(\sigma)\rangle\\
& = & \left\|\bar{S}_{C}(\sigma)\right\|^{2}
\end{array}
\label{sumunit}
\end{equation}

We may transform these simple relations into a couple of more fancy statements and definitions.

\begin{state}[\textbf{The wave function $a_{P}(p_{j},\sigma)$ is always normalized}]
We have $\sum_{j}|a_{P}(p_{j},\sigma)|^{2}=1$ for all $\sigma$ in the domain of $a_{P}$.
\label{normalized}
\end{state}

\begin{state}[\textbf{The contextual state vector $\bar{S}_{C}(\sigma)$ has unit norm}]
We have $\left\|\bar{S}_{C}(\sigma)\right\|=1$ for all $\sigma$ in the domain of the context family $C(\sigma)$.
\label{scnorm}
\end{state}

\begin{defi}[\textbf{The contextual evolution operator $u_{C}(\sigma)$}]
The operator $\bar{u}_{C}(\sigma)$ is defined by the relation $\bar{S}_{C}(\sigma)\equiv \bar{u}_{C}(\sigma)\bar{S}_{C}(0)$.
\label{cevol}
\end{defi}
Here $\bar{S}_{C}(0)$ is a limiting context such that the variation of the experimental setup that corresponds to a change of $\sigma$ cannot be pushed further. To give an example, we may consider the double slit experiment, and let variations in the position $y$ of the detector screen correspond to changes in $\sigma$ (Fig. \ref{Fig73}). Then we may let $\bar{S}_{C}(0)$ be the context in which the screen is pushed all the way up to the particle gun.

\begin{state}[\textbf{The contextual evolution operator $\bar{u}_{C}(\sigma)$ is unitary}]
For any family of contexts $C(\sigma)$ we have $\left\|\bar{S}_{C}(0)\right\|=1$ and $\left\|\bar{u}_{C}(\sigma)\bar{S}_{C}(0)\right\|=1$ for all $\sigma$ in the domain of $C(\sigma)$.
\label{unitary}
\end{state}

\begin{figure}[tp]
\begin{center}
\includegraphics[width=80mm,clip=true]{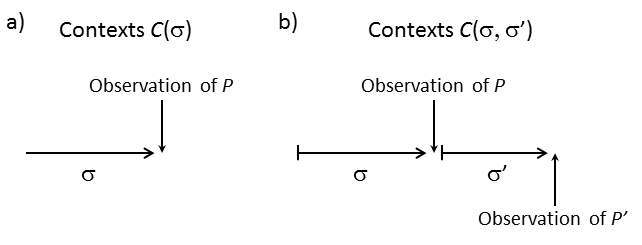}
\end{center}
\caption{a) A family of contexts $C(\sigma)$ in which one property $P$ is observed. b) A family of contexts $C(\sigma,\sigma')$ in which two properties $P$ and $P'$ are observed in succession.}
\label{Fig73a}
\end{figure}

The wave function $a_{P}(p_{i},\sigma)$ is no longer defined after the observation of $P$. We may, however, observe a second property $P'$ in the same context $C$. Then we may define a second wave function $a_{P'}(p_{j}',\sigma')$ that represents the contextual state just before $P'$ is observed. The evolution parameter $\sigma'$ is completely independent from $\sigma$. It describes variations in the experimental setup that affect the relational time $t_{PP'}$ that passes between the observations of $P$ and $P'$. We may thus define a family $C(\sigma,\sigma')$ of contexts in which two properties $P$ and $P'$ are observed (Fig. \ref{Fig73a}). The notation can obviously be generalized to contexts with three or more properties.

\begin{figure}[tp]
\begin{center}
\includegraphics[width=80mm,clip=true]{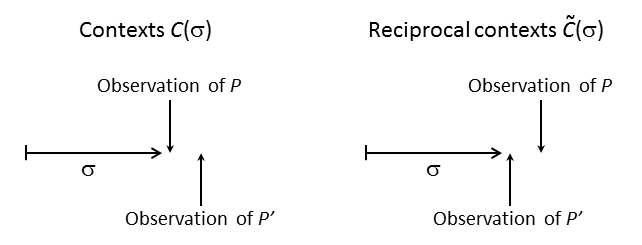}
\end{center}
\caption{a) A family of contexts $C(\sigma)$ in which two properties $P$ and $P'$ are observed. b) A family of reciprocal contexts $\tilde{C}(\sigma)$ where the same $\sigma$ is used. We may define the joint family of contexts $C\tilde{C}(\sigma)$.}
\label{Fig73b}
\end{figure}

There is one situation in which we can use a single evolution parameter $\sigma$ to describe contexts in which two properties are observed. Then we have to consider a pair of reciprocal contexts $C$ and $\tilde{C}$ (Definition \ref{reciprocalpair}). For a given $\sigma$, we just switch the order in which $P$ and $P'$ are observed, and assume that this can be done so that the contextual numbers are preserved according to Definition \ref{reciprocalcontext}. The fact that we use the same $\sigma$ in both $C$ and $\tilde{C}$ means that we consider the same variations in the experimental setup regardless which property $P$ or $P'$ is observed first. The situation is illustrated in Fig. \ref{Fig73b}. If we consider the members of the context pair together, without any predefined ordering, we can define the joint family of contexts $C\tilde{C}(\sigma)$ [compare Fig. \ref{Fig69d}(b)].  

\begin{defi}[\textbf{The joint contextual state vector} $\bar{S}_{C\tilde{C}}(\sigma)$]
Consider a pair $(C,\tilde{C})$ of reciprocal contexts described by the joint family $C\tilde{C}(\sigma)$. Then $\bar{S}_{C\tilde{C}}(\sigma)$ is defined in the joint Hilbert space $\mathcal{H}_{C\tilde{C}}$, so that we may write $\bar{S}_{C\tilde{C}}(\sigma)=\sum_{i}a_{P}(p_{i},\sigma)\bar{S}_{Pi}=\sum_{j}\tilde{a}_{P'}(p_{j}',\sigma)\bar{S}_{P'j}$, where $\bar{S}_{C}(\sigma)=\sum_{i}a_{P}(p_{i},\sigma)\bar{S}_{Pi}$ and $\bar{S}_{\tilde{C}}(\sigma)=\sum_{j}\tilde{a}_{P'}(p_{j}',\sigma)\bar{S}_{P'j}$. 
\label{jsdef}
\end{defi}

Simply put, we define the joint contextual state vector so that we are allowed to make a change of basis like in Fig. \ref{Fig69}(b) and still talk about `the same' state vector. It also allows us to apply the two different property operators $\bar{P}$ and $\bar{P}'$ to the same state vector.

\begin{equation}\begin{array}{lll}
\bar{P}\bar{S}_{C\tilde{C}} & = & \sum_{i}a_{P}(p_{i},\sigma)p_{i}\bar{S}_{Pi}\\
\bar{P}'\bar{S}_{C\tilde{C}} & = & \sum_{j}\tilde{a}_{P'}(p_{j}',\sigma)p_{j}'\bar{S}_{P'j}
\label{jointop}
\end{array}\end{equation}

In contrast, to the state vector $\bar{S}_{C}$ we can only apply the operator $\bar{P}$, and to the state vector $\bar{S}_{\tilde{C}}$ we can only apply the operator $\bar{P}'$.

\begin{equation}\begin{array}{lll}
\bar{P}\bar{S}_{C} & = & \sum_{i}a_{P}(p_{i},\sigma)p_{i}\bar{S}_{Pi}\\
\bar{P}'\bar{S}_{\tilde{C}} & = & \sum_{j}\tilde{a}_{P'}(p_{j}',\sigma)p_{j}'\bar{S}_{P'j}
\end{array}\end{equation}
The expressions $\bar{P}'\bar{S}_{C}$ and $\bar{P}\bar{S}_{\tilde{C}}$ are not defined.

Strictly speaking, we consider here the contextual property operators $\bar{P}_{C}$ that correspond to the contextual property $P_{C}$, or the discretized property $P^{M}$ (Definition \ref{contprop}). As discussed above, the property $P_{C}$ has just as many values $M$ as there are alternatives $\tilde{S}_{i}$, which means that the property values of the fundamental property $P$ are sometimes grouped together into $M$ bins. However, we drop the index $C$ to simplify notation.

The fact that we can express the same joint state vector $\bar{S}_{C\tilde{C}}$ in either basis $\{\bar{S}_{Pi}\}$ or $\{\bar{S}_{P'j}\}$ according to Definition \ref{jsdef} means that we can re-express Eq. [\ref{jointop}] as

\begin{equation}\begin{array}{lllll}
\bar{P}\bar{S}_{C\tilde{C}} & = & \sum_{i}(\bar{P}_{P}a_{P})\bar{S}_{Pi} & = & \sum_{j}(\bar{P}_{P'}\tilde{a}_{P'})\bar{S}_{P'j}\\
\bar{P}'\bar{S}_{C\tilde{C}} & = & \sum_{i}(\bar{P}_{P}'a_{P})\bar{S}_{Pi} & = & \sum_{j}(\bar{P}_{P'}'\tilde{a}_{P'})\bar{S}_{P'j}
\label{waveops}
\end{array}\end{equation}
for a quadruplet of operators $(\bar{P}_{P},\bar{P}_{P'},\bar{P'}_{P},\bar{P}_{P'}')$, where we have suppressed the arguments of the wave functions for clarity.

\begin{defi}[\textbf{The wave function property operators}]
In a joint context $C\tilde{C}$, the quadruplet of operators $(\bar{P}_{P},\bar{P}_{P'},\bar{P'}_{P},\bar{P'}_{P'})$ are defined by Eq. [\ref{waveops}]. We may say that $\bar{P}_{P}$ is the operator corresponding to property $P$ that acts on the wave function when it is expressed in terms of $P$, and that $\bar{P}_{P'}$ is the operator corresponding to $P$ that acts on the wave function when it is expressed in terms of $P$. The interpretation of $\bar{P'}_{P}$ and $\bar{P'}_{P'}$ is analogous.
\label{dwaveops}
\end{defi}

If we adopt array representations $[a_{1}(\sigma),a_{2}(\sigma),\ldots,a_{M}(\sigma)]^{T}$ and $[\tilde{a}_{1}(\sigma),\tilde{a}_{2}(\sigma),\ldots,\tilde{a}_{M}(\sigma)]^{T}$ of the wave functions $a_{P}(p_{i},\sigma)$ and $\tilde{a}_{P}(p_{j}',\sigma)$, respectively, we can express the wave function property operators as matrices. In particular, the operators $\bar{P}_{P}$ and $\bar{P}_{P'}'$ become diagonal.

\begin{equation}\begin{array}{cc}
\bar{P}_{P}=
\left(\begin{array}{cccc}
p_{1} & 0 & \ldots & 0\\
0 & p_{2} & & 0\\
\vdots & & \ddots & \vdots\\
0 & 0 & \ldots & p_{M}
\end{array}\right) &
\bar{P}_{P'}'=
\left(\begin{array}{cccc}
p_{1}' & 0 & \ldots & 0\\
0 & p_{2}' & & 0\\
\vdots & & \ddots & \vdots\\
0 & 0 & \ldots & p_{M}'
\end{array}\right)
\end{array}
\label{matrixops}
\end{equation}

From Statement \ref{propopfun} or \ref{propopnofun} we get the familiar expressions for the expectation value $\langle p\rangle$ of $P$ just before it is about to be observed.

\begin{equation}\begin{array}{lllll}
\langle p\rangle & = & \langle\bar{S}_{C},\bar{P}\bar{S}_{C}\rangle & = & \langle\bar{S}_{C\tilde{C}},\bar{P}\bar{S}_{C\tilde{C}}\rangle\\
\langle p'\rangle & = & \langle\bar{S}_{\tilde{C}},\bar{P}'\bar{S}_{\tilde{C}}\rangle & = & \langle\bar{S}_{C\tilde{C}},\bar{P}'\bar{S}_{C\tilde{C}}\rangle
\end{array}
\label{expectvalue}
\end{equation}

If we express the wave function in terms of $P$, we get from Eq. [\ref{waveops}]

\begin{equation}\begin{array}{lllll}
\langle p\rangle & = & \sum_{i}a_{P}^{*}(p_{i})[\bar{P}_{P}a_{P}](p_{i}) & = & \sum_{i}|a_{P}(p_{i})|^{2}p_{i}\\
\langle p'\rangle & = & \sum_{i}a_{P}^{*}(p_{i})[\bar{P}_{P}'a_{P}](p_{i}) & &
\end{array}
\label{wavemeans}
\end{equation}
Corresponding equations hold if we express the wave function in terms of $P'$. From the last equality we may identify the relation $\bar{P}_{P}a_{P}(p)=p_{i}a_{P}(p)$, which holds whenever the expression on the left hand side appears as a term in a sum. In such a situation we may therefore simply write

\begin{equation}
\bar{P}_{P}=p_{i}.
\end{equation}
This relation will hold for any property $P$, and can be used instead of the matrix representation [\ref{matrixops}].

\begin{defi}[\textbf{Eigenfunctions and eigenvalues to wave function operators}]
Let $\bar{S}_{P'j}$ be an eigenvector to the operator $\bar{P}'$ defined in some contextual Hilbert space $\mathcal{H}_{C}$, so that $\bar{P}'\bar{S}_{P'j}=p_{j}'\bar{S}_{P'j}$. Suppose that the property $P$ is observed in $C$, and that it has $D[\mathcal{H}_{C}]$ possible values $\{p_{i}\}$ with associated eigenvectors $\{\bar{S}_{Pi}\}$, so that we can write $\bar{S}_{P'j}=\sum_{i}a_{Pj}(p_{i})\bar{S}_{Pi}$. Then we call $a_{Pj}(p)$ an eigenfunction to the wave function operator $\bar{P}_{P}'$, with associated eigenvalue $p_{j}'$.
\label{eigenfunction}
\end{defi}

The terms `eigenfunction' and `eigenvalue' are motivated by the relation

\begin{equation}
\bar{P}_{P}'a_{Pj}(p)=p_{j}'a_{Pj}(p)
\end{equation}
that follows from Eq. [\ref{waveops}].

\begin{defi}[\textbf{Orthonormal eigenfunctions}]
Two eigenfunctions $a_{Pj}(p)$ and $a_{Pl}(p)$ to the same wave function operator $\bar{P}_{P}'$ are orthonormal if and only if $\sum_{i}a_{Pj}^{*}(p_{i})a_{Pl}(p_{i})=0$.
\label{orthoeigenfunction}
\end{defi}

We see that if the operator $\bar{P}'$ acting in $\mathcal{H}_{C}$ has two orthonormal eigenvectors $\bar{S}_{P'j}$ and $\bar{S}_{P'l}$, so that $\langle \bar{S}_{P'j},\bar{S}_{P'l}\rangle=0$, then the corresponding eigenfunctions $a_{Pj}(p)$ and $a_{Pl}(p)$ to $\bar{P}_{P}'$ are orthonormal according to Definition \ref{orthoeigenfunction}.

\begin{defi}[\textbf{A complete set of orthonormal eigenfunctions}]
Suppose that the wave function operator $\bar{P}_{P}'$ corresponds to an operator $\bar{P}'$ that acts in the contextual Hilbert space $\mathcal{H}_{C}$, in which property $P$ is observed with $D[\mathcal{H}_{C}]$ possible values $\{p_{i}\}$. If there are $D[\mathcal{H}_{C}]$ eigenfunctions $a_{Pj}(p)$ to $\bar{P}_{P}'$ such that $\sum_{i}a_{Pj}^{*}(p_{i})a_{Pl}(p_{i})=\delta_{jl}$ for all $1\leq (j,l)\leq D[\mathcal{H}_{C}]$, then $\bar{P}_{P}'$ has a complete set of orthonormal eigenfunctions.
\label{completeortho}
\end{defi}

Say that we express the wave function in terms of some property $P$. Then Eq. [\ref{waveops}] provides a one-to-one correspondence between a linear operator $\bar{P}_{P}'$ with real eigenvalues that acts on this wave function and a linear, self-adjoint operator $\bar{P}'$. Statement \ref{opisprop} asserts that to such an operator $\bar{P}'$ corresponds at least one property $P'$. We conclude the following.

\begin{state}[\textbf{Some wave function operators correspond to properties}]
Let $\bar{P}_{P}'$ be a wave function operator that acts on any wave function $a_{P}(p)$ that can be defined in the contextual Hilbert space $\mathcal{H}_{C}$. In this context $C$, the property $P$ is observed with $D[\mathcal{H}_{C}]$ possible values $\{p_{i}\}$. If $\bar{P}_{P}'$ has a complete set of orthonormal eigenfunctions, and real eigenvalues $\{p_{j}'\}$, then $\bar{P}_{P}'$ corresponds to at least one property $P'$ with $D[\mathcal{H}_{C}]$ possible values $\{p_{j}'\}$, in the sense described in Statement \ref{opisprop}. The property $C'$ referred to in that statement is one of the members in a pair of reciprocal contexts in which $P$ and $P'$ are observed.
\label{waveopisprop}
\end{state}

Suppose that we have a wave function expressed in terms of a property $P$ that we are about to observe. When we identify an operator $\bar{P}_{P}'$ acting on this wave function as a property $P'$, we should imagine that $P'$, together with $P$, are observed in a reciprocal pair of contexts $(C,\tilde{C})$. If we change basis and write down the wave function in terms of the new operator $P'$, then we should imagine that we switch perspective from $C$ to $\tilde{C}$, that we switch the property that we are about to observe from $P$ to $P'$ (Fig. \ref{Fig69d}).

The joint state vector allows us to define the commutator $[\bar{P},\bar{P}']$ as follows.

\begin{equation}
[\bar{P},\bar{P}']\bar{S}_{C\tilde{C}}\equiv (\bar{P}\bar{P}'-\bar{P}'\bar{P})\bar{S}_{C\tilde{C}}
\end{equation}

We used this commutator already in section \ref{propop} to formulate Statement \ref{communationrules}. The difference is that in connection with that statement we applied it to the contextual state vector just \emph{after} both $P$ and $P'$ have been observed. Here we apply it to the (joint) contextual state vector just \emph{before} the first of these two properties is observed. Of course both approaches lead to the same conclusion: $[\bar{P},\bar{P}']=\bar{0}$ if and only if $P$ and $P'$ are simultaneously knowable.

We may use Eq. [\ref{waveops}] to define wave function operator products $\bar{P}_{P}'\bar{P}_{P}$ according to

\begin{equation}\begin{array}{lllll}
\bar{P}'\bar{P}\bar{S}_{C\tilde{C}} & = & \sum_{i}(\bar{P}_{P}'\bar{P}_{P}a_{P})\bar{S}_{Pi} & = & \sum_{j}(\bar{P}_{P'}'\bar{P}_{P'}\tilde{a}_{P'})\bar{S}_{P'j}\\
\bar{P}\bar{P}'\bar{S}_{C\tilde{C}} & = & \sum_{i}(\bar{P}_{P}\bar{P}_{P}'a_{P})\bar{S}_{Pi} & = & \sum_{j}(\bar{P}_{P'}\bar{P}_{P'}'\tilde{a}_{P'})\bar{S}_{P'j}
\label{waveopsprod}
\end{array}\end{equation}
We see that $[\bar{P}_{P},\bar{P}_{P}']=[\bar{P}_{P'},\bar{P}_{P'}']=0$ if and only if $[\bar{P},\bar{P}']=0$, and conclude the following.

\begin{state}[\textbf{Simultaneous knowability of a property defined by a wave function operator}]
Consider the property $P$, observed in a context $C$, together with a property $P'$ specified by the operator $\bar{P}_{P}'$, as described in statement \ref{waveopisprop}. These two properties are simultaneously knowable if and only if $[\bar{P}_{P},\bar{P}_{P}']=0$. Then we also have $[\bar{P}_{P'},\bar{P}_{P'}']=0$.
\label{comwaveprop}
\end{state}

The above two statements will be essential when we analyze the evolution equation in section \ref{eveq}. 

Suppose that $P$ and $P'$ are simultaneously knowable. We restrict ourselves to contexts in which $M=M'$, so that the dimension of the relevant Hilbert spaces is $D[\mathcal{H}_{C}]=D[\mathcal{H}_{\tilde{C}}]=D[\mathcal{H}_{C\tilde{C}}]=M^{2}$. The representations $\bar{S}_{Pi}$ and $\bar{S}_{P'j}$ of the property value states will not be vectors but supspaces with dimension $D[\bar{S}_{Pi}]=D[\bar{S}_{P'j}]=M$, as discussed in connection with Fig. \ref{Fig68}. In context $C$, the state vector $\bar{S}_{C}$ will first be projected down to one of the supspaces $\bar{S}_{Pi}$ when $P$ is observed, and then further down to one of the vectors $\bar{S}_{ij}$ when $P'$ is observed. In the reciprocal context $\tilde{C}$, the state vector is first projected down to one of the subspaces $\bar{S}_{P'j}$ when $P'$ is observed, then, again, down to one of the vectors $\bar{S}_{ij}$ when $P$ is observed. We have

\begin{equation}\begin{array}{lll}
\bar{S}_{Pi} & = & \oplus_{j=1}^{M}\bar{S}_{ij}\\
\bar{S}_{P'j} & = & \oplus_{i=1}^{M}\bar{S}_{ij}
\end{array}\end{equation}

Clearly, $\{\bar{S}_{ij}\}$ is a simultaneous set of eigenvectors to $\bar{P}$ and $\bar{P}'$. We may therefore write

\begin{equation}\begin{array}{lll}
\bar{S}_{C}(\sigma) & = & \sum_{i}a_{P}(p_{i},\sigma)\bar{S}_{Pi},\\
\bar{S}_{\tilde{C}}(\sigma) & = & \sum_{j}\tilde{a}_{P'}(p_{j}',\sigma)\bar{S}_{P'j},\\
\bar{S}_{C\tilde{C}}(\sigma) & = & \sum_{ij}a_{PP'}(p_{i},p_{j}',\sigma)\bar{S}_{ij}.
\label{jointstates}
\end{array}\end{equation}
In the last row we have implicitly used the fact that we must have $\tilde{a}_{PP'}(p_{i},p_{j}',\sigma)=a_{PP'}(p_{i},p_{j}',\sigma)$ since $C$ and $\tilde{C}$ are reciprocals of each other (Definition \ref{reciprocalpair}).

The last row defines a two-dimensional wave function $a_{PP'}(p_{i},p_{j},\sigma)$.

\begin{defi}[\textbf{The combined wave function} $a_{PP'}(p_{i},p_{j}',\sigma)$]
Consider a pair $(C,\tilde{C})$ of reciprocal contexts described by the joint family $C\tilde{C}(\sigma)$ where the two simultaneously knowable properties $P$ and $P'$ are observed. Then $a_{PP'}(p_{i},p_{j}',\sigma)$ is defined according to the third row in Eq. [\ref{jointstates}]. It specifies the joint contextual state $S_{C\tilde{C}}$ just before the first property is observed in $C$ or in $\tilde{C}$.
\label{jwavedef}
\end{defi}

It will become necessary to consider a combined wave function when we try to motivate the Dirac equation in section \ref{eveq}. We will have to introduce a spin property $s$ that we consider together with the spatio-temporal four-position property $\mathbf{r}_{4}$ in order to find an evolution equation of the wave function. The spin and the four-position are simultaneously knowable, and thus we may say that the evolution equation applies to the combined wave function $a_{\mathbf{r}4s}(\mathbf{r}_{4i},s_{j},\sigma)$. Speaking about evolution, we thus have to consider a reciprocal pair of contexts in which two properties are observed, at least implicitly.

If we have more than two simultaneously knowable properties, we may define a multi-dimensional wave function that characterizes the initial state of the joint context, defined in this case as consisting of all permutations of the order in which the properties are observed.

\vspace{5mm}
\begin{center}
$\maltese$
\end{center}
\paragraph{}

Consider a relational property $P$ like the spatial position $x$, and idealize the context so that the number of teeth of the `detector comb' goes to infinity (Fig. \ref{Fig73}), that is $M\rightarrow\infty$. Then the set of possible values $x_{j}$ becomes dense along the property value axis of the corresponding funamental property $x$. Suppose that we cannot exclude that the possible values of $x$ are continuous (Definition \ref{continuousattribute}).

For such a property $P$, define $\Delta p_{j}$ so that the values of $P$ not excluded by the observation of value $p_{j}$ in a realistic context is contained in the interval $[p_{0j}, p_{0j}+\Delta p_{j})$. This expression reflects the fact that if we let the property values $p$ that are allowed by physical law form a continuous set, we are, in practice, always dealing with non-fundamental contexts of the kind illustrated in Figs. \ref{Fig69f} and \ref{Fig70}(c). In other words, the contextual property $P_{C}$ is a coarse-grained version of the fundamental property $P$ that we may identify with the discretized property $P^{M}$ discussed in connection with Statement \ref{repnofund} and Fig. \ref{Fig69f}. For this kind of contextual property we can define the familiar continuous wave function $\Psi$ according to

\begin{equation}
\Psi_{P}(p_{0j},\sigma)\Delta p_{j}\equiv a_{P}(p_{j},\sigma),
\label{psi}
\end{equation}
for
\begin{equation}
p_{j}\leftrightarrow[p_{0j}, p_{0j}+\Delta p_{j}).
\label{contpropv}
\end{equation}

\begin{defi}[\textbf{The continuous wave function} $\Psi_{P}(p,\sigma)$]
Suppose that $|[a_{P}(p_{j+1},\sigma)-a_{P}(p_{j},\sigma)]/[p_{j+1}-p_{j}]|<<1$ for all $\sigma$ and for all $j$, where $1\leq j\leq M-1$. Then there is a continuous wave function $\Psi_{P}(p,\sigma)$, which fulfils Eq. [\ref{psi}], and which, to a good approximation, can be used to represent the contextual state $S_{C}$ just before the observation of the contextual property $P_{C}$. The domain of $\Psi_{P}(p,\sigma)$ is $([p_{1},p_{M}],[0,\sigma_{\max}])$ for some arbitrary $\sigma_{\max}$ that depends on the parametrization, and the details of the family of contexts $C(\sigma)$.
\label{cwavedef}
\end{defi}

The state $S_{OS}$ of the specimen may be such that the relative volume $v[\tilde{S}_{j}]$ of some alternatives $\tilde{S}_{j}$ is zero for some $\sigma$, as illustrated in Fig. \ref{Fig73}. We may set $\Psi_{P}(p,\sigma)=0$ for values of $p$ that belong to the corresponding observed value $p_{j}$. In other words, the support of $\Psi_{P}(p,\sigma)$ may be smaller than its domain.

\begin{defi}[\textbf{The support $D_{P}(\sigma)$ of the continuous wave function}]
$D_{P}(\sigma)$ is the union of all intervals $[p_{0j}, p_{0j}+\Delta p_{j})$ such that $a_{P}(p_{j},\sigma)\neq 0$.
\label{csupportdef}
\end{defi}

\begin{figure}[tp]
\begin{center}
\includegraphics[width=80mm,clip=true]{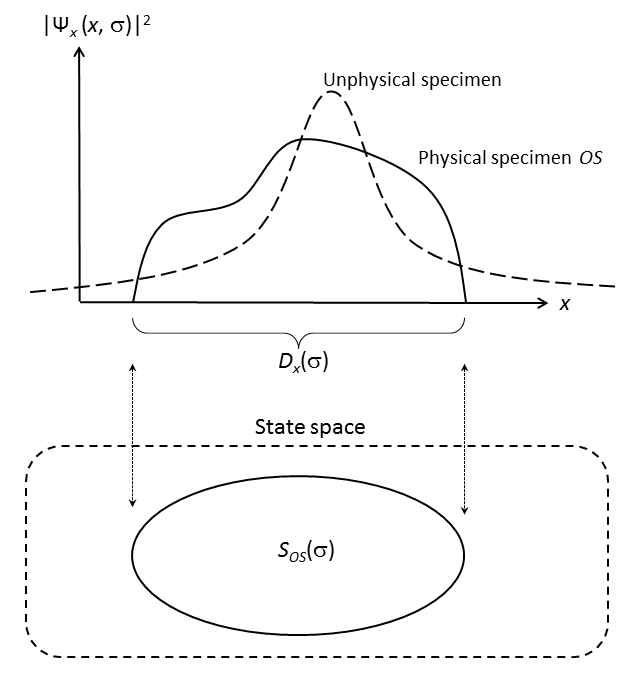}
\end{center}
\caption{Wave functions as defined here typically have a finite support $D(\sigma)$. In contrast, in the traditional picture the position of a particle is well described by a spatial wave function with a well defined hump, and with thin, but infinitely long tails. The spatial support $D_{x}(\sigma)$ can be seen as a projection of the state $S_{OS}$ along the $x$-axis. An infinite support would correspond to no knowledge at all about the position.}
\label{Fig75}
\end{figure}

Definition \ref{cwavedef} implies that we can choose $\Psi_{P}(p,\sigma)$ so that it is always differentiable with respect to $p$ inside its support $D_{P}(\sigma)$. However, it may be discontinuous at the boundary $\partial D_{P}(\sigma)$ (Fig. \ref{Fig75}). Let $a_{1}=a_{P}(p_{1})$ be the contextual numer that corresponds to the smallest possible value $p_{1}$ of property $P$, and let $a_{M}$ correspond to the largest possible value $p_{M}$. Then we have no reason to require that $|a_{1}|$ and $|a_{M}|$ are close to zero.

\begin{state}[\textbf{Continuous wave functions are piecewise differentiable}]
The continuous wave function $\Psi_{P}(p,\sigma)$ can be chosen to be at least piecewise differentiable with respect to $p$, and to be differentiable in the interior of $D_{P}(\sigma)$.
\label{psidiff}
\end{state}

We may use the continuous wave function to formulate a continuous state representation

\begin{equation}
\bar{S}_{C}(\sigma)=\int_{p\in D_{P}(\sigma)} \Psi_{P}(p,\sigma)dp\bar{S}_{P}(p),
\label{formalint}
\end{equation}
or

\begin{equation}
\bar{S}_{C}(\sigma)=\int_{-\infty}^{\infty} \Psi_{P}(p,\sigma)dp\bar{S}_{P}(p).
\label{formalint2}
\end{equation}
Remember that summation in Hilbert space corresponds to `OR' in knowledge space, as expressed in Table \ref{dictionary}. The meaning of the integration in the above equation is therefore just that all values of $P$ outside the support $D_{P}(\sigma)$ can be exluded as an outcome of the observation of $P_{C}$, based on the potential knowledge before the observation. The support $D_{P}$ can be seen as the projection of the state $S_{OS}$ down to the subspace in state space defined by property $P$. Consequently, it does not make sense to actually calculate the integral - there is nothing to add up. It should be seen as a purely formal representation of the contextual state just before an observation.

The vector $\bar{S}_{P}(p)$ in Eqs. [\ref{formalint}] and [\ref{formalint2}] is defined according to the relation
\begin{equation}
\bar{S}_{Pj}\equiv\Delta p_{j}^{-1/2}\int_{p_{0j}}^{p_{0j}+\Delta p_{j}}\bar{S}_{P}(p)dp.
\label{contstate}
\end{equation}
This relation expresses the fact that given the property value state $S_{Pj}$, where the property value $p_{j}$ corresponds to the interval given in Eq. [\ref{contpropv}], we cannot exclude any of the `continuous property value states' $S_{P}(p)$. Further, it is impossible in principle to define a context in which the exact continuous property value $p$ is determined. This means that we cannot define any meaningful relative volumes $v$ or contextual numbers $a$ to a corresponding set $S_{P}(p)$ in state space. In other words, there is no measure defined for an individual $p\in[p_{0j},p_{0j}+\Delta p_{j})$. Therefore we do not introduce any function $f(p)$ in the integrand. The factor $\Delta p_{j}^{-1/2}$ can be seen as a normalization constant, as we will see within short.

\begin{defi}[\textbf{The continuous property value space} $\mathcal{P}(p)$]
For a fundamental property $P$ with a continuous infinity of values $p$, the property value spaces $\mathcal{P}(p)$ partition the property space $\mathcal{P}$ in a continuous infinity of infinitely thin, distinct slices; $\mathcal{P}=\bigcup_{p}\mathcal{P}(p)$, and $\mathcal{P}(p)\cap \mathcal{P}(p')=\varnothing$ whenever $p\neq p'$.
\end{defi}

\begin{state}[\textbf{Representation of continuous property value spaces}]
The set $\{\mathcal{P}(p)\}$ of continuous property value spaces can be represented as an orthonormal basis $\{\bar{\mathcal{P}}(p)\}$ in a Hilbert space $\mathcal{H}_{P}$. Let the corresponding contextual property $P_{C}$ be such that the continuous wave function $\Psi_{P}$ can be defined. Then the contextual property value states $S_{Pj}$ can be expressed as in Eq. [\ref{contstate}], where we may identify $\bar{S}_{P}(p)=\bar{\mathcal{P}}(p)$. We have $\bar{P}_{C}\bar{S}_{Pj}=p_{j}\bar{S}_{Pj}$ and $\bar{P}\bar{S}_{P}(p)=p\bar{S}_{P}(p)$.
\label{contrep}
\end{state}

Equation [\ref{contstate}] implies
\begin{equation}\begin{array}{lll}
\delta_{jj'} & = & \langle\bar{S}_{Pj},\bar{S}_{Pj'}\rangle\\
& = & (\Delta p_{j}\Delta p_{j'})^{-1/2}\int_{p_{0j}}^{p_{0j}+\Delta p_{j}}\int_{p_{0j'}}^{p_{0j'}+\Delta p_{j}'}\langle\bar{S}_{P}(p),\bar{S}_{P}(p')\rangle dpdp',
\end{array}\end{equation}
which is fulfilled if and only if we identify
\begin{equation}
\delta(p-p')=\langle\bar{S}_{P}(p),\bar{S}_{P}(p')\rangle.
\label{deltarel}
\end{equation}
Note that the introduction of the Dirac delta function is needed only in the integral representation of the contextual state $\bar{S}_{C}$, and that this integral representation is, at best, a convenient approximation to the summation representation [\ref{sumrep}], which reflects the actual physical context, with its distinct alternatives. Thus the delta function plays no fundamental role and cause no conceptual trouble; it is just a convenient symbol in the manipulation of the integrals that may appear.

Since the relative volumes of all alternatives by definition add to one in any state, we have, in the same way as for the summation representation [\ref{sumrep}]:

\begin{equation}\begin{array}{lll}
1 & = & \langle\bar{S}_{C}(\sigma),\bar{S}_{C}(\sigma)\rangle\\
& = & \int\int \Psi_{P}(p,\sigma)^{*}dp\Psi_{P}(p',\sigma)dp'\langle\bar{S}_{P}(p),\bar{S}_{P}(p')\rangle\\
& = & \int\int \Psi_{P}(p,\sigma)^{*}dp\Psi_{P}(p',\sigma)dp'\delta(p-p')\\
& = & \int|\Psi_{P}(p,\sigma)|^{2}dp,
\end{array}
\label{intunit}
\end{equation}
for any $\sigma$. The expression on the right hand side is obviously an integral in the conventional sense, in contrast to that in Eq. [\ref{formalint}].

\begin{state}[\textbf{The continuous wave function $\Psi_{P}(p,\sigma)$ is always normalized}]
We have $\int|\Psi_{P}(p,\sigma)|^{2}dp=1$ for all $\sigma$ in the domain of $\Psi_{P}$.
\label{cnormalized}
\end{state}

If we are dealing with a family of contexts $C(\sigma,\sigma')$ according to Fig. \ref{Fig73a}, we may define a second continuous wave function $\Psi_{P'}(p_{j}',\sigma)$, which approximately specifies the contextual state just before observation of the second property $P_{C}'$, and fulfils the relation

\begin{equation}
\Psi_{P'}(p_{0j}',\sigma')\Delta p_{j}'\equiv a_{P'}(p_{j}',\sigma').
\label{psi2}
\end{equation}

At least, this is possible if the continuous approximation is good for the second property $P_{C}'$ also. Note that $\Psi_{P}(p,\sigma)$ is no longer defined at this stage in the context.

Let us now express some relations in the continuous representation in the case we have a reciprocal context $\tilde{C}$ and a joint family of contexts $C\tilde{C}(\sigma)$. We refer to the above discussion about the general wave function $a_{P}$ for the relevant definitions and the complete picture. We have

\begin{equation}\begin{array}{lllll}
\bar{S}_{C\tilde{C}} & = & \int\Psi_{P}(p,\sigma)dp\bar{S}_{P}(p) = \int\tilde{\Psi}_{P'}(p',\sigma)dp'\bar{S}_{P'}(p')\\
\bar{S}_{C} & = & \int\Psi_{P}(p,\sigma)dp\bar{S}_{P}(p) & &\\
\bar{S}_{\tilde{C}} & = & \int\tilde{\Psi}_{P'}(p',\sigma)dp'\bar{S}_{P'}(p') & &
\end{array}
\label{contbasechange}
\end{equation}
just before the observation of the first property in the context. The notation should be self-explanatory, and is analogous to the general case. In the same manner as in the general case we may define a quadruplet of operators $(\bar{P}_{P}^{\Psi},\bar{P}_{P'}^{\Psi},\bar{P'}_{P}^{\Psi},\bar{P'}_{P'}^{\Psi})$ by the following relations.

\begin{equation}\begin{array}{lllll}
\bar{P}\bar{S}_{C\tilde{C}} & = & \int(\bar{P}_{P}^{\Psi}\Psi_{P})dp\bar{S}_{P}(p) & = &
\int(\bar{P}_{P'}^{\Psi}\tilde{\Psi}_{P'})dp'\bar{S}_{P'}(p')\\
\bar{P'}\bar{S}_{C\tilde{C}} & = & \int(\bar{P'}_{P}^{\Psi}\Psi_{P})dp\bar{S}_{P}(p) & = &
\int(\bar{P'}_{P'}^{\Psi}\tilde{\Psi}_{P'})dp'\bar{S}_{P'}(p')
\end{array}
\label{psiops}
\end{equation}
where we have dropped the arguments of the wave functions for notational clarity.

\begin{defi}[\textbf{The continuous wave function property operators}]
In a joint context $C\tilde{C}$ where an integral representation of the contextual state $S_{C\tilde{C}}$ is possible, the quadruplet of operators $(\bar{P}_{P}^{\Psi},\bar{P}_{P'}^{\Psi},\bar{P'}_{P}^{\Psi},\bar{P'}_{P'}^{\Psi})$ are defined by Eq. [\ref{psiops}]. We may say that $\bar{P}_{P}^{\Psi}$ is the operator corresponding to $P$ that acts on the continuous wave function when it is expressed in terms of $P$, and that $\bar{P}_{P'}^{\Psi}$ is the operator corresponding to $P$ that acts on the continuous wave function when it is expressed in terms of $P'$. The interpretation of $\bar{P'}_{P}^{\Psi}$ and $\bar{P'}_{P'}^{\Psi}$ is analogous.
\label{dpsiops}
\end{defi}

From Eq. [\ref{wavemeans}] we see that the expected values of $P$ and $P'$ just before they are observed are

\begin{equation}\begin{array}{lllll}
\langle p\rangle & = & \int\Psi_{P}^{*}\bar{P}_{P}^{\Psi}\Psi_{P}dp & = & \int|\Psi_{P}|^{2}p dp\\
\langle p'\rangle & = & \int\Psi_{P}^{*}\bar{P'}_{P}^{\Psi}\Psi_{P}dp & &
\end{array}
\label{cwavemeans}
\end{equation}
where we have dropped all arguments for clarity, and we have used that

\begin{equation}
\bar{P}_{P}=p
\label{pisp}
\end{equation}
for any property $P$ that allows a continuous representation.

We may define

\begin{equation}
\bar{P}\bar{S}_{P}(p)\equiv p \bar{S}_{P}(p).
\label{conteigenstate}
\end{equation}
If we use this definition and apply $\bar{P}$ to both sides of Eq. [\ref{contstate}] we get $\bar{P}\bar{S}_{Pj}=\Delta p_{j}^{-1/2}\int_{p_{0j}}^{p_{0j}+\Delta p_{j}}p\bar{S}_{P}(p)dp$. This relation can be interpreted as to say that to the eigenvector $\bar{S}_{Pj}$ corresponds property values $p$ in the range $[p_{0j},p_{0j}+\Delta p_{j})$, and continuous property value states $\bar{S}_{P}(p)$ for $p$ in the same range. This is the picture we want to give, showing that Definition \ref{conteigenstate} is the proper way to describe the action of $\bar{P}$ on the continuous eigenstate.

\begin{defi}[\textbf{Eigenfunctions and eigenvalues to continuous wave function operators}]
Suppose that we can write $\bar{S}_{P'}(p')=\int_{-\infty}^{\infty}\psi_{P}(p',p)dp\bar{S}_{P}(p)$. Then $\bar{P'}_{P}^{\Psi}\psi_{P}(p',p)=p'\psi_{P}(p',p)$ according to Eq. [\ref{conteigenstate}] and Definition \ref{dpsiops}. We call $\psi_{P}(p',p)$ an eigenfunction to the continuous wave function operator $\bar{P'}_{P}^{\Psi}$ with associated eigenvalue $p'$.
\label{conteigenfunction}
\end{defi}

From Eq. [\ref{deltarel}] we see that

\begin{equation}
\int_{-\infty}^{\infty}\psi_{P}(p'',p)^{*}\psi_{P}(p',p)dp=\delta(p''-p').
\end{equation}
We may interpret this relation as to say that the eigenfunctions $\psi_{P}(p',p)$ for different fixed values of $p'$ are orthonormal to each other.

We see that if the operator $\bar{P}'$ acting in $\mathcal{H}_{C}$ has two orthonormal eigenvectors $\bar{S}_{P'j}$ and $\bar{S}_{P'l}$, so that $\langle \bar{S}_{P'j},\bar{S}_{P'l}\rangle=0$, then the corresponding eigenfunctions $a_{Pj}(p)$ and $a_{Pl}(p)$ to $\bar{P}_{P}'$ are orthonormal according to Definition \ref{orthoeigenfunction}.

\begin{defi}[\textbf{A complete set of orthonormal eigenfunctions in the continuous representation}]
Suppose that we can write $\Psi_{P}(p)=\int_{-\infty}^{\infty}\psi_{P}(p',p)dp'$ for any piecewise differentiable wave function $\Psi_{P}(p)$ that can be used in a continuous representation of any context $C$ in which property $P$ is observed. Then we say that the set of eigenfunctions $\psi_{P}(p',p)$ for different fixed values of $p'$ is complete.
\label{completconteortho}
\end{defi}

Using the above concepts we can formulate the continuous version of Statement \ref{waveopisprop}.

\begin{state}[\textbf{Some continuous wave function operators correspond to properties}]
Let $\bar{P'}_{P}^{\Psi}$ be a continuous wave function operator that acts on any wave function $\Psi_{P}(p)$ that can be used in a continuous representation of any context $C$ in which property $P$ is observed. If $\bar{P'}_{P}^{\Psi}$ has a complete set of orthonormal eigenfunctions, and real eigenvalues $\{p'\}$, then $\bar{P'}_{P}^{\Psi}$ corresponds to at least one fundamental property $P'$ with a continuous infinity of possible values $\{p'\}$.
\label{psiopisprop}
\end{state}

Note that we do not refer to any specific kind of context $C'$ in which the identified property $P'$ is to be observed, like in Statement \ref{waveopisprop}. Here we have been manipulating continuous wave functions and continuous wave function operators alone. The treatement has therefore been abstract, since these continuous objects only appear in approximative representations of actual contexts. This is the reason why we identify $P'$ as a fundamental property. Its infinity of possible values means that we always observe a corresponding contextual property $P_{C}'$ for which these values are grouped into $M$ bins, each of which correspond to one observed value $p_{j}'$ (Fig. \ref{Fig69f}).  

If $P$ and $P'$ are simultaneously knowable, and both properties allow a continuous representation, we may use Eq. [\ref{jointstates}] and write

\begin{equation}
\bar{S}_{C\tilde{C}}(\sigma)=\int\int\Psi_{PP'}(p,p',\sigma)dpdp'\bar{S}_{PP'}(p,p')
\label{cjointstates}
\end{equation}
for a two-dimensional continuous wave function $\Psi_{PP'}(p,p',\sigma)$. The `two-dimensional continuous property value state' $\bar{S}_{PP'}(p,p')$ is defined in analogy with Eq. [\ref{contstate}], and relates to other quantities in a way that is analogous to Statement \ref{contrep}.

\begin{defi}[\textbf{The combined continuous wave function} $\Psi_{PP'}(p,p',\sigma)$]
Consider a pair $(C,\tilde{C})$ of reciprocal contexts described by the joint family $C\tilde{C}(\sigma)$ where two simultaneously knowable properties $P$ and $P'$ are observed, which both allow a continuous representation. Then $\Psi_{PP'}(p,p',\sigma)$ is defined according to Eq. [\ref{cjointstates}]. It specifies the joint contextual state $S_{C\tilde{C}}$ just before the first property is observed in $C$ or in $\tilde{C}$.
\label{jpsidef}
\end{defi}

\vspace{5mm}
\begin{center}
$\maltese$
\end{center}
\paragraph{}

The wave functions $a_{P}$ or $\Psi_{P}$ that we have discussed above differ from the traditional ones in three respects.

\begin{enumerate}
\item They are only defined in certain experimental contexts.
\item Any given wave function is often just temporarily defined during the course of such an experiment.
\item The support $D_{P}(\sigma)$ is typically finite.
\end{enumerate}

Let us discuss this list of interpretational issues from back to front. A traditional wave function $\Psi_{x}(x,t)$, in the form of a localized wave packet with thin tails and infinite support that travels along a spatial $x$-axis as time $t$ passes, is most often unphysical from our point of view (Fig. \ref{Fig75}). This is so even if it can be normalized. It corresponds to an object whose position is completely unknown. It could be anywhere in the universe. However, if the wave function describes a property of a specific object - an identifiable specimen that we have identified in the past - then its position cannot become completely unknown in finite time. There will always be large chunks of state space that we can exclude. We can most often exlude faraway positions even if the specimen is not individually identified in the past. This is beacuse the specimens entering an observational context tend to have a known, localized source, like the particle gun in Fig. \ref{Fig73}.

\begin{figure}[tp]
\begin{center}
\includegraphics[width=80mm,clip=true]{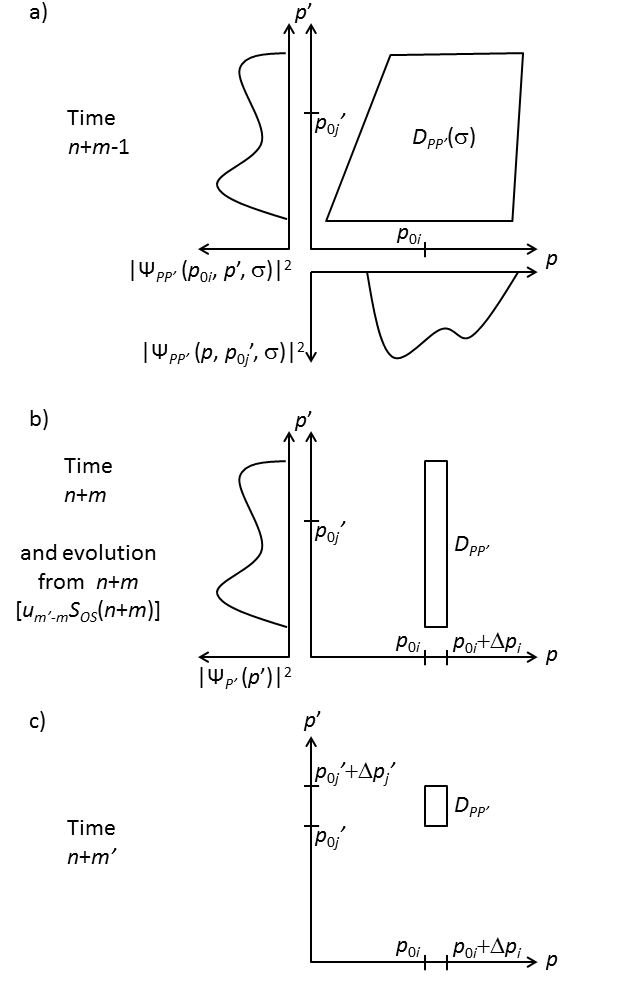}
\end{center}
\caption{A family of contexts $C(\sigma)$ in which $P$ is observed at time $n+m$, and then $P'$ at time $n+m'$. The properties are simultaneously knowable. At time $n+m-1$ we have $|\Psi_{PP'}(p_{0i},p')|^{2}\Delta p'=q(p'|p_{i})$ in context $C$ where $P$ is to be observed first, and where $p'=q(p'|p_{i})$ is the idealized conditional probability to find $p$ in the range $[p,p+\Delta p)$ given $p_{j}'$. Similarly, $|\Psi_{PP'}(p,p_{0j}')|^{2}\Delta p\approx q(p|p_{j}')$ in the reciprocal context $\tilde{C}$ where $P'$ is to be observed first. The wave function as a function of $p$ is no longer defined at time $n+m$ or later, neither is $\sigma$. The wave function as a function of $p'$ is no longer defined at time $n+m'$ or later. We see that the domain $D_{PP'}$ is a more fundamental object than the wave function. It is always defined since it is just a projection in state space of $S_{OS}$. That the domain $D_{PP'}$ is not rectangular to begin with means that there is conditional \emph{a priori} knowledge that excludes some value combinations $(p,p')$. Compare Fig. \ref{Fig76b}.}
\label{Fig76}
\end{figure}

\begin{figure}[tp]
\begin{center}
\includegraphics[width=80mm,clip=true]{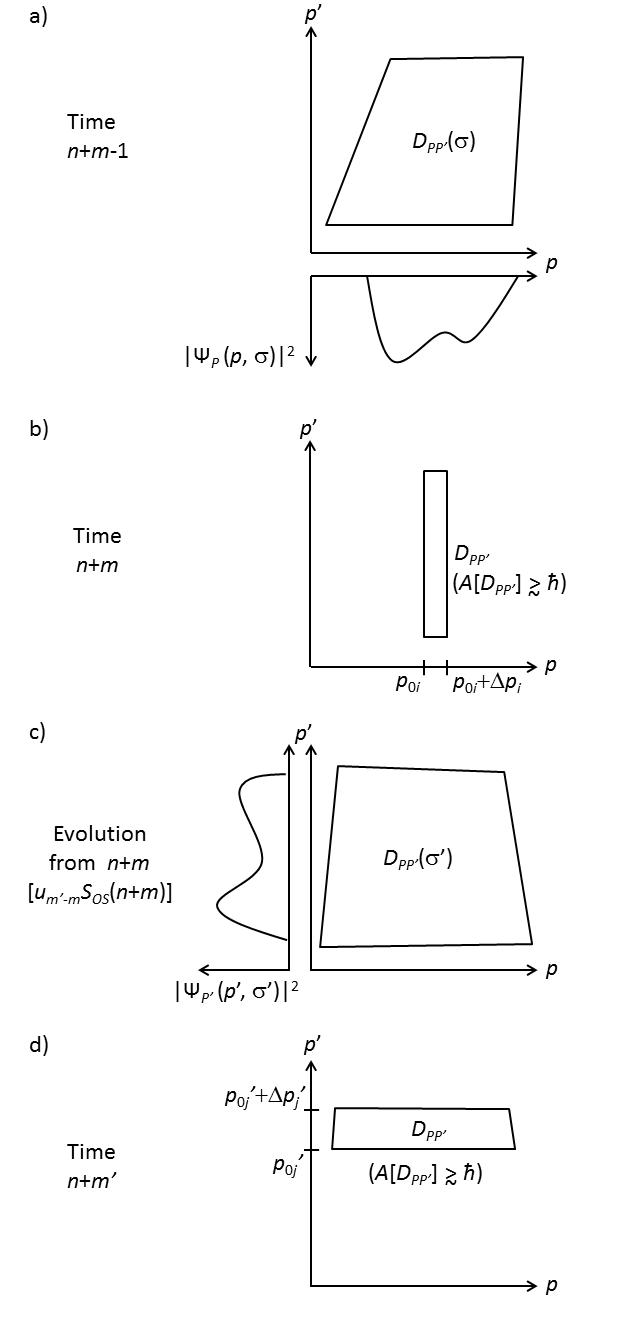}
\end{center}
\caption{A family of contexts $C(\sigma,\sigma')$ in which property $P$ is observed at time $n+m$, and $P'$ at time $n+m'$. The two properties are not simultaneously knowable. The wave function $\Psi_{P}(p,\sigma)$ is defined just before the observation of $p$, but not afterwards. The second wave function $\Psi_{P'}(p',\sigma')$ is defined just before the observation of $p'$, but not afterwards. If we interpret $P$ and $P'$ as position and momentum, there is an uncertainty relation $A[D_{PP'}]\approx\Delta p\Delta p'\gtrsim\hbar$, where $A[\ldots]$ denotes an area. This uncertainty refers to the physical state of the specimen rather than the wave function. Compare Figs. \ref{Fig76} and \ref{Fig76c}.}
\label{Fig76b}
\end{figure}

In other words, the essential measure of uncertainty in our description is the width $\Delta D_{P}$ of the support rather than the standard deviation $\Delta p=\sqrt{\mathrm{Var}\left[|\Psi_{P}(p)|^{2}\right]}$ of the probability distribution defined by the wave function. One may think that this causes trouble when we observe two not simultaneously knowable properties in succession, like position $x$ and momentum $p_{x}$. For suppose that we have measured $x$ to a fairly good precision, so that $\Delta D_{x}$ is small. Then we measure $p_{x}$ so that $\Delta D_{p_{x}}$ becomes small. To describe the state of the specimen after this event with a continuous wave function $\Psi_{x}(x)$ would require an infinite support $\Delta D_{x}$. This is so since the discontinuities in the derivatives of $\Psi_{p_{x}}(p_{x})$ that arise due to the finite $\Delta D_{p_{x}}$ can only be described by infinitely many eigenfunctions of position. This would mean that we have lost all the gained knowledge about position, that the specimen can suddenly be infinitely far away. This is clearly absurd, since the time difference $\Delta t$ between the two meaurements is finite, and the specimen cannot travel faster than the speed of light. The mistake in this reasoning is that we mix our own perspective (finite supports) with the traditional perspective (universally defined wave functions, which we can express in any basis we want to). More precisely, the infinitely many eigenfunctions of positions present in the wave function does not necessarily correspond to a set of possible outcomes of a second measurement of position, since this wave function is no longer defined at such a measurement. The point is illustrated in Fig. \ref{Fig76b}, and is further discussed in section \ref{evconsequences} in relation to Fig. \ref{Fig76c}. Here we touch upon the second item in the above list of interpretational issues.

After the observation of a property $P$ is made, the wave function as a function of $p$ and $\sigma$ is no longer defined. It provides the probabilities of the options in a complete set of future alternatives $\{\tilde{S}_{i}\}$ (Definition \ref{setfuturealt}), where the probability $q(p_{i})$ is the relative volume $v[\tilde{S}_{i},S_{O}(n)]$. After one alternative is realized, there are no longer any subjectively predefined alternatives, no relative volumes, and therefore no wave function. We can, of course, define new options for $P$. But that will give rise to a new, different, wave function, which should be expressed in terms of a different evolution parameter $\sigma'$ (Fig. \ref{Fig73a} and Eq. [\ref{psi2}]). These matters are illustrated in Figs. \ref{Fig76} and \ref{Fig76b}.

Even if there are predefined options $\tilde{S}_{i}$ for the value of a property $P$, it may nevertheless be impossible to define the wave function. This situation arises when the relative volumes $v[\tilde{S}_{i},S_{O}(n)]$ are not known at the time $n$ when the context is initiated, so that no contextual numbers $a_{i}$ can be associated with them, and therefore no corresponding wavefunction $a(p_{i})$. These matters were discussed in section \ref{probabilities}.

\begin{state}[\textbf{Collapse of the wave function}]
Suppose that we have a combined wave function of two or more property values $p,p',\ldots$ according to Statement \ref{jwavedef} or \ref{jpsidef}. Once property $P$ is observed, the wave function does no longer depend on property value $p$.
\label{collapse}
\end{state}

\begin{state}[\textbf{Loss of the wave function}]
Suppose that we have a wave function of a single property value $p$. Once property $P$ is observed, the wave function is no longer defined.
\label{loss}
\end{state}

\subsection{The evolution of a free specimen}
\label{eveq}

Equations [\ref{sumunit}] and [\ref{intunit}] express the unitarity of the evolution of the contextual state $\bar{S}_{C}$, as summarized in Statement \ref{unitary}. To determine the form of the evolution equation, we add two more requirements.

\begin{enumerate}
\item The form of the evolution equation should be relativistically invariant. From our perspective, this is a consequence of the assumptions of epistemic minimalism and (individual) epistemic invariance.
\item There should be a parametrization $\mathbf{r}_{4}=\mathbf{r}_{4}(\sigma)$ so that we can write $d\langle\mathbf{r}_{4}\rangle/d\sigma = \mathbf{v}_{0}+\beta\mathbf{f}(\sigma)$, where $\mathbf{r}_{4}=(\mathbf{x},ict)$ for $\mathbf{x}=(x,y,z)$, $\mathbf{v}_{0}$ is a constant vector, and $\beta$ is a scalar such that $\beta=0$ if and only if the specimen $OS$ does not interact with any other object.
\end{enumerate}

\begin{figure}[tp]
\begin{center}
\includegraphics[width=80mm,clip=true]{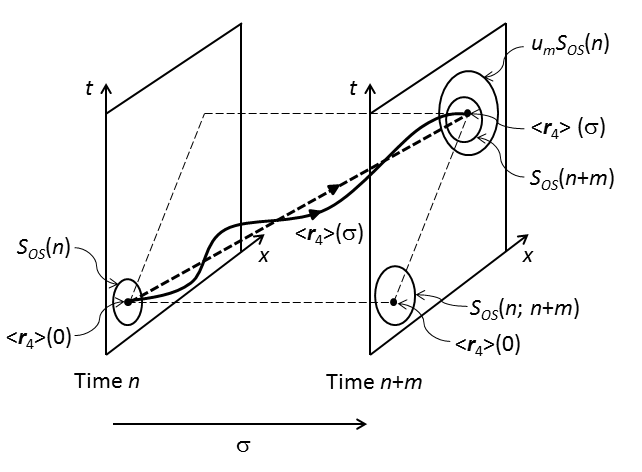}
\end{center}
\caption{Two parametrizations $\langle \mathbf{r}_{4}\rangle(\sigma)$, one straight and one winding, which are both contained in the tilted plane that defines all parametrizations that correspond to a free specimen with no net interactions that bend its expected trajectory $\langle x\rangle(\sigma)/\langle t\rangle(\sigma)$. The marked states $S_{OS}$ are really projections of these states onto space-time, but we suppress the projection operator for clarity. Recall from Fig. \ref{Fig52} that $S_{OS}(n;n+m)$ denotes the state that corresponds to the memory of $S_{OS}(n)$ at time $n+m$. Put differently, the state $S_{OS}(n+m)$ corresponds to an object $OS$ with presentness attribute $Pr=1$ at time $n+m$, whereas $S_{OS}(n;n+m)$ corresponds to an object with attribute value $Pr=0$. It is the memory of the state of the same object at a previous time $n$. The evolution equation relates the states of objects with presentness attribute $Pr=1$ at different sequential times $n$ and $n+m$.}
\label{Fig76d}
\end{figure}

The second condition is illustrated in Fig. \ref{Fig76d}. For a free specimen, any parametrization $\langle \mathbf{r}_{4}\rangle(\sigma)$ must be contained in the tilted plane between the space-times associated with times $n$ and $n+m$, respectively, so that $\langle x\rangle(\sigma)/\langle t\rangle(\sigma)$ does not depend on $\sigma$. Then there is a parametrization that follows the straight, diagonal line between $\langle \mathbf{r}_{4}\rangle(0)$ and $\langle \mathbf{r}_{4}\rangle(\sigma)$. (Recall from the discussion in Section \ref{knowstate} that in this setting $\sigma$ denotes the value of the evolution parameter just before the observation at time $n+m$. This means that when we vary $\sigma$ in Fig. \ref{Fig76d}, we vary the position of the spatio-temporal plane at time $n+m$ along the trajectory $\langle\mathbf{r}_{4}\rangle(\sigma)$.)

The possibility to use the concept of a straight trajectory as a starting point when we try to find a mathematical form for physical law is opened up by the discussion in Section \ref{statespaces}. There we list some basic relations between attribute values, relations which are taken to be primary from an epistemic point of view, impossible to reduce to something else. One of these relations is straightness (Definitions \ref{straightness} and \ref{straj}). A trajectory that seems straight to one subject may not seem straight to another. This means that a specimen that seems free as judged by the first subject may seem to be interacting as judged by the second. We return to these matterns in Sections \ref{eveqi} and \ref{gaugeprinciple}.

Consider a property $P$ and a physical state $S(n)$ such that we cannot exclude \emph{a priori} any exact value $p$ in some continuous interval. Relational attributes are properties of this kind. In such a situation, we can define a family of contexts $C(\sigma)$ such that the integral representation [\ref{formalint}] becomes an arbitrarily good approximation of the contextual state $\bar{S}_{C}$.

Let us focus on the evolution of the wave function $\Psi_{P}(p,\sigma)$ defined in this situation according to Eq. [\ref{psi}]. We have defined the family of contexts $C(\sigma)$ in such a way that $\Psi_{P}(p,\sigma')$ is uniquely given by $\Psi_{P}(p,\sigma)$ for any $\sigma'>\sigma$. We may therefore write

\begin{equation}
\frac{d}{d\sigma}\Psi_{P}(p,\sigma)=\bar{A}_{P}\Psi_{P}(p,\sigma),
\label{ev1}
\end{equation}
for some linear operator $\bar{A}$. That $\bar{A}$ has to be linear follows from the linearity of the evolution operator $u_{1}$ (Statement \ref{linearev}) and the definition of $\Psi_{P}$ (Eq. [\ref{formalint}]). The operator $\bar{A}$ cannot depend explicitly on $\sigma$, since $\sigma$ is not an attribute that describes the state of any object, and the evolution depends on nothing else but the physical state. Therefore we may also write $\Psi_{P}(p,\sigma)=\exp(\bar{A}_{P}\sigma)\Psi_{P}(p,0)$. We may identify the contextual evolution operator $\bar{u}_{C}(\sigma)=\exp(\bar{A}_{P}\sigma)$ in Statement \ref{unitary}. Therefore $\exp(\bar{A}_{P}\sigma)$ is unitary, so that we must have \begin{equation}
\bar{A}_{P}=i\bar{B}_{P},
\label{evbop}
\end{equation}
where $\bar{B}_{P}$ has real eigenvalues. We may interpret $\bar{B}_{P}$ as a continuous wave function operator according to Eq. [\ref{dpsiops}], which correspond to a self-adjoint operator $\bar{B}'$. We will see below that $\bar{B}_{P}$ is such that it has a complete set of eigenfunctions according to Definition \ref{completconteortho}. $\bar{B}_{P}$ therefore corresponds to a property $B$ according to Statement \ref{psiopisprop}. We may write

\begin{equation}
\Psi_{P}(p,\sigma)=\bar{u}_{C}(\sigma)\Psi_{P}(p,0)=\exp(i\bar{B}_{P}\sigma)\Psi_{P}(p,0).
\label{cbasicev}
\end{equation}

The above considerations apply for any property $P$. Let us find the form of $\bar{B}_{P}$ when $P$ is the spatio-temporal distance $\mathbf{r}_{4}$, and the specimen $OS$ moves freely without interacting with its environment before the observation of $\mathbf{r}_{4}$ takes place. To do so, we make use of requirement 2). Equations [\ref{expectvalue}] and [\ref{cwavemeans}] allow us to write

\begin{equation}
\langle \mathbf{r}_{4} \rangle=\langle\bar{S}_{C},\overline{\mathbf{r}_{4}} \bar{S}_{C}\rangle = \int\Psi_{\mathbf{r}_{4}}^{*}(\overline{\mathbf{r}_{4}})_{\mathbf{r}_{4}}\Psi_{\mathbf{r}_{4}}d\mathbf{r}_{4}.
\label{mean}
\end{equation}
These relations imply

\begin{equation}\begin{array}{lll}
\frac{d}{d\sigma}\langle\mathbf{r}_{4}\rangle & = & \frac{d}{d\sigma}\int\Psi_{\mathbf{r}_{4}}^{*}(\overline{\mathbf{r}_{4}})_{\mathbf{r}_{4}}\Psi_{\mathbf{r}_{4}}d\mathbf{r}_{4}\\
& = &  i\int\Psi_{\mathbf{r}_{4}}^{*}(\mathbf{r}_{4}\bar{B}_{\mathbf{r}_{4}}-\bar{B}_{\mathbf{r}_{4}}\mathbf{r}_{4})\Psi_{\mathbf{r_{4}}}d\mathbf{r}_{4}
\end{array}
\label{meanr}
\end{equation}
where we have used Eq. [\ref{ev1}], and the fact that $\bar{B}_{\mathbf{r}_{4}}=-i\bar{A}_{\mathbf{r}_{4}}$ corresponds to a self-adjoint operator $\bar{B}$ according to Eq. [\ref{dpsiops}]. We have suppressed the arguments in $\Psi_{\mathbf{r}_{4}}(\mathbf{r}_{4},\sigma)$ for clarity. We have also replaced the operator $(\overline{\mathbf{r}_{4}})_{\mathbf{r}_{4}}$, which corresponds to the property $\mathbf{r}_{4}=(\mathbf{r},ict)$ when the wave function is expressed in terms of the very same $\mathbf{r}_{4}=(\mathbf{r},ict)$, with its value $\mathbf{r}_{4}$, according to Eq. [\ref{pisp}]. (Strictly speaking, the four-position $\mathbf{r}_{4}$ is not a property as expressed above, since the fourth component is not real, but this is just a matter of notation.) 

Any wave function $\Psi_{\mathbf{r}_{4}}(\mathbf{r}_{4},\sigma)$ can be expressed as a Fourier integral

\begin{equation}
\Psi_{\mathbf{r}_{4}}(\mathbf{r}_{4},\sigma)=(2\pi)^{-5/2}\int_{-\infty}^{\infty}\tilde{\Psi}_{\mathbf{r}_{4}}(\tilde{\mathbf{r}}_{4},\tilde{\sigma})
e^{i(\tilde{\mathbf{r}}_{4}\cdot\mathbf{r}_{4}+\tilde{\sigma}\sigma)}d\tilde{\mathbf{r}}_{4}d\tilde{\sigma}.
\label{fexpansion}
\end{equation}
We may call the kernel $\tilde{\Psi}_{\mathbf{r}_{4}}(\tilde{\mathbf{r}}_{4},\tilde{\sigma})$ the `reciprocal wave function'. Analogously, we may call $\tilde{\mathbf{r}}_{4}$ the `reciprocal four-position', and $\tilde{\sigma}$ the `reciprocal evolution parameter'.

Inserting the expansion [\ref{fexpansion}] into Eq. [\ref{meanr}], we get

\begin{equation}\begin{array}{lll}
\frac{d}{d\sigma}\langle\mathbf{r}_{4}\rangle & = & (2\pi)^{-5}\int^{\infty}_{-\infty}\tilde{\Psi}_{\mathbf{r}_{4}}^{*}(\tilde{\mathbf{r}}_{4}',\tilde{\sigma}')\tilde{\Psi}_{\mathbf{r}_{4}}(\tilde{\mathbf{r}}_{4},\tilde{\sigma})\times\\
&  & \left\{\int_{\mathbf{r}_{4}\in D_{\mathbf{r}_{4}}}
e^{-i(\tilde{\mathbf{r}}_{4}\cdot\mathbf{r}_{4}+\tilde{\sigma}\sigma)}(\mathbf{r}_{4}\bar{B}_{\mathbf{r}_{4}}-\bar{B}_{\mathbf{r}_{4}}\mathbf{r}_{4})
e^{i(\tilde{\mathbf{r}_{4}}\cdot\mathbf{r}_{4}+\tilde{\sigma}\sigma)}d\mathbf{r}_{4}\right\}d\tilde{\mathbf{r}}_{4}d\tilde{\sigma}d\tilde{\mathbf{r}}_{4}'d\tilde{\sigma}'
\end{array}
\label{meanr2}
\end{equation}
 
Since we are considering a free specimen $OS$, the right hand side is a (real) constant, so that the integral over $\mathbf{r}_{4}$ within the curly brackets cannot depend on $\mathbf{r}_{4}$ or $\sigma$. The fact that it cannot depend on $\sigma$ implies $(\mathbf{r}_{4}\bar{B}_{\mathbf{r}_{4}}-\bar{B}_{\mathbf{r}_{4}}\mathbf{r}_{4})e^{i(\tilde{\mathbf{r}}_{4}\cdot\mathbf{r}_{4}+\tilde{\sigma}\sigma)}=f(\tilde{r}_{4},\tilde{\sigma},\mathbf{r}_{4})e^{i(\tilde{\mathbf{r}}_{4}\cdot\mathbf{r}_{4}+\tilde{\sigma}\sigma)}$, where the function $f$ does not depend on $\sigma$.

These considerations should hold for any wave function, in particular for an individual Fourier mode $\Psi_{\mathbf{r}_{4}}=e^{i(\tilde{\mathbf{r}}_{4}\cdot\mathbf{r}_{4}+\tilde{\sigma}\sigma)}$ for $\mathbf{r}_{4}\in D_{\mathbf{r}_{4}}$. In such a context $|\Psi_{\mathbf{r}_{4}}|^{2}$ is a positive constant for $\mathbf{r}_{4}\in D_{\mathbf{r}_{4}}$, and $|\Psi_{\mathbf{r}_{4}}|^{2}=0$ for $\mathbf{r}_{4}\not\in D_{\mathbf{r}_{4}}$, where $D_{\mathbf{r}_{4}}$ is the support (Definition \ref{csupportdef}). Since the specimen is assumed to be free, and since the spatio-temporal position $\mathbf{r}_{4}$ is basically a relational attribute whose value is defined in relation to an arbitrary reference frame of other objects, the constant $d\langle\mathbf{r}_{4}\rangle/d\sigma$ must be invariant under stiff translations of the region $D_{\mathbf{r}_{4}}$ in which we know that the specimen is located. (By a `stiff' translation we mean that the shape of the region does not change.) Therefore the function $f$ that describe the effect of the action of $\mathbf{r}_{4}\bar{B}_{\mathbf{r}_{4}}-\bar{B}_{\mathbf{r}_{4}}\mathbf{r}_{4}$ cannot depend on $\mathbf{r}_{4}$ either:

\begin{equation}
(\mathbf{r}_{4}\bar{B}_{\mathbf{r}_{4}}-\bar{B}_{\mathbf{r}_{4}}\mathbf{r}_{4})e^{i(\tilde{\mathbf{r}}_{4}\cdot\mathbf{r}_{4}+\tilde{\sigma}\sigma)}=f(\tilde{r}_{4},\tilde{\sigma})e^{i(\tilde{\mathbf{r}}_{4}\cdot\mathbf{r}_{4}+\tilde{\sigma}\sigma)}.
\end{equation}

Furthermore, the imaginary unit $i$ that appears in front of the integrals at the right hand side of Eq. [\ref{meanr2}] means that $f(\tilde{r}_{4},\tilde{\sigma})$ must be imaginary. These constraints imply

\begin{equation}\begin{array}{lll}
\bar{B}_{\mathbf{r}_{4}} & = & \sum_{k}b_{k}\frac{\partial^{2}}{\partial r_{k}^{2}}\\
& = & b_{x}\frac{\partial^{2}}{\partial x^{2}}+b_{y}\frac{\partial^{2}}{\partial y^{2}}+b_{z}\frac{\partial^{2}}{\partial z^{2}}-b_{t}\frac{1}{c^{2}}\frac{\partial^{2}}{\partial t^{2}},
\end{array}\end{equation}
for some array $(b_{x},b_{y},b_{z},b_{t})$ of real, scalar constants. Since the evolution equation [\ref{ev1}] must be relativistically invariant, we must have $b=b_{x}=b_{y}=b_{z}=b_{t}$, so that

\begin{equation}
\bar{B}_{\mathbf{r}_{4}}=-b\Box.
\label{dalembert}
\end{equation}

Equations [\ref{meanr}] and [\ref{dalembert}] imply that we may write

\begin{equation}\begin{array}{lll}
\frac{d}{d\sigma}\langle \mathbf{r}_{4}\rangle & = & -ib\int_{-\infty}^{\infty}
\Psi_{\mathbf{r}_{4}}^{*}\left(\mathbf{r}_{4}\Box-\Box\mathbf{r}_{4}\right)\Psi_{\mathbf{r}_{4}}d\mathbf{r}_{4}\\
& = & 2ib\int_{-\infty}^{\infty}
\Psi_{\mathbf{r}_{4}}^{*}\left(\frac{\partial}{\partial r_{1}},\frac{\partial}{\partial r_{2}},\frac{\partial}{\partial r_{3}},\frac{\partial}{\partial r_{4}}\right)\Psi_{\mathbf{r}_{4}}d\mathbf{r}_{4}.
\end{array}\end{equation}
If we insert the Fourier integral [\ref{fexpansion}] in the above expression, we get

\begin{equation}
\frac{d}{d\sigma}\langle \mathbf{r}_{4}\rangle = -2b\langle\tilde{\mathbf{r}}_{4}\rangle,
\label{meandr}
\end{equation}
where we have defined

\begin{equation}
\langle\tilde{\mathbf{r}}_{4}\rangle\equiv \int_{-\infty}^{\infty}\tilde{\Psi}_{\mathbf{r}_{4}}(\tilde{\mathbf{r}}_{4},\tilde{\sigma})^{*}\tilde{\mathbf{r}}_{4}\tilde{\Psi}_{\mathbf{r}_{4}}(\tilde{\mathbf{r}}_{4},\tilde{\sigma})d\tilde{\mathbf{r}}_{4}d\tilde{\sigma}
\end{equation}

The fact that time is directed means that $d\langle t\rangle/d\sigma$ has the same sign in all reference frames. It is natural to choose a parametrization so that the evolution parameter and relational time flow in the same direction:
\begin{equation}
d\langle t\rangle/d\sigma>0.
\label{directed}
\end{equation}
Equation [\ref{meandr}] implies that we may write

\begin{equation}
\frac{d}{d\sigma}\langle t\rangle = \frac{2i b}{c}\langle\tilde{r}_{4}\rangle,
\label{tct}
\end{equation}
since $r_{4}=ict$. Equation [\ref{directed}] should hold for all allowed values of $\tilde{\mathbf{r}}_{4}$. Equation [\ref{tct}] therefore implies that $\tilde{r}_{4}$ is imaginary, and that

\begin{equation}
i b\tilde{r}_{4}>0.
\label{signrule1}
\end{equation}
If we insert $\bar{A}_{\mathbf{r}}=-ib\Box$ in Eq. [\ref{ev1}], and express $\Psi_{\mathbf{r}_{4}}$ in terms of its Fourier integral [\ref{fexpansion}], we get $\tilde{\sigma}=-b\sum_{k}\tilde{r}_{k}^{2}$, or

\begin{equation}
-\tilde{r}_{4}^{2}=\frac{\tilde{\sigma}}{b}+\tilde{r}_{1}^{2}+\tilde{r}_{2}^{2}+\tilde{r}_{3}^{2}.
\label{protoeinstein}
\end{equation}
This relation must hold for all possible values of $\tilde{r}$, in particular when the reciprocal spatial positions are all zero; $\tilde{r}_{1}=\tilde{r}_{2}=\tilde{r}_{3}=0$. Since $\tilde{r}_{4}$ is imaginary, we must have

\begin{equation}
\frac{\tilde{\sigma}}{b}>0
\label{signrule2}
\end{equation}
for all $\tilde{\sigma}$. The sign of the parameter $b$ is arbitrary. In the following, let us use the convention

\begin{equation}
\begin{array}{lll}
b & < & 0,\\
i\tilde{r}_{4} & < & 0,\\
\tilde{\sigma} & < & 0.
\end{array}
\label{signconvention}
\end{equation}

We argued above that the wave function operator $\bar{B}_{P}$ corresponds to a property $B$ for any $P$, and therefore this is true in particular for $\bar{B}_{\mathbf{r}_{4}}$. The set of possible values of this property $B$ should be found among the set of eigenvalues $\tilde{\sigma}$ to $-i\frac{d}{d\sigma}$, and the corresponding property value states are described in terms of eigenfunctions as $\Psi_{\mathbf{r}_{4}}^{(\tilde{\sigma})}=\psi(\mathbf{r}_{4})e^{i\tilde{\sigma}\sigma}$, where the function $\psi(\mathbf{r}_{4})$ is arbitrary. In short,

\begin{equation}\begin{array}{c}
\Psi_{\mathbf{r}_{4}}^{(\tilde{\sigma})}=\psi_{\mathbf{r}_{4}}(\mathbf{r}_{4})e^{i\tilde{\sigma}\sigma}\\
-i\frac{d}{d\sigma}\Psi_{\mathbf{r}_{4}}^{(\tilde{\sigma})}=\bar{B}_{\mathbf{r}_{4}}\Psi_{\mathbf{r}_{4}}^{(\tilde{\sigma})}=\tilde{\sigma}\Psi_{\mathbf{r}_{4}}^{(\tilde{\sigma})}.
\end{array}
\label{eigensigma}
\end{equation}
In the same way (according to Statement \ref{psiopisprop}), there are properties $C$ and $D$ that correspond to the wave function operators

\begin{equation}\begin{array}{lll}
\bar{C}_{\mathbf{r}_{4}} & \equiv & b\left(\frac{\partial^{2}}{\partial x^{2}}+\frac{\partial^{2}}{\partial y^{2}}+\frac{\partial^{2}}{\partial z^{2}}\right)\\
\bar{D}_{\mathbf{r}_{4}} & \equiv & -b\frac{1}{c^{2}}\frac{\partial^{2}}{\partial t^{2}}.
\end{array}\end{equation}
In the same way as for $B$, the set of possible values of $C$ becomes the set of eigenvalues $\{\tilde{x}^{2}+\tilde{y}^{2}+\tilde{z}^{2}\}$. Likewise, the set of possible values of $D$ should be found among the set of eigenvalues $\tilde{t}^{2}$. We clearly have

\begin{equation}
\bar{B}_{\mathbf{r}_{4}}=\bar{C}_{\mathbf{r}_{4}}+\bar{D}_{\mathbf{r}_{4}}.
\end{equation}
This operator relation is closely related to the relativistic relation $E^{2}=\left(m_{0}c\right)^{2}+(cp)^{2}$, as we will discuss below.

We see in Eq. [\ref{eigensigma}] that for each eigenfunction $\Psi_{\mathbf{r}_{4}}^{(\tilde{\sigma})}$ to the operator $\bar{B}_{\mathbf{r}_{4}}$ with eigenvalue $\tilde{\sigma}$, there is another eigenfunction $\Psi_{\mathbf{r}_{4}}^{(-\tilde{\sigma})}$ with eigenvalue $-\tilde{\sigma}$. However, Eq. [\ref{signrule2}] tells us that all property values $\tilde{\sigma}$ must have the same sign (since $b$ is a fixed constant). This means that there are eigenvalues to $\bar{B}_{\mathbf{r}_{4}}$ that do not correspond to possible property values $\tilde{\sigma}$. Therefore $\bar{B}_{\mathbf{r}_{4}}=-id/d\sigma$ is not the unique operator associated with property $\tilde{\sigma}$ according to Statement \ref{psiopisprop}, since the match between the set of eigenvalues and the set of possible property values should be perfect.

Since $\tilde{\sigma}$ is a property with no positive values, we may define another property $M$ with values $\mu$ such that

\begin{equation}
\tilde{\sigma}=-\mu^{2}.
\end{equation}
Let us look for the appropriate operator $\bar{M}_{\mathbf{r}}$ (in the sense of Statement \ref{psiopisprop}) that corresponds to $M$.

In general, referring to Eq. [\ref{eigensigma}] we may write

\begin{equation}
\bar{B}_{\mathbf{r}_{4}}\psi_{\mathbf{r}_{4}}(\mathbf{r}_{4})=\tilde{\sigma}\psi_{\mathbf{r}_{4}}(\mathbf{r}_{4}),
\end{equation}
since $\bar{B}_{\mathbf{r}_{4}}$ does not act on $\sigma$. To ensure that all possible values of the property $B$ has the same sign, we may require

\begin{equation}
\bar{B}_{\mathbf{r}_{4}}=-\bar{W}_{\mathbf{r}_{4}}^{\dagger}\bar{W}_{\mathbf{r}_{4}},
\label{squareroot1}
\end{equation}
so that
\begin{equation}
\bar{B}_{\mathbf{r}_{4}}\psi_{\mathbf{r}_{4}}(\mathbf{r}_{4})=-\bar{W}_{\mathbf{r}_{4}}^{\dagger}\bar{W}_{\mathbf{r}_{4}}\psi_{\mathbf{r}_{4}}(\mathbf{r}_{4})=-\bar{W}_{\mathbf{r}_{4}}^{\dagger}\mu\psi_{\mathbf{r}_{4}}(\mathbf{r}_{4})=-\mu^{*}\mu\psi_{\mathbf{r}_{4}}(\mathbf{r}_{4}),
\end{equation}
and we get $\tilde{\sigma}=-|\mu|^{2}$, in agreement with the sign convention [\ref{signconvention}]. The wave function operator $\bar{M}_{\mathbf{r}}$ we look for should therefore be of the form [\ref{squareroot1}], and should correspond to a self-adjoint operator $\bar{M}$, so that $\mu$ becomes real. That is,

\begin{equation}
\bar{B}_{\mathbf{r}_{4}}=-\bar{M}_{\mathbf{r}_{4}}\bar{M}_{\mathbf{r}_{4}},
\label{squareroot2}
\end{equation}

Thus, in addition to the evolution equation [\ref{ev2}], we should add the constraint
\begin{equation}
\bar{M}_{\mathbf{r}_{4}}\psi_{\mathbf{r}_{4}}(\mathbf{r}_{4},\tilde{\sigma})=\mu\psi_{\mathbf{r}_{4}}(\mathbf{r}_{4},\tilde{\sigma})
\label{dirac1}
\end{equation}
for each function $\psi_{\mathbf{r}_{4}}(\mathbf{r}_{4},\tilde{\sigma})e^{i\tilde{\sigma}\sigma}$ in the general solution

\begin{equation}
\Psi_{\mathbf{r}_{4}}(\mathbf{r}_{4},\sigma)=\int_{-\infty}^{\infty}\psi_{\mathbf{r}_{4}}(\mathbf{r}_{4},\tilde{\sigma})e^{i\tilde{\sigma}\sigma}d\tilde{\sigma}
\end{equation}
to the evolution equation
\begin{equation}
\frac{d}{d\sigma}\Psi_{\mathbf{r}_{4}}(\mathbf{r}_{4},\sigma)=i\bar{B}_{\mathbf{r}_{4}}\Psi_{\mathbf{r}_{4}}(\mathbf{r}_{4},\sigma).
\label{ev2}
\end{equation}
(We know that the constraint [\ref{dirac1}] means that we have to work with spinors, but we stick to our single wave function notation here, and return to these matters below.)

The property $M$ is defined for any specimen $OS$, and is relativistically invariant according to the evolution equation [\ref{ev2}]. It may therefore be seen as an internal attribute. These qualities resemble those of the rest mass, or rest energy.

\begin{defi}[\textbf{Rest mass} $m_{0}$]
The property values of the rest mass are given by
\begin{equation}
m_{0}\equiv \frac{\hbar}{c\sqrt{-b}}\mu.
\end{equation}
The corresponding continuous wave function operator for a free specimen is
\begin{equation}
(\overline{m_{0}})_{\mathbf{r}_{4}}=\frac{\hbar}{c\sqrt{-b}}\bar{M}_{\mathbf{r}_{4}}
\end{equation}
when $\Psi$ is expressed in terms of the property $\mathbf{r}_{4}$.
\label{restmassdefi}
\end{defi}

The difference from the conventional notion of rest mass is that nothing in the above reasoning prevents $m_{0}$ from being negative.

\begin{defi}[\textbf{Squared rest mass} $m_{0}^{2}$]
The property values of the squared rest mass are given by
\begin{equation}
m_{0}^{2}\equiv -\frac{\hbar^{2}}{bc^{2}}\mu^{2}=\frac{\hbar^{2}}{bc^{2}}\tilde{\sigma}.
\end{equation}
A corresponding continuous wave function operator for a free specimen is
\begin{equation}
(\overline{m_{0}^{2}})_{\mathbf{r}_{4}}=\frac{\hbar^{2}}{c^{2}}\Box
\end{equation}
when $\Psi$ is expressed in terms of the property $\mathbf{r}_{4}$, or, equivalently,
\begin{equation}
\overline{m_{0}^{2}}=\frac{i\hbar^{2}}{bc^{2}}\frac{d}{d\sigma}.
\end{equation}
However, these operators do not correspond perfectly to $m_{0}^{2}$ in the sense of Statement \ref{psiopisprop}, since they have more eigenvalues than there are property values.
\label{masssquaredefi}
\end{defi}

We have effectively identified the reciprocal evolution parameter $\tilde{\sigma}$ with the squared rest mass, up to suitable constant of proportionality. In an analogous way, we can identify the reciprocal spatial position $(\tilde{x},\tilde{y},\tilde{z})$ with momentum, and the reciprocal temporal position $\tilde{r}_{4}$ with energy.

\begin{defi}[\textbf{Momentum} $\mathbf{p}$]
The property values of the momentum of a free specimen are given by
\begin{equation}
\mathbf{p}=\hbar\left(\tilde{x},\tilde{y},\tilde{z}\right).
\end{equation}
The corresponding continuous wave function operator is
\begin{equation}
\overline{\mathbf{p}}_{\mathbf{r}_{4}}=-i\hbar\left(\frac{\partial}{\partial x},\frac{\partial}{\partial y},\frac{\partial}{\partial z}\right)=-i\hbar\nabla
\end{equation}
when $\Psi$ is expressed in terms of the property $\mathbf{r}_{4}$.
\label{momentumdefi}
\end{defi}

\begin{defi}[\textbf{Energy} $E$]
The property values of the energy of a free specimen are given by
\begin{equation}
E=-ic\hbar\tilde{r}_{4}.
\label{energydef}
\end{equation}
A corresponding continuous wave function operator is
\begin{equation}
\overline{E}_{\mathbf{r}_{4}}=-c\hbar\frac{\partial}{\partial r_{4}}=i\hbar\frac{\partial}{\partial t}
\end{equation}
when $\Psi$ is expressed in terms of the property $\mathbf{r}_{4}$.
\label{energydefi}
\end{defi}

As a check that we have defined mass, momentum and energy appropriately, we straightforwardly deduce from Eq. [\ref{protoeinstein}] that

\begin{equation}
E^{2}=m_{0}^{2}c^{4}+c^{2}|\mathbf{p}|^{2}.
\label{einsteinrelation}
\end{equation}
We see in this equation that the fact that energy is non-negative does not exclude negative rest masses, since energy is related to the rest mass squared. Using the four-momentum $\mathbf{p}_{4}=(\mathbf{p},iE/c)$ this familiar relation reads

\begin{equation}
m_{0}^{2}c^{2}=-|\mathbf{p}_{4}|^{2}.
\end{equation}
This equation is fullfilled if we relate the four-momentum to the reciprocal spatio-temporal position $\tilde{\mathbf{r}}_{4}$ in the following way.

\begin{defi}[\textbf{Four-momentum} $\mathbf{p}_{4}$]
The property values of the four-momentum of a free specimen are given by
\begin{equation}
\mathbf{p}_{4}=\hbar\tilde{\mathbf{r}}_{4}.
\end{equation}
A corresponding operator is
\begin{equation}
(\overline{\mathbf{p}_{4}})_{\mathbf{r}_{4}}=-i\hbar\left(\frac{\partial}{\partial r_{1}},\frac{\partial}{\partial r_{2}},\frac{\partial}{\partial r_{3}},\frac{\partial}{\partial r_{4}}\right)
\end{equation}
when the wave function $\Psi$ is expressed in terms of the property $\mathbf{r}_{4}$.
\label{fourmomentumdefi}
\end{defi}

The parameter $b$ has no physical significance. Its numerical value depends on the way we parametrize the evolution with the help of the evolution parameter $\sigma$, which is not an observable property. We may contrast $b$ with $\hbar$. The latter parameter is a constant of proportionality that relates energy $E$ with the angular frequency $ic\tilde{r}_{4}$ with which the wave function oscillates as a function of $t$. Analogoulsy, $b^{-1}$ is a constant of proportionality that relates the squared rest mass $m_{0}^{2}$ with the angular frequency $\tilde{\sigma}$ with which the wave function oscillates as a function of the evolution parameter $\sigma$. Planck's constant $\hbar$ is a physical constant, since it relates the two observable attributes energy and time. On the other hand, $b$ is not a physical constant, since it relates the observable rest mass with the unobservable parameter $\sigma$.

We can fix the value of $b$ if we choose the natural parametrization

\begin{equation}
d\langle t\rangle/d\sigma=1.
\label{naturalp}
\end{equation}
This choice means that we can treat $\sigma$ as we treat time $t$ in classical mechanics, when we do not take into account that knowledge of relational time $t$ may be incomplete. Note, however, that the relation [\ref{naturalp}] is not relativistically invariant.

\begin{defi}[\textbf{The natural parametrization}]
The parametrization of the family of contexts $C(\sigma)$ is natural if and only if $d\langle t\rangle/d\sigma=1$ for all $\sigma\in[0,\sigma_{\max})$.
\label{natparadef}
\end{defi}

In the natural parametrization we get

\begin{equation}
b=-\frac{c^{2}\hbar}{2\langle E\rangle}.
\end{equation}
In this parametrization we can use Eq. [\ref{meandr}] and the definition of momentum (Definition \ref{momentumdefi}) to derive the counterpart of Ehrenfest's theorem:

\begin{equation}
\frac{d}{d\sigma}\langle\mathbf{r}\rangle=\frac{\langle\mathbf{p}\rangle}{\langle m\rangle},
\label{naturalehrenfest}
\end{equation}
where $\mathbf{r}=(x,y,z)$ and $m$ is the relativistic mass defined according to $E=mc^{2}$. Since Ehrenfest's theorem relates observable attributes or properties, it should not depend on a particular parametrization. The corresponding parametrization-independent formula is

\begin{equation}
\frac{d\langle\mathbf{r}\rangle}{d\sigma}/\frac{d\langle t\rangle}{d\sigma}=\frac{\langle\mathbf{p}\rangle}{\langle m\rangle}.
\label{ehrenfest}
\end{equation}

With the above definitions, we can write the evolution equation [\ref{ev1}] or [\ref{ev2}] in more familiar terms. 

\begin{state}[\textbf{The evolution equation}]
We have
\begin{equation}\begin{array}{lll}
i\hbar\frac{d}{d\sigma}\Psi_{\mathbf{r}_{4}}(\mathbf{r}_{4},\sigma) & = & \frac{-1}{2\langle E\rangle}(\overline{E_{0}^{2}})_{\mathbf{r}_{4}}\Psi_{\mathbf{r}_{4}}(\mathbf{r}_{4},\sigma)\\
& = & \frac{-1}{2\langle E\rangle}\left(\bar{E}_{\mathbf{r}_{4}}\bar{E}_{\mathbf{r}_{4}}-c^{2}\bar{\mathbf{p}}_{\mathbf{r}_{4}}\cdot\bar{\mathbf{p}}_{\mathbf{r}_{4}}\right)\Psi_{\mathbf{r}_{4}}(\mathbf{r}_{4},\sigma)
\end{array}
\end{equation}
in the natural parametrization, where $(\overline{E_{0}^{2}})_{\mathbf{r}_{4}}\equiv c^{4}(\overline{m_{0}^{2}})_{\mathbf{r}_{4}}$, or
\begin{equation}
i\hbar\frac{d}{d\sigma}\Psi_{\mathbf{r}_{4}}(\mathbf{r}_{4},\sigma)=
\frac{c^{2}}{2\langle E\rangle}\left[(\overline{\mathbf{p}_{4}})_{\mathbf{r}_{4}}\cdot(\overline{\mathbf{p}_{4}})_{\mathbf{r}_{4}}\right]\Psi_{\mathbf{r}_{4}}(\mathbf{r}_{4},\sigma).
\end{equation}
For a free specimen these relations correspond to
\begin{equation}
\frac{d}{d\sigma}\Psi_{\mathbf{r}_{4}}(\mathbf{r}_{4},\sigma)=\frac{ic^{2}\hbar}{2\langle E\rangle}\Box\Psi_{\mathbf{r}_{4}}(\mathbf{r}_{4},\sigma).
\label{freeeveq}
\end{equation}
\label{psieveq}
\end{state}

A plane wave solution to the evolution of a free specimen is

\begin{equation}
\Psi_{\mathbf{r}_{4}}(\mathbf{r}_{4},\sigma)\propto\exp[\frac{i}{\hbar}(\mathbf{p}\cdot\mathbf{r}-Et-\frac{E_{0}^{2}}{2\langle E\rangle}\sigma)],
\end{equation}
in the natural parametrization [\ref{naturalp}], or
\begin{equation}
\Psi_{\mathbf{r}_{4}}(\mathbf{r}_{4},\sigma)\propto\exp[\frac{i}{\hbar}(\mathbf{p}_{4}\cdot\mathbf{r}_{4}-\frac{E_{0}^{2}}{2\langle E\rangle}\sigma)].
\label{p4planewave}
\end{equation}
We write $\mathbf{r}=(x,y,z)$ as before, and introduce the rest energy $E_{0}=m_{0}c^{2}$. Even if $\langle E\rangle=E$ in a single plane wave, like the one above, we have to insert the average energy in the last denominator of each plane wave if we superpose several such waves with different energies, for instance in a Fourier integral.

\vspace{5mm}
\begin{center}
$\maltese$
\end{center}
\paragraph{}

The above formalism presupposes that the continuous wave function representation of the contextual state $S_{C}$ in Eq. [\ref{formalint}] is a valid approximation for the spatio-temporal position $\mathbf{r}_{4}$ that we supposedly observe in the context. As we have discussed above, this means in practice that the values of $\mathbf{r}_{4}$ we cannot exclude before the observation are dense in comparison to distance between the teeth of the `detector comb' that we introduced in Fig. \ref{Fig73}. The formalism also presupposes that the evolution of the contextual state can be represented by the evolution of a \emph{single} continuous wave function $\Psi_{\mathbf{r}_{4}}$. However, we recognize the Dirac equation in Eq. [\ref{dirac1}], and we know that the solutions to the Dirac equation are four-component spinors. Furthermore, we may want to consider the evolution when the wave function is not expressed in terms of $\mathbf{r}_{4}$, but some property whose values are inherently discrete, for which the integral representation never applies.

Let us therefore express the evolution equation in a more general form. Equations [\ref{ev1}] and [\ref{cbasicev}] hold equally well if we replace the continuous wave function with the general wave function:

\begin{equation}\begin{array}{lll}
\frac{d}{d\sigma}a_{P}(p_{i},\sigma) & = & i\bar{B}_{P}a_{P}(p_{i},\sigma)\\
& = & -\frac{i}{2\hbar\langle E\rangle}(\overline{E_{0}^{2}})_{P}a_{P}(p_{i},\sigma),
\end{array}
\end{equation}
where the last line holds in the natural parmetrization, or

\begin{equation}\begin{array}{lll}
a_{P}(p_{i},\sigma) & = & \exp(i\bar{B}_{P}\sigma)a_{P}(p_{i},0)\\
& = & \exp\left( -\frac{i}{2\hbar\langle E\rangle}(\overline{E_{0}^{2}})_{P} \right)a_{P}(p_{i},0).
\end{array}\end{equation}

As before, we let $M$ denote the number of alternatives for property $P$ within context, so that $1\leq i\leq M$. The solution to the general evolution equation can be written

\begin{equation}
a_{P}(p_{i},\sigma)=\int\alpha_{P}(p_{i},E_{0}^{2})\exp\left(\frac{-iE_{0}^{2}}{2\hbar\langle E\rangle}\sigma\right)dE_{0}^{2},
\end{equation}
where the stationary wave function $\alpha_{P}(p_{i},E_{0}^{2})$ can be seen as the eigenfunction to the continuous wave function operator $(\overline{E_{0}^{2}})_{P}$ with associated eigenvalue $E_{0}^{2}$, according to Definition \ref{eigenfunction}:

\begin{equation}
(\overline{E_{0}^{2}})_{P}\alpha_{P}(p_{i},E_{0}^{2})=E_{0}^{2}\alpha_{P}(p_{i},E_{0}^{2}).
\end{equation}

Let us use this more general notation to express and interpret the Dirac equation. The need to take the square root $\bar{M}_{\mathbf{r}}$ of the operator $\bar{B}_{\mathbf{r}}$ and use this operator $\bar{M}_{\mathbf{r}}$ to express a constraint [\ref{dirac1}] on the wave function, means that we cannot express the evolution of the state in terms of property $\mathbf{r}_{4}$ alone. We must express the wave function $a$ in terms of a combined property $P''$, which is the spatio-temporal position $\mathbf{r}_{4}$ \emph{together with} the degrees of freedom represented by the components of the spinor. This means that when we express the evolution of the contextual state $S_{C}$, we must implicitly consider a context $C$ in which both the spatio-temporal position and the spinor property are observed in succession. The spinor degrees of freedom do not refer to $\mathbf{r}_{4}$, and they are present even for a free specimen for which there are no other objects with which the specimen may interact. They should therefore be classified as internal rather than relational attributes. Let us denote the relational attribute $\mathbf{r}_{4}$ by $P$, and the internal spinor attributes by $P'$. The members of this property pair $P''=(P,P')$ will be simultanesously knowable, and we may therefore express the wave function $a_{P''}$ as a combined wave function

\begin{equation}
a_{PP'}(p,p',\sigma)=a_{\mathbf{r}_{4}s}((\mathbf{r}_{4})_{i},s_{j},\sigma)
\end{equation}
as discussed in relation to Definition \ref{jwavedef}. Here the property $s$ denotes the four components of the spinor, so that $1\leq j\leq 4$. In contrast, we have $1\leq i\leq M$, where the number $M$ of possible spatio-temporal positions within context is arbitrary. For large $M$ it becomes natural to choose the continuous representation of $\mathbf{r}_{4}$, so that the combined wave function can be written

\begin{equation}
a_{\mathbf{r}_{4}s}((\mathbf{r}_{4})_{i},s_{j},\sigma)\leftrightarrow \left(\begin{array}{c}\Psi_{\mathbf{r}_{4}1}(\mathbf{r}_{4},\sigma)\\\Psi_{\mathbf{r}_{4}2}(\mathbf{r}_{4},\sigma)\\\Psi_{\mathbf{r}_{4}3}(\mathbf{r}_{4},\sigma)\\\Psi_{\mathbf{r}_{4}4}(\mathbf{r}_{4},\sigma)\end{array}\right)
\label{spinorparts}
\end{equation}
as usual.

To formulate the Dirac equation properly, we should make it clear in the notation that we are dealing with the wave function expressed in terms of the combined property $P''=\mathbf{r}_{4}s$. Also, we should make it clear that the crucial operators act on this combined property, not just $\mathbf{r}_{4}$. Thus we make the following changes of notation: $\bar{M}_{\mathbf{r}_{4}}\rightarrow\bar{M}_{\mathbf{r}_{4}s}$ and $(\overline{E_{0}})_{\mathbf{r}_{4}}\rightarrow (\overline{E_{0}})_{\mathbf{r}_{4}s}$. 

\begin{state}[\textbf{The Dirac equation as an additional constraint on the wave function}]
Each wave function $\alpha_{\mathbf{r}_{4}s}((\mathbf{r}_{4})_{i},s_{j},E_{0}^{2})\exp\left(-\frac{iE_{0}^{2}}{2\hbar\langle E\rangle}\sigma\right)$ in the general solution
\begin{equation}
a_{\mathbf{r}_{4}s}((\mathbf{r}_{4})_{i},s_{j},\sigma)=\int_{-\infty}^{\infty}\alpha_{\mathbf{r}_{4}s}((\mathbf{r}_{4})_{i},s_{j},E_{0}^{2})e^{-\frac{iE_{0}^{2}}{2\hbar\langle E\rangle}\sigma}dE_{0}^{2}
\end{equation}
to the evolution equation
\begin{equation}\begin{array}{lll}
i\hbar\frac{d}{d\sigma}a_{\mathbf{r}_{4}s}((\mathbf{r}_{4})_{i},s_{j},\sigma) & = & \frac{-1}{2\langle E\rangle}(\overline{E_{0}^{2}})_{\mathbf{r}_{4}}a_{\mathbf{r}_{4}s}((\mathbf{r}_{4})_{i},s_{j},\sigma)\\
& = & \hbar\bar{M}_{\mathbf{r}_{4}s}\bar{M}_{\mathbf{r}_{4}s}a_{\mathbf{r}_{4}s}((\mathbf{r}_{4})_{i},s_{j},\sigma)\\
& = & \frac{c^{2}}{2\langle E\rangle}\left[(\overline{\mathbf{p}_{4}})_{\mathbf{r}_{4}}\cdot(\overline{\mathbf{p}_{4}})_{\mathbf{r}_{4}}\right]a_{\mathbf{r}_{4}s}((\mathbf{r}_{4})_{i},s_{j},\sigma)
\end{array}
\end{equation}
must be an eigenfunction to the continuous rest energy operator $(\overline{E_{0}})_{\mathbf{r}_{4}s}$:
\begin{equation}
(\overline{E_{0}})_{\mathbf{r}_{4}s}\tilde{\alpha}_{\mathbf{r}_{4}s}((\mathbf{r}_{4})_{i},s_{j},E_{0}^{2})=E_{0}\alpha_{\mathbf{r}_{4}s}((\mathbf{r}_{4})_{i},s_{j},E_{0}^{2}),
\end{equation}
where we define
\begin{equation}
(\overline{E_{0}})_{\mathbf{r}_{4}s}\equiv\sqrt{2\hbar\langle E\rangle}\bar{M}_{\mathbf{r}_{4}s}
\end{equation}
in the natural parametrization.
\label{diracconstraint}
\end{state}

Thanks to Dirac we know that we can more explicitly write
\begin{equation}
(\overline{E_{0}})_{\mathbf{r}_{4}s}=c(\overline{\mathbf{p}_{4}})_{\mathbf{r}_{4}}\cdot\bar{\mathbf{D}}_{s},
\end{equation}
where $\bar{\mathbf{D}}_{s}=(\bar{D}_{1},\bar{D}_{2},\bar{D}_{3},\bar{D}_{4})$ is a vector of $4\times 4$ matrices, closely related to the gamma matrices according to $D_{1}=i\gamma^{1}$, $D_{2}=i\gamma^{2}$, $D_{3}=i\gamma^{3}$, and $D_{4}=\gamma^{0}$.

We see that the Dirac equation can be interpreted as a stationary state equation, analogous the equation $\bar{H}\psi=E\psi$ in Schr\"odinger wave mechanics. We just replace the Hamiltonian $\bar{H}$ with the rest energy operator $\overline{E_{0}}$, and the energy eigenvalue $E$ with the rest energy eigenvalue $E_{0}$. The only difference is that the solution $\psi$ to the Dirac equation is a function of the relativistic spatio-temporal position $\mathbf{r}_{4}=(x,y,z,ict)$ instead of just the spatial position $\mathbf{r}_{3}=(x,y,z)$, as in the case of the stationary Schr\"odinger equation.

\subsection{The evolution of an interacting specimen}
\label{eveqi}

In the preceding section, we interpreted a `free specimen' to be a specimen that is expected to travel along a straight line. Conversely, a specimen that interacts with its environment should be expected to change direction in relation to the objects that surrounds it. Such a change of direction can come about in two ways. Either we observe that the specimen emits a second object and follows a new path afterwards [Fig. \ref{Fig42}(c)], or it follows curved path without knowably emitting another object. The latter case means that the chain of overlapping sets in Fig. \ref{Fig40} is bending.

When we speak about an interacting specimen, we assume that is is identifiable throughout the process, that it can be said to remain \emph{the same}, and that it preserves its perceived unity. This is so in the division process shown in Fig. \ref{Fig42}(c), but not in the object divisions shown in panels a) and b) in the same figure.  In these cases we may speak of \emph{specimen transformation} instead of \emph{specimen interaction}.

We may speak about a specimen that is expected to follow a curved path even if we do not continually observe it as its trajectory is bending. This is possible whenever the specimen is quasi-identifiable, so that it possesses an individual evolution (Definition \ref{objectevolution}). This is presupposed in any observational context where we define its wave function.

\begin{defi}[\textbf{A free specimen}]
An identifiable specimen $OS$ for which we can define a spatio-temporal wave function $a_{\mathbf{r}_{4}s}(\mathbf{r}_{4},s,\sigma)$ is free if and only if $d^{2}\langle\mathbf{r}_{4}\rangle/d\sigma^{2}=0$ in the entire interval $[0,\sigma_{\max}]$ in which the wave function is defined.
\label{freespec}
\end{defi}

\begin{defi}[\textbf{An interacting specimen}]
An identifiable specimen $OS$ for which we can define a spatio-temporal wave function $a_{\mathbf{r}_{4}s}(\mathbf{r}_{4},s,\sigma)$ is interacting if and only if $d^{2}\langle\mathbf{r}_{4}\rangle/d\sigma^{2}\neq 0$ for some $\sigma\in[0,\sigma_{\max}]$.
\label{interactingspec}
\end{defi}

\begin{defi}[\textbf{A knowably interacting specimen}]
A specimen $OS$ which is identifiable during the time period $[n,n+m]$ is knowably interacting during the same period if and only if it is concluded at an observation at time $n+m$ that it has divided at some time $n\leq n'\leq n+m$ in such a way that the identity of $OS$ could be upheld in the division process, like in Fig. \ref{Fig42}(c).
\label{knowintobject}
\end{defi}

In other words, there should be exactly one object that emerges from the division that can be identified with the original object. As an example of such a process, we may take an electron that knowably interacts electromagnetically with its environment by emitting an photon.

\begin{defi}[\textbf{A knowably transforming specimen}]
A specimen $OS$ which is identifiable during the time period $[n,n+m]$ is knowably transforming during the same period if and only if it is concluded at an observation at time $n+m$ that it has divided at some time  $n\leq n'\leq n+m$ in such a way that several or none of the identifiable objects emerging from the division can be identified with $OS$, like in Fig. \ref{Fig42}(a) or \ref{Fig42}(b).
\label{knowtransobject}
\end{defi}

\begin{figure}[tp]
\begin{center}
\includegraphics[width=80mm,clip=true]{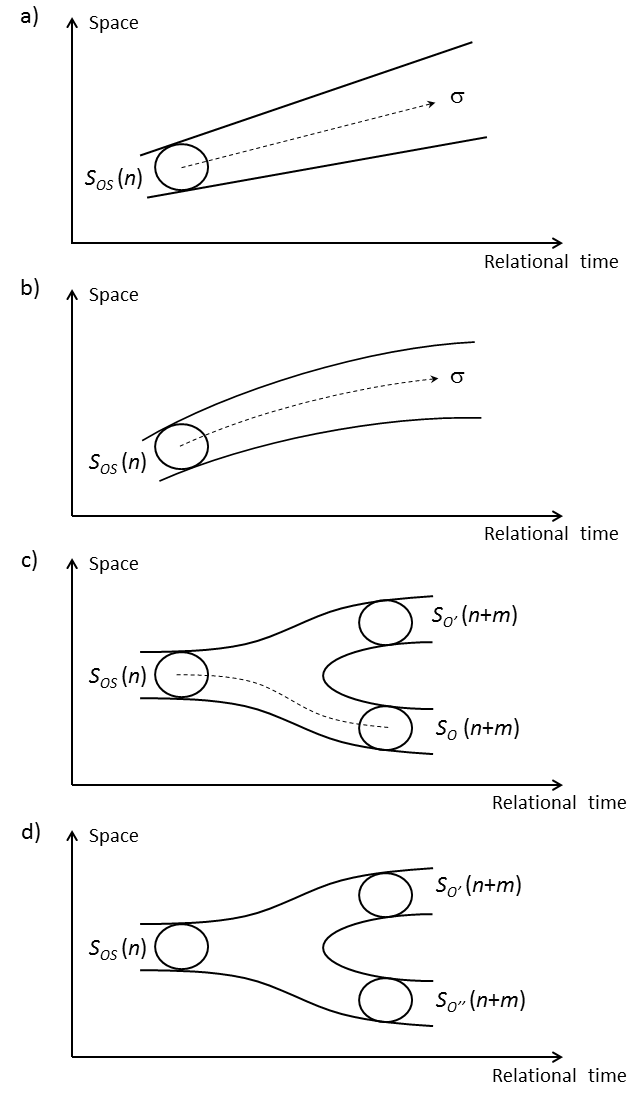}
\end{center}
\caption{Different types of evolution, shown in the projection of state space onto the spatio-temporal subspace. a) A free specimen $OS$ whose evolution is described by a wave function (Definition \ref{freespec}). b) An interacting specimen $OS$ whose evolution is described by a wave function (Definition \ref{interactingspec}). c) A knowably interacting specimen $OS$ emitting another object $O'$ (Definition \ref{knowintobject}). d) A knowably transforming specimen $OS$, dividing into two other objects $O'$ and $O''$ (Definition \ref{knowtransobject}).}
\label{Fig77}
\end{figure}

These four definitions are illustrated in Fig. \ref{Fig77}. To say that an object is knowably interacting or knowably transforming, we must actually observe it. It must be a specimen in a controlled experiment, in an observational context. We must measure a set of relational or internal attributes $P$ to determine whether it has changed internal nature or it has turned into two or several objects. A state reduction must take place. This means that we are not dealing with a context described by the simple spatio-temporal wave function $a_{\mathbf{r}_{4}s}$ discussed above, but with $a_{P}$, where the choice of $P$ determines which kinds of knowable interactions or transformations we are able to see. We may, however consider a specimen in \emph{the same} intial state $S_{OS}(n)$, but in another context, in which we just follow its spatio-temporal evolution. In that case we can describe its evolution with the ordinary evolution equation

\begin{equation}
\frac{d}{d\sigma}a_{\mathbf{r}_{4}s}(\mathbf{r}_{4},s,\sigma)=i\bar{B}_{\mathbf{r}_{4}s}a_{\mathbf{r}_{4},s}(\mathbf{r}_{4},s,\sigma).
\label{eveqgeneral}
\end{equation}

\begin{figure}[tp]
\begin{center}
\includegraphics[width=80mm,clip=true]{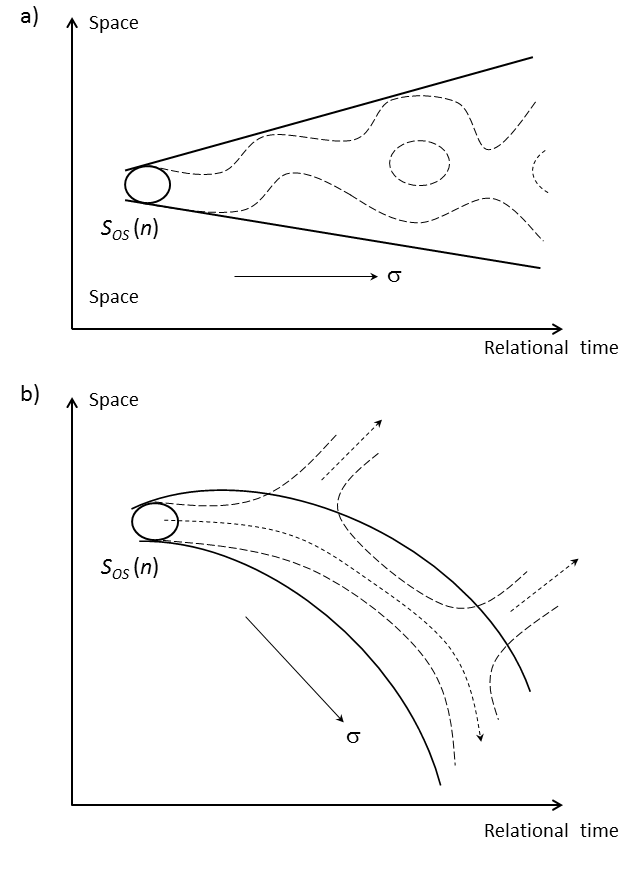}
\end{center}
\caption{a) A vast variety of object divisions and other processes may take place within a specimen $OS$ that is free according to the spatio-temporal evolution equation [\ref{eveqgeneral}]. But within the domain $[0,\sigma_{\max}]$ of the evolution parameter, we do not, by definition, observe the specimen to check what is going on inside the envelope of the wave function (Definition \ref{envelope}). b) In the same way, we cannot exclude that an interacting specimen is emitting other objects while preserving its identity. If such division processes were actually observed, the interacting specimen would be knowably interacting.}
\label{Fig78}
\end{figure}

Even if we do not explicitly observe it, we cannot \emph{exclude} that the specimen described by the above wave function transforms, divides, that the divided objects merge again, or that it knowably interacts with the environment. Therefore all such possibilities must be \emph{consistent} with the description of the evolution in terms of Eq. [\ref{eveqgeneral}]. This means that the trajectories of objects arising from $OS$ must all be contained in the enevolpe defined by the domain $D_{\mathbf{r}_{4}s}(\sigma)$ of the wave function (Fig. \ref{Fig78}).

Another way to see this is to refer to epistemic invariance (Definition \ref{epistemicinvariance}). The evolution of an object should be independent of the amount of knowledge we have about the object. More properly, the evolution of an object about which we have more knowledge should always be consistent with the evolution of the same object if we had less knowledge [Fig. \ref{Fig12}(b)]. This can be seen as an application of a generalized equivalence principle (Fig. \ref{Fig13}).

In section \ref{divideconserve} we argued that the concept of identifiability (section \ref{identifiability}) can be used to understand the discrete values of the internal attributes of minimal objects, and also the conservation laws of these values in processes where such objects divide or merge. Simply put, the sum of the values of a given attribute of all objects in the initial state should be the same as the corresponding sum in the final state. We may visualize this rule as in Fig. \ref{Fig47}, where an `attribute value substance' flows freely in the tubes that correspond to the objects that are present. This substance consists of parcels that correspond to a unit value of the attribute in question, and these parcels may choose whatever branch they like when an object divides.

This picture conforms with the idea that the evolution of the specimen $OS$ in a situation where we are ignorant of processes taking place in its interior is consistent with the evolution if we do have such knowledge. In fact, the conservation laws and the picture of a discrete attribute substance is necessary to make the idea work. This is so because consistent evolution at the two levels of knowledge relies on the fact that regardless how many objects are present within the specimen at a given time, the total value of the attribute contained in the envelope of $a_{\mathbf{r}_{4}s}(\mathbf{r}_{4},s,\sigma)$ is always the same.

\begin{defi}[\textbf{Envelope of the wave function}]
Suppose that the wave function $a_{\mathbf{r}_{4}s}(\mathbf{r}_{4},s,\sigma)$ is defined for $\sigma=[0,\sigma_{\max}]$. Then the envelope $UD$ of $a_{\mathbf{r}_{4}s}(\mathbf{r}_{4},s,\sigma)$ is $UD=\bigcup_{\sigma\in[0,\sigma_{\max}]}D_{\mathbf{r}_{4}}(\sigma)$, where $D_{\mathbf{r}_{4}}(\sigma)$ is the support of the wave function at the value $\sigma$ of the evolution parameter.
\label{envelope}
\end{defi}

The fact that no parcels of unit internal attribute value are allowed to leak across the boundary of the envelope does not exclude the possibility that objects may pass this border [Fig. \ref{Fig78}(b)]. However, such leaking objects must be the result of object divisions that corresponds to a knowably interacting object that preserves its identity throghout the series of divisions. Figuratively speaking, all parcels like those in Fig. \ref{Fig47} choose to follow the object that stays within the wave function envelope.

Leaking objects are therefore `empty', in a sense. Not necessarily in terms of relational attribute values such as momentum, but in terms of values of the internal attributes of the original specimen with state $S_{OS}(n)$, such as electric charge of baryon number. Actually, this is an oversimplification; the matter will be discussed further in section \ref{fermbos}. In that section we argue that such leaking objects are not potentially perceivable `real' objects, but are examples of `pseudoobjects'. We will try to identify massless elementary bosons with such pseudoobjects.

In Fig. \ref{Fig78}(b) we may, for example, interpret the identifiable specimen to be an electron, and the leaking pseudoobjects to be photons. At the level of less detailed knowledge, we may say that we have a single interacting electron that follows a curved trajectory given by some evolution equation for the wave function. At the level of more detailed knowledge, we may say that we have a knowably interacting electron, which interacts with the environment by emitting photons. At the latter level, we make a quantum field theory-like description of the evolution. At the former level, we make a quantum mechanical description of the same evolution, where interactions that bend trajectories must be described by classical fields. Following the ideas of Bohr, we may say that the two descriptions are complementary, suitable for different kinds of experminetal arrangements. The description in terms of classical fields is proper in contexts $C$ where we observe spatio-temporal positions, whereas the description in terms of quantum fields is proper in contexts $C'$ where we observe the number of objects and their internal attributes.

The main point is that these two descriptions should be mutually consistent. The aim of the present and the preceding section is to analyse evolution equations for spatio-temporal wave functions. This means that we focus here on the smooth quantum-mechanical evolution parametrized by $\sigma$. Then, by definition, no observations of what goes on inside the specimen are made during the time span $\Delta t$ that corresponds to the domain $[0,\sigma_{\max}]$ of the evolution parameter.

\vspace{5mm}
\begin{center}
$\maltese$
\end{center}
\paragraph{}

We argued in section \ref{minimalism} that there is no such thing as absolute acceleration. It is not possible to decide whether a given object is accelerating by measuring spatio-temporal attributes alone. Implicit epistemological minimalism (Assumption \ref{implicitepmin}) then implies that the form om physical law should not depend crucially on the existence of acceleration. In other words, if a specimen follows an accelerated or bended trajectory due to interaction with other objects, there should be an equivalent description in which the specimen follows a straight line, but in which the interaction looks different. In a given mathematical representation $\bar{S}_{C}$ of the contextual state, this equivalent description is reached via a change of variables that include the spatio-temporal coordinates. In the transformed representation $\bar{S}_{C}'$ we may, by definition, say that the net interaction is zero. The specimen appears to be free.

Conversely, if we would have a free specimen in the strong sense that we knew for sure that it did not interact at all with its environment, then a similar transformation of the variables that describe $\bar{S}_{C}$ would make the trajectory look accelerated or curved. This would mean, again by definition, that there is some interaction after all. We conclude that a specimen can never be truly free in the sense that it follows a straight line in all descriptions of its evolution allowed by physical law. We return to these matters in section \ref{gaugeprinciple}.

The same insights can be reached using the concept of epistemic invariance (Assumption \ref{epistemicinvariance}) rather than epistemic minimalism. Let us draw a hypothetical circle around the specimen, just like we did around Einstein's elevator in Fig. \ref{Fig13}. Even if we know about nothing else than the specimen inside the circle, having no idea whether there are any outside objects that are interacting with the specimen, its evolution must be consistent with the evolution of the specimen when we do know which outside objects there are, and how they interact with it. In the lack of such knowledge, there is no reason to assign an accelerated or curved trajectory to the specimen. Thus, again, we see that curved and straight trajectories are equally good in the description of the evolution of any specimen.

These considerations are local, meaning that it is not possible to measure three consecutive spatio-temporal positions $(\mathbf{r}_{1},ict_{1})$, $(\mathbf{r}_{2},ict_{2})$, and $(\mathbf{r}_{3},ict_{3})$ such that $t_{3}-t_{1}$ is a small number, and decide in an absolute sense that the trajectory $\mathbf{r}_{1}\rightarrow \mathbf{r}_{2}\rightarrow \mathbf{r}_{3}$ is straight or curved. That $t_{3}-t_{1}$ should be `small' just means that it should be small compared to the perceived curvature of the trajectory, so that we can exclude that it closes upon itself during the time period $[t_{1},t_{3}]$.

From a global perspective, the specimen can follow either a closed trajectory or not. These cases are not topologically equivalent, and it is only open trajectories that can be legitimately transformed into straight lines. Let us try to make these statements more precise.

\begin{defi}[\textbf{A bound state of a specimen}]
A specimen $OS$ is in a bound state if and only if 1) the domain of $\sigma$ can be extended to $D_{\sigma}=[0,\infty]$ in a natural parametrization, 2) $OS$ can be knowably divided into at least two objects $OS_{1}$ and $OS_{2}$, which are identifiable for all $\sigma\in D_{\sigma}$, 3) there is a finite upper bound $R_{\max}$ of the spatial distance $r_{12}$ between $OS_{1}$ and $OS_{2}$.
\label{boundspecimen}
\end{defi}

The existence of a bound $R_{\max}$ should be known at the time $n$ at which the observational context is initiated (corresponding to $\sigma=0$). The distance $r_{12}$ is defined as follows. Consider a context in which the spatio-temporal wave function $a_{\mathbf{r}_{4}s}(\sigma)$ is defined. Let $\mathbf{r}_{1}\in D_{\mathbf{r}1}(\sigma)$, where $D_{\mathbf{r}1}(\sigma)$ is the projection of the domain of $a_{\mathbf{r}_{4}s}(\sigma)$ to the spatial position of $OS_{1}$, and let $\mathbf{r}_{2}\in D_{\mathbf{r}2}(\sigma)$. Then $r_{12}=|\mathbf{r}_{2}-\mathbf{r}_{1}|$. Condition 3) means that for each $\sigma\in D_{\sigma}$ we have $r_{12}<R_{\max}$ for each possible pair of points $(\mathbf{r}_{1},\mathbf{r}_{2})$. To be physically meaningful, we note that the fulfilment of condition 3) should be invariant under a general coordinate transformation. That is, if we let $\mathbf{r}'=\mathbf{f}(\mathbf{r})$ where $\mathbf{f}$ is a diffeomorphism, then there is a finite $R_{\max}'$ such that $r_{12}'<R_{\max}'$ whenever $r_{12}<R_{\max}$.

\begin{defi}[\textbf{An unbound specimen}]
A specimen $OS$ that cannot knowably be identified with a component specimen $OS_{1}$ or $OS_{2}$ in a bound state acccording to Definition \ref{boundspecimen}.
\label{unboundspecimen}
\end{defi}

\begin{figure}[tp]
\begin{center}
\includegraphics[width=80mm,clip=true]{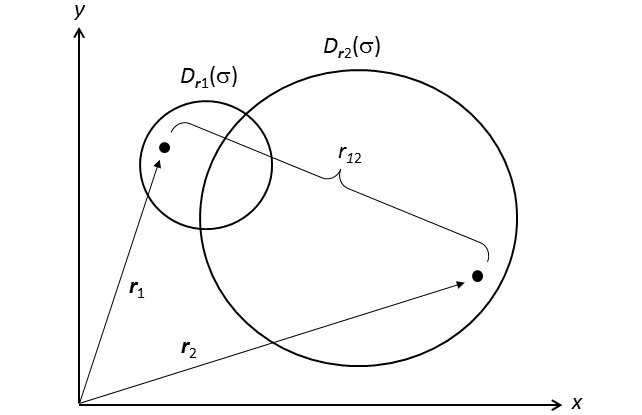}
\end{center}
\caption{Illustration of the concept of a bound specimen (Definition \ref{boundspecimen}). The wave function domains $D_{\mathbf{r}1}$ or $D_{\mathbf{r}2}$ correspond to the specimen states $S_{OS1}$ and $S_{OS2}$ according to Fig. \ref{Fig75}. An alternative choice of coordinates is to let the origin reside in one of the domains for all $\sigma\in[0,\infty]$. The most natural choice is then to place the origin within the smallest domain (in this case $D_{\mathbf{r}1}$). In atomic physics, this amounts to a coordinate system in which the origin is placed within the nucleus ($OS_{1}$) rather than anywhere in the cloud of electrons ($OS_{2}$).}
\label{Fig79}
\end{figure}

This definition is illustrated in Fig. \ref{Fig79}. In the following, we will analyze the evolution equation of an unbound, intercating specimen. We return to specimens in bound states in section \ref{reciprocaleq}.

We can consider the description of the free specimen in the preceding section to be a description of an interacting, unbound specimen for which the choice of variables is such that it appears to follow a straight trajcetory $\mathbf{r}_{4}$ in space-time. To make the form of the general evolution equation $\ref{eveqgeneral}$ invariant to a change of variables that transforms a straight trajectory into a bended trajectory and vice versa, we may write

\begin{equation}
\bar{B}_{\mathbf{r}_{4}s}=-b\Box+\bar{B}_{\mathbf{r}_{4}s}^{(int)}.
\label{generalb}
\end{equation}

We may want to express the interaction term $\bar{B}_{\mathbf{r}_{4}s}^{(int)}$ in terms of an interaction strength $\beta$ according to the definition given in the beginning of section \ref{eveq}:

\begin{equation}
d\langle\mathbf{r}\rangle/d\sigma = \mathbf{v}_{0}+\beta\mathbf{f}(\sigma),
\end{equation}
where $\mathbf{v}_{0}$ is a constant vector. Then we may write

\begin{equation}
\bar{B}_{\mathbf{r}_{4}s}=-b\Box+\beta\bar{B}_{\mathbf{r}_{4}s}^{(1)}
\label{firstorderb}
\end{equation}
for small $\beta$ in a first order expansion. 

The form [\ref{generalb}] of $\bar{B}_{\mathbf{r}_{4}s}$ follows from the discussion in the preceding section, where we argued that we must have $\bar{B}_{\mathbf{r}_{4}s}=-b\Box$ for $\beta=0$.

The operator $\bar{B}_{\mathbf{r}_{4}s}$ in Eq. [\ref{eveqgeneral}] that specifies the evolution cannot depend or act on $\sigma$ explicity, since the evolution of a physical state depends on this state only, that is, on the knowledge about its attributes. The evolution parameter $\sigma$ is no attribute, and is introduced just as a mathematical device to interpolate continuously between the evolution operators $u_{1}$, $u_{2}$, $u_{3}$, and so on, according to Eq. \ref{sigmadef}. Therefore $\bar{B}_{\mathbf{r}_{4}s}$ always has eigenfunctions

\begin{equation}
a_{\mathbf{r}_{4}s}^{(\tilde{\sigma})}(\mathbf{r}_{4},s,\sigma)=\alpha_{\mathbf{r}_{4}s}^{(\tilde{\sigma})}(\mathbf{r}_{4},s)e^{i\tilde{\sigma}\sigma}
\label{generaleigenf}
\end{equation}
such that

\begin{equation}
\bar{B}_{\mathbf{r}_{4}s}\alpha_{\mathbf{r}_{4}s}^{(\tilde{\sigma})}(\mathbf{r}_{4},s)e^{i\tilde{\sigma}\sigma}=
e^{i\tilde{\sigma}\sigma}\bar{B}_{\mathbf{r}_{4}s}\alpha_{\mathbf{r}_{4}s}^{(\tilde{\sigma})}(\mathbf{r}_{4},s)
\end{equation}
and

\begin{equation}
\bar{B}_{\mathbf{r}_{4}s}\alpha_{\mathbf{r}_{4}s}^{(\tilde{\sigma})}(\mathbf{r}_{4},s)=\tilde{\sigma}\alpha_{\mathbf{r}_{4}s}^{(\tilde{\sigma})}(\mathbf{r}_{4},s)
\label{generaleigenv}
\end{equation}
according to Eq. [\ref{eveqgeneral}]. Note that $a_{\mathbf{r}_{4}s}^{(\tilde{\sigma})}(\mathbf{r}_{4},s,\sigma)$ in Eq. [\ref{generaleigenf}] is an eigenfunction to $\bar{B}_{\mathbf{r}_{4}s}$ for any choice of function $\alpha_{\mathbf{r}_{4}s}^{(\tilde{\sigma})}(\mathbf{r}_{4},s)$ that makes the wave function normalized according to Statement \ref{normalized}. (In typical situations we also require that it has finite support $D_{\mathbf{r}_{4}}$, according to the discussion in relation to Fig. \ref{Fig75}.)

The general solution to Eq. [\ref{eveqgeneral}] can be written

\begin{equation}
a_{\mathbf{r}_{4}s}(\mathbf{r}_{4},s,\sigma)=\int_{-\infty}^{\infty}\alpha_{\mathbf{r}_{4}s}^{(\tilde{\sigma})}(\mathbf{r}_{4},s)e^{i\tilde{\sigma}\sigma}d\tilde{\sigma}.
\end{equation}

In the preceding section we identified $\bar{B}_{\mathbf{r}_{4}s}$ with a wave function operator that corresponds to a property, namely rest mass squared $m_{0}^{2}$. Can this identification be maintained in the general situation in which interactions are inherent? If so, we should always have

\begin{equation}
m_{0}^{2}\in\left\{\frac{\hbar^{2}}{bc^{2}}\tilde{\sigma}\right\},
\label{invariantmass}
\end{equation}
where $\tilde{\sigma}$ is an eigenvalue to $\bar{B}_{\mathbf{r}_{4}s}$, in accordance with Definition \ref{masssquaredefi}.

It will be possible in general to identify a squared rest mass $m_{0}^{2}$ with the eigenvalue $\tilde{\sigma}$ according to Eq. \ref{invariantmass} if this eigenvalue is invariant under all allowed changes of variables that turn an expected straight trajectory into an accelerated, bending trajectory. If the set of eigenvalues $\{\tilde{\sigma}\}$ would not be invariant under those transformations, we would have to abandon this identification, since rest mass is intended to be a physical property. As such, it cannot depend on a change of variables that leaves the physics invariant. The freedom in the choice of variables is simply a redundancy in the mathematical representation of the physical state. Which are the allowed variable transformations for which we should check eigenvalue invariance? First, we require invariance under any diffeomorphism

\begin{equation}
\mathbf{r}_{4}\rightarrow\mathbf{r}_{4}'=\mathbf{f}(\mathbf{r}_{4}).
\label{gravitytransformation}
\end{equation}
Second, we require eigenvalue invariance under any gauge transformation of a wave function $a\equiv a_{\mathbf{r}_{4}s}^{(\tilde{\sigma})}$ that is an eigenfunction associated to this eigenvalue (we drop the sub- and superscripts on the wave function for notational simplicity):

\begin{equation}
a(\mathbf{r}_{4},s,\sigma)\rightarrow a'(\mathbf{r}_{4},s,\sigma)=a(\mathbf{r}_{4},s,\sigma)e^{ig(\mathbf{r}_{4})}.
\label{electrictransformation}
\end{equation}
(We will discuss the gauge principle from the epistemic perspective in section \ref{gaugeprinciple}). Taking these two requirements together, we should check for invariance of $\{\tilde{\sigma}\}$ under any transformation

\begin{equation}
a(\mathbf{r}_{4},s,\sigma)\rightarrow a'(\mathbf{r}_{4}',s,\sigma)=a(\mathbf{r}_{4}',s,\sigma)e^{ig(\mathbf{r}_{4}')}.
\label{combinedtransformation}
\end{equation}

In short, the requirement is that the following implication holds true. 

\begin{equation}\begin{array}{rcl}
\bar{B}_{\mathbf{r}_{4}s}a(\mathbf{r}_{4},s,\sigma) & = & \tilde{\sigma}a(\mathbf{r}_{4},s,\sigma)\\
& \Downarrow & \\
\bar{B}_{\mathbf{r}_{4}'s}a'(\mathbf{r}_{4}',s,\sigma) & = & \tilde{\sigma}a'(\mathbf{r}_{4}',s,\sigma).
\end{array}
\end{equation}
But this is clearly so since Eq. [\ref{generaleigenf}] means that we can write

\begin{equation}
a(\mathbf{r}_{4},s,\sigma)=\alpha(\mathbf{r}_{4},s)e^{i\tilde{\sigma}\sigma}
\end{equation}
where $\alpha(\mathbf{r}_{4},s)$ is an arbitrary function that normalizes the eigenfunction $a(\mathbf{r}_{4},s,\sigma)$. That is, any transformation $\alpha(\mathbf{r}_{4},s)\rightarrow \alpha'(\mathbf{r}_{4}',s)$ leaves the eigenvalue $\tilde{\sigma}$ invariant. The check for invariance of $\tilde{\sigma}$ is therefore successful, since we have

\begin{equation}
a'(\mathbf{r}_{4}',s,\sigma)=\alpha'(\mathbf{r}_{4}',s)e^{i\tilde{\sigma}\sigma}
\label{invariantrestmass}
\end{equation}
for any transformation [\ref{combinedtransformation}].

This means that any unbound specimen, interacting or not, possesses a property with a value $m_{0}^{2}$ that is found in the set specified in Eq. [\ref{invariantmass}]. This set was shown in section \ref{eveq} to contain only non-negative numbers in the case of a non-interacting specimen (given the sign convention in Eq. [\ref{signconvention}] for the unphysical parameter $b$). Just as for non-interacting specimens this implies that we can write

\begin{equation}
\bar{B}_{\mathbf{r}_{4}s}=-\bar{M}_{\mathbf{r}_{4}s}\bar{M}_{\mathbf{r}_{4}s},
\label{squareroot3}
\end{equation}
in the general case also. This in turn means that the Dirac equation (Statement \ref{diracconstraint}) must hold for unbound, interacting specimens as well as for non-interacting specimens. The only change that we can do in the Dirac equation to make it generally applicable is to generalize the definition of four-momentum. In so doing, we can always express it in the following form.

\begin{defi}[\textbf{General four-momentum} $\mathbf{p}_{4}$]

A wave function operator that corresponds to $\mathbf{p}_{4}$ can be written
\begin{equation}
(\overline{\mathbf{p}_{4}})_{\mathbf{r}_{4}}=-i\hbar
\left(
\frac{\partial}{\partial r_{1}},\frac{\partial}{\partial r_{2}},\frac{\partial}{\partial r_{3}},\frac{\partial}{\partial r_{4}}
\right)+(\overline{\mathbf{p}_{4}})_{\mathbf{r}_{4}}^{(int)}
\end{equation}
The possible values of $\mathbf{p}_{4}$ is the set of eigenvalues to this operator.
\label{fourmomentumdefi2}
\end{defi}

We may use Eq. [\ref{generalb}] and Definition \ref{fourmomentumdefi2} to express

\begin{equation}
\bar{B}_{\mathbf{r}_{4}s}^{(int)}=\frac{b}{\hbar^{2}}\left(\{(\overline{\mathbf{p}_{4}})_{\mathbf{r}_{4}}^{(0)},(\overline{\mathbf{p}_{4}})_{\mathbf{r}_{4}}^{(int)}\}+(\overline{\mathbf{p}_{4}})_{\mathbf{r}_{4}}^{(int)}(\overline{\mathbf{p}_{4}})_{\mathbf{r}_{4}}^{(int)}\right),
\end{equation}
where $(\overline{\mathbf{p}_{4}})_{\mathbf{r}_{4}}^{(0)}=-i\hbar(\partial/\partial r_{1},\ldots,\partial/\partial r_{4})$, and we have introduced the anticommutator $\{(\overline{\mathbf{p}_{4}})_{\mathbf{r}_{4}}^{(0)},(\overline{\mathbf{p}_{4}})_{\mathbf{r}_{4}}^{(int)}\}=(\overline{\mathbf{p}_{4}})_{\mathbf{r}_{4}}^{(0)}(\overline{\mathbf{p}_{4}})_{\mathbf{r}_{4}}^{(int)}+(\overline{\mathbf{p}_{4}})_{\mathbf{r}_{4}}^{(int)}(\overline{\mathbf{p}_{4}})_{\mathbf{r}_{4}}^{(0)}$. 

\begin{defi}[\textbf{General momentum and energy}]
Momentum $\mathbf{p}$ is the vector of the first three components of $\mathbf{p}_{4}$, and energy $E$ is proportional to the fourth component, so that $\mathbf{p}_{4}=(\mathbf{p},iE/c)$.
\label{energymomentumdefi}
\end{defi}

Since the Dirac equation holds in the general case, so does the evolution equation (Statement \ref{psieveq}). If we use Definition \ref{energymomentumdefi} this in turn means that the Einstein energy-mass-momentum relation [\ref{einsteinrelation}] also holds generally for an interacting, but unbound specimen.

The transformations [\ref{gravitytransformation}] and [\ref{electrictransformation}] may be interpreted as gauge transformations associated with the gravitational and electro-magnetic fields, respectively (Section \ref{gaugeprinciple}). These transformations are specified by functions of $\mathbf{r}_{4}$, and arbitrary large distances $|\mathbf{r}|$ may be affected, distorting faraway trajectories and probabilities of alternatives. This fact conforms with the usual notion that the gravitational and electro-magnetic fields have infinite range. If the specimen is knowably interacting, we should use the quantum description of these fields in terms of emitted gravitons and photons, respectively (Fig. \ref{Fig78}). On the other hand, if they are `just' interacting, the classical descriptions in terms of general relativity and Maxwell's equations are appropriate.

What about the weak and strong forces? They are associated with an investigation of the internal attributes of the specimen, not the relational attribute $\mathbf{r}_{4}$ that is used to describe its trajectory. They determine what we will see if the specimen divides, or if we determine the value of an internal attribute precisely, given a fuzziness of our initial knowledge. This means that these forces may be associated with a knowably transforming specimen, rather than a (knowably) interacting specimen (Fig. \ref{Fig77}). We will discuss this interpretation briefly in the following section. Note that there can be no smooth classical description of the processes that lead to a knowable object transformation. Objects cannot be `just' transforming in the same way as they can be `just' interacting (Definitions \ref{interactingspec} to \ref{knowtransobject}). Therefore there is no alternative description of the weak and strong forces in terms of classical fields, like Einstein's or Maxwell's.

We argued above that the descriptions where a specimen is interacting and knowably interacting must be equivalent. The difference in the amount of knowledge about the system under study should not affect its evolution, according to the principle of epistemic invariance. A somewhat stronger statement is that a specimen that is knowably interacting must be considered free at those times it does not knowably interact. That is, between the emissions of `transfer quanta' according to Fig. \ref{Fig78}(b), the object is always expected to follow a straight line.  Put differently, all interactions (bending or accelerated trajectories in the given reference frame) must have an equivalent description in terms of knowable emissions of transfer quanta.

Interactions between two objects must encode information that makes it possible to identify which object interact with which; interactions between objects must be identifiable to some extent just like the objects themselves must be identifiable. If we are dealing with interacting minimal objects having discrete internal attributes, this implies that the information encoded in the interaction must be discrete. We identify the emitted `objects' in Fig. \ref{Fig78}(b) with such `transfer quanta'. If no such quanta are emitted, no interaction between the object and its environment takes place - it is free.

This conclusion makes it necessary to regard the emission of such quanta to depend on the choice of coordinates. This fact, togheter with other observations, make it clear that these quanta cannot be considered `real' objects; they can be neither directly perceived objects, nor deduced quasiobjects (Section \ref{fermbos}).

\begin{state}[\textbf{Objects are free between knowable interactions}]
Suppose that we, in a given coordinate system, choose to describe the evolution of an object in terms of knowable interactions and transformations only. Then the object must be considered free during any time interval $[n,n+m]$ in which no such interaction or transformation knowably occurs.
\label{freeinbetween}
\end{state}

\begin{figure}[tp]
\begin{center}
\includegraphics[width=80mm,clip=true]{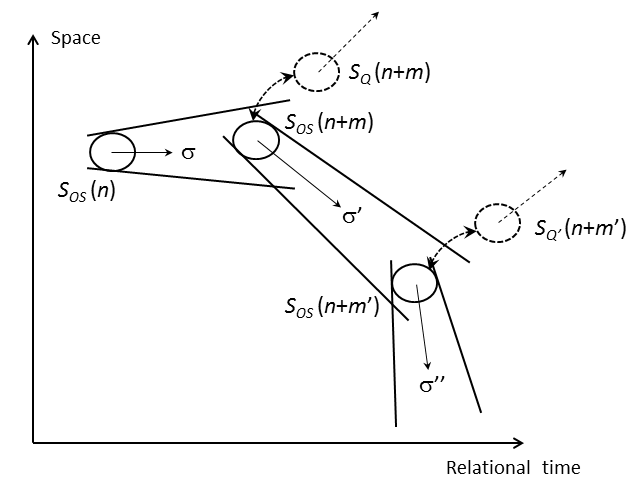}
\end{center}
\caption{The evolution of a knowably interacting specimen $OS$ as described according to Statement \ref{freeinbetween}. The evolution in between and after the observed emissions of transfer quanta $Q$ and $Q'$ can be described in a family of contexts $C(\sigma,\sigma',\sigma'')$ in terms of three consecutive wave functions $a_{\mathbf{r}_{4}s}(\mathbf{r}_{4},s,\sigma)$, $a_{\mathbf{r}_{4}s}'(\mathbf{r}_{4},s,\sigma')$, and $a_{\mathbf{r}_{4}s}''(\mathbf{r}_{4},s,\sigma'')$ [Fig. \ref{Fig73a}(a)]. Each of these describes a free specimen. The arrows in the middle of the three envelopes (Definition \ref{envelope}) represent the evolution of the expected spatio-temporal position $\langle\mathbf{r}_{4}\rangle$. It will follow a straight lines as the evolution parameters increase. The double-headed arrows that connect the specimen $OS$ with the transfer quanta $Q$ and $Q'$ symbolize the functional relationship between the attributes of the pairs of entities that makes it possible to say that the quanta are associated with the specimen [Fig. \ref{Fig42}(c)].}
\label{Fig80}
\end{figure}

This means that during such a time interval $[n,n+m]$ we can model the evolution of the object by treating it as a specimen in a family of contexts $C(\sigma)$, and describe its evolution between consecutive measurements of its state in terms of a free wave function $a_{\mathbf{r}_{4}s}(\mathbf{r}_{4},s,\sigma)$, as specified in section \ref{eveq}. In particular, the expected trajectory should be straight: $d\langle\mathbf{r}_{4}\rangle/d\sigma=\mathbf{v}_{0}$ for a constant vector $\mathbf{v}_{0}$ (Fig. \ref{Fig80}).

\vspace{5mm}
\begin{center}
$\maltese$
\end{center}
\paragraph{}

We have used the concepts of straight and curved trajectories as a point of departure in the foregoing discussion. We have allowed for the fact that there are equivalent descriptions in which a straight trajectories becomes bent, and vice versa, but we have nevertheless presupposed the subjective ability to distinguish \emph{straight} from \emph{bent} trajectories, in accordance with Definitions \ref{straightness} and \ref{straj}. From a strict epistemic perspective we should ask what aspect of physical law makes this subjective distinction possible. Only then do we achieve epistemic completeness (Assumption \ref{epcompleteness}).

We concluded in section \ref{minimalism} that we cannot use spatio-temporal attibutes alone to distinguish accelerated motion from motion with constant velocity. The relational nature of these attributes mean that if an object appears to accelerate, it may equally well be the objects of reference that accelerate. In the same way, we cannot decide in an absolute sense whether an elongated object like a stick is straight or bent. To answer the question we have to refer to a coordinate system, a grid of objects of reference. To decide whether the objects in this grid are equally spaced, we need another object, a measuring stick. But to decide whether this measuring stick keeps it length and shape as we check the spacings in the grid in consecutive order, we need still another coordinate system.

To escape this infinite regress, we have resorted to the \emph{feeling} of acceleration as the judge. However, since we can never decide in an absolute sense whether an object accelerates by means of the relational spatio-temporal attributes at hand, we also concluded in section \ref{minimalism} that there has to be another interpretation of this feeling, namely gravity.

But gravity in itself cannot account for this feeling. There has to be a force that opposes gravity. Otherwise we would always be falling freely. We would fall through the earth. The observer in Einstein's elevator would fall through the floor of the elevator without ever feeling the pressure against it. The question whether this feeling is due to gravity or acceleration would never arise. Since we would never be able to feel gravity, the concept would lose its epistemic meaning. The necessary opposing force is provided by electromagnetism.

\begin{state}[\textbf{The ability to feel acceleration implies gravity and electromagnetism}]
The existence of the feeling of acceleration and the requirement of epistemic closure (Assumption \ref{closure}) make gravity and electromagnetism necessary.
\label{accelectro}
\end{state}

The equivalence principle implies that electromagnetism must oppose any acceleration in exactly the same way as it opposes acceleration due to gravity. When we go around in a roundabout, the fluid in the semicircular canals in the ears is set in motion, which makes us feel dizzy. But in order for the nerve cells to detect this motion, the fluid has to be contained by the walls in these canals, exerting pressure on the hair cells from which the nerves emanate. Both the containment of the fluid, the pressure on the hair cells that set them in motion and the transmission of nerve signals is due to electromagnetism. That is, electromagnetism is necessary to account for the feeling of acceleration in this case also.

A world without the feeling of acceleration, without electromagnetism, would be truly bizarre in terms of spatio-temporal perceptions, almost impossible to imagine. There would simply be no subjective distinction between inertial and accelerated frames. We would not be able to speak about forces or interactions.

In order to define acceleration even in a \emph{relative} sense, we need knowably uniform yardsticks, and clocks that tick at knowably equal intervals. Without such devices we cannot tell whether two trains accelerate away from each other or move with steady relative speed. But we cannot define such yardsticks and clocks with the help of the perception of spatio-temporal attributes alone, as discussed above. We end up in infinite regress. Another kind of perception is needed to get out of it.

Another way to put it is to say that it is only possible to define proper yardsticks and clocks if the equality of the distance between the tick marks is invariant under all coordinate transformations of a certain type (linear), but is lost in another type of transformation (non-linear). Without the dubjective distinction between inertial and accelerated frames we could never in practice distinguish linear from non-linear transformations. Non-linear transformations that distort the yardsticks could be performed without anyone being able to tell that such a change of coordinates had actually taken place. The concept of acceleration would lose its meaning altogether.

\begin{state}[\textbf{Without electromagnetism, no metric space-time}]
Electromagnetism is necessary in order to measure distances, and thus to distinguish straight and curved trajectories subjectively.
\label{necessaryelectro}
\end{state}

Note, however, that we would still be able to speak about a space-time with ordered elements. By definition, all attribute values are possible to order sequentially, and we assume that spatial and temporal relations are attributes. We have discussed sequential time at length. We assume the ability to decide that event $B$ occurs after event $A$ but before event $C$. Likewise, we can always decide whether a point $B$ on a rope is placed between points $A$ and $C$ or not. These matters were discussed in Section \ref{statespaces}.

\section{Consequences of the evolution equation}
\label{evconsequences}

Statement \ref{freeinbetween} implies that the treatment of a free specimen in section \ref{eveq} is useful even if the aim is to describe interacting or transforming objects. More precisely, we can hope to be able to use the free wave function $a_{\mathbf{r}_{4}s}(\mathbf{r},s,\sigma)$ to calculate probabilities to see different kinds of interactions or transformations. We seek these probabilities as functions of the relational time $t$ passed since the last time $n$ at which we observed the state of the object. As discussed above, we do this by introducing a family of contexts $C(\sigma)$, using $\sigma$ as a proxy for the passage of relational time $t$. Since we argue in this way that the evolution equation for the free specimen is generally useful, we devote the present section to discuss some of its consequences.

The squared rest mass is given by $m_{0}^{2}=\frac{\hbar^{2}}{bc^{2}}\tilde{\sigma}$, according to Definition \ref{masssquaredefi}. This means that the possible values of $m_{0}^{2}$ are proportional to the eigenvalues $\tilde{\sigma}$ of the evolution operator $\bar{B}_{\mathbf{r}_{4}s}$, according to Eq. [\ref{generaleigenv}]. A specimen with a perfectly known rest mass is therefore described by the corresponding eigenfunction $a_{\mathbf{r}_{4}s}^{(\tilde{\sigma})}$, which, according to Eq. [\ref{generaleigenf}], always has the form

\begin{equation}
a_{\mathbf{r}_{4}s}^{(\tilde{\sigma})}(\mathbf{r}_{4},s,\sigma)=\alpha_{\mathbf{r}_{4}s}^{(\tilde{\sigma})}(\mathbf{r}_{4},s)e^{i\tilde{\sigma}\sigma}.
\end{equation}

We see that a specimen with a rest mass that is known to be zero would be described by a wave function of the form $a_{\mathbf{r}_{4}s}^{(\tilde{\sigma})}(\mathbf{r}_{4},s,\sigma)=\alpha_{\mathbf{r}_{4}s}(\mathbf{r}_{4},s)^{(\tilde{\sigma})}$, which does not depend on the evolution parameter $\sigma$. Then it cannot depend on relational time $t$ either, since $\sigma$ is a proxy for $t$. Such a wave function therefore corresponds to a static, timeless contextual state $S_{C}$. This fact cannot depend on the choice of spatio-temporal coordinates $\mathbf{r}_{4}$, and it should therefore remain true after a Lorentz transformation. [Since we restrict our interest to coordinate systems in which the specimen is free, we do not consider more general diffeomorphisms $\mathbf{r}_{4}'=\mathbf{f}(\mathbf{r}_{4})$.] This would possible only if the spatio-temporal part of the wave function were a plane wave

\begin{equation}
a_{\mathbf{r}_{4}}^{(0)}(\mathbf{r}_{4},\sigma)\propto e^{i\tilde{\mathbf{r}}_{4}\cdot\mathbf{r}_{4}}.
\label{masslesswavefunction}
\end{equation}
However, such a wave function is not normalizable according to Definition \ref{normalized}. Therefore there cannot be any specimen whose rest mass is observed within some context and is found to be zero, before some other properties of the specimen are observed.

The fact that we can never be sure that the rest mass is zero means that we can never exclude a non-zero rest mass. Can we go a step further and exclude the zero rest mass? Let us formulate the question more precisely. Let $\Sigma_{m_{0}=0}$ denote the hypothetical set in state space of all exact states $Z$ such that there is an observational setup $O$ with a specimen $OS$ with zero rest mass (using the concepts in Fig. \ref{Fig61c}). Can we be sure that $S_{OS}\cap \Sigma_{m_{0}=0}=\varnothing$ for all specimens that we can ever investigate?

The property $m_{0}$ that we have identified in the formalism presented in the preceding sections (Definition \ref{restmassdefi}) seems to conform precisely with the traditional concept of rest mass. We have assumed that any object can be represented in terms of a set of minimal objects. If we accept that $m_{0}$ shares all the qualitites of the rest mass, we can therefore write $m_{0}[OS]\geq \sum_{l}m_{0}[(O_{M})_{l}]$, where $\{(O_{M})_{l}\}$ is any set of minimal objects whose collective state can properly represent $S_{OS}$. This means that we have $m_{0}[OS]>0$ for all specimens $OS$ if there is no minimal object $(O_{M})_{l}$ whose rest mass can possibly be zero when it is free. We have concluded that we can never exclude the possibility that $m_{0}[(O_{M})_{l}]>0$. We have to argue that the two possibilites $m_{0}[(O_{M})_{l}]>0$ and $m_{0}[(O_{M})_{l}]=0$ are mutually inconsistent for a single minimal object.

Objects with zero rest mass always travel at the speed of light, i.e. $v=c$. In contrast, we have $v<c$ for any object with non-zero rest mass. These two possibilities are qualitatively different. We argued in the preceding section that $m_{0}$ is invariant under any diffeomorphism $\mathbf{r}_{4}'=\mathbf{f}(\mathbf{r}_{4})$ (Eq. [\ref{invariantrestmass}]). In particular, it is invariant under a Lorentz transformation. Therefore there is a valid reference frame in which any minimal object with non-zero rest mass is at rest. This possibility is indeed inconsistent with a minimal object that always travel at the speed of light. Since we cannot exclude $m_{0}[(O_{M})_{l}]>0$ we must exclude $m_{0}[(O_{M})_{l}]=0$, and therefore we can be certain that $m_{0}[OS]>0$ for any specimen $OS$. 

In these considerations we can equally well talk about `objects' instead of `specimens' and conclude that there cannot be any object with zero rest mass. This small semantic leap is possible since any object whose mass we would like to determine becomes a specimen within an observational context.

\begin{state}[\textbf{No object has zero rest mass}]
No object under study can be known to have zero rest mass. Furthermore, we can always exclude the possibility that the object has zero rest mass.
\label{nozeromass}
\end{state}

It is intuitively clear why the plane massless wave like that in Eq. [\ref{masslesswavefunction}] can never describe a specimen in any observational context: it is completely delocalized in space, and also in time! Since the evolution equation can be meaningfully constructed only in contexts with a specimen that has been actually observed at some previous time $n_{0}$, such solutions cannot arise for finite $\sigma$ if we define $S_{C}(n)=S_{C}(\sigma=0)$, where $n\geq n_{0}$ is the time at which the context is initiated, as usual. The reason is simply that the observation of the object at time $n_{0}$ means that far away parts of the possible values of the relational attributes can be excluded at this point in time, and also at time $n$.

Let us complicate the picture a little bit. The spatio-temporal support $D_{\mathbf{r}_{4}}$ of the wave function is obviously finite if the specimen is directly perceived at the initial time $n$. Then it will also remain finite for all finite $\sigma$. But the specimen may also be a quasiobject. It may, for instance, be a quantum of radiation that is expected to be emitted from a radioactive sample placed in the experimental setup at time $n$. In this case we may define a spatio-temporal coordinate system such that $t=0$ at sequential time $n$, and $\mathbf{r}=0$ at some point inside the radioactive sample. The context will then be such that all negative $t$ can be excluded as possible detection times for the radiated specimen, and we also know that $|\mathbf{r}|<R$ for some finite $R$ whenever the quantum of radiation is actially detected. However, in this case the support $D_{\mathbf{r}_{4}}$ nevertheless has infinite volume, since we cannot exlude that $t\rightarrow\infty$. We must be prepared to wait an arbitrarily long time before the first quantum is emitted from the sample.

The directionality of time, as expressed in Eq. [\ref{directed}], leads to Eq. [\ref{signrule1}]. This relation implies that the energy always has the same sign if we define it to be proportional to the reciprocal temporal position $\tilde{r}_{4}$, as we have done above. With the sign convention $b<0$ (Eq. [\ref{signconvention}]) and the choice of a positive constant of proportionality in Eq. [\ref{energydef}], we have $E\geq 0$. It follows from Statement \ref{nozeromass} that $m_{0}^{2}>0$. Thus we can use the relation [\ref{einsteinrelation}] between energy mass and momentum for a free specimen to conclude that $E>0$.

\begin{state}[\textbf{The energy of a free specimen is positive}]
Energy $E$, as defined in Eq. [\ref{energydef}], is either always negative or always positive for a free specimen. With our sign conventions it becomes positive.
\label{positiveenergy}
\end{state}

To ensure in the formalism that energy $E$ always stays non-negative, we could try to follow the same path as in the case of the squared rest mass $m_{0}^{2}$ and look for an operator $\bar{X}$ such that $\bar{E}=\bar{X}\bar{X}$. The eigenvalues of $\bar{X}$ would correspond to the square root of the energy, and this would become the basic energy measure. However, energy is not relativistically invariant, and neither would be the operator $\bar{X}$. Thus the additional constraint on the wave function that would arise (analogous to Eq. [\ref{dirac1}]) would not be relativistically invariant either. Therefore it cannot be part of fundamental physical law.

The issue of negative energies must be handled in terms of antimatter, as usual. The picture that antimatter are objects travelling backwards in time is particularly suitable from the present perspective. We found that $E>0$ if and only if $d\langle t\rangle/d\sigma>0$. In the same way, we may deduce that $E<0$ if and only if $d\langle t\rangle/d\sigma<0$. The solutions to the evolution equation with $E<0$

\begin{equation}
a_{\mathbf{r}_{4}s}(\mathbf{r},t,s,\sigma)^{-}=\int\tilde{a}_{\mathbf{r}_{4}s}(\mathbf{p},-E,s,E_{0}^{2})
e^{\frac{i}{\hbar}(\mathbf{p}\cdot\mathbf{r}+Et+\frac{E_{0}^{2}}{2\langle E\rangle}\sigma)}d\mathbf{p}dEdE_{0}^{2}.
\label{fexpansionplus}
\end{equation}
may thus be seen as specimens that travels backwards in time, but in all other respects are identical to the corresponding specimens with $E>0$ and wave function

\begin{equation}
a_{\mathbf{r}_{4}s}(\mathbf{r},t,s,\sigma)^{+}=\int\tilde{a}_{\mathbf{r}_{4}s}(\mathbf{p},E,s,E_{0}^{2})
e^{\frac{i}{\hbar}(\mathbf{p}\cdot\mathbf{r}-Et-\frac{E_{0}^{2}}{2\langle E\rangle}\sigma)}d\mathbf{p}dEdE_{0}^{2}.
\label{fexpansionminus}
\end{equation}
Let us clarify what we mean by `travelling backwards in time' in this setting. It means that as sequential time $n$ increases, the relational time $t$ decreases. The antimatter specimen $OS^{-}$ is assigned a time coordinate $t(n+m)$ when observed at time $n+m$ that is smaller than the time coordinate $t(n)$ it was assigned at the initiation of the context, at time $n$. To me, the meaning of the idea of objects travelling backwards in time becomes clearer in such a picture, in which we separate sequential and relational time. The separation makes it possible to say that the object travels backwards in time ($t$ decreases) in relation to something that travels forward by definition (sequential time $n$).

We will discuss the separation of sequential and relational time in connection with antimatter further in section \ref{antimatter}, where we suggest an epistemic interpretation of antimatter that demystifies the concept altogether. Since that interpretation is quite hard to swallow, we continue here with a more conventional narrative.

As usual, we may use the CPT-theorem to reinterpret the antimatter specimen $OS^{-}$ from being identical to the matter specimen $OS^{+}$, but travelling backwards in time, to being identical to $OS^{+}$ except from having inverted charge and parity, and travelling forwards in time. (Note that, strictly speaking, the CPT-theorem applies only in quantum field theory, which we do not consider here.)

From our epistemic perspective, the equivalence of this picture to that of the `Dirac sea' becomes apparent. This is so at least if we assume that all minimal objects with negative energy are quasiobjects, whose existence is deduced by physical law from the state of those objects that we actually perceive. The point is that we cannot exclude from the direct perception that there is a sea of such quasiobjects, given that there is one such object in each possible state. The reason is simply that the resulting infinite charge of the sea pulls any other charged object (with positive energy) equally forcefully in all directions, so that the net electromagnetic force is zero. In the same way, the rest mass of this sea of negative energy objects surrounds any other object equally in all directions so that the net gravitational pull becomes zero. Since we can never exclude neither the picture of antimatter as holes in the Dirac sea of negative energy objects, nor the picture of positive energy objects having opposite charge and parity, we should describe the total state $S$ as a union of those possibilites. That is, if we deduce the existence of antimatter at time $n$, then
\begin{equation}
S(n)=S_{hole}\cup S_{object}.
\end{equation}
However, since we can never discriminate between the two alternatives, we have no reason to uphold this superposition in our mathematical representations, but can choose whatever picture we like the most. More precisely, there never appears a set of future alternatives (Definition \ref{setfuturealt}) such that one picture or the other is the outcome of the observation.   

The introduction of the evolution parameter $\sigma$ introduces a third basic commutation relation, apart from the two familiar ones

\begin{equation}\begin{array}{rcl}
\left[x,\bar{p}_{x}\right] & = & i\hbar\\
\left[t,\bar{E}\right] & = & -i\hbar,
\end{array}\end{equation}
namely,

\begin{equation}
[\sigma,\overline{E_{0}^{2}}]=2i\hbar\langle E\rangle,
\label{newcom}
\end{equation}
in the natural parametrization, where we have defined $\overline{E_{0}^{2}}=c^{4}\overline{m_{0}^{2}}$. To each of these commutators it is possible to associate a corresponding uncertainty relation

\begin{equation}\begin{array}{rcl}
\Delta x\Delta p_{x} & \geq & \hbar/2\\
\Delta t\Delta E & \geq & \hbar/2\\
\Delta\sigma\Delta E_{0}^{2} & \geq & \hbar\langle E\rangle.
\label{uncertainrel}
\end{array}\end{equation}
The separation of relational and sequential time makes it possible to interpret the time-energy uncertainty relation in exactly the same way as the position-momentum relation. In contrast, in traditional quantum mechanics time $t$ is a parameter rather than an observable. The present description is an improvement since it respects the relativistic notion of space-time, in which the spatial and temporal coordinates are treated on equal footing and may be Lorentz-transformed into each other.

\begin{figure}[tp]
\begin{center}
\includegraphics[width=80mm,clip=true]{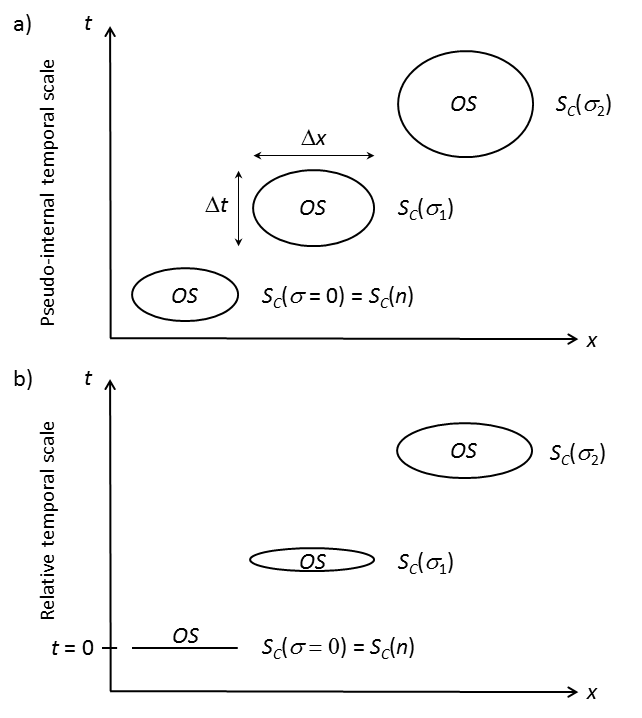}
\end{center}
\caption{In the present description, the interpretation of the temporal uncertainty relation $\Delta t\Delta E \geq \hbar/2$ is the same as that of $\Delta x\Delta p_{x} \geq \hbar/2$, respecting the symmetry between spatial an temporal coordinates inherent in relativistic space-time. a) Relational time $t$ in a coordinate system defined by objects external to the specimen $OS$. b) Relational time defined so that $t=0$ at the sequential time $n$ at which the experiment starts. In this case we must have $\Delta t\leq t$.}
\label{Fig74}
\end{figure}

Suppose that we observe a specimen $OS$ at sequential time $n$, and study its evolution as a function of $\sigma$ (Fig. \ref{Fig74}) in a family of contexts $C(\sigma)$, where time $n$ corresponds to $\sigma=0$. The evolution of $t$ may be measured in some external reference frame, in relation to some other event or object (panel a). It may also be measured in relation to the observation of $OS$ at time $n$ (panel b). In the first case, the uncertainty $\Delta t$ of $t$ is typically non-zero for all $\sigma\geq 0$. In the second case, the uncertainty grows from zero as $\sigma$ grows from zero. In the natural parametrization we get $\Delta t\leq \sigma$.

Let us look at the most common illustration of the time-energy uncertainty relation, namely the broadening of atomic spectral lines. Suppose that the electron was observed to be in the excited state at time $n$, and that the uncertainty of its energy at this time was $\Delta E$. The context is such that $\mathbf{r}_{4}=(\mathbf{r},ict)$ is observed at the moment $n+m$ defined by the deexcitation of the atom. This means that $t$ is part of the final observation. Then the uncertainty $\Delta t$ of the measured relational time $t$ that has passed before deexcitation fulfils the relation

\begin{equation}
\Delta t\geq \hbar/(2\Delta E).
\end{equation}

If we adopt the relative coordinate system in Fig. \ref{Fig74}(b), we get $\langle t\rangle \geq \Delta t$, so that

\begin{equation}
T\geq\hbar/(2\Delta E),
\end{equation}
where $T\equiv\langle t\rangle$ is the expected life-time of the excited state.

\begin{state}[\textbf{Uncertainty relations for spatio-temporal observations}]
Consider a context $C$ initiated at time $n$, and such that $\mathbf{r}_{4}=(\mathbf{r},ict)$ is observed at time $n+m$. Let $\Delta p_{x}$ and $\Delta E$ be the uncertainties of the momentum and energy of the specimen $OS$ in the evolved state $u_{m}S_{OS}(n)$. Let $\Delta x$ and $\Delta t$ be the uncertainties of the spatio-temporal distance between the specimen at times $n$ and $n+m$, as observed at time $n+m$. Then $\Delta x\geq\hbar/2\Delta p_{x}$ and $\Delta t\geq \hbar/2\Delta E$.
\label{uncertain1}
\end{state}

In traditional quantum mechanics the time-energy uncertainty relation is a bit awkward to interpret since $t$ is a parameter rather than an observable attribute. In the same way, the analogous relation $\Delta\sigma\Delta E_{0}^{2} \geq \hbar\langle E\rangle$ in the present description is a bit awkward to interpret. To give it a physical meaning, we note that it holds in the natural parametrization. In that parametrization we may set $\sigma=\langle t\rangle$. Then we get

\begin{equation}
\Delta\langle t\rangle\Delta E_{0}^{2}\geq\hbar\langle E\rangle.
\label{uncertainrel2}
\end{equation}

This inequality expresses the fact that it is not possible to know \emph{a priori} the expected relational time $t$ between the initiation of the context at time $n$ and the final observation at time $n+m$, unless the rest energy of the specimen $E_{0}^{2}$ is completely unknown. This situation does not occur in realistic contexts, and illustrates the conclusion expressed in Statement \ref{physicallaw2} that we can use $\langle t\rangle$ to parametrize the evolution only in idealized contexts.

\begin{state}[\textbf{Rest energy uncertainty}]
Consider a context $C$ initiated at time $n$, and such that $\mathbf{r}_{4}=(\mathbf{r},ict)$ is observed at time $n+m$. Let $\Delta E_{0}^{2}$ be the uncertainty of the squared rest energy of the specimen $OS$ in the evolved state $u_{m}S_{OS}(n)$. Let $\Delta \langle t\rangle$ be the uncertainty of the relational time passed between sequential times $n$ and $n+m$, and let $\langle E\rangle$ be the expected energy of the specimen. Then $\Delta\langle t\rangle\geq\hbar\langle E\rangle/\Delta E_{0}^{2}$.
\label{uncertainmass}
\end{state}

Suppose that the specimen at time $n$ is in an excited state with an expected life-time $T$. According to the above discussion we may identify $T=\langle t\rangle$, and conclude that $T$ cannot be exactly determined. We have $\langle t\rangle \geq \Delta\langle t\rangle$ and $E\geq E_{0}$, so that we may write

\begin{equation}
T\geq\hbar\langle E_{0}\rangle/\Delta E_{0}^{2}.
\label{masslifetime}
\end{equation}
  
Therefore, if a specimen is expected to undergo a distinct change from state $S_{OS}(n)$ to state $S_{OS}(n+m)\cap S_{OS}(n)=\varnothing$ within a time interval $T\approx\Delta\sigma$, then the rest energy of the original state $S_{OS}(n)$ has an uncertainty $\Delta E_{0}^{2}\geq \hbar\langle E_{0}\rangle/T$.

The uncertainty of an object's rest energy thus grows with its rest energy, and also grows if its expected lifetime becomes shorter. Conversely, if we manage to measure the rest energy of the specimen to arbitrary precision, then it cannot be expected to undergo any perceivable change in a time span $T$ that grows arbitrarily large. In other words, rest energies are only perfectly known in a static world. A static world is a world without time, since there will be no subjective change that can define the temporal update $n\rightarrow n+1$. We regain the conclusion that there can be no specimen in the real world which has a rest mass that is knowably zero (Statement \ref{nozeromass}).

\begin{state}[\textbf{The rest mass can never be exactly known}]
For all objects $O$ whose rest energy $E_{0}$ is observed in an experimental context, we have $\Delta E_{0}>0$ both before and after the observation.
\label{masswidth}
\end{state}

In the same way as in traditional quantum mechanics, we following statement follows from the fact that the evolution operator $\bar{B}_{\mathbf{r}_{4}s}$ commutes with the energy and rest energy operators.

\begin{state}[\textbf{The expected values of rest mass and energy are conserved}]
Consider any free specimen in a joint family of contexts $C\tilde{C}(\sigma)$ in which four-position $\mathbf{r}_{4}$ and four-momentum $\mathbf{p}_{4}$ are observed. Then we have $\frac{d}{d\sigma}\langle E_{0}^{2}\rangle=0$ and $\frac{d}{d\sigma}\langle E\rangle=0$
\label{conservedenergies}
\end{state}

\vspace{5mm}
\begin{center}
$\maltese$
\end{center}
\paragraph{}

The uncertainty relations we refer to above are derived in the standard manner from the properties of Fourier transforms. The position uncertainty $\Delta x$ is the standard deviation of the probability density function $|\Psi_{x}(x)|^{2}$, and the other uncertainties are defined analogously. These Fourier transforms are defined for continuous wave functions $\Psi_{P}(p,\sigma)$. In traditional quantum mechanics this is the basic layer of physical description. In contrast, in the present approach the basic layer of physical description are states $S_{O}$ of objects $O$. These states are sets in state space $\mathcal{S}$. All concepts and quantities should therefore be based on the physical state $S_{O}$, rather than on a particular mathematical representation $\bar{S}_{O}$ of this state.

We have argued that the representation of the contextual state $S_{C}$ of the specimen $OS$ in terms of a continuous wave function $\Psi_{P}(p,\sigma)$ with unbounded support $D_{p}$ is always an approximation (Definition \ref{cwavedef}). This is so for two reasons: first, the number of values $p_{j}$ that can be observed in any detector $OD$ is always finite. Second, there is always a maximum value $p_{\max}$ and a minimum value $p_{\min}$ that can be observed in any given context $C$. The basic problem is that observational contexts for relational attributes are never fundamental (Definition \ref{fundamentalcontext}).

\begin{figure}[tp]
\begin{center}
\includegraphics[width=80mm,clip=true]{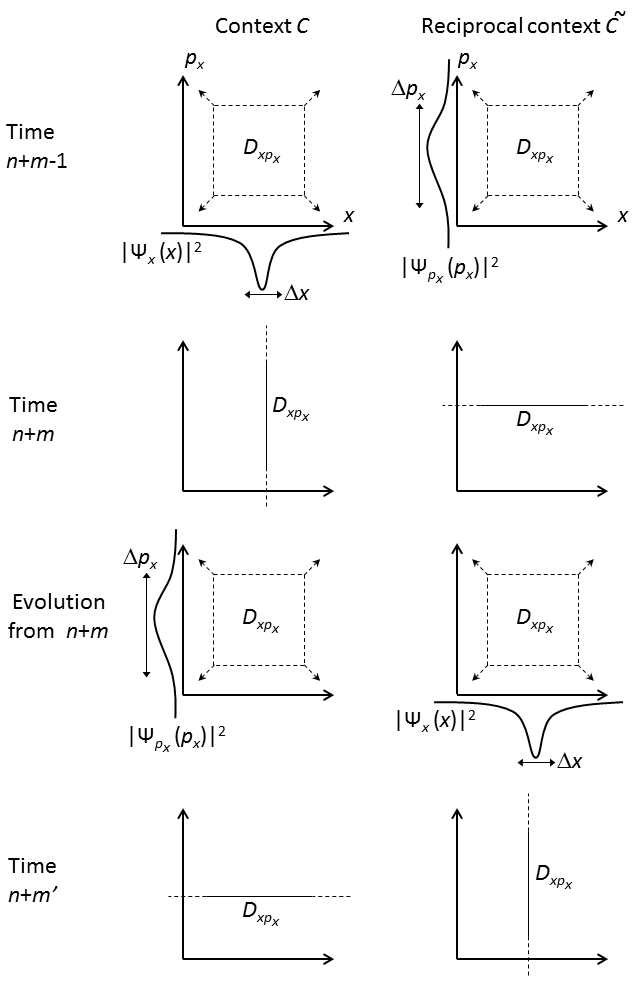}
\end{center}
\caption{An idealized context $C$ in which position $x$ and momentum $p_{x}$ are observed at times $n+m$ and $n+m'$, respectively. The observations are perfectly precise, and the detector is infinitely large, so that all values of $x$ and $p_{x}$ can be observed. The context $C$ is also neutral, so that the reciprocal $\tilde{C}$ can be defined. These conditions imply that the support $D_{xp_{x}}$ has no boundary just before an observation. No value $x$ is excluded by the potential knowledge just before we observe position, and no value $p_{x}$ is excluded afterwards. In the same way, no value $p_{x}$ is excluded before a momentum observation, and no value $x$ is excluded afterwards. It is only in this situation that the relation $\Delta x\Delta p_{x}\geq\hbar/2$ has a well-defined physical meaning. Compare the realistic context in Fig. \ref{Fig76b}.}
\label{Fig76c}
\end{figure}

For example, a position detector that were able to register positions arbitrarily far away would have to be infinitely large. The lack of registration of a position within a finite range $[x_{\min},x_{\max}]$ in a realistic, finite detector corresponds to the realization of the alternative `$x$ is larger than $x_{\max}$ or smaller than $x_{\min}$'. Such a `negative' alternative is necessary to make the set of future alternatives complete according to Definition \ref{setfuturealt}, but at the same time it makes it impossible to model the evolution of the contextual state $S_{C}$ with a continuous wave function in these faraway regions of property value space.

Another problem with the standard definition of the uncertainty relations is that the wave function is not universally defined, as discussed in the last part of section \ref{wavef}, and summarized in Statements \ref{collapse} and \ref{loss}. The physical interpretation of the Fourier transform as a change in the choice of which property is to be observed, is only possible if the observational context $C$ has a reciprocal $\tilde{C}$ (Definition \ref{reciprocalcontext}). Then we can change basis in the common Hilbert space $\mathcal{C\tilde{C}}$ at will (Fig. \ref{Fig69d}). However, the reciprocal context exists only if $C$ is neutral (Definition \ref{neutralcontext}). Neutral contexts are an idealization, just like the representation of the contextual state by a continuous wave function is an idealization.

The difference between a realistic and an idealized context in which we observe a pair of not simultaneously knowable relational properties is highlighted if we compare Figs. \ref{Fig76b} and \ref{Fig76c}. In the realistic context shown in Fig. \ref{Fig76b} we always have a finite support $D_{PP'}$, observations with limited resolution, and we have not shown any reciprocal context since there isn't any. In contrast, Fig. \ref{Fig76c} shows a corresponding idealized context in which we let $P$ be position $x$, and $P'$ be momentum $p_{x}$. It is only in this idealization that the standard deviations can be related by the uncertainty relation $\Delta x\Delta p_{x}\geq\hbar/2$.

Even so, the interpretation of this relation is a bit awkward, since the support $D_{xp_{x}}$ is unbounded. Any $x$ is a possible outcome of the observation of position, and any $p_{x}$ is a possible outcome of the observation of momentum. In terms of the widths $\Delta D_{x}$ and $\Delta D_{p_{x}}$ of $D_{xp_{x}}$, we always have

\begin{equation}
\Delta D_{x}\Delta D_{p_{x}}=\infty
\label{idealuncertainty}
\end{equation}
in the idealized context in Fig. \ref{Fig76c}. The support $D_{xp_{x}}$ is nothing else than the projection of the specimen state $S_{OS}$ onto the subspace of state space $\mathcal{S}$ defined by the position and momentum axes. Clearly, a proper definition of the uncertainty relations should be based upon the size $D_{xp_{x}}$, since this is the relevant physical quantity, in contrast to the abstract standard deviations $\Delta x$ and $\Delta p_{x}$. Therefore it is disappointing that the physical relation \ref{idealuncertainty} gives no useful information. However, the problem is just that we expressed the relation for an unphysical, idealized context. It does give useful information about the physical state in the realistic context in Fig. \ref{Fig76b}, in which case

\begin{equation}
0<\Delta D_{x}\Delta D_{p_{x}}<\infty.
\label{realuncertainty}
\end{equation}

The support $D_{PP'}$ does not need to be rectangular if there is conditional knowledge that exclude some combinations $(p,p')$. It is therefore more appropriate to refer to the area $A[D_{PP'}]$ of the support than the product of its two widths, if we want to formulate a general uncertainty relation.
We suggest the following.

\begin{state}[\textbf{A general uncertainty relation}]
Suppose that the property pair $(P,P')$ refers to $(x,p_{x})$, $(y,p_{y})$, $(z,p_{z})$ or $(t,E)$. Let $\Pi_{PP'}S_{O}$ be the projection $\Pi$ of the object state $S_{O}$ onto the two-dimensional subspace of $\mathcal{S}$ that is spanned by the axes defined by $P$ and $P'$. Also, let $A[\Pi_{PP'}S_{O}]$ be the area of this projected set in a given coordinate system that define units for the values $p$ and $p'$. Then we have $A[\Pi_{PP'}S_{OS}]\geq\hbar/2$, where $\hbar$ is expressed in these units.
\label{generaluncertainty}
\end{state}

The fact that $\hbar$ is very small in coordinate systems defined by units that relate to everyday experience means that effects due to incomplete potential knowledge (quantum effects) are very hard to discern in everyday experience. The state space $\mathcal{S}$ is not in itself equipped with a coordinate system and an associated set of units (Section \ref{statespaces}). However, the uncertainty relation \ref{generaluncertainty} defines a minimal area $A\sim\hbar$ of certain two-dimensional slices of state space. This minimal area is therefore the most natural area unit for such slices, in concordance with the assignment $\hbar=1$ in the commonly used `natural units'. The area $A$ translates to the property `action', and $\hbar$ is often called `the quantum of action'. In our terminology we may equally well call it `a quantum of state space area'.

\begin{figure}[tp]
\begin{center}
\includegraphics[width=80mm,clip=true]{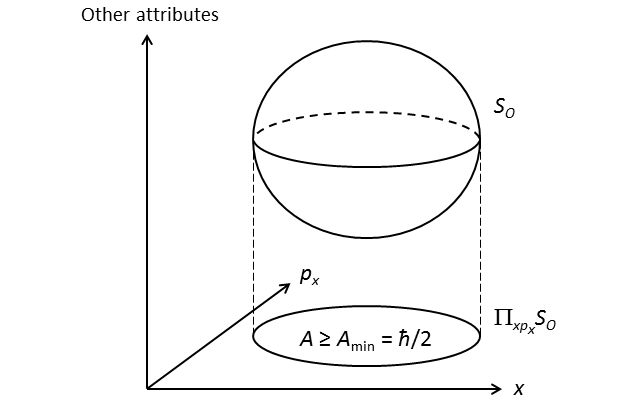}
\end{center}
\caption{Planck's constant may be defined as (twice) the minimum state space area $A$ of the projection of the object state $S_{O}$ onto the plane spanned by the $x$-components of position and momentum. The state space $\mathcal{S}$ is no vector space, so the fact that the attribute axes are drawn perpendicular has no particular meaning. The projection $\Pi_{xp_{x}}S_{O}$ of the state $S_{O}$ is just the union of the value pairs $(x(Z),p_{x}(Z))$ of all exact states $Z\in S_{O}$.}
\label{Fig80b}
\end{figure}

Let us motivate Statement \ref{generaluncertainty}. In so doing, we turn it around so that it becomes more like a definition of Planck's constant. We defined the properties momentum and energy in a formal way in section \ref{eveq}, as proportional to the reciprocal position $\tilde{\mathbf{r}}$ and the reciprocal time $\tilde{t}$ in the Fourier expansion of the continuous wave function $\Psi_{\mathbf{r}_{4}}(\mathbf{r},t)$ (Eq. [\ref{fexpansion}]). We were able to identify these expansion coefficients as values of properties because they appear as eigenvalues of self-adjoint wave function operators (Statement \ref{psiopisprop}). From this perspective, Planck's constant is just an arbitrary constant of proportionality that relates the expansion coefficients $\tilde{\mathbf{r}}$ and $\tilde{t}$ with the corresponding properties $\mathbf{p}$ and $E$. The wave function operators for the components of $(\mathbf{p},E)$ do not commute with the wave function operators for the components of $(\mathbf{r},t)$. Therefore the same is true for the corresponding property operators (Statement [\ref{comwaveprop}]), and thus the four property pairs listed in Statement \ref{generaluncertainty} are not simultaneously knowable (Statement \ref{comwaveprop}). This means that the area $A[\Pi_{PP'}S_{O}]$ has to be larger than zero, where $\Pi_{PP'}S_{O}$ is the projection of the object state $S_{O}$ onto the plane spanned by the axes that correspond to the property pair $(P,P')$, and we let this property pair correspond to any of the four pairs listed in Statement \ref{generaluncertainty}. The notation is illustrated in Fig. \ref{Fig80b}. We may write

\begin{equation}
A[\Pi_{PP'}S_{O}]\geq A_{\min}(\mathbf{r},t,\mathbf{p},E)>0,
\end{equation}
allowing in general that the minimum area $A_{\min}$ may depend on the values of the properties at which this area is located in state space $\mathcal{S}$. However, the translational invariance, together with the Lorentz invariance that apply to appriopriate combinations of these properties means that there can be no such dependence, so that

\begin{equation}
A[\Pi_{PP'}S_{O}]\geq A_{\min}>0,
\end{equation}
where $A_{\min}$ is a positive constant. Then we may simply define Planck's constant according to

\begin{equation}
\hbar\equiv 2A_{\min}.
\label{planckdef}
\end{equation}

If we use this definition as a starting point, we may say that we introduce $\hbar$ in the constants of proportionality that define $\mathbf{p}$ and $E$ from the reciprocal variables $\tilde{\mathbf{r}}$ and $\tilde{t}$ in such a way that the conventional uncertainty relations $\Delta x\Delta p_{x}\geq\hbar/2$ and $\Delta t\Delta E\geq\hbar/2$ resemble the more fundamental relation in Statement \ref{generaluncertainty} as much as possible.

Alternatively, we could use the minimum area $A_{\min}$ directly in all equations instead of $\hbar$, including those that define momentum, energy and rest mass. Then we would get $\Delta x\Delta p_{x}\geq A_{\min}$, $[x,\bar{p}_{x}]=2i A_{\min}$ and correspondingly for the other pairs of conjugate variables. The plane wave solution [\ref{p4planewave}] to the evolution equation would read $\Psi_{\mathbf{r}}(\mathbf{r},\sigma)\propto\exp[\frac{i}{A_{\min}}(\mathbf{p}_{4}\cdot\mathbf{r}-\frac{E_{0}^{2}}{2\langle E\rangle}\sigma)]$.

\vspace{5mm}
\begin{center}
$\maltese$
\end{center}
\paragraph{}

In the following paragraphs we discuss the rest energy or rest mass of minimal objects, which we identify with elementary fermions. We argue that the rest masses of these elementary fermions might be possible to determine from the eigenvalue equation of the rest energy operator, acting in a suitable Hilbert space, corresponding to a suitable choice of observational context. The discussion will be superficial, tentative and somewhat meandering.

The fact that the rest mass cannot be exactly known (Statement \ref{masswidth}) must be true for minimal objects also, of course.  

\begin{state}[\textbf{The rest mass of a minimal object cannot be exactly known}]
Let a minimal object species $M_{l}$ be defined by the values of an array of discrete internal attributes. Suppose that $M_{l}$ is an element in a minimal set $\mathcal{M}$ (Definition \ref{minimalset}) with more than one member. Then we have $\Delta E_{0}[O_{Ml}]>0$ in all physical object states $S_{O_{Ml}}$, where $O_{Ml}$ is a minimal object of species $M_{l}$.
\label{masswidth2}
\end{state}

In section \ref{minimal} we discussed minimal objects in some detail, and tried to identify the concept with that of elementary fermions. We argued that one or several arrays $\upsilon$ of internal attributes may correspond to the same minimal object species $M_{l}$ (Eq. [\ref{completeminimal}]). According to the discussion in section \ref{divideconserve} we expect that the internal attributes take discrete values. We are therefore talking about a finite set $\mathcal{M}$ of $m$ minimal objects species $M_{l}$, which each correspond to a finite set of arrays $\upsilon$, which all have the same finite length, and in which each array component can take a finite number of discrete values.

The rest energy $E_{0}$ must also be considered to be an internal attribute. It is defined by the evolution equation for any specimen, including minimal objects $O_{Ml}$. But it has a different status as compared to the discrete internal attributes that are connected to the conservation laws that apply when minimal objects divide and merge (section \ref{divideconserve}). In particular, there is nothing in the evolution equation that makes it possible to exclude \emph{a priori} any positive value of $E_{0}$, just as there is nothing \emph{a priori} that makes it possible to exclude any positive value of spatio-temporal distance. In other words, the rest mass $E_{0}$ of an object is chosen from a continuous set of possible values.

In the following, we do not include the rest energy in the specification of a minimal object species $M_{l}$; it is defined exclusively by the values of its discrete internal attributes. This is a matter of choice, but the notation becomes simpler in this way. We know that at least different rest energies are possible for each minimal object, corresponding to the three known generations of elementary fermions. In our notation there is then, apparently, just four minimal object species (so that $m=4$), namely two quarks with different electric charges, the electron and the neutrino.

Suppose that we construct an observational context $C$ in which it is determined which minimal object we are dealing with, if we know to begin with that it belongs to a minimal set $\mathcal{M}$ of distinguishable minimal object species $M_{1},M_{2},\ldots,M_{m}$. The identification of the minimal object $O_{Ml}$ corresponds to the observation of the property value $p_{l}\leftrightarrow M_{l}$ of a contextual property $P_{C}$, according to Definition \ref{contprop}. The fact that several arrays $\upsilon$ of attribute values may be consistent with a single minimal object corresponds to the fact that the contextual property $P_{C}$ does not need to be fundamental (Statement \ref{propopnofun}).

Suppose further that we introduce a family $C(\sigma)$ of such contexts. Then we may express the evolution equation

\begin{equation}
\frac{d}{d\sigma}a_{\mathcal{M}}(M_{l},\sigma)=i\bar{B}_{\mathcal{M}}a_{\mathcal{M}}(M_{l},\sigma),
\label{eveqminimal}
\end{equation}
with general solution

\begin{equation}\begin{array}{lll}
a_{\mathcal{M}}(M_{l},\sigma) & = & \int_{-\infty}^{\infty}\alpha_{\mathcal{M}}^{(\tilde{\sigma})}(M_{l})e^{i\tilde{\sigma}\sigma}d\tilde{\sigma}\\
& = & \int_{-\infty}^{\infty}\alpha_{\mathcal{M}}^{(E_{0}^{2})}(M_{l})e^{-\frac{iE_{0}^{2}}{2\hbar\langle E\rangle}\sigma}dE_{0}^{2},
\end{array}\label{minimalsolution}\end{equation}
where the second row holds only in the natural parametrization (Eq. [\ref{naturalp}]). Defining the wave function operator for the squared rest energy as $(\overline{E_{0}^{2}})_{\mathcal{M}}=2\hbar\langle E\rangle\bar{B}_{\mathcal{M}}$, we get the eigenvalue equation

\begin{equation}
(\overline{E_{0}^{2}})_{\mathcal{M}}\alpha_{\mathcal{M}}^{(E_{0}^{2})}(M_{l})=E_{0}^{2}\alpha_{\mathcal{M}}^{(E_{0}^{2})}(M_{l}).
\label{masseigenvalues}
\end{equation}

We see that the contextual Hilbert space is seemingly $m$-dimensional, as it is seemingly spanned by the property value state vectors $\bar{S}_{Ml}$ corresponding to the set of $m$ minimal object species $M_{l}$. This would mean that there are $m$ eigenvectors

\begin{equation}
\alpha_{\mathcal{M}}^{(E_{0}^{2})}(M_{l})=(\alpha_{1},\alpha_{2},\ldots,\alpha_{m}),
\label{masseigenfunctions}
\end{equation}
and $m$ corresponding eigenvalues $E_{0}^{2}$. If this would be so, we could associate one squared rest energy $(E_{0}^{2})_{l}$ and one eigenvector $(\alpha_{1l},\alpha_{2l},\ldots,\alpha_{ml})$ to each minimal object species $M_{l}$.

The rest energy and the identity of the associated minimal object would then be simultaneously knowable if and only if $\alpha_{il}=\delta_{il}$. Otherwise there would be several minimal objects that cannot be excluded given an observed squared rest mass $(E_{0}^{2})_{l}$ for a specimen that is known to belong to the minimal set $\mathcal{M}$.

But we concluded in Statement \ref{masswidth2} that the rest energy of a minimal object can never be exactly known. Therefore we cannot have $\alpha_{il}=\delta_{il}$, since that would mean that we could deduce the precise value $(E_{0}^{2})_{l}$ as the rest energy of the observed minimal object $O_{Ml}$ as soon as we have observed its internal attributes and concluded that it belong to species $M_{l}$.

This means that in the tentative $m$-dimensional Hilbert space introduced above, the rest energy and the identity of the associated minimal object are not simultaneously knowable properties. In terms of commutators, we would then write $[\bar{M}_{l},(\overline{E_{0}^{2}})_{\mathcal{M}}]\not\equiv 0$, where $\bar{M}_{l}$ is the minimal object species operator.

If the species $M_{l}$ is known for $\sigma=0$, the initial state of the wave function can be written $a_{\mathcal{M}}(M_{l},0)=(a_{1}(0),\ldots,a_{l}(0),\ldots,a_{m}(0))=(0,\ldots,exp(i\phi_{l}),\ldots,0)$. The fact that $[\bar{M}_{l},(\overline{E_{0}^{2}})_{\mathcal{M}}]\not\equiv 0$ then would mean that as $\sigma$ increases there appear non-zero amplitudes $a_{j}(\sigma)$ for $j\neq l$. That is, even if we would know that we start out with one minimal object of species $M_{l}$, we would be able to find that this one minimal object has species $M_{j}$ at a later time. This possibility violates the conservation laws for the discrete internal attributes of minimal objects as discussed in Section \ref{divideconserve} and summarized in Statement \ref{notransform}. Therefore the simple $m$-dimensional Hilbert space is insufficient to describe the evolution of minimal objects in a family of contexts $C(\sigma)$ in which the nature of the minimal object is investigated.

What we have to do is to take into account that minimal objects can divide into other minimal objects as time goes, as $\sigma$ increases. Put differently, the fact that the rest energy of an object can never be exactly known implies that minimal objects must be able to divide and merge, that their number is not conserved. This conclusion is the same as that reached from general epistemic arguments in Section \ref{limits}, in relation to Definition \ref{minimalset}.

Therefore we let each set $SM_{l}$ of minimal objects that may arise via division from one minimal object $O_{Ml}$ of species $M_{l}$ be one possible outcome of the observation performed in the family of contexts $C(\sigma)$. We may specify the set $SM_{l}$ by a vector 

\begin{equation}
SM_{l}\leftrightarrow(N_{1l},N_{2l},\ldots,N_{ml}),
\label{minimalcollection}
\end{equation}
where $N_{kl}$ is the number of minimal objects of species $M_{k}$ in the collection of objects that have arisen from the division of the initial object of species $M_{l}$. There are, of course, many possible sets $SM_{l}$, so that we have to introduce an index $j$ when we speak in general about one example of such a set, writing $SM_{jl}$. In a proper Hilbert space representation we should allow one basis vector or eigenstate $\overline{SM}_{jl}$ for each such set. We have $1\leq l\leq m$, whereas the range of $j$ is possibly infinite. The dimension of $\mathcal{H}_{C}$ is therefore possibly infinite. We may write

\begin{equation}
\mathcal{H}_{C}=\mathcal{H}_{C1}\oplus\ldots\oplus\mathcal{H}_{Cl}\oplus\ldots\oplus\mathcal{H}_{Cm},
\label{directsumhilbert}
\end{equation}
where each subspace $\mathcal{H}_{Cl}$ is spanned by the set $\{\overline{SM}_{jl}\}_{j}$. These subspaces are invariant under the evolution

\begin{equation}
\frac{d}{d\sigma}a_{\mathcal{SM}}(SM_{jl},\sigma)=i\bar{B}_{\mathcal{SM}}a_{\mathcal{SM}}(SM_{jl},\sigma),
\label{eveqminimalb}
\end{equation}
meaning that

\begin{equation}
a_{\mathcal{SM}}(SM_{jl},\sigma_{1})\in \mathcal{H}_{Cl}\Rightarrow a_{\mathcal{SM}}(SM_{jl},\sigma_{2})\in \mathcal{H}_{Cl}
\label{invariantevo}
\end{equation}
for any $\sigma_{2}\geq \sigma_{1}$. The subscript $\mathcal{SM}$ of the evolution operator $\bar{B}_{\mathcal{SM}}$ indicates that we are now dealing with a family of contexts $C(\sigma)$ in which we do not just observe the values of the overall set of $m$ discrete internal attributes associated with each minimal object species $M_{l}$, but also how these attributes are distributed across a set of $SM_{jl}$ other minimal objects, in accordance with the proper conservation laws.

Instead of Eqs. [\ref{minimalsolution}] and [\ref{masseigenvalues}] we should write

\begin{equation}\begin{array}{lll}
a_{\mathcal{SM}}(SM_{jl},\sigma) & = & \int_{-\infty}^{\infty}\alpha_{\mathcal{SM}}^{(\tilde{\sigma})}(SM_{jl})e^{i\tilde{\sigma}\sigma}d\tilde{\sigma}\\
& = & \int_{-\infty}^{\infty}\alpha_{\mathcal{SM}}^{(E_{0}^{2})}(SM_{jl})e^{-\frac{iE_{0}^{2}}{2\hbar\langle E\rangle}\sigma}dE_{0}^{2},
\end{array}\label{minimalsolutionb}\end{equation}
and

\begin{equation}
(\overline{E_{0}^{2}})_{\mathcal{SM}}\alpha_{\mathcal{SM}}^{(E_{0}^{2})}(SM_{jl})=E_{0}^{2}\alpha_{\mathcal{SM}}^{(E_{0}^{2})}(SM_{jl}),
\label{masseigenvaluesb}
\end{equation}
where $(\overline{E_{0}^{2}})_{\mathcal{SM}}=2\hbar\langle E\rangle\bar{B}_{\mathcal{SM}}$.

The fact that the squared rest mass cannot be negative means that we have to require that the eigenvectors $\alpha_{\mathcal{SM}}^{(E_{0}^{2})}(SM_{jl})$ obey a Dirac constraint, in addition to Eq. [\ref{masseigenvaluesb}], just like in the case of the spatio-temporal wave functions discussed above (Statement \ref{diracconstraint}). More precisely, there must be a square-root operator $\sqrt{(\overline{E_{0}^{2}})_{\mathcal{SM}}}$ such that

\begin{equation}
\sqrt{(\overline{E_{0}^{2}})_{\mathcal{SM}}}\alpha_{\mathcal{SM}}^{(E_{0}^{2})}(SM_{jl})=E_{0}\alpha_{\mathcal{SM}}^{(E_{0}^{2})}(SM_{jl}).
\label{dmeigenvalues}
\end{equation}

Regarding the rest energy eigenvalues $E_{0}$, we deduce from Statement \ref{nozeromass} that we cannot have $E_{0}=0$, that is, the rest masses must all be greater than zero.

The eigenvectors $\alpha_{\mathcal{SM}}^{(E_{0}^{2})}(SM_{jl})$ of the self-adjoint operator $\sqrt{(\overline{E_{0}^{2}})_{\mathcal{SM}}}$ are expected to span $\mathcal{H}_{C}$, and we therefore expect equally many eigenvalues as there are configurations $SM_{jl}$, even though some of the eigenvalues may be degenerate. Each such eigenvalue corresponds to a rest mass. We should then be able to write each vector $\overline{SM}_{jl}\in\mathcal{H}_{C}$ corresponding to the configuration $SM_{jl}$ as a linear combination

\begin{equation}
\overline{SM}_{jl}=c_{\nu}(\alpha_{\mathcal{SM}})_{\nu}+c_{\nu'}(\alpha_{\mathcal{SM}})_{\nu'}+c_{\nu''}(\alpha_{\mathcal{SM}})_{\nu''}+\ldots,
\end{equation}
so that we can identify the incomplete knowledge of the rest energy $(E_{O})_{jl}$ of $SM_{jl}$ with the knowledge that it belongs to the set of eigenvalues $\{(E_{0})_{\nu},(E_{0})_{\nu'},(E_{0})_{\nu''},\ldots\}$. These rest energy values are ideally bounded from below and from above.

The question is whether we can, with this method, associate such a rest mass $(E_{O})_{jl}\in\{(E_{0})_{\nu},(E_{0})_{\nu'},(E_{0})_{\nu''},\ldots\}$ to all configurations $SM_{jl}$, including those consisting of a single minimal object. It is the rest energy of these that is of primary interest, of course. One problem is that a single minimal object of the lightest species $M_{1}$ cannot spontaneoulsy divide into another configuration $SM_{j1}$ containing more than one object. The evolutionary invariance of $\mathcal{H}_{1}$ according to Eq. [\ref{invariantevo}] then means that $\overline{SM}_{1}$ corresponds to an eigenvector of the rest energy operator, so that the rest mass becomes precisely known. This is impossible whenever such a minimal object is actually observed at a given time, according to the uncertainty relation involving rest mass and time (Statement \ref{uncertainmass}). Such an observation is necessary to identify the minimal object in the first place. Note that this is true even if the lightest minimal object is stable, in contrast to the analogous case in which the energy of a stable ground state of a bound composite system can be precisely determined, at least in principle.

To overcome this problem, we could alter the Hilbert space $\mathcal{H}_{C}$ by considering subspaces $\mathcal{H}_{Cl}$ according to Eq. [\ref{directsumhilbert}] that correspond to the initial existence of \emph{two} minimal objects of species $M_{l}$ instead of just one. In that case we can consider collisions between these two minimal objects, so that other minimal objects may appear even when two minimal objects of the lightest species collide and transform, converting kinetic energy into the required rest mass of the new minimal objects.

Let us emphasize two main points at the basis of the above discussion.

\begin{state}[\textbf{The species of a minimal object and its rest energy are not simultaneously knowable}]
The species of a minimal object can be known exactly, but its rest energy cannot. This means that species and rest mass are two properties that are not simultaneously knowable. The same goes for sets of minimal objects according to Eq. [\ref{minimalcollection}].
\label{nosimknowspeciesmass}
\end{state}

According to Statement \ref{comwaveprop} this means that the two corresponding wave function property operators do not commute.

\begin{state}[\textbf{The minimal object species configuration operator} $\overline{SM}_{l}$ \textbf{and rest energy operator} $(\overline{E_{0}^{2}})_{\mathcal{SM}}$ \textbf{do not commute}]
The wave function operators $\overline{SM}_{l}$ and $(\overline{E_{0}^{2}})_{\mathcal{SM}}$ act on the wave functions $a_{\mathcal{SM}}$ in Eq. [\ref{eveqminimalb}], being defined in the Hilbert space $\mathcal{H}_{C}$ according to Eq. [\ref{directsumhilbert}]. We have $[\overline{SM}_{l},(\overline{E_{0}^{2}})_{\mathcal{SM}}]\not\equiv 0$.
\label{nocommutespeciesmass}
\end{state}

Finally, we emphasize that all minimal objects must have non-zero rest mass, since this is true for any object (Statement \ref{nozeromass}).

\begin{state}[\textbf{All minimal object species have positive rest mass}]
We can exclude the possibility that a minimal object has zero rest mass.
\label{nozerominimalmass}
\end{state}

In section \ref{divideconserve} we discussed briefly the role of the generation quantum number as an internal attribute of minimal objects. There is no conservation law for generation number in elementary particle physics, despite the fact that we argued that such conservation laws are necessary to uphold the identifiability of minimal objects. Let us call the generation number $g$.

One possibility is that $g$ emerges from the eigenvalue equation Eq. [\ref{dmeigenvalues}] in a way that is analogous to the quantum number $n$ in atomic physics. In the hydrogen atom the state of the electron is specified by the attribute value triplet $(n,l,m_{z})$, where $l$ is the angular momentum and spin direction quantum numbers, respectively. There is no conservation law for $n$ just as there is no conservation law for $g$. To each triplet $(n,l,m_{z})$ is associated one electronic energy level $E$. This means that we could equally well specify the electronic state by the triplet $(E,l,m_{z})$, since $E=f(n,l,m_{z})$. The number $n$ is more dispensable than $l$ and $m_{z}$ since the latter two corresponds to observable physical properties that are independent from $E$. In contrast, $n$ is just a number that is used to encode different energy levels. In this way we see that the non-conservation of $n$ is no more mysterious than the non-conservation of $E$, which just means that the potential energy stored in the atom may change by the emission or absorption of a photon.

To be a little bit more concrete, the eigenvalue equation Eq. [\ref{dmeigenvalues}] may over-determine the rest energies of the $m$ minimal object species, since there is potentially an infinite set of eigenvalues $E_{O}$, given that the number of configurations $SM_{jl}$ of minimal objects is possibly infinite, in which case the dimension of $\mathcal{H}_{C}$ becomes infinite. This opens up for the possibility that several distinct rest energies can be associated to a single species $M_{l}$. These distinct rest energies can then be indexed by a generation quantum number $g$. However, most of the rest energies that emerge as eigenvalues are associated with configurations containing several minimal objects, so that the over-determination of the rest energies of single minimal objects may be avoided. Nevertheless, the circumstance that there are several invariant subspaces $\mathcal{H}_{Cl}$ according to Eq. \ref{directsumhilbert} suggests that the rest-masses may be over-determined after all, if the they can be dedcued from the eigenvalue problem in each invariant subspace spearately. In that case we might naïvely expect as many generations as there are basic classes of elementary fermions, namely four.

\section{Antimatter}
\label{antimatter}

Many readers may find some claims and arguments in this text strange or misguided. In this section, a perspective on antimatter will be put forward that even I find hard to swallow. I ask myself: can it really be that simple, or am I swept away by a silly idea that does not hold water?

As discussed in section \ref{evconsequences}, for all families of contexts $C(\sigma)$ described by the evolution equation [\ref{psieveq}] we have

\begin{equation}\begin{array}{rcl}
d\langle t\rangle/d\sigma>0 & \Leftrightarrow & E>0\\
d\langle t\rangle/d\sigma<0 & \Leftrightarrow & E<0.
\end{array}
\label{posnegenergies}
\end{equation}
Basically, this is a consequence of the directionality of sequential time, as parametrized by $\sigma$, together with the \emph{definition} of energy $E$ as proportional to the Fourier expansion coefficient $\tilde{\mathbf{r}}_{4}$ (Definition \ref{energydefi}).

Say that the event that marks the beginning of the context $C$ is assigned sequential time $n$, corresponding to $\sigma=0$. Let us choose a relative scale for relational time $t$ according to Fig. \ref{Fig74}(b), so that we have $t(n)=0$.

In a family of contexts $C(\sigma)$ which is parametrized so that $d\langle t\rangle/d\sigma>0$ we will have $t(n+1)>t(n)=0$. Suppose that the specimen at time $n$ belongs to the present, so that $Pr_{OS}(n)=1$, where $Pr$ is the presentness attribute introduced in section \ref{time} (Definition \ref{presentness}). Then it will also belong to the present the next time we observe at time $n+m$, since $t(n+m)>t(n)$ and there is not as yet any later observation of $OS$ of which the observation at time $n+m$ can be a memory. That is, $Pr_{OS}(n+m)=Pr_{OS}(n)=1$.

\begin{state}[\textbf{Parametrization of observations of the present}]
Let $C^{+}(\sigma)$ be a family of contexts such that the final observation at time $n+m$ is an observation of the present state of the specimen $OS$ at this time, so that $Pr_{OS}(n+m)=1$. Then the parametrization must be such that $d\langle t\rangle/d\sigma>0$. Conversely, if we consider such a parametrization, then we consider such a family $C^{+}(\sigma)$.
\label{deducepresent}
\end{state}

Consider instead a family of contexts $C(\sigma)$ parametrized so that $d\langle t\rangle/d\sigma<0$. What does this mean? Suppose again that $Pr_{OS}(n)=1$. Since $t(n+m)<t(n)$ we must conclude that $Pr_{OS}(n+m)=0$. That is, the observation at time $n+m$ must be a deduction about a past state of the specimen.

\begin{state}[\textbf{Parametrization of observations that are deductions about the past}]
Let $C^{-}(\sigma)$ be a family of contexts such that the final observation at time $n+m$ is an observation of a past state of the specimen $OS$ at this time, so that $Pr_{OS}(n+m)=0$. Then the parametrization must be such that $d\langle t\rangle/d\sigma<0$. Conversely, if we consider such a parametrization, then we consider such a family $C^{-}(\sigma)$.
\label{deducepast}
\end{state}

An object that we deduce at time $n+m$ to belong to the past is necessarily a quasiobject. Only objects with $Pr=1$ that we see here and now are directly perceived objects.

\begin{state}[\textbf{The specimen in a context} $C^{-}$ \textbf{is a quasiobject}]
Suppose that we follow the evolution of specimen $OS$ in a family of contexts $C^{-}(\sigma)$, as defined in Statement \ref{deducepast}. Then $OS$ is a quasiobject.
\label{pastquasiobject}
\end{state}

However, any quasiobject $\tilde{O}$ that belongs to the past must be in one-to-one correspondence to a directly perceived object $O$ that belongs to the present, as discussed in relation with Definition \ref{quasiobjectdefi}. The role of deduced quasiobjects of the past is elaborated upon in connection with Eqs. [\ref{deduction1}] and [\ref{deduction2}]. Here we skip the formal subtleties, for instance the fact that the deduction itself is an event that corresponds to an object and a temporal update.

\begin{defi}[\textbf{The perceived object} $POS$ \textbf{that corresponds to the specimen in contexts} $C^{-}$]
Consider a context $C^{-}$ as defined in Statement \ref{deducepast}. Then $POS$ is the perceived object $O$ which defines the observation within context at time $n+m$. The specimen $OS$ is the quasiobject $\tilde{O}$ that corresponds to $POS$ according to Definition \ref{quasiobjectdefi} and Statement \ref{pastquasiobject}.
\label{posdefi}
\end{defi}

\begin{figure}[tp]
\begin{center}
\includegraphics[width=80mm,clip=true]{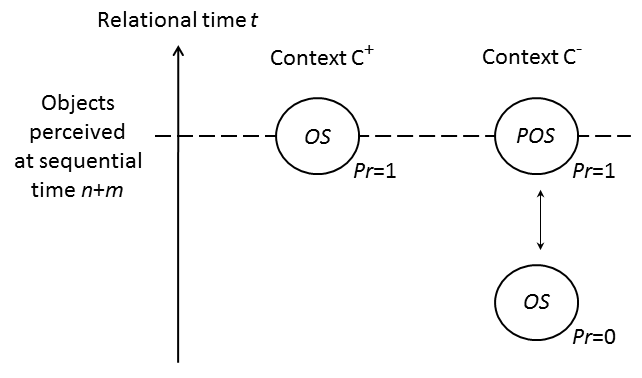}
\end{center}
\caption{In a context $C^{+}$ (Statement \ref{deducepresent}), the specimen $OS$ is a present object ($Pr=1$) at the time $n+m$ of the final observation. The energy of $OS$ is positive. In a context $C^{-}$ (Statement \ref{deducepast}), $OS$ is a deduced quasiobject belonging to the past at the time $n+m$ of the final observation. The energy of $OS$ is negative. The object that corresponds to the actual observation at time $n+m$ is called $POS$ (Definition \ref{posdefi}).} 
\label{Fig82a}
\end{figure}

This definition is illustrated in Fig. \ref{Fig82a}. Now we arrive at the identification that is the main hypothesis in this section.

\begin{defi}[\textbf{Antimatter}]
If the perceived object $POS$ in a context $C^{-}$ is real (Definition \ref{realobjects}), then $POS$ is an anti-object.
\label{antimatterdef}
\end{defi}

Consider a family contexts $C^{-}(\sigma)$ that corresponds to deductions about the past according to Statement \ref{deducepast}. For this family we may define an anti-natural parametrization that mirrors the natural parametrization defined for familes $C^{+}(\sigma)$ according to Definition \ref{natparadef}.

\begin{defi}[\textbf{The anti-natural parametrization}]
The parametrization of the family of contexts $C(\sigma)$ is anti-natural if and only if $d\langle t\rangle/d\sigma=-1$ for all $\sigma\in[0,\sigma_{\max})$.
\label{antinatparadef}
\end{defi}

\begin{figure}[tp]
\begin{center}
\includegraphics[width=80mm,clip=true]{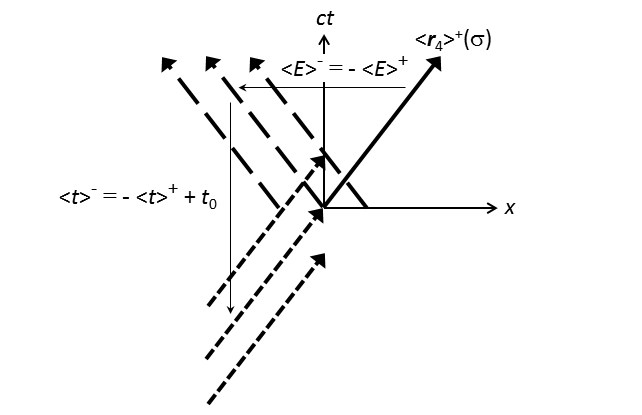}
\end{center}
\caption{If the energy of a free specimen changes sign, so does the expected velocity. A negative energy also means that the direction of the expected relative time $t$ changes sign in relation to $\sigma$. The arrows show the direction of motion of the object.} 
\label{Fig81b}
\end{figure}

Assume that the specimen is free in the family $C^{+}(\sigma)$, and that we choose a coordinate system so that $\langle\mathbf{r}_{4}\rangle^{+}(\sigma)=\mathbf{v}_{0}\sigma$ (Fig. \ref{Fig81b}). The expected energy $\langle E\rangle^{+}$ will not depend on $\sigma$. Introduce a corresponding family $C^{-}(\sigma)$ such that $\langle E\rangle^{-}=-\langle E\rangle^{+}$, in accordance with Eq. [\ref{posnegenergies}]. We have

\begin{equation}
\langle E\rangle^{-}=-\langle E\rangle^{+}\Rightarrow\langle\mathbf{r}\rangle^{-}=-\langle\mathbf{r}\rangle^{+}+\mathbf{r}_{0}
\label{spacemirror}
\end{equation}
according to Ehrenfest's theorem (Eq. [\ref{naturalehrenfest}] or [\ref{ehrenfest}]). In the natural and anti-natural parametrizations we get

\begin{equation}
\langle t\rangle^{-}(\sigma)=-\langle t\rangle^{+}(\sigma)+t_{0},
\label{timemirror}
\end{equation}
where $t_{0}$ is an arbitrary constant. These transformations are illustrated in Fig. \ref{Fig81b}.

\begin{defi}[\textbf{A pair of corresponding context families} $(C^{+}(\sigma),C^{-}(\sigma))$]
Suppose that for each exact state $Z^{+}\in S_{OS}^{+}(n+m)$ in which the specimen $OS^{+}$ has energy $E>0$ in a context $C^{+}$, there is exactly one exact state $Z^{-}\in S_{OS}^{-}(n+m)$ in which the specimen $OS^{-}$ has energy $-E$ in the context $C^{-}$, but in which all other attributes of the two specimens are the same. Then the contexts $C^{+}$ and $C^{-}$ is a corresponding pair. If this relation between the contexts $C^{+}(\sigma)$ and $C^{-}(\sigma)$ holds for each $\sigma\in[0,\sigma_{\max})$, then $(C^{+}(\sigma),C^{-}(\sigma))$ is a pair of corresponding context families in the domain $\sigma\in[0,\sigma_{\max})$.
\label{correspondingfamilies}
\end{defi}

The following statement is a simple consequence of this definition.

\begin{state}[\textbf{All contexts can be grouped in corresponding pairs}]
The set $\{C\}$ of all contexts allowed by physical law has the same elements as the set $\{(C^{+},C^{-})\}$ of all pairs of corresponding contexts.
\label{contextpairing}
\end{state}

Suppose that the free specimen $OS^{+}$ is observed at time $t=0$ in the context $C^{+}$, and that the free specimen $OS^{-}$ is observed in the corresponding context $C^{-}$. (Recall that we use the relative temporal time scale, as expressed in Fig. \ref{Fig74}.) The state $S_{OS}^{-}(n+m)$ of the specimen $OS^{-}$ in the family $C^{-}$ can be interpreted as the deduced history at the expected time $-\langle t\rangle$ of the specimen $OS^{+}$ that is observed at expected time $\langle t\rangle$ in the corresponding context $C^{+}$.

This is so since the constants $\mathbf{r}_{0}$ in Eq. [\ref{spacemirror}] and $t_{0}$ in Eq. [\ref{timemirror}] are arbitrary. That is, we cannot \emph{exclude} that $\mathbf{r}_{0}=\mathbf{0}$ and $t_{0}=0$ so that the dashed arrow in Fig. \ref{Fig81b} ends at the origin, where the solid arrow starts. Further, the transformations [\ref{spacemirror}] and [\ref{timemirror}] leave $d\langle \mathbf{r}\rangle/d\langle t\rangle$ invariant, so that the two arrows can be interpreted as a single world line of a free specimen, stretching from the past into the future. If we cannot exclude that two objects are different, then they should be considered the same, as discussed in section \ref{identifiability}. Therefore we may identify $OS^{-}$ with $OS^{+}$.

\begin{state}[\textbf{The specimens in a pair of corresponding context families are the same object, projected into the future and into the past}]
Consider a pair of corresponding context families $(C^{+}(\sigma),C^{-}(\sigma))$ in which the specimens $OS^{+}$ and $OS^{-}$ are free. Then these specimens are the same object $OS$. The context family $C^{+}(\sigma)$ enables the direct observation of future states of $OS$, whereas $C^{-}(\sigma)$ enables the observation of past states of $OS$.
\label{samespecimen}
\end{state}

The state of any object can be expressed in terms of minimal objects. Minimal objects are identifiable (Statement \ref{allminimalidentity}); they do not suddenly appear or disappear. In other words, the state of potential knowledge must always be consistent with the hypothesis that these minimal objects have existed as far back in time as the world itself has existed, and will exist as far away in the future as the world itself will exist. In this sense, every object has a history as well as a future at any given moment $n$.

This statement has epistemic meaning if and only if there is a context in which a future state of the object (or specimen) can be observed, as well as a corresponding context in which a past state of the object can be observed. Such a pair of contexts must exist for each pair of expected relational times $(\langle t\rangle,-\langle t\rangle)$ for the observations of the future and past states, respectively. (Of course, physical law may restrict which values of $t$ are possible. In that case, we assume that the value $-t$ is allowed if and only if $t$ is allowed.)

\begin{state}[\textbf{For each future state of an object, there is a corresponding past state}]
Suppose that physical law allows an observation of a future state of an object at the expected relative temporal distance  $\langle t\rangle$ from the present time. Then physical law must also allow an observation of a past state of the same object at the expected relative temporal distance  $-\langle t\rangle$. The reverse of this statement is also true.
\label{futurepaststates}
\end{state}

Observations of past states of an object $O$, with $Pr[O]=0$, may either be a recollection of a memory, or a deduction from observations of other objects $O'$ belonging to the present, with $Pr[O']=1$. In the first case we have $Dp[O]=1$, and in the second case we have $Dp[O]=0$ (Table \ref{objectinterpretation}).

Note the perfect congruence between the epistemic Statement \ref{futurepaststates} and the formal Statement \ref{contextpairing}. The fact that each object has both a future and a past is matched perfectly by the evolution equation and the Dirac equation, for which there is exactly one solution with energy $-E$ for each solution with energy $E$. These two solutions correspond to the corresponding pair of contexts $(C^{+},C^{-})$. To me, this is one of many observations that supports the idea that all proper mathematical expressions of physical law match the basic epistemic conditions of perception.

This symmetry between looking forwards and backwards in time does not imply that there are equal amounts of matter and antimatter, however. Recall that we required that the present object $POS$ is \emph{real} (Definition \ref{realobjects}) in the definition of antimatter (Definition \ref{antimatterdef}). This is not necessarily the case.

Consider, for example, a context $C^{-}$ in which the observation at time $n+m$ is a recollection of a memory. Imagine that you are watching a cloud passing by in the sky. If you keep watching, you a realizing a context $C^{+}$. If you close your eyes for a moment and reacall where the cloud came from, you are realizing a corresponding context $C^{-}$. Then $POS$ is the recalled image of the cloud $OS$. The image $POS$ is not real, according to the discussion in section \ref{contextualstates}. It must be described in terms of more or less stationary minimal objects in the brain, rather than minimal objects passing by in the sky.

We may express this conclusion in another way. To account for the memory of the cloud, it is not necessary to postulate the existence of real minimal objects passing by in the sky at the moment of recollection. The image encoded by the state of minimal objects in the brain is sufficient.

Sometimes it is necessary to postulate such minimal objects, however. Imagine that a ball suddenly appears out of nowhere in a doorway. It comes bouncing into the room where you are standing. You have no memory of it, and it turns out to be impossible to deduce its past existence indirectly, say, from sounds of bounces from the other room that you heard before it turned up before our eyes. Simply put, the prior existence of the ball is outside potential knowledge. Since you should have heard such bounces, the present knowledge even contradicts the prior existence of the ball. In such a situation, at the same time as the ball appears, there has to appear an object that makes it possible to account for its history.

A more scientific example of the same phenomenon is the sudden appearance of a track of an electron in the middle of a cloud chamber. The absence of a prior track means that we have no evidence of the history of the new electron. Furhermore, we should have seen such a prior track entering the cloud chamber if it existed beforehand. To make it possible to account for the history of the new electron, a track of a positron must appear at the same time.  

The reason is the requirement that minimal objects should be identifiable. They cannot appear out of nowhere, without a history. At each moment $n$ at which the existence of the minimal object is part of potential knowledge, a possible history of the minimal object that is consistent with its present state should also be part of potential knowledge. If the past existence of the minimal object is completely unknown and even contradicts the existing evidence, then the appearance of a new type of minimal object has to be postulated, the role of which is to make it possible to account for the history of the original minimal object. This new type of minimal object is the anti-object.

\begin{state}[\textbf{Anti-objects must sometimes appear to uphold identifiability}]
Suppose that the potential knowledge $PK(n)$ contains a new minimal object $O_{Ml}$, observed at time $n+m$ as a (part of a) specimen $OS$ in a context $C$. Also, $PK(n+m)$ excludes the existence of $O_{Ml}$ at time $n+m-1$, in the sense that it does not allow an object state $S_{O_{Ml}}(n+m-1)$ such that $S_{O_{Ml}}(n+m-1)\cap S_{O_{Ml}}(n+m)\neq\varnothing$. Then it must be possible to observe an anti-object in $C$ at time $n+m$.
\label{necessaryanti}
\end{state}

\begin{figure}[tp]
\begin{center}
\includegraphics[width=80mm,clip=true]{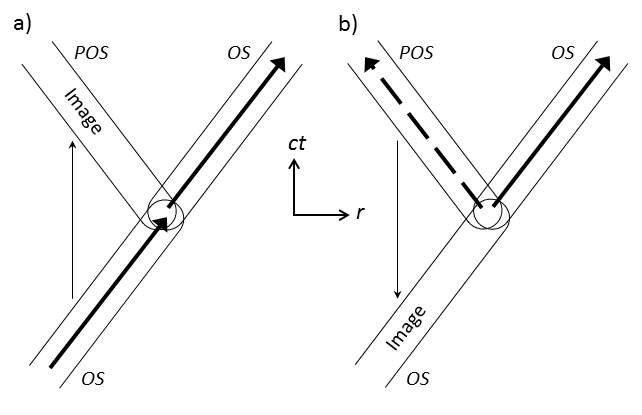}
\end{center}
\caption{In a context $C^{-}$, either the directly percevied present object $POS$, or the history of the specimen $OS$ is real. a) The history of $OS$ is real, and $POS$ is just an image of this history. b) $POS$ is real, which means that it is an anti-object, and the history of $OS$ is just an image of $POS$. In both cases, the $OS$ has both a history and a future, as required by the requirement of identifiability. The world tubes represent the spatio-temporal positions that are not excluded by potential knowledge (Fig. \ref{Fig50}). The pseudo-internal temporal scale is used (Fig. \ref{Fig74}).} 
\label{Fig81c}
\end{figure}

It may seem paradoxical to say that an anti-object appear to account for the history of an object, at the same time as we say that the evidence exludes the existence of such a history. The solution to this paradox is that in this case the history is not a real object. This matter is discussed further in connection with Fig. \ref{Fig81c} and Statement \ref{singlerealobject}.

Statement \ref{necessaryanti} expresses a situation where an anti-object must appear. There are other situations where an anti-object may or may not appear.

\begin{state}[\textbf{Anti-objects may sometimes appear to account for an unknown history}]
Suppose that the potential knowledge $PK(n+m)$ is greater than the preceding potential knowledge $PK(n+m-1)$ in the sense that $PK(n+m)$ contains a new minimal object $O_{Ml}$ observed as a (part of a) specimen $OS$ in a context $C$, whereas $PK(n+m-1)$ does not contain this particular minimal object. Then it may be possible to observe an anti-object in $C$.
\label{possibleanti}
\end{state}

Returning to the ball, we may have a situation where the ball is again suddenly appearing in the doorway. This time it is silently rolling into the room where you are standing. Again, there is no way to deduce the previous existence of the ball from your present potential knowledge. But this time this knowledge does not exclude its previos existence. There are no bounces that you should have heard if it were there.

The corresponding situation in the cloud chamber is an electron track that is present already at the chamber boundary. You cannot know for sure that the electron existed before it entered the chamber, its prior existence is outside potential knowledge, but you cannot exclude it either. In this case you may see a positron track beside the electron track, making it possible to interpret this positron as a representation of the history of the electron, but it is not necessary in order to uphold the identifiablity of the electron.

If there is such a positron track, it must be present already at the chamber boundary, of course, just like the associated electron track. This makes it possible to see such a positron track without the associated electron track. In such a case it is not possible to identify the electron whose history the positron is supposed to represent. It may be far away from the cloud chamber. According to our interpretation of antimatter, it should still be possible to uphold the notion that it represents the history of \emph{some} electron.

Since minimal objects may divide and transform into each other, we have to define what we mean by a `new minimal object' $O_{Ml}$ in Statements \ref{necessaryanti} and \ref{possibleanti}.

\begin{defi}[\textbf{New minimal object}]
If the conservation law \ref{conservationtypes} is not fulfilled in a division of a minimal object, and this happens in such a way that the sum of the values of some internal attribute is greater after the division than before, then a new minimal object is created in the division. This situation also occurs if a minimal object is apparently created out of nothing. Then we may say that the prior absence of the object corresponds to a set of internal attributes having zero value.
\label{newminimal}
\end{defi}

Apart from upholding identifiability, the appearance of the anti-object according to Statement \ref{necessaryanti} is also necessary to make sure that the conservation laws \ref{conservationtypes} are, in fact, fulfilled, provided that we take account of the internal attribute values of the anti-object. This means, for example, that the anti-object that appears together with a new electron must have the electric charge $+e$, so that it can be identified with a positron.

Any minimal object can, in principle, be created out of nothing, provided that a corresponding anti-minimal object is also created at the same time to account for its history (Statement \ref{necessaryanti}). To uphold the conservation laws in object division, the values of all internal attributes of anti-minimal object must have the same magnitude but the opposite sign as those in the corresponding ordinary minimal object. This observation was made already in connection with Statement \ref{newminimal}.

\begin{state}[\textbf{To each minimal object species} $M_{l}$ \textbf{corresponds an anti-minimal object species} $M_{l}^{-}$]
Suppose that $M_{l}$ is specified by the set $\{\Upsilon_{1l},\ldots,\Upsilon_{ml}\}$ of internal attributes. Then $M_{l}^{-}$ is specified by the array $\{-\Upsilon_{1l},\ldots,-\Upsilon_{ml}\}$. The rest masses of $M_{l}$ and $M_{l}^{-}$ are the same.
\label{minimalobjectpairs}
\end{state}

Why are the rest masses the same? In the solution to the evolution equation (and to the Dirac equation) that represents to the anti-object, we just flip the sign of $E$, but do not change $m_{0}$. Also, the interpretation of $M_{O}^{-}$ as a history of $M_{O}$ means that the rest masses must be the same. The rest mass is not an internal attribute on the same footing as the array of attributes with discrete values, as discussed in section \ref{evconsequences}. There is no conservation laws for rest masses in object division that forces the rest mass of $M_{O}^{-}$ to change sign.

In the present interpretation of antimatter, there is no reason to expect that there are as many minimal anti-objects as minimal objects. Since any anti-object is interpreted to be an imposed history of a present object, there cannot be more anti-objects than objects. Some objects does not have an assiciated anti-object since the first knowledge of this object may have appeared at a time when the existence of the object did not contradict its existence even further back in time. Without going into quantitative details, we may say that the forced creation of matter and antimatter in equal amounts according to Statement \ref{necessaryanti} should be a rare event which requires carefully designed experimental contexts. 

\begin{hypo}[\textbf{There is more matter than antimatter in the world}]
There cannot be more anti-objects than objects in the world. There is no symmetry principle that requires equal numbers of minimal anti-objects and minimal objects. Knowledge of new objects may appear without any associated anti-objects.
\label{lessantimatter}
\end{hypo}

A condition for the appearance of new minimal objects is that the corresponding increase of the potential knowledge about the past fulfils epistemic consistency (Definition \ref{epconsistency2}). The inferred past existence of the new object must be such that it would not have caused a different present potential knowledge $PK(n)$ if it would have been part of $PK(n-m)$ for some positive integer $m$. Simply put, the new object must be so small and light that its existence does not make any knowable difference for the state of its surroundings.

Thus it is very hard to create massive and composite new objects like balls or houses close to aware subjects apparently out of nothing - together with their associated anti-object. Such an event would almost certainly violate epistemic consistency. If not, the large new object must have been extremely well isolated in its (unknown) past. This is roughly the same line of reasoning that explains why it is hard to observe interference between two potential evolutions of a massive and composite object (say in a two-slit experiment). It is then very hard to keep the experimental setup isolated so that the choice of path is not imprinted in its `walls'. If so, the path becomes part of potential knowledge before the interference pattern is created. The appearance of interference in such a situation would violate epistemic consistency. We would both have and not have path information at the same time.

Say that a new minimal object $OS$ appears out of nowhere at the time $n+m$. In other words, we know that it did not exist at its present location the moment before (Statement \ref{necessaryanti}). Let the expected location of its appearance define the origin in space-time. We may then say that it follows the world line towards the upper right in Fig. \ref{Fig81b}. In this situation, the mirrored world line towards the upper left must be interpreted as a real anti-object $POS$. The lower dashed arrow pointing towards the origin must be interpreted as the history of $OS$ that the appearance of the anti-object $POS$ makes it possible to deduce. Because of the identifiability condition, there can be no temporal gaps in the ability to deduce the history of $OS$. Therefore the lower dashed arrow must point approximately at the origin, so that the deduced historical state $S_{OS}^{-}(\sigma)$ overlaps the present state $S_{OS}^{+}(\sigma)$ as $\sigma\rightarrow 0$. (We consider here the pseudo-internal assignment of spatio-temporal coordinates, as described in Fig. \ref{Fig74}.) This means that we must choose $\mathbf{x}_{0}\approx 0$ and $t_{0}\approx 0$ in Eqs. [\ref{spacemirror}] and [\ref{timemirror}].

\begin{state}[\textbf{Pair production}]
Suppose that the family of contexts $C^{+}(\sigma)$ is such that a new free minimal object $OS=O_{M}^{(n)}$ appears  out of nowhere at time $n+m$ (Statement \ref{necessaryanti}), and that we observe it the next time $n+m'$ at the position $\mathbf{r}_{4}[OS]$. Then, to uphold the identifiablity of minimal objects, $C^{+}(\sigma)$ must also be such that the position of a real anti-object $POS$ is observed at time $n+m'$. The position of $POS$ must fulfil $\mathbf{r}_{4}[POS]=-\mathbf{r}_{4}[OS]$ in a fundamental context.
\label{mustanti}
\end{state}
Recall that a fundamental context (Definition \ref{fundamentalcontext}) is such that there is one alternative for each value allowed by physical law, so that the observed value of $\mathbf{r}_{4}$ is exact.

We see that in a context $C^{+}$ such as that in Statement \ref{antimatterdef}, we observe the new present object $OS$ at the same time as we are able deduce its past via $POS$. Since $POS$ is real, it can be described by a set of minimal objects that `move along' with it. If the new object $OS$ is an electron, then $POS$ must be described as a single positron, as discussed above.

\begin{state}[\textbf{Corresponding contexts coincide if we observe pair production}]
Suppose that we have a context $C^{+}$ such as that described in Statement \ref{mustanti}. Consider the corresponding context $C^{-}$ in the pair $(C^{+},C^{-})$ according to Definition \ref{correspondingfamilies}. We have $C^{+}=C^{-}$.
\label{correspondingsame}
\end{state}

A pair-producing family of contexts is shown in Fig. \ref{Fig81c}(b). If $POS$ is real (an anti-object), then the history of $OS$ cannot be real. We cannot have two different real minimal objects that represent the same thing. One of them must be an `image' of the real thing. In a pair production event, it is the history of $OS$ that is the image of the real object $POS$. In a recollection of a memory, or a deduction of a historical fact, it is the present object $POS$ that is an image of the real historical object $OS$. This is illustrated in Fig. \ref{Fig81c}(a).

\begin{state}[\textbf{The reality of the past state of} $OS$ \textbf{and the present state of} $POS$ \textbf{is mutually exclusive}]
Consider a context $C^{-}$. If $POS$ is real according to Definition \ref{realobjects}, then $OS$ is not real, and vice versa.
\label{singlerealobject}
\end{state}

\begin{figure}[tp]
\begin{center}
\includegraphics[width=80mm,clip=true]{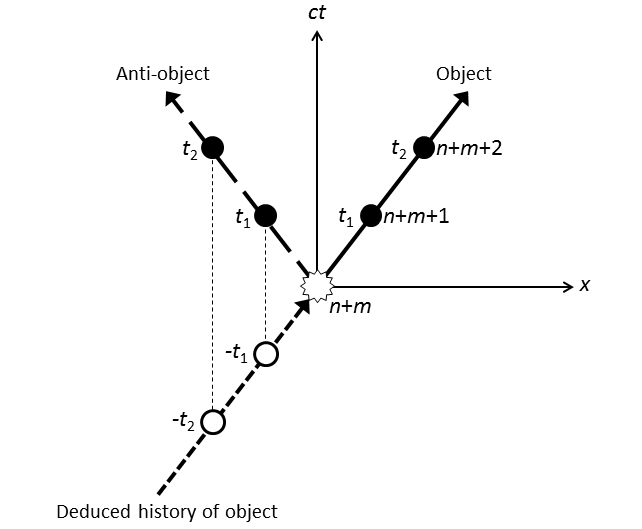}
\end{center}
\caption{A context in which the temporal update $n+m-1\rightarrow n+m$ is defined by the appearance of a new object. This object and the corresponding anti-object is tracked in a series of observations. The observation of real objects is marked by filled circles. The history of the object, as deduced from the anti-object, is not real. The corresponding observations are marked by hollow circles. Compare Fig. \ref{Fig81c}.} 
\label{Fig82}
\end{figure}

Instead of considering a family of contexts $C(\sigma)$ in which we observe a new object at time $n+m$ and check that there is an object as well as an anti-object at time $n+m'$, we may track the object and the anti-object during a sequence of observations in a family $C(\sigma_{1},\sigma_{2},\ldots)$, corresponding to observations at times $n+m,n+m+1,n+m+2,\ldots$. These sequential times will correspond to relational times $t(n+m)=0,t(n+m+1)=t_{1},t(n+m+2)=t_{2},\ldots$. If the new object appearing at time $n+m$ is free, the outcome of such a context will look like that in Fig. \ref{Fig82}.

Since the anti-object is assumed to be real according to Definition \ref{realobjects}, it can be described in terms of a set of minimal objects that share the relative attributes with the anti-object (it `moves along' with it, as compared to the surroundings). Since all minimal objects are identifiable, they can, in principle, persist indefinitely. This means that we may have contexts in which we study an anti-object which were created a long time ago, so that it exist already at the start of the context at time $n$. It also means that the anti-object does not have to be accompanied by a corresponding ordinary object within context. Since these anti-minimal objects can be studied independently from ordinary minimal objects, they should be given the same status of reality.

We have talked about pair production. Let us talk a bit about the reverse of this process, namely pair annihilation. The picture of pair production that we have presented is the following: at a certain time $n+m$, a new object and its history appears. The present and the past side of the coin appear simultaneosly because of the identifiability requirement. In plain language, this just the requirement that the amount of matter is preserved. We may look at pair annihilation in the same way. At a certain time $n+m$, an object and its history disappear simultaneoulsy. At the next time instant $n+m+1$, the object is missed by nobody, so that the requirement that matter is preserved is \emph{apparently} fulfilled.

From the epistemic perspective, this \emph{apparently} is all there is. Recall that \emph{all} potential knowledge at time $n+m$ is contained in $PK(n+m)$, including knowledge about the past and possible deductions about the future. At time $n+m$ we cannot peer back at $PK(n+m-1)$ and say: "hey, one object is missing!" Pair annihilation is the perfect crime: you kill the victim - the object - and erase all evidence that it ever existed - the anti-object. Then no one will miss it - the preservation of matter is apparently upheld. Since no one will ever know that you broke the law, you did not actually break it, from the epistemic point of view. In the same way, in a pair production, you add an uninvited guest to the party who behaves in such a way that everyone thinks she belongs there. 

\begin{figure}[tp]
\begin{center}
\includegraphics[width=80mm,clip=true]{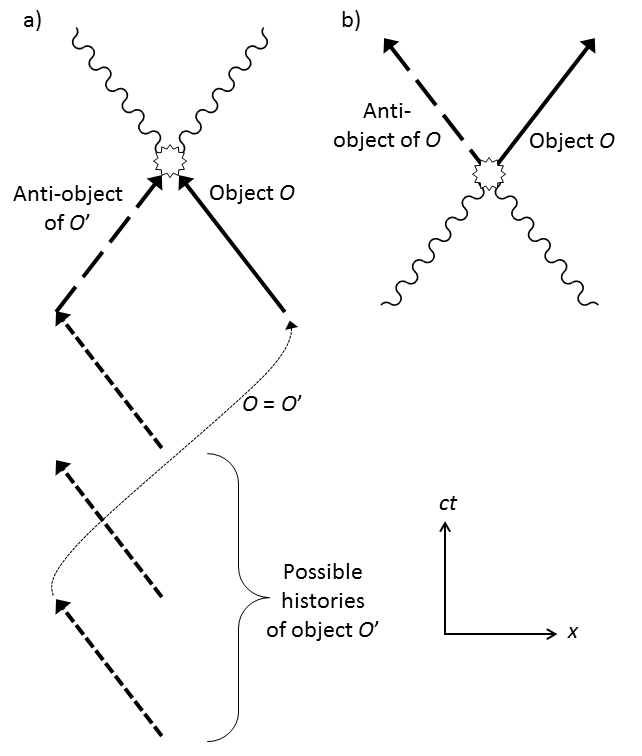}
\end{center}
\caption{a) Pair annihilation of object $O$ and the anti-object that represents the history of object $O'$. The freedoom in the choice of how far back in time we should place the history $O'$ means that $O'$ and $O$ can be identified. The object $O$ and its history disappear together. b) Pair production. The object $O$ and the anti-object that represents its history appear together.} 
\label{Fig82c}
\end{figure}

Is this allegoric description really consistent with the formalism? Consider Fig. \ref{Fig82c}(a). The natural point of view is that the annihilating object $O$ and the anti-object $O'$ are two different objects. But the constant $t_{0}$ in Eq. \ref{timemirror} is arbitrary (Fig. \ref{Fig81b}). We can move the history of the object $O'$ that is deduced from its anti-object back in time so far that we cannot exclude the possibility that it is, in fact, the history of $O$. That is, we cannot exclude that $O'$ and $O$ are the same object. Then the interpretation of the annihilation conforms with the allegoric description above. According to our view on identifiability, if the identity of two objects cannot be excluded, they should be regarded to be the same. Thus $O'=O$. This conclusion may seem far-fetched, but reflect the basic difficulty in keeping track of individual minimal objects (elementary fermions). Note that, in the same way, we have arbitrarily moved the dashed line segments between $-t_{2},-t_{1}$ and $t=0$ in Fig. \ref{Fig82} up and down along the temporal axis to make them form a connected historical world line of a single object.

The photons associated with pair production and annihilation may be seen as book-keeping symbols that take care of the conservation of the expected momentum and energy as time passes, in the absense of the objects themselves. To realize a pair production, there has to be enough energy to withdraw from your account to buy the rest energy of these objects. Conversely, if you let two objects annihilate, the amount of energy in your account grows, energy that may be exchanged for new objects in the future.

Since the rest mass of a photon is strictly zero, we cannot, according to the discussion in section \ref{evconsequences}, see them as minimal objects, as building blocks of other objects. In fact, if we insist on looking at photons as objects, we have to wait an infinitely long time before we can actually observe them, according to the uncertainty relation [\ref{masslifetime}]. They cannot be localized in time and space according to Eq. [\ref{masslesswavefunction}]. The only information they carry is energy and momentum, as already stated. The interpretation of elementary bosons is discussed further in section \ref{fermbos}.

\begin{figure}[tp]
\begin{center}
\includegraphics[width=80mm,clip=true]{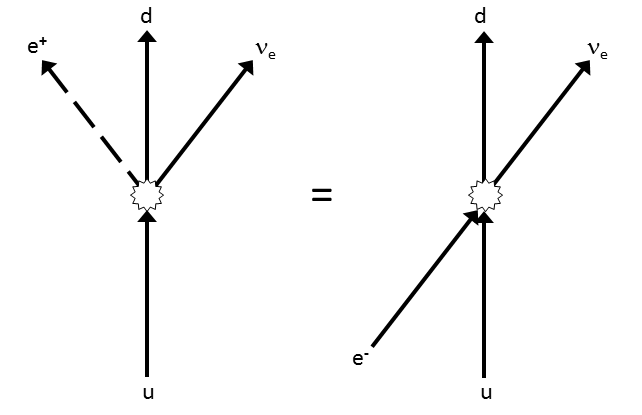}
\end{center}
\caption{Two equivalent descriptions of beta plus decay. A process which is knowably of the second kind is called K-capture. In that case we can deduce that the process took place in an atomic nucleus, and that the electron was taken from the K-shell of the atom. Otherwise we can only deduce the history of the electron via the apparent presence of a positron.} 
\label{Fig82d}
\end{figure}

Of course, there are other processes involving anti-objects in which objects merge or divide than pair production and annihilation. Figure \ref{Fig82d} shows beta plus decay. If we insist on the interpretation that anti-objects always represent the unknown history of other objects, the two processes shown in the figure should be seen as one and the same. A positron may be emitted if the existence of the electron that went into the reaction was outside potential knowledge before the reaction. Otherwise, the process should be described as in the right panel. Indeed, this process occurs in the atomic nucleus exactly when we have a previously knowable electron in the K-shell of the atom that is captured by the $u$-quark in the nucleus. Then there is no need to introduce a positron to account for the history of the electron - it is already known. The occurrence of electron capture can be deduced via the associated emission of X-rays or Auger electrons. If we check for such particles but do not see them, then we have to see a positron emission.

To summarize, the basic difference between the traditional and the present view on antimatter is that instead of speaking about particles travelling backwards in time (whatever that means), we speak here about particles whose histories we \emph{track} backwards in time. This tracking sometimes requires the appearance of anti-particles.

The reasoning behind this conclusion rests upon the epistemic perspective on time. We use the distinction between present and past as an input in the physical description, and get as output the distinction between matter and antimatter. In other words, the present intepretation of antimatter is an application of epistemic completeness (Assumption \ref{epcompleteness}). The structure of the evolution equation and the Dirac equation honors this distinction in the sense that they allow pairs of distinct solutions with energies $E$ and $-E$.

\section{The reciprocal evolution equation}
\label{reciprocaleq}

The purpose of the present and the next section is to explore the symmetries between the equations that hold for the position and the momentum properties of a specimen, respectively, and to discuss some consequences of these symmetries.

From Eqs. [\ref{ev1}], [\ref{evbop}] and [\ref{dalembert}], we see that we can express the evolution equation for the continuous wave function for a free specimen as

\begin{equation}
\frac{d\Psi_{\mathbf{r}_{4}}}{d\sigma}=-ib\Box_{\mathbf{r}_{4}}\Psi_{\mathbf{r}_{4}}
\label{eveqfree}
\end{equation}
for some real constant $b$, with

\begin{equation}
\Box_{\mathbf{r}_{4}}\equiv-\sum_{k=1}^{4}\frac{\partial^{2}}{\partial r_{k}^{2}}
=\frac{1}{c^{2}}\frac{\partial^{2}}{\partial t^{2}}-\frac{\partial^{2}}{\partial x^{2}}-\frac{\partial^{2}}{\partial y^{2}}-\frac{\partial^{2}}{\partial z^{2}}.
\end{equation}

To derive this equation, we assumed that the parametrization is such that $d\langle\mathbf{r}_{4}\rangle/d\sigma$ is a constant. However, once we have established Eq. [\ref{eveqfree}], we may lift this condition, and make an arbitrary re-parametrization $\sigma'=f(\sigma)$. This is so since the general evolution operator $\bar{A}_{\mathbf{r}_{4}}=i\bar{B}_{\mathbf{r}_{4}}$ does not depend explicitly on the evolution parameter $\sigma$ (see Eq. [\ref{cbasicev}]). The wave function evolution operator cannot have such a dependence since $\sigma$ is just an unphysical parameter that interpolates between the repeated action of the basic evolution operator $u_{1}$ according to Eq. [\ref{sigmadef}]. We may thus write

\begin{equation}
\frac{d\Psi_{\mathbf{r}_{4}}}{d\sigma}=-ib(\sigma)\Box_{\mathbf{r}_{4}}\Psi_{\mathbf{r}_{4}}
\label{eveqfree2}
\end{equation}
for an arbitrary continuous, real function $b(\sigma)$. It is easily checked that the energy-momentum relation [\ref{einsteinrelation}] still holds for this equation, if we generalize the definition of rest mass (Definition \ref{masssquaredefi}) to read $m_{0}^{2}\equiv-\hbar\tilde{\sigma}/b(\sigma)c^{2}$.

We may expand the wave function as a Fourier integral

\begin{equation}
\Psi_{\mathbf{r}_{4}}(\mathbf{r}_{4},\sigma)=(2\pi)^{-5/2}\int_{-\infty}^{\infty}\tilde{\Psi}_{\mathbf{r}_{4}}(\tilde{\mathbf{r}}_{4},\tilde{\sigma})
e^{i(\tilde{\mathbf{r}}_{4}\cdot\mathbf{r}_{4}+\tilde{\sigma}\sigma)}d\tilde{\mathbf{r}}_{4}d\tilde{\sigma}
\label{fexpansioncopy}
\end{equation}
with inverse

\begin{equation}
\tilde{\Psi}_{\mathbf{r}_{4}}(\tilde{\mathbf{r}}_{4},\tilde{\sigma})=(2\pi)^{-5/2}\int_{-\infty}^{\infty}\Psi_{\mathbf{r}_{4}}(\mathbf{r}_{4},\sigma)
e^{-i(\tilde{\mathbf{r}}_{4}\cdot\mathbf{r}_{4}+\tilde{\sigma}\sigma)}d\mathbf{r}_{4}d\sigma.
\label{fexpansioninverse}
\end{equation}

Consider the evolution equation [\ref{eveqfree}]. By mere symmetry, we may expect that the reciprocal wave function $\tilde{\Psi}_{\mathbf{r}_{4}}(\tilde{\mathbf{r}}_{4},\tilde{\sigma})$ fulfils the equation

\begin{equation}
\frac{d\tilde{\Psi}_{\mathbf{r}_{4}}}{d\tilde{\sigma}}=-i\tilde{b}(\tilde{\sigma})\Box_{\tilde{\mathbf{r}}_{4}}\tilde{\Psi}_{\mathbf{r}_{4}}
\label{reciprocaleveqfree}
\end{equation}
for some continuous, real function $\tilde{b}(\tilde{\sigma})$, with

\begin{equation}
\Box_{\tilde{\mathbf{r}}_{4}}\equiv-\sum_{k=1}^{4}\frac{\partial^{2}}{\partial \tilde{r}_{k}^{2}}.
\end{equation}
But is this really so? We motivated the evolution equation for a free specimen from the two conditions at the beginning of section \ref{eveq}. The condition $d\langle\tilde{\mathbf{r}}_{4}\rangle/d\tilde{\sigma} = const.$ that is `reciprocal' to the second of these conditions cannot be used to motivate Eq. [\ref{reciprocaleveqfree}], since it does not (yet) have any physical interpretation. Recall that we \emph{defined} momentum, energy and rest energy from the Fourier expansion coefficients $(\tilde{\mathbf{r}}_{4},\tilde{\sigma})$, and that we should not use any other knowledge about the behaviour of these physical quantities to motivate an equation that the expansion coefficients may fulfil.

Another way forward is to note that the evolution equation [\ref{eveqfree2}] implies Eq. [\ref{protoeinstein}], which we repeat here:

\begin{equation}
-\tilde{r}_{4}^{2}=\frac{\tilde{\sigma}}{b(\sigma)}+\tilde{r}_{1}^{2}+\tilde{r}_{2}^{2}+\tilde{r}_{3}^{2}.
\label{protoeinsteincopy}
\end{equation}
This relation holds for all allowed quantity pairs $(\tilde{\mathbf{r}}_{4},\tilde{\sigma})$. In the same way, Eq. [\ref{reciprocaleveqfree}] implies

\begin{equation}
-r_{4}^{2}=\frac{\sigma}{\tilde{b}(\tilde{\sigma})}+r_{1}^{2}+r_{2}^{2}+r_{3}^{2}
\label{reciprocaleinstein}
\end{equation}
for all allowed quantity pairs $(\mathbf{r}_{4},\sigma)$. In fact, the implication goes both ways, so that Eq. [\ref{reciprocaleinstein}] implies Eq. [\ref{reciprocaleveqfree}]. We may interpret Eq. [\ref{reciprocaleinstein}] as the definition of the Lorentz distance

\begin{equation}
l^{2}=c^{2}t^{2}-x^2-y^{2}-z^{2}
\end{equation}
if we let

\begin{equation}
l^{2}=\frac{\sigma}{\tilde{b}(\tilde{\sigma})}.
\label{spara}
\end{equation}
This interpretation is allowed if and only if
\begin{equation}
\tilde{b}(\tilde{\sigma})>0
\end{equation}
for all $\tilde{\sigma}$, since, by definition, the evolution parameter $\sigma$ is non-negative. In this way we get $l^{2}\geq 0$, and the distance element $\sqrt{l^{2}}$ in Minkowski space-time becomes invariant under Lorentz transformations, as required. We conclude that Eq. [\ref{reciprocaleveqfree}] follows from special relativity.

Note that the parametrization [\ref{spara}] is different from the natural parametrization [\ref{naturalp}] that we used to express the evolution equation (Statement \ref{psieveq}) with the Dirac constraint (Statement \ref{diracconstraint}). The former parametrization means that $\sigma = t +\sigma_{0}$ for some constant $\sigma_{0}$, whereas the latter parametrization means that $\sigma\propto t^{2}+\sigma_{0}'$ for some other constant $\sigma_{0}'$. We are allowed to use a different parametrization to motivate Eq. [\ref{reciprocaleveqfree}] than that we used to motivate Eq. [\ref{eveqfree}] since the parametrization is arbitrary, as expressed in Eq. [\ref{eveqfree2}].

\vspace{5mm}
\begin{center}
$\maltese$
\end{center}
\paragraph{}

To interpret Eq. [\ref{reciprocaleveqfree}] physically, it is natural to try to identify the reciprocal wave function $\tilde{\Psi}_{\mathbf{r}_{4}}$ for a context $C$ in which we are about to observe $\mathbf{r}_{4}$, with the wave function $\tilde{\Psi}_{\mathbf{p}_{4}}$ in the reciprocal context $\tilde{C}$ in which we are about to observe $\mathbf{p}_{4}$. It is the latter wave function that has physical meaning. Let us discuss how to do this.

Let us form the scalar product $\langle\bar{S}_{\mathbf{r}_{4}}(\mathbf{r}_{4}),\bar{S}_{C\tilde{C}}\rangle$ using Eq. [\ref{contbasechange}], with the choice of properties $P=\mathbf{r}_{4}$ and $P'=\mathbf{p}_{4}$. We get

\begin{equation}
\Psi_{\mathbf{r}_{4}}(\mathbf{r}_{4},\sigma)=\int_{-\infty}^{\infty}\tilde{\Psi}_{\mathbf{p}_{4}}(\mathbf{p}_{4},\sigma)d\mathbf{p}_{4}
\langle\bar{S}_{\mathbf{r}_{4}}(\mathbf{r}_{4}),\bar{S}_{\mathbf{p}_{4}}(\mathbf{p}_{4})\rangle,
\label{crcrel}
\end{equation}
since $\langle\bar{S}_{\mathbf{r}_{4}}(\mathbf{r}_{4}),\bar{S}_{\mathbf{r}_{4}}(\mathbf{r}_{4}')\rangle=\delta(\mathbf{r}_{4}-\mathbf{r}_{4}')$ according to Eq. [\ref{deltarel}].

In the definition of four-momentum (Definition \ref{fourmomentumdefi}) we identified the corresponding wave function operator to be

\begin{equation}
(\overline{\mathbf{p}_{4}})_{\mathbf{r}_{4}}=i\hbar\Box_{\mathbf{r}_{4}}.
\end{equation}
This expression holds for a free specimen in the context $C$ in which we are about to observe $\mathbf{r}_{4}$, before we are going to observe $\mathbf{p}_{4}$. Consider a context $C$ such that the value of the four-momentum is exactly known to be $\mathbf{p}_{4}'$ before the observation of $\mathbf{r}_{4}$ takes place. Then, according to Definition [\ref{conteigenfunction}], the contextual state $\bar{S}_{C}$ is described by an eigenfunction to the continuous wave function operator $(\overline{\mathbf{p}_{4}})_{\mathbf{r}_{4}}$ at this moment, that is

\begin{equation}
\psi_{\mathbf{r}_{4}}(\mathbf{r}_{4})\propto e^{\frac{i}{\hbar}\mathbf{p}_{4}'\cdot \mathbf{r}_{4}}.
\label{eigendiff}
\end{equation}

The contextual state $\bar{S}_{\tilde{C}}$ of the reciprocal context $\tilde{C}$ will, on the other hand, be described by the wave function

\begin{equation}
\tilde{\psi}_{\mathbf{p}_{4}}(\mathbf{p}_{4})=\beta\delta(\mathbf{p}_{4}-\mathbf{p}_{4}')
\label{eigensame}
\end{equation}
just before the observation of $\mathbf{p}_{4}$, for some complex constant $\beta$ such that $|\beta|=1$. This relation follows from Definition \ref{conteigenfunction} in the case $P=P'$.

If we insert the relations [\ref{eigendiff}] and [\ref{eigensame}] in Eq. [\ref{crcrel}] we see that

\begin{equation}
\langle\bar{S}_{\mathbf{r}_{4}}(\mathbf{r}_{4}),\bar{S}_{\mathbf{p}_{4}}(\mathbf{p}_{4})\rangle=(2\pi)^{-2}e^{\frac{i}{\hbar}\mathbf{p}_{4}\cdot \mathbf{r}_{4}}.
\end{equation}

In the general case when we do not assume that the initial state in $C$ or $\tilde{C}$ is such that the four-momentum is exactly known, we may insert the above relation in Eq. [\ref{crcrel}] to find the following relation between the wave functions $\Psi_{\mathbf{r}_{4}}$ and $\tilde{\Psi}_{\mathbf{p}_{4}}$:

\begin{equation}
\Psi_{\mathbf{r}_{4}}(\mathbf{r}_{4},\sigma)=(2\pi)^{-2}\int_{-\infty}^{\infty}\tilde{\Psi}_{\mathbf{p}_{4}}(\mathbf{p}_{4},\sigma)
e^{\frac{i}{\hbar}\mathbf{p}_{4}\cdot \mathbf{r}_{4}}d\mathbf{p}_{4}.
\label{wavefrec}
\end{equation}
The factor $(2\pi)^{-2}$ appearing in the two equations above arises because of the condition that all wave functions should always be normalized (Statement \ref{cnormalized}).

Let us define the reciprocal wave function $\tilde{\Psi}_{\mathbf{r}_{4}}(\mathbf{p}_{4},\sigma)$ as the kernel that appears in the Fourier expansion

\begin{equation}
\Psi_{\mathbf{r}_{4}}(\mathbf{r}_{4},\sigma)=(2\pi)^{-2}\int_{-\infty}^{\infty}\tilde{\Psi}_{\mathbf{r}_{4}}(\tilde{\mathbf{r}}_{4},\sigma)
e^{i\tilde{\mathbf{r}}_{4}\cdot\mathbf{r}_{4}}d\tilde{\mathbf{r}}_{4}.
\label{recwavef}
\end{equation}
Comparing Eqs. [\ref{wavefrec}] and [\ref{recwavef}] we see that

\begin{equation}
\tilde{\Psi}_{\mathbf{p}_{4}}(\mathbf{p}_{4},\sigma)=\frac{1}{\hbar}\tilde{\Psi}_{\mathbf{r}_{4}}(\frac{\mathbf{p}_{4}}{\hbar},\sigma).
\label{recwfiswfrec}
\end{equation}
That is, the reciprocal wave function associated with the context $C$ can be identified with the wave function associated with the reciprocal context $\tilde{C}$.

We may now define a wave function $\tilde{\Psi}_{\mathbf{p}_{4}}(\mathbf{p}_{4},\tilde{\sigma})$ associated with a family $\tilde{C}(\tilde{\sigma})$ of reciprocal context $\tilde{C}$ as follows:

\begin{equation}
\tilde{\Psi}_{\mathbf{p}_{4}}(\mathbf{p}_{4},\tilde{\sigma})\equiv(2\pi)^{-1/2}\int_{-\infty}^{\infty}
\tilde{\Psi}_{\mathbf{p}_{4}}(\mathbf{p}_{4},\sigma)e^{-i\tilde{\sigma}\sigma}d\sigma.
\label{recwfiswfrec2}
\end{equation}
We see that a given reciprocal context in the family $\tilde{C}(\tilde{\sigma})$ does not correspond to a given reciprocal context in the family $\tilde{C}(\sigma)$, but is a function of the entire family. In other words, if the reciprocal evolution parameter $\tilde{\sigma}$ has a given exact value, then the value of the evolution parameter $\sigma$ is completely undetermined. Of course, the same statements hold if we let $\tilde{\sigma}$ and $\sigma$ change roles.

If we consider Eqs. [\ref{fexpansioncopy}], the inverse of [\ref{recwavef}], [\ref{recwfiswfrec}], and [\ref{recwfiswfrec2}], we see that

\begin{equation}\begin{array}{lll}
\tilde{\Psi}_{\mathbf{p}_{4}}(\mathbf{p}_{4},\tilde{\sigma}) & = & \frac{1}{\hbar}\tilde{\Psi}_{\mathbf{r}_{4}}(\frac{\mathbf{p}_{4}}{\hbar},\tilde{\sigma})\\
& &\\
& = & \frac{(2\pi)^{-5/2}}{\hbar}\int_{-\infty}^{\infty}\Psi_{\mathbf{r}_{4}}(\mathbf{r}_{4},\sigma)
e^{-\frac{i}{\hbar}(\mathbf{p}_{4}\cdot\mathbf{r}_{4}+\hbar\tilde{\sigma}\sigma)}d\mathbf{r}_{4}d\sigma.
\end{array}
\label{cwfp}
\end{equation}

Let us insert this relation in Eq. [\ref{reciprocaleveqfree}]. We get

\begin{equation}
\frac{d\tilde{\Psi}_{\mathbf{p}_{4}}}{d\tilde{\sigma}}=-i\tilde{b}(\tilde{\sigma})\hbar\Box_{\mathbf{p}_{4}}\tilde{\Psi}_{\mathbf{p}_{4}},
\label{reciprocaleveqfree2}
\end{equation}
with

\begin{equation}
\Box_{\mathbf{p}_{4}}\equiv-\sum_{k=1}^{4}\frac{\partial^{2}}{\partial p_{k}^{2}}
=c^{2}\frac{\partial^{2}}{\partial E^{2}}-\frac{\partial^{2}}{\partial p_{x}^{2}}-\frac{\partial^{2}}{\partial p_{y}^{2}}-\frac{\partial^{2}}{\partial p_{z}^{2}}.
\label{pdalembert}
\end{equation}
We call this equation \emph{the reciprocal evolution equation}. It is an equation that relates physical quantities and therefore expresses physical law, in contrast to Eq. [\ref{reciprocaleveqfree}], in which the abstract kernel $\tilde{\Psi}_{\mathbf{r}_{4}}(\mathbf{r}_{4},\tilde{\sigma})$ in the Fourier expansion [\ref{fexpansioncopy}] appears. We may say that Eq. [\ref{reciprocaleveqfree2}] is the physical interpretation of Eq. [\ref{reciprocaleveqfree}], where we relate the Fourier expansion coefficients $\mathbf{r}_{4}$ with four-momentum, and the kernel $\tilde{\Psi}_{\mathbf{r}_{4}}(\mathbf{r}_{4},\tilde{\sigma})$ with the wave function that describes the reciprocal context $\tilde{C}$.

\begin{state}[\textbf{The reciprocal evolution equation}]
Equation [\ref{reciprocaleveqfree2}] holds whenever we parametrize the evolution so that $\sigma$ fulfils Eq. [\ref{spara}]. It specifies the probabilities for the possible outcomes of an observation of $\mathbf{p}_{4}$ in the family of contexts $\tilde{C}(\tilde{\sigma})$, defined by Eq. [\ref{recwfiswfrec2}], at some time $n+m$ given the contextual state at initial time $n$.
\label{recistate}
\end{state}

Note that the reciprocal evolution equation makes a statement about temporal evolution even though the reciprocal evolution parameter $\tilde{\sigma}$ is not related to temporal (or spatial) distances. The statement concerns sequential time, no matter if the parameter that defines the family of contexts refers to relational time or not. Below we will relate $\tilde{\sigma}$ to the expected energy of the specimen before we actually observe it (as part of the observation of $\mathbf{p}_{4}$).

For an exactly known four-position $\mathbf{r}_{4}'$ we have $\Psi_{\mathbf{r}_{4}}(\mathbf{r}_{4})=\beta\delta(\mathbf{r}_{4}-\mathbf{r}_{4}')$ for some constant $\beta$ such that $|\beta|=1$. We see from Eq. [\ref{cwfp}] that in this situation we have

\begin{equation}
i\hbar\left(\frac{\partial}{\partial p_{1}},\frac{\partial}{\partial p_{2}},\frac{\partial}{\partial p_{3}},\frac{\partial}{\partial p_{4}}\right)
\tilde{\Psi}_{\mathbf{p}_{4}}=\mathbf{r}_{4}'\tilde{\Psi}_{\mathbf{p}_{4}}.
\end{equation}
That is, the known position $\mathbf{r}_{4}'$ is the eigenvalue of the self-adjoint operator on the left hand side. This operator can therefore be interpreted as the continuous wave function operator $(\overline{\mathbf{r}_{4}})_{\mathbf{p}_{4}}$.

\begin{state}[\textbf{The wave function operator for four-position}]
The continuous wave function operator for four-position $\mathbf{r}_{4}$ for a free specimen is

\begin{equation}
(\overline{\mathbf{r}_{4}})_{\mathbf{p}_{4}}=
i\hbar\left(\frac{\partial}{\partial p_{1}},\frac{\partial}{\partial p_{2}},\frac{\partial}{\partial p_{3}},\frac{\partial}{\partial p_{4}}\right).
\end{equation}
when the wave function $\Psi$ is expressed in terms of four-momentum $\mathbf{p}_{4}$.
\label{posop}
\end{state}

This statement should be compared with Definition \ref{fourmomentumdefi}.

The reciprocal evolution equation as given by Statement \ref{recistate} contains one arbitrary parameter, namely $\tilde{b}$. In the ordinary evolution equation, we got rid of the corresponding arbitrary parameter $b$ by choosing the natural parametrization

\begin{equation}
\frac{d\langle t\rangle}{d\sigma}=1.
\label{natpara}
\end{equation}
By symmetry, the natural parametrization for $\tilde{\sigma}$ becomes

\begin{equation}
\frac{d\langle E\rangle}{d\tilde{\sigma}}=1.
\label{rnatpar}
\end{equation}
The relation that corresponds to Eq. [\ref{meandr}] in the reciprocal picture is

\begin{equation}
\frac{d\langle\mathbf{p}_{4}\rangle}{d\tilde{\sigma}}=-2\tilde{b}\langle\mathbf{r}_{4}\rangle.
\label{meandp}
\end{equation}
This relation can be verified straightforwardly by an evaluation of the left hand side, using the reciprocal evolution equation. We may use Eq. [\ref{meandp}] to express a `reciprocal Ehrenfest theorem' (compare Eq. [\ref{ehrenfest}]). Namely, if we again define $m$ as the relativistic mass according to $E=mc^{2}$, we get

\begin{equation}
\frac{d\langle\mathbf{p}\rangle}{d\tilde{\sigma}}/\frac{d\langle m\rangle}{d\tilde{\sigma}}=\frac{\langle\mathbf{r}\rangle}{\langle t\rangle}.
\label{rehrenfest}
\end{equation}
This relation simplifies to $d\langle\mathbf{p}\rangle/d\langle m\rangle=\langle\mathbf{r}\rangle/\langle t\rangle$ in the natural parametrization $[\ref{rnatpar}]$. If we apply this parametrization to the fourth component of Eq. [\ref{meandp}] we get

\begin{equation}
\tilde{b}=-\frac{1}{2c^{2}\langle t\rangle}.
\end{equation}

In the pair of natural parametrizations [\ref{natpara}] and [\ref{rnatpar}], we may therefore express the pair of reciprocal evolution equations as follows.

\begin{equation}\begin{array}{rcl}
d\Psi_{\mathbf{r}_{4}}/d\langle t\rangle & = & \frac{ic^{2}\hbar}{2\langle E\rangle}\Box_{\mathbf{r}_{4}}\Psi_{\mathbf{r}_{4}}\\
& & \\
d\tilde{\Psi}_{\mathbf{p}_{4}}/d\langle E\rangle & = & \frac{i\hbar}{2c^{2}\langle t\rangle}\Box_{\mathbf{p}_{4}}\tilde{\Psi}_{\mathbf{p}_{4}}
\end{array}
\label{receqpair}
\end{equation}

Here we have replaced the derivatives on the left hand sides with respect to $\sigma$ and $\tilde{\sigma}$ with derivatives with respect to the expected time and energy, in order to get equations that relate physical properties only. We have to remember, however, that the expected time or energy is never exactly known in any real context (compare the disscussion in section \ref{knowlaw}).

We may say that the ordinary equation expresses the change of the probabilites for different outcomes in the family of contexts $C(\sigma)$ as we increase the expected value of the relational time to be measured at sequential time $n+m$. The reciprocal evolution equation, on the other hand, expresses the change of the probabilites in the family $\tilde{C}(\tilde{\sigma})$ as we increases the expected energy of the specimen to be measured at sequential time $n+m$.

In the reciprocal family $\tilde{C}(\tilde{\sigma})$, the expected values of the spatio-temporal properties $\mathbf{r},t,l^{2}$ do not change as the expected energy increases, or when $\mathbf{p}_{4}$ is finally observed. They are `passive passengers' in the family. In the same way, in the family $C(\sigma)$, the expected values of the momentum-energy properties $\mathbf{p},E,E_{0}^{2}$ do not change as the expected passed time increases, or when $\mathbf{r}_{4}$ is finally observed. The are passive passengers in the family $C(\sigma)$.

We may express the general solutions to Eqs. [\ref{receqpair}] as follows. 

\begin{equation}\begin{array}{rcl}
\Psi_{\mathbf{r}_{4}}(\mathbf{r}_{4},\langle t\rangle) & = & (2\pi)^{-5/2}\int\tilde{\Psi}_{\mathbf{p}_{4}}(\mathbf{p}_{4},E_{0}^{2})\exp[\frac{i}{\hbar}(\mathbf{p}_{4}\cdot\mathbf{r}-\frac{E_{0}^{2}}{2\langle E\rangle}\langle t\rangle)]d\mathbf{p}_{4}dE_{0}^{2}\\
& & \\
\tilde{\Psi}_{\mathbf{p}_{4}}(\mathbf{p}_{4},\langle E\rangle)& = & \frac{(2\pi)^{-5/2}}{\hbar}\int\Psi_{\mathbf{r}_{4}}(\mathbf{r}_{4},l^{2})\exp[\frac{i}{\hbar}(\mathbf{p}_{4}\cdot\mathbf{r}-\frac{l^{2}}{2c^{2}\langle t\rangle}\langle E\rangle)]d\mathbf{r}_{4}dl^{2}
\end{array}
\label{recsolutions}
\end{equation}

Note that the function $\Psi_{\mathbf{r}_{4}}$ that appears in the first equation is not the same as that appearing in the second equation; they are the result of different parametrizations, as discussed above. In the first equation we use $\sigma\propto\langle t\rangle$, and in the second equation we use $\sigma\propto l^{2}$. Likewise, the functions $\tilde{\Psi}_{\mathbf{p}_{4}}$ that appear in the first and second equations are the result of the different parametrizations $\tilde{\sigma}\propto E_{0}^{2}$ and $\tilde{\sigma}\propto\langle E\rangle$, respectively.

Put differently, $\Psi_{\mathbf{r}_{4}}(\mathbf{r}_{4},l^{2})$ represents the state in a hypothetical context $C$ such that the spatio-temporal distance $l^{2}$ between the initial and final positions of the specimen is exactly known just before the observation of $\mathbf{r}_{4}$ is made. On the other hand, no such exact \emph{a priori} knowledge is assumed in any realistic family of contexts $C(\langle t\rangle)$ described by $\Psi_{\mathbf{r}_{4}}(\mathbf{r}_{4},\langle t\rangle)$. Likewise, $\tilde{\Psi}_{\mathbf{p}_{4}}(\mathbf{p}_{4},E_{0}^{2})$ describes a hypothetical context $\tilde{C}$ in which the rest energy is precisely known just before the observation of $\mathbf{p}_{4}$, whereas $\tilde{\Psi}_{\mathbf{p}_{4}}(\mathbf{r}_{4},\langle E\rangle)$ describes a realistic family of contexts $\tilde{C}(\langle E\rangle)$ in which we have no such \emph{a priori} knowledge.

It may seem confusing to have both the value of a property and its expected value appearing in the same equation. In the evolution equation both $t$ and $\langle t\rangle$ appear, and in the reciprocal evolution equation both $E$ and $\langle E\rangle$ appear. The interpretation is that the evolution equations determine the probabilities for different property values ($t$ or $E$) given their expected values \emph{just before} they are actually observed.

It is clearly seen in Eq. [\ref{recsolutions}] that the squared spatio-temporal distance $l^{2}$ plays the same role in the reciprocal evolution equation as does the squared rest mass $m_{0}^{2}$ or rest energy $E_{0}^{2}$ in the evolution equation. They are symmetrically chosen as

\begin{equation}\begin{array}{rcl}
l^{2} & = & \sigma/\tilde{b}(\tilde{\sigma})\\
m_{0}^{2}c^{2}/\hbar & = & \tilde{\sigma}/b(\sigma)
\end{array}
\end{equation}
in the pair of reciprocal evolution equations, reflecting their analogous roles in the relations

\begin{equation}\begin{array}{rcccl}
l^{2}& = & -|\mathbf{r}_{4}|^{2} & = & c^{2}t^{2}-|\mathbf{r}|^{2}\\
m_{0}^{2}c^{2} & = & -|\mathbf{p}_{4}|^{2} & = & c^{-2}E^{2}-|\mathbf{p}|^{2}.
\end{array}
\end{equation}

\begin{figure}[tp]
\begin{center}
\includegraphics[width=80mm,clip=true]{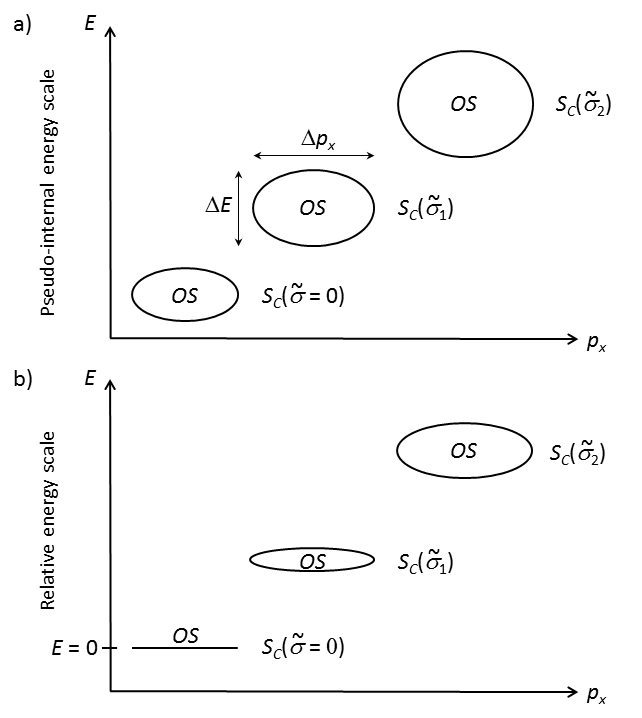}
\end{center}
\caption{Two possible coordinate systems in reciprocal space that are analogous to the two possible coordinate systems in ordinary space shown in Fig. \ref{Fig74}. a) Energy $E$ in a coordinate system defined so that the reciprocal evolution parameter $\tilde{\sigma}$ has no particular relation to $E$, except for the requirement $E>0$. b) Energy defined in relation to a parametrization such that $\tilde{\sigma}=0$ corresponds to $E=0$. In this case we must have $\Delta E\leq E$ because of the condition $E>0$.}
\label{Fig74b}
\end{figure}

In other words, we may say that $l^{2}$ is conjugate to $m_{0}^{2}$ on the same way as $\mathbf{r}_{4}$ is conjugate to $\mathbf{p}_{4}$, or, dividing the four-vectors into their Newtonian parts:

\begin{equation}\begin{array}{rcl}
\mathbf{r} & \leftrightarrow & \mathbf{p}\\
t & \leftrightarrow & E\\
l^{2} & \leftrightarrow & m_{0}^{2}
\end{array}
\end{equation}

This list of conjugate properties naturally leads us back to the commutators and uncertainty relations discussed in section \ref{evconsequences}. To be precise, we should reformulate Statement \ref{uncertain1} as follows, since we are now dealing with observations of $\mathbf{p}_{4}$ rather than $\mathbf{r}_{4}$.

\begin{state}[\textbf{Uncertainty relations for momentum-energy observations}]
Consider a context $\tilde{C}$ initiated at time $n$, and such that $\mathbf{p}_{4}=(\mathbf{p},iE/c)$ is observed at time $n+m$. Let $\Delta x$ and $\Delta t$ be the uncertainties of the position $(x,t)$ in space and time of the specimen $OS$ in the evolved state $u_{m}S_{OS}(n)$. Let $\Delta p_{x}$ and $\Delta E$ be the uncertainties of the momentum and energy observed at time $n+m$. Then $\Delta p_{x}\geq\hbar/2\Delta x$ and $\Delta E\geq \hbar/2\Delta t$.
\label{uncertain2}
\end{state}
Note that here we use the pseudo-internal coordinate system shown in Fig. \ref{Fig74}(b) in space and time, whereas we use a relational coordinate system for energy and momentum, as illustrated in Fig. \ref{Fig74b}(b).

We may re-express the commutator [\ref{newcom}] in the natural parametrization as

\begin{equation}
[\langle t\rangle,(\overline{E_{0}^{2}})_{\mathbf{r}_{4}}]=2i\hbar\langle E\rangle,
\label{newcom2}
\end{equation}
where we have added an index to the squared rest energy operator to indicate that it applies to the family of contexts $C(\sigma)$ in which $\mathbf{r}_{4}$ is observed. The corresponding uncertainty relation is

\begin{equation}
\Delta\langle t\rangle\Delta E_{0}^{2}\geq\hbar\langle E\rangle.
\end{equation}

By symmetry, we find the following corresponding relations in the reciprocal family of contexts $\tilde{C}(\tilde{\sigma})$ in which $\mathbf{p}_{4}$ is observed.

\begin{equation}
[\langle E\rangle,(\overline{l^{2}})_{\mathbf{p}_{4}}]=2ic^{2}\hbar\langle t\rangle,
\label{newcom3}
\end{equation}
and

\begin{equation}
\Delta\langle E\rangle\Delta l^{2}\geq c^{2}\hbar\langle t\rangle.
\end{equation}

\begin{state}[\textbf{Lorentz distance uncertainty}]
Consider a context $\tilde{C}$ initiated at time $n$, and such that $\mathbf{p}_{4}=(\mathbf{p},iE/c)$ is observed at time $n+m$. Let $\Delta l^{2}$ be the uncertainty of the squared Lorentz position $l^{2}=c^{2}t^{2}-r^{2}$ of the specimen $OS$ in the evolved state $u_{m}S_{OS}(n)$. Let $\Delta\langle E\rangle$ be the uncertainty of its expected energy $\langle E\rangle$ observed at sequential time $n+m$, and let $\langle t\rangle$ be the expected passed time. Then $\Delta\langle E\rangle\geq c^{2}\hbar\langle t\rangle/\Delta l^{2}$.
\label{uncertaindist}
\end{state}

Since $\langle E\rangle\geq \Delta \langle E\rangle$ and $\langle t\rangle\geq\langle l\rangle$ we may express the following counterpart to Eq. [\ref{masslifetime}]:

\begin{equation}
\langle E\rangle\geq c^{2}\hbar\langle l\rangle/\Delta l^{2}.
\label{distenergychange}
\end{equation}

Simply put, this means that in a context in which a small (positive) energy is measured, the uncertainty of the spatio-temporal position of the specimen with this energy must be large. Conversely, if the position is precisely known, the measured energy must be large.

\begin{figure}[tp]
\begin{center}
\includegraphics[width=80mm,clip=true]{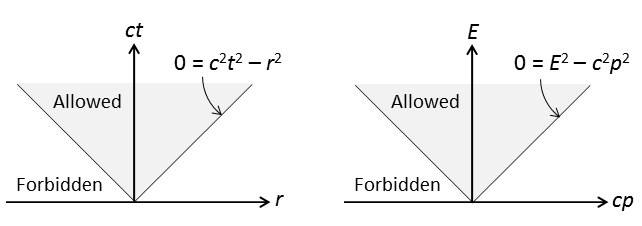}
\end{center}
\caption{Symmetry of the $\mathbf{r}_{4}$- and $\mathbf{p}_{4}$-spaces. Consider a context $C^{+}$ in which the present state of a free specimen is observed (section \ref{antimatter}). Then we always have $E>0$. Since sequential time is directed by definition, we may define $t(n+m)\geq t(n)$, so that we always have $t\geq 0$ in the relative coordinate system (Fig. \ref{Fig74}). The fact that $c$ is an upper speed limit implies that $c^{2}t^{2}-r^{2}\geq 0$. The Dirac equation implies that $E^{2}-c^{2}p^{2}\geq 0$. Compare Fig. \ref{Fig87}} 
\label{Fig83}
\end{figure}

The symmetry of the physical relations between variables in ordinary and reciprocal space can be illustrated as in Fig. \ref{Fig83}. Let us explore this symmetry in order to motivate a reciprocal Dirac equation.

The ordinary Dirac equation (Statement \ref{diracconstraint}) can be seen as a relation that guarantees that $E^{2}-c^{2}p^{2}\geq 0$. This property can therefore be set equal to the square of another property - the rest energy - whose allowed values are the eigenvalues given by the Dirac equation. Equation [\ref{protoeinstein}] is the reason why we have to make sure that $E^{2}-c^{2}p^{2}\geq 0$. This equation implies that the reciprocal evolution parameter $\tilde{\sigma}$ always has the same sign - which we chose to be negative with our sign convention [\ref{signconvention}]. This motivates the introduction of a rest mass according to $m_{0}^{2}\propto-\tilde{\sigma}$. Basically, the reason why $\tilde{\sigma}$ always must have the same sign is that time is directed, and that the derivative in the evolution equation with respect to the evolution parameter $\sigma$ is first order, whereas they are second order with respect to the components of the four-vector $\mathbf{r}_{4}$.  

In a similar way, the fact that $c$ is an upper speed limit means that we have to make sure that $c^{2}t^{2}-r^{2}\geq 0$. This condition is not automatically fulfilled by the reciprocal evolution equation [\ref{reciprocaleveqfree2}]. To fulfil it, we have to introduce the additional constraint on the reciprocal wave function $\tilde{\Psi}_{\mathbf{p}_{4}}$ given by a reciprocal Dirac equation. The lines of reasoning are analogous to those in section \ref{eveq}, where we arrived at the ordinary Dirac equation from the requirement $E^{2}-c^{2}p^{2}\geq 0$.

The symmetry of the equations means that the solutions to the reciprocal Dirac equation are spinors, just as the solutions to the ordinary Dirac equation. Therefore we have to abandon the single continuous reciprocal wave function $\tilde{\Psi}$ and work with a general reciprocal wave function $\tilde{a}$.

\begin{state}[\textbf{The reciprocal Dirac equation as an additional constraint on the reciprocal wave function}]
Each reciprocal wave function $\tilde{\alpha}_{\mathbf{p}_{4}s}((\mathbf{p}_{4})_{i},s_{j},l^{2})\exp\left(-\frac{il^{2}}{2\hbar c^{2}\langle t\rangle}\tilde{\sigma}\right)$ in the general solution
\begin{equation}
\tilde{a}_{\mathbf{p}_{4}s}((\mathbf{p}_{4})_{i},s_{j},\tilde{\sigma})=\int_{-\infty}^{\infty}\tilde{\alpha}_{\mathbf{p}_{4}s}((\mathbf{p}_{4})_{i},s_{j},l^{2})e^{-\frac{il^{2}}{2\hbar c^{2}\langle t\rangle}\tilde{\sigma}}dl^{2}
\end{equation}
to the reciprocal evolution equation in the natural parametrization
\begin{equation}\begin{array}{lll}
i\hbar\frac{d}{d\tilde{\sigma}}\tilde{a}_{\mathbf{p}_{4}s}((\mathbf{p}_{4})_{i},s_{j},\tilde{\sigma}) & = & \frac{1}{2c^{2}\langle t\rangle}(\overline{l^{2}})_{\mathbf{p}_{4}}\tilde{a}_{\mathbf{p}_{4}s}((\mathbf{p}_{4})_{i},s_{j},\tilde{\sigma})\\
& = & \frac{-1}{2 c^{2}\langle t\rangle}\left[(\overline{\mathbf{r}_{4}})_{\mathbf{p}_{4}}\cdot(\overline{\mathbf{r}_{4}})_{\mathbf{p}_{4}}\right]\tilde{a}_{\mathbf{p}_{4}s}((\mathbf{p}_{4})_{i},s_{j},\tilde{\sigma})
\end{array}
\end{equation}
must be an eigenfunction to the continuous Lorentz distance operator $\overline{l}_{\mathbf{p}_{4}s}$:
\begin{equation}
\overline{l}_{\mathbf{p}_{4}s}\alpha_{\mathbf{p}_{4}s}((\mathbf{p}_{4})_{i},s_{j},l^{2})=l\tilde{\alpha}_{\mathbf{p}_{4}s}((\mathbf{p}_{4})_{i},s_{j},l^{2}),
\end{equation}
where $(\overline{l^{2}})_{\mathbf{p}_{4}}=\overline{l}_{\mathbf{p}_{4}s}\overline{l}_{\mathbf{p}_{4}s}$.
\label{recdiracconstraint}
\end{state}

\begin{figure}[tp]
\begin{center}
\includegraphics[width=80mm,clip=true]{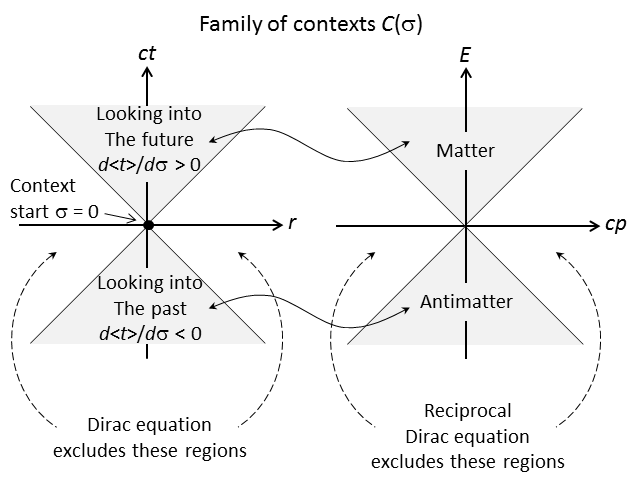}
\end{center}
\caption{More symmetries of the $\mathbf{r}_{4}$- and $\mathbf{p}_{4}$-spaces. The association of contexts in which $d\langle t\rangle/d\sigma>0$ with matter, and contexts with $d\langle t\rangle/d\sigma<0$ with antimatter, is discussed in detail in section \ref{antimatter}. The present illustration is simplified. See text for further explanation.} 
\label{Fig87}
\end{figure}

With both Dirac equations in place, we may continue the exploration in Fig. \ref{Fig83} of the symmetries between ordinary space $\{\mathbf{r}_{4}\}$ and reciprocal space $\{\mathbf{p}_{4}\}$. Figure \ref{Fig87} includes both contexts $C^{+}$ looking at the present, and contexts $C^{-}$ looking into the past, so that a complete light cone is created. This light cone is mirrored by a cone in reciprocal space. We associate the past part of this cone with antimatter, as discussed in section \ref{antimatter}. (We should keep in mind, however, that not all past contexts $C^{-}$ contains antimatter.) The two Dirac equations create the two cones in the sense that they exclude the space-like parts of ordinary space, and the `momentum-like' parts of reciprocal space.

\begin{figure}[tp]
\begin{center}
\includegraphics[width=80mm,clip=true]{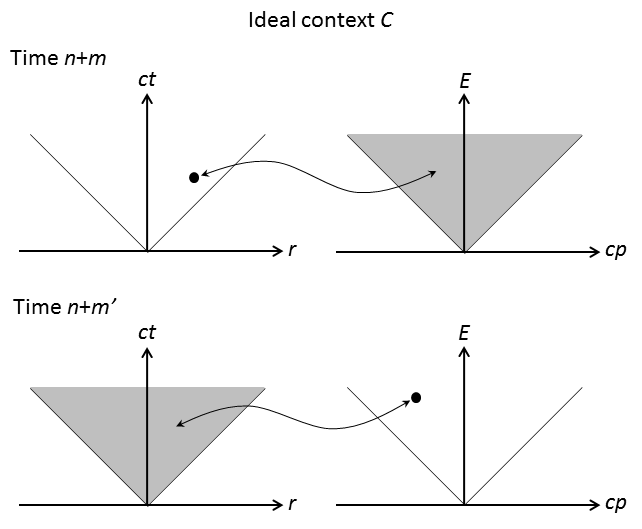}
\end{center}
\caption{Projections of the specimen state onto ordinary and reciprocal space in an idealized context such as that in Fig. \ref{Fig76c}. The projections correspond to the black points and the grey areas. The symmetry between the projections onto the two spacs can be seen as an expression of Bohr's complementarity principle. Compare Fig. \ref{Fig85}.} 
\label{Fig84}
\end{figure}

The uncertainty relations (Statements \ref{uncertain1} and \ref{uncertain2}) can also be used to highlight the symmetries between ordinary space $\{\mathbf{r}_{4}\}$ and reciprocal space $\{\mathbf{p}_{4}\}$. Consider an idealized, fundamental context $C$ such as that in Fig. \ref{Fig76c}. The four-position $\mathbf{r}_{4}$ is observed at time $n+m$ and the four-momentum $\mathbf{p}_{4}$ is subsequently observed at time $n+m'$. If we plot the projection $\Pi_{\mathbf{r}_{4}}S_{OS}$ of the specimen state $S_{OS}$ onto ordinary space and the projection $\Pi_{\mathbf{p}_{4}}S_{OS}$ onto reciprocal space at these time instants, we get the picture in Fig. \ref{Fig84}. The fact that the position measurement is perfectly precise means that the position in reciprocal space is completely undetermined at sequential time $n+m$. Conversely, the precise momentum measurement at time $n+m'$, means that the position in reciprocal space is perfectly known at this time, whereas the position in ordinary space has become completely undetermined. This can be seen as an illustration of Bohr's complementarity principle.

At each time instant we know, however, that the state is contained in the upper parts of the two light cones. We have assumed that we are dealing with a context $C^{+}$ that tracks the present attributes of the specimen forwards in time. Note also that since all objects have non-zero mass (Statement \ref{nozeromass}), the boundaries of the two cones defined by $ct=|\mathbf{r}|$ and $E=c|\mathbf{p}|$, respectively, are never part of the projected states $\Pi_{\mathbf{r}_{4}}S_{OS}$ and $\Pi_{\mathbf{p}_{4}}S_{OS}$.

\begin{figure}[tp]
\begin{center}
\includegraphics[width=80mm,clip=true]{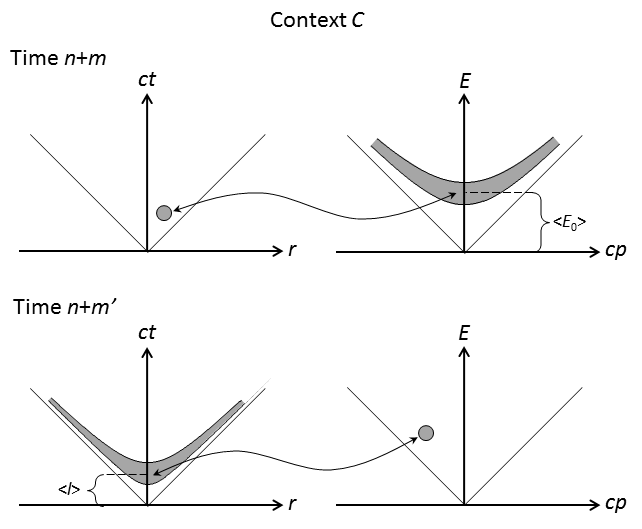}
\end{center}
\caption{Projections of the specimen state onto ordinary and reciprocal space in a real context such as that in Fig. \ref{Fig76b}. The projections correspond to the grey areas. The width of the grey belt in reciprocal space at time $n+m$ represents the uncertainty of the rest energy $E_{0}$. The width of the grey belt in ordinary space at time $n+m'$ represents the uncertainty of the Lorentz position $l$. Compare Fig. \ref{Fig84}.} 
\label{Fig85}
\end{figure}

No real context in which position or momentum is observed is fundamental. If these properties are observed in succession, we are dealing with contexts such as that in Fig. \ref{Fig76b} rather than that in Fig. \ref{Fig76c}. Then we get projected specimen states like those in Fig. \ref{Fig85} rather than the idealized ones in Fig. \ref{Fig84}.

At time $n+m$, the grey ball in ordinary space represents the possible values of four-position just after the measurement of this property, whereas the grey `belt' in reciprocal space represents the possible values of four-momentum. The diameter of the ball is inversely proportional to the squared width of the belt according to Statement \ref{uncertainmass}.

Analogously, at time $n+m'$, the grey ball in reciprocal space represents the possible values of four-momentum just after the measurement of this property, whereas the grey `belt' in ordinary space represents the possible values of four-position. The diameter of the ball is inversely proportional to the squared width of the belt according to Statement \ref{uncertaindist}.


\section{Bound states and the spectrum of space-time}
\label{boundstates}

In a bound state, the energies allowed by physical law form a discrete set. In unbound states, on the other hand, no values of the energy can be excluded \emph{a priori}. The set of allowed energies is continuous. Given the extensive symmetries between ordinary space and the reciprocal space - and the corresponding evolution equations - one may ask if this is true for spatio-temporal distances also. Are they discrete in bound states, and continuous in unbound states?

\begin{figure}[tp]
\begin{center}
\includegraphics[width=80mm,clip=true]{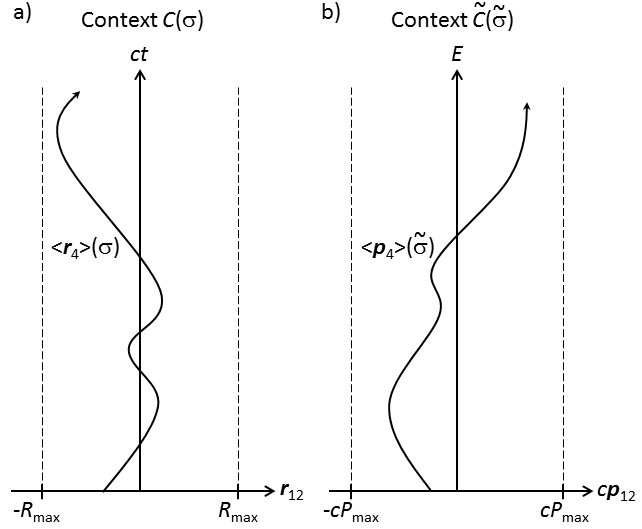}
\end{center}
\caption{A bound state of a specimen $OS$ with two parts $OS_{1}$ and $OS_{2}$, expressed with the help of a) a family of contexts $C(\sigma)$ in ordinary space, and b) a family of contexts $\tilde{C}(\tilde{\sigma})$ in reciprocal space. In the contexts $C$, the spatial position $\mathbf{r}_{12}$ is defined so that the origin is the spatial position of $OS_{1}$. The time $t$ is the relational time passed between the initiation of the context and the sequential time $n$ at which $\mathbf{r}_{4}$ is observed. In the contexts $\tilde{C}$, the momentum $\mathbf{p}_{12}$ is the relative momentum of the two parts $OS_{1}$ and $OS_{2}$. The energy $E$ is measured so that $\tilde{\sigma}=0$ corresponds to $E=0$.}
\label{Fig88}
\end{figure}

To answer this question, we must first make clear what a bound state means in terms of the properties that span reciprocal space. Definition \ref{boundspecimen} expresses what is meant by a bound state in terms of the properties that span ordinary space. Finding the corresponding expression in repciprocal space can be seen as a step in our ongoing exploration of the symmetries between the two spaces.

Loosely speaking, we defined a bound state of two parts $OS_{1}$ and $OS_{2}$ of a specimen to be such that the spatial distance between these parts remains bounded as the evolution parameter $\sigma$ goes to infinity together with time $t$ [Fig. \ref{Fig88}(a)]. A direct translation of this condition to reciprocal space would read as follows: a bound state is such that the relative momentum of the two parts remains bounded as the reciprocal evolution parameter $\tilde{\sigma}$ goes to infinity together with energy $E$ [Fig. \ref{Fig88}(b)].

Is this statement reasonable? Yes, it means that increasing energy in the composite specimen is absorbed as increasing rest energy, rather than momentum. This is exactly what we would expect in a bound state.

\begin{state}[\textbf{A bound state of a specimen in reciprocal space}]
A specimen $OS$ is in a bound state according to Definition \ref{boundspecimen} if and only if 1) the domain of $\tilde{\sigma}$ can be extended to $D_{\tilde{\sigma}}=[0,\infty]$ in a natural parametrization, 2) $OS$ can be knowably divided into at least two objects $OS_{1}$ and $OS_{2}$, which are identifiable for all $\tilde{\sigma}\in D_{\tilde{\sigma}}$, 3) there is a finite upper bound $P_{\max}$ of the magnitude of the relative momentum $p_{12}$ of $OS_{1}$ and $OS_{2}$.
\label{boundspecimen2}
\end{state}

\begin{figure}[tp]
\begin{center}
\includegraphics[width=80mm,clip=true]{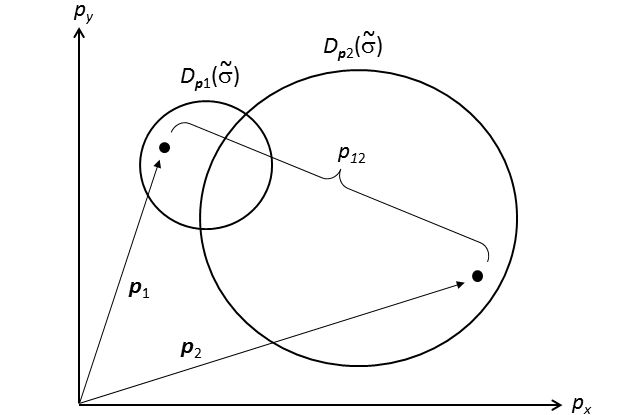}
\end{center}
\caption{Illustration of the concept of a bound specimen in reciprocal space (Statement \ref{boundspecimen2}). The reciprocal wave function domains $D_{\mathbf{p}1}$ or $D_{\mathbf{p}2}$ correspond to the specimen states $S_{OS1}$ and $S_{OS2}$ according to Fig. \ref{Fig75}. Compare Fig. \ref{Fig79}.}
\label{Fig89}
\end{figure}

The existence of a bound $P_{\max}$ should be known at the time $n$ at which the observational context is initiated (corresponding to $\tilde{\sigma}=0$). The magnitude of the momentum $p_{12}$ is defined as follows (Fig. \ref{Fig89}). Let $\mathbf{p}_{1}\in D_{\mathbf{p}1}(\tilde{\sigma})$, where $D_{\mathbf{p}1}(\tilde{\sigma})$ is the domain of the part of the reciprocal wave function that describes $OS_{1}$, and let $\mathbf{p}_{2}\in D_{\mathbf{p}2}(\tilde{\sigma})$. Then $p_{12}=|\mathbf{p}_{2}-\mathbf{p}_{1}|$. Condition 3) means that for each $\tilde{\sigma}\in D_{\tilde{\sigma}}$ we have $p_{12}<P_{\max}$ for each possible pair of points $(\mathbf{p}_{1},\mathbf{p}_{2})$.

According to the formalism of quantum mechanics, there is a ground state with the greatest possible binding energy in any bound state. This means that there is a state $S_{OS}$ that has the smallest possible rest energy. We let $E_{0}^{(G)}$ denote this `ground state rest energy'. In a context $C$ in which $\mathbf{r}_{4}$ is about to be observed, the ground state is described by a wave function $\alpha_{\mathbf{r}_{4}s}^{(G)}$ that is an eigenfunction to the rest energy operator according to the Dirac equation (Statement \ref{diracconstraint}) that is associated to the eigenvalue $E_{0}^{(G)}$.

\begin{equation}
(\overline{E_{0}})_{\mathbf{r}_{4}s}\alpha_{\mathbf{r}_{4}s}^{(G)}=E_{0}^{(G)}\alpha_{\mathbf{r}_{4}s}^{(G)}
\end{equation}

From the symmetry of the Dirac equation and its reciprocal (Statement \ref{recdiracconstraint}), we see that there is another ground state that is associated with a smallest possible value $l^{(G)}$ of the Lorentz distance $l$.

\begin{equation}
\overline{l}_{\mathbf{p}_{4}s}\tilde{\alpha}_{\mathbf{p}_{4}s}^{(G)}=l^{(G)}\tilde{\alpha}_{\mathbf{p}_{4}s}^{(G)}.
\end{equation}

What distance are we talking about? It is the Lorentz distance between an event corresponding to the observation of $OS_{1}$ and another event corresponding to the observation of $OS_{2}$. This means that there is a smallest Lorentz distance between any pair of bound objects that is possible to observe. Physical law forbids the observation of any smaller distance. 

\begin{figure}[tp]
\begin{center}
\includegraphics[width=80mm,clip=true]{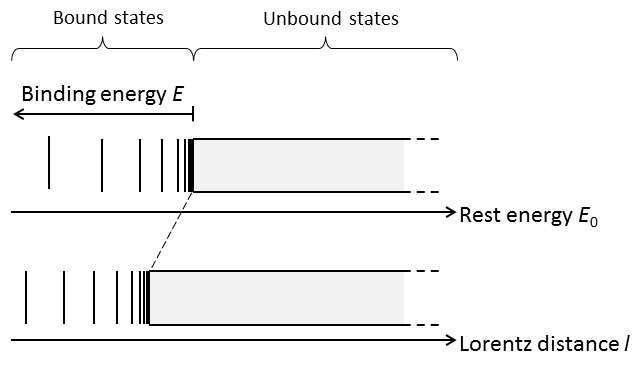}
\end{center}
\caption{In a bound state of a composite specimen, the spectra of rest energy and Lorentz distance are discrete. In an unbound state, these spectra are continuous.}
\label{Fig91}
\end{figure}

Note that there is one fundamental asymmetry between the two eigenvalues equations given above, amid all the symmetries between ordinary and reciprocal space. To define the Lorentz distance we need a composite object with two parts. The rest mass, on the other hand, is defined for all objects, composite or not. The structure of the distance and energy spectra of a given composite specimen is sketched in Fig. \ref{Fig91}.

\begin{figure}[tp]
\begin{center}
\includegraphics[width=80mm,clip=true]{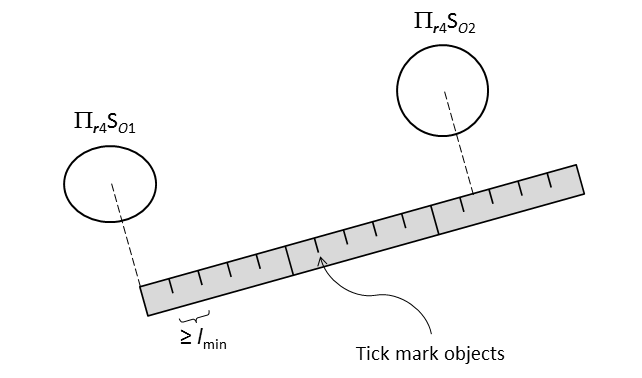}
\end{center}
\caption{To measure a distance, we need a ruler. To make the scale of the ruler fixed and known \emph{a priori}, any pair of tick marks must correspond to a bound state of two objects with a given relative Lorentz distance. Since there is a smallest such distance $l_{\min}$, there is a smallest scale we can ever be able to measure. This is true even if we measure the distance between objects $O_{1}$ and $O_{2}$ that are \emph{not} bound to each other. The distance between such objects can, in principle, take any value.}
\label{Fig92}
\end{figure}

Given a finite set of minimal objects, and a given set of interactions, there will be a finite set of specimens composed of these minimal objects. For each such composite specimen, there is one smallest ground state Lorentz distance. Among these ground state distances, there will be a smallest one. This distance will be the smallest distance $l_{\min}$ that is possible to measure (Fig. \ref{Fig92}). In the present superficial discussion, we do not attempt to determine this distance, but just point out its existence.

\begin{state}[\textbf{There is a smallest observable distance}]
There is a Lorentz distance $l_{\min}$ such that for any two objects or events $O_{1}$ and $O_{2}$, it is never posible to know that the Lorentz distance between them is smaller than $l_{\min}$.
\label{minimumdistance}
\end{state}

We may use the terminology in section \ref{presfut} to express the same thing, saying that it is not a realizable property to have $l<l_{\min}$ (Definition \ref{realizableprop}). A property that is not realizable cannot correspond to a present or future alternative (Definitions 
\ref{presentalt} and \ref{futurealt}). These realizable alternatives are the "nodes" in the directed network of alternatives that define a context (Fig. \ref{Fig65}), and contribute to superpositions and amplitudes in quantum mechanical calculations. Therefore we should not treat two states which differ less than $l_{\min}$ in the spatio-temporal position as two terms that contribute separately to the amplitude in such a calculation. In other words, we get a natural cutoff scale.

\begin{state}[\textbf{The ultraviolet cutoff}]
Consider a Hilbert space context representation $\bar{u}_{C}\bar{S}_{C}(n)=\sum_{j}a_{j}\bar{S}_{Pj}$ according to Statement \ref{hilbertrep}, where $\bar{S}_{Pj}$ is the property value state that corresponds to property value $p_{j}$. Let $P$ be the Lorentz distance between two parts $OS_{1}$ and $OS_{2}$ of a composite specimen. Consider the representation of $p_{j}$ given in Eq. \ref{contpropv}. Then we must have $\Delta p_{j}\geq l_{\min}$ for all $j$.
\label{cutoff}
\end{state}

\begin{figure}[tp]
\begin{center}
\includegraphics[width=80mm,clip=true]{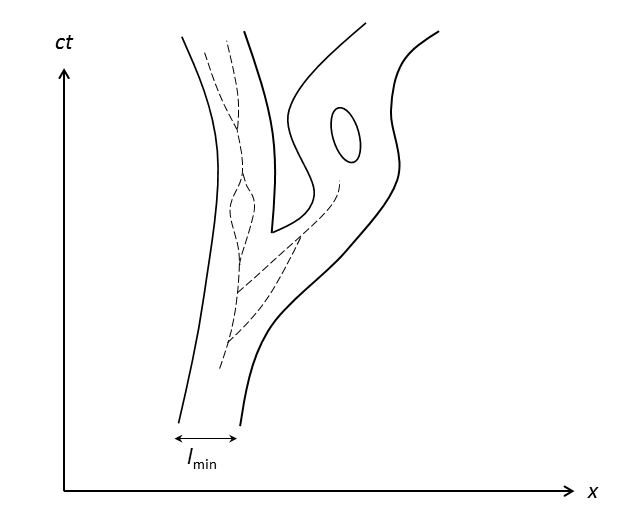}
\end{center}
\caption{Processes taking place on shorter scales than $l_{\min}$ cannot be resolved. Therefore we should not treat object divisions and mergings on such small scales as realizable alternatives. The hypothetical processes marked by dashed lines can be ignored when we construct superpositions of alternatives, and the path of a minimal object can be treated as a world-tube in space-time (compare Fig. \ref{Fig47}).}
\label{Fig95}
\end{figure}

In the rest frame of the specimen, we get a maximum temporal resolution of two events $t\geq l_{\min}/c$. This statement should be relevant in particle physics, for example. Calculating the total amplitude of a certain reaction, we should not include contributing loops that occurs on a shorter spatial and temporal scale (Fig. \ref{Fig95}).

It is important to stress, however, that this finite spatio-temporal resolution, the lack of absolute locality, does not mean that space-time is fundamentally discrete. In bound states, it is indeed discrete, but not so in unbound states. In those states we cannot \emph{exclude} any distance between two objects or events, however small, but we cannot \emph{confirm} any distance shorter than $l_{\min}$.

The truly discrete nature of space-time in bound states should have consequences when it comes to the calculation of energy levels in such states. This is so since it involves finding the eigenfunctions $\alpha_{\mathbf{r}_{4}s}(\mathbf{r}_{4},s,E_{0}^{2})$ with the corresponding (rest) energy eigenvalue $E_{0}$. Traditionally, we let a continuous spatio-temporal eigenfunction $\psi_{\mathbf{r}_{4}k}(\mathbf{r},t,E_{0}^{2})$ represent each component in the spinor (compare Eq. \ref{spinorparts}). In bound states, this is not quite correct, not even in idealized, fundamental contexts, since the eigenfunction $\psi$ is defined only for arguments $\mathbf{r}_{4}$ that are allowed by physical law (Fig. \ref{Fig91}). In the same way, we should only include values of $\mathbf{p}_{4}$ that correspond to the discrete spectrum of binding energies $E$ as arguments in the reciprocal four-momentum component $\tilde{\psi}_{\mathbf{p}_{4}k}(\mathbf{p},E,l^{2})$ of the eigenfunction $\tilde{\alpha}_{\mathbf{p}_{4}s}(\mathbf{p}_{4},s,l^{2})$, obeying the reciprocal Dirac equation (Statement \ref{recdiracconstraint}).

To determine the spectrum of bound Lorentz distances we therefore need the spectrum of binding energies, and to determine the spectrum of binding energies we need the spectrum of bound Lorentz distances. We get a pair of eigenvalue equations that have to be solved self-consistently.

\begin{state}[\textbf{Self-consistent spectra of energy and distance}]
Suppose that we want to find the set $\{(\mathbf{p}_{4})_{i}\}$ of allowed four-momenta in a bound state of a specimen (Definition \ref{boundspecimen} and Statement \ref{boundspecimen2}), together with the set $\{(\mathbf{r}_{4})_{j}\}$ of allowed four-distances. These sets are determined self-consistently from the Dirac equations specified in Statements \ref{diracconstraint} and \ref{recdiracconstraint}.

\begin{equation}\begin{array}{rcl} \overline{l}_{\mathbf{p}_{4}s}\tilde{\alpha}_{\mathbf{p}_{4}s}((\mathbf{p}_{4})_{i},l_{j}^{2}) & = & l_{j}\tilde{\alpha}_{\mathbf{p}_{4}s}((\mathbf{p}_{4})_{i},l_{j}^{2})\\ (\overline{E_{0}})_{\mathbf{r}_{4}s}\alpha_{\mathbf{r}_{4}s}((\mathbf{r}_{4})_{j},(E_{0})_{i}^{2}) & = & (E_{0})_{i}\alpha_{\mathbf{r}_{4}s}((\mathbf{r}_{4})_{j},(E_{0})_{i}^{2})
\end{array}\end{equation}
\label{selfconsistency}
\end{state}

We have argued that there is no minimum Lorentz distance $l^{2}=c^{2}t^{2}-|\mathbf{r}_{4}|^{2}$ in unbound states. However, there is always a smallest possible time difference $t$ between any two objects that belong to different sequential times $n$ and $n+1$. This is true regardless whether the objects are in a bound state or not. In contrast, if the events or objects belong to the same sequential time $n$, there is no such minimum time difference. In conventional terminology, there is a minimum time difference $t_{\min}$ between events with time-like separation, but there is no such minimal time difference between events with space-like separation. Let us argue why this is so, using the method of \emph{reductio ad absurdum}.

At the beginning of section \ref{evolutionparameter} we argued that relational time $t$ is a meaningful attribute because physical states $S$ that belong to adjacent sequential times $n$ and $n+1$ tend to be more similar than are physical states separated by more than one time step. Therefore, to give meaning to the hypothesis that $t$ is continuous and can take arbitrary small values, we have to allow the difference between $S(n+1)$ and $S(n)$ to become arbitrarily small.

\begin{figure}[tp]
\begin{center}
\includegraphics[width=80mm,clip=true]{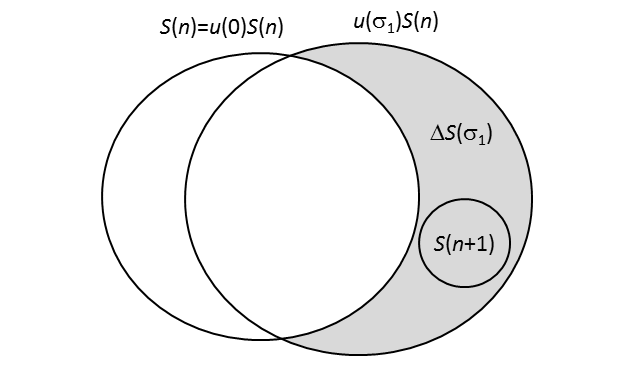}
\end{center}
\caption{The reason why there has to be a smallest relational time difference $t_{\min}$ between sequential states $S(n)$ and $S(n+1)$. The incompleteness of knowledge means that the volume of $S(n+1)$ is non-zero. This means that the grey region $\Delta S(\sigma_{1})$ must have non-zero volume. This volume grows from zero as $\sigma_{1}$ grows from zero. Thus $\sigma_{1}$ and the corresponding time difference $t$ must be non-zero.}
\label{Fig93}
\end{figure}

We may use the continuous evolution parameter $\sigma$ to parametrize the difference between these two states. Let us write

\begin{equation}
S(n+1)\subseteq u_{1}S(n)=u(\sigma_{1})S(n),
\label{paratime}
\end{equation}
where the parametrization is such that

\begin{equation}
S(n)=u(0)S(n)
\end{equation}
and let

\begin{equation}
\Delta S(\sigma)\equiv u(\sigma)S(n)/S(n)
\end{equation}
according to Fig. \ref{Fig93}. Further, let

\begin{equation}
V(\sigma)=V[\Delta S].
\end{equation}
If $t$ is continuous, we are allowed to take the limit $t\rightarrow 0$. In this limit, $\sigma_{1}\rightarrow 0$, leading to $V(\sigma_{1})\rightarrow 1$ according to Eq. [\ref{paratime}]. This necessarily means that $\Delta S\rightarrow Z$, for some exact state $Z$. In that case we get $S(n+1)=Z$, since we must have $S(n+1)\subseteq\Delta S$ because of the condition $S(n+1)\cap S(n)=\varnothing$. But the incompleteness of knowledge means that this cannot happen. Therefore there has to be a minimum state space volume

\begin{equation}
V(\sigma_{1})\geq V_{\min}>1,
\end{equation}
leading to $\sigma_{1}\geq\sigma_{\min}$ (where $\sigma_{\min}>0$ is parametrization dependent), and a minimum time difference

\begin{equation}
t\geq t_{\min}>0.
\end{equation}

Clearly, this non-continuity of relational time is a consequence of the incompleteness of knowledge (Statement \ref{incompleteknowledge}).

\begin{state}[\textbf{There is a smallest relational time between sequential events}]
Consider two objects $O_{1}\in PKN(n)$ and $O_{2}\in PKN(n+1)$, directly perceived by a subject $j$. Let $t(n)$ and $t(n+1)$ be the relational times that $j$ assign to $O_{1}$ and $O_{2}$, respectively. Then $t(n+1)-t(n)\geq t_{\min}$ for some real constant $t_{\min}>0$.
\label{smallesttime}
\end{state}

\begin{figure}[tp]
\begin{center}
\includegraphics[width=80mm,clip=true]{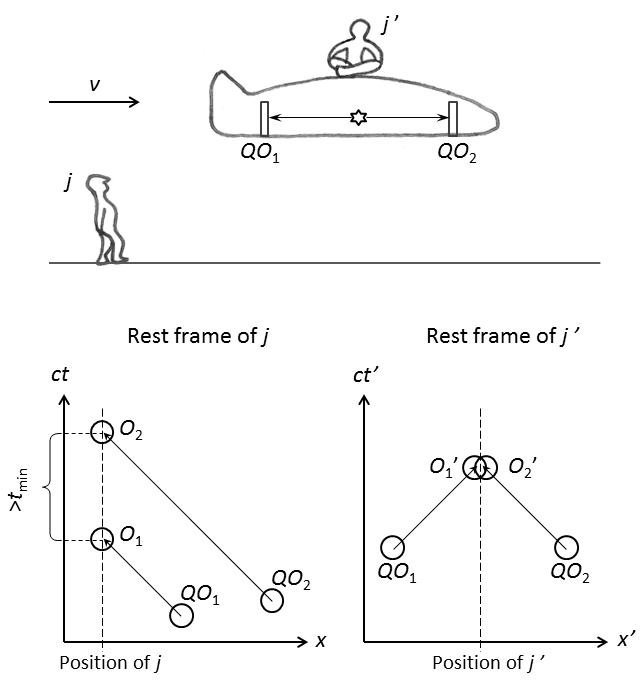}
\end{center}
\caption{The time difference between two directly perceived objects is either greater than $t_{\min}$ (like that between $O_{1}$ and $O_{2}$), or zero (like that between $O_{1}'$ and $O_{2}'$). All subjects agree on these statements. The fact that the temporal ordering of the quasiobjects $QO_{1}$ and $QO_{2}$ is ambiguous has no primary importance in our epistemic approach.} 
\label{Fig97}
\end{figure}

Is the existence of $t_{\min}$ consistent with Lorentz invariance? Recall from section \ref{time} that sequential time $n$ is updated to $n+1$ if and only if the two events or objects that define these two time instants have a time-like separation (Definition \ref{temporalupdates} and Fig. \ref{Fig96}). Otherwise they are judged to belong to the same time $n$. Further, a given subject $j$ assigns the same relational time $t^{(j)}(n)$ to all events or objects that belong to sequential time $n$. This is the `personal klock' of subject $j$, which may be said to belong to the rest frame of $j$ (Fig. \ref{Fig97}). Thus, if $j$ measures the time difference $t(n+1)-t(n)=t^{(j)}\geq t_{\min}$ between the events occurring at time $n$ and $n+1$, then all other subjects $j'$ will measure a time difference $t^{(j')}\geq t^{(j)}\geq t_{\min}$ due to time dilation. Thus $ t_{\min}$ is Lorentz invariant.

The requirement that the two events or objects $O_{1}$ and $O_{2}$ are \emph{directly perceived} by subject $j$ is the crucial fact that makes an unambiguous temporal ordering possible, and makes $t_{\min}$ Lorentz invariant. Both these beneficial qualities dissappear if we consider \emph{deduced} quasiobjects $QO_{1}$ and $QO_{2}$. If these quasiobjects have space-like separation, two subjects $j$ and $j'$ may assign different temporal ordering to them, and the time difference may be arbitrarily small. Note, however, that it is sufficient to consider directly perceived objects in our epistemic approach. They make it possible to specify the physical state $S(n)$ completely, as discussed in relation to Fig. \ref{Fig24b}, and also the physical law that governs its evolution.

In any case, it may be a good idea to illustrate the difference between our `subjective' minimal time difference $t_{\min}$ and Lorentz transformations of ordinary time differences $t$, which may apply to quasiobjects. We use the familiar example where two light pulses are sent in opposite directions inside an aircraft or spaceship (Fig. \ref{Fig97}). The events that the light rays hit the rear and front walls of the vessel are two quasiobjects $QO_{1}$ and $QO_{2}$ with space-like separation. The corresponding directly perceived objects are the events when information about these quasievents reach some subject. Suppose that this information reaches a subject via light signals that are emitted immediately after the quasievents have occurred. A subject $j'$ located in the middle of the vessel receive the information about $QO_{1}$ and $QO_{2}$ at the same time. These events correspond to two objects $O_{1}'$ and $O_{2}'$ that belong to the same sequential time $n$. Another subject $j$ at the ground will agree that these two \emph{perceived} events are indeed simultaneous. This subject will judge, however, that the quasievents $QO_{1}$ and $QO_{2}$ are not simultaneous. He will receive information that $QO_{1}$ has occurred before he receives information that $QO_{1}$ has occurred. These events correspond to two objects $O_{1}$ and $O_{2}$ that belong to different sequential times, say $n$ and $n+1$. The relational time difference must be greater than $t_{\min}$. The subject $j'$ inside the vessel wll agree that $O_{1}$ occurs before $O_{2}$ and that the time difference is greater than $t_{\min}$.

\begin{state}[\textbf{Subjective and objective time differences}]
Let $t^{(j)}$ be the time difference that subject $j$ measures between any two objects $O_{1}$ and $O_{2}$ that she directly perceives. Then $t^{(j)}=0$ or $t^{(j)}\geq t_{\min}$. Any other subject $j'$ agrees with these statements. Let $t$ be the deduced time difference between two quasiobjects $QO_{1}$ and $QO_{2}$ in a given reference frame. Then $t$ can take any value.
\label{subjectiveobjectivetime}
\end{state}

To be precise, we note that from the perspective of subject $j'$, the objects $O_{1}$ and $O_{2}$ are quasiobjects. When he agrees that $t^{(j)}\geq t_{\min}$, it is a deduced conclusion.

\begin{figure}[tp]
\begin{center}
\includegraphics[width=80mm,clip=true]{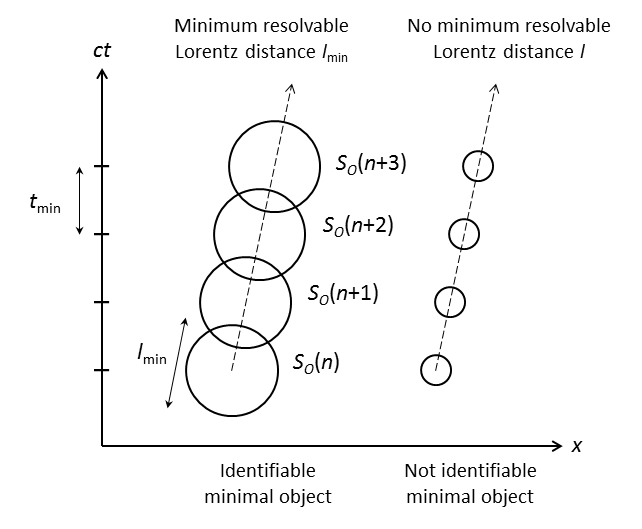}
\end{center}
\caption{Suppose that we track a minimal object $O$ during a sequence of times $n,n+1,n+2,\ldots$. Let $S_{O}(n')$ be the projection of the state of $O$ on space-time. The existence of $t_{\min}$ makes it necessary to have a minimum resolvable Lorentz distance $l_{\min}$ that is at least as large. Otherwise we cannot be sure that all successive states $S_{O}$ overlap. The minimal object would lose its identifiability (section \ref{identifiability}).}
\label{Fig94}
\end{figure}

There is a fundamental connection between the existence of a smallest distance $t_{\min}$ and a smallest \emph{resolvable} $l_{\min}$ on the one hand, and the identifiability of minimal objects on the other. Namely, the existence $t_{\min}$ makes it necessary to have a minimum resolvable Lorentz distance $l_{\min}$ of the same order of magnitude to uphold the identifiability of minimal objects. Figure \ref{Fig94} makes it clear why.

\begin{state}[\textbf{The minimum resolvable Lorentz distance is necessary for identifiability}]
Given $t_{\min}$ according to Statement \ref{smallesttime}, the existence of identifiable minimal objects (Statement \ref{allminimalidentity}) makes it necessary to have a minimum resolvable Lorentz distance $l_{\min}$ (Statement \ref{minimumdistance}) which fulfils $l_{\min}\geq t_{\min}$.
\label{lorentzidenti}
\end{state}

We have argued that Lorentz distances are discrete in bound states, and that we can never resolve any shorter distances. We have also argued that there is a smallest time difference between perceptions that can ever be measured. Do these observations mean that space-time must be considered discrete? No, we cannot treat the discreteness or the continuity of space-time as two mutually exclusive possibilites. What we can say is that in some aspects and circumstances the continuous description of space-time is inadequate. This ambiguity is no mystery if we stick to the basic ingredients of physical states: objects, their internal and relational attributes, and the changing potential knowledge about these. The mystery appears only if we confuse the landscape with the map, where the map is the smooth manifold in which we represent the set of relational attributes. This map seems to be redundant at small scales, just as it is redundant in the sense that translations and rotations of the map has no epistemic meaning - a fact sometimes described as the homogeneity and isotropy of space. If a sailor finds an old map from the 15th century and uses it to navigate, he should not be frightened when he approaches a position where a sea monster is painted on the map.

We end this conceptual discussion with another conceptual note. The discreteness of space-time in bound states might shed some light on the paradoxes that arise in connection with black holes and space-time singularities.

\section{The orientability of space}
\label{orient}

We argued in section \ref{minimalism} that the dependence of physical law on distances and angles is not enough to give epistemic meaning to the parity operation $\mathbf{r}\rightarrow -\mathbf{r}$. In other words, there has to be something else that makes it possible to distinguish between left and right at a fundamental level. In this section we discuss the idea that it is the directionality of time that makes space orientable in this sense. We will motivate this statement in two ways. First, we try to relate temporal directionality and spatial orientability directly, and second, we try to relate the two qualities via the spin degree of freedom.

\begin{figure}[tp]
\begin{center}
\includegraphics[width=80mm,clip=true]{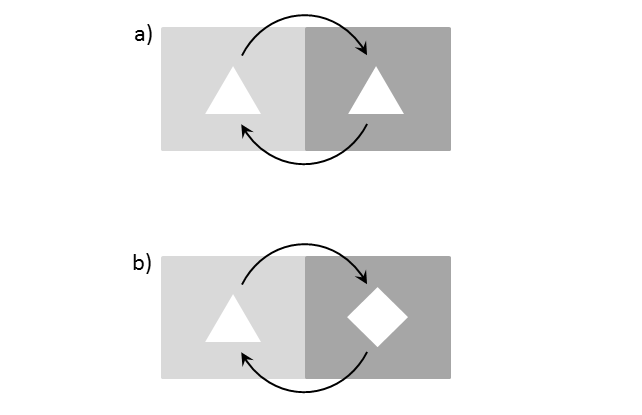}
\end{center}
\caption{a) The parity operation $\mathbf{r}\rightarrow -\mathbf{r}$ corresponds to the spatial interchange of identical objects $O_{1}$ and $O_{2}$ that have different environments. These different environments correspond to a coordinte system. b) We do not need the parity operation to describe the effect of the interchange of two non-identical objects.}
\label{Fig150}
\end{figure}

To give meaning to the parity operation, there must be a way to identify a spatial interchange of two objects $O_{1}$ and $O_{2}$ that are identical except for their spatio-temporal position $\mathbf{r}_{4}$. We also need a coordinate system with an origin. Such a coordinate system requires external objects, so that the state of knowledge that we consider must contain more objects than $O_{1}$ and $O_{2}$.

If these two objects are not identical in such a situation, it is trivial that the state after their interchange is different from the state before (Fig. \ref{Fig150}). Only if we investigate the effect of an interchange of identical objects do we investigate the orientability of space itself. If we find an identifiable effect, we have a vector space. Otherwise it is sufficient to describe physical space as a metric space in which distances and angles between objects are defined, but in which there is no other structure.

\begin{defi}[\textbf{Orientable space}]
Consider two objects $O_{1}$ and $O_{2}$ that have identical states $S_{O1}$ and $S_{O2}$, but different spatio-temporal positions $\mathbf{r}_{4}(O_{1})$ and $\mathbf{r}_{4}(O_{2})$. Physical space is orientable if and only if the spatial interchange of $O_{1}$ and $O_{2}$ can lead to a new physical state. More precisely, there should exist two such objects $O_{1}$ and $O_{2}$, with spatial positions $\mathbf{r}(O_{1})$ and $\mathbf{r}(O_{2})$, such that

\begin{equation}
\left.\begin{array}{rrr}
\mathbf{r}'(O_{1}) & = & \mathbf{r}(O_{2})\\
\mathbf{r}'(O_{2}) & = & \mathbf{r}(O_{1})
\end{array}\right\}
\Rightarrow S'\neq S.
\end{equation}

\label{orientable}
\end{defi}

Say that the temporal positions $t(O_{1})$ and $t(O_{2})$ of the identical objects $O_{1}$ and $O_{2}$ differ, as well as the spatial positions. We treat the directed nature of time as fundamental (Assumption \ref{timeconcept}). This means that the \emph{temporal} interchange of these objects does make a knowable difference. That is, we have, by assumption,

\begin{equation}
\left.\begin{array}{rrr}
\mathbf{t}'(O_{1}) & = & \mathbf{t}(O_{2})\\
\mathbf{t}'(O_{2}) & = & \mathbf{t}(O_{1})
\end{array}\right\}
\Rightarrow S'\neq S.
\label{timeswitch}
\end{equation}
whenever $O_{1}$ and $O_{2}$ are placed in a spatial coordinate system that corresponds to the different environments in Fig. \ref{Fig150}(a). Such a temporal interchange is identical to a spatial interchange according to Definition \ref{orientable}, so that space is indeed orientable.

We might argue that we should not involve time when we discuss the orientability of space. That is, we should not allow different spatio-temporal positions $\mathbf{r}_{4}$ of $O_{1}$ and $O_{2}$ in Definition \ref{orientable}, just different spatial positions $\mathbf{r}$. But relativity has taught us that the spatial and temporal parts of space-time cannot be separated. An object should be described as an event, and two events that are judged to be simultaneous by one observer are not simultaneous as judged by another. 

\begin{figure}[tp]
\begin{center}
\includegraphics[width=80mm,clip=true]{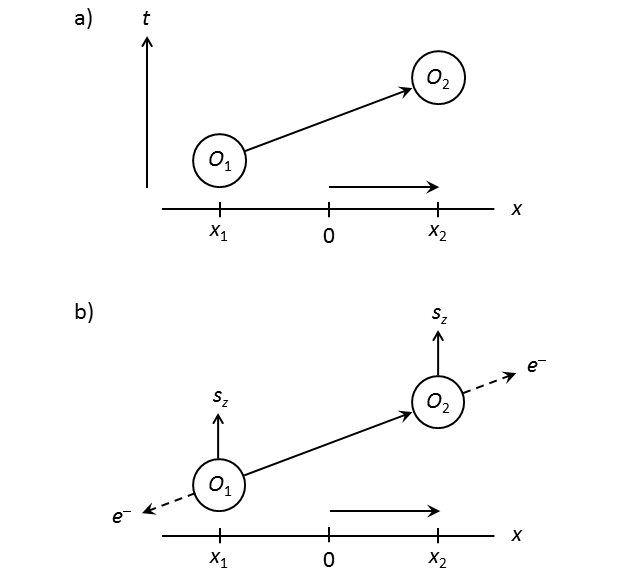}
\end{center}
\caption{We say that space is orientable whenever we may know the difference when two identical objects $O_{1}$ and $O_{2}$ change positions (Definition \ref{orientable}). Then we may define a directed spatial arrow that starts at $O_{1}$ and ends at $O_{2}$. The parity operation $x\rightarrow-x$ acquires meaning if we introduce an origin halfway between $O_{1}$ and $O_{2}$, and let $-x\equiv x_{1}$ and $x\equiv x_{2}$. a) A temporal difference between the objects allows us to let the arrow start at the early object. b) Spin dependent weak interactions allow us to let the arrow start at the object from which the intensity of electrons emitted along the corresponding dashed arrow is the highest.}
\label{Fig103}
\end{figure}

The fact that we allow temporal differences makes it possible to attach a spatial arrow that points from the early object to the late object [Fig. \ref{Fig103}(a)]. We may let the arrow start at the origin halfway between the two objects rather than at the position of the early object. Let us denote such an arrow by $\mathbf{r}$. If we interchange the spatial position of the two identical objects, the direction of the arrow is reversed, whereas the starting point at the origin is preserved. We may denote the reverse arrow by $-\mathbf{r}$. We have given physical meaning to the parity operation.

Let us change focus and approach the orientability of space via the Dirac equations and the spin inherent in all objects. The fact that wave functions have to obey the Dirac equations, in addition to the evolution equations, can be seen as a direct consequence of the directed nature of time. Let us discuss why this is so.

Equation [\ref{directed}] is a consequence of the directionality of time. This equation leads to Eq. [\ref{signrule1}], which in turn leads to Eq. [\ref{signrule2}], stating that the reciprocal evolution parameter $\tilde{\sigma}$ always has the same sign. This condition is not fulfilled by the evolution equation $\frac{d}{d\sigma}\Psi_{\mathbf{r}_{4}}(\mathbf{r}_{4},\sigma)=i\bar{B}_{\mathbf{r}_{4}}\Psi_{P}(\mathbf{r}_{4},\sigma)$ itself (Eq. [\ref{ev2}]. It can only be fulfilled if we require that $\bar{B}_{P}$ is the square of another operator $\overline{M}_{\mathbf{r}_{4}}$ according to Eq. [\ref{squareroot2}], and that each stationary state $\psi(\mathbf{r}_{4},\tilde{\sigma})$ to the evolution equation is an eigenfunction to this square root of the evolution operator. This is the Dirac equation (Statement \ref{diracconstraint}).

We defined energy and momentum from the Fourier expansion coefficients of the wave function $\Psi_{\mathbf{r}_{4}}$. This definition implies the familiar relation $E^{2}-c^{2}p^{2}\geq 0$. If we consider this inequality as given beforehand, we can derive the Dirac equation from this relation.
The corresponding `reciprocal' relation is $c^{2}t^{2}-r^{2}\geq 0$. When we motivated the reciprocal Dirac equation (Statement \ref{recdiracconstraint}), we considered this relation as given beforehand from special relativity, and used it to derive the reciprocal Dirac equation. The relation $c^{2}t^{2}-r^{2}\geq 0$ can be regarded as a condition that all observers agree on the direction of time. Therefore we can see the reciprocal Dirac equation as a consequence of the directionality of time in the same way as the ordinary Dirac equation.

\begin{state}[\textbf{Directed time implies the Dirac equations}]
Both the ordinary and the reciprocal Dirac equations are constraints on the corresponding wave functions that are forced upon us by the directed nature of time.
\label{dirtimedirac}
\end{state}

This means that the directionality of time implies an internal (spin) degree of freedom of each minimal object that couples to space-time. According to the principle of explicit epistemic minimalism, there has to be a knowable distinction between the two possible states of this degree of freedom in relation to a given spatial $z$-direction. Since this spin degree of freedom is independent of other attributes, such a distinction must be possible to make even if we have two minimal objects which are identical except for the spin direction in relation to the chosen $z$-direction. Further, since spin is an internal atribute in the sense that is defined without reference to other objects, the distinction should be possible to make even if we disregard interactions with other objects. For this reason we disallow the separation of objects with different spin directions in an external magnetic field as the basic knowable distinction that we are looking for.

If we do not allow reference to other objects when we look for a fundamental distinction, we have to look for a distinction in relation to space itself. In this way we arrive at the conclusion that there has to be a knowable transformation of the object itself (Definition \ref{knowtransobject}) that depends on the spin direction. Of course, we know that there are such object transformations, caused by weak interactions. These makes it possible to associate the binary spatial operation $z\rightarrow -z$ with the knowable difference between the two spin values in relation to $z$. In this way we have arrived at the orientability of space via the Dirac equations.

\begin{state}[\textbf{Parity and spin}]
The binary parity operation $\mathbf{r}\rightarrow-\mathbf{r}$ is epistemically well-defined if there is a binary (spin) degree of freedom associated with minimal objects defined in relation to a spatial direction, and there are also knowable object transformations (Definition \ref{knowtransobject}) that depend on the value of this spin.
\label{parityspin}
\end{state}

\begin{figure}[tp]
\begin{center}
\includegraphics[width=80mm,clip=true]{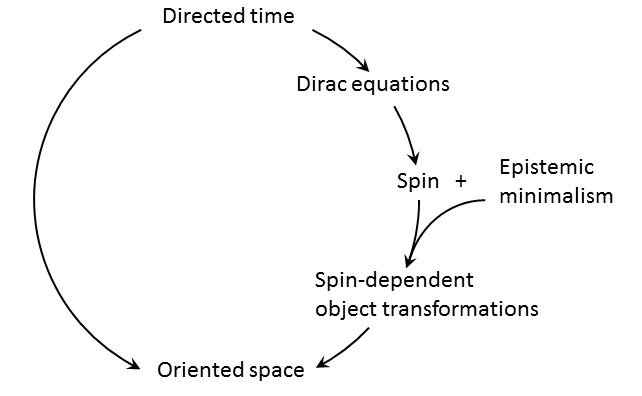}
\end{center}
\caption{We argue along two different lines that an oriented space follows from the directionality of time. Compare Fig. \ref{Fig103}. See text for further explanation.}
\label{Fig104}
\end{figure}

Looking at Fig. \ref{Fig103}(b), we see that we can use such a spin-dependent object transformation to tell the difference when the spatial locations of two identical (minimal) objects $O_{1}$ and $O_{2}$ are interchanged, assuming that the spin directions are also identical. In weak interactions, the intensity of electrons emitted from $O_{1}$ and $O_{2}$ along the two dashed lines may be different. In this way, we can define a spatial arrow between the two objects, going from the object with the higher emission intensity to the object with lower emission intensity. This is analogous to the case shown in Fig. \ref{Fig103}(a), where we used a temporal ordering of the objects to define the arrow. The conclusion is the same in both cases: the ability to define such an arrow corresponds to the orientability of space.

A further clarification might be in order here. When we interchange the positions of $O_{1}$ and $O_{2}$ in Fig. \ref{Fig103}(b), we should imagine that we interchange the spinning (minimal) objects \emph{together with} the entire observational setups with which we observe electron emissions (Fig. \ref{Fig61c}). This means that object $O_{1}$ is moved from position $x_{1}$ to position $x_{2}$ together with the dashed arrow that starts at $O_{1}$ and defines the direction in which the apparatus detects the emitted electrons. Consequently, after the objects are interchanged, the dashed arrows point towards each other rather than away from each other.

The reason why we have to interchange the entire experimental setups is that the minimal objects that decay in spin-dependent weak interactions are quasiobjects. The setups are necessary to deduce the existence of the minimal objects and their electron emissions. Without the setups the interchange of $O_{1}$ and $O_{2}$ would not be epistemically well-defined.

The reader may object that $O_{1}$ and $O_{2}$ are no longer identical if we include the observational setups, since the two dashed arrows point in different directions in relation to the spin direction - the angle between them is different. But the spin direction is not defined \emph{a priori}; if that were the case we would know beforehand that space is orientable. Rather, we should see the marked spin direction that forms different angles in relation to the dashed arrows just as an indicator of the fact that the interchange of identical object in this case indeed makes a knowable difference - the intensity of the electron emission at position $x_{2}$ increases, and the intensity at $x_{1}$ decreases.

Figure \ref{Fig104} summarizes in a flow diagram how we can associate an orientable space with a directed time along the two paths discussed above. 

\begin{state}[\textbf{Directed time implies oriented space}]
The fact that each set of relational times $\{t_{k}\}$ can be ordered in a directed sequence $\{t_{1},t_{2},\ldots\}$ gives epistemic meaning to the parity operation $\mathbf{r}\rightarrow -\mathbf{r}$.
\label{directedoriented}
\end{state}

\section{Fermions and bosons}
\label{fermbos}

In sections \ref{eveq} and \ref{reciprocaleq} we have argued that any object obeys the Dirac equation and its reciprocal. This means that all objects should be described as particles with spin quantum number $s=1/2$. How does this go together with the fact that the spin of composite objects may be larger than $1/2$, and that some elementary particles are bosons with integer spin?

The first question is resolved if all objects can be described as being composed of minimal objects with spin $1/2$. Note that, in deriving the evolution equation and the Dirac equation, we considered a fundamental context in which a \emph{single} four-position $\mathbf{r}_{4}$ was observed. The assumed fundamentality of the context means that the description of the specimen in terms of a single position vector $\mathbf{r}_{4}$ should hold to arbitrary high spatio-temporal resolution. Therefore the Dirac equation is not exactly valid for composite specimens. In the same way, the reciprocal Dirac equation is not valid for composite specimens, since it presupposes the observation of a single four-momentum $\mathbf{p}_{4}$ to arbitrarily high resolution in reciprocal space.

A further limitation of the applicability of the Dirac equations was discussed in section \ref{boundstates}. They do not hold exactly in bound states, where the spectra of ordinary and reciprocal space become discrete. Then the differential operators are not properly defined.

\begin{state}[\textbf{Applicability of the Dirac equations}]
The Dirac equation (Statement \ref{diracconstraint}) and its reciprocal (Statement \ref{recdiracconstraint}) hold exactly if and only if the specimen is an unbound minimal object.
\label{diracapplies}
\end{state}

Any minimal object that can be found in a bound state can also break away from this state. To be able to say that any object can be described as a set of minimal objects that are bound together, there must be a way to verify this statement by breaking the bindings and observe the minimal objects one by one, or in new constellations, in other bound states. If the minimal objects can be observed one by one, they obey the Dirac equations according to Statement \ref{diracapplies}. If this is not possible, as for quarks, they may nevertheless  be excited to bound states in which the spectra of space-time and the reciprocal momentum space are continuous to arbitrarily good approximation. This means that all minimal objects obey the Dirac equation in some situations. The Dirac equations imply $s=1/2$. Therefore all minimal objects always have spin $1/2$, since this attribute is internal and does not depend on the relational attributes that defines whether the minimial object is bound or unbound.

\begin{state}[\textbf{All minimal objects have spin $1/2$}]
All minimal objects are fermions with spin quantum number $s=1/2$.
\label{allminimalfermions}
\end{state}

This statement excludes elementary bosons from the family of objects and quasiobjects. If we believe in the reasoning that leads to this conclusion, what role can be given to such bosons?

Let us first discuss the notion of `pseudoobject' that was introduced in section \ref{eveqi}. These are objects that are emitted by a knowably interacting object according to Definition \ref{knowintobject} and Fig. \ref{Fig77}. Since the object that is knowably interacting preserves its identity throughout the division process, the emitted pseudoobject can have no internal attributes. Its role is to make sure that the conservation laws of relational attributes are fulfilled in the division. This means that the pseudoobject must carry momentum and angular momentum.

\begin{defi}[\textbf{Pseudoobject}]
Consider an object $O$ that is knowably interacting during the time period $[n,n+m]$ and divides at some time $n\leq n'\leq n+m$. The entity $O'$ that is emitted in the division according to Fig. \ref{Fig77} is a pseudoobject.
\label{pseudoobject}
\end{defi}

\begin{defi}[\textbf{Minimal pseudoobject}]
If the object $O$ in Definition \ref{pseudoobject} is a minimal object, then $O'$ is a minimal pseudoobject.
\label{minimalpseudo}
\end{defi}

The preservation of angular momentum, and the fact that all minimal objects have total spin quantum number $s=1/2$, imply that a minimal pseudoobject should be able to carry spin angular momentum $s_{z}=-1$, $s_{z}=0$ or $s_{z}=1$ along a given $z$-direction. In contrast, it should not be able to carry any half integer spin.

\begin{state}[\textbf{Minimal pseudoobjects have integer total spin}]
The spin quantum number $s$ of a pseudoobject is a non-negative integer.
\label{spinone}
\end{state}

Any object that follows a curved trajectory can be equivalently described as a knowably interacting object where the change of velocity is accounted for by the emission of pseudoobjects that carry a certain amount of momentum. This is discussed in section \ref{eveqi}, and it is a consequence of epistemic invariance. In particular, such a description is possible regardless the energy (or energy) change of the object that is accelerating. To uphold energy conservation in the division process, we must therefore require that pseudoobjects are massless, if we are to assign any specific rest mass to them at all.

\begin{state}[\textbf{Pseudoobjects are massless}]
The rest mass $m_{0}$ of a pseudoobject is zero.
\label{massless}
\end{state}

The masslessness of pseudoobjects implies that they cannot be located, according to the discussion in section \ref{evconsequences}.

\begin{state}[\textbf{Pseudoobjects are not localized}]
Consider the projection $\Pi_{\mathbf{r}_{4}}S_{PO}$ of the state $S_{PO}$ of a pseudoobject $PO$ onto space-time. This projection has no boundary $\partial\Pi_{\mathbf{r}_{4}}S_{PO}$.
\label{nopseudolocation}
\end{state}

Since pseudoobjects cannot be located, they cannot be observed in the same sense as objects or quasiobjets can. This fact motivates the label `pseudo' in the name chosen for these entities. But if they cannot be observed, what role can they possibly play in a physical model? The answer is, of course, that they can deliver attribute values from one object to another. This role is fulfilled when a pseudoobject emitted from an object $O_{1}$ is absorbed by another object $O_{2}$ in a merging (Section \ref{objectmerging}). Since a pseudoobject carries relational attributes only, the receiving object $O_{2}$ preserves its identity in the merging process, just as $O_{1}$ preserves its identity in the division process.

\begin{state}[\textbf{A knowable interaction takes place between two objects}]
When a knowably interacting object $O_{1}$ emits a pseudoobject, this pseudoobject is always absorbed by another knowably interacting object $O_{2}$.
\label{twototango}
\end{state}

We argued in section \ref{minimalism} that the existence of a unique upper speed limit $c$ can be regarded as a consequence of epistemic minimalism and epistemic invariance. Further, we concluded that there must be some objects that travels at this speed, and that all observers must agree on this fact. However, we concluded in section \ref{evconsequences} that all objects and quasiobjects are massive and therefore travel at a lower speed $v<c$. Further, different observers may disagree on the value of $v$. This means that the only kind of `objects' that may travel at the speed $v=c$ are pseudoobjects. This goes together well with the conclusion that they are massless (Statement \ref{massless}).

Further, the information carried by pseudoobjects cannot travel \emph{slower} than $c$. The role of the pseudoobject is to associate the two interacting objects. This can only be done if their attributes change in well-defined manner that may be interpreted as an attribute transfer. Only in that case can we give meaning to the statement that this particular pair of objects are interacting. (We will elaborate further on these matters below.) A pseudoobject that travels slower than $c$ would mean that different observers would disagree about the relational time difference between the events where $O_{1}$ changed state (emitted the pseudoobject), and $O_{2}$ changed state (absorbed the pseudoobject). This would make the association between $O_{1}$ and $O_{2}$ ill-defined. We could never be sure which pairs of objects to associate in an interaction. The concept of a knowable interaction would lose its meaning. Any concept or entity that is an essential part of physical law must be possible to express in a Lorentz invariant way.


\begin{state}[\textbf{pseudoobjects travel at the speed of light}]
All pseudoobjects travel at the universal upper speed limit $c$.
\label{pseudospeed}
\end{state}

It does not make sense to talk about pseudoobjects that divide into other pseudoobjects, or two or more pseudoobjects that merge into one. This is so since they do not possess any internal attributes apart from the spin quantum number $s=1$. This attribute is of the type 2) in Eq. [\ref{additivity}], just labeling the type of entity involved in the division process. Such attributes cannot be used in themselves to define a division. To do so, we must have at least one internal attribute that can be divided into two or more `bags', according to the additive rule 1) in Eq. [\ref{additivity}].

\begin{state}[\textbf{Pseudoobjects are linear}]
Pseudoobjects do not divide or merge. In other words, they do not interact with each other.
\label{linearpseudo}
\end{state}

We argued in section \ref{eveqi} that the evolution of each object should be consistent with a description in which it is interacting as well as a description in which it is knowably interacting. This means that to each force that gives rise to an interaction there should be an associated pseudoobject.

\begin{state}[\textbf{Pseudoobjects are force carriers}]
To each force that is able to bend the trajectory of an object without changing its identity is associated a minimal pseudoobject.
\label{pseudoforce}
\end{state}

The forces of this type that are known today are electromagnetism and gravity. Going through the above list of qualities of pseudoobjects, we see that they are consistent with those of photons.

\begin{state}[\textbf{The photon}]
The photon is the minimal pseudoobject associated with electromagnetism.
\label{pseudophoton}
\end{state}

Statement \ref{pseudoforce} implies that there must be a pseudoobject associated with gravity, and the necessary qualities of gravitons are consistent with those of pseudoobjects.

\begin{state}[\textbf{The graviton}]
The graviton exists. It is the minimal pseudobject associated with gravity.
\label{pseudograviton}
\end{state}

To get a better understanding of pseudoobjects, let us analyse their role in the interaction of objects from a purely epistemic perspective. To give operational meaning to the statement that two objects $O_{1}$ and $O_{2}$ are interacting, there must be a connection between these two objects that is possible in principle to decide by means of observations. It must be possible to say that these two objects are potentially interacting, independently from other objects. The only way this can be done is to say that $O_{1}$ and $O_{2}$ are potentially interacting when the attribute values of $O_{1}$ change by the amount $\Delta \upsilon$, and the attribute values of $O_{2}$ change by the opposite amount $-\Delta \upsilon$.

\begin{figure}[tp]
\begin{center}
\includegraphics[width=80mm,clip=true]{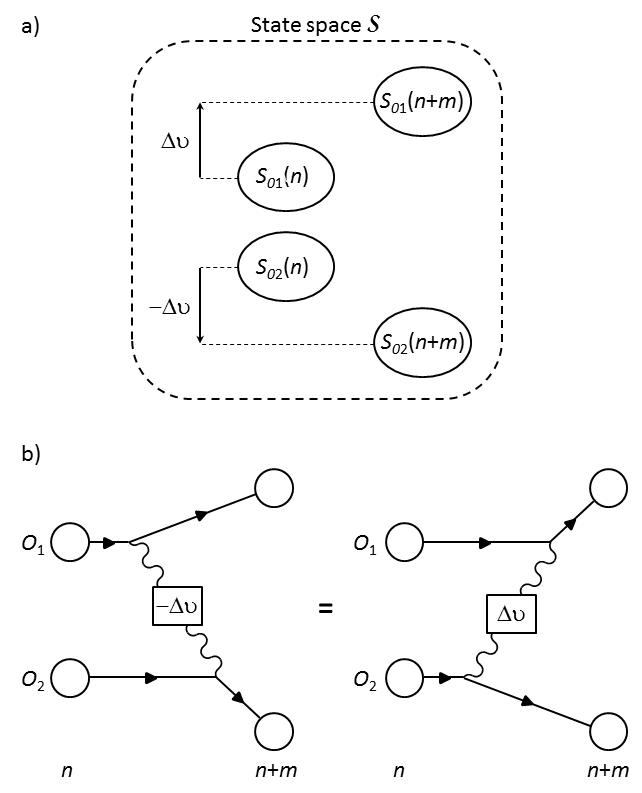}
\end{center}
\caption{A quasi-identifiable interaction between objects $O_{1}$ and $O_{2}$. a) The attribute change $\Delta\upsilon$ of the two objects match. b) This fact can be interpreted as a potential interation between them, mediated by a transfer quantum that carries the attributes $\Delta\upsilon$. There are no observations that can be used to assign a direction of motion of the transfer quantum. The potential knowledge about the distance between $O_{1}$ and $O_{2}$, and the time difference between $n$ and $n+1$, has to be consistent with information transfer at the speed of light. Compare Fig. \ref{Fig98b}.}
\label{Fig98}
\end{figure}

In effect, we are saying that if the balance of one bank account decreases by a certain amount, and the balance of another account increases by the same amount, then a potential transfer has taken place between these two accounts. Note that the balance of the attributes is all the information that we have about objects. There are no account numbers, and there is no one giving transfer orders, so we can never be sure that a transfer has actually occurred between these particular accounts.

The idea is illustrated in Fig. \ref{Fig98}. The states of two objects $O_{1}$ and $O_{2}$ are observed at times $n$ and $n+m$. It is found that the change of internal and relational attributes $\Delta \upsilon$ of $O_{1}$ balances the change $-\Delta \upsilon$ of the same set of attributes of $O_{2}$. This means that we can imagine an `transfer quantum' that carries the set of attributes $\Delta \upsilon$ from $O_{1}$ to $O_{2}$. Equivalently, we can imagine an transfer quantum that carries the attributes $-\Delta \upsilon$ from $O_{2}$ to $O_{1}$. The direction of motion of the quantum has no epistemic meaning since, by definition, it fulfils the imagined journey from one object to the other \emph{between} two subsequent observations of the involved objects.

\begin{figure}[tp]
\begin{center}
\includegraphics[width=80mm,clip=true]{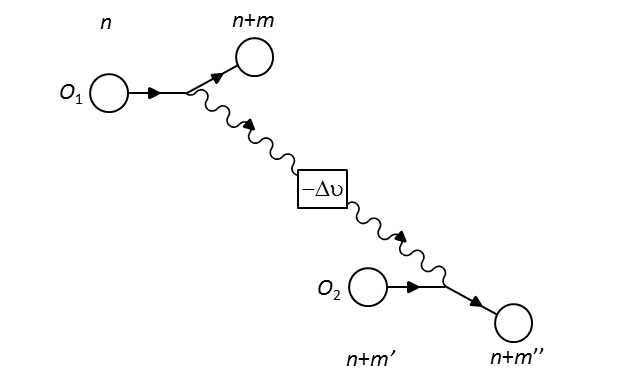}
\end{center}
\caption{A directed quasi-identifiable interaction between objects $O_{1}$ and $O_{2}$, to be compared with the interaction shown in Fig. \ref{Fig98}. In this case, the observed change of object $O_{2}$ occurs after the observed change of $O_{1}$, so that a direction of motion of the transfer quantum from $O_{1}$ to $O_{2}$ can be assigned. The potential knowledge about the distance between $O_{1}$ and $O_{2}$, and the time difference between $n$ and $n+m$, has to be consistent with information transfer at the speed of light.}
\label{Fig98b}
\end{figure}

We may generalize this picture according to Fig. \ref{Fig98b}. Here the update of sequential time $n+m''-1\rightarrow n+m''$ that is defined by the observed change of the state of $O_{2}$ may occur later than the update $n+m-1\rightarrow n+m$ that is defined by the observed change of the state of $O_{1}$. This situation occurs if $m'>m$. In this case the directional symmetry of the interaction is broken. If an interaction is actually taking place between $O_{1}$ and $O_{2}$, the transfer quantum must transfer the bag of attributes $-\Delta \upsilon$ from $O_{1}$ to $O_{2}$.

However, the qualifier `actually' in the above sentence has no epistemic content. All we can ever know is that the attribute changes of $O_{1}$ and $O_{2}$ match when conservation laws are taken into account. This fact may be interpreted as a potential pairwise interaction. Consequently, the imagined quantum that carries the attributes $\Delta \upsilon$ between the two objects has no physical reality. It is neither an object, nor a quasiobject. It is just a book-keeping device that keeps track of attribute changes.

For the same reason, there are no conservation laws that apply to the imagined vertices in Figs. \ref{Fig98} and \ref{Fig98b}. Conservation laws apply to observed changes of objects at sequential time instants $n,n+1,n+2,\ldots$, whereas these vertices are placed in an imaginary world in between these time instants. They do not correspond to object divisions or mergings, but are just graphical illustrations of our book-keeping efforts.

In section \ref{identifiability}, we defined a quasi-identifiable object as an object that can be \emph{modelled} by identifiable minimal objects, even though we cannot be \emph{sure} that the object is the same when we come back and look at it a later time (Definition \ref{quasiidentifiable}). In a similar way, we can define a quasi-identifiable interaction between two objects as a change of the states of these objects that can be \emph{modelled} by a  pair-wise interaction of the kind we have discussed above, mediated by a transfer quantum, even if we cannot be \emph{sure} that these two objects are actually interacting.

We introduced pseudoobjects as force carriers, entities that bend trajectories of identifiable objects. After that, we introduced transfer quanta from the epistemic consideration that an interaction can only be deduced indirectly from matching attribute changes of two objects. Obviously, we can identify pseudoobjects with transfer quanta. This provides another condition for which objects can be part of a quasi-identifiable interaction, since we concluded that all pseudoobjects have to travel at the speed of light (Statement \ref{pseudospeed}). Apart from the condition that the changes of the attribute accounts of two potentially interacting objects have to balance each other, we must therefore require that the distance between the two objects is such that the attribute transfer can be modelled as taking place at the speed of light.

\begin{defi}[\textbf{Quasi-identifiable interaction}]
Let $\upsilon=(\upsilon_{1},\upsilon_{2},\ldots)$ be an array of exact values of a set of attributes $\{A_{1},A_{2},\ldots\}$ that specify object $O_{1}$ as well as object $O_{2}$. Let $Z_{1}=(\upsilon_{11},\upsilon_{21},\ldots)$ be an exact state of object $O_{1}$, and let $Z_{2}=(\upsilon_{12},\upsilon_{22},\ldots)$ be an exact state of object $O_{2}$. To any pair of exact states $Z_{1}^{(n)}\in S_{O1}(n)$ and $Z_{1}^{(n+m)}\in S_{O1}(n+m)$ corresponds an attribute value change $\Delta \upsilon=Z_{1}^{(n+m)}-Z_{1}^{(n)}$. Suppose that there is a pair of exact states $Z_{2}^{(n)}\in S_{O2}(n)$ and $Z_{2}^{(n+m)}\in S_{O2}(n+m)$ such that $\Delta \upsilon=Z_{2}^{(n)}-Z_{2}^{(n+m)}$. Suppose also that there is such a pair that fulfils $\mathbf{r}(Z_{2}^{(n+m)})-\mathbf{r}(Z_{1}^{(n)})\leq c[t(n+m)-t(n)]$ or $\mathbf{r}(Z_{1}^{(n+m)})-\mathbf{r}(Z_{2}^{(n)})\leq c[t(n+m)-t(n)]$. Then there is a quasi-identifiable interaction between $O_{1}$ and $O_{2}$ during the time interval $[n,n+m]$.
\label{identinteraction}
\end{defi}

This definition corresponds to the intuitive process shown in Fig. \ref{Fig98}. The following definition formalizes the generalized process shown in Fig. \ref{Fig98b}.

\begin{defi}[\textbf{Directed quasi-identifiable interaction}]
To any pair of exact states $Z_{1}^{(n)}\in S_{O1}(n)$ and $Z_{1}^{(n+m)}\in S_{O1}(n+m)$ corresponds an attribute value change $\Delta \upsilon=Z_{1}^{(n+m)}-Z_{1}^{(n)}$. Suppose that there is an integer $m'>m$ and a pair of exact states $Z_{2}^{(n+m')}\in S_{O2}(n+m')$ and $Z_{2}^{(n+m'')}\in S_{O2}(n+m'')$ such that $\Delta \upsilon=Z_{2}^{(n+m')}-Z_{2}^{(n+m'')}$. Suppose also that there is such a pair that fulfils $\mathbf{r}(Z_{2}^{(n+m')})-\mathbf{r}(Z_{1}^{(n+m)})\leq c[t(n+m)-t(n)]$ and $\mathbf{r}(Z_{2}^{(n+m'')})-\mathbf{r}(Z_{1}^{(n)})\geq c[t(n+m)-t(n)]$. Then there is a directed quasi-identifiable interaction from $O_{1}$ to $O_{2}$ during the time interval $[n,n+m'']$.
\label{didentinteraction}
\end{defi}

A quasi-identifiable interaction between two objects may be interpreted as a truly identifiable interaction if we can exclude the existence of any other pair of objects that fulfils one of the above definitions. Or, as Sherlock Holmes put it: "Eliminate all other factors, and the one which remains must be the truth." \cite{sherlock}

\begin{defi}[\textbf{Identifiable interaction}]
An interaction between a pair of objects $(O_{1},O_{2})$ is identifiable if and only if the pair fulfil the conditions in Definition \ref{identinteraction} or \ref{didentinteraction}, and if there are no other pairs $(O_{1}',O_{2})$ or $(O_{1},O_{2}')$ that fulfil the same conditions.
\label{identinter}
\end{defi}

\begin{figure}[tp]
\begin{center}
\includegraphics[width=80mm,clip=true]{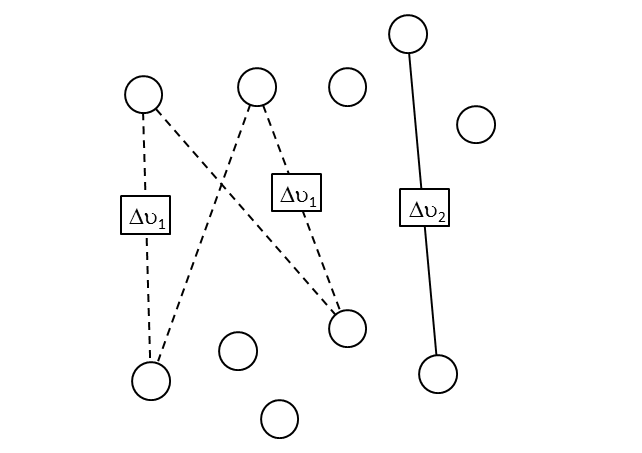}
\end{center}
\caption{The difference between a quasi-identifiable and an identifiable interaction. Two objects in the upper group change their attributes by the same amount $\Delta\upsilon_{1}$, and two objects in the lower group change their attributes by the opposite amount $-\Delta\upsilon_{1}$. This situation can be described in terms of exchanges of pseudoobjects along any of the dashed lines. Here we choose a particular model with two such exchanges. The freedom of choice means that the interaction is quasi-identifiable. One object change its attributes by the amount $\Delta\upsilon_{2}$, and one object change its attributes by the amount $-\Delta\upsilon_{2}$. We are forced into a model where a pseudoobject transfer the attributes between these two objects. The interaction is identifiable.}
\label{Fig101}
\end{figure}

The difference between a quasi-identifiable interaction and an identifiable interaction is illustrated in Fig. \ref{Fig101}.

\begin{figure}[tp]
\begin{center}
\includegraphics[width=80mm,clip=true]{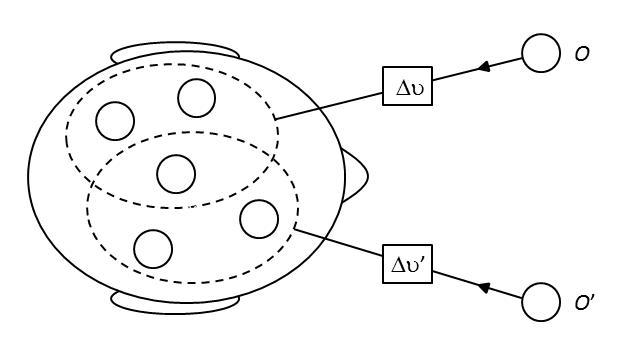}
\end{center}
\caption{The perception of external objects must consist of a one-to-one link between the observed object and a group of quasiobjects in the body whose change of state corresponds to the perception. Such a link can only be achieved by an identifiable interaction.}
\label{Fig100}
\end{figure}

There is one basic situation in which we have to say that the interaction is identifiable, almost by definition. It is the interaction that takes place between the outside world and the body of the subject, making perceptions possible.

We can perceive two different objects $O_{1}$ and $O_{2}$ if and only if two different transfer quanta $\Delta\upsilon_{1}$ and $\Delta\upsilon_{2}$ carry information about these objects to two objects that belong to the body (Fig. \ref{Fig100}). These objects are groups of quasiobjects according to the discussion in section \ref{bwstates} (see Fig. \ref{Fig41a1}). The possibility to link a given external object to a given (composite) object in the body is necessary to uphold detailed materialism and intertwined diualism. Without such an explicit identifiablity there would be no well-defined relation between the body and the world. Consequently, there would be no well-defined relation between subjective perceptions and the objects that are perceived.

But are we allowed to speak about a truly identifiable interaction in this case? We argued above that in general we can only speak about quasi-identifiable interactions. Two objects may be potentially linked by an interaction, but we can never be sure that an interaction is actually taking place between these two particular objects. A quantum of information $\Delta\upsilon$ carried to the body could in principle have come from several different external object that changed their attributes in the same way.

However, this is an empty statement from the epistemic point of view, since the transfer of the attribute values $\Delta\upsilon$ to the subject is the only way to define an external object. We cannot say that there may be several different external objects that may have transferred these attributes to the body since we have to allow different transfers $\Delta\upsilon,\Delta\upsilon',\ldots$ to speak about different external objects in the first place.

\begin{state}[\textbf{Interactions between the external world and the body are identifiable}] 
All interactions between the external world and the body are directed and identifiable. They are mediated by pseudoobjects that link a given external object to a given composite quasiobject that belong to the body.
\label{worldbodyint}
\end{state}

We judge that all interactions between the body and the objects that are perceived are directed, since the flow of information has to go from the observed objects to the perceiving subject, not the other way around.

\begin{figure}[tp]
\begin{center}
\includegraphics[width=80mm,clip=true]{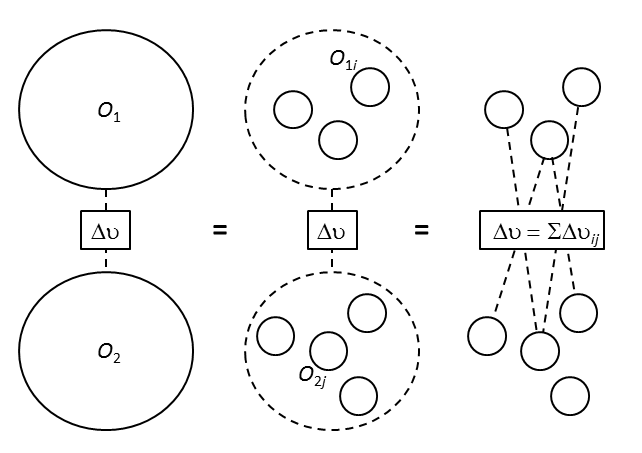}
\end{center}
\caption{A pseudoobject mediates the quasi-identifiable interaction between two objects $O_{1}$ and $O_{2}$. These objects can be decomposed into two sets of minimal objects $\{O_{1i}\}$ and $\{O_{2j}\}$. Analogously, the interaction can be decomposed into pairwise interactions between these minimal objects, mediated by minimal pseudoobjects $PO_{ij}$. These carry the attributes $\Delta\upsilon_{ij}$ from $O_{1i}$ to $O_{2j}$.}
\label{Fig99}
\end{figure}

We have noted the similarity between quasi-identifiable objects and quasi-identifiable interactions. In section \ref{identifiability} we stated that all quasi-identifiable objects can be decomposed into identifiable minimal objects (Statement \ref{allminimalidentity}). In an analogous manner, we will state here that all interactions between two objects can be decomposed into knowable, pairwise interactions between the minimal objects that constitute the two objects. These pairwise interactions define minimal pseudoobjects (Fig. \ref{Fig99}).

The very fact that any object can be described in terms of minimal obects means that any interaction can be described in terms of interactions between minimal objects. One may imagine that these decomposed interactions are not necessarily pairwise decomposable, but redistributed attributes between two groups of minimal objects. However, this cannot be the case since all pseudoobjects are linear according to Statement \ref{linearpseudo}. Interactions that cannot be decomposed into pairwise interations would have to be modelled by pseudoobjects that interact with themselves along the way between the two hypothetical groups of minimal objects. Therefore any knowable interation between two minimal objects can be modelled as a directed, identifiable interaction mediated by a minimal pseudoobject.

\begin{state}[\textbf{All interactions are decomposable}]
Consider any knowable interaction between objects $O_{1}$ and $O_{2}$, where the attributes $-\Delta \upsilon$ are carried from $O_{1}$ to $O_{2}$. These two objects can be decomposed into two sets of identifiable minimal objects $\{O_{1i}\}$ and $\{O_{2j}\}$. In this description, the interaction can be modelled as mediated by minimal pseudoobjects traveling between pairs of minimal objects $(O_{1i},O_{2j})$. Each of these carries the attributes $\Delta\upsilon_{ij}$, and these attributes filfil $-\Delta\upsilon=\sum_{ij}\Delta\upsilon_{ij}$. 
\label{decompinteraction}
\end{state}

We have painted a picture in which all minimal objects are elementary fermions with spin $1/2$. We have also introduced minimal pseudoobjects, corresponding to photons and gravitons. These are the elementary bosons in our description. There is a clear distinction between these fermions and bosons. The former are objects, whereas the latter represent interactions between objects.

But doesn't these mediators of interactions have some characteristics of actual objects after all? Is the distinction that clear? We have stated that pseudoobjects have no location. But the diffraction and interference of light follows the same quantum mechanical principles as the diffraction and interference of electrons. When the electro-magnetic field interacts with itself after having passed the two slits in a double-slit experiment, doesn't that represent a kind of locality? We are tempted to say that a given part of the wave actullay passes a given slit.

There is a basic difference between the interference of electrons and photons, though. We can never investigate which slit a photon actually passed. We must always treat this as unknowable; we always have an experimental setup where these two alternatives have knowability level 1 (Table \ref{levels}). The reason is simply that in order to investigate whether a photon is present, it must be absorbed by an object. Then \emph{another} photon is emitted that may reach our eyes, or another detector. The first photon is no longer defined. A photon can never be tracked and at the same time preserve its identity. It is only the object that absorbs the photon that can be localized. Therefore we have to stick to a more abstract interpretation of the electro-magnetic field. In a double-slit experiment it simply provides the probability that an object at a given position in the detector screen will change its attributes by the amount $\Delta\upsilon$, given that the object corresponding to the radiation source has changed its attributes by the opposite amount $-\Delta\upsilon$. In that case there is a quasi-identifiable interaction between them, encoded as the exchange of a photon. We get an apparent interference pattern since the photon is an inherently non-local entity. Physical law must therefore contradict a description in which we assign a definite path to it, according to explicit epistemic minimalism, just as we get interference in a double slit experiment arranged so that it is forever unknowable which slit the object actually passed, making the object non-local in relation to the slits.

The fact that interactions travel at a given speed $c$ may be regarded as another trait which makes the carriers of these interactions similar to objects. But this speed can be interpreted in a more abstract way, just as the electro-magnetic field itself. We should regard the speed of light simply as a number that encapsulates another condition that has to be fulfilled by two objects that are involved in a quasi-identifiable interation. Namely, the distance between them divided by the observed relational time passed between the attribute changes of the first and the second object should equal $c$. In this way we avoid one of the paradoxes of special relativity - that observers that move in relation to one another nevertheless mesure the same speed $c$ of a passing object. The photon simply is no object.

\vspace{5mm}
\begin{center}
$\maltese$
\end{center}
\paragraph{}

Pseudoobjects are identified as elementary bosons. These elementary bosons are described as book-keeping devices that are useful to express knowable interactions between identifiable objects. The other kind of knowable change an object can undergo is that of transformation (Definition \ref{knowtransobject}). The basic difference is that in a transformation the object changes identity. Do we need to introduce another kind of elementary boson than pseudoobjects in order to represent knowable transformations?

The basic reason why we need pseudoobjects is that two objects that interact are most often separated spatio-temporally. We can make a subjective, spatial distinction between Jupiter and Mars, or between the comb and the hair that is attracted to it by static electricity. The projections of their states onto space-time do not overlap. To describe such a situation as a knowable, quasi-identifiable interaction, we have to create a hypothetical link between the two objects. We have given these links the name pseudoobjects.

In contrast, in an object transformation there is no knowable spatio-temporal separation between the objects that exist just before the transformation and the objects that exist after the transformation. The projections of the states of the ingoing and outgoing objects onto space-time overlap in these processes, as illustrated in Figs. \ref{Fig46} and \ref{Fig47}. They can be identified and linked without the help of entities similar to pseudoobjects.

\begin{figure}[tp]
\begin{center}
\includegraphics[width=80mm,clip=true]{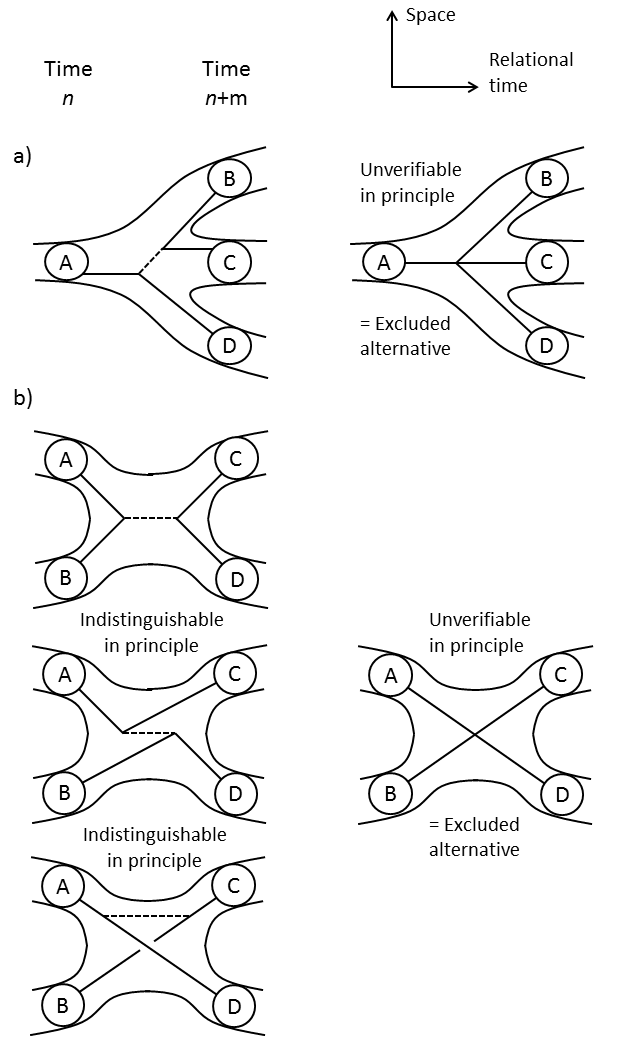}
\end{center}
\caption{Three-legged vertices can and should be used to describe all knowable transformations, since we can never verify that a transformation corresponding to a vertex with four or more legs has actually taken place. a) An object observed at time $n$, and three objects observed at time $n+m$, deduced to be the result of a division of the first object. Such a $1\rightarrow 3$ transformation can be the result of more or less complex sequences of divisions and mergings that are not observed within context. The simplest allowed alternative that describes a $1\rightarrow 3$ transformation can be represented by a graph that consists of two three-legged vertices. b) The same goes for $2\rightarrow 2$ transformations. The three shown graphs of this type must be treated as a single alternative since they cannot be distinguished in any context.}
\label{Fig114}
\end{figure}

However, to be able to represent such processes mathematically, there is still a problem that needs to be resolved. Exactly \emph{how} should we link the incoming and outgoing objects? To discuss these questions, we use the following definition.

\begin{defi}[\textbf{A} $q\rightarrow r$ \textbf{transformation}]
Consider a knowable transformation in which $q$ objects observed at time $n$ or earlier can be identified with $r$ objects observed at time $n+m$ or later in a knowable transformation. Suppose that no member in the set of involved objects $\mathcal{O}_{qr}\equiv\{O_{1},\ldots,O_{q+r}\}$ is observed at any intermediate time $n<n'<n+m$. Then we are dealing with a $q\rightarrow r$ transformation.
\label{qrtrans}
\end{defi}

Figure \ref{Fig114} illustrates $1\rightarrow 3$ and $2\rightarrow 2$ transformations. At time $n+m$ there is a state reduction

\begin{equation}
u_{m}S_{\mathcal{O}_{i}}(n)\rightarrow S_{\mathcal{O}_{f}}(n+m)\subset u_{m}S_{\mathcal{O}_{i}}(n),
\end{equation}
where $S_{\mathcal{O}_{i}}(n)$ is the initial state, $S_{\mathcal{O}_{f}}(n+m)$ is the final state, $\mathcal{O}_{i}\equiv\{O_{1},\ldots,O_{q}\}$ and $\mathcal{O}_{f}\equiv\{O_{q+1},\ldots,O_{r}\}$. Since the objects are not observed between times $n$ and $n+m$, we cannot know exactly when and how the objects transform. In a family of contexts $C(\sigma)$ we can speak about a continuous evolution of the world tube from the initial state $S_{\mathcal{O}_{i}}(n)$, but it should be noted that this world tube does not divide or merge until time $n+m$. Only then do we know that a transformation has actually taken place. The illustrations in Fig. \ref{Fig114} are therefore a bit misleading. This observation justifies the statement that a state reduction has to take place at time $n+m$; we cannot have $S_{\mathcal{O}_{f}}(n+m)=u_{m}S_{\mathcal{O}_{i}}(n)$, since there is no transformation at all in such a situation.

Before the state reduction, all alternatives that are not excluded by the initial state should be included in the state representation $\bar{u}_{m}\bar{S}_{\mathcal{O}_{i}}$ in a superposition. These alternatives can be grouped into topological classes according to those divisions and mergings that occur. These classes can be defined by a graph with directed edges embedded in space-time. A few such graphs are shown in Fig. \ref{Fig114}. Only those topologies or graphs that can in principle be observed should be included. That is, there should exist another family of context $C'(\sigma)$ such that the topology in question can be verified in the sense that no simpler graph is consistent with the sequence of observations made within $C'(\sigma)$. Only in that case is the corresponding alternative realizable. This condition excludes, for example, graphs with edges shorter than the minimum resolvable Lorentz distance $l_{\min}$ (Fig. \ref{Fig95}).

Explicit epistemic minimalism (Assumption \ref{explicitepmin}) provides the reason why we should only include graphs that correspond to realizable alternatives. Physical law should be inconsistent with models that incorporate processes that cannot be verified even in principle. The inclusion of such entities should give the wrong answer when such a model is used to predict the outcome of an experiment.

For the same reason we should demand that all internal vertices have degree three, that is, three legs. Since there is a smallest observable Lorentz distance $l_{\min}$, we can never exclude the possibility that a vertex that appears to contain an internal vertex with a degree larger than three in reality corresponds to a sequence of transformations that each correspond to a vertes with degree three. Examples of such decompositions of vertices with degree four are shown in Fig. \ref{Fig114}.

\begin{defi}[\textbf{Irreducible transformation graph}]
Let $G_{rp}$ be a transformation graph that is consistent with a $r\rightarrow p$ transformation. Let $G_{rp}^{-}$ be a graph constructed from $G_{rp}$ by removing edges. $G_{rp}$ is irreducible if and only if there is a context $C'$ such that $G_{rp}$ is consistent with a sequence of observations that can be made within $C'$, but no graph $G_{rp}^{-}$ is consistent with this particular sequence of observations.
\label{verigraph}
\end{defi}

\begin{figure}[tp]
\begin{center}
\includegraphics[width=80mm,clip=true]{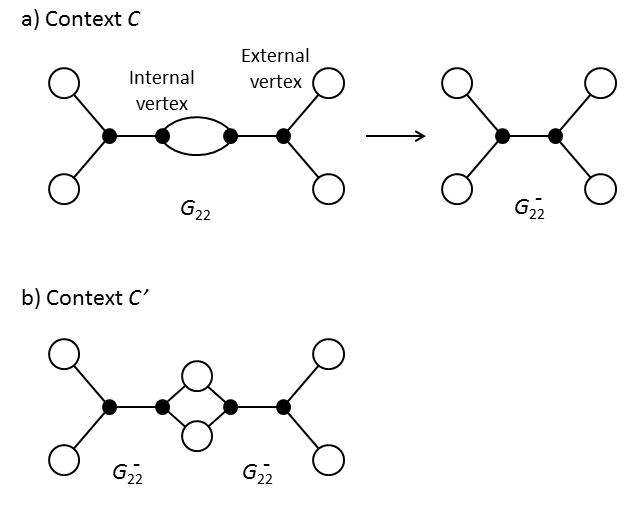}
\end{center}
\caption{Graphs $G_{qr}$ that describe possible topologies in $q\rightarrow r$ transformations. a) A graph $G_{22}$, and another graph $G_{22}^{-}$, created from $G_{22}$ by removing one edge. In so doing, two internal vertices lose their meaning and can be removed [compare Fig. \ref{Fig114}(c)]. These graphs represent two classes of realizable alternatives in a given $2\rightarrow 2$ transformation. b) Another context $C'$ in which a sequence of five observations realizes the alternative represented by $G_{22}$.}
\label{Fig115}
\end{figure}

The graph $G_{22}$ in Fig. \ref{Fig115} is irreducible by this criterion, since the context $C'$ makes it possible to detect two objects in an intermediate observation, which correspond to the internal loop in $G_{22}$.

\begin{state}[\textbf{Realizable transformation alternatives}]
An alternative consistent with a given $q\rightarrow r$ transformation that may be the outcome of a context family $C(\sigma)$ should be included in a state representation $\bar{u}(\sigma)\bar{S}_{\mathcal{O}_{i}}$ if and only if it corresponds to an irreducible transformation graph in which all internal vertices have degree three.
\label{realtrans}
\end{state}

The rule that all internal vertices must have degree three implies that each object transformation $q+r>3$ should be represented by transformation graphs in which there is at least one edge that does not correspond to the world tube of an observed object. These edges are dashed in Fig. \ref{Fig114}.

The addition rule for angular momenta means that exactly one of the three edges must carry integer spin. Since all observed minimal objects have spin $1/2$ according to the Dirac equations, it is the dashed internal edges that must be ascribed this integer spin value. This observation also implies that $1\rightarrow 2$ transformations are forbidden.

\begin{defi}[\textbf{Cryptoobject}]
A cryptoobject is an edge in a transformation graph that connects two internal vertices $v_{1}$ and $v_{2}$. The edge is such that all other edges that are connected to $v_{1}$ and $v_{2}$ correspond to objects.
\label{crypto}
\end{defi}

Figure \ref{Fig114} shows the cryptoobjects as dashed lines.

\begin{state}[\textbf{Cryptoobjects mediate all object transformations}]
There is at least one cryptoobject associated with each knowable object transformation.
\label{cryptotrans}
\end{state}

We argued above that cryptoobjects must be ascribed integer spin. Conservation of the invariant mass implies that they must be ascribed non-zero rest mass.

\begin{state}[\textbf{Cryptoobjects are massive elementary bosons}]
An integer total spin and a non-zero rest mass can be associateed with each cryptoobject.
\label{massivecrypto}
\end{state}

The picture that emerges is that all elementary massive gauge bosons can be interpreted as cryptoobjects. These cryptoobjects are as `unreal' as pseudoobects in the sense that they can never be observed. They appear only as internal edges in transformation or interaction graphs that properly describes the outcome of an observational context. Their introduction as elements in the mathematical representation of physical law is necessary in order to fulfil two basic requirements in our epistemic approach to physics: epistemic minimalism in the case of cryptoobjects, and identifiability in the case of pseudoobjects.

\begin{figure}[tp]
\begin{center}
\includegraphics[width=80mm,clip=true]{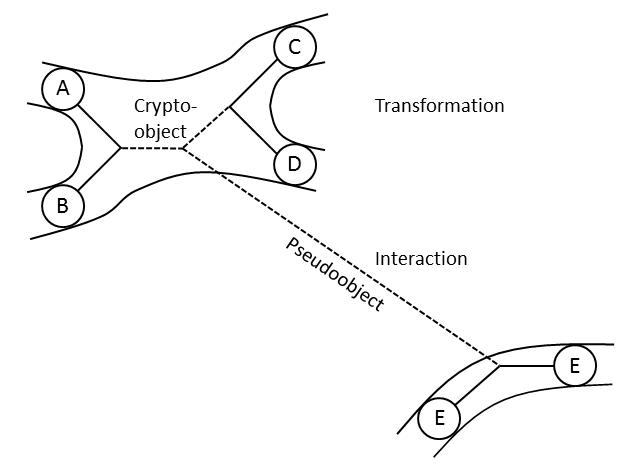}
\end{center}
\caption{Cryptoobjects, which mediate transformations, can be modelled as being able to emit pseudoobjects, which mediate interactions. However, this model is indistinguishable from a model in which the pseudoobject emanate from any of the four objects involved in the transformation (solid lines). These possibilities should therefore be treated as a single realizable alternative.}
\label{Fig116}
\end{figure}

The fact that cryptoobjects and pseudoobjects cannot be observed means that there is no irreducible graph (Definition \ref{verigraph}) such that these elementary bosons form loops among themselves. All loops involve edges that correspond to observable objects (Fig. \ref{Fig115}). Elementary bosons may transform among themselves in a restricted way, though. Cryptoobjects carry mass and are localized in the sense that they are confined to the known region in space-time in which the transformation takes place. We may therefore have an irreducible graph corresponding to a quasi-identifiable interaction in which the cryptoobject emit a pseudoobject, which is absorbed later by an object that undergo a knowable interaction. In this way the concepts of transformations and interactions can be mixed (Fig. \ref{Fig116}).

However, the `transformation-interaction' in Fig. \ref{Fig116} is indistinguishable from a process in which the transformation is separate from the interaction, and the pseudoobject is emitted from one of the vertices that correspond to one of the four objects involved in the transformation. These two possibilities are therefore just two representations of the same alternative. As such, they should be assigned the same amplitude $a$ in a wave function, and they should not be double-counted when the wave function is normalized. The same goes for the three indistinguishable possibilites shown in Fig. \ref{Fig114}(b).

\begin{state}[\textbf{The use of indistinguishable, irreducible graphs in wave functions}]
Suppose that there are $m$ irreducible graphs $\{G_{1},\ldots,G_{m}\}$ that conform with a set of transformations or interactions observed within some context $C$. Suppose further that there is no other context $C'$ in which the same set of transformations or interactions are observed, such that some of the $m$ graphs are excluded. Then, if graphs are used to define alternatives $\bar{S}_{j}$ in a wave function $a(j)$ that are defined for $C$, the entire set of graphs $\{G_{1},\ldots,G_{m}\}$ corresponds to a singe alternative $\bar{S}_{j}$.   
\label{indigraphs}
\end{state}

\section{The spin-statistics theorem}
\label{spinstatistics}

In section \ref{identifiability} we defined an identifiable object $O$ to be such that its state $S_{OO}(n)$ at time $n$ overlap its state $S_{OO}(n+1)$ at time $n+1$. Since we cannot distinguish the two states, we have to say that they describe the same object. We simply cannot justify a claim that the object we observe at time $n+1$ is different from that we observed at time $n$.

\begin{figure}[tp]
\begin{center}
\includegraphics[width=80mm,clip=true]{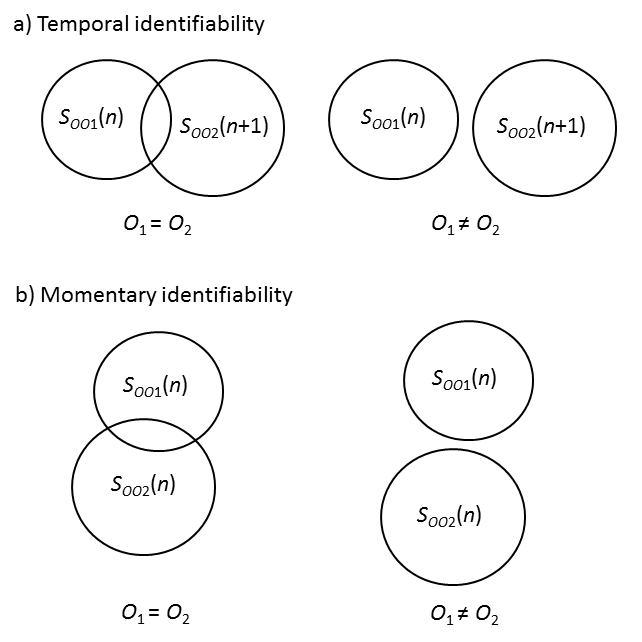}
\end{center}
\caption{Identifiability and the Pauli exclusion principle. a) If the states of two objects $O_{1}$ and $O_{2}$ at subequent times overlap, the objects must be considered to be the same. We speak about an \emph{identifiable} object that we can track as time passes. b) If the states of two objects $O_{1}$ and $O_{2}$ at the same time overlap, they must also be considered to be one and the same object, if the states are represented in object state space $\mathcal{S}_{O}$. They cannot be distinguished. From an epistemic point of view, the statement that the states of two objects overlap is paradoxical. If the overlapping states cannot be reduced to non-overlapping states by future observations, then we can exclude the possibility that the object will divide and turn out to be two objects at a later time. See Fig. \ref{Fig90b} for further explanation.}
\label{Fig90}
\end{figure}

Consider the states $S_{OO1}(n)$ and $S_{OO2}(n)$ of two objects $O_{1}$ and $O_{2}$, which we observe at the same time $n$. Since we have given the objects different names, there must be a way to distinguish them. This means that the state $S_{OO1}(n)$ does not overlap the state $S_{OO2}(n)$ (Fig. \ref{Fig90b}). If they do overlap, on the other hand, we cannot justify the claim that the two objects are different, in the same way as we cannot justify that the identifiable object $O$ at time $n+1$ is different from that at time $n$. For the same reason, we must judge that $O_{1}$ and $O_{2}$ is one and the same object. We may say that these two objects fulfil \emph{momentary identifiability}, and that the objects described by the two states $S_{OO}(n)$ and $S_{OO}(n+1)$ fulfil \emph{temporal identifiability}. These considerations are illustrated in Fig. \ref{Fig90}.

Even if we cannot distinguish two objects at the present moment, it may be possible to observe them more closely in the future, in order to decide that we are indeed dealing with a composite object with two parts. In other words, we may observe object division. We have to distinguish between the cases when this is possible in principle, and when it is not. In the first case it is meaningful to speak about the potential existence of two distinct objects, and in the second case it is not.

In the latter case explicit epistemic minimalism requires that a proper representation of a contextual state should not contain terms with two factors that represents a specimen consisting of two objects, since it is impossible in principle to decide whether two objects are really present. In the former case the representation should contain such terms, since they represent possible states of potential knowledge.

We may compare these considerations to those discussed in section \ref{boundstates}, in relation to the discreteness of the set of possible states in bound states. The Lorentz distances are discrete in such states, meaning that it is impossible in principle to measure any other distance. Therefore we should not include terms in the contextual state that corresponds to distances that do not belong to the discrete spectrum. This means that it is not quite correct to use a continuous spatial wave function (Statement \ref{selfconsistency}). Similarly, the impossibility to measure arbitrarily short distances means that we should not include processes on shorter scales as terms in representations of possible particle reactions (Statement \ref{cutoff}).

In general, we may express as follows a contextual state in which the alternatives $S_{j}$ are realized with probability $|a_{j}|^{2}$, and in which the specimen is possibly composite (see section \ref{compspec}):

\begin{equation}
\bar{S}_{C}=\sum_{j}a_{j}\sum_{m}\delta(j,m)\prod_{kl}^{(m)}\bar{S}_{Pkl}.
\label{comprep}
\end{equation}
This can be seen as an expression of the form [\ref{collectivecontext}], where we observe a collective property $cP$ with property value states

\begin{equation}
\bar{S}_{cPj}=\sum_{m}\delta(j,m)\prod_{kl}^{(m)}\bar{S}_{Pkl}.
\end{equation}
The property value states $\bar{S}_{Pkl}$, on the other hand, refer to the individual objects $O_{l}$ that the specimen is potentially made of. In a fundamental description, these are minimal objects. In the following, we let the individual property $P$ refer to the array of \emph{all} attributes of $O_{l}$ that can be known simultaneously with the collective property $cP$, which defines the alternatives $S_{j}$. We let $k$ be an index that numbers all the property values of the array $P$ of simultaneously knowable individual attributes. The state $\bar{S}_{Pkl}$ thus corresponds to a state of maximum potential knowledge about a minimal object that is part of the specimen $OS$, given that the value of the collective property $cP$ is also known.

\begin{equation}
\bar{S}_{Pkl}\leftrightarrow state\;of\;maximum\;knowledge\;about\;object\;O_{l},
\label{maximumol}
\end{equation}
This means that the product corresponds to a state of maximum knowledge about the entire specimen $OS$:

\begin{equation}
\prod_{kl}^{(m)}\bar{S}_{Pkl}\leftrightarrow state\;of\;maximum\;knowledge\;about\;specimen\;OS,
\label{maximumos}
\end{equation}
where $m$ is an index that points to a given such state of maximum knowledge among all possibilities. These states may contain a varying number of minimal objects $O_{M}$, so that the product may contain a varying number of factors.

The binary function $\delta(j,m)$ equals one if the collective maximum knowledge state $\prod_{kl}^{(m)}\bar{S}_{Pkl}$ is consistent with the collective property value $cP_{j}$ that is observed when alternative $S_{j}$ is realized, and $\delta(j,m)$ equals zero otherwise.

\begin{state}[\text{The Pauli exclusion principle}]
Consider a representation of a contextual state of the form \ref{comprep}. We have $\delta(j,m)=0$ whenever the product $\prod_{kl}^{(m)}\bar{S}_{Pkl}$ contains a factor $\bar{S}_{Pkl}\bar{S}_{Pkl'}$.
\label{pauli}
\end{state}

Such a factor $\bar{S}_{Pkl}\bar{S}_{Pkl'}$ would correspond to identical property value states of two different objects, without any hope to tell the objects apart via further observations. This would be paradoxical from the epistemic point of view, as discussed at the beginning of this section.

Consider the classical illustration of the Pauli exclusion principle, the fact there is only room for two electrons in each atomic shell, corresponding to one electron with spin $s_{z}=1/2$, and one electron with spin $s_{z}=-1/2$. We let $\bar{S}_{C}$ represent a context in which the specimen $OS$ is the entire atom. Say that we know which element we are dealing with, but we don't know which isotope. Thus the number of minimal objects $O_{M}$ that are part of the product [\ref{maximumos}] is not fixed by our knowledge \emph{a priori}. Let the context be such that the rest energy of the atom is measured. This is the collective property $cP$ whose possible values define the alternatives $S_{j}$. It is possible to measure the spin of an individual electron in the atom along any given $z$-direction at the same time as the rest energy of the atom; this particular individual property and the collective property $cP$ are simultaneuously knowable. Therefore the spin direction of an electron is an element in the array of individual attributes $P$. According to Statement \ref{pauli} we should thus exclude all maximum knowledge states [\ref{maximumos}] that contain more than one individual maximum knowledge state [\ref{maximumol}] that represents an electron with a given spin direction. Therefore, in any proper representation of $\bar{S}_{C}$ there are at most two electrons which have identical quantum numbers, if we disregard the spin direction. Note that this is true even if we do not explicitly determine the spin directions of all the eletrons. The only property that we actually observe in our example is the atomic rest mass, and therefore the different possible outcomes of this observation are the only property values that are assigned probabilities $|a_{j}|^{2}$.

What about elementary bosons, that is, minimal pseudoobjects and cryptoobjects? Let us repeat the main messages of the preceding section. Elementary bosons are book-keeping devices, and as such they have no individuality when it comes to actual perceptions. We cannot expect the Pauli exclusion principle to hold. Since both pseudoobjects and cryptoobjects must be assigned integer spin, we get a primitive version of the spin-statistics theorem.

Let us discuss the case of pseudoojects in more detail for the sake of illustration. They identify pairs of potentially interacting minimal objects. According to Statement \ref{decompinteraction}, all interactions are decomposable into such pairs. It may be possible to pin down exactly which elementary object is interacting with which, in which case the interaction is identifiable. The interactions may also be quasi-identifiable. In this case we can just say that a there is a group of $N$ minimal objects, each of which may be interacting with any the minimal objects in another group of $N$ minimal objects. Each object in the first group changes its attributes by the same amount $\Delta\upsilon$, and each object in the second group changes its attributes by the opposite amount $-\Delta\upsilon$.

In Fig. \ref{Fig101}, there are two such groups containing $N=2$ objects each. Attributes can be modelled to be transferrred from one group to another along any of the dashed lines. Even if we cannot determine exactly which paths they take, we know that the number of pseduoobjects involved in the interaction must be $N$.

The main point to be made in the present section is that all these $N$ pseduoobjects can be described as being identical, carrying the same array of attributes $-\Delta\upsilon$. To say that there is a group of $N>1$ identical pseudoobjects is just another way to say that we have an interaction that is observed to be quasi-identifiable rather than identifiable. There is no epistemic reason to rule out this situation, since each of the $2N$ objects we observe to deduce that there are $N$ identical pseudoobjects is distinct from all the other objects. Therefore the Pauli exclusion principle does not hold for pseduoobjects.

\begin{state}[\text{The spin-statistics theorem for elementary fermions and bosons}]
Minimal objects with spin quantum number $s=1/2$ obey the Pauli exclusion principle, but minimal pseudoobjects with integer spin quantum number do not.
\label{minispinstat}
\end{state}

One may argue that it is not meaningful to construct a contextual state like that in Eq. [\ref{comprep}] when we are dealing with a collection of pseudoobjects, since we cannot compose a specimen out of pseudoobjects. It may nevertheless be done, if we reinterpret the meaning of some of the symbols.

Consider a context in which we observe the way in which two objects $O_{1}$ and $O_{2}$ interact, assuming that they actually do. The specimen is the composite object consisting of $O_{1}$ and $O_{2}$. The alternatives $S_{j}$ correspond to different attribute changes $\Delta\upsilon_{j}$ and $-\Delta\upsilon_{j}$ of these two objects. Then we may write

\begin{equation}
\bar{S}_{C}=\sum_{j}a_{j}\sum_{m}\delta(j,m)\prod_{k}^{(m)}N_{k}\overline{\Delta S}_{Pk}.
\label{comprep2}
\end{equation}

The `differential state' $\overline{\Delta S}_{Pk}=\Delta\upsilon_{k}$ represents the state of a mimimal pseudoobject whose attributes are determined to the maximum precision that is possible to know simultaneously with $\Delta\upsilon_{j}$. These states are distinct in the sense that $\overline{\Delta S}_{Pk}\cap \overline{\Delta S}_{Pk'}=\varnothing$ whenever $k\neq k'$. The product $\prod_{k}^{(m)}N_{k}\overline{\Delta S}_{Pk}$ represents a specific decomposition of the interaction between $O_{1}$ and $O_{2}$ in terms of pairwise interactions between minimal objects belonging to these two objects. The number $N_{k}$ is the number of pairwise interaction with attribute transfer $\Delta\upsilon_{k}$ in this particular decomposition. The binary function $\delta(j,m)$ has a similar interpretation as in Eq. [\ref{comprep}] and has to be introduced to make sure that the condition $\Delta\upsilon_{j}=\sum_{k}N_{k}\Delta\upsilon_{k}$ is fulfilled (Fig. \ref{Fig99}).

Looking at Fig. \ref{Fig101} from this point of view, we may interpret the upper group of objects as a composite object $O_{1}$, the lower group as $O_{2}$, and all object as the specimen $OS$. We have observed the overall attribute change $\Delta_{j}=\Delta\upsilon_{1}+\Delta\upsilon_{2}$, and we have $N_{1}=2$ and $N_{2}=1$.

The total number of minimal objects that change their attributes may be impossible to determine within context. This corresponds to the statement that the total number of observed pseudoobjects cannot be predicted; there is no conservation law for the number of elementary bosons.

\vspace{5mm}
\begin{center}
$\maltese$
\end{center}
\paragraph{}

The spin-statistics theorem, as stated above, holds only for elementary fermions and bosons. More precisely, it holds for minimal objects on the one hand, and minimal pseudo- and cryptoobjects on the other. The traditional formulation of the spin-statistics theorem is more general. It accounts for the statistics of composite fermions and bosons as well. An object that can be decomposed into an even number of minimal objects obeys Bose-Einstein statistics. Several such objects can occupy the same state, at least if we consider collective properties of the entire object. From our perspective, it may seem strange that a composite object can behave as if it were not an object at all, but a pseudo- or cryptoobject. To be able to account for this behaviour we will express ourselves in terms of evolving, spatio-temporal wave functions $\Psi_{\mathbf{r}_{4}}(\sigma)$ rather than the more fundamental and general property value states $\bar{S}_{P}$.

We start by considering property value states, however. Let $P_{l}$ be the array of all properties of object $O_{l}$ that is simultaneously knowable with its position $\mathbf{r}_{4}$. Then we can express a state of maximum knowledge about $O_{l}$ as $\bar{S}_{Pkl}(\mathbf{r}_{4})$ (compare Eq. [\ref{maximumol}]). A state of maximum knowledge about the entire specimen can be written

\begin{equation}
\bar{S}_{Pk1}(\mathbf{r}_{4})\bar{S}_{Pk'2}(\mathbf{r}_{4}')\bar{S}_{Pk''3}(\mathbf{r}_{4}'')\ldots.
\label{maximumospos}
\end{equation}

\begin{figure}[tp]
\begin{center}
\includegraphics[width=80mm,clip=true]{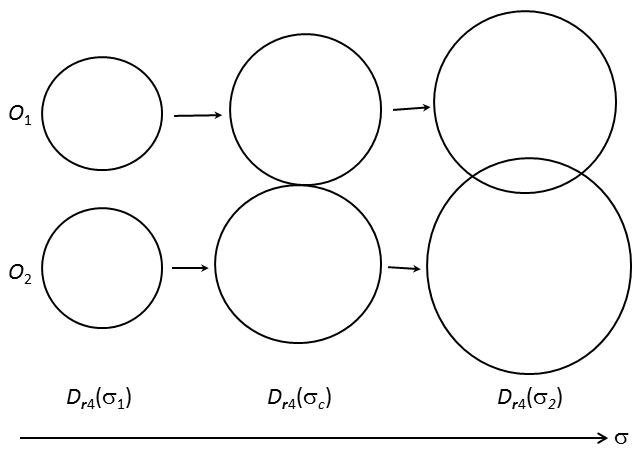}
\end{center}
\caption{A context in which the spatio-temporal positions of two objects $O_{1}$ and $O_{2}$ are observed. The support $D_{\mathbf{r}_{4}}$ of their wave functions start to overlap as these evolve. Since they do not overlap for $\sigma<\sigma_{c}$ we can decide that we are dealing with two objects even if they share the same values of all the other attributes. For $\sigma>\sigma_{c}$ we may find the two objects in the same position. However, this is forbidden by the Pauli exclusion principle if all the other attributes are also equal. This fact motivates the choice of anti-symmetric wave functions.}
\label{Fig102}
\end{figure}

For simplicity, consider a specimen that consists of just two minimal objects $O_{1}$ and $O_{2}$. These are observed in a family of contexts $C(\sigma)$ such that the projections $D_{\mathbf{r}_{4}1}$ and $D_{\mathbf{r}_{4}2}$ of the two object states onto space-time do not overlap at the time $n$ when the context is initiated, corresponding to $\sigma=0$ (Fig. \ref{Fig102}). The context $C(\sigma)$ is assumed to be idealized or fundamental, and to be such that the exact positions $r_{4}$ and $r_{4}'$ of the two objects are observed. These two positions are simulataneously knowable, so that a combined wave function

\begin{equation}
\Psi_{\mathbf{r}_{4}}(\mathbf{r}_{4},\mathbf{r}_{4}',\sigma)
\label{twowave}
\end{equation}
exists according to Definition \ref{jwavedef}. At the intial time $n$ we can separate the objects spatio-temporally, order them in a list, and associate the first argument in the wave function to the position of $O_{1}$ and the second argument to the position of $O_{2}$. Such a wave function continues to de defined even above a critical value $\sigma_{c}>0$ of the evolution parameter at which $D_{\mathbf{r}_{4}1}$ and $D_{\mathbf{r}_{4}2}$ starts to overlap. The knowledge that there are two objects cannot be erased by the evolution.

Let us assume that the contexts $C(\sigma)$ are such that $P$ is already known at time $n$, so that the specimen is in a state of maximum knowledge of the form [\ref{maximumospos}] after the position observations. Then, if there is a critical value $\sigma_{c}>0$ as described above, the Pauli principle excludes the collective maximum knowledge state 

\begin{equation}
\bar{S}_{Pk1}(\mathbf{r}_{4})\bar{S}_{Pk2}(\mathbf{r}_{4})
\label{maximumospos2}
\end{equation}
for $\sigma\geq\sigma_{c}$. For these $\sigma$ we must add the condition

\begin{equation}
\Psi_{\mathbf{r}_{4}}(\mathbf{r}_{4},\mathbf{r}_{4},\sigma)=0
\label{idwave}
\end{equation}
by hand if the two objects are identical, meaning that $k'=k$. For $\sigma<\sigma_{c}$ we know before observation that $\mathbf{r}_{4}\neq\mathbf{r}_{4}'$, so that this condition is not necessary. The left hand side of Eq. [\ref{idwave}] is nevertheless well-defined, since the wave function [\ref{twowave}] still exists for $\sigma\geq\sigma_{c}$.

Next, let us relate the meaning of expression [\ref{twowave}] to the meaning of the same expression with interchanged arguments:

\begin{equation}
\Psi_{\mathbf{r}_{4}}(\mathbf{r}_{4}',\mathbf{r}_{4},\sigma).
\end{equation}

The former expression denotes the probability amplitude for the event that the object with property array $p_{k}$ is found at position $\mathbf{r}_{4}$, and the object with property array $p_{k'}$ is found at position $\mathbf{r}_{4}'$. Naturally, the latter expression then denotes the probability amplitude for the event that the object with properties $p_{k}$ is found at $\mathbf{r}_{4}'$, and that the object with properties $p_{k'}$ is found at $\mathbf{r}_{4}$.

Let us again assume that the two minimal objects are identical, so that $k=k'$ and $p_{k}=p_{k'}$, and study the effect of such object swapping in this case. Since nothing knowably changes in the operation, we must have

\begin{equation}
|\Psi_{\mathbf{r}_{4}}(\mathbf{r}_{4},\mathbf{r}_{4}',\sigma)|^{2}=|\Psi_{\mathbf{r}_{4}}(\mathbf{r}_{4}',\mathbf{r}_{4},\sigma)|^{2}
\end{equation}
for all position pairs $(\mathbf{r}_{4},\mathbf{r}_{4}')$ and for all $\sigma$ in the domain $[0,\sigma_{\max}]$. If we denote the object swapping operation by $\pi$, we may therefore write

\begin{equation}\begin{array}{rcl}
|\pi\Psi_{\mathbf{r}_{4}}(\mathbf{r}_{4},\mathbf{r}_{4}',\sigma)|^{2} & = & |\Psi_{\mathbf{r}_{4}}(\mathbf{r}_{4},\mathbf{r}_{4}',\sigma)|^{2}\\
\pi^{2}\Psi_{\mathbf{r}_{4}}(\mathbf{r}_{4},\mathbf{r}_{4}',\sigma) & = & \Psi_{\mathbf{r}_{4}}(\mathbf{r}_{4},\mathbf{r}_{4}',\sigma).
\end{array}
\end{equation}
The only solutions to these equations are $\pi=I$ or $\pi=-I$, where $I$ is the identity operator. That is,

\begin{equation}
\Psi_{\mathbf{r}_{4}}(\mathbf{r}_{4}',\mathbf{r}_{4},\sigma)=\Psi_{\mathbf{r}_{4}}(\mathbf{r}_{4},\mathbf{r}_{4}',\sigma),
\label{sympsi}
\end{equation}
or

\begin{equation}
\Psi_{\mathbf{r}_{4}}(\mathbf{r}_{4}',\mathbf{r}_{4},\sigma)=-\Psi_{\mathbf{r}_{4}}(\mathbf{r}_{4},\mathbf{r}_{4}',\sigma).
\label{asympsi}
\end{equation}

The evolution parameter $\sigma$ is introduced to allow a continuouos description of the evolution between discrete events. Therefore we should require that the wave function changes continuously with $\sigma$. This means that if the wave function is symmetric according to Eq. [\ref{sympsi}] or anti-symmetric according to Eq. [\ref{sympsi}] for a given $\sigma$, it sticks to the same kind of symmetry for all $\sigma$. Similarly, the wave function should change continuously with the position arguments.

This means that the symmetry type of a wave function that describes a pair of identical minimal objects is a global quality that does not depend on the arguments. To ensure that Eq. [\ref{idwave}] is always fulfilled, we have to choose anti-symmetric wave functions, regardless whether $\sigma<\sigma_{c}$ or $\sigma\geq\sigma_{c}$. In this way we have arrived at the traditional defining characteristic of a fermionic wave function.

Note that we are not allowed to use this characteristic as a starting point for the discussion in this section, since wave functions are not fundamental entities in the epistemic formalism. Rather, the fundamental entity is the physical state $S$. Consequently, we started out with a discussion about object states that were overlapping or non-overlapping.

Note also that to we have to consider wave functions to be able to introduce the minus-sign that defines the anti-symmetry that characterizes fermions. In other words, we have to consider contexts in which the positions of the fermions are actually measured, so that there are probability amplitudes for all the possible outcomes, amplitudes that may be negative. In the general expression [\ref{comprep}] of the contextual state in terms of property value states $\bar{S}_{Pkl}$, the plus-sign is defined according to the discussion on section \ref{compspec}, but we have given no meaning to the minus-sign. Therefore minus-signs cannot appear in such expressions.

It is not essential to use spatio-temporal wave functions to arrive at the conclusion that they must be anti-symmetric. We may consider any property $P$, and argue along the same lines that for any pair of identical minimal objects we must have

\begin{equation}
\Psi_{P}(p',p,\sigma)=-\Psi_{P}(p,p',\sigma),
\label{asympsi2}
\end{equation}
if we assume that all other attributes $\breve{P}$ that are simultaneously knowable with $P$ are indeed precisely known at the initiation of the context at time $n$. Put another way, we assume that the initial state $S_{Ol}$ of each minimal object fulfils $S_{Ol}(n)\subseteq S_{\breve{P}kl}$, where $S_{\breve{P}kl}$ is a state of maximum knowledge about $O_{l}$ according to Eq. [\ref{maximumol}]. 

\begin{defi}[\textbf{The co-property} $\breve{P}$ \textbf{to} $P$]
Consider any property $P$ that describes an object $O_{l}$. A co-property $\breve{P}$ associated with $P$ is an array of properties that is simultaneously knowable with $P$, and such that there is no simultaneously knowable property that is independent from the array of properties $(P,\breve{P})$, in the sense that the value of any property $P'$ is a function of the values of $(P,\breve{P})$. All properties in this array are assumed to be independent from the others.  
\label{coproperty}
\end{defi}

Let us give a couple of examples, focusing on relational attributes. If $P=t$, then $\breve{P}$ may be chosen as $(\mathbf{r},s_{z})$, where $s_{z}$ is the spin along some $z$-direction. If $P=E$, then $\breve{P}$ may be chosen as $(\mathbf{p},s_{z})$.

Above we have considered a specimen that consists of two minimal objects only. The same reasoning holds for any specimen, as long as the number of minimal objects in the specimen is known at the initiation of the context (at time $n$).

\begin{state}[\textbf{Anti-symmetric wave functions for identical minimal objects}]
Consider a family of fundamental contexts $C(\sigma)$ in which the specimen is known to contain $N$ minimal objects $O_{l}$ at the initial time $n$. The individual property $P$ of each of these minimal objects is observed, whereas the value $\breve{p}_{kl}$ of the co-property $\breve{P}$ is already exactly known at time $n$ for each object $O_{l}$. Suppose that the minimal objects are identical, meaning that $\breve{p}_{l}=\breve{p}_{l'}$ for all object pairs $(O_{l},O_{l'})$. Suppose also that for each such object pair, there is a $\sigma>0$ for which $p_{l}=p_{l'}$ is a possible outcome of the observation. Then

\begin{equation}
\Psi_{P}(p_{1},\ldots,p_{l'},\ldots,p_{l},\ldots,p_{N},\sigma)=-\Psi_{P}(p_{1},\ldots,p_{l},\ldots,p_{l'},\ldots,p_{N},\sigma)
\end{equation}
for all such object pairs, for all property values $(p_{1},\ldots,p_{l'},\ldots,p_{l},\ldots,p_{N})$ allowed by physical law, and for all $\sigma$ in its domain $[0,\sigma_{\max}]$.
\label{spinstatevol}
\end{state}

Let us finally discuss the case of composite objects. Suppose that $N$ is divisible by $M$, and that we know at time $n$ that all minimal objects in the specimen can be grouped in bound states which contain $M$ minimal objects each. In the case $M=2$ we can represent this knowledge in the wave function as follows:

\begin{equation}
\Psi_{P}=\Psi_{P}(p_{1},p_{2};p_{3},p_{4};\ldots;p_{N-1},p_{N}),
\label{twobound}
\end{equation}
where the semi-colons separate minimal objects belonging to different bound states. Now we cannot assume that all minimal objects are identical, just that each group of $M$ objects is identical to each other such group. This means that we can order the minimal objects so that $\breve{p}_{l}=\breve{p}_{l+M}=\breve{p}_{l+2M}=\ldots$. In Eq. [\ref{twobound}] this means that $\breve{p}_{1}=\breve{p}_{3}=\ldots=\breve{p}_{M-1}$ and $\breve{p}_{2}=\breve{p}_{4}=\ldots=\breve{p}_{M}$.

We may use Statement \ref{spinstatevol} in this situation if we restrict argument swapping to the sets of arguments that correspond to identical minimal objects. In the case $M=2$ we get, for example,

\begin{equation}\begin{array}{rcl}
\Psi_{P}(p_{3},p_{2};p_{1},p_{4};\ldots;p_{N-1},p_{N}) & = & -\Psi_{P}(p_{1},p_{2};p_{3},p_{4};\ldots;p_{N-1},p_{N})\\
\Psi_{P}(p_{1},p_{4};p_{3},p_{2};\ldots;p_{N-1},p_{N}) & = & -\Psi_{P}(p_{1},p_{2};p_{3},p_{4};\ldots;p_{N-1},p_{N}).\\
\end{array}
\label{twoswap}
\end{equation}

If we perform these two swappings sequentially, we get

\begin{equation}
\Psi_{P}(p_{3},p_{4};p_{1},p_{2};\ldots;p_{M-1},p_{M})=\Psi_{P}(p_{1},p_{2};p_{3},p_{4};\ldots;p_{N-1},p_{N}).
\label{twoswap2}
\end{equation}

The bound states correspond to composite objects. Let us call the first composite object in the argument list $O_{c1}$, the second $O_{c2}$, and so on. We may define a collective version $cP$ of the observed property $P$. For instance, if $P=\mathbf{r}_{4}$ in the case $M=2$, then $cP$ may be the center of mass position of the two bound minimal objects. The contexts are assumed to be fundamental, so that the value $p$ of $P$ is exactly known for each minimal object after the observations. Then the same is true for the value $cp$ of $cP$. We may therefore define a wave function which has the collective property values as arguments. For $M=2$ we may write

\begin{equation}
\Psi_{cP}=\Psi_{P}(cp_{1},cp_{2},\ldots,cp_{N/2}),
\label{twobound2}
\end{equation}
with

\begin{equation}\begin{array}{rcl}
cp_{1} & = & f(p_{1},p_{2})\\
cp_{2} & = & f(p_{3},p_{4})\\
& \vdots &\\
cp_{N/2} & = & f(p_{N-1},p_{N})
\end{array}
\end{equation}
Equation \ref{twoswap2} then implies

\begin{equation}\begin{array}{c}
\Psi_{cP}(cp_{1},\ldots,cp_{l'},\ldots,cp_{l},\ldots,cp_{N/2},\sigma)=\\
\Psi_{cP}(cp_{1},\ldots,cp_{l},\ldots,cp_{l'},\ldots,cp_{N/2},\sigma)
\end{array}\end{equation}
in the case $M=2$.

The symmetry of this collective wave function means that the Pauli exclusion principle does not exclude that a set of composite objects that contains two minimal objects each are found in the same state, meaning that $cP$ and $c\breve{P}$ have the same values for all composite objects in this set.

Generalizing to an arbitrary number $M$ of minimal objects in each bound state, we see that the collective wave function becomes symmetric for even $M$ and anti-symmetric for odd $M$:

\begin{equation}\begin{array}{c}
\Psi_{cP}(cp_{1},\ldots,cp_{l'},\ldots,cp_{l},\ldots,cp_{N/M},\sigma)=\\
(-1)^{M}\Psi_{cP}(cp_{1},\ldots,cp_{l},\ldots,cp_{l'},\ldots,cp_{N/M},\sigma).
\end{array}\end{equation}

The same line of reasoning as for the collective property $cP$ can be followed in order to define a collective co-property $c\breve{P}$ (Definition \ref{coproperty}) such that the values $c\breve{p}$ are exactly known \emph{a priori} for all composite objects whenever the individual values $\breve{p}$ are known \emph{a priori}. This means that we can formulate a collective version of Statement \ref{spinstatevol}.

\begin{state}[\textbf{Wave function symmetries for identical composite objects}]
Consider a family of fundamental contexts $C(\sigma)$ in which the specimen is known to contain $N/M$ composite objects $O_{l}$ at the initial time $n$, containing $M$ minimal objects each. The collective property $cP$ of each of these composite objects is observed, whereas the value $\breve{cp}_{kl}$ of the co-property $\breve{cP}$ is already exactly known at time $n$ for each object $O_{l}$. Suppose that the composite objects are identical, meaning that $\breve{cp}_{l}=\breve{cp}_{l'}$ for all object pairs $(O_{l},O_{l'})$. Suppose also that for each such object pair, there is a $\sigma>0$ for which $cp_{l}=cp_{l'}$ is a possible outcome of the observation. Then

\begin{equation}\begin{array}{c}
\Psi_{cP}(cp_{1},\ldots,cp_{l'},\ldots,cp_{l},\ldots,p_{N/M},\sigma)=\\
(-1)^{M}\Psi_{cP}(cp_{1},\ldots,cp_{l},\ldots,cp_{l'},\ldots,cp_{N/M},\sigma)
\end{array}\end{equation}
for all such object pairs, for all property values $(cp_{1},\ldots,cp_{l'},\ldots,cp_{l},\ldots,cp_{N/M})$ allowed by physical law, and for all $\sigma$ in its domain $[0,\sigma_{\max}]$.
\label{cspinstatevol}
\end{state}

Note that the arguments leading to Statement \ref{cspinstatevol} rely on the fact that each minimal object in the specimen can be attributed to a given composite object $O_{l}$. This means that the composite objects have to be weakly interacting, so that they hardly exchange or share any minimal objects. This condition excludes, for instance, specimens in which the composite objects $O_{l}$ are chosen to be atoms in molecules, or in metals.

\begin{state}[\textbf{The spin-statistics theorem for identical composite objects}]
Consider a family of context such as that described in Statement \ref{cspinstatevol}, in which the specimen consists of independent or weakly interacting composite objects $O_{l}$. Assume that all these composite objects are identical, meaning that $\breve{cp}_{l}=\breve{cp}_{l'}$ for all object pairs $(O_{l},O_{l'})$ Suppose that each $O_{l}$ contains an even number of minimal objects, and has an integer collective spin quantum number $cs$. Then the Pauli exclusion principle does not exclude the possibility to observe $cP_{l'}=cP_{l}$ for some pairs of objects $(O_{l},O_{l'})$. Suppose instead that each $O_{l}$ contains an odd number of minimal objects, and has a half-integer collective spin quantum number $cs$. Then the Pauli exclusion principle excludes the possibility to observe $cP_{l'}=cP_{l}$ for all pairs of objects $(O_{l},O_{l'})$.
\label{spinstatcompobj}
\end{state}

\section{Symmetries and redundancies}
\label{symmetries}

Recall the notion of a mathematical representation $\bar{S}$ of a physical state $S$, first introduced in section \ref{state}. We may write

\begin{equation}
\bar{S}\hookrightarrow S.
\end{equation}

The state $S$ is a set in state space $\mathcal{S}$, whereas $\bar{S}$ is a string of symbols with arithmetic, algebraic or analytic meaning.

\begin{defi}[\textbf{Proper family of state representations}]
Let $\{\bar{S}(\nu)\}$ be a family of mathematical representations defined for all $\nu\in D_{\nu}$, where $\nu$ is some array of numbers. If any family member $\bar{S}(\nu)$ is a representation of exactly one state $S(\nu)$ for each $\nu\in D_{\nu}$, then $\{\bar{S}(\nu)\}$ is a proper family of representations with domain $D_{\nu}$ of the family of states $\{S(\nu)\}\subseteq\mathcal{PS}$. Here, $\mathcal{PS}$ is the power set of state space $\mathcal{S}$. 
\label{repfam}
\end{defi}

If a transformation $R$ is applied to $\bar{S}$, and $R\bar{S}$ represents the same state $S$ as does $\bar{S}$, then we call $R$ a redundancy transformation. That is, for a redundancy transformation $R$ we have

\begin{equation}
\bar{S}\hookrightarrow S\Rightarrow R\bar{S}\hookrightarrow S.
\label{redeq}
\end{equation}

We use Defintion \ref{repfam} to define a redundancy transformation that applies not only to a single representation $\bar{S}$, but to an entire family $\bar{S}(\nu)$ within some domain $D_{\nu}$.

\begin{defi}[\textbf{Redundancy transformation}]
Consider a family $\{\bar{S}(\nu)\}$ of proper state representations. $R$ is a redundancy transformation with domain $D_{\nu}$ if and only if $\bar{S}(\nu)\hookrightarrow S(\nu)\Rightarrow R\bar{S}(\nu)\hookrightarrow S(\nu)$ for each $\nu\in D_{\nu}$, but for no $\nu\notin D_{\nu}$. 
\label{redundancy}
\end{defi}

Familiar examples are translations of all spatial coordinates in the entire known world $\mathbf{r}\rightarrow\mathbf{r}+\mathbf{r}_{0}$, and rotation of all coordinates in the world any given angle. It has no epistemic meaning to say that such coordinate transformations correspond to physcial translations or rotations of the world itself, since it is not possible to decide whether such operations have been carried out or not. To do that we need a point of reference. We will return this this matter shortly.

Let us first discuss the effect on a representation of physical law of a redundancy transformation. Physical law is embodied in the evolution operator $u_{1}$, defined so that $S(n+1)\subseteq u_{1}S(n)$. We may write

\begin{equation}
S_{u}(n)\equiv u_{1}S(n).
\label{evostate}
\end{equation}
It may seem unnecessary to re-write the evolved state in this way, but I think it will make the following notation a little bit less confusing.

We may say that $\bar{u}_{1}$ is a mathematical representation of $u_{1}$ whenever we may write $\bar{S}_{u}(n)=\bar{u}_{1}\bar{S}(n)$ for all states $S(n)$, where $\bar{S}(n)$ and $\bar{S}_{u}(n)$ are representations of $S(n)$ and $S_{u}(n)$, respectively. Then we have

\begin{equation}
R\bar{S}_{u}(n)=R\bar{u}_{1}\bar{S}(n).
\label{revo}
\end{equation}
This means that if $R\bar{u}_{1}$ acts on $\bar{S}(n)$, which represents $S(n)$, we get $R\bar{S}_{u}(n)$, which represents $S_{u}(n)$ according to Eq. [\ref{redeq}]. Thus $R\bar{u}_{1}$ is also a representation of $u_{1}$. We may write

\begin{equation}
\bar{u}_{1}\hookrightarrow u_{1}\Rightarrow R\bar{u}_{1}\hookrightarrow u_{1}.
\end{equation}

\begin{state}[\textbf{Physical law is invariant under a redundancy transformation}]
If $\bar{u}_{1}$ represents the evolution operator $u_{1}$, then so does $R\bar{u}_{1}$.
\label{redundlaw}
\end{state}

In other words, $\bar{u}_{1}$ and $R\bar{u}_{1}$ represent the same physical law. We would also like to express physical law in such a way that its form $\bar{u}_{1}$ does not depend on any parameter $r$ of any redundancy transformation $R(r)$ that we may choose to apply to $\bar{S}$. Only in that case can we regard $\bar{u}$ as a `generally valid' or `true' representation. The line of reasoning is similar to that which made us define the evolution $u_{1}$ so that is does not depend on the state $S$ that we apply it to. It should also be generally valid.

\begin{defi}[\textbf{A proper evolution representation} $\bar{u}_{1}$]
A representation $\bar{u}_{1}$ of the evolution $u_{1}$ is proper if and only if $R(r)\bar{S}_{u}(n)=\bar{u}_{1}R(r)\bar{S}(n)$ for any redundancy transformation $R$. The evolved state $S_{u}(n)$ is defined by Eq. [\ref{evostate}].
\label{properevorep}
\end{defi}

Combining the defining property of a proper evolution representation with Eq. [\ref{revo}], we immediately conclude the following.

\begin{state}[\textbf{A redundancy transformation commutes with a proper evolution representation}]
For any redundany transformation $R$, and any proper evolution representation $\bar{u}_{1}$, we have $[R,\bar{u}_{1}]=0$.
\label{rucommute}
\end{state}

Another way to express the same thing is, of course, to say that $R^{-1}\bar{u}_{1}R=\bar{u}_{1}$ for a proper evolution representation $\bar{u}_{1}$. If we make a change of coordinates in the description of a state that does not change its epistemic content, evolve the transformed state, and then undo the coordinate change, we should get the same evolved state as if we did not fiddle with the coordinates in the first place.

\vspace{5mm}
\begin{center}
$\maltese$
\end{center}
\paragraph{}

Let us turn from transformations of the physical state $S$ of the entire world $\Omega$ to transformations of the state $S_{O}$ of an object within the world (Definition \ref{objectstate}). This is a different matter at a qualitative level, since $\Omega$ is not an object according to Statement \ref{worldnoobject}.

A transformation of the state $S_{O}$ is never the same thing as a transformation of $S$, since there is always a possible environment $\Omega_{O}$ to any object $O$ that can act as a point of reference when $S_{O}$ is transformed. In contrast, there is never such an external point of reference when $S$ is transformed. The environment $\Omega_{O}$ may not be actually perceived or known (Statement \ref{kworldmayobject}), but it enters any representation of the world $\Omega$ as exact states $Z$ that cannot be \emph{excluded} by our knowledge, and therefore are elements in $S$.

The basic point is that if we transform the state of an object, we can change its relation to the environment even if we do not change the state of the object itself. If you rotate an apple in your hand we change its relation to the background we see behind the apple even if the internal state of the apple remains the same. If there were no such background, and you were not there holdning it, it would be impossible to tell whether the apple has actually been rotated. The operation would be a redundancy transformation.

Let us try to distinguish symmetry transformations $T$ that act on object states, as opposed to redundancy transformations $R$ acting on such states. In a symmetry transformation, something stays the same while something else knowably changes. In a redundancy transformation, nothing knowably changes. The only thing that changes is the representation of the object state.

More precisely, in a symmetry transformation something in the \emph{relation} between the object and the environment stays the same, while something else in this relation changes. (The object itself, seen in isolation, does not change.) To have one thing in the object-environment relation changing and another remaining the same, either the object $O$ or the environment $\Omega_{O}$ must have two parts that react differently to the transformation.

\begin{figure}[tp]
\begin{center}
\includegraphics[width=80mm,clip=true]{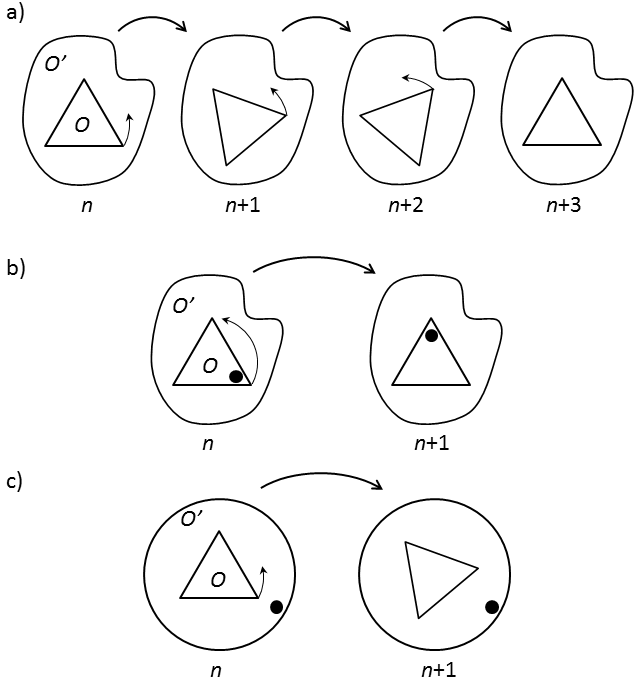}
\end{center}
\caption{Symmetry transformations of an object $O$. The object $O'$ belongs to its environment $\Omega_{O}$. a) The final state of $O$ is the same as the initial one, but the final state of the world is different since it contains memories of intermediate rotations. b) A marker dot in $O$ makes it possible to decide that a rotation has taken place. c) A marker dot in $O'$ makes it possible to decide that a rotation has taken place.}
\label{Fig106}
\end{figure}

This reasoning is exemplified in Fig. \ref{Fig106}. The object we transform is $O$, whereas object $O'$ belongs to the environment; $O'\subset\Omega_{O}$. In panel a) we rotate $O$ in three consequtive steps ($40^{\circ}$ each time) to finally find it back in its original position. In this case it is the environment that has two parts that react differently to the symmetry transformation. Object $O'$ stays the same throughout the process, but there have also appeared new objects $O''$ and $O'''$ at time $n+3$ in the form of memories of the intermediate steps at time $n+1$ and $n+2$. The existence of these objects makes the state $S(n+3)$ different from the state $S(n)$. This is so even though $S_{O}(n+3)=S_{O}(n)$, $S_{O'}(n+3)=S_{O'}(n)$, and the relation between $O$ and $O'$ is the same. Note that if memories were not formed, or the relation between $O$ and $O'$ in the intermediate steps where not knowably different than at the beginning and in the end, then the transformation would be reduced to a redundancy transformation.

In Fig. \ref{Fig106}(b) we give an example of the opposite situation, in which it is the object $O$ itself that has two parts that react differently to the transformation, rather than the environment. Here, $O$ consists of the triangle contour together with the black dot, which acts as a marker when $O$ is rotated $120^{\circ}$ in a single transformation. The relation between the triangle and the environmental object $O'$ does not change, but the relation between the marker and $O'$ does. If the marker were removed, nothing would knowably change, and the symmetry transformation would reduce to a redundancy transformation.

\begin{figure}[tp]
\begin{center}
\includegraphics[width=80mm,clip=true]{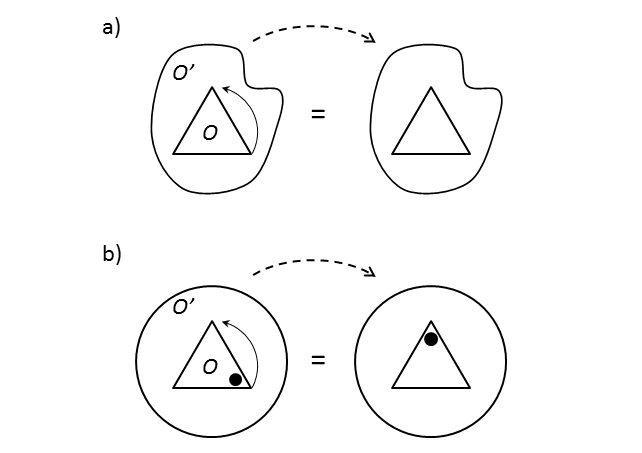}
\end{center}
\caption{Redundancy transformations of an object $O$. a) The state of $O$ does not change when it is rotated $120^{\circ}$ in a single transformation. The environment, as represented by the object $O'$, also stays the same, as well as the relation between $O$ and $O'$. No temporal update $n\rightarrow n+1$ can be defined. b) Even if we introduce a marker in $O$, nothing knowably changes if the relation between the environment and this marker is invariant under the transformation.}
\label{Fig107}
\end{figure}

This fact can also be expressed as follows. If the marker were removed, the transformation would not define a temporal update $n\rightarrow n+1$, since such updates are defined by the appearance of a knowable change. This observation further strengthens the picture that redundancy transformations are `unphysical'.

This is illustrated in Fig. \ref{Fig107}(a), where the arched arrow that indicates the execution of the transformation is dashed, and we have not labelled the initial and final states with different sequential times. Figure \ref{Fig107}(b) shows another situation in which the symmetry reduces to a redundancy. Here, the circular environment is invariant with respect to all rotation transformations $T(\phi)$, where $\phi$ is the rotation angle. This means that the relation between $O$ and $O'$ cannot change for any $\phi$ when $T$ is applied to $O$. Of course, when we look at the picture we can indeed see a change, but this is only because we perceive additional environmental objects $O'',O''',\ldots$, which are \emph{not} invariant with respect to all rotations. We need only mention the text and drawings that surround $O$ and $O'$, and the frame of the paper (or the computer screen) on which we look at the figure.

At this point it should be noted that a symmetry transformation $T$ that acts on an object can be equivalently described as a symmetry transformation that acts on the environment $\Omega_{O}$ `in the opposite direction'. We may denote this transformation $T^{-1}$. We may therefore say that a necessary condition for $T$ to be a symmetry transformation is that the environment is \emph{not} invariant with respect to $T$.

The possibility to let the object and the environment change roles in the transformation is highlighted in Fig. \ref{Fig106}(c). Here the `marker dot' is moved from the object to the environment, from $O$ to $O'$. In this case any rotation $T(\phi)$ becomes a symmetry transformation since the circular environmental object $O'$ is invariant to any inverse rotation $T^{-1}(\phi)\Omega_{O}$ or $T^{-1}(\phi)O'$. The role of the marker is to make the rotation angle $-\phi$ of the circle knowable. The symmetry of the circle would reduce to a redundancy if the marker were removed. This symmetry belongs to the same class as that in Fig. \ref{Fig106}(a) in the sense that the environment has two parts, one which changes its relation to $O$, whereas the other does not. The changing parts of the environment in Figs. \ref{Fig106}(a) and \ref{Fig106}(c) are the appearing memories of intermediate steps and the marker, respectively.

\begin{figure}[tp]
\begin{center}
\includegraphics[width=80mm,clip=true]{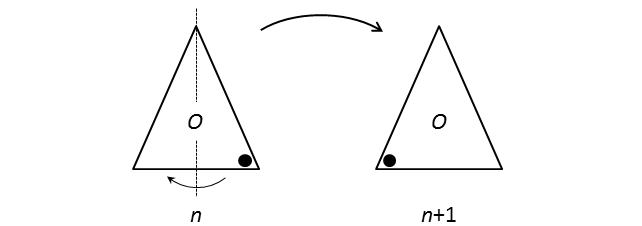}
\end{center}
\caption{Mirror symmetry. The parity-breaking parts of physical law corresponds to the asymmetric environment $O'$ in Fig. \ref{Fig106}, which makes it possible to decide that something has changed (the position of the marker), while something else stays the same (the shape of the triangle).}
\label{Fig108}
\end{figure}

Let us discuss the peculiar case of mirror symmetry in the context of the environment not being invariant to $T$. As discussed in section \ref{orient}, parity breaking physical law is necessary to even \emph{define} parity or mirror transformations $M$. This means that the environment $\Omega_{O}$ to any object $O$ is inherently asymmetric with respect to $M$. The state $MS_{\Omega_{O}}$ is different from $S_{\Omega_{O}}$ since the parity breaking weak force causes them to evlove differently. To be able to evolve differently, the states themselves have to be different, according to our definition of the evolution operator $u_{1}$. Even more, this asymmetry holds for any object $O'\subset\Omega_{O}$. We may say that all background objects are asymmetric with respect to $M$ just like the object $O'$ in Figs. \ref{Fig106} and \ref{Fig107}(a). 

Assume that we apply $M$ to an object $O$ (Fig. \ref{Fig108}). Since the background itself is mirror-asymmetric, we do not need to show any particular background object $O'$ in Fig. \ref{Fig108} as a point of reference. Recall that in a symmetry transformation $T$ of an object $O$, either $O$ or $\Omega_{O}$ have to contain two parts that react differently to $T$. One part remains the same and the other changes. We have just argued that all background objects $O'\subset\Omega_{O}$ reacts in the same way to a mirror symmetry $M$. Therefore, to make $M$ a symmetry transformation, $O$ itself has to contain two parts that react differently to $M$. That is, it has to contain a `marker', just as in Fig. \ref{Fig106}(b).

\begin{state}[\textbf{Mirror symmetry}]
The fact that physical law is parity breaking makes it possible for mirror transformations $M$ to be symmetry transformations.
\label{mirrorsym}
\end{state}

Let us formalize the discussion about symmetry transformations.

\begin{defi}[\textbf{Symmetry transformation} $T$]
Let $T$ be a transformation that is an evolution operator $u_{m}$ for some state $S$ and some $m\geq 1$. Then $S(n+m)=TS(n)\neq S(n)$. Suppose that there is such a state $S$ which contains an object $O$ with complement $\Omega_{O}$ so that $TS_{O}=S_{O}$. Then $T$ is a symmetry transformation if and only if at least one of the following conditions is fulfilled.

\begin{enumerate}
\item $T$ gives rise to a new object $O''\subset\Omega_{O}$ such that $T\Omega_{O}\neq \Omega_{O}$, but $TS_{O}\cap (TS_{\Omega_{O}})/S_{O''}=S_{O}\cap S_{\Omega_{O}}$.

\item There are two objects $O_{A}\subseteq O$ and $O_{B}\subseteq O$ such that $T(S_{O_{A}}\cap\Omega_{O})=S_{O_{A}}\cap\Omega_{O}$ whereas $T(S_{O_{B}}\cap\Omega_{O})\neq S_{O_{B}}\cap\Omega_{O}$.

\item There are two objects $O_{A}'\subset \Omega_{O}$ and $O_{B}'\subset \Omega_{O}$ such that $T(S_{O}\cap S_{O_{A}'})=S_{O}\cap S_{O_{A}'}$ whereas $T(S_{O}\cap S_{O_{B}'})\neq S_{O}\cap S_{O_{B}'}$.

\end{enumerate}
\label{symtrans}
\end{defi}

The first condition corresponds to the symmetry transformation shown in Fig. \ref{Fig106}(a), the second condition corresponds to Fig. \ref{Fig106}(b), and the third to Fig. \ref{Fig106}(c).

As indicated above, we may split the three cases into two classes in two different ways. Cases 1. and 3. belong to the same class in the sense that it is the environment that has two parts, one of which canges its relation to the transformed object whereas the other does not. Cases 2. and 3. belong to the same class in the sense that one turns into the other if we let the object and the environment change roles.

If we have $TS_{O}=S_{O}$, but none of the three conditions are fulfilled and and we have $TS=S$, then, of course, the symmetry reduces to a redundancy. Such a redundancy would apply to an object rather than the entire world. Like any redundancy, such an object redundancy transformation means that the representation of the state changes, but not the state itself.

If we have a representation $\bar{S}$ of a state which contains an object $O$. We may write

\begin{equation}
\bar{S}=\bar{S}(\bar{S}_{O},\bar{S}_{\Omega_{O}},\overline{PK}_{R},\overline{PK}_{C}),
\label{relrep}
\end{equation}
where $\overline{PK}_{R}$ and $\overline{PK}_{C}$ represent the potential knowledge about the relations and the conditions, respectively, that link $O$ to its environment (see the discussion in relation to Statement \ref{relobjectstate}).

We use this notation to formalize the idea of an object redundancy.

\begin{defi}[\textbf{Object redundancy transformation} $R_{O}$]
Consider a transformation $R_{O}$ such that $\bar{S}_{O}\hookrightarrow S_{O}\Rightarrow R_{O}\bar{S}_{O}\hookrightarrow S_{O}$. Then $R_{O}$ is an object redundancy transformation if and only if $\bar{S}(\bar{S}_{O},\bar{S}_{\Omega_{O}},\overline{PK}_{R},\overline{PK}_{C})\hookrightarrow S\Rightarrow \bar{S}(R_{O}\bar{S}_{O},\bar{S}_{\Omega_{O}},R_{O}\overline{PK}_{R},R_{O}\overline{PK}_{C})\hookrightarrow S$.
\label{oredundancy}
\end{defi}

If the last sufficient and necessary condition is not fulfilled, then $R_{O}$ may or may not represent a symmetry transformation $T$. Note again the difference between `physical' and 'unphysical' transformations. To define a physical symmetry transformation $T$, we let it act directly on a physical object state $S_{O}$. To define an object redundancy transformation $R_{O}$, we have to let it act on the \emph{representation} $\bar{S}_{O}$ of the object state, since nothing physical changes.

Let us discuss some concrete examples of object redundancies and (object) symmetries.

Suppose that an electron moves within an infinite, perfect lattice of atoms. If we let it move exactly one lattice spacing, then the operation corresponds to an object redundancy transformation $R_{O}$. The translation does not change the state of the electron $O$, so that $\bar{S}_{O}\hookrightarrow S_{O}\Rightarrow R_{O}\bar{S}_{O}\hookrightarrow S_{O}$. Nothing knowably changes in the overall state of the lattice with the moving electron, so that $\bar{S}(\bar{S}_{O},\bar{S}_{\Omega_{O}},\overline{PK}_{R},\overline{PK}_{C})\hookrightarrow S\Rightarrow \bar{S}(R_{O}\bar{S}_{O},\bar{S}_{\Omega_{O}},R_{O}\overline{PK}_{R},R_{O}\overline{PK}_{C})\hookrightarrow S$.

In this case $R_{O}$ corresponds to a transformation $\mathbf{x}\rightarrow\mathbf{x}+\mathbf{x}_{L}$ of the spatial coordinates of the electron, without making the same transformation of the lattice coordinates. If we did make such an overall coordinate change, we would have a redundancy transformation $R$ rather than an object redundancy transformation $R_{O}$.

If the transformation $\mathbf{x}\rightarrow\mathbf{x}+\mathbf{x}_{L}$ does not take the electron to an identical position in another lattice cell, then $\bar{S}(R_{O}\bar{S}_{O})\hookrightarrow S'\neq S$. We have induced a physical change. This physical transformation is not a symmetry transformation, however.

The movement of the electron to an identical lattice position in a the lattice becomes a symmetry transformation if there are other (finite) objects in the environment than the lattice. Then one aspect of the relation between the electron and the environment stays the same (the position of the electron in relation to the lattice), whereas another aspect changes (the position of the electron in relation to the other objects). We have a symmetry of the type shown in Fig. \ref{Fig106}(c), where the circle $O'$ corresponds to the lattice, and the marker dot to the other environmenal objects. Formally speaking, condition 3 in Definition \ref{symtrans} is fulfilled.

The rotation of an atom with a spherically symmetric electron distribution is clearly an object redundancy transformation. Nothing knowably changes in the relation between the object and its environment. If the electron distribution is not spherically symmetric, then we get a physical transformation that is not a symmetry transformation if we let it rotate an arbitrary angle along an axis that is not a symmetry axis of the electron cloud. If we let the atom rotate $360^{\circ}$ around such an axis in a gradual way, we get a symmetry transformation of the type shown in Fig. \ref{Fig106}(a). Formally, condition 1 in Definition \ref{symtrans} is fulfilled.

Since rotations of spherically symmetric objects are always (object) redundancies, we may use explicit epistemic minimalism (Assumption \ref{explicitepmin}) to conclude that such states cannot have any angular momentum. Physical law cannot allow any angular momentum since there is no way to distinguish different values of this attribute.

\begin{state}[\textbf{Spherical symmetry implies zero angular momentum}]
Any object whose state is invariant under all rotations has zero angular momentum.
\label{zeroangular}
\end{state}

Physical law is invariant under object redundancy transformations, of course. Formally,

\begin{equation}\begin{array}{rcl}
\bar{u}_{1}\bar{S}(\bar{S}_{O},\bar{S}_{\Omega_{O}},\overline{PK}_{R},\overline{PK}_{C}) & \hookrightarrow & u_{1}S(S_{O},S_{\Omega_{O}},PK_{R},PK_{C})\\
& \Downarrow & \\
\bar{u}_{1}\bar{S}(R_{O}\bar{S}_{O},\bar{S}_{\Omega_{O}},R_{O}\overline{PK}_{R},R_{O}\overline{PK}_{C}) & \hookrightarrow & u_{1}S(S_{O},S_{\Omega_{O}},PK_{R},PK_{C}).
\end{array}
\label{invevo}
\end{equation}

In contrast, physical law is never invariant under symmetry transformations, in the sense that the evolution of $O$, as well as of the environment $\Omega_{O}$, will be different after the transformation $T$. Even if $TS_{O}=S_{O}$ and $TS_{\Omega_{O}}=S_{\Omega_{O}}$ the evolution of these two parts of the world will be different since they are part of the same world and their interactions depend on their relation, which knowably changes under $T$. If no part of the relation would change, we would have anobject redundancy rather than a symmetry. We may write

\begin{equation}
u_{1}S(S_{O},S_{\Omega_{O}},PK_{R},PK_{C})\neq u_{1}S(TS_{O},S_{\Omega_{O}},TPK_{R},TPK_{C})
\end{equation}
in a notation that should be self-explanatory, given the corresponding notation above for the mathematical representation of the states and the potential knowledge.

Evidently, the unavoidable relation between object $O$ and its environment is the reason for the inequality in the above equation (Statement \ref{relobjectstate}). In the idealized situation where $O$ is isolated (Definition \ref{isoobjectstate}) we may write

\begin{equation}
u_{1}S(S_{O},S_{\Omega_{O}})=S(u_{1}S_{O},u_{1}S_{\Omega_{O}}).
\label{isoevo}
\end{equation}
This is the approximation often made in scientific experiments. We want to isolate the observed object $O$ as much as possible from outside perturbations. Ideally, the reaction of the specimen $OS$ to a manipulation of the experimenter $OB$ is a function of the state $S_{O}$ only (Fig \ref{Fig61c}).

If Eq. [\ref{isoevo}] is fulfilled, then physical law is, in fact, invariant under symmetry transformations $T$. We may trivially write

\begin{equation}
u_{1}S(S_{O},S_{\Omega_{O}})=u_{1}S(TS_{O},S_{\Omega_{O}}),
\label{isoinvevo}
\end{equation}
since $TS_{O}=S_{O}$. This equation may be approximately fulfilled in carefully designed experiments.

If we can isolate an experiment, then we have to get the same results in any environment, for example in any spatial surroundings. This can be interpreted as a statement about the homogeneity and isotropy of space. However, such language is misplaced in an epistemic approach, in which space should not be given any properties in itself. 

\begin{state}[\textbf{Space is homogeneous and isotropic}]
If an experimental context $C$ including the body of an observer $OB$ and an observed object $O$, with given states $S_{OB}$ and $S_{O}$, can be isolated from the environment at two different locations, then the expected outcome of the context is the same at these two locations. This means that the relative volume $v[\tilde{S}_{j}]$ of each future alternative $\tilde{S}_{j}$ is the same.
\label{homospace}
\end{state}

This is merely a tautology from the epistemic point of view, since the two locations are nothing more than two different environments, and we have assumed that the system is isolated from the environment. One might try to escape the tautological nature of the conclusion by formulating a practical version of Statement \ref{homospace}: the expected outcome is \emph{almost} the same at the two locations if the experimental context can be \emph{almost} isolated at both places. However, there is no way to define the meaning of the phrase `almost isolated' other than to refer to `almost the same outcome' or `almost the same evolution'. That is, we may say that $S_{O}$ is almost isolated from $\Omega_{O}$ if and only if the evolution of $S_{O}$ is almost the same regardless $S_{\Omega^{s}}$ in the sense of Eq. [\ref{isoevo}]. We are stuck with the tautology.

\section{The gauge principle}
\label{gaugeprinciple}

The gauge principle is often expressed in the following way. Suppose that the mathematical description of a physical system contains redundant degrees of freedom, which do not correspond to different physical states. Then, in a proper description, there have to be \emph{global} transformations associated with these redundancies, which leave the form of the equations invariant. That these transformations are global means that they affect all parts of the system in the same way.

Suppose that we require that the form of the equations should also be invariant under the corresponding \emph{local} transformations. That the transformations are local means that they may affect each part of the system individually. To achieve such local invariance, an additional term must be introduced in the equations, which absorbs the change introduced by the local transformation. This term can be interpreted as an interaction between the parts of the system. All known interactions in nature can be interpreted in this way.

In this section, we want to express these ideas as clearly as possible in terms of the concepts we have previously introduced, in particular those discussed in the preceding section \ref{symmetries}. We also want to motivate from the epistemic point of view why we must require invariance under a local symmetry transformation whenever there is a corresponding global symmetry. In other words, we want to motivate why gauge forces are inevitable in any mathematical representation of physical law that contains redundancies.

\begin{figure}[tp]
\begin{center}
\includegraphics[width=80mm,clip=true]{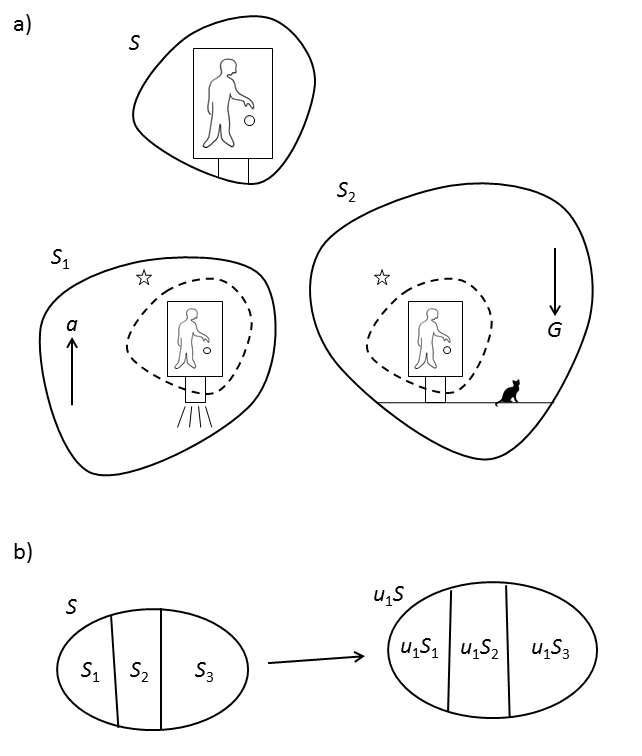}
\end{center}
\caption{Einstein's elevator revisited (compare Fig. \ref{Fig13}). a) The state $S$ corresponds to the absense of knowledge about the outside. In the state of expanded knowledge $S_{1}$, the elevator is accelerating in empty space. In the state $S_{2}$ it is standing on a gravitating body. To make the list of alternatives complete, we add a state $S_{3}$ where the perceptions inside the elevator are due to a combination of acceleration and gravity. b) Einstein's equivalence principle corresponds to the linearity of the evolution.}
\label{Fig109}
\end{figure}

Let us first return to Einstein's elevator, discussed in relation to Fig. \ref{Fig13}. Let $S$ be the physical state that corresponds to the state of knowledge where we have no idea what is outside the elevator [Fig. \ref{Fig109}(a)]. Further, let $S_{1}$ be the union of all physical states where the elevator is accelerating through space without any gravitating body in sight. Correspondingly, let $S_{2}$ be the union of all states where the elevator is instead standing on such a body. We should also consider states where the perceptions inside the elevator are the result of a combination of acceleration and gravitation. Let us denote by $S_{3}$ the union of all possible such states. We clearly have

\begin{equation}
S=S_{1}\cup S_{2}\cup S_{3},
\end{equation}
as illustrated in [Fig. \ref{Fig109}(b)]. Clearly, these alternatives are mutually exclusive, so that $S_{i}\cap S_{j}=\varnothing$ for $i\neq j$.

According to epistemic invariance (Assumptions \ref{epistemicinvariance} and \ref{epistemicinvariance2}), the evolution of the elevator in these three alternative external worlds should be consistent with the evolution of the state $S$, in which we have no knowledge at all about the external world. This condition implies that evolution is linear (Statement \ref{linearev}):

\begin{equation}
u_{1}S=u_{1}S_{1}\cup u_{1}S_{2}\cup u_{1}S_{3}.
\end{equation}
This is Einstein's equivalence principle.

Let us now choose particular mathematical representations of the evolution operator and the involved states. We may then write

\begin{equation}
\bar{S}=\bar{S}_{1}+\bar{S}_{2}+\bar{S}_{3},
\end{equation}
and

\begin{equation}
\bar{u}_{1}\bar{S}=\bar{u}_{1}\bar{S}_{1}+\bar{u}_{1}\bar{S}_{2}+\bar{u}_{1}\bar{S}_{3},
\end{equation}
according to the general recipe in which we formally translate unions of sets in state space to addition of mathematical representations in algebraic space (Table \ref{dictionary}).

We have $R^{-1}\bar{u}_{1}R=u_{1}$ for any redundancy transformation $R$ according to Statement \ref{rucommute}, provided $\bar{u}_{1}$ is a `proper' or `generally valid' representation of the evolution $u_{1}$. Speaking about Einstein, it is natural to use the Lorentz transformation $L(\mathbf{v})$ as an example. Say that the elevator is set to move with speed $\mathbf{v}$ relative its original state of motion. From inside the elevator it will be impossible to decide whether anything has changed, since the apparent speed of light is the same due to Lorentz invariance. Therefore $L(\mathbf{v})$  is a redundancy transformation when applied to $\bar{S}$:

\begin{equation}
\bar{S}\hookrightarrow S\Rightarrow L(\mathbf{v})\bar{S}\hookrightarrow S.
\label{lred}
\end{equation}

We may apply the same reasoning to representations of the states $S_{1}$, $S_{2}$, or $S_{3}$. In the first case we describe the accelerating elevator and all the surrounding objects (such as the star indicated in Fig. \ref{Fig109}(a)) with a set of spatio-temporal coordinates in a representation $\bar{S}_{1}$. If we apply a Lorentz transformation $L(\mathbf{v}_{1})$ to such a representation, we may write

\begin{equation}
\bar{S}_{1}\hookrightarrow S_{1}\Rightarrow L(\mathbf{v}_{1})\bar{S}_{1}\hookrightarrow S_{1},
\end{equation}
just as we did for the representation of the `entire' state $\bar{S}$ in Eq. [\ref{lred}]. Note that in both cases, the Lorentz redundancy transformation is global in the sense that we apply it to \emph{all} objects in the corresponding state.  In the case of $\bar{S}$, all objects within the elevator are set to move with velocity $\mathbf{v}$. In the case of $\bar{S}_{1}$, the star, the elevator, and all the objects within it, are set to move with velocity $\mathbf{v}_{1}$.

Consider the transformed representation

\begin{equation}
L(\mathbf{v}_{1},\mathbf{v}_{2},\mathbf{v}_{3})\bar{S}\equiv L(\mathbf{v}_{1})\bar{S}_{1}+L(\mathbf{v}_{2})\bar{S}_{2}+L(\mathbf{v}_{3})\bar{S}_{3}.
\end{equation}
Since each term $L(\mathbf{v}_{j})\bar{S}_{j}$ on the right hand side represents the corresponding alternative $S_{j}$, we clearly have

\begin{equation}
\bar{S}\hookrightarrow S\Rightarrow L(\mathbf{v}_{1},\mathbf{v}_{2},\mathbf{v}_{3})\bar{S}\hookrightarrow S.
\label{gaugered}
\end{equation}

Also, since each $L(\mathbf{v}_{j})$ is a redundancy transformation and $\bar{u}$ is assumed to be a proper evolution representation, Statement \ref{rucommute} implies

\begin{equation}
\begin{array}{rcl}
\bar{u}_{1}L(\mathbf{v}_{1},\mathbf{v}_{2},\mathbf{v}_{3})\bar{S} & \equiv & \bar{u}_{1}L(\mathbf{v}_{1})\bar{S}_{1}+\bar{u}_{1}L(\mathbf{v}_{2})\bar{S}_{2}+\bar{u}_{1}L(\mathbf{v}_{3})\bar{S}_{3}\\
& = & L(\mathbf{v}_{1})\bar{u}_{1}\bar{S}_{1}+L(\mathbf{v}_{2})\bar{u}_{1}\bar{S}_{2}+L(\mathbf{v}_{3})\bar{u}_{1}\bar{S}_{3}\\
& \equiv & L(\mathbf{v}_{1},\mathbf{v}_{2},\mathbf{v}_{3})\bar{u}_{1}\bar{S}.
\end{array}
\label{lgpdef}
\end{equation}

That is,

\begin{equation}
[L(\mathbf{v}_{1},\mathbf{v}_{2},\mathbf{v}_{3}),\bar{u}_{1}]=0.
\label{lgprinciple}
\end{equation}

This is an example of the gauge principle. We can choose the velocity $\mathbf{v}_{j}$ in the Lorentz transformation of each alternative state representation $\bar{S}_{j}$ \emph{independently}. The above commutator should still be zero. We are not restricted to the common choice $\mathbf{v}=\mathbf{v}_{1}=\mathbf{v}_{2}=\mathbf{v}_{3}$ that corresponds to

\begin{equation}
[L(\mathbf{v}),\bar{u}_{1}]=0.
\end{equation}

Equation [\ref{lgprinciple}] can be regarded as a condition that further restricts the possible forms a proper evolution representation $\bar{u}_{1}$ can have. It must fulfil Eq. [\ref{lgprinciple}] for any division of $S$ into realizable alternative states $\{S_{j}\}$, and for any set of velocities $\{\mathbf{v}_{j}\}$ that specifies a set of independent Lorentz transformations.

\begin{figure}[tp]
\begin{center}
\includegraphics[width=80mm,clip=true]{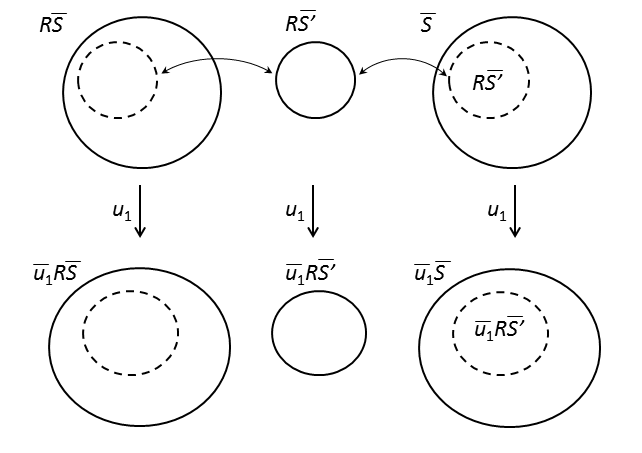}
\end{center}
\caption{Schematic illustration of the idea behind the Gauge principle. If we can make a redundancy transformation $R$ of the state representation $\bar{S}$ without affecting its evolution representation $\bar{u}_{1}$, we should be able to do the same on the representation $\bar{S}'$ of \emph{any} specific part $S'\subset S$ of the original state $S$, leaving the representation of the other parts $S\setminus S'$ of $S$ unchanged. See text for further explanation.}
\label{Fig110}
\end{figure}

The basic reason for this is the assumed epistemic invariance. If we can do a redundancy transformation of a state representation $\bar{S}$ without affecting the evolution, we should be able to do it on the representation $\bar{S}'$ of any specific part $S'\subset S$ of this state, since this part can be seen as a state in its own right, regardless the surrounding parts $S\setminus S'$ (Fig. \ref{Fig110}). The evolution of a part of the state should be consistent with the evolution of the whole state. Therefore, if the evolution representation is unaffected by a redundancy transformation $R$ applied to $\bar{S}$, it should be unaffected by any redundancy transformation applied specifically to $\bar{S}'$.

The only requirement is that $S_{j}$ is a realizable alternative, so that the evolution $u_{1}$ can be applied to it in the first place. In other words, $S_{j}$ should be an observable state (Statement \ref{evolutiondomain}). This exludes redundancy transformations applied to representations $\bar{Z}$ of exact states $Z$.

Let us generalize the example with the Lorentz transformations of Einstein's elevator to a general statement of the Gauge principle.

\begin{defi}[\textbf{Family of redundancy transformations} $R(r)$]
Say that $R$ is a redundancy transformation of a certain type such that we need to specify a parameter $r$ in order to define it completely. Let $D_{r}$ be the domain of values of $r$ for which $R$ is defined. Then $R(r)$ is a family of redundancy transformation with domain $D_{r}$. The parameter $r$ may be a scalar or a vector.
\label{redfam}
\end{defi}

Apart from the Lorentz transformation where we need to specify the velocity $r=\mathbf{v}$, we may take as an example a rotation $R$ along a given axis, where we need to specify one angle $r=\phi$. We have $D_{v_{x}}=D_{v_{y}}=D_{v_{z}}=(-c,c)$, and $D_{\phi}=[0,2\pi)$.

\begin{defi}[\textbf{Proper state partition}]
The partition $S=\bigcup_{j=1}^{M}S_{j}$ is proper if and only if $S_{j}\cap S_{j'}$ for all $j\neq j'$, and $S_{j}$ is an observable state for all $j$. An observable state can be defined according to Statement \ref{evolutiondomain} as a state to which the evolution operator $u_{1}$ can be applied.
\label{statepart}
\end{defi}

\begin{defi}[\textbf{Gauge transformation}]
Let $\bar{S}=\sum_{j=1}^{M}\bar{S}_{j}$ be a representation of a proper state partition. Further, let $R(r_{j})$ be a redundancy transformation in a family $R(r)$, where $r_{j}\in D_{r}$. Then $R(r_{1},\ldots,r_{M})\bar{S}\equiv \sum_{j=1}^{M}R(r_{j})\bar{S}_{j}$ is a gauge transformation. The gauge transformation is defined for any array of parameters $R(r_{1},\ldots,r_{M})$ such that $r_{j}\in D_{r}$ for each $j$.
\label{gaugetrans}
\end{defi}

Note that any gauge transformation is a redundancy transformation. This is a direct generalization of Eq. [\ref{gaugered}]. 

The following statement generalizes Eq. [\ref{lgprinciple}].

\begin{state}[\textbf{The gauge principle}]
For any gauge transformation $R(r_{1},\ldots,r_{M})$, and for any proper evolution representation $\bar{u}_{1}$ (Definition \ref{properevorep}), we have $[R(r_{1},\ldots,r_{M}),\bar{u}_{1}]=0$.
\label{gprinciple}
\end{state}

\vspace{5mm}
\begin{center}
$\maltese$
\end{center}
\paragraph{}

The discussion of the gauge principle so far may seem a bit abstract. To recast it in more familiar terms, let us consider how it can be applied in experimental contexts $C$ in which a wave function $\Psi$ is defined. (To make the notation familiar and simple, we write $\Psi(p,\sigma)$ in this section, even though a general wave function is denoted $a_{P}(p_{j},\sigma)$ in section \ref{wavef}.)

\begin{defi}[\textbf{The field} $\Phi$]
The field $\Phi$ is a representation of the potential knowledge at initial time $n$ about the complement $\Omega_{OS}$ to the observed specimen $OS$ in an experimental context $C$, and of those attributes of $OS$ that are not observed within context. Specifically, it is a representation of the part of this knowledge that affects the evolution of $OS$ in a way that depends on $\sigma$.
\label{field}
\end{defi}

\begin{figure}[tp]
\begin{center}
\includegraphics[width=80mm,clip=true]{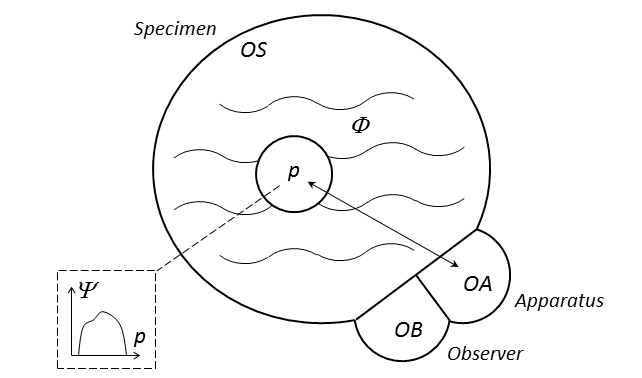}
\end{center}
\caption{Elaborated version of Fig. \ref{Fig61c}, showing the necessary parts of an experimental setup. $OS$ is the observed specimen, and $OA$ is the apparatus used to measure the value $p$ of the observed property $P$. The observer $OB$ is required to define alternative outcomes, and is thus necessary to define the wave function $\Psi(p,\sigma)$. The choice to observe property $P$, rather than another property $P'$, may be described as the choice to observe a certain aspect or part of $OS$ (smaller circle inside $OS$). This requires an interaction between this part and the apparatus (double arrow). The other parts of $OS$ that are not observed may influence the probabilities $|\Psi(p,\sigma)|^{2}$ of different values $p$ of $P$ in a way that depends on the evolution parameter $\sigma$. These influences are modelled by the field $\Phi(p,\sigma)$.}
\label{Fig113}
\end{figure}

Figure \ref{Fig61c} shows the schematic division of an experimental setup into the specimen $OS$, the apparatus $OA$, and the body $OB$ of the observer. This picture is elaborated in Fig. \ref{Fig113}. In an idealized context $C$, the experimental setup is isolated from the environment, and the initial states $S_{OA}(n)$ and $S_{OB}(n)$ are chosen so that the influence of the apparatus and the observer does not depend on $\sigma$. In that case, we just need to include in $\Phi$ the potential knowledge at time $n$ of those attributes of the specimen $OS$ that are \emph{not} observed within context.

In the terminology of Fig. \ref{Fig61c}, the object $O$ is the specimen $OS$ together with the apparatus $OA$. We may represent the state of this object as follows:

\begin{equation}
\bar{S}_{O}(\sigma)=\left[\begin{array}{c}\Psi(p,\sigma)\\\Phi(\sigma)\end{array}\right],
\label{phistate}
\end{equation}
where $\Psi$ is the wave function that describes the possible observed property values $p$ of the specimen. It is necessary to include both the specimen and the apparatus in this representation, since the set of alternative outcomes that define the wave function $\Psi$ depends on the state of the apparatus. 

The two parts of this state representation can `trade content' with each other. Referring to section \ref{eveqi}, a specimen with a given state $S_{OS}$ may be free in one context, whereas it is interacting in another. In the former case the field is zero, while it is non-zero in the latter. In general, the evolution can be expressed as

\begin{equation}
\left[\begin{array}{c}\Psi(p,\sigma)\\\Phi(p,\sigma)\end{array}\right]
=\bar{u}_{1}(\sigma)\left[\begin{array}{c}\Psi(p,0)\\\Phi(p,0)\end{array}\right].
\label{phiev}
\end{equation}

Note that, in this evolution equation, we let the field depend both on the value $p$ of the property $P$ that we observe within context, apart from the evolution parameter $\sigma$. The field encapsulates the initial knowledge about those aspects of $OS$ that affect the outcome $p$ of the observation of $P$. The way these aspect affect the outcome depends on the actual outcome $p$. If we toss a ball so that it bounces down along stony slope, the way in which the topography of the slope affects the course of the ball depends on which course it actually takes. This may not be known until its position $p$ at the bottom of the slope is observed.

In contrast to the conventional understanding of fields, $\Phi$ is only defined within a context $C$, just like the wave function $\Psi$. It does not exist \emph{per se}. From the epistemic perspective it is meaningless metaphysical baggage to say that there are fields floating around in the universe when we do not perform any experiments that are affected by these fields. They just represent those environmental attributes known at the start of the experiment that affect the probabilities for different outcomes, as collected in $\Psi$. These environmental attributes are not themselves observed within context. Therefore the knowledge about them is not updated during the course of the experiment.

Another way to put it is to recall from section \ref{wavef} that the evolution of $\Psi$ with respect to $\sigma$ does not correspond to an actual history of the specimen between the initial time $n$ and the time of observation $n+m$. It is just a parametrization of a family of experiments $C(\sigma)$ in which different relational times $t$ are expected to pass between the start and the end of the experiment. In the same way, the interplay between the wave function and the field in the evolution equation \ref{phiev} does not represent an actual, continuous interaction.

It all comes down to the basic fact that physical law does not say anything about what happens in between the observations at times $n$ and $n+1$. It just provides the operator $u_{1}$ that maps the system from time $n$ to time $n+1$. Everything is fixed when the experiment set sails. To model an actual influence from the environment (the field) on the evolution of the observed attribute $P$ of the specimen between the times $n$ and $n+1$ is therefore inappropriate, or at least unnecessary.

Formally, we may write $\bar{u}_{1}(\sigma)=\exp(\bar{A}\sigma)$, so that we can recast Eq. [\ref{phiev}] in the differential form

\begin{equation}
\frac{d}{d\sigma}\left[\begin{array}{c}\Psi(p,\sigma)\\\Phi(p,\sigma)\end{array}\right]
=\bar{A}\left[\begin{array}{c}\Psi(p,\sigma)\\\Phi(p,\sigma)\end{array}\right].
\label{dphiev}
\end{equation}

Comparing with Eqs. [\ref{ev1}] and [\ref{evbop}], we see that

\begin{equation}
\bar{A}\left[\begin{array}{c}\Psi\\\Phi\end{array}\right]\equiv\left[\begin{array}{c}\bar{A}_{\Psi}\Psi\\\bar{A}_{\Phi}\Phi\end{array}\right]=\left[\begin{array}{c}i\bar{B}_{P}\Psi\\\bar{A}_{\Phi}\Phi\end{array}\right].
\end{equation}

It does not make sense to let the evolution of $\Phi$ depend on $\Psi$ in the second component of the right hand side of the above equation. The probabilities for different outcomes is a function of the field, but not vice versa. This is the role we have given to the field by definition.
Therefore we can always write

\begin{equation}\begin{array}{rcl}
\bar{B}_{P} & = & \bar{B}_{P}(\Phi)\\
\bar{A}_{\Phi} & \neq & \bar{A}_{\Phi}(\Psi).
\end{array}
\end{equation}

Consider a context in which the spatio-temporal position (and spin) is observed. Since the differential evolution operator $\bar{B}_{\mathbf{r}_{4}s}$ can always be written $\bar{B}_{\mathbf{r}_{4}s}=-b\Box+\bar{B}_{\mathbf{r}_{4}s}^{(int)}$ according to Eq. [\ref{generalb}], we may identify

\begin{equation}
\Phi=\bar{B}_{\mathbf{r}_{4}s}^{(int)}
\label{field1}
\end{equation}
so that

\begin{equation}
\bar{B}_{\mathbf{r}_{4}s}(\Phi)=-b\Box+\Phi.
\end{equation}

For such contexts we therefore have
\begin{equation}
\frac{d}{d\sigma}\left[\begin{array}{c}\Psi(p,\sigma)\\\Phi(p,\sigma)\end{array}\right]
=\left[\begin{array}{c}-ib(\Box+\Phi(p,\sigma))\Psi(p,\sigma)\\\bar{A}_{\Phi}\Phi(p,\sigma)\end{array}\right].
\label{fieldevo}
\end{equation}

We may equally well use Definition \ref{fourmomentumdefi2} to write

\begin{equation}
\bar{B}_{\mathbf{r}_{4}s}=\frac{b}{\hbar^{2}}((\overline{\mathbf{p}_{4}})_{\mathbf{r}_{4}}^{(0)}+(\overline{\mathbf{p}_{4}})_{\mathbf{r}_{4}}^{(int)})^{2},
\end{equation}
where $(\overline{\mathbf{p}_{4}})_{\mathbf{r}_{4}}^{(0)}=-i\hbar(\partial/\partial r_{1},\ldots,\partial/\partial r_{4})$, so that another valid representation of the field $\Phi$ is

\begin{equation}
\Phi'=(\overline{\mathbf{p}_{4}})_{\mathbf{r}_{4}}^{(int)}.
\label{field2}
\end{equation}

Then the evolution equation [\ref{fieldevo}] transforms to

\begin{equation}
\frac{d}{d\sigma}\left[\begin{array}{c}\Psi(p,\sigma)\\\Phi'(p,\sigma)\end{array}\right]=
\left[\begin{array}{c}\frac{ib}{\hbar^{2}}((\overline{\mathbf{p}_{4}})_{\mathbf{r}_{4}}^{(0)}+\Phi'(p,\sigma))^{2}\Psi(p,\sigma)\\\bar{A}_{\Phi}'\Phi'(p,\sigma)\end{array}\right].
\label{fieldevo2}
\end{equation}

In general, the field $\Phi$ must be described as an operator. Sometimes it can be described as a function of the evolution parameter $\sigma$ and the attribute value $p$ that is observed within context. If the field representation $\Phi'$ in Eq. \ref{fieldevo2} is a function, we can identify it with the electro-magnetic or the gravitational potential.

Consider a proper partition of the initial state $S_{O}(n)$ of a context $C$ into a set of states $S_{j}(n)$ (Definition \ref{statepart}). Let the partition be such that it divides $S_{O}$ of the wave function exactly along those lines that defines the set of property values $\{p_{k}\}$ that can be observed within context. In this way we know that the partition is proper, since each value $p_{k}$ corresponds to an observable state, by the definition of a wave function. Thus we know that the evolution operator $u_{1}$ can be applied to each state $S_{j}$. Such a partitioned initial state means that we have defined a new experimental context $C_{j}$.

\begin{figure}[tp]
\begin{center}
\includegraphics[width=80mm,clip=true]{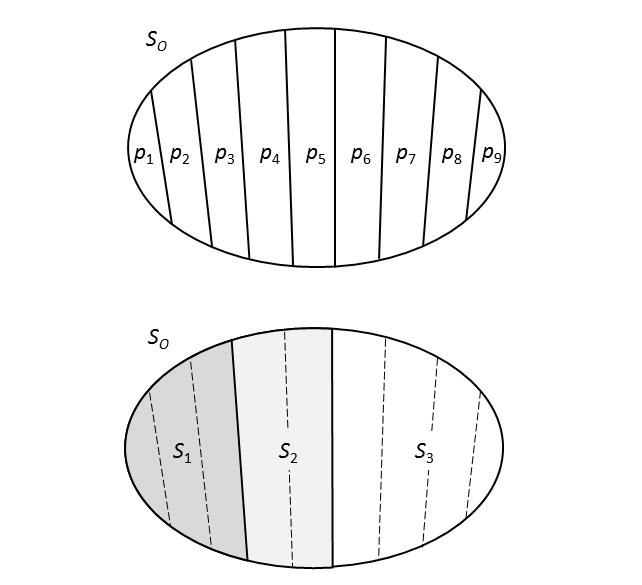}
\end{center}
\caption{A state $S_{O}$ that corresponds to an experimental context $C$ in which property $P$ is observed. The probability of each value $p$ is encoded in a wave function $\Psi(p)$ with domain $D=\{p_{1},\ldots,p_{9}\}$. A partition of $S_{O}$ into states $S_{j}$ which corresponds to contexts $C_{j}$. The probabilities of each value in these contexts are encoded in wave functions $\Psi_{j}(p)$ with domains $D_{1}=\{p_{1},p_{2},p_{3}\}$, $D_{2}=\{p_{4},p_{5}\}$ and $D_{1}=\{p_{6},\ldots,p_{9}\}$. We have $\Psi_{j}(p)=\nu_{j}\Psi(p)$ whenever $p\in D_{j}$, where the normalization constant $\nu$ is given in Eq. [\ref{normj}]. Compare Fig. \ref{Fig109}.}
\label{Fig111}
\end{figure}

Further, assume that for each state $S_{j}(n)$ of increased initial knowledge, nothing is knowably changed in the experimental arrangement, as compared to $S_{O}(n)$. The only change is an increase in the knowledge about the initial state of the specimen $OS$. Some alternative outcomes $p_{k}$ are ruled out right from the start. The domain $D$ of the wave function shrinks to $D_{j}\subset D$. This means that the state space volumes $V[p_{k}]$ of those alternatives that are still possible is the same in each context $C_{j}$ as in the original context $C$. We may therefore define a partitioned wave function $\Psi_{j}(p)$ such that

\begin{equation}
\Psi_{j}(p)=\frac{V[S_{O}]}{V[S_{j}]}\Psi(p),\ \forall p\in D_{j},
\label{normj}
\end{equation}
The real constant $V[S_{O}]/V[S_{j}]$ is needed to keep $\Psi_{j}$ normalized.

\begin{defi}[\textbf{Neutral context partition}]
Consider a context $C$ in which property $P$ is observed. Let $D$ be the domain of the wave function. Suppose that we increase the initial state of potential knowledge $PK_{OS}(n)$ about the specimen $OS$ in such a way that only a subset $D_{j}\subset D$ of the property values $p\in D$ are now possible to see in $C$ at time $n+m$. There is no other change of $PK_{OS}(n)$, no change of $PK_{OA}(n)$, and no change of the potential knowledge that relates $OS$, the apparatus $OA$ and the body $OB$ of the observer. These prescriptions define a new context $C_{j}$. A set $\{C_{1},C_{2},\ldots,C_{M}\}$ of such contexts is a netural partition of $C$ if and only if $D=\bigcup_{j}^{M}D_{j}$.
\label{neutralpart}
\end{defi}

Let us apply an object redundancy transformation $R_{O}$ (Definition \ref{oredundancy}) to a representation $\bar{S}_{O}$ of the state $S_{O}$ that corresponds to the experimental context $C$. Say that we have defined a neutral partition $\{C_{1},\ldots,C_{M}\}$ of $C$. Suppose that $R_{O}$ belongs to a family $R_{O}(r)$ according to Definition \ref{redfam}. The gauge principle (Statement \ref{lgprinciple}), together with Eqs. [\ref{phistate}] and [\ref{phiev}], makes it possible to write

\begin{equation}\begin{array}{rcl}
R_{O}(r_{1},\ldots,r_{M})\left[\begin{array}{c}\Psi(p,\sigma)\\\Phi(p,\sigma)\end{array}\right] & = & \bar{u}_{1}(\sigma)R_{O}(r_{1},\ldots,r_{M})\left[\begin{array}{c}\Psi(p,0)\\\Phi(p,0)\end{array}\right]\\
&&\\
& = & \sum_{j=1}^{M}\bar{u}_{1}(\sigma)R_{O}(r_{j})\left[\begin{array}{c}\Psi_{j}(p,0)\\\Phi_{j}(p,0)\end{array}\right]\\
&&\\
& = & \sum_{j=1}^{M}\bar{u}_{1}(\sigma)\left[\begin{array}{c}R_{\Psi}(r_{j})\Psi_{j}(p,0)\\R_{\Phi}(r_{j})\Phi_{j}(p,0)\end{array}\right]
\end{array}
\label{phievg}
\end{equation}
The second line follows because $\bar{S}_{j}=[\Psi_{j}(p,0),\Phi_{j}(p,0)]^{T}$ is a representation of $S_{j}$ (compare Eq. [\ref{lgpdef}]). We write $\Phi_{j}$ rather than $\Phi$ in this representation since it is only necessary to consider those parts of the field that affect the reduced set of possible values $p\in D_{j}$.

What does Eq. [\ref{phievg}] say? We assume that $R_{O}$ is a redundancy transformation in the sense that $R_{O}[\Psi(p,0),\Phi(p,0)]$ represents the same experimental setup as $[\Psi(p,0),\Phi(p,0)]$. Then the equation says that we can break up the domain $D$ of $\Psi$ (and $\Phi$) into pieces in whatever way we like, apply $R_{O}$ with an individual parameter choice $R_{O}(r_{j})$ for each piece, and the collection of `transformed pieces' still represents the same experimental setup.

Having performed this partition, we may glue the transformed pieces together again, forming an overall transformed wave function $R_{\Psi}(r_{1},\ldots,r_{M})\Psi(p,0)$. Recall that the piecewise wave functions $\Psi_{j}$ are normalized individually according to Eq. [\ref{normj}], so that the bare sum $\sum_{j=1}^{M}R_{\Psi}(r_{j})\Psi_{j}$ is not a normalized transformed wave function defined in the entire domain $D=\bigcup D_{j}$. This is merely a technicality, and we can regain the proper normalization if we replace $\Psi_{j}(p)$ with $\Psi(p\in D_{j})$, where the latter expression is understood as the piece of an overall wave function defined in $D_{j}$ (Eq. [\ref{normj}]).
In this way each individual state representation $\bar{S}_{j}$ we glue together obtains the appropriate weight in terms of relative state space volume (contemplate Fig. \ref{Fig111}). These considerations amount to the statement

\begin{equation}\begin{array}{rcl}
R_{O}(r_{1},\ldots,r_{M})\left[\begin{array}{c}\Psi(p,\sigma)\\\Phi(p,\sigma)\end{array}\right] & = & \sum_{j=1}^{M}\bar{u}_{1}(\sigma)\left[\begin{array}{c}R_{\Psi}(r_{j})\Psi(p,0)\\R_{\Phi}(r_{j})\Phi(p\in D_{j},0)\end{array}\right]\\
&&\\
& \equiv & \bar{u}_{1}(\sigma)\left[\begin{array}{c}R_{\Psi}(r_{1}\ldots,r_{M})\Psi(p,0)\\R_{\Phi}(r_{1},\ldots,r_{M})\Phi(p,0)\end{array}\right]
\end{array}
\label{phievg2}
\end{equation}
The second row merely defines a short-hand notation for the expression in the first row. In this notation, the technical, somewhat pedantic, discussion in the preceding paragraphs just motivates the movement of the redundancy transformations inside the square bracket: $R_{O}(r_{1},\ldots,r_{M})[\Psi(p,0),\Phi(p,0)]^{T}=[R_{\Psi}(r_{1},\ldots,r_{M})\Psi(p,0),R_{\Phi}(r_{1},\ldots,r_{M})\Phi(p,0)]^{T}$. Note also that the normalization problem does not apply to the field $\Phi$, so that we gladly exchange $\Phi_{j}(p)$ with $\Phi(p\in D_{j})$ in the gluing together of the pieces.

\begin{figure}[tp]
\begin{center}
\includegraphics[width=80mm,clip=true]{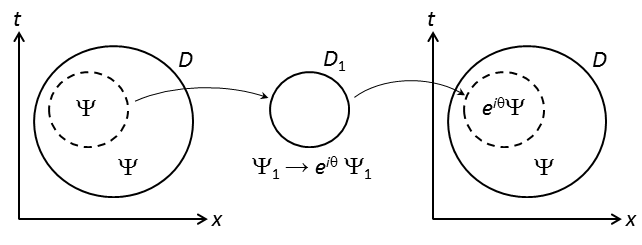}
\end{center}
\caption{The gauge principle for wave functions. Here we consider a context $C$ described by $\Psi$ in which the position $\mathbf{r}_{4}$ is observed. We may choose an arbitrary piece of the spatio-temporal domain $D$, cut it out and treat it as the initial state of another context $C_{1}$. We may  perform an overall change of phase of the corresponding wave function $\Psi_{1}$. Since this is a redundancy transformation, we can glue the transformed piece back into $\Psi$. Compare Fig. \ref{Fig110}.}
\label{Fig105}
\end{figure}

A simple illustration of the process is given in Fig. \ref{Fig105}. We first break up the domain $D$ in two parts, look at the pieces as two independent states of the experimental setup, apply a phase transformation to one piece, and then glue the pieces together again. In this example, the redundancy transformation is chosen to be an overall change of phase of one of the two individual wave functions $\Psi_{j}$. 

Let us consider a neutral context partition $\{C_{1},\ldots,C_{M}\}$ such that each state $S_{j}$ corresponds to exactly one property value $p_{j}$. In the example shown in Fig. \ref{Fig111} this means that $S_{1}$ is the part of $S_{O}$ that defines property value $p_{1}$, $S_{2}$ is the part that defines $p_{2}$, and so on. In short, $S_{j}\leftrightarrow p_{j}$. Then we may write

\begin{equation}
R_{O}(r_{j})=G(r(p_{j})).
\end{equation}
We may spell out the meaning of $G(r(p_{j}))$ as a gauge transformation $G$ that is specified by a parameter $r$, the value of which depends on the property value $p_{j}$. In this notation, and for a context partition such that $S_{j}\leftrightarrow p_{j}$, we may rewrite Eq. [\ref{phievg2}] as

\begin{equation}\begin{array}{rcl}
\left[\begin{array}{c}G_{\Psi}(r(p))\Psi(p,\sigma)\\G_{\Phi}(r(p))\Phi(p,\sigma)\end{array}\right]
=\bar{u}_{1}(\sigma)\left[\begin{array}{c}G_{\Psi}(r(p))\Psi(p,0)\\G_{\Phi}(r(p))\Phi(p,0)\end{array}\right].
\end{array}
\label{phievg3}
\end{equation}
Here, the meaning is that for each $p\in D$, the Gauge transformation $G$ is performed, as specified by the parameter $r(p)$ that can be independently chosen for each $p$. The differential form of Eq. [\ref{phievg3}] is

\begin{equation}
\frac{d}{d\sigma}\left[\begin{array}{c}G_{\Psi}(r(p))\Psi(p,\sigma)\\G_{\Phi}(r(p))\Phi(p,\sigma)\end{array}\right]=\bar{A}\left[\begin{array}{c}G_{\Psi}(r(p))\Psi(p,\sigma)\\G_{\Phi}(r(p))\Phi(p,\sigma)\end{array}\right].
\label{dphiev2}
\end{equation}
This can be seen as a generalization of Eq. [\ref{dphiev}], which corresponds to the case $G(r(p))=I$.

For contexts in which the sptio-temporal position $\mathbf{r}_{4}$ is observed, we may choose one of the expressions [\ref{field1}] or [\ref{field2}] of the field to narrow down the form of the evolution equation further. The most common choice is $\Phi=(\overline{\mathbf{p}_{4}})_{\mathbf{r}}^{(int)}$. Adopting this field representation we get

\begin{equation}\begin{array}{c}
\frac{d}{d\sigma}\left[\begin{array}{c}G_{\Psi}(r(p))\Psi(p,\sigma)\\G_{\Phi}(r(p))\Phi(p,\sigma)\end{array}\right]=\\
\\
=\left[\begin{array}{c}\frac{ib}{\hbar^{2}}\left((\overline{\mathbf{p}_{4}})_{\mathbf{r}}^{(0)}+G_{\Phi}(r(p))\Phi(p,\sigma)\right)^{2}G_{\Psi}(r(p))\Psi(p,\sigma)\\\bar{A}_{\Phi}G_{\Phi}(r(p))\Phi(p,\sigma)\end{array}\right].
\end{array}
\label{gfieldevo}
\end{equation}
with $(\overline{\mathbf{p}_{4}})_{\mathbf{r}}^{(0)}=-i\hbar(\partial/\partial r_{1},\ldots,\partial/\partial r_{4})$. To avoid cluttered notation, we have kept the general property value symbol $p$ in Eq. [\ref{gfieldevo}], keeping in mind that $p=(\mathbf{r}_{4},s)$, where $s$ is the spin. If the field is `static' in the sense that it does not depend on the evolution parameter $\sigma$, we get

\begin{equation}
\frac{d}{d\sigma}G_{\Psi}(r(p))\Psi(p,\sigma)=\frac{ib}{\hbar^{2}}\left((\overline{\mathbf{p}_{4}})_{\mathbf{r}}^{(0)}+G_{\Phi}(r(p))\Phi(p)\right)^{2}G_{\Psi}(r(p))\Psi(p,\sigma).
\label{gfieldevo2}
\end{equation}

The parameter $r$ that specifies the gauge transformation must somehow be present in the wave function. It may correspond to a property value that describes the specimen, but it may also be an abstract parameter that appears in the state representation.

An example of the latter case is when $r$ is chosen to be the phase $\theta$ of the complex amplitudes in the wave function of a context in which the spatio-temporal position $\mathbf{r}_{4}$ is observed (Fig. \ref{Fig105}). Then

\begin{equation}\begin{array}{rcl}
G_{\Psi}(\theta(\mathbf{r}_{4}))\Psi & = & e^{i\theta(\mathbf{r}_{4})}\Psi\\
G_{\Phi}(\theta(\mathbf{r}_{4}))\Phi & = & \Phi-\hbar\left(\frac{\partial\theta}{\partial r_{1}},\ldots,\frac{\partial\theta}{\partial r_{4}}\right),
\end{array}
\end{equation}
and Eq. [\ref{gfieldevo2}] expresses the gauge invariance of the evolution equation of an object moving in the electromagnetic potential $\Phi(\mathbf{r}_{4})$. In other words, we have demonstrated that the electromagnetic interaction follows from the fact that $\exp(i\theta)$ is a redundancy transformation when applied to a spatio-temporal wave function:

\begin{equation}
\left[\begin{array}{c}\Psi_{\mathbf{r}_{4}}\\\Phi\end{array}\right]\hookrightarrow S_{O}\Rightarrow
\left[\begin{array}{c}e^{i\theta}\Psi_{\mathbf{r}_{4}}\\\Phi\end{array}\right]\hookrightarrow S_{O}
\end{equation}
for any constant $\theta\in[0,2\pi)$. This is a conventional line of reasoning, of course. The new thing is that we have used epistemic invariance to motivate why we are allowed to take the step from a `global' redundancy transformation to a `local' gauge transformation.

\begin{state}[\textbf{The gauge principle for wave functions}]
Let $C(\sigma)$ be a family of contexts in which property $P$ is observed, and let $O$ be the specimen $OS$ together with the apparatus $OA$. Suppose that $\bar{S}_{O}=[\Psi(p),\Phi(p)]^{T}\hookrightarrow S_{O}$, and that $\bar{S}_{O}=\bar{S}_{O}(r)$ depends on the parameter $r$ in such a way that $R_{O}(r)$ is a family of object redundancy transformations. Then $[\bar{u}_{1}(\sigma),G(r(p))]=0$ for any `appropriate' function $r(p)$, where $G(r(p))\bar{S}_{O}\equiv[G_{\Psi}(r(p))\Psi,G_{\Phi}(r(p))\Phi]^{T}$. The meaning of the operators $G_{\Psi}(r(p))$ and $G_{\Phi}(r(p))$ is defined in relation to Eq. [\ref{phievg3}].
\label{wavegauge}
\end{state}

Note that $G(r(p))$ is an object redundancy transformation as well as a gauge transformation. The operator $G(r(p))$ is just a special case of a general gauge transformation according to Definition \ref{gaugetrans}, and we argued that all such transformations are redundancy transformations. That is,

\begin{equation}
\left[\begin{array}{c}\Psi(p)\\\Phi(p)\end{array}\right]\hookrightarrow S_{O}\Rightarrow
\left[\begin{array}{c}G_{\Psi}(r(p))\Psi(p)\\G_{\Phi}(r(p))\Phi(p)\end{array}\right]\hookrightarrow S_{O}.
\end{equation}
Therefore, the fact that the wave function and the field both changes in a gauge transformation does not mean that the physical state changes. 

We need to elaborate on the vague phrase `for any appropriate function $r(p)$' in Statement \ref{wavegauge}. If $r$ is an abstract parameter in the state representation, and the set $D_{p}$ of allowed property values $p$ is discrete, then $r(p)$ may be any function $r(p):D_{p}\rightarrow D_{r}$. The same is true if $r$ corresponds to a property value, and $D_{p}$ is still discrete. If $P$ is a property that can take a continuity of values $p$ in principle, and we are dealing with an idealized context (Fig. \ref{Fig76c}), then we can restrict our interest to differentiable functions $r(p)$. This is so since any meaningful idealized context is constructed as an approximation to an actual context. These actual contexts always have a finite number of alternatives, to which a differentiable function $r(p)$ can always be matched in the idealized approximation.

If $r$ is a property, it may correspond either to a relational or to an internal attribute, for example color charge. The three possible values of color charge are cyclic, and a translation of their values $r$ is therefore the same thing as a rotation. The strong interaction is invariant under such rotations, so that $R(r)$ is a redundancy transformation. We may therefore perform the corresponding `local' gauge transformation $G(r(p))$.

\begin{figure}[tp]
\begin{center}
\includegraphics[width=80mm,clip=true]{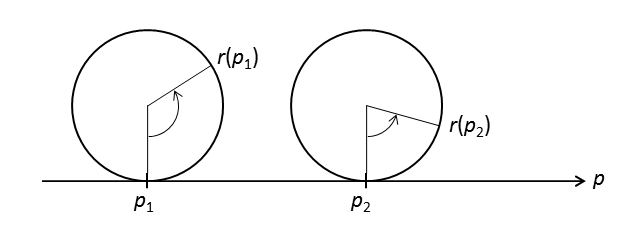}
\end{center}
\caption{A gauge transformation $G(r(p))$ derived from a family of redundancy transformations $R_{O}(r)$ in which the parameter $r$ is cyclic. It may, for example, correspond to the phase $\theta$ of the wave function $\Psi$, or the color charge of the observed part of the specimen. That the gauge symmetry of the evolution representation $\bar{u}_{1}$ is `local' means that we can rotate the value of $r$ independently at each value $p$ that is possible to observe within context. $\bar{u}_{1}$ stays the same nevertheless.}
\label{Fig112}
\end{figure}

The procedure is illustrated in Fig. \ref{Fig112}. Note that an independent rotation of a cyclic internal attribute $r$ at each possible value $p$ of the observed property is analogous to the independent rotation of the abstract wave function phase $\theta$ at each such value $p$.

Note that the term `local' gains its normal meaning only if we let $p$ be the spatio-temporal position, i.e. $p=\mathbf{r}_{4}$. In other words, saying that a gauge symmetry is local pressupposes that we are dealing with a context in which $\mathbf{r}_{4}$ is observed.

If $r$ is a property value, it may or may not be observed within context. Suppose that we are dealing with a context in which both the values $r$ and $p$ are observed, and that the corresponding properties are simultaneously knowable. Then we should express the wave function as $\Psi(p,r)$ (Definition \ref{jwavedef}). The transformed wave function $G_{\Psi}(r(p))\Psi(p,r)$ is then typically different than $\Psi(p,r)$. On the other hand, if the property that has values $r$ is \emph{not} observed within context, then the gauge transformation cannot act on $\Psi(p)$, since $r$ is not a variable that appears in this wave function. We get $G_{\Psi}(r(p))\Psi(p)=\Psi(p)$.

The field $\Phi$ may depend on the propery value $r$ that defines the gauge transformation even if $r$ is not observed. That is, we typically have $G_{\Phi}(r(p))\Phi(p,r)\neq\Phi(p,r)$.

A special case occurs when $r=p$. Then we may write $G(r(p))=G(p'(p))$, where $p'$ is the transformed value of the property $P$. We have assumed that the possible values of any attribute can be ordered (Definition \ref{attridefi} and Assumption \ref{knowattri}). This means that if $v_{2}$ is found between $v_{1}$ and $v_{3}$ in one coordinate system, this should remain true in any other appropriate coordinate system. Therefore, a gauge transformation $G(p'(p))$ cannot be allowed to change this ordering. We have to require that the function $p'(p)$ is invertible, that it is monotonically increasing or decreasing.

Let us discuss the case when $r=p=\mathbf{r}_{4}$. The underlying redundancy transformation $R_{O}(p)$ may correspond to a spatial translation or rotation of the entire experimental setup. The derived gauge transformation $G(p'(p))$ then becomes a local spatial translation or rotation that is applied independently to each point $\mathbf{r}_{4}$. The requirement that $p'(p)$ is invertible means that space-time can be stretched and compressed in an arbitrary manner, but that it is not allowed to be folded.

Since we can approximate any invertible function $p'(p)$ in any actual context (offering a finite number of alternative values $p_{j}$) by a diffeomorphism in a corresponding idealized context, we may, without loss of generality, restrict our interest to those transformations of space-time coordinates that are, indeed, diffeomorphisms. We arrive a the diffeomorphism-invariance assumed in general relativity. This is true at least in the idealized description in which space-time is treated as a continuum.

However, we argued in section \ref{boundstates} that there is a smallest distace that can ever be measuerd, even if we cannot exclude a continuum of larger distances between unbound objects. On the one hand, this fact validates the restriction to diffeomorphisms: since we cannot measure arbitrary small distances, it has no epistemic meaning to speak about cusps and tears in the transformed fabric of space-time, defined as points at which $\frac{d}{dp}p'(p)$ is not defined. On the other hand, this fact makes it clear that the treatment of space-time as a continuum is, indeed, an idealization.

\begin{state}[\textbf{All monotonic coordinate changes are allowed in gauge transformations}]
Let $C$ be an actual context in which one of the values $p_{j}$ in the finite set $\{p_{j}\}$ is observed. Suppose that $r=p$ in a gauge transformation $G(r(p))$, so that we may write $G(p'(p))$. Then any monotonically increasing or decreasing function $p_{j}'(p_{j})$ is `appropriate', in the sense that $[\bar{u}_{1}(\sigma),G(p_{j}'(p_{j}))]=0$ (Statement \ref{wavegauge}).
\label{invertrans}
\end{state}

In the above statement we have presupposed that the attribute $p$ for which we are allowed to change coordinates (as long as we respect the attribute value ordering) is a parameter $r$ in a family of redundancy transformations $R_{O}(r)$. Is this always so? In other words, are coordinate translations $p\rightarrow p+r$ always redundancy transformations, for any attribute $P$?

For relative attributes, like distance and momentum, it is easy to conclude that this must be so. To fix a symbol, a number, a position, to a single object in order to describe its relation to another inevitably introduces a redundancy. It is only the relation between two such symbols that matters.

What about internal attributes? Actually, these are also relational in a sense. They determine how the object interact with other objects. Electric charge is one example. To find out its value we have to let another object with a known charge interact with the first object. If the charges are the same the two objects repel each other, and if they are different they attract each other. Again, the value of the charge \emph{per se} does not matter, what matters is the relation between the charge values of the two interacting objects.

Therefore I would like to answer the question affirmatively: yes, attribute value translations are redundancy transformations for all attributes. If the set of possible attribute values is compact, this means that they have to be arranged in a circular manner (Definition \ref{circularvalues}), so that the translation can also be described as a rotation. This is true for the three color charges, and it may also said to be true for the two fundamental electric charges. (In the latter case the rotation degenerats into an interchange.) Note that we have to exclude the zero charge from these considerations. That is, we should rather say that an object lacks the attribute \emph{charge} than saying that the value of its charge is zero.

If all translations indeed are redundancies, we can use the gauge principle to constrain the form of the evolution operator representation in a specific way for each attribute. Such a constraint can be seen as an interaction or transformation (section \ref{eveqi}). This is to say that to each degree of freedom corresponds a gauge force.

We used the relational nature of all attributes to come to this conclusion. But to define a relation we need an interaction. Therefore, it would be a satisfying epistemic closure if the above conclusion holds true. Such a closure would also mean that all interactions or transformations are possible to derive from the gauge principle.

It may seem disturbing to be able to derive the laws of nature from redundancies alone. Redundancies are not the real thing - rather the opposite. But here we mean derivations of \emph{mathematical representations} of the laws of nature. In such representations, we cannot do without coordinates. And as soon as we have coordinates, we have redundancies, as we argued above. Then comes the gauge principle, and the gauge forces. Therefore there is no arbitrariness in the use of redundancies to derive equations that describe physical law, even if these redundancies are unphysical in themselves. The set of attributes defines the set of possible redundancy transformations; we cannot choose them freely. At the basic level, we may say that it is from the set of degrees of freedom of our perceptions that we derive the form of physical law, rather than from the set of redundancy transformations that follows from any attempt to represent these degrees of freedom symbolically.

We close this section with a basic observation. It is the finite support of the wave functions (Fig. \ref{Fig75}) in the present epistemic approach that makes it possible to motivate the gauge principle in the way we have done. Only if supports are typically finite it is possible to cut out any given patch of the wave function and treat it as the entire wave function in another context. Only then can a phase change in this patch be treated as a global phase change in the other context, allowed because because it is a redundancy transformation.

\section{Entropy}
\label{entropy}

In this section we define the entropy of a state as the logarithm of the state space volume of this state. The state space volume is given by Definition \ref{voldef}. In this way entropy becomes closely related to probability, in the sense of the word introduced in section \ref{probabilities}. The relative volume of an alternative as compared to the present state gives the probability of the alternative, whereas the absolute volume of the alternative gives its entropy. We will discuss similarities and differences between the present and the conventional notion of entropy.

\begin{defi}[\textbf{Entropy}]
The entropy of the physical state $S$ is $\mathcal{E}[S]\equiv \log(V[S])$. The entropy of the object state $S_{O}$ is $\mathcal{E}[S_{O}]\equiv \log(V[S_{O}])$. The object entropy of $S_{O}$ is $\mathcal{E}_{O}[S_{O}]\equiv \log(V_{O}[S_{O}])$. Here $V$ is the volume in state space $\mathcal{S}$, and $V_{O}$ is the volume in object state space $\mathcal{S}_{O}$.
\label{entrodef}
\end{defi}

The incompleteness of knowledge (Statement \ref{incompleteknowledge}) means that we always have $\mathcal{E}[S]>1$. Since it is impossible in practice to relate a state space volume to the unit volume of an exact state, we have to resort to relative volumes, and consequently to entropy differences $\Delta \mathcal{E}=\mathcal{E}[S_{2}]-\mathcal{E}[S_{1}]=\log(V[S_{2}]/V[S_{1}])$.

According to Definition \ref{entrodef}, the entropy of the entire world $\mathcal{E}[S]$ is always smaller than the entropy of each object in the world. Let $O$ and $O'$ be two objects. We have $S\subseteq S_{O}\cap S_{O'}$, so that $V[S]<V[S_{O}]$ and $V[S]<V[S_{O}']$. These relations hold provided $O$ and $O'$ are not identical, which they cannot be if they can be distinguished and be given different labels or names.

\begin{state}[\textbf{The entropy of the world is bounded from above by the entropy of its objects}]
Suppose that the world contains the two distinct objects $O$ and $O'$. Then $\mathcal{E}[S]<\mathcal{E}[S_{O}]$ and $\mathcal{E}[S]<\mathcal{E}[S_{O'}]$.
\label{entropybound}
\end{state}
This statement means that if we increase the knowledge of the objects in the world, if we sharpen our perception by putting on better glasses, we may force the total entropy to decrease. This means that we cannot expect any strict adherence to the law that entropy must increase with time.

Nevertheless, Definition \ref{entrodef} is, on the surface, very similar to the conventional definition of entropy: $\mathcal{E}=k\log(\Omega)$, where $\Omega$ is the number of microstates consistent with a given macrostate, and $k$ is Boltzmann's constant. We simply replace `microstate' with `exact state', and `macrostate' with `physical state'. The problem with the conventional definition is that the terms `microstate' and `macrostate' are poorly defined. In particular, the evolution of the macroscopic constraints such as pressure, temperature and energy that specify the macrostate is not clearly defined in terms of fundamental physical law. This fact obscures the relation between physical law and the evolution of entropy.

In contrast, in our conceptual framework, the physical state $S$ is the fundamental object on which the evolution operator $u_{1}$ acts. The evolution of the entropy is therefore well-defined; there is no bridge to cross between the microscopic and the macroscopic descriptions. On the other hand, the evolution of exact states $Z$ is not defined at all. These states do not belong to the domain of the operator $u_{1}$. In the conventional picture, physical law acts on the entity analogous to the exact state, namely the `microstate'. We may therefore say that we turn the coin upside down: we let physical law act on the `macrostate' rather than the `microstate'.

\begin{figure}[tp]
\begin{center}
\includegraphics[width=80mm,clip=true]{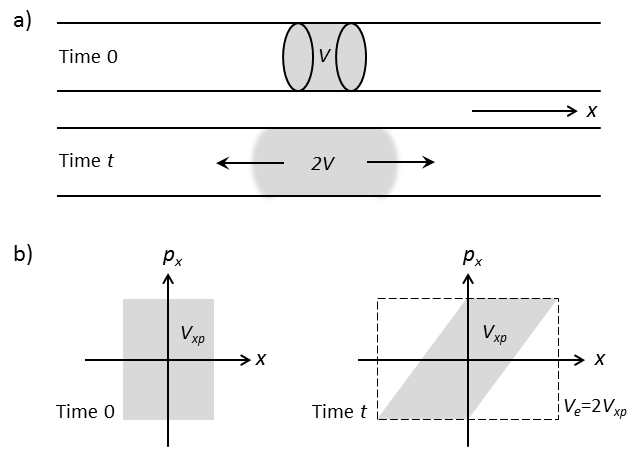}
\end{center}
\caption{A perfect gas is confined by two partitions to a small region of a long tube, after which the partitions are removed. a) The free expansion of the gas corresponds to an increasing entropy. b) This conclusion relies on the fact that we disregard the known evolution from time $0$ to time $t$, which makes it possible to exclude some regions in the classical phase space envelope volume $V_{e}$.}
\label{Fig118}
\end{figure}

The fact that the evolution of entropy is not clearly defined in the conventional picture makes the second law of thermodynamics ill-defined in this picture. Consider a perfect gas contained by removable walls in a small segment of a long tube (\ref{Fig118}). When the walls are removed, the gas diffuses in both directions. After some time $t$ the volume $V$ of the region in which gas molecules can be found doubles: $V(t)=2V(0)$. Then the number of microstates consistent with the confinement also doubles: $\Omega(t)=2\Omega(0)$. We get $\mathcal{E}(t)=log(2)+\mathcal{E}(0)$.

However, this conclusion is a comparison of two \emph{static} macroscopic states; it does not take into account the \emph{evolution} from the first state to the second. Our knowledge of the initial state of the gas makes it possible to exclude some microscopic states at time $t$. Consider Fig. \ref{Fig118}(b). Let us assume the knowledge that the position in phase space of all molecules at time $0$ is confined to the rectangular gray region with volume $V_{xp}$. We assume no knowledge about the distribution of positions within this rectangle. At time $t$, this rectangle is deformed so that the volume $V_{e}$ of the smallest envelope rectangle that covers the gray region is twice as large as that of the initial gray rectangle: $V_{e}(t)=2V_{xp}(t)$. This fact corresponds to the increase of entropy.

\begin{figure}[tp]
\begin{center}
\includegraphics[width=80mm,clip=true]{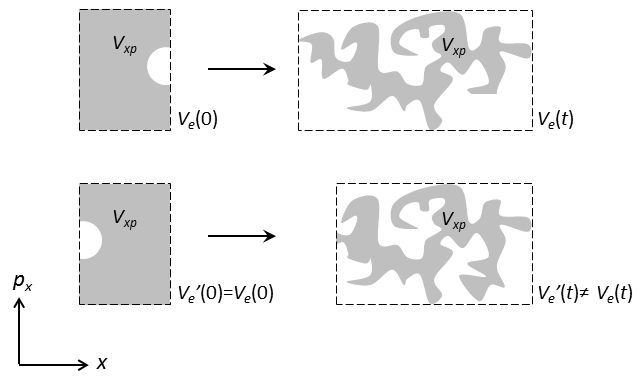}
\end{center}
\caption{Entropy, in its conventional sense, is a function of variables that define the macrostate. It will be a function of the volume $V_{e}$ of the envelope in phase space that covers the ensemble of microscopic states (gray regions). Microsopic details of the initial state may then affect the evolution of the entropy.}
\label{Fig119}
\end{figure}

If we take into account our initial knowledge and knowledge about physical law, however, we can exclude the white regions within the envelope rectangle. The volume $V_{xp}$ of the gray region is a constant of motion in classical mechanics. Therefore we can argue that the entropy should stay constant. If we nevertheless insist that the entropy should depend only on the macroscopic contraints that define the envelope, we face the fact that the result depends on the details of the initial state. This is illustrated in Fig. \ref{Fig119}.

In what sense then does the entropy increase in the conventional picture? Clearly, it is not related to our actual knowledge about the state of the gas, but to the evolution of a set of pre-chosen intervals of macroscopic variables that constrains this state. The product of these widening intervals define the entropy, regardless the dynamics of the gas. If the dynamics is non-linear, we may end up with very complex gray regions in phase space (Fig. \ref{Fig119}). The volume $V_{xp}$ is still the same, but the region must be covered with a much larger envelope rectangle. One may argue that the shape of the gray microscopic state ensemble in phase space typically becomes so complex after a while that the covering box becomes a good approximation of our `knowledge in practice'. However, this is a hand-waving phrase, and we should not rely on such vague concepts if we want to use entropy to say something fundamental about Nature. Even if the complicated shape of the ensemble becomes harder and harder to compute by us humans as time goes, it is nevertheless a function of our initial knowledge.

To make the notion of an evolving, increasing entropy more precise, we may focus on the growing complexity of the ensemble in phase space rather than its envelope. This notion comes closer, it seems to me, to the everyday illustration of the second law of thermodynamics: that things tend to become more and more disordered as time passes. A common picture is that of a vase that falls from a table and breaks into a thousand pieces. It is hard to decide whether the number of microscopic states consistent with the sad mess we see on the floor is smaller or larger than the number of such states consistent with the intact vase, but it is certainly harder to describe the mess in its details - it is more complex.

There is a sense in which the entropy of the vase actually decreases when it breaks. The shape of the ensemble in phase space is often very sensitive to the initial conditions, to the boundaries of our initial knowledge. If this sensititvity is present at all times, the dynamics is chaotic. Then, as the ensemble grows more complex, we learn more and more about the initial conditions; the sensitivity acts as a magnifying glass. Fewer and fewer initial conditions become consistent with what we observe. This gain of knowledge may be said to correspond to a decreased entropy of the initial state.

The vase might have had invisible cracks along which it breaks. These weaknesses may become knowable only after the breaking. We are able to conclude afterwards that fewer microscopic states are consistent with the intact vase we remember with regret. On the other hand, if the vase does not break when it falls, we learn nothing new about its initial state. Its entropy does not change.

In our terminology, these considerations become almost self-evident from the definition of the physical state $S$, its evolution $u_{1}S$, and the concept of a state reduction: $u_{1}S(n)\rightarrow S(n+1)\subset u_{1}S(n)$.

\begin{state}[\textbf{Knowledge cannot decrease as a result of an observation}]
Suppose that a knowable change of object $O$ defines the temporal update $n\rightarrow n+1$, so that $S(n)\cap S(n+1)=0$. Then $u_{1}^{-1}S_{O}(n+1)\subseteq S_{O}(n)$. Correspondingly, $u_{1}^{-1}S(n+1)\subseteq S(n)$.
\label{knowincrease}
\end{state}

If the invisible cracks in the vase was outside potential knowledge, it is tempting to say that we gain potential knowledge about the vase in retrospect, as we examine the fragments. However, this is an incorrect statement in a formal sense. We can seldom gain new potential knowledge of the past. This tends to violate epistemic consistency (Assumptions \ref{epconsistency}, \ref{epconsistency1} and \ref{epconsistency2}, illustrated in Fig. \ref{Fig25c}). What we can say without any risk of contradiction is that if we apply the inverse evolution operator to the state of the broken vase, we get a state which is a subset of the state of the intact vase. We get $u_{1}^{-1}S_{O}(n+1)\subset S_{O}(n)$, in accordance with Statement \ref{knowincrease}.

Sometimes the increase of retrodicted knowledge may be very small when an observation is made, or even non-existent. It is non-existent if and only if no state reduction of the object state takes place, meaning that $S_{O}(n+1)=u_{1}S_{O}(n)$. We claimed above that we do not learn anything new if the vase does not break as it falls. But this is not quite so. We may learn that is had no critical cracks to begin with. However, if we observe the vase anew while it is still standing, we can be more confident that we learn nothing new.

Let us discuss another situation in which little is learned by an observation. Suppose that the opening of a vault of a gas tube at time $n$ allows high pressure gas to fill a large empty chamber. Let the object $O$ be the chamber with the gas and the tube. Say that we open the chamber at time $n+m$ and finds that it is evenly filled with gas. The gain of knowledge in this observation is extremely small, meaning that $V[u_{m}S_{O}(n)]\approx V[S_{O}(n+m)]$. This is so because a very, very small fraction of the unknown microstates compatible with $S_{O}(n)$ leads to subsequent states $S_{O}(n+m)$ where the gas is unevenly distributed on a macroscopic scale, or is even still hiding inside the tube. Only a very small fraction of the state space volume $V[S_{O}(n)]$ is thus excluded by the observation.

We may convert this conclusion to a reformulation of the law of thermodynamic equilibrium. Let us assume that we have no prior potential knowledge about the state $S_{O}$ of an object $O$, except that it can be divided into a very large number of objects. We also now the values of a set $\{cP\}$ of collective properties that describe the entire object $O$, such as mass or temperature. Suppose that we partition the object $O$ into $N$ different pieces in a context $C$, and observe the property $P$ in each piece $O_{l}$ at time $n+m$. That is, we observe a vector-valued property 

\begin{equation}
\mathbf{P}=(P_{1},P_{2},\ldots,P_{N}).
\label{vectorp}
\end{equation}
Just like $O$, we assume that the piece $O_{l}$ can be divided into a very large number of objects. Suppose finally that

\begin{equation}
cp=p_{1}+p_{2}+\ldots+p_{N}
\label{sump}
\end{equation}
is the value of one of the collective properties $cP$ known \emph{a priori} (at time $n$), where $p_{l}$ is the observed value of $P_{l}$. Then we will find 

\begin{equation}
p_{1}\approx p_{2}\approx\ldots\approx p_{N}\approx cp/N
\label{equalp}
\end{equation}
with probability $q\approx 1$. We express this statement sloppily since it the result of conventional reasoning in statistical mechanics. The number $V[S_{1}]$ of exact states $Z$ that conform with Eq. [\ref{equalp}], given the knowledge at time $n$ expressed in Eq. [\ref{sump}], is very much larger than the number $V[S_{2}]$ of exact states that do not conform with this equation.

Note that, to reach this conclusion, we do not have to assume that a thermalization process takes place before the observation at time $n+m$. The lack of prior potential knowledge about $\mathbf{P}$ means that we start with \emph{tabula rasa}, and can apply the principle of \emph{a priori} equal probabilities (Section \ref{aprioriequal}). Thus, even if the reasoning behind Eq. [\ref{equalp}] is conventional, the meaning of the statement and its range of applicability is not conventional, since we have used our own concepts as input, rather than the usual thermodynamic ones.

We may write $q_{1}=V[S_{1}]/(V[S_{1}]+V[S_{2}])$ and $q_{2}=V[S_{2}]/(V[S_{1}]+V[S_{2}])$, so that we may re-express our conclusion as

\begin{equation}
q_{1}\gg 1-q_{1}.
\end{equation}

\begin{state}[\textbf{Law of expected thermodynamic equilibrium}]
Suppose that an object $O$ is divided in $N\gg 1$ pieces $O_{l}$, each of which can be divided into $M_{l}\gg 1$ smaller pieces. Suppose that the value $p$ of property $P$ of $O$ is known, but that we have no potential knowledge about the corresponding value $p_{l}$ of any piece $O_{l}$, except that $p_{l}$ is consistent with $p$. Let $S_{1}$ be the alternative that Eq. [\ref{equalp}] is fulfilled, and let $q_{1}$ be the probability that this alternative comes true in a context $C$ in which property $\mathbf{P}$ is observed, as defined in Eq. [\ref{vectorp}]. Then $q_{1}\gg 1-q_{1}$.
\label{thermoeq}
\end{state}

\begin{figure}[tp]
\begin{center}
\includegraphics[width=80mm,clip=true]{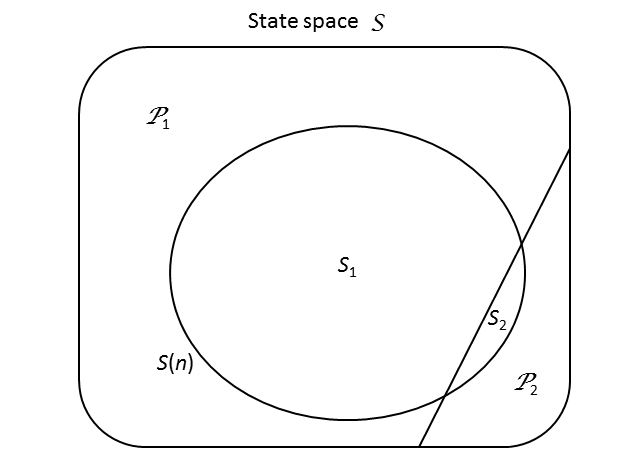}
\end{center}
\caption{Illustration of Statement \ref{matterdist}. $S(n)$ is the state just before we decide whether the distribution of matter and radiation in the universe is isotropic and homogeneous or not. $\mathcal{S}/S(n)$ is the part of state space $\mathcal{S}$ that consists of exact states not consistent with local observations with naked eye on a planet with intelligent life. $\mathcal{P}_{1}$ is the part of $\mathcal{S}$ that consists of exact states with homogeneous and isotropic large scale matter distribution, whereas $\mathcal{P}_{2}$ correponds to inhomogeneity or anisotropy. The corresponding two alternatives are denoted $S_{1}$ and $S_{2}$. We get $V[S_{2}]/V[S_{1}]<<1$ by the usual line of reasoning in statistical mechanics.}
\label{Fig121}
\end{figure}

This statement can be applied to the large scale distribution of matter (and radiation) in outer space. Consider Fig. \ref{Fig121}. Suppose that the first astronomic observation that makes it possible to decide whether this distribution is isotropic and homogeneous takes place at time $n+m$. This observation may not necessarily have been done by humans on earth. Even before that observation, at time $n$, some properties of the matter distribution follows by physical law from local observations with the naked eye (or less sophisticated astronomical instruments). For example, we can deduce by simple means that there has to be a countless number of stars, exoplanets, gas clouds, galaxies, and so on. In this way we can exclude at time $n$ parts of the state space $\mathcal{S}$ that correspond to exact states that lack such deduced properties. We are left with the state $S(n)$.

Independently, we can divide the entire state space in two parts: in one part $\mathcal{P}_{1}$ the matter distribution is fairly homogeneous on sufficiently large scales, in the other part $\mathcal{P}_{2}$ it is not. We ratio $V[\mathcal{P}_{2}]/V[\mathcal{P}_{1}]$ will be extremely close to zero. (Figure \ref{Fig121} is misleading in this respect for the sake of illustration.) Therefore we will have $q_{1}\equiv V[S_{1}]/(V[S_{1}]+V[S_{2}])\approx 1$, where $S_{1}$ and $S_{2}$ are the two alternatives that define the question about the matter distribution answered at time $n+m$. In other words, the probability that we would have seen an anisotropic matter distribution was essentially zero.

\begin{state}[\textbf{Matter distribution in the universe}]
The chance that we would have seen anything else than an isotropic and homogeneous distribution of matter and radiation on sufficiently large scales was essentially zero.
\label{matterdist}
\end{state}

If we accept the choice of fundamental concepts used in the present work, the above statement is almost self-evident. The prize we pay, though, is that we deprive the universe any definite matter distribution before the first aware beings cared to investigate it. However, this is no different from the conclusion that the test object used in a double slit experiment does not pass any definite slit unless we actually investigate which slit it passes. In the language used here, there will be interference between these two alternatives if the path information is outside potential knowledge, just as we assume above that the large scale matter distribution is outside potential knowledge until time $n+m$.

There has to be such a critical time $n+m$ at which the question about the matter distribution is settled. Otherwise we have to assume that an isotropic and homogeneous distribution is necessary for the emergence of aware organisms, so that potential knowledge about this distribution is present as soon as the first aware beings emerge. This seems highly unlikely. (The relation between the emergence of aware beings and the emergence of the universe will be discussed below.)

Let us turn to the evolution of entropy. Recall from section \ref{law}, and in particular from Definition \ref{exactstate}, that a state evolves deterministically if and only if it is exact. We often illustrate exact states $Z$ as points with volume $V[Z]=1$, and actual states $S$ as hollow circles with volume $V[S]>1$ containing a set $\{Z\}$ of such points. However, if such a `hollow' state evolves without any state reductions according to $S(n+m)=u_{m}S(n)$ for all $m$, then it becomes exact by definition and should rather be illustrated as a point. We therefore concluded in Statement \ref{reductionsoccur} that state reductions must occur.

\begin{figure}[tp]
\begin{center}
\includegraphics[width=80mm,clip=true]{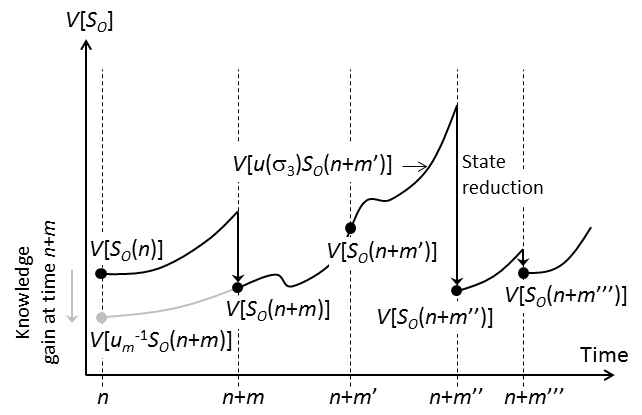}
\end{center}
\caption{The evolution of the state space volume or entropy of an object state $S_{O}$. In a context family $C(\sigma_{1},\sigma_{2},\ldots)$ the continuous evolution operator $u(\sigma)$ is expected to make volumes increase, even though exceptions may occur. State reductions decrease the volume, and enables knowledge gain of the preceding state via the repeated application of $u_{1}^{-1}$. They occur regularly, counterbalancing the effect of $u(\sigma)$, so that the sequence $(V[S_{O}(n)],V[S_{O}(n+m)],\ldots)$ does not have to be growing.}
\label{Fig120}
\end{figure}

In fact, the must occcur regularly; the state volume $V[S]$ must decrease by a substantial amount at finite time intervals. Otherwise we cannot distinguish the evolution of $S(n)$ from that of a deterministic state (Definition \ref{determinism3}) in any knowable sense. That is, to give the statement that the evolution is non-deterministic epistemic meaning we must require that $\langle V[S(n)]-V[u_{m}^{-1}S(n+m)]/V[S(n)]\rangle_{n}$ differ substantially from zero for large enough $m$, where $\langle\ldots\rangle_{n}$ is an average over all sequential times $n$. The situation is illustrated in Fig. \ref{Fig120} for an object state $S_{O}$ and for $m=1$.

\begin{state}[\textbf{Substantial state reductions occur regularly}]
There is a constant $g_{A}[\mathcal{T}_{S}]>0$ such that $\langle V[S(n)]-V[u_{1}^{-1}S(n+1)]/V[S(n)]\rangle_{n}\geq g_{A}[\mathcal{T}_{S}]$ for all trajectories $\mathcal{T}_{S}=(S(1),S(2),\ldots,S(n),S(n+1),\ldots)$ given any initial state $S(1)$.
\label{regularreductions}
\end{state}

We may reformulate this statement as follows: The expected knowledge gain in the sense of Statement \ref{knowincrease} at each time instant is a positive constant $g$. In a long sequence of observations, this means that the knowledge about the initial state will grow forever. This can in turn be interpreted as a law of ever increasing entropy. This interpretation is justified by the idea mentioned above that entropy may be said to increase when an ensemble in classical phase space grows forever more complex, requiring a forever larger envelope (Fig. \ref{Fig119}).

The sensitivity of such an ensemble $\Sigma(t)$ to its initial condition $\Sigma(0)$ is the source of the knowledge growth; it acts as a magnifying glass that zooms in more and more on the initial ensemble as time passes. (However the image of the initial condition becomes more and more distorted in the process.) This situation occurs when the dynamics is chaotic, in classical mechanical sense of the word. In that context, our \emph{increasing} entropy corresponds to the notion of \emph{positive} metric or topological entropy \cite{Ott}.

\begin{state}[\textbf{Law of increasing entropy A}]
As sequential time passes, the retrodicted potential knowledge about the initial physical state $S(1)$ grows without bound in the following sense: $V[u_{n-1}^{-1}S(n)]\rightarrow 1$ as $n\rightarrow\infty$.
\label{entropylawA}
\end{state}

In other words, the retrodicted knowledge approaches a state of potential knowledge that corrsponds to an exact physical state $Z$. This statement does not contradict the fact that knowledge is always incomplete, since the retrodiction of knowledge via the inverse evolution operator $u_{n-1}^{-1}$ does not correspond to actual knowledge at the initial time $1$.

Statement \ref{entropylawA} expresses the evolution of entropy in terms of state space volume changes associated with observations, with temporal updates $n\rightarrow n+1$. What can we say about the volume changes associated with the application of the evolution operator $u_{1}$?

Given Statement \ref{regularreductions}, we see that the expected value of $V[u_{1}S]$ must be substantially larger than $V[S]$. Otherwise we expect $V[S(n)]$ to decrease steadily towards $V[S(n)]\rightarrow 1$ as $n\rightarrow\infty$. This would contradict the perpetual incompleteness of knowledge.

\begin{state}[\textbf{Evolution increases volumes substantially}]
There is a constant $g_{B}[\mathcal{T}_{S}]>0$ such that $\langle V[u_{1}S(n)]-V[S(n)])/V[S(n)]\rangle_{n}\geq g_{B}[\mathcal{T}_{S}]$ for all trajectories $\mathcal{T}_{S}=(S(1),S(2),\ldots,S(n),S(n+1),\ldots)$ given any initial state $S(1)$.
\label{regularincrease}
\end{state}

Just as in Statement \ref{regularreductions}, $\langle\ldots\rangle_{n}$ is an average along the trajectory of the state starting with $S(1)=S$. To avoid that $V[S(n)]\rightarrow 1$ as $n\rightarrow\infty$, we must require

\begin{equation}
\frac{g_{B}[\mathcal{T}_{S}]}{g_{B}[\mathcal{T}_{S}]+1}\geq g_{A}[\mathcal{T}_{S}],
\label{opposefactors}
\end{equation}
for each initial condition $S(1)$. This inequality follows from Statements \ref{regularreductions} and \ref{regularincrease}, together with the observation that $\langle V[u_{1}S(n)]/V[S(n+1)]\rangle_{n}=\langle V[S(n)]/V[u_{1}^{-1}S(n+1)]\rangle_{n}$.

The overall situation is illustrated in Fig. \ref{Fig120} in the case of an object state $S_{O}$ that is observed within a context family $C(\sigma_{1},\ldots,\sigma_{n},\sigma_{n+1},\sigma_{n+2},\ldots)$. The presence of the continuous evolution parameter make it possible to depict a continuously growing state space volume between the observations. The volume does not have to be growing monotonically as a function of $\sigma_{n}$. The requirement expressed in Statement \ref{regularincrease} just means that it has to grow in the mean. In a similar way, Statement \ref{regularreductions} does not mean that a discernible state reduction takes place at each observation, just that such reduction take place at regular intervals. (In Fig. \ref{Fig120}, no discernible state reduction takes place at time $n+m'$.)

An important observation in Fig. \ref{Fig120} is that there is no clear trend in the sequence of volumes $(V[S_{O}(n)],V[S_{O}(n+m)],V[S_{O}(n+m')],\ldots)$. We cannot exclude the possiblity that the increasing volumes caused by the evolution, and the decreasing volumes caused by observation counterbalance perfectly in the long run. In that case we have $g_{B}[\mathcal{T}_{S}]/(g_{B}[\mathcal{T}_{S}]+1)= g_{A}[\mathcal{T}_{S}]$ in Eq. [\ref{opposefactors}]. Entropy does \emph{not} have to increase in the sense that the volumes in  a sequence $(V[S(n)],V[S(n+m)],V[S(n+m')],\ldots)$ grow steadily.

\begin{figure}[tp]
\begin{center}
\includegraphics[width=80mm,clip=true]{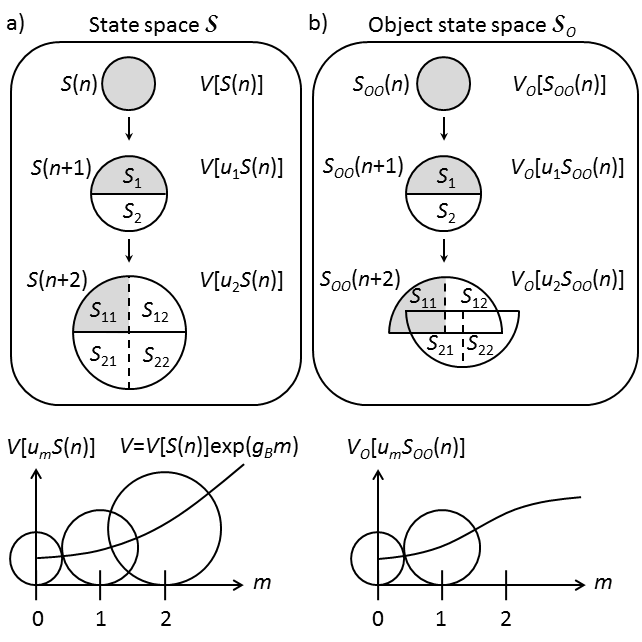}
\end{center}
\caption{The evolution operator $u_{1}$ can be applied to each possible alternative, even if they are not realized. a) For a state $S$ in state space $\mathcal{S}$ this means that $V[u_{m}S]$ grow exponentially, since distinct alternatives cannot overlap at a later time. b) In contrast, for an object state $S_{O}$ represented in object state space $\mathcal{S}_{O}$, the alternatives may start to overlap, halting the exponential volume increase.}
\label{Fig122}
\end{figure}

There is another sense, though, in which entropy grows, apart from that expressed in Statement \ref{entropylawA}. In a state reduction at time $n+1$ where one of several alternatives $S_{1},S_{2},\ldots$ is realized, the expected volume growth due to evolution `between' times $n+1$ and $n+2$ expressed in Statement \ref{regularincrease} can be applied to each of these alternative, not only that which is realized (compare Fig. \ref{Fig30} and Fig. \ref{Fig122}(a)). This means that we do expect a steadily growing trend in the sequence of volumes $(V[S(n)],V[u_{1}S(n)],V[u_{2}S(n)],\ldots)$.

\begin{state}[\textbf{Law of increasing entropy B}]
We have $\langle V[u_{m}S]\rangle_{S}>\langle V[u_{m-1}S]\rangle_{S}$ for each $m\geq 1$.
\label{entropylawB}
\end{state}

Here we let $u_{0}\equiv I$, and let $\langle\ldots\rangle_{S}$ be an average over all states $S\in\mathcal{PS}$ that belong to the domain of $u_{1}$.
With this interpretation of the concept of entropy, it stays the same, becoming a constant of motion, if and only if the evolution is deterministic, meaning that $S(n+1)= u_{1}S(n)$.

In what way does the sequence $(V[S(n)],V[u_{1}S(n)],V[u_{2}S(n)],\ldots)$ grow? As noted above, a growth factor related to $g_{B}$ applied at each sequential time can be applied to all alternatives at each time step, regardless whether these alternatives are realized or not. In Fig. \ref{Fig122}(a), alternative $S_{1}$ is realized at time $n+1$, meaning that $S(n+1)=S_{1}$. This state evolves according to $u_{1}S(n+1)=S_{11}\cup S_{12}$, where $S_{11}$ and $S_{12}$ are two alternatives that can be realized at time $n+2$. However, the other alternative $S_{2}$ also belongs to the domain of $u_{1}$, so that we may write $u_{1}S_{2}=S_{21}\cup S_{22}$, where $S_{21}$ and $S_{22}$ are two alternatives that could have been realized at time $n+2$, if $S_{2}$ were realized at time $n+1$. We conclude that

\begin{equation}
\langle V[u_{m}S]\rangle_{S}\propto V[S]e^{gm},
\label{expgrowth}
\end{equation}
for some average growth factor $g>0$, where it is understood that $u_{0}\equiv I$, where $I$ is the unit operator.

This conclusion relies heavily on the fact that the evolution $u_{1}$ is unique and invertible (Assumption \ref{uniqueu1}), as illustrated in Fig. \ref{Fig31b}. If distinct alternatives $S_{1}$ and $S_{2}$ defined at time $n$ may evolve so that $u_{m}S_{1}$ and $u_{m}S_{2}$ overlap at some later time $n+m$, then the exponential volume increase may come to a halt.

This possibility arises if we consider object states $S_{OO}$ represented in object state space $\mathcal{S}_{O}$, rather than the full physical state $S$ itself [Fig. \ref{Fig122}(b)]. In this case two alternatives $S_{1}$ and $S_{2}$ can equally well be interpreted as two different \emph{objects}. As such they can merge (Section \ref{objectmerging}). This means that two distinct states start to overlap, the reverse of the process shown in Fig. \ref{Fig42}.

Recall that in the full state space $\mathcal{S}$ the states $S_{O1}$ and $S_{O2}$ of two different objects $O_{1}$ and $O_{2}$ must always overlap; otherwise there would be no physical state $S$ consistent with the existence of both objects, as we perceive them (Fig. \ref{Fig90b}). On the other hand, in the object state space $\mathcal{S}_{O}$ two different object states $S_{OO1}$ and $S_{OO2}$ \emph{cannot} overlap - if they do the objects cannot be perceived as different (Fig. \ref{Fig90}).

\begin{figure}[tp]
\begin{center}
\includegraphics[width=80mm,clip=true]{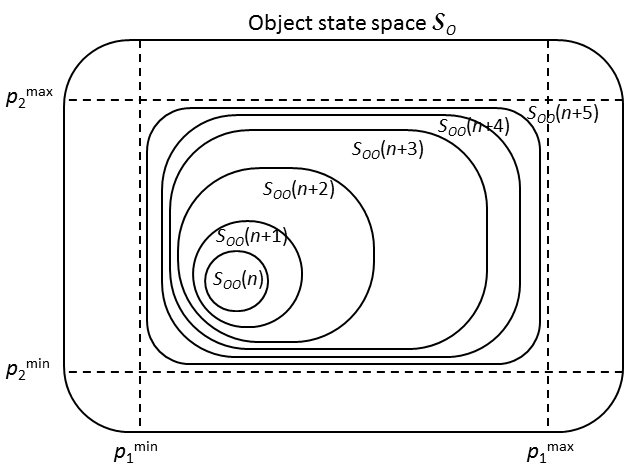}
\end{center}
\caption{The evolution in object state space $\mathcal{S}_{O}$ of an object state $S_{OO}$, which we do not observe during a time interval $[m,M]$. Even if we do not observe it during this time, we know that it must comply with certain constraints; there is a set of properties $\{P_{k}\}$ such that $p_{k}^{\min}\geq p_{k}(n)\geq p_{k}^{\max}$ for each $k$ whenever $m\leq n \leq M$. The object may be a light ball that bounces inside a chamber with massive walls that we cannot see through. It may also be a gas that expands within the same chamber. In the former case the state expansion is reversible, and in the latter case it is irreversible.}
\label{Fig117}
\end{figure}

This means that the exponential volume increase may be interrupted, as illustrated in Fig. \ref{Fig122}(b). This is typically the case, since there are most often known overall constraints that limit the growth of the object state (Fig. \ref{Fig117}). These constaints may correspond to knowlege that the object is located somewhere in a physical container. More generally, this corresponds to knowledge that the object is in a bound state. We may also have knowledge that macroscopic properties such as pressure or temperature stays within given limits.

In contrast, it does not make sense to talk about such constraints on the physical state $S$, on the world as a whole. By definition, there is nothing outside the universe that is able to constrain its evolution.

The foregoing considerations are similar to a discussion about sensitivity to initial conditions, non-linearity and chaos in dynamical systems theory. To explore the analogies we identify classical phase space with our state space ($\mathcal{S}$ or $\mathcal{S}_{O}$). The fact that the volume of any alternative tends to increase as it evolves correponds to the sensitivity to initial conditions - most phase space ensembles are stretched in some direction. The growth factor $g$ in Eq. \ref{expgrowth} is similar to a Lyapunov exponent $\lambda$. That different evolving alternatives represented in $\mathcal{S}_{O}$ eventually may overlap corresponds to the folding of the stretched phase space ensemble. This process is necessary to keep all variables finite, to contain their values inside some large box. Folding can occur in non-linear systems only, and is the hallmark of chaos. No overlap analogous to this folding can occur for states represented in the full state space $\mathcal{S}$. The dynamics resembles that of a linear system with sensitivity to initial conditions, where any phase space ensemble explodes exponentially along some direction in phase space. For the physical state $S$ this means that the boundary $\partial S$ approaches infinite values of some set of attributes at exponential speed. We return to these matters in Section \ref{expansion}.

The expansion of the object state $S_{OO}$ in object state space $\mathcal{S}_{O}$ can be either reversible or irreversible. Suppose that the object $O$ whose state is shown in Fig. \ref{Fig117} represents a light ball that bounces inside a chamber that constrains its motion. If we do not look inside the chamber during a long time interval $[n,M]$, all we can say after that is that the ball is somewhere in the chamber. In other words, the state $S_{OO}(n+M)$ fills the entire chamber. If we open the chamber at time $n+M+1$, we will find the ball somewhere. The object state immediately reduces to its original size:

\begin{equation}
V_{O}[S_{OO}(n+M+1)]\approx V_{O}[S_{OO}(n)]\ll V_{O}[S_{OO}(n+M)].
\end{equation}
The expansion is clearly reversible. Note that this state reduction is guaranteed only if it is part of potential knowledge at time $n$ that the ball will stay intact at least during the time interval $[n,M]$, and so will the chamber that encloses it.

The situation is different if we replace the ball with gas that is released from a small region defined by $S_{OO}(n)$ at time $n$, and is allowed to expand to fill the entire chamber. If we open the chamber at time $n+M+1$ and measure the density of the gas in different regions of the chamber, we will almost certainly find that it is the same everywhere (Statement \ref{thermoeq}). If we let $O$ represent an individual gas molecule, we will not be able to see any substantial state reduction at all:

\begin{equation}
V_{O}[S_{OO}(n+M+1)]\approx V_{O}[S_{OO}(n+M)] \gg V_{O}[S_{OO}(n)].
\end{equation}
The state expansion is clearly irreversible.

\begin{figure}[tp]
\begin{center}
\includegraphics[width=80mm,clip=true]{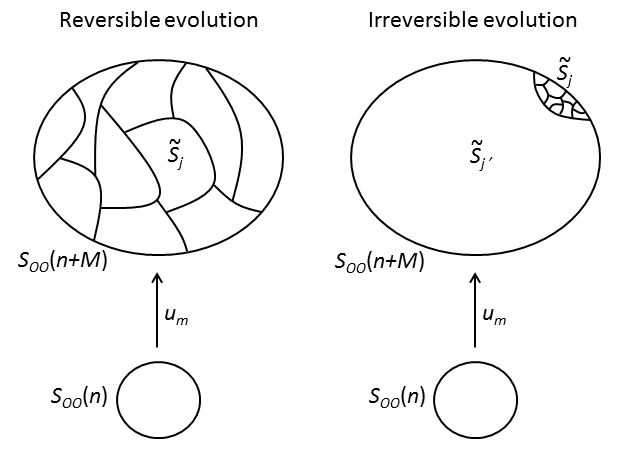}
\end{center}
\caption{Reversible and irreversible object evolution. In a reversible evolution we can reduce the volume of the object state to its original size $V_{O}[S_{OO}(n)]$ in an observational context $C$ with a set of possible outcomes $\tilde{S}_{j}$ of comparable volumes $V_{O}[\tilde{S}_{j}]\approx V_{O}[\tilde{S}_{j'}]\lesssim V_{O}[S_{OO}(n)]$. In an irreversible evolution no such context $C$ is possible. There is typically one alternative $\tilde{S}_{j}$ with $V_{O}[\tilde{S}_{j}]\gg V_{O}[S_{OO}(n)]$, which is realized with overwhelming probability. A ball bouncing in the chamber in Fig. \ref{Fig117} illustrates the reversible case, and the gas expanding in the same chamber illustrates the irreversible case. The dominating alternative $S_{j'}$ correponds to the observation of the gas evenly distributed in the chamber.}
\label{Fig128}
\end{figure}

Of course, we could see the gas spontaneously recollected in a small region in the chamber, but the probablity that this happens is essentially zero.
Probabilites are volumes of a complete set $\{\tilde{S}_{j}\}$ of future alternatives. Therefore we can distinguish between reversible and irreversible expansion of object states in the following way.

\begin{defi}[\textbf{Reversible and irreversible object evolution}]
Suppose that the object state $S_{OO}$ evolves without observation between times $n$ and $n+M$ with $M\gg 1$, but that $O$ is observed at time $n+M+1$. Suppose also that $V_{O}[S_{OO}(n+M)]\gg V_{O}[S_{OO}(n)]$. Then the evolution is reversible if and only if it is possible to construct an observational context $C$ with alternative outcomes $\{\tilde{S}_{j}\}$ such that we have $V_{O}[\tilde{S}_{j}]\lesssim V_{O}[S_{OO}(n)]$ for some $j$ such that there is no $j'$ for which $q_{j'}\gg q_{j}$. Here, $q_{j}=V[\tilde{S}_{j}]/V[S_{O}(n+M)]$ and $q_{j'}=V[\tilde{S}_{j'}]/V[S_{O}(n+M)]$ are the probabilities for the two alternatives.
\label{reversible}
\end{defi}

The idea behind this definition is illustrated in Fig. \ref{Fig128}. The existence of irreversible object evolution is what makes us expect that the entropy $\mathcal{E}[S]=\log(V[S])$ of the entire world tends to increase with time. However, according to Statement \ref{entropybound} it is sufficient that \emph{some} objects evolve reversibly in order to ensure that $\mathcal{E}[S]$ stays small. 

\vspace{5mm}
\begin{center}
$\maltese$
\end{center}
\paragraph{}

In this connection, we noted above that we cannot say anything definitive about long term trends among the volumes in the sequence $(V[S(n)],V[S(n+1)],\ldots)$. This fact seemingly contradicts the conventional notion that entropy always increases as time passes. However, we can restore this conclusion if we introduce a coarse-grained entropy, or take the incomplete knowledge about the physical state into account, together with the fact that the evolution is irreducible.

\begin{figure}[tp]
\begin{center}
\includegraphics[width=80mm,clip=true]{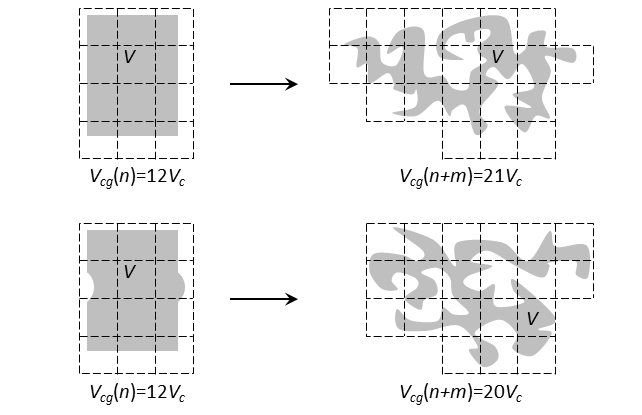}
\end{center}
\caption{Two similar states that evolve into two complex states that have no details in common. This behaviour can arise due to the irreducible nature of the evolution operator. In contrast, the evolution is pointwise in classical phase space, so that identical segments of the boundary of different ensembles remain identical as they evolve (Fig. \ref{Fig119}). The coarse-grained state space volume $V_{cg}$ is calculated in units of the volume $V_{C}$ of the coarse-grained cells in state space. $V_{cg}$ increases more rapdly with time than the actual state space volume $V$.}
\label{Fig123}
\end{figure}

Paul and Tanya Ehrenfest introduced the concept of coarse-grained entropy in 1911 \cite{ehrenfest}. We may partition the classical phase space into compartments and count the number $V_{cg}(t)$ of such compartments that intersects an evolving ensemble of points in this phase space (Fig. \ref{Fig123}). Even if the volume $V(t)$ of these points stays constant due to Liouville's theorem, $V_{cg}(t)$ tend to increase since the shape of the ensemble becomes more and more complex.

In classical mechanics, this procedure can be used to deduce that entropy increases, under an assumption of ergodicity. The problem is, however, that it can also be used to deduce that entropy increses in the past, as $-t$ grows, from a given present value at $t=0$ where it is at a minimum.

This will not be a problem in our description, with our different treatment of time. (However, we will discuss an analogous problem below, in relation to Fig. \ref{Fig127}, concerning the evolution of the support of wave functions.) We replace $t$ with sequential time $n$, and $V(t)$ with $V[S(n)]$. We have not proven any counterpart to Liouville's theorem for the state space volumes $V[S(n)]$, but we know that they cannot tend to zero as $n\rightarrow\infty$ because of the incompleteness of knowledge. There is a minimum volume $V_{\min}$ such that

\begin{equation}
V[S(n)]\geq V_{\min}
\end{equation}
for all $n$ and all initial conditions $S(1)$. (The projection of $V_{\min}$ onto $(x,p_{x})$-space in the case of object states is associated with Planck's constant $\hbar$ in Fig. \ref{Fig80b}).

In the classical case we expect $V_{cg}(t)\rightarrow\infty$ as $t\rightarrow\infty$ if the evolution is ergodic and unconstrained. By analogy, we expect the following hypothesis to hold true.

\begin{hypo}[\textbf{The coarse-grained entropy increases without bound}]
Consider a collection of covering sets $\Sigma(n)\equiv\{\Sigma_{Ck}(n)\}$ such that $\Sigma_{Ck}(n)\cap\Sigma_{Ck'}(n)=\varnothing$ and $S(n)\subseteq\bigcup_{k}\Sigma_{Ck}(n)$. Suppose that the state space volume of each covering set is the same, and that it is independent of $n$; $V[\Sigma_{Ck}(n)]=V_{C}$. Let $N[\Sigma(n)]$ be the smallest possible number of elements in such a collection $\Sigma(n)$. For each fixed $V_{C}$, and for each initial condition $S(1)$, we have $N[\Sigma(n)]\rightarrow\infty$ as $n\rightarrow\infty$.
\label{cgincrease}
\end{hypo}

In classical physics, the counterpart of this hypothesis is at the basis of the second law of thermodynamics. In this picture we have to assume that the initial condition of the universe is very special in the sense that the ensemble in phase space is `simple', like the square in Figs. \ref{Fig119} and \ref{Fig123}. Then it becomes more and more complex as time passes.

This fact does not mean that physical law in itself has to be time-asymmetric. It may just reflect the fact that there are more complex ensembles than simple ones, and that a simple ensemble typically evolves into a more complex one, regardless whether you apply the evolution rule forwards or backwards. Put another way, the set $S_{sf}$ of simple ensembles in phase space that stays simple when we evolve them forwards is very small as compared to the set $S_{s}$ of all simple ensembles. The same goes for the set $S_{sb}$ of simple ensembles that stays simple when we evolve them backwards. The intersection $S_{sf}\cap S_{Sb}$ is even smaller. This means that, to get the second law, we just have to choose a \emph{typical} simple initial ensemble $S_{i}$. Then we can be almost sure that it grows more complex as time passes, so that the coarse-grained entropy increasees.

The situation is different in the present description. Statement \ref{entropylawB} means that physical law is inherently time-asymmetric. If we apply the evolution operator $u_{1}$ to a state, its volume tends to increase, whereas it tends to decrease if we apply the inverse evolution operator $u_{1}^{-1}$. The only way to regain symmetry would be to let $u_{1}$ and $u_{1}^{-1}$ act one exact states $Z$, so that $V[u_{1}^{-1}Z]=V[Z]=V[u_{1}Z]=1$ and we achieve a precise trajectory $Z(n)$ in state space. We know, however, that exact states are not in the domain of $u_{1}$ or $u_{1}^{-1}$ (Statement \ref{irreduciblelaw}).

\begin{state}[\textbf{Physical law is temporally asymmetric}]
The evolution operator $u_{1}$ is different from its inverse $u_{1}^{-1}$.
\label{timeasymmetry}
\end{state}

The coarse-grained entropy is helpful in the sense that it couples entropy to physical law; it makes the concept dynamical. However, from the present  epistemic perspective, it is not satisfying to depend on a collection of covering sets $\Sigma_{Ck}$ in order to say something fundamental about the evolution of the apparent volume of $S(n)$. What do these covering sets represent in terms of knowledge? They cannot be treated as a complete set of realizable alternatives with equal probabilities $1/N$ (Statement \ref{cgincrease}), since their volume $V[\Sigma_{Ck}]$ in part represent regions of state space \emph{outside} the physical state $S(n)$.

Instead, we may consider the fundamental unknowability of the details of the boundary $\partial S$ of the physical state (Section \ref{knowstate}). This means that the evolution of $S$ becomes fuzzy. Let $\{_{1}S,\,_{2}S,\,_3{S},\ldots\}$ be a set of possible physical states $S$, given the potential knowledge $PK$. Then we may define a fuzzy state 

\begin{equation}
_{f}S(n)\equiv\,_{1}S(n)\cup\,_{2}S(n)\cup\,_3{S}(n),\ldots
\end{equation}
such that $S\subseteq\,_{f}{S}$. We may define 

\begin{equation}
_{fm}u[_{f}S(n)]\equiv u_{m}[_{1}S(n)]\cup u_{m}[_{2}S]\cup u_{m}[_3{S}]\cup\ldots.
\end{equation}

In a mathematical representation $_{f}\bar{S}$, we may want to replace $\cup$ with $+$ according to Table \ref{dictionary}. This algebraic notation has limited value, however, since there are no alternatives associated with the possible states $_{j}S$ - we can never decide which state is the true one. Therefore there are no amplitudes $_{j}a$ and no Hilbert space.

\begin{figure}[tp]
\begin{center}
\includegraphics[width=80mm,clip=true]{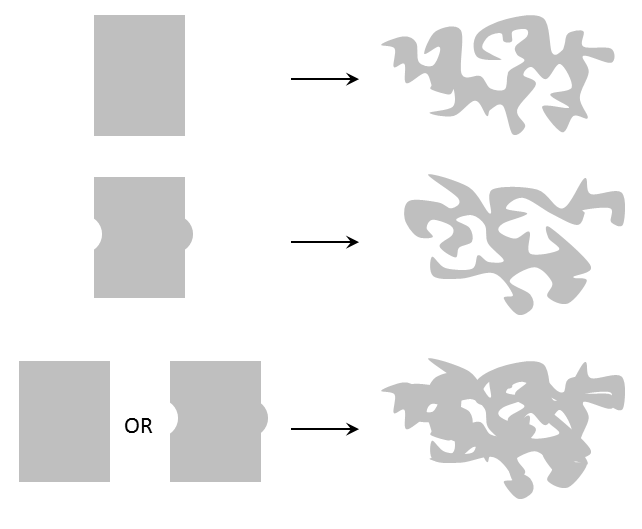}
\end{center}
\caption{Two similar states that evolve into two complex states that have no details in common. If none of the original states can be excluded as a representation of the potential knowledge, none of the exact states in the superposed complex state can be excluded as a representation of the evolved potential knowledge. The superposed state space volume increases more rapidly with time than that of an individual state. Compare Fig. \ref{Fig123}.}
\label{Fig125}
\end{figure}

The crucial point is that in general we have

\begin{equation}
_{f1}u[_{f}S]\neq u_{1}[_{f}S]
\end{equation}
because of the irreducibility of physical law, making the state boundary act like a membrane (Statement \ref{membrane}). The evolution of the state does not have to equal the union of the evolution of its parts. The idea is illustrated in Fig. \ref{Fig125}. Even if the two states at the left are very similar, they may evolve into quite different states, which share only global features such as the level of complexity and the approximate volume of the envelope (Fig. \ref{Fig119}). This means that the volume of the superposed evolved state may be much larger than the volume of each individual evolved state. We may express this fact as the possibility that

\begin{equation}
V[_{f1}u[_{f}S]]> V[u_{1}[_{f}S]].
\label{fastergrowth}
\end{equation}

This situation may be contrasted with that in classical mechanics, where the evolution of an ensemble in phase space is pointwise. In Fig. \ref{Fig119}, there is a small difference between the two simple emsembles to the left, leding to a small difference between the evolved ensembles to the right. The removal of a small part of the rectangle to the left causes the removal of a small part of the complex ensemble to the right. If we superpose the two complex ensembles we regain the ensemble that correspond to the evolution of the intact rectangle. The volume of union of the two perturbed evolved ensembles is exactly the same as the volume of the intact rectangle.

What can be said about the evolution of the sequence of volumes $(V[_{f}S(n)],V[_{f}S(n+1)],\ldots)$? We note first that we may have a state reduction

\begin{equation}
_{f1}u[_{f}S(n)]\rightarrow\,_{f}S(n+1)\subset\,_{f1}u[_{f}S(n)]
\end{equation}
at time $n+1$ of the evolved fuzzy state, which takes place if and only if there is a state reduction $u_{1}[_{j}S(n)]\rightarrow\,_{j}S(n+1)\subset\,u_{1}[_{j}S(n)]$ of each individual possible state $_{j}S$.

As discussed above, the existence of state reductions of makes it impossible to decide in general whether the sequence $(V[_{j}S(n)],V[_{j}S(n+1)],\ldots)$ is growing, even though $(V[u_{1}[_{j}S(n)]],V[u_{2}[_{j}S(n)]],\ldots)$ is indeed growing (Fig. \ref{Fig120}). All we can say is that the perpetual incompleteness of knowledge implies that there is a constant $g_{\min}>1$ such that

\begin{equation}
g_{\min}\leq V[_{j}S(n)]
\label{lbe}
\end{equation}
for each $n$ and for each possible state $_{j}S$. It may also be the case that $V[_{j}S(n)]$ stays bounded so that 

\begin{equation}
g_{\min}\leq V[_{j}S(n)]\leq g_{\max}
\label{lube}
\end{equation}
for each $n$ and for each possible state $_{j}S$. This situation occurs if the volume increase caused by evolution is exactly balanced in the long run by the volume decrease caused by state reductions.

If we accept the hypothesis that states $_{j}S$ tend to become more and more complex in analogy with ergodic classical ensembles in phase space, then $V[_{fm}u[_{f}S]]$ will grow faster than $V[u_{m}[_{j}S]]$ with $m$, as expressed in Eq. [\ref{fastergrowth}] and illustrated in Fig. \ref{Fig125}. Suppose that that there is no upper bound $[_{j}S(n)]\leq g_{\max}$ according to Eq. [\ref{lube}]. Then there cannot be an upper bound $V[_{f}S(n)]\leq g_{\max}$ either. Suppose, on the other hand, that the bound $[_{j}S(n)]\leq g_{\max}$ exists. Then, if such a bound exists for $[_{f}S(n)]$ also, we would have to conclude that Eq. [\ref{lbe}] is not fulfilled as $n\rightarrow\infty$, because of the different growth rates of $V[_{fm}u[_{f}S]]$ and $V[u_{m}[_{j}S]]$. Therefore there can never be any bound $V[_{f}S(n)]\leq g_{\max}$ as $n\rightarrow\infty$.

\begin{defi}[\textbf{Fuzzy entropy} $_{f}\mathcal{E}$]
Suppose that we choose a set of physical states $\{_{j}S(1)\}$ with more than one member, such that each state $_{j}S(1)$ is consistent with some initial state of potential knowledge $PK(1)$. Then $_{f}\mathcal{E}(n)=\log\left(V\left[\bigcup_{j}\,_{j}S(n)\right]\right)$.
\label{knowentro}
\end{defi}

\begin{hypo}[\textbf{Law of increasing entropy C}]
The fuzzy entropy increases without bound. Suppose that aware subjects exist for arbitrarily long time, so that the limit $n\rightarrow\infty$ is defined. Then, for a typical set of possible states $\{_{j}S(1)\}$ with more than one member, we have $_{f}\mathcal{E}\rightarrow\infty$ as $n\rightarrow\infty$.
\label{entropylawC}
\end{hypo}

This hypothesis is vague in the sense that it is not specify what is meant by a `typical' initial set of states $\{_{j}S(1)\}$. It could mean that almost all states of potential knowledge $PK$ can be described by a set $\{_{j}S(1)\}$ that fulfils the hypothesis.

If we accept detailed materialism (assumption \ref{localmaterialism}), we accept that gradual biological evolution is necessary to create aware beings. Then there must be an initial sequential time $n=1$ at which the first aware perceptions emerge, and these perceptions must be rudimentary, corresponding to a small $PK(1)$ and a large physical state $S(1)$ with $V[S(1)]\gg 1$ (Fig. \ref{Fig3}).

As the biological evolution goes on, and as the number of aware beings increase, one may expect that the potential knowledge $PK(n)$ grows steadily. However, Hypothesis \ref{entropylawC} implies that such a development eventually must come to a halt. Just like in the conventional picture, thermodynamic equilibirum finally settles, making meaningful distinctions impossible, so that potential knowledge dissipates, approaching the `naked' awareness in Fig. \ref{Fig5b}, before nothingness sets in at some time $n_{final}$.

\begin{figure}[tp]
\begin{center}
\includegraphics[width=80mm,clip=true]{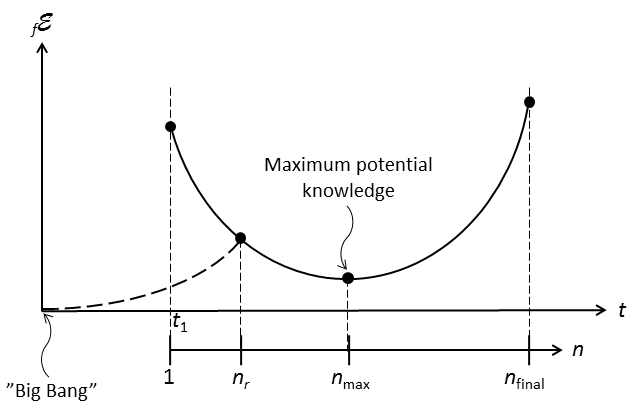}
\end{center}
\caption{A model of the evolution of the fuzzy entropy $_{f}\mathcal{E}$. The birth of the universe corresponds to the appearance of the first aware subject at sequential time $n=1$. The entropy $_{f}\mathcal{E}$ decreases as more advanced species develop and they become more numerous. The potential knowledge peaks at some time $n_{\max}$, after which the entropy $_{f}\mathcal{E}$ increases again, as the universe approaches thermodynamic equilibrium. The universe comes to an end as the last subject closes her eyes at time $n_{final}$. The Big Bang is regarded to be nothing more than an extrapolation backwards in time, using the inverse evolution operator. In terms of the relational time $t$, it has to be positioned before the time $t_{1}$ that corresponds to $n=1$.}
\label{Fig124}
\end{figure}

These considerations mean that there ought to be a time $n_{\max}$ at which the potential knowledge $PK$ peaks, and the fuzzy entropy reaches a minimum. A possible evolution of the fuzzy entropy between the initial time $n=1$ and $n=n_{final}$ is shown in Fig. \ref{Fig124}.

\begin{hypo}[\textbf{There is a time at which the potential knowledge peaks}]
If aware life persist for sufficiently long time, so that $n_{final}$ is large enough, there will be a time $1<n_{\max}< n_{final}$ such that $_{f}\mathcal{E}(n_{\max})$ is at a global minimum.
\label{maxknowledge}
\end{hypo}

Suppose that we are living at some time $n_{r}$ before time $n_{\max}$. We look at the skies and try to retrodict the history of the cosmos. What we do is to apply the inverse evolution operator $u_{1}^{-1}$ repeatedly. The state space volume of each possible retrodicted state $u_{m}^{-1}S(n_{r})$ is expected to decrease steadily with $m$. In fact, according to Eq. [\ref{expgrowth}]

\begin{equation}
\langle V[u_{m}^{-1}S(n_{r})]\rangle_{S}\propto V[S(n_{r})]e^{-gm}.
\label{expshrink}
\end{equation}
This means that the fuzzy entropy also shrinks towards one as we let $m$ increase. We get the usual conclusion that entropy becomes extremely small as we approach Big Bang (Fig. \ref{Fig124}). One attribute value that pops out of the application of $u_{m}^{-1}$ is the expected relational time: We may write $\langle t\rangle(m)$, where $\langle t\rangle$ decreases as $m$ increases.

It is perfectly allowed to let $m>n_{r}$, meaning that we retrodict the state of the universe to an epoch before the emergence of the first aware beings. According to the strict epistemic perspective adopted here, such states have no actual existence. The proper birth of the universe should be taken to be the sequential time $n=1$. Retrodiction towards the Big Bang becomes an exercise in the repeated application of $u_{1}^{-1}$ rather than a reconstruction of a real situation. The extremely small entropy associated with the initial state of the universe ceases to be a puzzle, as it can be regarded as a mirage, a mathematical consequence of Statement \ref{entropylawB}.

Let us elaborate a bit on this point. We base the concept of entropy on the present knowledge, which can be represented as a physical state. Then we extrapolate backwards in time using physical law. The conventional line of reasoning is the other way around. The initial condition of the universe is regarded as the primary physical state. Then we extrapolate forwards in time using physical law. The conclusion is that the initial condition must be extremely fine-tuned in order to produce our perceptions today. We have an apparent mystery.

What we have done, in effect, is to replace the mystery of the extremely fine-tuned initial state with the mystery of our present existence. The primary wonder is that we are aware, and that we can differentiate between well-defined objects. We may extrapolate this wonder backwards in time using the inverse physical evolution $u_{1}^{-1}$, getting an extremely fine-tuned initial condition that enables such an existence, but this is a derived, secondary wonder. It is a consequence, not a cause.

From our strict epistemic perspectie we have claimed that the proper birth of the universe is the sequential time $n=1$ when the first awareness emerges. The gradual decrease of the fuzzy entropy from that time can be described as a process in which the universe becomes more and more aware of itself, learning like a child. In fact, since the size of physical states and the size of the state of potential knowledge are reciprocal quantities (Fig. \ref{Fig25}), it is natural to define $_{f}\mathcal{E}^{-1}$ as a measure of the amount of potential knowledge. Figure \ref{Fig124} then expresses a cosmological model in which potential knowledge grows from zero, peaks at time $n_{\max}$ an then falls back towards zero.

\begin{figure}[tp]
\begin{center}
\includegraphics[width=80mm,clip=true]{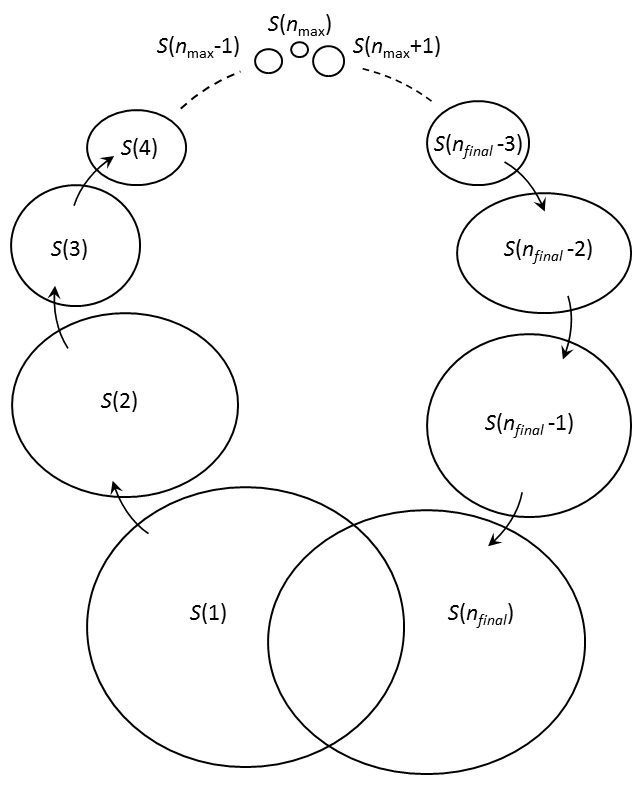}
\end{center}
\caption{A cosmological model in which the initial and the final physical states overlap. This would mean that the end becomes indistinguishable from the beginning.}
\label{Fig129}
\end{figure}

Since both the initial state $S(1)$ and the final sate $S(n_{final})$ ought to have a very large volume, and presumably covers a substantial part of state space $\mathcal{S}$, one may ask if they overlap, so that

\begin{equation}
S(1)\cap S(n_{final})\neq \varnothing.
\label{recurrence}
\end{equation}
If this is indeed the case we get a cosmological model like that in Fig. \ref{Fig129}. It is tempting to call such a model cyclical, but this is not really the case. The initial state $S(1)$ must correspond to a state of knowledge $PK(1)$ without memories, without objects having presentness attribute $Pr=0$ (Definition \ref{presentness}). If $S(1)$ and $S(n_{final})$ do overlap, $PK(n_{final})$ must also be a state of knowledge without memories. We get a picture of a universe that loses its memory like an old person. Now, to say that we are dealing with a cyclical cosmology, there has to be a way to distinguish the different cycles from each other and to order them. This is not possible in the present model because of the lack of memory. The model rather corresponds to a breakdown of time at the beginning and at the end, which makes the meanings of the words `beginning' and `end' coincide. In this interpretation, Eq. \ref{recurrence} does not contradict Statement \ref{norecurrence}. 

\vspace{5mm}
\begin{center}
$\maltese$
\end{center}
\paragraph{}

So far, we have discussed the entropy of physical states $S$ and object states $S_{O}$ very generally, without referring to any experimental context $C$, or to any specific mathematical representations $\bar{S}$ and $\bar{S}_{O}$ of these states. Let us check that the conceptual framework can be properly applied in those circumstances.

Consider a context $C$ in which we have a complete set $\{\tilde{S}_{j}\}$ of future alternatives defined at the intial time $n$, corresponding to a set $\{p_{j}\}$ of possible values of property $P$. Let the observation of $P$ in this context define the temporal update $n+m-1\rightarrow n+m$. We may write

\begin{equation}
u_{1}S_{O}(n+m-1)=\bigcup_{j}u_{1}S_{j}=\bigcup_{j}u_{1}S_{O}(n+m-1)\cap\mathcal{P}_{j},
\end{equation}
where $\{S_{j}\}$ is the corresponding set of present alternatives defined at time $n+m-1$ according to Definition \ref{presentalt}. Consequently,

\begin{equation}
S_{O}(n+m)=u_{1}S_{j}=u_{1}S_{O}(n+m-1)\cap\mathcal{P}_{j}
\end{equation}
for some $j$.

To the observation of property $P$ of object $O$ at time $n+m$ we can associate a change $\Delta\mathcal{E}_{O}(P)$ of the object entropy of $O$:
\begin{equation}
\Delta\mathcal{E}_{O}(P)\equiv \log(V_{O}[S_{O}(n+m)])-\log(V_{O}[u_{1}S_{O}(n+m-1)]).
\end{equation}
The probability $q_{j}$ to observe the value $p_{j}$ can be expressed as

\begin{equation}
q_{j}=v[\tilde{S}_{j},S_{O}(n)]=v[u_{1}S_{j},u_{1}S_{O}(n+m-1)]=\frac{V_{O}[u_{1}S_{j}]}{V_{O}[u_{1}S_{O}(n+m-1)]},
\end{equation}
so that $\Delta\mathcal{E}_{O}(P)=\log(q_{j})$ (Section \ref{probabilities}). The expected entropy change of object $O$ associated with the observation of $P$ becomes

\begin{equation}
\langle\Delta\mathcal{E}_{O}(P)\rangle=\sum_{j}q_{j}\log(q_{j}),
\end{equation}
so that it equals the Shannon entropy $H$ with a minus sign:

\begin{equation}
\langle\Delta\mathcal{E}_{O}(P)\rangle=-H(P).
\end{equation}
In other words, the expected entropy decrease of object $O$ equals the expected information gain associated with the observation of a discrete random variable $P$, as conventionally defined.

The Shannon entropy can be used to express the difference between reversible and irreversible evolution in a more concise way than in Definition \ref{reversible}. Suppose that the state $S_{O}$ of object $O$ evolves with an average growth factor $e^{g}>1$, that is, $\langle V_{O}[u_{1}S_{O}]\rangle=e^{g}\langle V_{O}[S_{O}]\rangle$. Consider contexts $C$ which are initiated at time $n$ and in which $O$ is not observed again until time $n+m$. Then the evolution is reversible if and only if we, for any $m\geq 1$, can construct such a context $C$ for which

\begin{equation}
mg\lesssim H(P),
\end{equation}
where $P$ is the property of $O$ that we observe within context at time $n+m$. In short, we should be able to regain at least as much information about object $O$ as we lost when we did not keep track of it. 

\begin{figure}[tp]
\begin{center}
\includegraphics[width=80mm,clip=true]{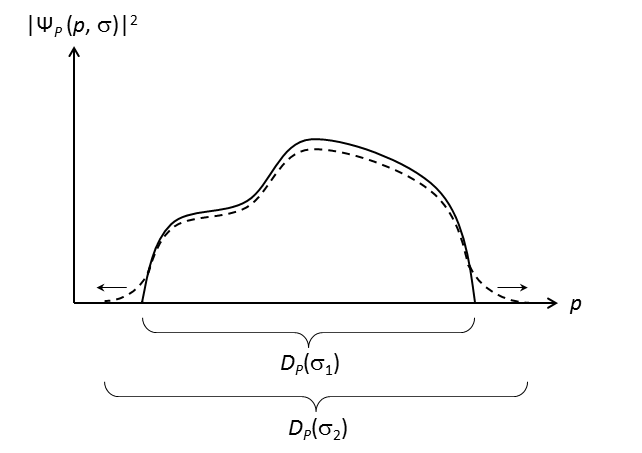}
\end{center}
\caption{A proper evolution equation for $\Psi_{P}(p,\sigma)$ is such that the width of the support $D_{P}(\sigma)$ typically becomes wider as $\sigma$ increases. Ordinary evolution equations like Eq. [\ref{freeeveq}] cannot be tested in this way since they cannot be applied at a sharp boundary $\partial D_{P}(\sigma)$ where the second derivative $\partial^{2}\Psi_{P}(p,\sigma)/\partial p^{2}$ is undefined. We have to use a smoothed wave function and calculate the evolution of the standard deviation $\Delta p(\sigma)$ as an indicator of the true behavior. Compare Fig. \ref{Fig75}.}
\label{Fig126}
\end{figure}

\begin{figure}[tp]
\begin{center}
\includegraphics[width=80mm,clip=true]{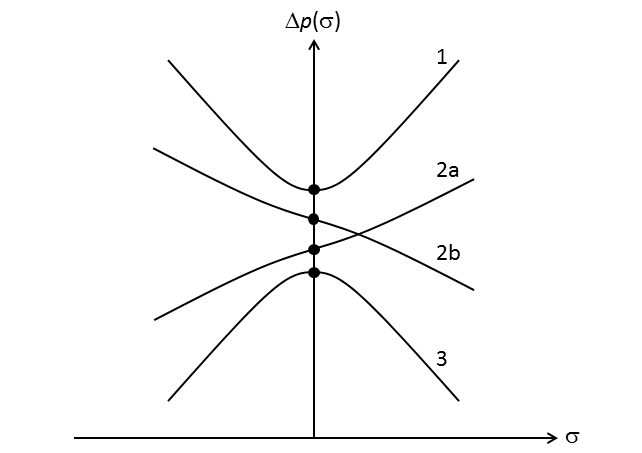}
\end{center}
\caption{A typical choice of wave function $\Psi(p,\sigma)$ at $\sigma=0$ evolves so that its standard deviation $\Delta p(\sigma)$ increases regardless whether we evolve it forwards or backwards with respect to $\sigma$ (evolution type 1). Any deviation from this behavior is expected to occur with zero probability if we pick $\Psi(p,0)$ randomly from a function space with a prescribed value of $\Delta p(0)$. The evolution types 2a and 2b are equally (im)probable because of the invariance under $\sigma\rightarrow -\sigma$, whereas evolution type 3 is even more improbable. More precisely, given that $d\Delta p/d\sigma<0$ for $\sigma>0$, we argue that the probability is zero that $d\Delta p/d\sigma>0$ for $\sigma<0$, and given that $d\Delta p/d\sigma>0$ for $\sigma<0$, we argue that the probability is zero that $d\Delta p/d\sigma<0$ for $\sigma>0$. }
\label{Fig127}
\end{figure}

Let us next consider a family $C(\sigma)$ of contexts in which $P$ is observed in the same manner. Then we can define a wave function $a_{P}(p_{j},\sigma)$ according to Definition \ref{wavedef}. Since we expect that the evolution operator expands volumes in the object state space (Statement \ref{entropylawB}), we expect that the support $D_{P}(\sigma)$ of the wave function expands when the evolution parameter $\sigma$ grows. Any correct evolution equation $da_{P}/d\sigma=\bar{A}a_{P}$ should reproduce this behavior (Fig \ref{Fig126}).

Equation [\ref{freeeveq}] gives the evolution of a free specimen in a fundamental context in which the spatio-temporal position $\mathbf{r}_{4}$ is observed. The problem with this equation is that it cannot be used to predict the evolution of a finite support $D_{\mathbf{r}_{4}}(\sigma)$, or, equivalently, of the boundary $\partial D_{\mathbf{r}_{4}}(\sigma)$. It relies on the existence of the second derivative $\partial^{2}\Psi_{\mathbf{r}_{4}}/\partial r_{k}^{2}$ for $k=1,2,3,4$, whereas these derivatives are not defined at a finite boundary $\partial D_{\mathbf{r}_{4}}$.

This is not a fundamental problem, since it is impossible in principle to arrange fundamental contexts in which Eq. [\ref{freeeveq}] holds exactly. The use of continuous wave functions is always an approximation, as discussed in section \ref{wavef}. Therefore the use of evolution equations for continuous wave functions is always an approximation, too. We are therefore allowed to soften the sharp boundary $\partial D_{\mathbf{r}_{4}}(\sigma)$, allowing a small but nonzero value of $|\Psi_{\mathbf{r}_{4}}(\mathbf{r}_{4},\sigma)|$ for $\mathbf{r}_{4}\notin D_{\mathbf{r}_{4}}(\sigma)$. Then we can use the evolution of the standard deviation

\begin{equation}
\Delta\mathbf{r}_{4}(\sigma)\equiv\sqrt{\langle|\Psi_{\mathbf{r}_{4}}(\mathbf{r}_{4},\sigma)|^{2}\rangle-\langle|\Psi_{\mathbf{r}_{4}}(\mathbf{r}_{4},\sigma)|\rangle^{2}}
\end{equation}
as an \emph{indicator} of the evolution of the size of $\partial D_{\mathbf{r}_{4}}(\sigma)$. It is well known that $\Delta\mathbf{r}_{4}$ typically evolves as

\begin{equation}
\Delta\mathbf{r}_{4}(\sigma)\propto\sqrt{1+\alpha\sigma^{2}},
\label{widthevolution}
\end{equation}
indicating that the support $D_{\mathbf{r}_{4}}(\sigma)$ is expected to expand, as required.

The problem to determine the evolution of the boundary $D_{\mathbf{r}_{4}}(\sigma)$ is similar to the problem to determine the Heisenberg uncertainty relations, as discussed in section \ref{evconsequences}. From our epistemic perspective, we define such a relation as a minimum area of a projection of the object state on a two-dimensional subset of object state space, a subset spanned by two properties that are not simultaneously knowable (Statement \ref{generaluncertainty}). However, to derive these relations we tend to rely on continuous wave functions that describe idealized, fundamental contexts. (Compare Figs. \ref{Fig76b} and \ref{Fig76c}.) In that way we get inequalities relating the product of two standard deviations, like $\Delta x\Delta p_{x}\geq\hbar/2$. These inequalitites should be seen as nothing more than indicators of the true relations, just like the evolution of $\Delta\mathbf{r}_{4}(\sigma)$ expressed in Eq. [\ref{widthevolution}] is just an indicator of the evolution of the true size of $D_{\mathbf{r}_{4}}(\sigma)$.

The basic reason why $\Delta p$ typically increases with $\sigma$ is that there are less complex-valued functions $\Psi(p,\sigma)$ that are well localized than those that are poorly localized. This statement can be understood in terms of the Fourier expansion

\begin{equation}
\Psi(p,\sigma)=(2\pi)^{-1/2}\int_{-\infty}^{\infty}\tilde{\Psi}(\tilde{p})e^{i(\tilde{p}p+\tilde{\sigma}\sigma)}d\tilde{p}
\label{freefourier}
\end{equation}
of the wave function $\Psi(p,\sigma)$ of an object that has a precisely known rest mass, which is a constant of motion. If you pick at random a Fourier transform $\tilde{\Psi}(\tilde{p})$ that gives rise to a wave function $\Psi(p,0)$ with a given standard deviation $\Delta p(0)$, then we will almost certainly have $d\Delta p/d\sigma>0$ for $\sigma>0$. This is so since we have to fine-tune the phases $\tilde{p}p$ in the Fourier expansion in order to keep $\Psi(p,0)$ localized. This fine-tuning is almost certainly gradually lost as $\sigma$ grows from zero and the phase becomes $\tilde{p}p+\tilde{\sigma}\sigma$.

Three notes are in order here.

First, there are exceptions. We may happen to pick a transform $\tilde{\Psi}(\tilde{p})$ that makes the wave function become more and more localized as $\sigma$ grows, but this happens with zero probability. The spontaneous localization may happen in two ways (Fig. \ref{Fig127}). Either $\Psi(p)$ become more localized also as $\sigma$ decreases to negative values from zero (evolution type 3), or it becomes less localized during such a reverse evolution (evolution type 2b).

Second, the reasoning is invariant with respect to a reflection $\sigma\rightarrow -\sigma$ of the evolution parameter. This invariance is captured in Eq. \ref{widthevolution}. More generally, we typically have $d\Delta p/d\sigma>0$ for $\sigma>0$, but $d\Delta p/d\sigma<0$ for $\sigma<0$. This does not contradict the fact that there are choices of $\tilde{\Psi}(\tilde{p})$ for which $d\Delta p/d\sigma>0$ for $\sigma>0$ and $d\Delta p/d\sigma>0$ for $\sigma<0$ also. The probability to pick such a transform by chance is zero, but they arise naturally as the evolution of a wave function $\Psi'(p)$ that has a deviation $\Delta p'<\Delta p(0)$, since we typically have $d\Delta p/d\sigma>0$, as discussed above. In this case we may write $\tilde{\Psi}(p,0)=\tilde{\Psi}'(\tilde{p})e^{i\tilde{\sigma}\sigma'}$ for some $\sigma'>0$. The invariance with respect to the reflection $\sigma\rightarrow -\sigma$ is expressed by the fact that to each such choice of $\tilde{\Psi}(\tilde{p})$ there corresponds one choice for which $d\Delta p/d\sigma<0$ for $\sigma>0$ and $d\Delta p/d\sigma<0$ for $\sigma<0$. This means that it is equally (im)probable to pick a $\tilde{\Psi}(\tilde{p})$ that gives rise to evolution of the types 2a and 2b shown in Fig. \ref{Fig127}.

Third, the wave function [\ref{freefourier}] is quasi-periodic, so that it returns arbitrarily close to its original shape if we let $\sigma$ grow sufficiently large. This means we must finally enter a region of values of $\sigma$ in which $\Delta p(\sigma)$ shrinks. This situation typically happens after such a long time that it has no practical consequences; in any realistic context family $C(\sigma)$ the maximum evolution parameter $\sigma_{\max}$ for which the context is defined is much smaller than the recurrence value $\sigma_{\mathrm{recurrence}}$.

Several aspects of this discussion about the evolution of wave functions are similar to commonly discussed aspects of the evolution of entropy. To see the similarities, we regard the width $\Delta p$ of the wave function and the entropy as analogous quantitites. To derive the second law of thermodynamics, the \emph{typicality} of the present state of the world is often assumed. That is, given the macroscopic variables that specify our actual knowledge of the state of the world, it is assumed that the microscopic state is typical given these known constraints. More precisely, it is assumed that the microscopic state, which determines the evolution, is picked at random from the subset of phase space which is consistent with the macroscopic constraints. Then the probability that entropy increases with time is essentially one, since there are many more microscopic states that evolve into macroscopic states with larger entropy than the present macroscopic state, than there are microscopic states that evolve into macroscopic states with equal or smaller entropy. The problem with this argument is that it can be applied backwards in time as well as forwards. A typical present microscopic state which is evolved backwards in time by a physical law that is invariant under time reflections leads to an entropy that increases in both temporal directions, just like the wave function evolution of type 1 in Fig. \ref{Fig127}. In that case, the typical choice of a microscopic state in phase space that conforms with the macroscopic knowledge corresponds to a typical choice of wave function in function space that conforms with the given width $\Delta p(0)$.

In the present approach to entropy, we avoid these difficulties. By construction, physical law is not invariant under time reflection (Statement \ref{timeasymmetry}). To change sign of the evolution parameter $\sigma$ does not mean that we change the direction of time. It just means that we change the parametrization of physical law. The evolution equation [\ref{freeeveq}] is invariant under time reflection in the same way as the Schr\"odinger equation with a real Hamiltonian: if we let $\sigma'\equiv-\sigma$ then

\begin{equation}
\frac{d}{d\sigma'}\Psi_{\mathbf{r}_{4}}'(\mathbf{r}_{4},\sigma)=\frac{ic^{2}\hbar}{2\langle E\rangle}\Box\Psi_{\mathbf{r}_{4}}'(\mathbf{r}_{4},\sigma).
\label{freeeveqprime}
\end{equation}
with $\Psi'=\Psi^{*}$. We start with an equation expressed in the natural parametrization $d\langle t\rangle/d\sigma=1$, and we get back the same equation expressed in the anti-natural parametrization $d\langle t\rangle/d\sigma=-1$ (see Definitions \ref{natparadef} and \ref{antinatparadef}). This is an expression of the fact that physical law is parametrization independent; it is not a statement about physical law itself.

It simply has no meaning to evolve a wave function backwards in time to make retrodictions about the past. A wave function is constructed from a known present state of a specimen, together with a complete set of alternatives that apply to this specimen, one of which is known to be realized at some time in the future. It does not make sense to have a complete set of alternatives for the past. All that can be said about the past is already known. An alternative represents something outside potential knowledge that may become known. That something `may become known' can be regarded, by definition, as a statement of a possible \emph{future} event.

The realization of an alternative corresponds to a state reduction. Basically, it is the existence of these state reductions that makes physical law asymmetric with respect to time reflections. This is neatly illustrated in Fig. \ref{Fig120}. A reversal of the direction of sequential time $n$ makes the evolution of the state space volume $V[S_{O}]$ look fundamentally different.

All of this means that we do not have to bother about the fact that the typical choice of wave function become wider both as $\sigma$ grows from zero and as it decreases from zero, as indicated in Fig. \ref{Fig127}. In this figure, $\sigma=0$ is taken to correspond to a \emph{given} width $\Delta p$. A given width means a \emph{known} width, so that it corresponds to the \emph{present} state of the specimen. (Recall that the present physical state is nothing more than a representation of the potential knowledge.) Since the role of $\sigma$ is to interpolate the evolution from the present time $n$ to the next time $n+1$, we should only follow $\Delta p(\sigma)$ in one direction from $\sigma=0$. Which one is a matter of the choice of parametrization. In either case we do obtain the correct behavior: a free specimen typically increases its entropy as it evolves, when this evolution is modelled by Eq. [\ref{freeeveq}].

\section{Expansion of state space}
\label{expansion}

Equation [\ref{expgrowth}] asserts that the repeated application of the evolution operator $u_{1}$ typically causes the state space volume of a state $S$ to increase exponentially. This conclusion can be translated to a statement about the expansion of the state space $\mathcal{S}$ itself. The evolved state $u_{1}S(n)$ defines what elbow room must be present at time $n+1$ in the state space in which $S$ lives, since any subset of $u_{1}S(n)$ can potentially be a part of $S(n+1)$. This follows from the minimality condition used to define $u_{1}$ (Definition \ref{evolutionu1}).

Suppose that we partition the state space into a set of states

\begin{equation}
\hat{\mathcal{S}}\equiv\{\hat{S}_{k}\}.
\label{hatspace}
\end{equation}
That is, we cover $\mathcal{S}$ with sets $\hat{S}_{k}$ such that $\hat{S}_{k}\cap\hat{S}_{k'}=\varnothing$ whenever $k\neq k'$, such that

\begin{equation}
\bigcup_{k}\hat{S}_{k}=\mathcal{S},
\label{cover}
\end{equation}
and such that the evolution operator is defined for each covering set (assuming that this is possible). This means that the set $u_{1}\hat{S}_{k}\in\mathcal{PS}$ is uniquely defined by physical law for each $k$.

It follows that we can define the evolved set of covering states

\begin{equation}
u_{1}\hat{\mathcal{S}}\equiv\{u_{1}\hat{S}_{k}\}.
\label{evhatspace}
\end{equation}

We used the invertibility of physical law (Assumption \ref{uniqueu1}) to conclude that the volume of a physical state typically grows exponentially (Eq. [\ref{expgrowth}]). We can use it again to conclude that we must have $u_{1}\hat{S}_{k}\cap u_{1}\hat{S}_{k'}=\varnothing$ whenever $k\neq k'$. We must also have

\begin{equation}
\bigcup_{k} u_{1}\hat{S}_{k}=\mathcal{S}.
\label{evcover}
\end{equation}

The last statement is an apparent contradiction, since we ought to be able to write

\begin{equation}
V[\mathcal{S}]=V[\bigcup_{k} \hat{S}_{k}]=\sum_{k} V[\hat{S}_{k}]
\end{equation} 
because of Eq. [\ref{cover}], as well as

\begin{equation}
V[\mathcal{S}]=V[\bigcup_{k} u_{1}\hat{S}_{k}]=\sum_{k} V[u_{1}\hat{S}_{k}]=e^{g} \sum_{k} V[\hat{S}_{k}]
\end{equation} 
for some growth factor $e^{g}>1$, because of Eq. [\ref{evcover}]. We seemingly get

\begin{equation}
V[\mathcal{S}]=e^{g}V[\mathcal{S}]>V[\mathcal{S}].
\end{equation}

\begin{figure}[tp]
\begin{center}
\includegraphics[width=80mm,clip=true]{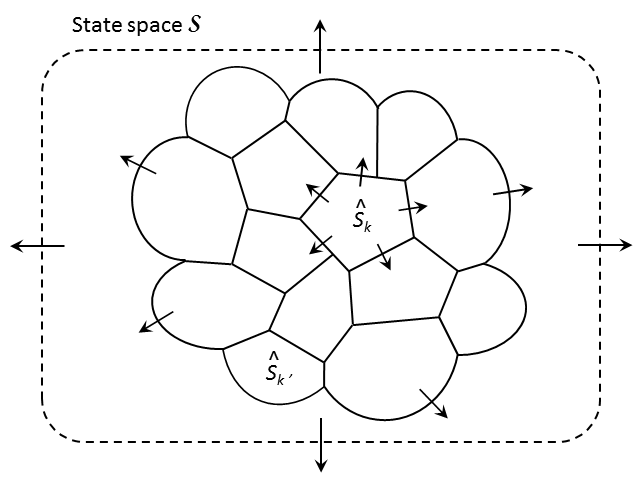}
\end{center}
\caption{We may cover state space $\mathcal{S}$ with states $\hat{S}_{k}$, like cells in a honeycomb. The volume of each such cell is expected to grow when the evolution operator $u_{1}$ is applied. Therefore we may say that the entire state space grows with a factor $e^{g}$, defined as the average of the growth factor of each cell.}
\label{Fig130}
\end{figure}

The paradox can be resolved simply by saying that the volume of the entire state space $\mathcal{S}$ is not defined. The measure $V$ can be given meaning only as a means to compare the size of two physical states $S$ and $S'$, or, equivalently, the amount of knowledge contained in two states of potential knowledge $PK$ and $PK'$. It is true that we have defined an absolute scale for $V$ by the assignment $V[Z]=1$ for all exact states $Z$, but this expression has a formal meaning only. Since physical states cannot be exact we are unable to count the number of exact states contained in a physical state, or in the entire state space. Since we can use $V$ to compare two states $S$ and $S'$, one might argue that we should be able to use it to compare two state spaces $\mathcal{S}$ and $\mathcal{S}'$. But the physical state $S$ is the only tool we have in our hands to explore the world. To say something about the state space itself, we have to use a state $S=\mathcal{S}$. Such a state corresponds to no knowledge at all. A comparison of the size of two state spaces is therefore a comparison of two zeros, of two `states' of nothingness. An equivalent statement would be that the evolution operator $u_{1}$ cannot be applied to $\mathcal{S}$. The state space is not in the domain of $u_{1}$, just as the exact states are not in this domain.

In contrast, we \emph{can} define the evolution of $\hat{S}_{k}$, as expressed in Eq. [\ref{evhatspace}]. Since the choice of covering states is arbitrary, this set can be used to define the evolution of the size of state space, in a restricted sense. We may define a growth factor $e^{g}[\hat{\mathcal{S}}]$ as follows.

\begin{equation}
e^{g}[\hat{\mathcal{S}}]\equiv\frac{V[u_{1}\hat{\mathcal{S}}]}{V[\hat{\mathcal{S}}]}\equiv\left\langle\frac{V[u_{1}\hat{S}_{k}]}{V[\hat{S}_{k}]}\right\rangle_{k}.
\end{equation}

\begin{state}[\textbf{The size of state space increases with time}]
We define $G$ by $e^{G}\equiv\langle e^{g}[\hat{\mathcal{S}}]\rangle_{\hat{\mathcal{S}}}>1$, where $\langle\ldots \rangle_{\hat{\mathcal{S}}}$ is an average over all possible state space coverings $\hat{\mathcal{S}}$ according to Eq. [\ref{hatspace}]. Since $G$ is not a functional of the covering, we may write $Size[u_{1}\mathcal{S}]=e^{G}Size[\mathcal{S}]$.
\label{sgrows}
\end{state}

The statement is illustrated in Fig. \ref{Fig130}. We may, at least formally, say that the size of state space increases exponentially when the evolution operator is repeatedly applied:

\begin{equation}
Size[u_{m}\mathcal{S}]=e^{Gm}Size[\mathcal{S}]
\end{equation}

\begin{figure}[tp]
\begin{center}
\includegraphics[width=80mm,clip=true]{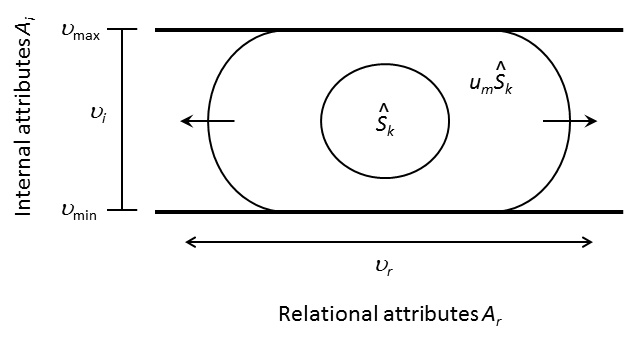}
\end{center}
\caption{The continued expansion of the covering states $\hat{S}_{k}$ in Fig. \ref{Fig130} must be effected by an increased uncertainty $\Delta \upsilon_{r}$ of the values of relational attributes $A_{r}$. The uncertainty of the values of the internal attributes $A_{i}$ that specify a minimal object is bounded by the allowed range $[\upsilon_{\min},\upsilon_{\max}]$ of these values. A continued expansion of the uncertainty $\Delta\mathbf{r}_{4}$ of spatio-temporal distance $\mathbf{r}_{4}$ as $u_{1}$ is repeatedly applied means that space-time itself expands indefinitely.}
\label{Fig131}
\end{figure}

To say that the size of state space increases means the same as to say that the uncertainty $\Delta \upsilon$ of the values $\upsilon$ of some attributes $A$ increases. To make this possible, the range $[\upsilon_{\min},\upsilon_{\max}]$ of possible values of these attributes must become wider, since we must have $[\upsilon_{\min},\upsilon_{\max}]\geq \Delta \upsilon$. However, not all attribute ranges can expand in this way.

We have assumed that all objects can be represented as a composition of minimal objects (Assumption \ref{finitedepth}), and we have concluded that the evolution can always be expressed as the evolution of a physical state expressed in terms of such minimal objects (Statement \ref{evred}). To make such a representation meaningful, the minimal objects have to be identifiable (Statement \ref{allminimalidentity}); otherwise it would not make any sense to speak about the evolution of these minimal objects. Such identifiability relies upon the existence of fixed, discrete sets of possible values of the internal attributes that specify the minimal objects, as discussed in Section \ref{divideconserve}. Therefore, the ranges $[\upsilon_{\min},\upsilon_{\max}]$ of the possible values $\upsilon_{i}$ cannot change as we apply $u_{1}$. This means that set of possible minimal objects stays fixed, as does their respective rest masses. Also, The Dirac equation holds generally, so that the range of possible spin projection values $[-1/2,1/2]$ must stay fixed.

Even if the range of possible values of the internal attributes (and the spin projection) of a minimal object is fixed, the number $N$ of minimal objects can vary, and also the uncertainty $\Delta N$ of this number. The physical state $S$ is supposed to describe the entire world, and therefore we can never know the total value $\upsilon_{i}(S)$ of an internal attribute $A_{i}$ that describes a minimal object. By the `total value' of $A_{i}$ we mean the sum the values of this attribute of all the $N$ minimal objects in the world.

Suppose first that we know that the universe is spatially closed, having a finite volume. Let $\Upsilon_{i}(S)$ be the set of possible total values of $\upsilon_{i}(S)$ that are not excluded by the potential knowledge encoded in such a physical state $S$. Then, for each $A_{i}$, there is a pair of numbers $(\upsilon_{i}^{\min},\upsilon_{i}^{\max})$ such that

\begin{equation}
\Upsilon_{i}(u_{m}S)\equiv[\upsilon_{i}^{\min}(u_{m}S),\upsilon_{i}^{\max}(u_{m}S)]\subseteq [\upsilon_{i}^{\min},\upsilon_{i}^{\max}]
\label{ivaluefint}
\end{equation}
for each $m\geq 0$ (defining $u_{0}\equiv I$). The finite range $[\upsilon_{i}^{\min},\upsilon_{i}^{\max}]$ of possible `total values' is a simple consequence the finite volume and the assumption of finite depth of knowledge (Assumption \ref{finitedepth1}), leading to a finite total number $N$ of minimal objects. It is clear that a hypothetical initial widening of $\Upsilon_{i}(u_{m}S)$ for small $m$ cannot contribute to the exponential expansion of state space in the long run because of the bound $\Upsilon_{i}(S)\in[\upsilon_{i}^{\min},\upsilon_{i}^{\max}]$ for all $S$.

Suppose next that the physical state $S$ is such that it is impossible in principle to exlude the possibility that $N=\infty$. This is true whenever it cannot be excluded that the universe is spatially infinite. For some internal attributes that can be both positive and negative, like electric charge, this means that we cannot exclude \emph{any} total value. We have

\begin{equation}
\Upsilon_{i}(S)=(-\infty,\infty)
\label{ivalueinf}
\end{equation}
for any $S$. Clearly, such a set cannot change volume under the application of $u_{1}$ and cannot contribute to the exponential expansion of state space.

We may also have internal attributes $A_{i}$ for which the sign of $\upsilon_{i}(S)$ is known, for example the baryon or lepton number (see Hypothesis \ref{lessantimatter}). Then it is of interest that we can set a lower bound $N^{\min}(S)$ of $N$. The size of this bound depends on the size of the visible universe, and is therefore a function of $S$. Supposing that the sign of $\upsilon_{i}(S)$ is positive, this means that we can associate to $S$ a set $\Upsilon_{i}(S)$ of possible total values of $A_{i}$ according to

\begin{equation}
\Upsilon_{i}(S)=[\upsilon_{i}^{\min}(S),\infty).
\label{ivalueint}
\end{equation}
It is possible that $\Upsilon_{i}(S)\neq \Upsilon_{i}(u_{m}S)$, and we may imagine that $\upsilon_{i}^{\min}(u_{m}S)$ decreases exponentially with $m$. However, such a process would not contribute to the exponential increase of $V[u_{m}S]$ in the long run, since we hit rock bottom when $\upsilon_{i}^{\min}(u_{m}S)=0$. The value of $m$ at which this happens may be very large, however. Let us therefore argue in more detail why a dependence of $\upsilon_{i}^{\min}(u_{m}S)$ on $m$ is irrelevant.

The only way to compare two state space volumes $V$ and $V'$ is to calculate their ratio $V/V'$; they cannot be calculated separately in an absolute sense. The ratio of the lengths of the intervals $[\upsilon_{i}^{\min}(S),\infty)$ and $[\upsilon_{i}^{\min}(u_{m}S),\infty)$ is one for all $m$ regardless the form of the function $\upsilon_{i}^{\min}(m)$. This is why a hypothetical change of $\upsilon_{i}^{\min}$ cannot give rise to an exponentially increasing volume $V[u_{m}S]$.

More formally, consider the attribute value space $S(A_{i},\upsilon_{i})$ according to Definition \ref{valuespacedef}. We have

\begin{equation}\begin{array}{rcl}
\frac{V[\bigcup_{\upsilon_{i}\in [\upsilon_{i}^{\min}(u_{m}S),\infty)}S(A_{i},\upsilon_{i})]}{V[\bigcup_{\upsilon_{i}\in [\upsilon_{i}^{\min}(S),\infty)}S(A_{i},\upsilon_{i})]} & = & \frac{\sum_{\upsilon_{i}\in [\upsilon_{i}^{\min}(u_{m}S),\infty)}V[S(A_{i},\upsilon_{i})]}{\sum_{\upsilon_{i}\in [\upsilon_{i}^{\min}(S),\infty)}V[S(A_{i},\upsilon_{i})]}\\
& & \\
& = & \frac{\sum_{\upsilon_{i}\in [\upsilon_{i}^{\min}(u_{m}S),\upsilon_{i}^{\min}(S))}V[S(A_{i},\upsilon_{i})]}{\sum_{\upsilon_{i}\in [\upsilon_{i}^{\min}(u_{m}S),\infty)}V[S(A_{i},\upsilon_{i})]}+1\\
& & \\
& = & 1
\end{array}
\end{equation}
according to Definition \ref{voldef}.

To conclude, to account for the exponential state space volume increase, we are left with changes of the ranges of possible values of purely relational attributes (Fig. \ref{Fig131}), spin projections excluded. We argued in Section \ref{minimalism} that we should consider angles as a relational attribute that relates three objects $O_{1}$, $O_{2}$ and $O_{3}$, and that these angle should be treated as independent attributes in the spirit of the Riemannian geometry of general relativity. In other words, they are able to vary independently of the three distances $\mathbf{r}_{4}^{(12)},\mathbf{r}_{4}^{(13)},\mathbf{r}_{4}^{(23)}$ that relate $O_{1}$, $O_{2}$ and $O_{3}$. However, it follows from the definition of the angle that the range of possible values is always $[0,2\pi)$. It cannot be altered by the application of $u_{1}$.

The only attributes whose range of possible values \emph{can} expand are, as far as I understand, the four-position $\mathbf{r}_{4}$ and the four-momentum $\mathbf{p}_{4}$. Let us call the ranges of possible values of these attributes $\Delta\mathbf{r}_{4}$ and $\Delta\mathbf{p}_{4}$, respectively. These ranges correspond to uncertainties of the attribute values in question. $\Delta\mathbf{r}_{4}$ and $\Delta\mathbf{p}_{4}$ are tightly connected. It is easy to see that one of these ranges expand if and only if the other range expand. This means that both of them has to expand to account for the expansion of state space.

Let us focus on the spatio-temporal attributes. Instead of speaking about positions, which are `pseudo-internal' attributes according to the terminology of Section \ref{divideconserve}, we will speak about distances between two objects $O_{1}$ and $O_{2}$. By Lorentz invariance we see that the uncertainty $\Delta r_{12}$ of the spatial distance $r_{12}$ between the objects will expand exponentially under the repeated application of $u_{1}$ if and only if the uncertainty $\Delta t$ of the temporal distance $t_{12}$ expands exponentially. We concluded above that they do indeed expand in this way. However, the spatial distance $r_{12}$ and the temporal distance $t_{12}$ need not both behave in this way. Let us discuss each of them in turn. This separation means that we break Lorentz invariance of the discussion, and choose a specific reference frame, distances measured by a specific observer.

Consider first temporal distances $t_{12}$ between two objects $O_{1}$ and $O_{2}$ that both belong to the past at some time $n$. We say that they both have presentness attribute zero: $Pr[O_{1}]=Pr[O_{2}]=0$. Say that $O_{1}$ occurred first, and that this event is assigned the time $t[O_{1}]=0$ in a given reference frame. Then $t[O_{2}]=t_{12}>0$.

\begin{figure}[tp]
\begin{center}
\includegraphics[width=80mm,clip=true]{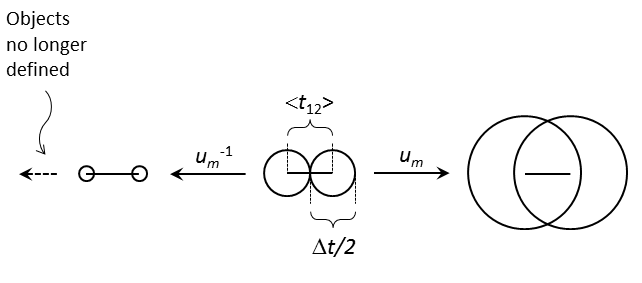}
\end{center}
\caption{Schematic illustration of the evolution of the temporal distance $t_{12}$ between two objects or events $O_{1}$ and $O_{2}$. The expected distance $\langle t_{12}\rangle$ cannot change, but the uncertainty $\Delta t_{12}$ must change according to the general exponential expansion of the spatio-temporal part of state space $\mathcal{S}$ under repeated application of $u_{1}$. For large enough $m$, the temporal ordering between the events $O_{1}$ and $O_{2}$ in the evolved state $u_{m}S$ becomes ambiguous. The inverse evolution $u_{m}^{-1}$ can only be applied for small enough $m$. If we exceed a given $m$, we retrodict a state before $O_{1}$ or $O_{2}$ occurred. Compare Fig. \ref{Fig133}.}
\label{Fig134}
\end{figure}

Clearly, $O_{1}$ and $O_{2}$ are pushed further and further back into history as $u_{1}$ is repeatedly applied to $S(n)$. In this process, $\langle t_{12}\rangle$ must stay fixed (Fig. \ref{Fig134}). This follows from the very idea of relational time. It does not make sense to define it so that the time difference between a pair of events that have already occurred increases with time. This time difference should characterize a relation between the two events, not the temporal distance from which we look at them.

This means that the ordering between the events may become ambiguous in the state $u_{m}S$ for large enough $m$, as indicated in Fig. \ref{Fig134}. Going in the opposite direction, the expected time difference $\langle t_{12}\rangle$ becomes more an more sharply defined when $m$ increases in the retrodicted state $u_{m}^{-1}S(n)$. This process comes to a halt when we reach a value of $m$ for which $O_{2}$ is no longer defined. We have reached a point in the retrodicted history before this event took place.

Let us philosophize a bit about our conclusions. Suppose that we would like to define relational time $t$ so that temporal distances expand under the application of the evolution operator. It would not be possible to give such an expansion epistemic meaning, since it would not be possible to measure it.

As discussed in Section \ref{statespaces}, to measure a distance essentially means to place a number of reference objects between the two objects $O_{1}$ and $O_{2}$ whose distance from each other we want to determine. The number of reference objects that can be fitted between $O_{1}$ and $O_{2}$ is the measured distance. In the case of temporal distances, this means to count the number of reference events that takes place between $O_{1}$ and $O_{2}$. These reference events may be heartbeats or ticks of an artificial clock. But a general expansion of the time scale under $u_{1}$ would mean that the temporal distance betwen \emph{all} events increase, including the reference events. However, the \emph{number} of reference events fitted between $O_{1}$ and $O_{2}$ would not change. Therefore the \emph{measured} distance would not change.

We may extend this reasoning to argue that it does not make sense to say that the temporal distance between successive sequential times $n$ and $n+1$ vary with $n$. Actually, $t(n+1)-t(n)$ cannot be measured at all, since, by definition, we can put no reference objects between the two events that follow imediately after each other. If we would like to, we can define a basic unit of relational time $t_{0}$ and say that

\begin{equation}
\langle t(n+1)-t(n)\rangle=t_{0}\ \ \mathrm{for}\;\mathrm{all}\;n,
\label{timeunit}
\end{equation}
and also

\begin{equation}
t(n+1)-t(n)\geq t_{\min}\ \ \mathrm{for}\;\mathrm{all}\;n.
\label{minitime}
\end{equation}

Note that such a basic unit of temporal distance has to be defined in a reference frame that is at rest with respect to the two events $O_{1}$ and $O_{2}$ that defines the temporal updates $n-1\rightarrow n$ and $n\rightarrow n+1$, respectively. If it is the same subject $k$ who experience both these events, the reference frame of interest is the rest frame of this observer (compare the discussions in relation to Figs. \ref{Fig96} and \ref{Fig97}). We have also concluded that we are free to choose a smallest observable time distance $t_{\min}$. The necessity for such a minimum distance reached from a slightly different point of view in Section \ref{boundstates}, in the form of Statement \ref{smallesttime}.

We can reformulate Eq. [\ref{timeunit}] to a statement about the uniform flow of time under the repeated application of the evolution operator. That is, we can without loss of generality define relational time $t$ so that the following claim holds true.

\begin{state}[\textbf{Time evolves uniformly}]
For each pair of physical states $S$ and $u_{m}S$ we have $\langle t[O_{m}]-t[O_{0}]\rangle=m t_{0}$ and $t[O_{m}]-t[O_{0}]\geq m t_{\min}$, where $O_{m}$ is a present object in $u_{m}S$, and $O_{0}$ is a present object in $S$.
\label{uniformtime}
\end{state}

Let us turn from the non-existent expansion of time to the hypothetical expansion of space. We do not run into the same problem of lack of a reference distance when it comes to measurement of the expansion of space. The reason is the existence of bound states. If these are insensitive to the expansion, they can be used as a fixed ruler to put in between the growing distance between unbound objects (compare Fig. \ref{Fig92}). This means that it may be possible to have an epistemically well-defined exponential expansion of space.

We consider now the spatial distance $r_{12}$ between pairs of objects $O_{1}$ and $O_{2}$ that are present at some reference time $n$, having presentness attribute $Pr[O_{1}]=Pr[O_{2}]=1$. Suppose that they are identifiable, so that we can track them as $u_{1}$ (or its inverse $u_{1}^{-1}$) is repeatedly applied. This means that $S_{O1}\cap u_{1}S_{O1}\neq\varnothing$ and $S_{O2}\cap u_{1}S_{O2}\neq\varnothing$ when the object states are represented in object state space $\mathcal{S}_{O}$ (Section \ref{statespaces}). Regardless how many times we apply $u_{1}$, we always consider the present version of the objects, so that we consider $u_{m}S_{O1}(n)$ and $u_{m}S_{O1}(n)$ when we speak about the distance between $O_{1}$ and $O_{2}$ in the evolved state $u_{m}S(n)$.

Let the expected distance between $O_{1}$ and $O_{2}$ in the state $S(n)$ be $\langle r_{12}\rangle(n)$ with uncertainty $\langle \Delta r_{12}\rangle(n)$. Then the evolved state $u_{1}S(n)$ must be such that it allows an uncertainty $e^{g}\langle \Delta r_{12}\rangle(n)$, for some growth factor $e^{g}$, which is expected to be greater than one.

\begin{figure}[tp]
\begin{center}
\includegraphics[width=80mm,clip=true]{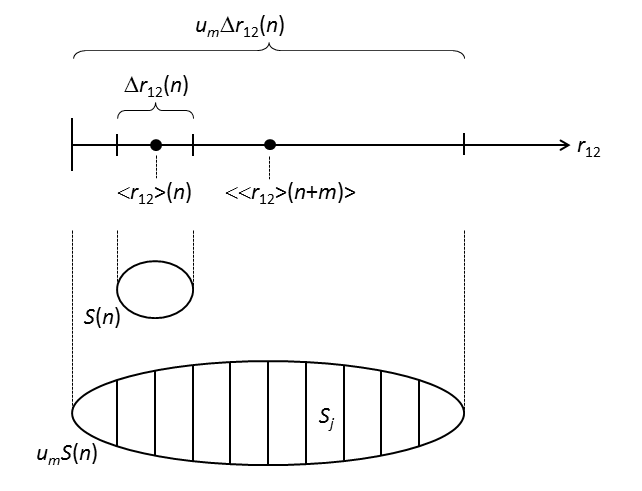}
\end{center}
\caption{When the uncertainty $\Delta r_{12}(n)$ of the distance $r_{12}(n)$ between two objcts $O_{1}$ and $O_{2}$ increases due to the evolution $u_{m}$, the central value $\langle r_{12}\rangle(n)$ of those allowed by the present physical state also has the potential to increase. This is expected to happen if we have no additional knowledge at time $n$ that tells us whether small or large values are preferred in an observation at time $n+m$ with alternatives $S_{j}$. The expected value of $\langle r_{12}\rangle(n+m)$ is denoted $\langle\langle r_{12}\rangle(n+m)\rangle$.}
\label{Fig132}
\end{figure}

For some $m\geq 1$ the evolved state $u_{m}S(n)$ will be such that

\begin{equation}
u_{m}\Delta r_{12}(n)=e^{gm}\Delta r_{12}(n)\gg \langle r_{12}\rangle(n).
\label{expdist}
\end{equation}
As noted above, this means that $u_{m}S(n)$ must allow for the elbow room to have

\begin{equation}
\langle r_{12}\rangle(n+m)\approx e^{gm}\langle r_{12}\rangle(n).
\end{equation}
It does not mean, however, that the expected distance between the object \emph{has to be} increasing exponentially in this way. This will be the case only if there is no additional potential knowledge at time $n$ which tells us which of all the possible distances within the range of uncertainty are more likely. In other words, it will be the case only if all alternative distances have a symmetric probability distribution, upon observation of $\langle r_{12}\rangle$ at time $n+m$ in a context $C$ (Fig. \ref{Fig132}). In a fundamental context this would mean that the square modulus $|\Psi_{r}(r_{12})|^{2}$ of the wave function is symmetric with respect to the middle point of its support $D_{r_{12}}$.

Such an observation causes a state reduction that resets $\Delta r_{12}(n+m)$ to a smaller value. If $\langle r_{12}\rangle(n)$ is defined to be at the centre of the possible range $[r_{\min}(n),r_{\max}(n)]$ of distances at time $n$, then we may let $\langle\langle r_{12}\rangle(n+m)\rangle$ be the expected centre just before the observation at time $n+m$ of the possible range of distances just after the observation, given the probabilities specified by $|\Psi_{r}(r_{12})|^{2}$. The quantity may also be interpreted as the average value of $\langle r_{12}\rangle(n+m)$ after a large number of similar observations. We conclude that without additional knowledge encoded in the state $S(n)$ about the evolution of $r_{12}$ we have

\begin{equation}
\langle\langle r_{12}\rangle(n+m)\rangle=e^{gm}\langle r_{12}\rangle(n).
\label{expsep}
\end{equation}

Such an exponential expansion of distances can occur only if objects $O_{1}$ and $O_{2}$ are not in a bound state (Defintion \ref{boundspecimen}). If we have knowledge at time $n$ that they are indeed bound to each other, and also knowledge about the nature of the forces that keep them together, then nothing is expected to change between times $n$ and $n+m$. We have

\begin{equation}
\langle\langle r_{12}\rangle(n+m)\rangle=\langle r_{12}\rangle(n).
\label{constsep}
\end{equation}

If $O_{1}$ and $O_{2}$ are not bound to each other, but there are nevertheless known intercations or forces that attract them to each other, then $\langle\langle r_{12}\rangle(n+m)\rangle$ will increase with time, but the separation may be slower than exponential. In cosmology, this happens becuase of the gravitational pull exerted on each object by all the other objects. Still, since $O_{1}$ and $O_{2}$ are not bound to each other by assumption, they are expected to fly farther and farther apart, so that finally the graviational force becomes negligible. Then the exponential separation according to Eq. [\ref{expsep}] sets in.

\begin{state}[\textbf{Exponential expansion of space}]
If two objects $O_{1}$ and $O_{2}$ are in an unbound state, and if they are not attracted to or repelled from each other by any known force, then the expected distance between them increases exponentially with time. This situation occurs in unbound states after sufficiently long time.
\label{spaceexpand}
\end{state}

The statement starts with an "If". We have to assume that there are objects in unbound states, that two objects farther apart than a certain large observable distance are never bound to each other.

Suppose that all states are bound at some time $n$. For such states both $\langle r_{12}\rangle$ and $\langle\Delta r_{12}\rangle$ are stationary. In a situation described by the continuous evolution equation [\ref{psieveq}], such stationary states are specified by the Dirac equation [\ref{diracconstraint}] that do not depend on the evolution parameter $\sigma$ and have a fixed rest mass $m_{0}$. More generally, $\langle r_{12}\rangle$ and $\langle\Delta r_{12}\rangle$ are invariant under $u_{1}$. If all pairs of objects were in bound states, there would be no uncertainty $\Delta r_{12}$ that increases when $u_{1}$ is applied to $S$. Since this is necessary to account for the exponential expansion of state space, there would be no such expansion. The entire world could be described as a stationary, bound state, so that $u_{1}S=S$. This is forbidden by the definition of $u_{1}$, as illustrated in Fig. \ref{Fig29}. We conclude that there are indeed pairs of objects $O_{1}$ and $O_{2}$ at any time $n$ that are not bound to each other.

This means that at each time $n$ there are distances $r_{12}$ that are expected to increase with time according to Fig. \ref{Fig132}. We can remove the "If" in Statement \ref{spaceexpand} and conclude the following.

\begin{state}[\textbf{The cosmological constant is positive}]
There is a sequential time $n$ at which the expansion of space starts to accelerate. The acceleration becomes exponential for large enough $n$, meaning that $\langle r_{12}\rangle(n)\propto e^{gn}$ for some constant $g>0$, for large enough $n$, and for large enough distances $\langle r_{12}\rangle(n)$ between two identifiable objects $O_{1}$ and $O_{2}$ which have presentness attribute $Pr[O_{1}]=Pr[O_{2}]=1$ in the state $S(n)$.
\label{cosmoconst}
\end{state}

\begin{figure}[tp]
\begin{center}
\includegraphics[width=80mm,clip=true]{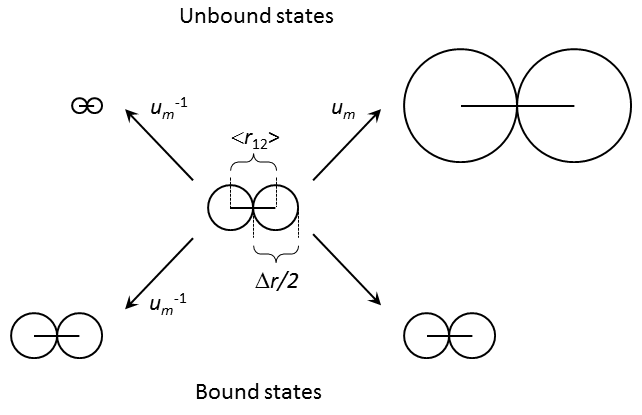}
\end{center}
\caption{Schematic illustration of the evolution of the expected spatial distance $r_{12}$ between two objects $O_{1}$ and $O_{2}$. In unbound states, both $\langle r_{12}\rangle$ and $\Delta r_{12}$ increase when $u_{1}$ is applied. Consequently, both these quantities decrease when the inverse evolution $u_{1}^{-1}$ is applied. In bound states, neither the expected distance $\langle r_{12}\rangle$ nor the uncertainty $\Delta r_{12}$ of this distance can change when $u_{1}$ is applied. Compare Fig. \ref{Fig134}.}
\label{Fig133}
\end{figure}

The above reasoning can be reversed. At each time $n$ there are distances $r_{12}$ and uncertainties $\Delta r_{12}$ between unbound states that are expected to decrease under the application of the inverse evolution operator $u_{1}^{-1}$. There may also be distances and uncertainties relating bound objects that stay constant. The evolution of $\langle t_{12}\rangle$ and $\Delta t_{12}$ was schematically illustrated in Fig. \ref{Fig134}. Figure \ref{Fig133} provides the analogous illustration of the evolution of  $\langle r_{12}\rangle$ and $\Delta r_{12}$.

The existence of distances $r_{12}$ in each state $S$ that decrease under the application of $u_{1}^{-1}$ means that there are distances $r_{12}$ between unbound, identifiable and present objects in the state $u_{m}^{-1}S$ such that

\begin{equation}\begin{array}{rcl}
\lim_{m\rightarrow\infty}u_{m}^{-1}\langle r_{12}\rangle & = & 0,\\
\lim_{m\rightarrow\infty}u_{m}^{-1}\Delta r_{12} & = & 0,
\end{array}
\label{zerodist}
\end{equation}
provided that the limit exists. To reach this conclusion, we must exclude the possibility that each distance $r_{12}$ converge in a decreasing sequence $(\langle r_{12}\rangle,u_{1}^{-1}\langle r_{12}\rangle,\ldots,u_{m}^{-1}\langle r_{12}\rangle,\ldots)$ to a non-zero value $\lim_{m\rightarrow\infty}u_{m}^{-1}\langle r_{12}\rangle=r_{12}^{(-\infty)}>0$. This is forbidden since it would correspond to the entire world being in a bound limit state $S^{(-\infty)}=\lim_{m\rightarrow\infty}u_{m}^{-1}S$ with $u_{1}S^{(-\infty)}=S^{(-\infty)}$. Such a fixed point state contradicts the rule $u_{1}S\cap S=\varnothing$. In plain language, that something evolves means that something changes.

Since distances between \emph{unbound} objects go to zero according to Eq. [\ref{zerodist}], we cannot have distances between \emph{bound} states that stay constant in this limit. All bound states have to be compressed and destroyed in this limit, if it exists.

Define

\begin{equation}
\Sigma_{BB}(S)\equiv\lim_{m\rightarrow\infty}u_{m}^{-1}S.
\label{singularlimit}
\end{equation}

\begin{figure}[tp]
\begin{center}
\includegraphics[width=80mm,clip=true]{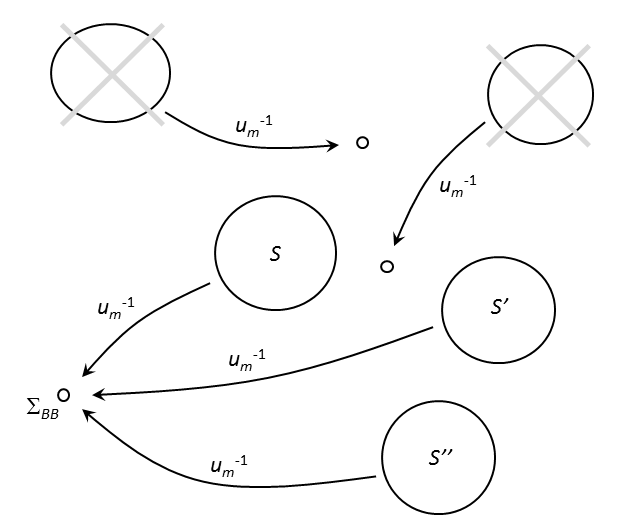}
\end{center}
\caption{The state space volume $V[S]$ decreases as the inverse evolution operator $u_{m}^{-1}$ is repeatedly applied. Many states $S,S',S'',\ldots$ have to converge towards the same limit set $\Sigma_{BB}$ in a trajectory through state space $\mathcal{S}$.}
\label{Fig136}
\end{figure}

Since $u_{1}$ is assumed to be invertible (Assumption \ref{uniqueu1}), it may seem natural to assume that to each sequence $(S,u_{1}^{-1}S,u_{2}^{-1}S,\ldots)$ corresponds exactly one set $\Sigma_{BB}$. However, $\Sigma_{BB}$ is a limiting set, and we cannot be sure that it belongs to the domain of $u_{1}$ (we will return to this matter below). Therefore, the condition of invertibility may not apply to the limit expressed in Eq. [\ref{singularlimit}], and the hypothetical one-to-one correspondence may be broken. In fact, this must be the case (Fig. \ref{Fig136}).

What attributes can be used to characterize $\Sigma_{BB}$? Can there be many such limit sets? Clearly they all fulfil Eq. [\ref{zerodist}], so that spatial distances play no role. Neither do temporal distances since $t\rightarrow -\infty$ according to Statement \ref{uniformtime}. What about momenta or velocities? Let $v_{12}$ be the relative velocity between two present, identifiable objects $O_{1}$ and $O_{2}$. Then we may write, in a notation that should be self-explanatory,

\begin{equation}
u_{m}^{-1}v_{12}\equiv\frac{\Delta r_{12}(m)}{\Delta t(m)}\equiv\frac{u_{m}^{-1}r_{12}-u_{m+1}^{-1}r_{12}}{u_{m}^{-1}t-u_{m+1}^{-1}t}\leq \frac{u_{m}^{-1}r_{12}-u_{m+1}^{-1}r_{12}}{\Delta t_{\min}},
\end{equation}
where the inequality follows from Eq. [\ref{minitime}]. Equation [\ref{zerodist}] then implies that
 
\begin{equation}\begin{array}{rcl}
\lim_{m\rightarrow\infty}u_{m}^{-1}\langle v_{12}\rangle & = & 0,\\
\lim_{m\rightarrow\infty}u_{m}^{-1}\Delta v_{12} & = & 0.
\end{array}
\label{zerovdist}
\end{equation}

Clearly, momenta can play no role either to distinguish different sets $\Sigma_{BB}(S)$ and $\Sigma_{BB}(S')$, since they are all zero, just like distances. This means that specific values of angular momenta and spin projections cannot be defined either. Therefore the sets $\Sigma_{BB}(S)$ can be a function only of the purely internal values of minimal objects.

Since spatial distances contract to zero as $m\rightarrow\infty$, all minimal objects must undergo object merging (Section \ref{objectmerging}) until there is only one object left for high enough values of $m$. This primordial ball will be characterized by the sets $\Upsilon_{i}$ of possible total values of the internal attributes $A_{i}$ that were discussed above (see Eqs. [\ref{ivaluefint}], [\ref{ivalueinf}], and [\ref{ivalueint}]). This is so beacuse of the additive conservation law for internal attribute values in object division and merging that is expressed in Eq. [\ref{additivity}]. We may write

\begin{equation}
\Sigma_{BB}(S)=\Sigma_{BB}(\{\Upsilon_{i}\}).
\end{equation}
This means that all states $S$ with given a given set $\{\Upsilon_{i}\}$ of allowed total values of the internal attibutes $A_{i}$ converge to the same limit set. Since there is a continuous infinity of possible values of the relational attributes that characterize a physical state $S$, we conclude that the basin of attraction $B(\Sigma_{BB}(\{\Upsilon_{i}\}))\subseteq\mathcal{PS}$ contains a continuous infinity of elements $S$.

We have postponed the discussion about the existence of the limit $\lim_{m\rightarrow\infty}u_{m}^{-1}S$. If we cannot make any sense of this limit, we cannot make any sense of $\Sigma_{BB}$. Recall that the evolution operator $u_{1}$ is not defined for exact states $Z$, but only for physical states $S$ that encode a state of potential knowledge $PK$ that can actually be realized by some group of subjects (Statement \ref{irreduciblelaw}). Already at this point we conclude that the limit cannot exist. There cannot be any aware observers in $\Sigma_{BB}$. One basic reason for this is that there has to be more than one object in any physical state that contains observers, since there has to be a correspondence between the observed objects, and the objects of the body of the observer that reacts to the observation (Section \ref{bwstates}).

Heisenberg's uncertainty relations (Statement \ref{generaluncertainty}) provide another reason why $\Sigma_{BB}$ cannot be considered to be a physical state to which we can apply $u_{1}$ or $u_{1}^{-1}$. Namely, Eqs. [\ref{zerodist}] and [\ref{zerovdist}] show that $\Delta r_{12}\Delta v_{12}=0$, contradicting the condition $\Delta x\Delta p_{x}\geq\hbar/2$ that must be fulfilled for any object $O$ with state $S_{O}$ that is part of $S$.

\begin{figure}[tp]
\begin{center}
\includegraphics[width=80mm,clip=true]{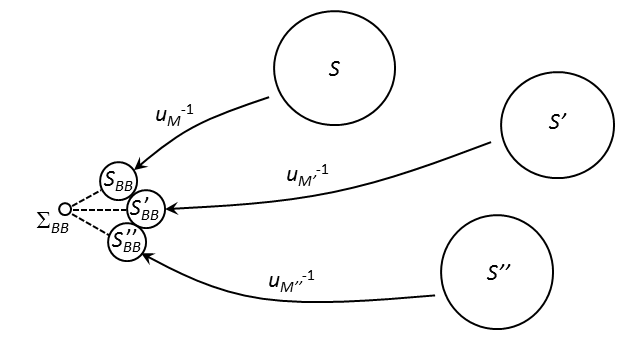}
\end{center}
\caption{A model of the Big Bang, as described in Statement \ref{bigbang}. The Big Bang states $S_{BB}(S)$ are the first physical states, reached after $M(S)$ applications of the inverse evolution operator $u_{1}^{-1}$ to the present state $S$. The singular Big Bang set $\Sigma_{BB}$ cannot be reached via physical law, but correponds to an abstract extrapolation.}
\label{Fig137}
\end{figure}

Instead, we must resort to the following picture. When we apply the inverse evolution operator $u_{1}^{-1}$ repeatedly to a state $S$, we finally reach a state $S_{BB}(S)=u_{1}^{-M}S$ such that $u_{1}^{-1}S_{BB}(S)$ is not defined. The states $S_{BB}$ are perfectly normal in the sense that we can apply $u_{1}$ repeatedly to them and reach back to $S$. They also have to fulfil $S_{BB}(S)\cap S_{BB}(S')=\varnothing$ whenever $S\cap S'=\varnothing$ like any other state.

It is natural to depict the states $S_{BB}(S)$ as lying very close to each other in a schematic sketch of state space $\mathcal{S}$, since they are similar in the sense that they are characterized by very small space-times (Fig. \ref{Fig137}). Actually, the space-time of $S_{BB}$ cannot contain any past objects, no time intervals $\Delta t_{12}>0$. Therefore the space-time of $S_{BB}$ may be said to be degenerated.

\begin{state}[\textbf{The Big Bang}]
For each physical state $S$ there is a positive integer $M(S)$ such that $S_{BB}(S)\equiv u_{M(S)}^{-1}S$ is a physical state, but $u_{1}^{-1}S_{BB}(S)$ is not defined. $S_{BB}$ is the `Big Bang state' associated with $S$. A set of Big Bang states can be associated with a single `Big Bang set' $Z_{BB}$. We can define an invertible operator $w_{1}$ such that $\lim_{m\rightarrow\infty}w_{m}^{-1}S_{BB}(S)=Z_{BB}$ whenever $S\in B(\Sigma_{BB})$, where $B(\Sigma_{BB})$ is a basin of attraction that contains a continuous infinity of states $S$. However, $w_{1}\neq u_{1}$. The Big Bang state $S_{BB}$ is abstract in the sense that it is a retrodiction via physical law that has not been a physical state $S(n)$ at any sequential time $n\geq 1$. That is, $S(n)\neq S_{BB}(S)$ for all $n\geq 1$. The singular Big Bang set $\Sigma_{BB}$ is even more abstract in the sense that we have to go outside physical law to reach it.
\label{bigbang}
\end{state}

\chapter{\normalfont{SUMMARY AND CONCLUSIONS}}
\label{summaryconclusions}

In this section, I discuss some basic physical concepts and facts on which the present approach may shed some new light. They are discussed item by item. In this way the discussion takes the form of a summary of the main conclusions reached in this study. I will focus more on the physical questions than on the philosophical ones. In so doing, I will connect more clearly with the standard vocabulary used by physicists than I have done in the main body of this text. An example is the concept of \emph{entanglement}, which I have not yet mentioned at all. Nevertheless, I start with a brief summary of the philosophical perspective, and of the crucial assumptions that makes it possible to use it as a tool.

Before that, I state my philosophical motto: to apply common sense. I have tried to motivate the structure of physical law as a reflection of the structure of our perceptions. Many physicists take the opposite road. They start with the structure of physical law and argue that we should not try to relate it to our everyday perceptions. Trying to do so is considered narrow-minded, and purportedly leads to Newtonian mechanics, we we know is incorrect at a fundamental level. We are told that concepts such as entanglement and superpositions cannot be understood in an ordinary sense, but should be regarded exclusively as abstract, mathematical structures. I reject this view. In my view the lack of everyday understanding of such formal concepts is due to the lack of a proper dictionary of interpretation. I have tried to contribue to such a dictionary. 

\subsubsection{Philosophical perspective}

The basic assumption is what I call \emph{intertwined dualism}. Subject and object are two indispensable aspects of the world, but it makes no sense to talk about them separately. From this idea follows that the physical state can be described as a state of knowledge. This knowledge can be seen as the tie between the subjective and the objective aspect. This tie is assumed to exist at every level of detail, meaning that there is no little object that is not rooted in a subjective perception, and there is no subjective perception that is not rooted in the state of some objects. This is what I call the assumption of \emph{detailed materialism}.

To be able to talk about knowledge, we have to add interpretatations to the perceptions. Therefore a fundamental subjective ability to distinguish between \emph{proper and improper interpretations} has to be assumed, an ability that cannot be analyzed further. To make the concept of knowledge a solid foundation of a well-defined scientific world-view, we also have to distinguish between the knowledge that we are aware of at a given moment, and the \emph{potential knowledge} that is possible to be aware of in principle. It is the latter that corresponds to the physical state. Another requirement to make such a physical state well-defined is that we let it correspond to the \emph{collective potential knowledge} of all perceiving subjects, rather than the potential knowledge of some random individual. Implicit in this requirement is the assumption that the world is such that it allows the existence of several subjects. The alternative is solipsism.

Regarding the structure of knowledge, it is assumed that all knowledge can be expressed in terms of \emph{internal and relational attributes} of a set of distinguishable \emph{objects}. An internal attribute refers to a single object, whereas a relational attribute relates two or more objects. Gain or loss of knowledge may cause objects to divide or merge. It is assumed that any given object can only be divided a finite number of times, until we reach the level of \emph{minimal objects}. These minimal objects can be identified with elementary particles. Minimal objects may not be directly perceivable; their existence may have to be deduced from the observation of other objects. Such deduced objects are called \emph{quasiobjects}.

We may have \emph{complete} or \emph{incomplete knowledge} about the values of the attributes that describe the objects. In the latter case there is a set of attribute values (with more than one element) that cannot be excluded as the true one. Any knowledge about the state of an object exludes \emph{some} attribute values, however. This exclusion may take the form of \emph{conditional knowledge}. This simply means the knowledge contained in an implication $A\rightarrow B$. If $A$ is a value of one attribute, and $B$ is a value of another, then we can exclude all attribute value arrays of the form $(A,\neg B)$. Since the physical state is assumed to correspond to a state of knowledge, any proper representation of the physical state should allow the representation of incomplete and conditional knowledge at the fundamental level.

To extract physics from the knowledge-based physical state we have to choose an approach to the concept of time, so that we can express physical law appropriately. Here we treat time as an attribute that is fundamentally directed and takes discrete values. That is, we treat it as a directed sequence of instants $n$. Any change of perception experienced by a subject corresponds to a temporal update from one instant to the next. If two subjects perceive changes, these changes correspond to different temporal updates if and only if the separation between the bodies of these subjects is time-like in the relativistic sense. The physical state at any given time instant contains objects which belong to the present, as well as those which belong to the past (like memories). Thus we introduce a binary \emph{presentness attribute} that applies to all objects. Any two objects can be related with a measure of temporal distance $t$. If both the objects have the value \emph{present} of the presentness attribute, we always have $t=0$. The relational attribute $t$ is analogous to the spatial distance $r$ between any two objects, and the two can be transformed into each other in a Lorentz transformation in the usual way. This means that we define an entire space-time for \emph{each} time instant $n$. In effect, what we do is to separate \emph{sequential time} $n$ from \emph{relational time} $t$. We use sequential time to define physical law as the mapping of the physical state from one instant to the next. 

A central concept in this study is that of \emph{identifiability} and the associated notion of \emph{identity}. As time passes, we need a criterion to tell whether two objects observed at different time instants is the same. This is necessary in order to express physical law as the dynamics of objects. As indicated above, we have adopted the usual reductionistic idea that all objects, and their dynamics, can be modelled by minimal objects, and their interactions. That is, we assume that the dynamics of all objects can be described in terms minimal objects which preserve their identity as time passes. More precisely, we are never able to \emph{exclude} a model in which all individual minimal objects follow individual trajectories. If two objects are observed at two different times, and the second object can be modelled as the evolved state of the set of minimal objects that model the first object, then the two objects are said to be one and the same. The basic idea is that if we cannot exclude that they are the same, we have to judge that they are the same. We say that the objects are \emph{quasi-identifiable}.

To say that two directly observed objects are \emph{identifiable} is a stronger claim. This means that their states cannot be subjectively told apart at two successive time instants, that they look the same, and therefore are judged to be the same. The world preserves its identity as time passes if and only if there are some objects that are identifiable in each temporal update. At each such update, there has to some other objects that are not identifiable, that subjectively look different. Otherwise we would be unable to define the passage of time. In the same manner, a subject is said to preserve her identity as time passes if and only if some of the objects she perceives stay the same while she perceives the change of others. The objects that stays the same may be internal as well as external; they may be related to her state of mind, her memories, as well as things she sees. For example, a stem of a tree may stay the same as she watches one of its leaves fall to the ground.

\subsubsection{Philosophical assumptions}

The above ideas are elements of a philosophical world-view, but cannot be used to find the form of pysical law. To this end, we need another set of assumptions, which constrains the form of such laws.

The most basic assumption is that of \emph{epistemic consistency}. This simply means that the potential knowledge at one time instant should not contradict the potential knowledge at another, given the laws of physcis that determine the evolution of the physical state from the first instant to the second. This assumption may look like a tautology that adds no insight, provided the laws of physics are always obeyed. However, it is no tatutology if we accept the idea that the physical state is a state of knowledge, and if we assume that knowledge is inherently incomplete.

Namely, in that case we may deduce knowledge about a previous time instant at a later time, and this knowledge changes the physical state at the previous time, in a sense. In general we have to assume that two different physical states at a previous time may evolve into two different physical states at a later time. Therefore, the knowledge deduced afterwards could lead to an evolution that creates a physical state at the later time that contradicts the state we actually experience. An example is the interference pattern in a double slit experiment. If we have seen such a pattern, it should be impossible to deduce afterwards which slit the particle actually passed. Such knowledge would mean that no interference pattern shows up, which contradicts the memory of this pattern in the present state of knowledge. In essence, the condition of epistemic consistency means that we cannot gain `too much' knowledge about a situation afterwards.    

A second important assumption is that of \emph{explicit epistemic minimalism}. This means that a proper representation of physical law does not allow entities or processes that cannot be observed. A weaker form of this condition is that of \emph{implicit epistemic minimalism}. This means that the representation of physical law does not \emph{require} such entitites or processes. Another way to put epistemic minimalism is to say that we should dismiss any expression of physical law that makes use of distinctions that does not correspond to distinctions that can be made subjectively, in the perception of the world and of ourselves. When we write down physical law we are not allowed to make use of any fantasies, to make anything up. All epistemic dead weight must be thrown overboard.

On the other hand, any proper expression of physical law must take into account all entities and processes than \emph{can} be observed in principle. We should dismiss any expression of physical law that does \emph{not} make use of all distinctions that can be made subjectively, in the perception of the world and of ourselves. We must take all epistemic baggage on board. We call this principle \emph{epistemic completeness}.

We may combine the conditions of epistemic minimalism and epistemic completeness to the the assumption of \emph{epistemic closure}. Everything physical must have an epistemic root, and everything epistemic must have a physical representation. This principle embodies the assumption of intertwined dualism. The subjective and objective aspects of the world are equally fundamental, and cannot be separated.

Another constraint we put on physical law is that of \emph{epistemic invariance}. This is the idea that the evolution of an object should not depend on the amount of knowledge we have about it. That is, gain or loss of knowledge about its initial state cannot lead to a knowably different final state. We cannot affect the evolution of an object just by taking on or off our glasses when we look at it. Just like epistemic consistency, this principle is meaningful as a guide to physical law only in the present epistemic approach, where we identify the physical state with a state of knowledge, and this knowledge may be more or less complete.

We also assume \emph{individual epistemic invariance}. Just as the evolution cannot depend on how much knowledge we have about the evolving object, it cannot depend \emph{who} is having this knowledge. Physical law must be insensitive to a change of perspective from one person to another when we look at a given object. Again, this principle would not be necessary if the physical state did not correspond to a state of knowledge, and this knowledge were not rooted in a set of individual subjects.

\subsubsection{Ontology}
Even an epistemic approach to physics requires an ontology. Someone has to exist who possesses the knowledge, as well as something which is the target of the knowledge. Many people feel that quantum mechanics does not provide a viable or coherent ontology. Therefore, these people argue, we must seek a more fundamental model of the world. The present approach to physics does not represent an alternative to quantum mechanics. Rather, it is an attempt to motivate it from a set of philosophical assumptions. Here, we scrutinize these assumptions to find their ontic basis. If the reader finds our motivation of quantum mechanics convincing, she might consider the following short list of four `beables' to be a draft of the ontic basis of quantum mechanics.

First, we have emphasized that we must add an interpretation to get knowledge from a subjective perception. This interpretation can be proper or improper. The ability to distinguish between such interpretations must be assumed in a knowledge-based approach to physics. Otherwise we do not get off the ground. Further, in this approach, this ability must be seen as fundamental; it cannot be explained in terms of something else. This means, for example, that we must assume the ability to distinguish dreams from reality. In terms of the physical interpretation, this means the ability to locate the objects that correspond to the subjective perception correctly: they belong either to our own body, or to the external world. The existence of this fundamental ability does not mean that we always make the correct interpretation. This fact gives meaning to the word \emph{mistake}. It just means that the distinction between correct and incorrect interpretations exists as a basic quality of the world. To put it solemnly: the truth is out there.

Second, the idea of intertwined dualism means that we assume the existence of something outside our own body, the latter being a proper subset of the physical world. This `something' matches the proper interpretation of some objects as belonging to the external world, as discussed above. However, the knowledge is incomplete about these external objects (as well as the knowledge about the internal objects), as discussed under the next heading. Therefore they should not be assigned precisely defined attribute values in proper representations of the physical state, assuming that the physical state corresponds to a state of knowledge. It should be stressed that these external objects and their attributes are \emph{not} assumed to exist objectively, regardless whether they are perceived by any subject or not. Using the language of Kant, both the objects that are properly interpreted to belong to the body and those who belong to the external world are things-as-they-appear-to-us rather than things-in-themselves.


Third, physical law exists objectively. It is a quality of the world that transcends ourselves in the sense that it cannot be changed or broken at will. It also transcends our perceptions in the sense that it maps a state at time $n$ to a state at the \emph{different} time $n+1$, whereas all our perceptions belong to the \emph{same} present time, even if some of these perceptions may correspond to memories of past times. Under the heading \emph{epistemic closure} we argue that physical law is a reflection of the categories of perception that we are given, the distinctions that we are able to make. This means that it should be possible in principle to deduce the form of physical law from an analysis of such catergories and distinctions. Much of the present text is devoted to this task. For example, we treat time and its directed nature as fundamental categories of perception, and argue that the Dirac equation emerges as a consequence. Such a connection between pure thought an physics is viewed with suspicion by many, since it downplays the role of experiment. However, it is unavoidable if we take materialism seriously. Then our logic and the categories of perception are functions of the action of the brain, which is a product of the action of physical law. 

Fourth, there exist several subjects. You are not alone. This is acknowledged by the fact that we treat the collective potential knowledge of all subjects taken together as the entity that corrresponds to the physical state of the world. The knowledge of each subject contribute to this knowledge, but is not sufficient as a basis to construct a physical state to which physical law can be applied unambiguously. Below we argue that the invariance of physical law under Lorentz transformations is an expression of the fact that the world is arranged to accommodate several subjects. If this is indeed so, the world continues to exist after our individuality is lost, when we die.

\subsubsection{The incompleteness of knowledge}
I argue that potential knowledge has to be incomplete, since the bodies of the set of all subjects is a proper subset of all objects in the world. Therefore the bodies can never encode the exact state of the entire world. To reach this conclusion we make use of the abovemnetioned assumption of detailed materialism. The incompleteness of knowledge must be expressed in all representations of the physical state and of physical law. This is the most basic conclusion in this study. It opens the door for the philosophical assumptions discussed above, gives them power, makes them useful as tools to chisel out the form of physical law.

\subsubsection{Physical states} 
We introduce the physical state $S$ as the set of all states of complete knowledge that are not excluded by our actual incomplete knowledge. This means that $S$ always has more than one element $Z$. From this fact follows that physical law is not deterministic. If it were indeed deterministic despite the incompleteness of knowledge, it would be impossible in principle to gain more knowledge at a later time about those attribute values for which the knowledge is incomplete at the moment. It would be impossible to uphold the notion that these attributes were independent variables. In effect, $S$ would correspond to an exact state $Z$, and the knowledge would be complete.

\subsubsection{Trajectories of objects}
Since two successive time instants correspond to a knowable change, we have $S(n)\cap S(n+1)=\varnothing$. We may also introduce the state $S_{O}$ of an object $O$. This object is \emph{identifiable} if and only if $S_{O}(n)\cap S_{O}(n+1)\neq\varnothing$. The object we look at at times $n$ and $n+1$ is \emph{the same}. Naturally, there has to be at least one object $O'$ such that $S_{O'}(n)\cap S_{O'}(n+1)\neq\varnothing$, so that we get the required knowable change.

An identifiable object does not have to be static, as illustrated in Fig. \ref{Fig40}. In contrast, in a world with complete knowledge, an object is identifiable only if its state does not change: $S_{O}(n)=S_{O}(n+1)=S_{O}(n+2)=\ldots$. This means that the existence of a well-defined, non-trivial notion of identifiablity relies on a knowledge-based physical state in a world where knowledge is incomplete. Traditional discussions about the foundations of physics do not address the problem of identifiability even though the notion of objects that we follow through time is central to almost all attempts to formulate physical law.

\subsubsection{Conservation laws}
The concept of identifiability is central to our analysis of conservation laws in particle reactions. If the nature of the incoming and outgoing particles is not the same, we cannot use an overlap of their object states $S_{O}$ during a sequence of time instants that contains the reaction as a defining criterion for the statement that the outgoing particles are produced from the incoming ones, that they should be associated to each other. However, the assumed difference in their nature means that there is no such overlap.

Instead, the association of the incoming particles with the outgoing ones has to be accomplished by a relation between the internal attributes that define their nature. We argue that the only way to create an unambiguous association is to postulate discrete values of the internal attributes, and conservation laws that stipulate that the sum of their values in the set of incoming particles should be the same as the sum of their values in the set of outgoing particles.

The conservation laws for energy and momentum has a different source. They are simple consequences of the relational nature of these attributes. It has no epistemic meaning to say that the total energy or momentum of a set of particles isolated from the environment changes with time, since such an environment is necessary to define the change. Since there is no basis for a distinction between the values of these attributes before an after a reaction among the set particles, the values have to be considered the same in any proper numerical representation of their state and its evolution. Of course, in practice there is always an environment to any system under observation. However, in a carefully designed experiment we minimize the interaction between the system and the environment. This minimization corresponds to a minimzation of the deviation from the conservation laws that hold in the idealized limit of complete isolation.

\subsubsection{Statistics of identical particles}
It is epistemically meaningless to treat a permutation of a pair of identical particles as a new state. Therefore the assumption of explicit epistemic minimalism implies that we get the wrong answer to physical questions if we do treat such permutations as different. This is indeed the case, since no gas obeys Maxwell-Boltzmann distribution perfectly, and the deviation from this distribution grows when the hypothetical effect of such permutations becomes more important.

\subsubsection{Pauli's exclusion principle}
It does not make epistemic sense to say that two objects are found in the same state. To say that we have two of something, we have to able to separate what we are talking about into two parts. To do that the two parts must differ in some respect. In our vocabulary, there has to be an attribute whose value is knowably different in the two objects. Explicit epistemic minimalism then requires that we get wrong physical answers if we treat a theoretical arrangement with several objects in the same individual state as a possible collective physical state. This is Pauli's exclusion principle, which leads to Fermi-Dirac statistics. What about bosons and Bose-Einstein statistics? We claim that all objects are indeed composed of elementary fermions. Elementary bosons are different kinds of entities, for which the above argument does not apply. We return to this matter below.

\subsubsection{Angular momentum and spherical symmetry}
If an object is perfectly spherically symmetric, then it is impossible to tell whether it is rotating or not. There is no marker that makes it possible to trace the rotation. It would go against epistemic minimalism to assign a non-zero angular momentum to such an object, since it is a value that cannot be checked. Indeed, quantum mechanics always assign zero angular momentum to an object in a bound state with a spherically symmetric spatial probability distribution.

\subsubsection{Relativity and the existence of many subjects}
We argue that the finite speed of light can also be seen as a consequence of explicit epistemic minimalism. Namely, since the idea of absolute speed has no epistemic meaning, physical law should make it impossible to uphold this idea. As long as the addition law for velocities (Eq. [\ref{additionlaw}]) always hold, we can indeed uphold such a notion, by choosing an arbitrary object as the origin at rest in a universal spatial coordinate system. To exclude this possibility, we have to introduce a finite maximum velocity - the speed of light. To give such a speed epistemic meaning, all subject must agree on the value of this velocity.

To arrive at the Lorentz invariance of physical law from such an invariance of the speed of light, we have to make use of a thought experiment in which two observers who are moving in relation to one another measure the spatial and temporal distance between a given pair of events. We also have to make use of the assumption that the results of both measurements are equally valid. We have observer democracy. In our fancy terminology, we may say that we apply the assumption of individual epistemic invariance.

This argument relies on the possibility that two subjects observe \emph{the same} object or event. This corresponds to the prescence of several subjects in the same objective world. That is, Lorentz invariance means that the world is arranged to accomodate many subjects. In other words, the fact that physical law seems to be Lorentz invariant is an argument against solipsism. If you were the only aware observer, there would be no need for Lorentz invariance to make physical law consistent.

We may say that Lorentz invariance is an example of the fact that the content of the physical state $S$ is individual, whereas the form of physical law which acts on $S$ is collective. By the `content' of the state we mean the measured values of attributes, for instance the spatio-temporal distance $\mathbf{r}_{4}$ between two events. These are always based on individual perceptions. By the statement that the form of physical law is `collective' we mean that it is neutral to any discrepancies of measurements made by different individuals.  

Another example of the interplay between the individual and the collective in representations of physical law is the concept of straightness (Definition \ref{straightness}). We regard the individual ability to distinguish the straight from the curved as a fundamental assumption, and we try to use it as a tool to derive physical law (see below). Without the use of straightness as a fundamental quality it is impossible to talk about forces and interactions that bend trajectories. Nevertheless, since acceleration cannot be defined in an absolute sense, the judgement that something is straight cannot be transcended from the individual to the collective realm. Different individuals may judge straightness differently. Again, the form of physical law must be neutral to any such discrepancies. This is the equivalence principle. We may therefore say that both special and general relativity gives support to the idea that the world is arranged to host both individual subjects and collections of subjects.

To conclude, the evolution of a world in which many individuals live cannot depend on attribute values on which these individuals may disagree. Then the world perceived by the different individuals would not stay the same, contrary to assumption. Conversely, the fact that physical law is neutral to such disagreements supports our identification of the physical state with the \emph{collective} state of potential knowledge, and also the common sense idea that we are many people living in the same world.

As a side note, if the evolution of the world indeed depends on the collective state of knowledge only, it does not lead to any epistemic inconsistency if an individual subject transcends her personal knowledge and gains direct access to the knowledge of other subjects, or to the `pool' of collective knowledge. Nothing really changes in that case. In other words, the present approach to physics does not \emph{exclude} ESP, like remote viewing. However, any hypothetical knowledge gained in such an unconventional way must still be within the collective potential knowledge at the time it is acquired, meaning that it belongs to the personal potential knowledge of \emph{someone}. Otherwise we do risk inconsistencies.

In summary, the fact that different aspects of physical law can be successfully represented in a way that takes into account the interplay between the individual and the collective can be seen as a hint that the subjective aspect of the world should be divided into several individual parts.

\subsubsection{State representations, quasiobjects and reductionism}
The definition of the physical state $S$ given above is purely conceptual. To do physics, we need a mathematical representation $\bar{S}$ of $S$. There may be several proper representations of the same state $S$. This means that two such representations $\bar{S}$ and $\bar{S}'$ may encode the same state of knowledge. For example, a rigid translation of all spatial coordinates changes the representation, but not the physical state. The distinction between the state $S$ and its representation $\bar{S}$ is used at several places in this study. It provides the basis for the motivation of the gauge principle, for instance (see below).

A quasiobject is an object that is not directly perceived, but is deduced from the observation of other objects. Elementary particles are examples of quasiobjects. It is sufficient to consider the directly perceived objects to specify the physical state $S$, but its representation $\bar{S}$ may contain quasiobjects as symbolic elements. The advantage of using quasiobjects in $\bar{S}$ is that physical law can be efficiently expressed in terms of interactions between the small parts of the perceived objects. If we divide an observed object into sufficiently small parts, these inevitably become quasiobjects.

It is the principle of reductionism that allows us to express physical law in terms of such small parts, ultimately elementary particles. We may look at reductionism as a consequence of epistemic invariance. The evolution of a state of enhanced knowledge attainable in principle should be consistent with the evolution of the actual state of lesser knowledge. Loss or gain of knowledge should not in itself enable different behavior.

An enhanced state of knowledge may correspond to the division of each perceived object into smaller parts. Knowledge increases in such an imagined process since the total number of perceived objects increase. Therefore we may account for the behavior of the macroscopic objects we actually perceive in terms of its microscopic constituents, even if we are not directly aware of these constituents. However, we must first check in a spot test the extent to which the perceived objects can actually be divided. If we try to express the physical state and physical law in terms of even smaller parts, we get an improper representation, which should be contradicted by experiment according to the assumption of explicit epistemic minimalism.

\subsubsection{The superposition principle and the linearity of evolution}
What we are saying is that the evolution $u_{1}$ of a state $S$ of lesser knowledge, where we do not actually see the smaller parts of the observed object, is consistent with the evolution of each possible state of greater knowledge, where we do see the detailed composition of this object. Let us assume that there are only two microscopic alternatives $S_{1}$ and $S_{2}$ that are consistent with what we see. Then we may write $S=S_{1}\cup S_{2}$ and conclude that $u_{1}S=u_{1}S_{1}\cup u_{1}S_{2}$. If we choose particular representations $\bar{S}$, $\bar{S}_{1}$ and $\bar{S}_{2}$ of these states, we may formally write $\bar{S}=\bar{S}_{1}+\bar{S}_{2}$ and $\bar{u}_{1}\bar{S}=\bar{u}_{1}\bar{S}_{1}+\bar{u}_{1}\bar{S}_{2}$. These relations express the superposition principle and the linearity of evolution. They are both consequences of epistemic invariance.

\subsubsection{Observations and measurements}
Any perceived change can be regarded as an observation. Such observations are the driving force of the flow of time, since they are responsible for each temporal update $n\rightarrow n+1$.

A measurement is a particular kind of observation, in which the observed system is prepared at time $n$ so that it is known at that time that one of $M$ prepared alternative states $S_{j}$ of the system will be observed at some later time.

\subsubsection{Schr\"odinger's cat}
A superposition of two alternatives $S_{1}$ and $S_{2}$ means that they can both be realized in principle, but that it is currently unknowable which is true. Therefore a cat in a closed box, who dies if a radioactive nucleus decays, is not dead and alive at the same time. A superposition of the two alternatives rather means that nobody knows whether it is dead or alive. If the cat is a conscious being, it may judge for itself. Someone knows whether the cat is dead or alive - the cat itself. The perceptions of the cat contribute to the state of collective potential knowledge in such a way that there cannot be any superposition between its life and death. This resolution of the paradox relies on the assumption that it is the \emph{collective} state of potential knowledge that determines the physical state, rather than the knowledge of a given individual, like the cruel experimenter.

If the cat is not a conscious being, then there is no mystery either. Then it is just a big lump of matter that is in a superposed state. This may also be called a mysterious state of affairs, however. How can a big object be in a superposition between two very different macroscopic states? From the epistemic perspective, this is not strange at all. We just have to isolate the object well enough so that no aware being is able to decide, now or later, which is the true current state of the object. Of course, it is harder to achieve the necessary isolation if the object becomes bigger, or the observers gain access to more sensitive instruments. But these are practical difficulties, in the end. We may have an entire galaxy in a superposition between the alternatives that it is spiral or elliptical. This is possible if there are no intelligent observers inside the galaxy or within a certain very large distance.

\subsubsection{Probability}
In a measurement, we may or may not know the probability of each alternative $S_{j}$. Since we identify states of knowledge with physical states, probabilities are defined only if they are known beforehand.

If the probability of an alternative state $S_{j}$ of the observed system $O$ exists, then it can be identified with the volume $V[S_{j}]$ of  $S_{j}$, divided by $V[S_{O}]$, where $S_{O}$ is the state of $O$. The volume of a state is the number of exact states $Z$ of complete knowledge that is consistent with the actual knowledge.

Exact states $Z$ cannot be counted. They are abstract entities with no individual physical meaning, since, by the incompleteness of knowledge, they cannot be observed. This means, for example, that the evolution of an exact state is not defined. You may say that they are introduced just to dress the statement "we know that there is something we cannot know anything about" in set-theoretical clothing. Consequently, only relative volumes can be known, like the probability $V[S_{j}]/V[S_{O}]$.

\subsubsection{Born's rule}
Consider a measurement such as that in Fig. \ref{Fig59}(c). It is known at the start of the experiment at time $n$ that the specimen $OS$ that is going to be observed realizes one of a given number of predefined alternatives in an intermediate process before it is actually observed. However, it is also known at time $n$ that it is forever outside potential knowledge which of these alternatives is realized. Then explicit epistemic minimalism says that we get wrong predicitions for the final measurement performed on the specimen $OS$ if we treat these intermediate alternatives as if we could actually know which of them was chosen by $OS$. This means that the multiplication law for probabilities (Eq. [\ref{normalprob}]) cannot be used. We argue in Section \ref{bornrules} that the only alternative to that multiplication law that is generally applicable is to assign complex probability amplitudes to each alternative and use Born's rule to calculate probabilities to the alternatives that define the final outcome of the measurement.

We cannot assign probabilitites to the intermediate alternatives, since the outcome of these are never checked, and such probabilities therefore would have no epistemic meaning. In contrast, none of our philosophic assumptions prohibit the use of complex probability amplitudes.

To motivate the introduction of probability amplitudes and Born's law, we could equally well appeal to epistemic completeness as epistemic minimalism. Namely, there is a knowable, fundamental distinction between what is part of potential knowledge and what is not part of this knowledge. Since knowledge is incomplete we know that this demarcation line must exist, even if it may be impossible to locate it exactly (Fig. \ref{Fig3}). The existence of this distinction or demarcation line must be reflected in the physical representation. Such a distinction is introduced in the formalism if we treat the calculation of probabilities differently in the two cases.

\subsubsection{Properties and operators}
A property $P$ is a statement about attributes of objects. We may define a set $\mathcal{P}$ that contains all those exact states $Z$ such that the property can be defined among the objects that are present in this state. Likewise, we may define sets $\mathcal{P}_{j}\subset\mathcal{P}$ containing all those exact states for which the value of property $P$ is $p_{j}$. The sets $\mathcal{P}$ and $\mathcal{P}_{j}$ resemble the object state $S_{O}$ and the alternative outcomes $S_{j}$ of a measurement, respectively. However, there is a fundamental difference. The sets associated with properties are abstract. They do not refer to any physical state of a system, or to the possible results of an experiment performed on this system.

Property values may or may not have an inherent ordering, even if the values of all attributes are ordered. Property values that are not ordered are those whose values are instances of a categorical property, like the species of birds. In either case, we may formally label each attribute value with a number $p_{j}$.

We may formally identify each property $P$ with a linear, self-adjoint operator $\bar{P}$ in the following sense. To each set $\mathcal{P}_{j}$ we associate an eigenvector $\bar{P}_{j}$ in a Hilbert space $\mathcal{H}_{P}$. The corresponding eigenvalue is $p_{j}$. This construction is regarded as the basis for the association in quantum mechanics between observable properties and self-adjoint operators.

\subsubsection{The evolution parameter}
Let us say that an experiment is started at time $n$, and that the measurement is made at time $n+m$. The value of $m$ is variable. We may introduce a continuous evolution parameter $\sigma$ that interpolates between the integer values of $m$. An increase of $\sigma$ means that the expected relational time $t$ passed between the start of the experiment and the measurement increases. In practice, a variation of $\sigma$ may correspond to a movement of the detector in the experimental setup, as illustrated in Fig. \ref{Fig72}.

The evolution parameter is no attribute, it is just an abstract parameter introduced in order to express continuous evolution equations. In contrast, relational time $t$ is an attribute. It may be one of the observable properties measured in the experiment. The essential difference between our approach to evolution equations and the conventional one is that we release relational time from its duty to evolve the state. It becomes an observable distance on equal footing with spatial distances $x$. This liberation of $t$ is enabled by the separation of sequential and relational time, and makes it possible to express evolution equations with full relativistic symmetry.

\subsubsection{The wave function}
There is a finite number $M$ of possible outcomes of the measurement in any experiment, corresponding to an alternative $S_{j}$. We may associate a complex probability amplitude $a_{j}$ to each of these outcomes. If we consider a family of experiments parameterized by $\sigma$, we may write $a_{j}=a_{P}(p_{j},\sigma)$. The latter function is the wave function. It is defined as soon as an experiment is started at time $n$ in which the value $p_{j}$ of property $P$ is known to be measured at som later time $n+m$. Further, the probabilities $|a_{P}(p_{j},\sigma)|^{2}$ should be part of potential knowledge already at time $n$. As soon as the measurement is performed at time $n+m$, the wave function is no longer defined.

Continuous wave functions $\Psi_{P}(p,\sigma)$ are just convenient approximations to the actual, discrete wave function $a_{P}(p_{j},\sigma)$. They can be used to simplify calculations whenever property $P$ can take more or less continuous values, and the number $M$ of alternative outcomes is large.

Apart from their finite life time, the wave functions used here differ from the conventional ones in another respect: they typically have finite support. A spatial wave function that can take non-zero values arbitrariy far away correponds in the present approach to a specimen for which we have absolutely no idea where it is located at the start of the experiment. This is seldom the case in a controlled experiment. We may, for example, consider a specimen which is emitted from a source with a known position, like an electron from an electron gun. In general, we may regard the support of the wave function as the projection of the state $S_{OS}$ onto the plane in state space defined by the property $P$ we are about to observe (Fig. \ref{Fig75}). 

\subsubsection{The state space and Hilbert spaces}
We see that the wave function has a very limited range of validity in the present approach to physics. It is only defined in carefully designed experimental situations during a limited amount of time. Even so, it is used in exactly the same manner as in conventional quantum mechanics to extract probabilities for the outcomes of such experiments.

This means that the Hilbert space is not the fundamental state space. It ceases to exist as soon as the corresponding experiment is completed. We can define a new Hilbert space in another experiment. This Hilbert space may look different from the first one. This happens, for example, if we measure a different property $P'$, or use a detector with a different resolution, leading to a different number $M'$ of possible outcomes.

Rather, the fundamental state space is the set $\mathcal{S}$ of all possible states $Z$ of exact knowledge in which the physical states $S$ live. The fundamental temporal evolution $S(n)\rightarrow S(n+1)$ can be expressed as a mapping from the power set of $\mathcal{S}$ to itself.

\subsubsection{State reductions and wave function collapse}
By a state reduction we mean a temporal update $n\rightarrow n+1$ such that the observed change that defines the update is not completely determined by physical law, as expressed in the evolution operator $u_{1}$. Symbolically, $S(n+1)\subset u_{1}S(n)$. The physical state suddenly shrinks.

The Hilbert space in which one of the $M$ possible values $p_{j}$ is measured is $M$-dimensional, with a basis $\{\bar{S}_{j}\}$, where $\bar{S}_{j}$ is a representation of the corresponding alternative $S_{j}$. This representation takes the form of a vector that is orthogonal to all the other vectors $\bar{S}_{j'}$ in the set $\{\bar{S}_{j}\}$. The wave function is an element in this Hilbert space, as usual.

When $p_{j}$ is measured, we might say that the wave function collapses. Such a collapse always corresponds to a state reduction, since $S_{j}\subset u_{1}S_{O}$ whenever there are more than one ralizable alternative. However, after the measurement, the wave function is no longer defined at all if no further measurements will take place in the experiment. The wave function not only collapses, it is completely lost.

We may, however, define an experiment such that we know from the outset at time $n$ that two or more measurements of different properties will take place in succession. In such an experiment, the continued existence of the wave function can be motivated. Then there is a pair of integers $(m,m')$ such that $m'>m$, and such that the measurment of property $P$ takes place at time $n+m$ and the measurement of property $P'$ takes place at time $n+m'$. Then we may say that the wave function collapses in the ordinary sense at time $n+m$, and is lost at time $n+m'$ (Fig. \ref{Fig76}).

In summary, we may say that a wave function collapse is a state reduction after the first measurement in a well-defined experiment in which two properties are measured in succession. Clearly, the state reduction is the fundamental concept.

\subsubsection{Decoherence and wave function collapse}
Some physicists seem to argue that state reductions or wave function collapses are not necessary as a fundamental ingredient in proper physical models. They are considered to be just an apparent effect of decoherence. It is true that the measureable effects of a wave function collapse may look the same as those of decoherence. In a double slit experiment, the interference pattern disappears in a similar way if we gain knowledge about which slit the electron passed, as if the electron is entangled with contaminations in the experimental setup.

But the \emph{causes} of the loss of interference pattern are very different in the two cases. In the first case it disappears since we \emph{gain} knowledge about the electron path. In the second case it disappears because we \emph{lose} knowledge about the electron phase. Decoherence and de-superposition are different things, and I can see a conceptually important role to play only for the latter. The fact that we lose the chance to confirm experimentally that the two paths are superposed when decoherence destroys the interference pattern does not mean that the paths are no longer superposed. That there is a mess of footprints in the mud, obscuring the footprints of the murderer, does not mean that the murderer did not pass.

\subsubsection{Intention and choice}
Given a choice of property $P$ to be measured in an experiment, standard quantum mechanics provides the probabilities for different outcomes. The theory does not address how the choice of $P$ comes about, which basis is chosen in Hilbert space. Nevertheless, this choice determines the future state of the world, since it determines in advance from which set of orthogonal state vectors the future state is to be picked. Quantum mechanics does not try to reduce this choice to someting else, to a consequence of some other component of the theory. In a sense, it therefore accepts free will as a starting point. Strangely, this fact is seldom discussed. Most often, physicists say that free will is not relevant to quantum mechanics since we cannot influence the probabilities for the alternative outcomes of a given experiment. However, the word \emph{given} is crucial in that sentence.

One may say that the choice of property $P$ is the outcome of another `experiment' taking place in the experimenter's head, and that probabilities are defined for this meta-experiment. However, we argue that we end up in infinite regress if we continue along this road. In particular, we argue that probabilities must be knowable to someone in advance to deserve that name. This condition cannot be fulfilled for such meta-experiments.

Just like many people have done before me, I define free will negatively: as choices that cannot be predicted by anyone, neither deterministically, nor probabilistically. The existence of such choices or events corresponds to a `white hole' in physical law, at least if the physical sate is identified with a state of knowledge. I argue that intentionality can flow out of such white holes, from the subjective to the objective aspect of the world, leading to objective effects without knowable cause. I try to define individuals as aware entities capable of independent intentions and choices. It is important to stress that a clear-cut distinction between the individual and the collectice is necessary in the present approach to physics. The interplay between the individual and the collective is crucial in several places of the construction.  

That said, my attempts to describe individuality, intention and choice with the help of the epistemic formalism should be seen as provisional (Sections \ref{bwstates}, \ref{aic}, \ref{individualsubjects} and \ref{indtime}). The proposed definitions and statements are not heavily used in the subsequent sections where the focus is on more traditional physics. I include my attempts merely to stress that since I consider the subjective aspect of the world to be as fundamental as the objective aspect, it should be possible to formalize the structure of the subjective aspect in an analogous manner as the structure of the objective aspect is formalized in traditional science. I do not see these matters very clearly myself. However, even if some or all of the proposed ideas along this line are misguided, there should be a way forward.

\subsubsection{Entanglement}
Consider a specimen $OS$ such that we know at the start of the experiment at time $n$ that it has two parts $O_{1}$ and $O_{2}$ with physical object states $S_{O1}(n)$ and $S_{O2}(n)$, respectively. Suppose that the object states $S_{O1}(n)$ and $S_{O2}(n)$ are related by conditional knowledge. This may mean that we know at the start of the experiment that if we learn something new about $O_{1}$ at time $n+m$, then we may automatically learn something new about $O_{2}$, without actually looking at it. Alternatively, it may mean that if we learn something new about $O_{2}$ at time $n+m$, then we may automatically learn something new about $O_{1}$. In traditional language we say that the states of the two parts are entangled.

In the first case there is an exact state $Z_{1}$ of part $O_{1}$ and an exact state $Z_{2}$ of part $O_{2}$ such that $Z_{1}\in S_{O1}(n)$ and $Z_{2}\in S_{O2}(n)$. However, if $Z_{1}\in S_{O1}(n+m)$ then $Z_{2}\not\in S_{O2}(n+m)$. In the second case we may use the same pair of exact states to say that if $Z_{2}\in S_{O2}(n+m)$ then $Z_{1}\not\in S_{O1}(n+m)$. The pair of exact object states $(Z_{1},Z_{2})$ is excluded in the state $S_{OS}$ of the entire specimen $OS$ even though $Z_{1}$ and $Z_{2}$ are not excluded individually as possible exact states of its parts $O_{1}$ and $O_{2}$, respectively. Formally, the condition for the presence of conditional knowledge or entanglement at time $n$ becomes $S_{OS}(n)\subset S_{O1}(n)\cap S_{O2}(n)$. In contrast, if there is no conditional knowledge or entanglement between the parts of the specimen, then $S_{OS}(n)=S_{O1}(n)\cap S_{O2}(n)$. These matters are discussed in more detail in Section \ref{compspec}.

In experiments in which measurements are made on both parts $O_{1}$ and $O_{2}$ of the specimen, this definition of entanglement translates to the usual condition that the combined wave function for the two parts does not factorize to a product of two wave functions that describe each part individually.

In essence, we regard entaglement to be nothing more than the representation in the physical state of the knowledge contained in a logical implication $A\rightarrow B$, in the absence of any knowledge whether $A$ is true. The pair of logical states $(A,\neg B)$ is excluded just as the pair of exact physical object states $(Z_{1},Z_{2})$. Of course, this identification is made possible by our epistemic approach, in which the physical state corresponds to a state of potential knowledge. In this way, all the spookiness contained in the concept of entanglement evaporates. If we consider entaglement spooky, we should also consider the conclusion $B$ spooky, after having learned $A$. In a way it is - it is a sudden leap of logic over an abyss that we cannot look down into, that cannot be analyzed further, using as a bridge the arrow of implication.

The reader might complain that the abstract objects in logical reasoning which might have properties $A$ and $B$ are not material objects whose behavior can be described by physics. Therefore they cannot be `entangled', the reader might argue. However, when we defined an object at the beginning of this text, we did it in a very general way, as something we perceive that can be distinguished from everything else we perceive. This quality is possessed by logical objects as well, of course. Nowhere do we make use of any other qualities of an object.

We have made one additional assumption though. We suppose that all perceived objects corresponds to other perceivable objects in our bodies, and we gave this assumption the name `detailed materialism'. In fact, the claimed correspondence between entanglement and implications means that our logical categories of thought are perfectly rooted in the material world, in accordance with such a materialistic world-view.

We may even say that the existence of entangled physical states makes the concept of logical implications meaningful. If there were no entanglement, there would be no knowledge of $A\rightarrow B$ without the additional knowledge whether $A$ is true. The implication iself would lose its independent meaning. The concept of physical law would also lose its meaning, since physical law can be seen as a set of logical implications: "If conditions $A$ are fulfilled at time $n$, the physical state will fulfil conditions $B$ at time $n+1$."

The existence of implications deduced from physical law are essential to virtually all scientific experiments. When we observe a change in a macroscopic detector, we can conclude that this corresponds to a well-defined event in a microscopic specimen only by means of knowledge \emph{a priori} about the entaglement between the states of the detector and the specimen.

\subsubsection{Evolution equations}

Consider an experiment such that probabilites of the alternative outcomes are knowable beforehand. Then there is a Hilbert space and a wave function. In that case we should be able to express an evolution equation for the wave function. We can use the abstract evolution parameter $\sigma$ in order to express a differential evolution equation

\begin{equation}
\frac{d\Psi_{P}}{d\sigma}=i\bar{B}_{P}\Psi_{P},
\label{generalevo}
\end{equation}
where $\bar{B}_{P}$ has real eigenvalues, using the fact that the evolution must be unitary since the probabilites of all alternatives must add to one regardless the value of $\sigma$.

It should be stressed that differential evolution equations of this type are always approximations. They apply exactly only in idealized experiments in which the spatio-temporal resolution is perfect, so that the number $M$ of alternative outcomes becomes infinite. They also presupposes that an infinite support of the wave function is allowed, corresponding to an detector that is infinitely big.

In Section we \ref{eveq} motivate the form of such an evolution equation in the case where the measured property $P$ is the spatio-temporal distance $\mathbf{r}_{4}$ traveled by the specimen $OS$ between the start of the experiment and the measurement. We make use of the fundamental subjective distinction between straight and curved trajectories. We may say that a specimen that is expected to travel with constant speed along a straight trajectory is \emph{free}; it does not interact with any other objects as judged by the individual who performs the experiment. For such a specimen we arrive at

\begin{equation}
\bar{B}_{\mathbf{r}_{4}}=\frac{c^{2}\hbar}{2\langle E\rangle}\Box.
\label{summaryeveq}
\end{equation}

In so doing, we use the additional requirements that the evolution equation must be Lorentz invariant and that the differential evolution operator $\bar{B}_{P}$ cannot depend explicitly on $\sigma$. This is so since $\sigma$ is no attribute or property - a change of its value does not change the physical state, whereas the evolved physical state depends only on the previous physical state. Apart from Lorentz invariance, we also use the self-evident fact that the equation must be invariant under stiff coordinate translations $\mathbf{r}_{4}\rightarrow \mathbf{r}_{4}+\mathbf{c}$, since we are dealing with relational attributes.

Furthermore, to arrive at Eq. [\ref{summaryeveq}], we make use of what we call the \emph{natural parametrization} for which $d\langle t\rangle/d\sigma=1$. This means that we choose an evolution parameter that mimics, in an averaged sense, the conventional evolution equations in which the relational attribute $t$ is used as evolution parameter. Here $\langle t\rangle$ is the time that is expected at the start of the experiment to pass until the measurement of $\mathbf{r}_{4}$ (including $t$) is performed.

\subsubsection{Construction of momentum and energy}

The quantity $\langle E\rangle$ in Eq. [\ref{summaryeveq}] can be identified as the expected energy of the specimen. This is a constant that does not depend on $\sigma$, and it is not measured in the experiment.

In our approach, momentum, energy and rest mass are not properties of objects that are defined in the Newtonian sense via spatio-temporal relations such as velocities and accelerations, in combination with the concept of force. Rather, they are defined via the fact that certain operators always correspond to properties. In our approach a property is defined simply as a region $\mathcal{P}$ in state space $\mathcal{S}$ which can be divided into disjoint subsets $\mathcal{P}_{j}$. These subsets correspond to the property values $p_{j}$, which must be observable in the sense that there must be an object $O$ whose state $S_{O}$ can fulfil $S_{O}\subseteq\mathcal{P}_{j}$ for each $j$.

We argue in Sections \ref{propop} and \ref{wavef} that a self-adjoint operator that acts on a wave function can be associated with a property in this sense. The eigenvalues are the property values $p_{j}$. Performing a Fourier decomposition of the wave function $\Psi_{\mathbf{r}_{4}}$ and using the evolution equation [\ref{summaryeveq}], we see that the triplet of eigenvalues $(p,p',\mathbf{p}'')$ of the three self-adjoint operators

\begin{equation}\begin{array}{rcl}
\bar{P}_{\mathbf{r}_{4}} & = & i\hbar\frac{\partial}{\partial t}\\
& & \\
\bar{P}_{\mathbf{r}_{4}}' & = & -\frac{2i\hbar\langle E\rangle}{c^{4}}\frac{d}{d\sigma}\\
& & \\
\bar{P}_{\mathbf{r}_{4}}'' & = & -i\hbar\nabla
\end{array}
\label{operatorlist}
\end{equation}
that share the same plane wave eigenfunctions fulfils the relation

\begin{equation}
p^{2}=p'c^{4}+|\mathbf{p}''|^{2}c^{2}.
\label{protoein}
\end{equation}
This equation is fulfilled for any choice of triplet $(p,p',\mathbf{p}'')$. Since we recognize Einstein's relativistic relation between energy, rest mass and momentum, we can identify

\begin{equation}\begin{array}{rcl}
p & = & E\\
p' & = & m_{0}^{2}\\
\mathbf{p}'' & = & \mathbf{p}.
\end{array}
\end{equation}

We may combine the first and third operator in Eq. [\ref{operatorlist}] to a four-momentum operator for a free specimen. By generalizing its definition according to Eq. [\ref{fourmomentumdefi2}], we may retain the form of the evolution equation according to Statement [\ref{psieveq}] even if the specimen is interacting, following a curved trajectory in space-time.

\subsubsection{The Dirac equation}
Since Eq. [\ref{protoein}] is fulfilled for any eigenvalue triplet $(p,p',\mathbf{p}'')$, we must have $p'\geq 0$. However, there is nothing in the evolution equation [\ref{summaryeveq}] that guarantess that this is the case. Indeed, for each wave function that solves the equation with $p'=\alpha$ there is another solution with $p'=-\alpha$. The only way to ensure that $p'\geq 0$ is to say that the corresponding property $P'$ is the square of another property. This requirement is fulfilled by the identification of $P'$ with the square of the rest mass.

In the same way as we argued that any self-adjoint operator acting on the wave function can be associated with a property, any property can be associated with such a self-adjoint operator. This means that we can write

\begin{equation}
\bar{P}_{\mathbf{r}_{4}}'=\bar{P}_{\mathbf{r}_{4}}'''\bar{P}_{\mathbf{r}_{4}}''',
\end{equation}
where $\bar{P}_{\mathbf{r}_{4}}'''$ is the operator associated with the rest mass, having eigenvalues $m_{0}$.

Just as in conventional quantum mechanics, we can decompose a general wave function into a sum of stationary states. In our approach this means stationary wave functions $\psi_{\mathbf{r}_{4}}(\mathbf{r}_{4},p')$ that does not depend on $\sigma$. To ensure that we always have $p'\geq 0$, we must make sure that this is true for each such stationary state. This is done by requiring

\begin{equation}
\bar{P}_{\mathbf{r}_{4}}'''\psi_{\mathbf{r}_{4}}=p'''\psi_{\mathbf{r}_{4}}=m_{0}\psi_{\mathbf{r}_{4}}.
\end{equation}

This is the Dirac equation. It is a constraint on stationary wave functions that has to be fulfilled in addition to the defining condition 

\begin{equation}
\bar{P}_{\mathbf{r}_{4}}'\psi_{\mathbf{r}_{4}}=p'\psi_{\mathbf{r}_{4}}m_{0}^{2}\psi_{\mathbf{r}_{4}},
\end{equation}
given by the evolution equation.

In essence, the Dirac equation follows from the directed nature of sequential time $n$. This nature is reflected in the fact that the derivative with respect to $\sigma$ is first order in the evolution equation. In contrast, the derivatives with respect to $x$ and $t$ are second order, reflecting the undirected nature of space and relational time. It follows that $p$ and $p''$ are squared in Eq. [\ref{protoein}], but not $p'$. The absence of the last square means that we have to make sure `by hand' that $p'$ stays positive, so that Eq. [\ref{protoein}] becomes generally valid.

\subsubsection{The structure of space-time}
The release of $t$ from the burden to evolve the wave function, from the double task of being both a parameter and an observable, releases the full symmetry between space-time and four-momentum space. A major conclusion along this line is that the spectrum of space-time has both continuous and discrete parts, just as the spectrum of four-momentum. The discrete parts of space-time appear in bound states, just as the energy levels become discrete in such states. More precisely, if two objects are not bound to each other, we cannot exclude \emph{\emph{a priori}} any spatio-temporal distance between them. In contrast, if they are are bound to each other, we can exclude small enough distances even before we check them.

We need a bound state as a ruler when we measure distances (Fig. \ref{Fig92}). The discrete spectrum of space-time in bound states means that there should be a minimum Lorentz distance $l_{\min}$ that can ever be measured. This means that we can never \emph{confirm} smaller distances in unbound states, but we cannot \emph{exclude} them either. It is therefore improper to say that space-time is discrete \emph{per se}.

The basic relation between space-time and momentum space is given by the Fourier transform, of course. We may expect that the Fourier transform of the wave function $\Psi_{\mathbf{r}_{4}}$ fulfils a reciprocal evolution equation with an evolution operator that has the same form as Eq. [\ref{summaryeveq}], with the space-time derivatives replace by four-momentum derivatives. To show that this is indeed the case, we reverse the reasoning that led to the ordinary evolution equation above. In so doing, we are able to derive the Einstein relation 

\begin{equation}
E^{2}=m_{0}^{2}c^{4}+|\mathbf{p}|^{2}c^{2},
\end{equation}
provided we make the accurate association between these properties and the corresponding operators. In deriving the reciprocal evolution equation, we go the other way around, assuming the relativistic relation

\begin{equation}
c^{2}t^{2}=l^{2}+|\mathbf{r}|^{2}.
\end{equation}
In this way we arrive at the reciprocal evolution equation, as displayed in Eq. [\ref{receqpair}] together with the ordinary evolution equation, in order to make their symmetry manifest.

It must be stressed that the reciprocal evolution equation does not express the evolution with respect to time, but rather to energy. In the ordinary evolution equation a change of the evolution parameter  $\sigma$ represents a change in the experimental setup such that the expected passed time $\langle t\rangle$ until measurement changes. In the natural parametrization we may say sloppily that $\sigma=\langle t\rangle$. In contrast, in the reciprocal evolution equation we imagine a continuous change in the experimental setup such that we change the expected energy $\langle E\rangle$ at the moment just before it is measured.

The conclusion mentioned above that space-time becomes discrete in bound states is based on the possibility to make conjugate definitions of what a bound state means. Using spatio-temporal coordinates it means that the distance between two objects is bound from above as $t\rightarrow\infty$. Using momentum coordinates it means that their relative momentum is bound from above as $E\rightarrow\infty$. In other words, the energy increase is absorbed as an increasing rest mass. The discrete spectrum of space-time follows from symmetry of the conjugate pair of evolution equations and their corresponding stationary states.

We see in these considerations the conjugate roles played by $t$ and $E$. This is conventional wisdom, just like the conjugate roles played by $\mathbf{r}$ and $\mathbf{p}$. What is a little bit new, I think, is to see the Lorentz distance $l$ and the rest mass $m_{0}$ as a conjugate pair of variables. The association makes sense since both are relativistically invariant.

Just as we derived the Dirac equation from the fact that the rest mass squared must be non-negative, we may derive a reciprocal Dirac equation from the fact that the Lorentz distance squared must be non-negative (Statement \ref{recdiracconstraint}). We argued that, in essence, the Dirac equation follows from the directed nature of time. This becomes even more clear in the case of the reciprocal Dirac equation. A negative squared Lorentz distance would mean that we leave the light cone, so that the temporal ordering between events become ambiguous. The roles of the pair of conjugate Dirac equations as constraints on space-time and momentum space are illustrated in Fig. \ref{Fig87}.

\subsubsection{Heisenberg's uncertainty relations}
The conventional uncertainty relations $\Delta x\Delta p_{x}\geq \hbar/2$ and $\Delta t\Delta E\geq \hbar/2$ can be derived in the usual manner from the wave function, using general qualities of Fourier transforms. The only difference is that the time-energy relation can now be interpreted in exactly the same simple way as the position-momentum relation, since $t$ has been released from its assignment as evolution parameter. It can therefore be treated like any other observable, and may have an incompletely known value.

Its successor $\sigma$ is matched with the rest mass or rest energy in a conjugate pair. Consequently, we get a new uncertainty relation $\Delta\sigma\Delta E_{0}^{2}\geq\hbar\langle E\rangle$. In the natural parametrization we may write $\Delta\langle t\rangle\Delta E_{0}^{2}\geq\hbar\langle E\rangle$. This relation can be interpreted as to say that the uncertainty of the expected time until measurement is inversely proportional to the uncertainty of the rest mass of the specimen on which we make the measurement. This fact has consequences regarding the possible masses of elementary particles, as discussed below.

Since wave functions are not always defined, the conventional derivation of the uncertainty relations via Fourier transforms cannot be the fundamental one. Even when the wave function is indeed defined, we should keep in mind that it is never truly continuous, and that it typically has a finite support. That is, the conventional treatment is an approximation at best.

To get to the bottom of things, we should look directly at the state space. The fact that knowledge is always incomplete means that there are pairs of attributes whose values cannot be known simultaneously. The conjugate attribute pairs above are examples of such pairs. We may consider the projection of the state of an object onto the plane in state space spanned by the members of such a pair (Fig. \ref{Fig80b}). Basically, Planck's constant is the minimum area of such a projection. The existence of such a minimum area, a quantum of action, reflects the fact that the state can never shrink to a point, to an exact state $Z$.

\subsubsection{The equivalence principle}
So far, we have mainly talked about the evolution of a free specimen, which is expected to travel along a straight line at constant speed. We have mentioned, however, that the quality of straightness is not absolute. To uphold the distinction between straight and curved trajectories at least at the individual, subjective level, we need forces or interactions. These can be derived from the gauge principle. Let us first discuss Einstein's equivalence principle as an example of the underlying reasoning.

The assumption of epistemic invariance means that it should be possible to account for the perceptions of a subject locked into an elevator without referring to specific unknown objects outside this elevator. The evolution of her perceptions should be consistent with many different relations between the elevator and the outside world. The amount of potential knowledge should not matter. A feeling of being pressed against the floor should be consistent with both gravity and acceleration. The existence of this subjective feeling is treated as fundamental, whereas its interpretation as gravity or acceleration is not. This approach conforms with the fact that it is impossible to define absolute acceleration without referring to such a feeling. We fail if we refer to spatio-temporal attributes alone; it is impossible in that way to judge whether a body is accelerating or its entire environment is accelerating in the opposite direction.

\subsubsection{The gauge principle}
Let us choose symbolic representations $\bar{S}_{1}$ and $\bar{S}_{2}$ of each of the two possible environments to the elevator. This means, among other things, that we choose two coordinate systems, one in which the elevator is gravitating towards an outside body, another in which it is accelerating in empty space. To each of these coordinate systems we may apply a global Lorentz transformation. Such a pair of transformations just change the coordinate systems, not the physical states themselves. They are redundancy transformations. We have seen that a representation of a state of lesser knowledge can be represented as the sum of the representations of each possible state of greater knowledge consistent with the former state. We may therefore apply an individual Lorentz transformation $\bar{L}_{1}=\bar{L}(\mathbf{v}_{1})$ to $\bar{S}_{1}$ and another Lorentz transformation $\bar{L}_{2}=\bar{L}(\mathbf{v}_{2})$ to $\bar{S}_{2}$ without changing the physics, that is, the evolution of $S=S_{1}\cup S_{2}$. This may be interpreted as the invariance of physical law under a local symmetry transformation, that is, as the invariance under a gauge transformation. Therefore the gauge principle can be seen as a consequence of epistemic invariance, since this assumption is at the heart of the superposition principle and the linearity of evolution.

We may say that the gauge principle makes it possible to derive gravity as a necessary alternative explanation of the feeling of acceleration, given that acceleration cannot be defined in any absolute sense. It is well known that electromagnetism can be derived in an analogous way. The physical invariance of a global phase change in the wave function corresponds to the physical invariance of a Lorentz transformation applied to a global coordinate system. They are both redundancy transformations. The superposition principle, as stated above, means that each small part of the wave function can be seen as a representation of a state with more exact knowledge about the position of the particle. A local phase change in this part of the wave function is therefore allowed, since it corresponds to a global phase change in a possible state of larger knowledge. Electromagnetism follows from the requirement that the evolution is invariant under any such a local phase change.

We put forward the idea that the values of all attributes are inherently ordered, and that any numerical translation of the entire set of ordered values of each attribute is a redundancy transformation. That is, the numerical values of the attributes have no physical meaning, only their ordering.
If this is so, we can use the gauge principle to constrain the form of physical law in a specific way for each attribute, for each degree of freedom. Such a constraint can be seen as an interaction or transformation. This is to say that to each degree of freedom corresponds a gauge force. An example of such a degree of freedom is the feeling of acceleration. The values of this degree of freedom are ordered in the sense that we an decide subjectively the strength of this feeling. The corresponding gauge force is gravity.

When we say that a rigid translation of the set of values of each attribute is a redundancy transformation, we speak about attributes that are assigned to individual objects. The reason why such translations are always redundancy transformations is that numerical values of attributes only have meaning in relation to other objects. In other words, the relation between two objects is represented by one mathematical object pointing at one object, and another mathematical object pointing at the other object. This is obvious when it comes to relational attributes such as distance between two objects, which is replaced by one position for each object. But the same situation occurs for the internal attributes, giving rise to internal redundancies such as invariance with respect to rotations in color space. In these cases, it is only the relation of the values of the internal attributes of two interacting objects that is relevant: if we rotate color space of all present baryons, the interactions between them stay the same. The same invariance holds if we switch sign of the electric charges of all particles in a system. The forces between them stay the same, and it cannot be epistemically decided that the charge transformation has taken place.

The set of attributes or degrees of freedom that can be assigned to individual objects defines the set of possible redundancy translations; we cannot choose them freely. At the basic level, we may therefore say that it is from the set of degrees of freedom of our perceptions that we derive the form of physical law, rather than from the set of redundancy transformations that follow from any attempt to represent these degrees of freedom numerically.

The statement that the values of each attribute are inherently ordered means that they have an inherent relation. But to define a relation we need an interaction. We may say that we need a force to activate a degree of freedom, just like we need to exert some force to open a chinese fan. Therefore, it would be a satisfying epistemic closure if there is a one-to-one correspondence between a force and a degree of freedom. Such a closure would mean that all interactions or transformations are possible to derive from the gauge principle. At an even more basic level, we may say that there is an epistemic closure between the existence of forces and the ability, at the individual level, to distinguish between straight and curved lines.

One may ask the following question: Since interactions are the result of the redundancies that are created when the relation between several objects are described by attribute values pointing at each individual object, wouldn't the interactions go away if we use a strictly relational representation of the physical state? No, nothing would change, since interactions are changes in the states of those individual objects we perceive, something which can only be defined in the traditional, object-based representation. We naively associate a position to each object in our field of vision, and define forces according to this representation. In a strictly relational state representation, we would not be able to use the gauge principle to derive forces, but on the other hand, the notion of force would lose its intuitive meaning. In short, an object-based state representation that deals with the resulting redundancies is the most natural way to represent the physical state and physical law.

\subsubsection{Classical and quantum description of forces}
As long as we do not observe the trajectory of a specimen, we can always describe it as if it is bending in response to an external force. This means that we can use a description according to classical mechanics, or treat it according to old-school quantum mechanics, where the evolution of the wave function is affected by a potential. We get a continuous change of the expected direction of motion as a function of the evolution parameter $\sigma$.

When we actually observe a change of direction, we can say that the difference between its initial and final momentum has been `emitted' from the specimen, finally to be `absorbed' by another object. This is merely a game of words as long as we allow arbitrarily small changes. But there is an inherent discontinuity in the very concept of \emph{change}. We must actually perceive a difference. This is why we treat sequential time as a discrete sequence of instants $n$, rather than a continuous temporal axis with an arrow. We may therefore speak about `quanta of change'. In the case of an observed change in the state of motion of a specimen, we may speak in a meaningful way of a `quanta of interaction' that are emitted from from the specimen. They carry a given amount of momentum, and possibly a given amount of other attributes such as angular momentum, in order to fulfil the conservation laws for relational attributes.

This discrete description is more reminiscent of quantum field theory. The continuous (classical) and discrete (quantum) descriptions are complementary. Which description is appropriate depends on the experimental setup, the observations we are about to make, the detail of our knowledge. According to the assumption of epistemic invariance the two descriptions cannot contradict each other, since the amount of knowledge should not matter when we use physical law to predict the future.

\subsubsection{Quantum mechanics and general relativity}
The separation between sequential time $n$ and relational time $t$ means that the so called `problem of time' disappears in attempts reconcile quantum mechanics and general relativity. There is no longer any conceptual contradiction between the evolution equations in quantum mechanics and the static trajectories in the space-time of general relativity. For each $n$ there is an entire space-time $(\mathbf{r},t)$, possibly curved.

The static trajectories in space-time can be parametrized by the evolution parameter $\sigma$ according to $(\mathbf{r}(\sigma),t(\sigma))$ to get motion into the relativistic picture. For each sequential time instant $n$ we get a snapshot $(\mathbf{r}(\sigma_{n}),t(\sigma_{n}))$. Strictly speaking, at each instant $n$, the points $(\mathbf{r}(\sigma_{n-1}),t(\sigma_{n-1}))$, $(\mathbf{r}(\sigma_{n-2}),t(\sigma_{n-2}))$ represent the \emph{memory} at time $n$ of the coordinates of the previous part of the trajectory. Different observers may disagree about the value of the distance $\mathbf{r}_{4}=(\mathbf{r}(\sigma_{n-1}),t(\sigma_{n-1}))-(\mathbf{r}(\sigma_{n-2}),t(\sigma_{n-2}))$ between the two events that correspond to the observation of the object along the trajectory at times $n-1$ and $n-2$. Some observers may judge that the trajectory is straight, others may judge that is curved.

Taking quantum mechanics into account just means that we allow for incomplete knowledge about the position $(\mathbf{r}(\sigma_{n}),t(\sigma_{n}))$ of the observed object. It becomes a four-dimensional sphere rather than a point. We should also allow for the memory of an event to get fuzzier and fuzzier as time goes, so that the volume of the sphere that correponds to all points $(\mathbf{r}(\sigma_{n-2}),t(\sigma_{n-2}))$ that are not excluded by the potential knowledge at time $n$ is expected to be greater than that of all points $(\mathbf{r}(\sigma_{n-1}),t(\sigma_{n-1}))$ not excluded by potential knowledge at the same time.

The points $(\mathbf{r}_{4})_{j}$ that define such a sphere should be read as follows: "The position of the object is $(\mathbf{r}_{4})_{1}$, or $(\mathbf{r}_{4})_{2}$, or $(\mathbf{r}_{4})_{3}$, or..." In an algebraic representation, the word `or' is formally translated to `$+$', so that the state of incomplete knowledge is represented by a superposition. Some people reads the quantum-mechanical `$+$' as `and', which leads to all sorts of nonsensical interpretations of the formalism. This is the reading at the heart of the `many-worlds' interpretation, as well as the idea that the shape of space-time undergoes wild fluctuations at tiny scales, creating a `space-time foam'.

To observe the detailed structure of space-time, to determine whether it fluctuates wildly or not, we need a massive apparatus. Only then are we allowed to define a wave function that represents a superposition of more or less fluctuating alternative space-times. The smaller scales we wish to probe, the more energy we need to focus on this small area. This, in itself, means that space-time will be wildly deformed, according to general relativity. But this is a result of our actions, of our assembly of a powerful apparatus, our focusing of a high energy beam on a small spot, rather than a property of space-time itself (whatever that would mean). There is no storm in the sea of space-time if we do not start blowing the wind ourselves. 

\subsubsection{Elementary fermions and bosons}
We have motivated the Dirac equation from the directed nature of time without making any assumptions about the nature of the specimen to which it applies. How does this go together with the fact that there are bosons with integer spin? Are they not invited to the party? Are they not allowed as objects of experimental inquiry?

Let us first distinguish between elementary and composite fermions. Implicit in the derivation of the Dirac equation is the assumption that the specimen is not spatially extended. This is so since we are considering a continuous wave function specified by differential operators. Such an approximation is justified only if the position of the specimen can be specified to arbitrary precision with a single 4-vector $\mathbf{r}_{4}$. This means that it can apply to elementary fermions only.

We can compose bosons out of an even number of elementary fermions. Such composite bosons can therefore be used as specimens in experiments, and they can be observed as ordinary objects by the naked eye if they are large enough. Therefore, we need to explain only why the elementary bosons should not be regarded as proper objects.

What we try to do is to account for all such bosons as bookkeeping devices without independent existence. We regard massless elementary bosons, such as photons or gravitons, as abstract lists of attributes that make it possible to say that object $O_{1}$ is interacting with object $O_{2}$. This is possible only if the relational attributes of $O_{1}$ changes by the amount $\Delta\upsilon$ and the relational attributes of $O_{2}$ changes by the opposite amount $-\Delta\upsilon$, given the conservation laws for such attributes (Fig. \ref{Fig98}). The elementary boson is just the list $\Delta\upsilon$ together with a hypothetical link between $O_{1}$ and $O_{2}$, indicating that these objects \emph{may} have been interacting. We call such an entity a \emph{pseudoobject}. If there is only one pair of objects whose attributes are observed to change by the opposite amount, then the interaction is \emph{identifiable}. If there are many candidate object pairs, then we say that the interaction is \emph{quasi-identifiable}.

The basic reason why we need such pseudoobjects is that two interacting objects are most often separated spatio-temporally. We can make a subjective, spatial distinction between Jupiter and Mars, at the same time as we say that they interact gravitationally. The projections of their states onto space-time do not overlap. An additional link is necessary to identify them as possibly interacting.

By definition, two interacting objects preserve their identity in the process. Their rest masses are not changing. Their energies may change, but does not have to. If we are to assign any specific rest mass to a pseudoobject, we must therefore set it to zero.

We have argued that implicit epistemic minimalism implies a finite upper speed limit about which all subjects agree. What kind of entity can travel at this speed? Any object is potentially perceivable, and by continuous observation we can follow its trajectory. It may appear to travel very fast. But velocities are relative, and we may equally well imagine that the object is at rest and that you, the observer, is travelling equally fast in the opposite direction. A second observer who is moving in relation to yourself will always judge the speed of the object differently than you do. It is clear that the upper speed limit cannot be associated with any object.

On the other hand, the conclusion that two objects are potentially interacting cannot depend on the state of motion of the observer. All subjects must agree about that, according to individual epistemic invariance. If different subjects had different opinions about the interactions of objects, universal physical law would break down. We have two problems with one solution: it is the pseudoobjects which travel at the speed of light, making the judgement which objects are interacting with which unambiguous and universal.

In fact, it is not appropriate to say that pseudoobjects are travelling at all, since they are no objects. Rather, we should express ourselves as follows. Consider two events $A$ and $B$, where $A$ is the event where object $O_{1}$ changes its attributes by the amount $\Delta\upsilon$, and $B$ is the event where object $O_{2}$ changes its attributes by the opposite amount $-\Delta\upsilon$. Objects $O_{1}$ and $O_{2}$ should be associated in an identifiable (or quasi-identifiable) interaction if only if the spatial distance between $A$ and $B$ divided by the corresponding temporal distance equals $c$.

What about the massive elementary bosons? They may be associated with object transformations. In this case there is no need to introduce an explicit link to identify the involved objects, as in the case of interactions between spatially separated objects. The states of the transforming objects overlap in space-time, as indicated in Figs. \ref{Fig46} and \ref{Fig47}, so that the identifiaction is automatic. However, we can refer to explicit epistemic minimalism in order to argue that at a fundmental level, we should consider only transformations in which one entity divides into two, or two entities merge into one. That is, we should decompose a more involved transformation into a set of such simpler transformations with `three legs'. We regard the massive bosons as a way to associate the different members of such a set of simpler transformations with each other in this kind of decomposition. We call the corresponding links \emph{cryptoobjects}. 

The reason why we allow three-legged transformations only is that it is impossible in principle to \emph{verify} the occurrence of transformation graphs with degree four or higher. To be able to say that four or more legs meet at exactly one point in space-time means that we must \emph{exlude} all small-scale decompositions into sets of vertices with three legs. This is impossible due to the presumed existence of a minimum measurable Lorentz distance $l_{\min}$. Explicit epistemic minimalism states that any attempt to introduce physical processes that in principle cannot be observed, or distinguished from other processes, should lead to erroneous physical answers.

The fact that a given reaction involving elementary fermions can sometimes be decomposed in several ways with the help of cryptoobjects means that the cross sections of some fundamental reactions have to be equal, as exemplified in Fig. \ref{Fig114}. The decompositions are equally good, and their difference is imaginary. Cryptoobjects can be assigned rest masses in a formal sense if we apply the conservation law for the invariant mass to the corresponding link. At the imagined time during which the link exists, the ingoing objects exist no more, and the outging objects have not yet come into existence.

In summary, we look at elementary bosons as tools that help us represent physical law efficiently. The physical state on which physical law acts can be efficiently represented in terms of elementary fermions and their relational attributes. Elementary bosons and fermions cannot be mixed in superpositions, meaning that our state of knowledge cannot be such that we know there is an entity around, but we do not know whether it is an elementary boson or fermion. Such a situation can arise only if we give up the distinction between physical law and physical state. 

\subsubsection{The spin-statistics theorem}
We have already motivated Pauli's exclusion principle, which applies to any directly perceived object, or any indirectly perceived quasi-object. Therefore it applies to all elementary fermions of which these objects are composed. Thus all elementary fermions (with spin $1/2$) obeys Fermi-Dirac statistics.

Pseudoobjects and cryptoobjects must be assigned integer spin, if we apply the conservation of angular momentum to the vertices at which they are `emitted' and `absorbed' by ordinary objects. We must bear in mind, however, that these processes are purely imaginary bookkeeping exercises that are introduced to express physical law in a pictorial way. Since they are imaginary entities, there is no reason why they should obey Pauli's exclusion principle. For example, two pseudoobjects being in the same state just means that there are two objects which change their attibutes by the same amount $\Delta \upsilon$, and two other objects which change their attibutes by the opposite amount $-\Delta \upsilon$. This is perfectly allowed, of course, as long as the four involved objects can be distinguished both before and after the change, that is, if they are all in different states both before and afterwards (Fig. \ref{Fig101}). In other words, all elementary bosons obey Bose-Einstein statistics.

The spin-statistics theorem applies to composite bosons as well as to elementary bosons. One may argue that a boson composed of an even number of fermions is an object just like any other, and should therefore abide by Pauli's exclusion principle. However, we present an argument in Section \ref{spinstatistics} that aims to show that such composite bosons escape Pauli's exclusion principle at least in a controlled experiment in which a property that applies to the composite boson \emph{as a whole} is observed, and a wave function for a collection of such composite bosons can be defined. We also assume that the composite bosons in this collection are identical and weakly interacting.

\subsubsection{Particle masses}
In principle, it should be possible to determine the (squared) rest masses of all elementary fermions as eigenvalues of an evolution operator $\bar{B}_{P}$ (Eq. [\ref{generalevo}]). These matters are discussed superficially in Section \ref{evconsequences}. The specimen $OS$ in a corresponding experiment is an elementary fermion, and the observed property $P$ is its species. If we assume that such a fermion species is specified by $m$ internal attributes (like charge), then $\bar{B}_{P}$ becomes an $m\times m$-matrix. Since the rest mass cannot be known exactly according to the Heisenberg relation for rest masses discussed above, the matrix is not diagonal. This means that the operator that corresponds to the particle species property $P$ does not commute with the evolution operator. In other words, elementary fermions need not be stable, and may be found to transform into each other upon repeated observation. Since rest masses are always non-negative, we get a Dirac equation, meaning that the (non-squared) rest masses must be eigenvalues to the square-root operator $\sqrt{\bar{B}_{P}}$.

The fact that rest masses of objects cannot be determined precisely according to a Heisenberg uncertainty relation means that there cannot be any fermion with zero rest mass. Such a mass is special, and a particle with zero mass can in principle be distinguished from massive particles by its qualitatively different behavior. We cannot form a superposition between a massless and a massive particle. Simply put, all elementary fermions are massive.

We have already argued above, from our conceptual perspective, why photons and gravitons must be massless, whereas gauge bosons involved in weak and strong transformations must be massive. We called these entities pseudoobjects and cryptoobjects, respectively.

\subsubsection{The second law of thermodynamics}
As discussed above, we identify probability with a \emph{relative} volume, with the state space volume $V[S_{j}]$ of an alternative $S_{j}$ divided by the volume of the state $S_{O}$ of the system $O$ to which the alternative applies, before we learn which alternative comes true. Similarly, we relate the entropy $\mathcal{E}$ of a state $S$ or $S_{O}$ to the \emph{absolute} volume of this state, according to $\mathcal{E}=\log(V[S])$ or $\mathcal{E}_{O}=\log(V[S_{O}])$.

The physical state $S$ corresponds to a state of potential knowledge in such a way that $V[S]$ increases when the knowledge shrinks, and vice versa. Therefore we expect the entropy of the entire world to have been \emph{decreasing} in the process during which intelligent life developed. However, we argue in Section \ref{entropy} that it will finally start to increase again, according to Fig. \ref{Fig124}. The detailed argument is a little bit involved, but basically it is the same as the usual one. After sufficiently long time we approach thermodynamic equilibrium, so that the distinctions of perception can no longer be upheld. The knowledge shrinks and the entropy increases without bound, as the boundary $\partial S$ of the state $S$ recedes towards infinity.

We may look at the personal death in the same way. The knowledge of each of us can be encoded in a personal physical state $S^{k}$. The physical state of the world is the intersection of the personal states of all subjects $S=\bigcap_{k}S^{k}$. When we are about to die we may say that $\partial S^{k}$ expands until its boundaries disappear. In the absence of the distinctions and contrasts of perception, nothing is excluded. I come to think of the first few rows in the poem \emph{There is a lake and nothing ever more} by the Swedish poet Hjalmar Gullberg:
\footnote{Translated by Judith Moffett, Poetry Magazine (June 1976). The poem was first published in Hjalmar Gullberg's collection \emph{D\"{o}dsmask och lustg\r{a}rd} (P. A. Norstedt \& S\"{o}ners F\"{o}rlag, Stockholm, 1952). The Swedish original reads: Det finns en sj\"{o} och sedan aldrig mer / och floden sl\"{a}tas ut i den gr\r{a} spegeln / som ingen strand inramar, inga roddarslag / ska spr\"{a}cka.}

\begin{quote}
\emph{There is a lake and nothing ever more,\\
and the river is smoothed out in the gray mirror\\
no strand frames, no oar strokes\\
will ever crack}
\end{quote}

This is the opposite to the distinctions of life expressed so vividly in the song by Violeta Parra cited in the preface.

In any case, we see that the second law of thermodynamics is not valid in an absolute sense in our conceptual framework. There is a slightly different sense in which it is upheld more strictly, though. State reductions $S(n)\rightarrow S(n+1)\subset u_{1}S(n)$ occur quite regularly. They correspond to moments when we learn something new, when something happens that is not bound to happen. Repeated such reductions are necessary to avoid determinism, which is not consistent with the deduced incomplete knowledge. On the other hand, the same incomplete knowledge never goes away, meaning that the volume $V[S(n)]$ cannot decrease steadily as $n\rightarrow\infty$; it is bounded from below. We have $V[u_{1}S(n)]> V[S(n+1)]$ whenever a state reduction takes place. This volume decrease must therefore be countered by an expected volume increase caused by the evolution $u_{1}$:

\begin{equation}
\langle V[S]\rangle < \langle V[u_{1}S]\rangle
\label{evolincrease}
\end{equation}
This expected volume increase corresponds to an expected entropy increase, of course.

We may also turn our eyes backwards, and conclude from the above equation that whenever we use physical law to retrodict the past from the present physical state $S(n)$, we are expected to see an entropy that decreases towards zero. This fact explains the minuscule entropy at the Big Bang that puzzles som many people. It does not correspond to the entropy of an actual physical state, just to the entropy of the deduced state that results after applying the inverse evolution operator $u_{1}^{-1}$ many times to $S(n)$. In fact, there cannot be an actual physical state that corresponds to the Big Bang, since there cannot be any aware subjects in such a state. There is simply no place for intertwined dualism in a singularity. In other words, the actual creation of the world, at which subject and object emerge together, occur at later time than that assigned to the Big Bang. We may say that we trade the mystery of the extremely low entropy just after the Big Bang for the mystery of the appearance of aware beings who are able to make distinctions.

\subsubsection{The postulate of \emph{a priori} equal probabilities}
While talking about statistical mechanics, we may take the opportunity to discuss briefly the perspective given by the present approach to the postulate of \emph{a priori} equal probabilities. Loosely speaking, the postulate states that each microstate compatible with given macrostate should be assigned equal probabilities when statistical weights are calculated.

We may translate a macrostate to the physical state $S_{O}$ of a system of interest, and a microstate to an exact state $Z$ that is an element of the set $S_{O}$ in state space. However, we should avoid explicit reference to exact states, since they are not observable. Instead, we may associate a microstate with a well-defined group $\Sigma$ of exact states that can in principle be observed as a state of a microscopic object that is part of the system of interest. These groups $\Sigma$ can be chosen to be equivalent in the sense that they contain the same number of exact states. They provide a unit in which the volume $V[S_{O}]$ can be measured.

Now, in the present approach we cannot assign probabilities to microstates, but only to macrostates that correspond to an alternative $S_{j}$ in a future observation. The alternative $S_{j}$ is a division of $S_{O}$ in a limited number of `slices', and corresponds to a directly perceived macrostate just as much as $S_{O}$ does. Different alternatives $S_{j}$ may have different volumes (different probabilities), but it has no meaning to assign different volumes (different `probabilities'), to different exact states $Z$ or groups of such states $\Sigma$. They are the units in which volumes (probabilities) are measured, by definition.

In other words, the postulate of equal \emph{a priori} probabilities is a self-evident consequence of the way in which we have defined probability to be proportional to state space volume.

\subsubsection{Conservation of information}
In traditional approaches to physics, we can translate the reversibility or invertibility of physical law at the fundamental level to the statement that information is conserved. If a physical state can be evolved forward from time $t_{1}$ to time $t_{2}$, then the information contained in the state at time $t_{2}$ is inherent already in the state at time $t_{1}$. No new information can be obtained. If the physical state can also be evolved backwards from $t_{2}$ to $t_{1}$, then the information contained in the state at time $t_{1}$ is still there in the state at time $t_{2}$. No information is lost. In short, information is conserved.

In classical mechanics, physical law evolves a point in phase space to another point in this space along a well-defined trajectory. In quantum mechanics, physical law evolves a vector of unit length in Hilbert space to another vector of unit length. Its tip traces out a well-defined trajectory just as in classical mechanics. In both cases the evolution is point-wise. The only difference is that the evolving point belongs to different kinds of state spaces. Information is conserved in both cases according to the general consideration above.

In the present approach we abandon this pointwise description of the physical evolution, and therefore information does not have to be conserved. Nevertheless, we keep the reversibility or invertibility of physical law. As we see it, the evolution $u_{1}$ can be applied only to physical states $S$ that correspond to actual states of knowledge, or states that in principle can be realized as states of knowledge. Since knowledge is incomplete, such states $S$ are always sets of several exact states $Z$. These states $Z$ are the elements or `points' of state space $\mathcal{S}$. Since the elements $Z$ are not themselves in the domain of $u_{1}$, the evolution is not point-wise. 

Of course, we are not allowed to deviate from quantum mechanics when it comes to physical predictions, even if we do not see it as the fundamental level of description. How is it possible then to come to different conclusions as regards information conservation? As we see it the conservation of information inherent in the formalism of unitary quantum mechanical evolution is devoid of physical content; it is a conclusion that stems from a misunderstanding of the meaning of the formalism.

We have argued above that the Hilbert space and the wave function are defined during a well-defined experiment $C$ only. The evolution of the wave function is taken care of by an abstract parameter $\sigma$. We may see a change of $\sigma$ as a proxy for a change of the experimental setup that changes the expected time difference between the start of the experiment and the final measurement. We may, for example, move a detector farther away from an electron gun. In this way we define a \emph{one-parameter family} of experiments $C(\sigma)$. In this light, the conservation of information just means that if we first move the detector farther away and then back to its original position, we get the same probabilities for different alternative outcomes as if we did not move it at all.

If we accept that the present approach allows a changing information content in the physical state, in what way does this information actually change? We may identify the information with the inverse entropy $\mathcal{E}^{-1}=1/\log(V[S])$. According to the above discussion about the second law of thermodynamics we may therefore say that the evolution $u_{1}$ tends to decrease the information content of the physical state, a tendency that is balanced by the information gain in a state reduction. There is no general law that tells us for an arbitrary state $S(n)$ whether the information increases, decreases or stays constant as time $n$ goes.

By abandoning point-wise evolution in state space, we enable information gain or loss at the same time as we keep the reversibility of physical law.
A crucial point is that we define the information and entropy via the same object $S$ as we use to define the fundamental evolution $u_{1}$. No strain between `microscopic' reversibility (information conservation) and `macroscopic' irreversibility (information loss) can therefore arise.

\subsubsection{The cosmological constant}
The state space volume expansion expressed in Eq. [\ref{evolincrease}] cen be expected to hold for any state $S$. We may cover the entire state space $\mathcal{S}$ with a large number of such states. Therefore we may say that the entire state space expands when the evolution $u_{1}$ is applied. On the average, we can expect the same expansion factor each time $u_{1}$ is applied, so that the expansion becomes exponential. The increase of $V[S]$ means that the \emph{uncertainty} of the value of some attributes increase. Not all attribute values can become more and more uncertain without limit. We argue in Section \ref{expansion} that only spatial distances can. The exponential expansion of state space $\mathcal{S}$ is therefore taken care of by an exponential increase of the uncertainty of spatial distances. If the forces that attract objects to each other are weak, for example if they are located far away from each other, the exponential increase of the \emph{distance uncertainty} translates to an exponential increase of the \emph{expected distance} in an actual measurement. In this way we try to motivate a positive cosmological constant from general principles.

\chapter{\normalfont{OUTLOOK}}

In this chapter I let my thoughts wander, extrapolating from the material treated in the main body of this study. I will not make any definite claims, just discuss some possibilities.

\section{The neurological basis of intention and choice}
\label{neurological}
I have argued repeatedly that there seems to be a `white hole' in physical law, that allows intentions and choice to `flow out', but no physical influences to `flow in' (Sections \ref{twoways}, \ref{individualsubjects} and \ref{probabilities}). In other words, it seems that we must allow for choices that have an effect but no cause. By saying that the choice has no cause I mean that there is nothing in the physical state that precedes it that makes it possible to deduce its appearance by means of physical law, neither in a deterministic, nor in a probabilistic sense.

The assumption of detailed materialism (Assumptions \ref{localmaterialism} and \ref{localmaterialism2}) means that every subjective change corresponds to a change in the state of some physical object (possibly a deduced quasiobject). The appearance of an intention in a subject therefore means that the state of some object in her body changes. At the same time, according to the discussion above, sometimes there is no cause of this change of state. How is this possible?

It is self-evident that the potential knowledge of the body of any subject is always incomplete, just like the knowledge of the entire world. Every elementary particle in our bodies obeys the Heisenberg uncertainty relation. This provides the room for a subjective change with no known cause, a state reduction of the body that cannot be predicted.

But exactly where in the body does the crucial lack of knowledge reside when it comes to intention? In this case the lack of microscopic knowledge must somehow couple to macroscopic choices, just like in the Schr\"{o}dinger's cat \emph{gedankenexperiment}. In that case the lack of knowledge corresponds to uncertainty when a single radioactive nucleus will decay. At the moment it decays, when the state of the nucelus reduces, it triggers the death of the cat via some ingenious mechanism involving a gun or some poison. Returning to intention, we should look inside the brain to find a corresponding magnifying mechanism, of course. An intention followed by a choice must have its origin in the organ in which the nerve signals that cause bodily action starts. In contrast to the case of the decaying nucleus, it should not even be possible to assign probabilities for the appearance of different intentions at the crucial location.

The existence of such a place in the brain where an unknowable seed of intention can sprout is essential in the bodies of all subjects, at least if the web of defintions that I have weaved is accepted. Namely, I have defined individuals by their ability to form independent intentions, as discussed in Section \ref{individualsubjects} (Definitions \ref{individuals}, \ref{intentsubject} and Assumption \ref{separatesubjects}).

It seems to me that the appearance of traveling action potentials across the cell membrane of neurons cannot be the basic layer in the process where the physical change brought about by an intention is magnified. Neurons are very comlex entities with a rich internal landscape, like most other cell types. Furthermore, they are fairly large objects, whereas the structure in which the seed for intention is formed has to be small and `well defended', so that it cannot be effectively probed from the outside to determine its state. (Of course, if the probe is forceful enough it can penetrate the defences of any structure inside which the seed may be formed, but if structure and seed are small and delicate, this means that they are destroyed in the process. No actual knowledge is gained. This is the same argument as that of the Heisenberg microscope, leading to the conclusion that the position of the studied object becomes more and more uncertain the smaller it gets, or the more energetic light is used as a probe.)

\begin{figure}[tp]
\begin{center}
\includegraphics[width=80mm,clip=true]{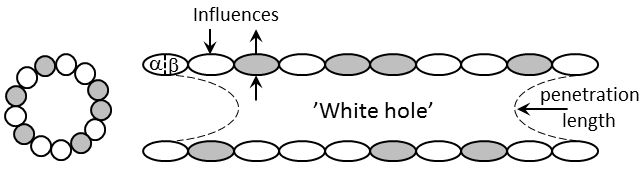}
\end{center}
\caption{Microtubules are composed of tubulin molecules, which have two distinct parts $\alpha$ and $\beta$. The relative orientation of these parts can change, creating (at least) two conformations of the tubulin molecule (white and grey). Environmental influences may affect the state of the microtubule, and so may probably the water that fills it. One may speculate that the state of this water is largely unaffected by the environment, at least far away from the two ends.}
\label{Fig142}
\end{figure}

When I first read Roger Penrose's account of microtubules in his book \emph{Shadows of the mind} \cite{penrose} I was very intrigued. The idea that these structures constitute a basic layer of information processing in the brain have been around since the seventies \cite{atema}, and Penrose argued persuasively for the case. I am just an interested layman, and I will not try to review the morphology and function of the microtubules in any detail. I would just distort the facts. Instead, the reader is referred to Penrose's book, or a recent review by Hameroff and Penrose \cite{hameroff}.

Nevertheless, I will try to describe some of their features very superficially (Fig. \ref{Fig142}). I think that microtubules might fulfil the requirements for a structure inside which unknowable seeds of intention may have their origin, and I want to be able to discuss why this might be so. Microtubules are composed of the protein \emph{tubulin}. The circular circumference of the tube is composed of thirteen tubulin molecules, whereas its length is arbitrary and may consist of a very large number of molecules. The tubulin molecule is somewhat peanut-shaped and can be divided into two parts, called $\alpha$ and $\beta$. There are at least two discrete conformations of the molecule. These are related to the relative orientation of $\alpha$ and $\beta$. A conformational change possibly corresponds to a discrete change of the electric dipole moment.

These qualities suggest that the state of a tubulin molecule may correspond to a bit of information in a computation that is carried out on its surface. In that case the cylindrical tubulin lattice acts as a cellular automaton. Signals might be processed and transmitted from one end of the microtubule to another. This opens up the possibility that the computational power of the brain is vastly bigger than that suggested by the number of neurons, since each neuron contain about $10^{9}$ tubulin molecules \cite{georgiev}, which may change state $10^{6}$ times faster than a nerve signal is transmitted from one neuron to another.

The microtubule interior is mostly filled with water, but may contain some other molecules as well. One may speculate that the state of this water is unknowable to a large extent, since it is shielded from the outside by the regularly arranged tubulin lattice. This lattice may act like a kind of armor that protects influences from the outside to reach the inside. This might be so because of the rigid structure formed by the tubulin molecules, which only allows a discrete number of collective states of the entire microtubule surface, corresponding to the different possible combinations of tubulin conformations. The number of possible states of the water inside the tube is probably much higher, maybe forming a continous set. If this is really the case, some information about the inside is necessarily lost at the surface.

Since microtubules are often very long, the amount of information about its interior that can be tapped from its ends is probably much smaller than the information stored in an exact knowledge of its interior state. One might speak about a `penetration length', the maximum distance that outside influences may travel towards the center of the tube via the ends (Fig. \ref{Fig142}). In the vocabulary used in this text, we would say that there might be very little conditional knowledge that relates the state of the microtubule interior with the state of the outside world. In conventional quantum mechanical language, we would say that there is little entanglement between the interior and the outside world. If there is substance in these speculations, there might indeed be white holes hiding inside microtubules.

Hameroff and Penrose speculate that there may be quantum computations going on in the tubulin lattice. This would require that the tubulin lattice itself is disentangled from the environment, so that `clean' superpositions between different lattice conformations can be upheld. In the vocabulary of this text, the conformation of the lattice would be unknowable. It should be emphasized that my own perspective is different. I am arguing that it is the state of the interior of the microtubules that is unknowable, not the state of its lattice surface. In the conventional vocabulary, many such interior states coexist in a quantum superposition.

Recent investigations of the microtubules isolated \emph{in vitro} indicate anomaluous optical and electronic behavior. This may be seen as an indication that quantum effects are indeed at play \cite{sahu1,sahu2}. It seems that the presence of water in the microtubule interior is crucial for the appearance of these effects.

If it is indeed true that the microtubule interior is a white hole whose state is largely unknowable, then it is impossible to predict how this interior affects the conformation of the tubulin surface lattice. We get the effect without any knowable cause that we are looking for. The effect, the change of the tubulin lattice conformation, may act as input for a computation or a signal that is transmitted along the microtubule. The output of this process may then be amplified to a larger scale, affecting the entire organism and its environment. We get an action $\mathcal{A}$ such as that indicated in Fig. \ref{Fig24}.

This picture does not exclude, of course, that a large part of those physiological processes that affect our aware state and determine choices do have a knowable origin. It is well known that processes outside the microtubulues determine their formation and dynamics. Such outside process may affect the hypothetical computations and signal processing that goes on in the tubulin lattice to a larger degree than the unknowable input from the interior. This fact is indicated in Fig. \ref{Fig142} as an arrow of influence that point at the microtubule from the outside. Such influences would correspond to those aspects of the behavior of aware beings that are dictated by physical law. 
 
The purported unknowability of the microtubule interior means that its action on the surface may have non-local features. Conditional knowledge may connect the interior state at one point $A$ to another distant point $B$. In conventional language, faraway points in the microtubule interior may be entangled. The influence of the interior on the tubulin lattice that causes a conformation change at point $A$ may then dictate an immediate change at point $B$. Since microtubules in brain cells are thought to reach lenghts of millimeters or even centimeters, we may get non-local quantum effects that connect different parts of the brain at macroscopic distances.

\begin{figure}[tp]
\begin{center}
\includegraphics[width=80mm,clip=true]{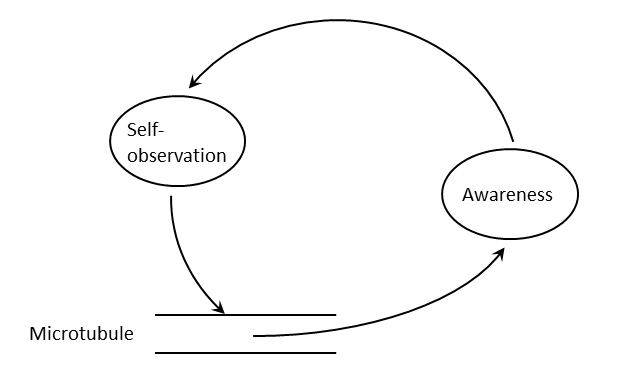}
\end{center}
\caption{G\"odelian microtubules? The tubulin lattice that forms the surface of microtubules might be seen as an armor that prevents perfect self-awareness, both in terms of the state of our own bodies and of our intentions. If this picture is correct, and if the physical state can be identified with a state of knowledge, then the appearance of some ideas, images, associations and conclusions may lack a computational basis - a cause dictated by physical law acting on a symbolic representation of the physical state.}
\label{Fig143}
\end{figure}

The idea that the microtubule surface acts as an armor behind which the interior can hide may be seen as a mechanism that prevents perfect self-reference (Fig. \ref{Fig143}). Such a mechanism is necessary at two levels. First, the knowledge about the body that is possible to deduce by observation of the objects it consists of must be smaller than the knowlede of the body inherent in the state of subjective perception via the assumption of detailed materialism. If this were not so, the knowledge about the body would become a proper subset of itself whenever we are aware of more things than the deduced knowledge about our own bodies. Second, an indeterministic world requires an unknowable seed of intention, as discussed above, for which not even probabilities can be assigned. Detailed materialism then demands a physical armor around this seed, since it is assumed to be rooted in the physical world and cannot originate `in the heavens'. This means that if microtubules cannot fulfil this role, there has to be some other small scale structure in the brain that does the job.

\section{Many worlds}
Several subjects can perceive the same object. One subject can perceive several objects. These basic facts are illustrated in Fig. \ref{Fig7b}. Can we replace the word `object' with `world' in these statements? It is clear that several subjects can perceive the same world. But can one subject perceive several worlds? From the epistemic perspective, the idea of a multiverse is meaningful if and only if the answer to this question is positive.

\begin{figure}[tp]
\begin{center}
\includegraphics[width=80mm,clip=true]{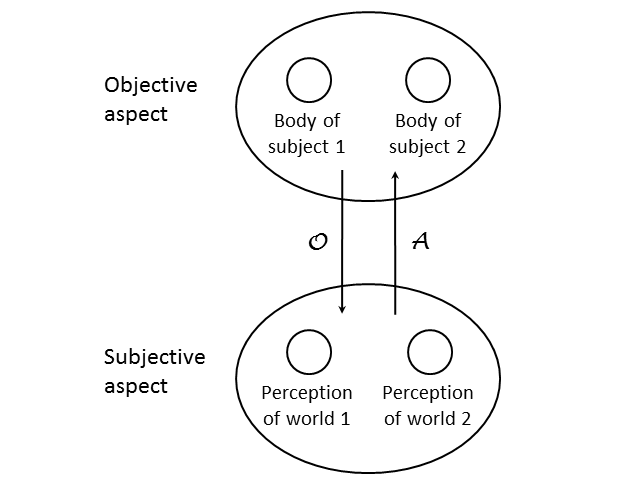}
\end{center}
\caption{Possible symmetries between the subjective and objective aspects of existence. There may be several subjects. These are associated to several bodies in the objective aspect, according to detailed materialism. Conversely, we speculate that there there may be several worlds. In that case these are associated to several independent sets of perceptions in the subjective aspect, according to the epistemic approach to physics. The subjective and objective aspects may also influence each other symmetrically, as discussed in relation to Fig. \ref{Fig24}.}
\label{Fig151}
\end{figure}

This possibility is attractive from an aestethic point of view, since it would make the roles of the subjective and objective aspects of the world more symmetric. In the previous section we discussed the possible symmetry sketched in Fig. \ref{Fig24}. The subjective aspect may influence the objective aspect via the apperance of intention an choice, and the objective aspect influences the subjective aspect by observations that are determined (in part) by physical law. These different kinds of influences in opposite directions are equally fundamental; one of them cannot be explained in terms of the other. We could further add to this symmetry by saying that just as one objective world may contain many subjects, one subject may `contain' many objective worlds (Fig. \ref{Fig151}).

But what might it mean to say that a subject can perceive several worlds? Clearly, we must make a clear distinction between the perception of two objects that belong to the same world, and two objects that belong to different worlds. We may say that two objects $O_{1}$ and $O_{2}$ are part of the same world $W$ if and only if they are associated by some relational attribute $r_{12}$. Physical law couple the evolution of two objects if and only if there is such a relational attribute. This would mean that the statement that two objects belong to different worlds means that the state $S_{O1}$ of $O_{1}$ does not influence any future state $S_{O2}$ of $O_{2}$, and vice versa.

It may seem that this fact makes the notion that one and the same subject can perceive different worlds inconsistent. To make such a notion meaningful, the subject must be able to remember objects from world 1 at the same time as she perceives objects from world 2, so that she can decide in practice that there are indeed two worlds. If she cannot do that, she cannot preserve her identity in the movement from one world to another, and the entire discussion loses its meaning. This means that she can carry influences from one world to another. We have introduced a possible relation between objects from different worlds, in conflict with their definition.

Indeterminism provides a loophole. A subject can perform an action in world 2 `inspired' by an experience in world 1 without presuming an influence between the two worlds that is dictated by physical law. This is possible because of the `white hole' in physical law that allows intentions, choices and actions that are dictated neither by necessity, nor probability. We should introduce a relational attribute that associate two objects only if physical law acts on this attribute, changing its value as time goes. If influences between two worlds go exclusively via a perceiving subject who makes use of the `white hole', then we can uphold the distinction between two objects belonging to the same world, and two objects belonging to different worlds.

Let us formalize the discussion a bit, playing with indices. Let $(S_{Ol})_{W}^{k}$ denote the state of an object $O_{l}$ perceived by subject $k$ in world $W$. The index $k$ can vary independently for given $W$ and $l$, meaning that the notion that different subjects perceive the same object in the same world is well-defined. Likewise, the index $l$ can vary independently for given $W$ and $k$, meaning that the notion that a given subject perceive different objects in the same world is well-defined. In contrast, the notion that a given subject perceive the same object in different worlds is nonsensical. We cannot let the index $W$ vary independently for given $k$ and $l$.

\begin{figure}[tp]
\begin{center}
\includegraphics[width=80mm,clip=true]{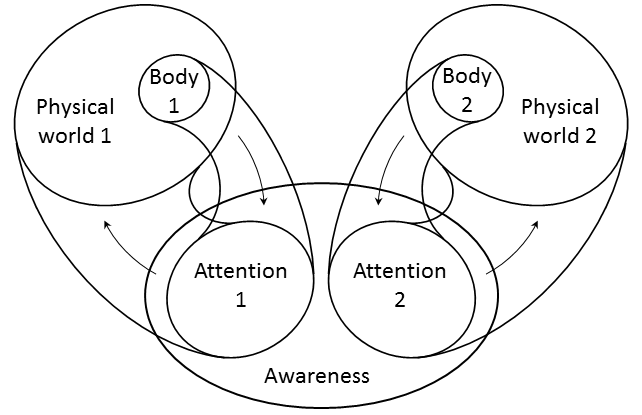}
\end{center}
\caption{A sketch of intertwined duality in the case a given subject may perceive different worlds as her attention changes. Note that we must say that she carries different bodies when she visits different worlds. Compare Fig. \ref{Fig149}.}
\label{Fig148}
\end{figure}

What about letting the indices $W$ and $l$ vary jointly for a given $k$? That is, can a given subject perceive different worlds? Is there a pair of objects $O_{l}$ and $O_{l'}$ with states $(S_{Ol})_{W}^{k}$ and $(S_{Ol'})_{W'}^{k}$? If there is, and the world $W'$ obeys the same laws of physics as $W$, then $W'$ allows several subjects $k$ and $k'$. Then there is also a pair of objects with states $(S_{Ol})_{W}^{k}$ and $(S_{Ol'})_{W'}^{k'}$. Note that we can imagine such a pair even if there is no pair of states $(S_{Ol})_{W}^{k}$ and $(S_{Ol'})_{W'}^{k}$. In that case the presumed existence of the pair $[(S_{Ol})_{W}^{k},(S_{Ol'})_{W'}^{k'}]$ just corresponds to the speculation that there may be other worlds in which different beings live, which we cannot know anything about.

If a pair $[(S_{Ol})_{W}^{k},(S_{Ol'})_{W'}^{k}]$ exists, then subject $k$ must have a different body $\mathcal{B}'$ when she experiences the object $O_{l'}$ in world $W'$ than the body $\mathcal{B}$ that she has when she experiences object $O_{l}$ in world $W$. This is so since the body of a subject must be associated by relational attributes to the objects the body lets her perceive. The body must belong to the same world $W$ as the objects perceived in $W$. Since bodily objects in different worlds are not associated to each other by any relational attribute, by definition of different worlds, the body itself must be considered different, rather than being different parts of the same body. The latter expression presupposes a relation between the different parts.

\begin{figure}[tp]
\begin{center}
\includegraphics[width=80mm,clip=true]{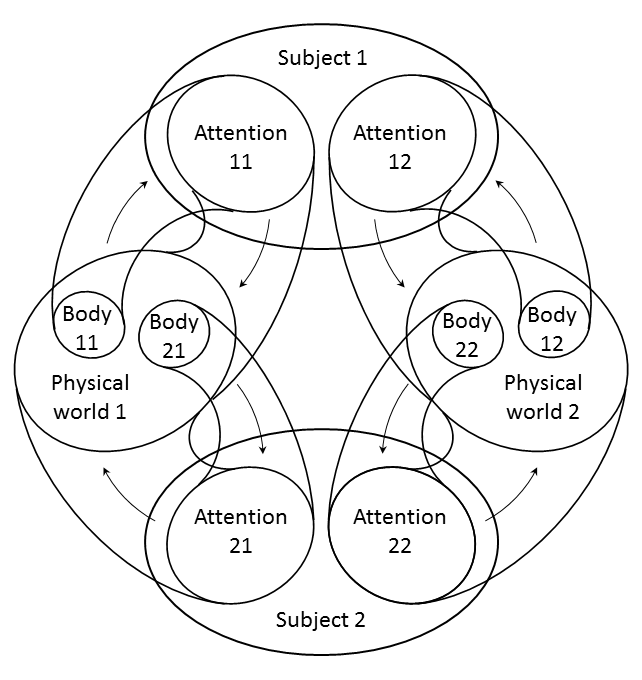}
\end{center}
\caption{If both the worlds indicated in Fig. \ref{Fig148} allow several subjects, like our normal world does, then that sketch has to be further elaborated in order to capture all the relations that are involved.}
\label{Fig149}
\end{figure}

These considerations are illustrated in Fig. \ref{Fig148}. We call the perceptions of different worlds different `attentions'. The awareness of a single subject may transcend a given attention. We should also allow different subjects in this picture. In that case the relations between the concepts that we have used may be illustrated as in Fig. \ref{Fig149}.

It should be stressed that I am not arguing in favor of the existence of different worlds in the abovse sense. I am just playing with the possibilities provided by the structure of relations between different aspects of existence that has been used as assumptions in this study.

\section{Different worlds}
The discussion above about the structure of a hypothetical multiverse offers no conceptual difficulties as long as all the worlds have the same basic characteristics as our own. By this I mean that the same categories of perception are valid in all the worlds, as well as the same conditions for knowledge. In such a situation, there is time in all the worlds, time is inherently directed everywhere, and at each instant each subject in each world may distinguish between mental images that belong to the past, the present, and the future. Also, knowledge is incomplete in all the worlds since the bodies of all subjects are a proper subset of the world itself. This means that a clear distinction can always be made between subject and object, between the observers and the observed.

It is a lot harder to conceive different worlds which have different basic characteristics in this sense. We are stuck in the world we are born into and cannot imagine a completely different kind of existence, for example one without time. This is so, at least, if we accept detailed materialism. Then the degrees of freedom of imagination are determined by the degrees of freedom of the brain, which are determined by the physical law that applies to the world we live in. This physical law is, in turn, expressed in terms of the fixed set of categories and degrees of freedom that applies to this world.

Is it meaningful, then, to consider the possibility for different kinds of worlds in a multiverse? Such a multiverse makes epistemic sense if and only if a given subject can perceive these qualitatively different worlds and remember them as her attention shifts from one such world to another. But how can this be done if the categories of perception are different? If we assume epistemic closure in each of these worlds, the structure of perception would vary from world to world. For example, the very notion of `remembering' relies on the same category of time that we are used to. On the other hand, some `less severe' variations might be acceptable, for example, worlds with and without the distinction between left and right, or worlds with and without the subjective distinction between straight and curved lines.

In this connection, it may be appropriate to turn the perspective around and discuss the possible perceptions of different subjects in the same world, rather than the hypothetical perceptions of the same subject in different worlds. That different subjects inhabit the same world means that they all possess bodies that are part of this world. It must therefore be possible to represent all these bodies by means of the same set of minimal objects (elementary particles) subject to the same physical law. I have tried to derive the form of physical law from epistemic principles that seem reasonable to me, and hopefully to other human beings as well. If this project is judged to be successful, then we have established a direct link between the form and limitations of our perceptions and the workings of the objective world. In so doing, we have treated all subjects in this world as equal; we have not given the preference to any particular kind of aware being when it comes to whose subjective perceptions are to be linked to the workings of the world. This picture can only be consistent if all aware beings that can acquire knowledge have the same type of perceptions, regardless whether they are humans or octopuses. For example, they must all have the same conception of time.

We may imagine a pair of subjects that meet each other in two separate worlds, as illustrated in Fig. \ref{Fig149}. In so doing, they must see two different bodies of their friend in the two worlds, as discussed above, even if these bodies belong to the same subject. Apart from this oddity, there is nothing conceptually troubling in such a situation. Again, problems arise only if the two worlds are qualitatively different in terms of physical laws and forms of subjective perceptions.

\section{The subjective world, the objective world, and the world of reason}

\begin{figure}[tp]
\begin{center}
\includegraphics[width=80mm,clip=true]{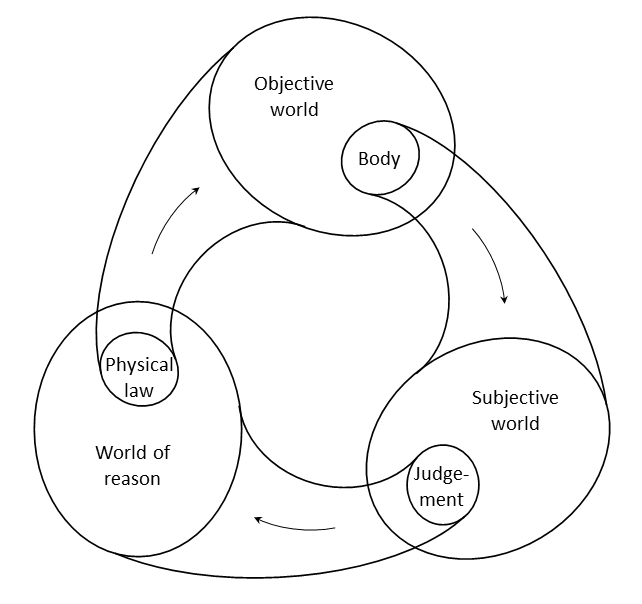}
\end{center}
\caption{Roger Penrose's three intertwined worlds. He does not seem to be dogmatic about the details of the relations between these worlds, but the basic picture is the following. The subjective world emerges from the objective world via the body, which is a subset of the objective world. This is an expression of materialism. The world of reason emerges from a subset of the subjective world, acknowledging the fact that there are other facets of perception than the systematization of experiences and the application of logic. The objective world emerges from a subset of the world of reason, meaning that not all mathematical objects and models have relevance to the categories and laws of physics. Compare Fig. \ref{Fig147}}
\label{Fig152}
\end{figure}

Roger Penrose's \cite{penrose} picture of 'three worlds' (Fig. \ref{Fig152}) has been a great inspiration for me when formulating the idea of intertwined dualism. The basic idea common to the `three worlds-picture' and `intertwined dualism' is, as I see it, that there are several aspects of the world, none of which is more fundamental than any other, each of which emerges from the others. The difference is that there are only two such aspects in my picture of intertwined dualism, whereas there are three aspects in Penrose's picture of three worlds. My departure from Penrose's picture may be compared to the departures discussed by Hut, Alford and Tegmark \cite{hut}.

\begin{figure}[tp]
\begin{center}
\includegraphics[width=80mm,clip=true]{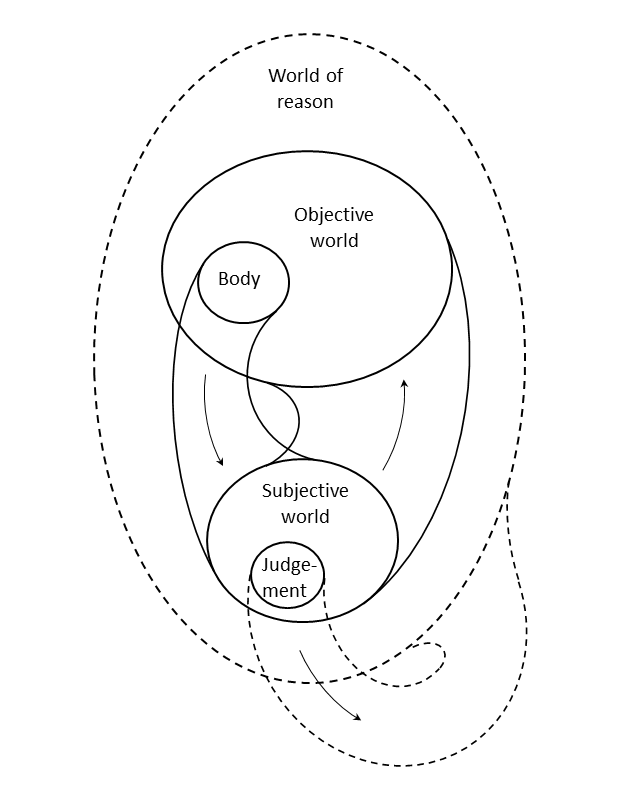}
\end{center}
\caption{The relationships between the subjective and obective aspects of the world, as sketched in Fig. \ref{Fig1}, are defined within the world of reason only. This world emerges from the analytical mind of proper interpretations according to Fig. \ref{Fig2}. We would like to leave room for non-analytical interpretations of perceptions. These are neither proper, nor improper, and are not the subject of this text. Compare Fig. \ref{Fig152}.}
\label{Fig147}
\end{figure}

The reason why I remove the world of reason from the triangle is that I think the very construction of such a triangle relies on reason. What we do, after all, is to try to sort out the logical and set-theoretical relations between abstract aspects or categories. Figuratively speaking, the entire triangle in Fig. \ref{Fig152} baths in a sea of reason. Therefore, I would like to let the world of reason emerge from the subjective judgement in Fig. \ref{Fig152} to embrace both the subjective and the objective worlds, as illustrated in Fig. \ref{Fig147}.

We may look at the world of logic that emerges from the subject as a tool of interpretation, a tool that transforms bare perceptions to knowledge (Fig. \ref{Fig2}). One may speculate about other tools of interpretation that emerge from other parts of the subject. This means that we cannot exclude \emph{a priori} that there is something outside this `plane of reason' in which we place our intertwined world, that we can perceive and acquire knowledge by other means than science. The structure of the world we see in the plane of reason should then be seen as a projection in a wider space of possible interpretations. This would mean that whenever we apply reason to our perceptions we see patterns that conform with the structure in this plane, including physical law. Other interpretations would never contradict these structures, however, partly because the term contradiction would be undefined. Poetically speaking, this intertwined dualism may be seen as a projection of the predicament of existence onto the logical, symbolic subspace of knowledge. In short, I would like to open up for the possibility that the world of reason is not as fundamental as the subjective and the objective worlds.

\chapter{\normalfont{DISCUSSION}}
\label{discuss}

\section{The philosophical heritage}

In this section I will discuss some ideas of a few thinkers who have inspired me. I include some quotes from them that are relevant to this study. Without shame, I choose quotes that express views similar to my own. The aim is not to try to prove my points by saying "look, these great authorities support me, therefore I must be right". Rather, the aim is just to make it clear that my perspective on physics is part of a long tradition, and that many thinkers have been troubled by the same blind spots and inconsistencies in the naive scientific world view as I have been.

Since I was a teenager, I have had great respect for Kant. In fact, he is the only metaphysician that I have ever found interesting. One might say that his basic insight was that a proper ontology does not concern the general qualities of those things that exist, but the general conditions for our knowledge about those things. He introduced the strict epistemic perspective in philosophy that I have tried to apply to physics. Such an epistemic perspective does not deny ontology, it just changes its target.

It is popular among physicists to say that the aim of their subject is not to unravel the ultimate nature of the world, just to ask well-posed questions to nature, and to order the answers they are able to apprehend systematically. That is, physics tries to describe what we can say about the world, not the world itself. This attitude is Kantian, in a sense. Many of these physicists does not seem to draw the philosophical consequences of the approach, however. Most often it is no more than an modest attitude. Kant did draw the philosophical consequences when it comes to metaphysics. He realized that a detailed analysis of what can be said about our ability to form knowledge tells us something fundamental about the structure of the world itself.

A well-known catchphrase in Kant's philosophy is: "We cannot know anything about the thing in itself". Every act of knowledge acquisition has a subjective aspect; it is our intuition about the object that is the only possible target for our knowledge, not the thing itself, detached from the observer. Nevertheless, Kant is not an idealist in the sense that he denies the \emph{existence} of the `thing in itself', he just states that its nature is beyond our knowledge; it cannot be \emph{intuited} or \emph{cognized}. He motivates this conclusion at several places in his \emph{Critique of pure reason} \cite{kant}. The most concise version of his arguments is possibly the following.

\begin{quote}
\emph{[E]ven if we cannot \emph{cognize} these same objects as things in themselves, we at least must be able to \emph{think} them as things in themselves. For otherwise there would follow the absurd proposition that there is an appearance without anything that appears.}
\end{quote}

In the philosophical construction presented in this study, we have taken for granted that the outside world exists in the sense that we have assumed the fundamental ability to distinguish internal objects from external objects. In other words, we are able to know that `something is out there'. The picture of intertwined dualism needs this ability, but it does not need the properties of this `something', seen as things in themselves. It just needs the properties of these things as they appear to us.

Since the only thing we can know about the thing in itself is that it exists, it is not the subject of scientific inquiry, according to Kant. If this proposition is accepted, it may seem that the only alternative is to say that all scientific knowledge is empirical. But this is not so, according to Kant. There are certain forms of perception and reason that must be given \emph{\emph{a priori}} to make sense of empirical observations.

\begin{quote}
\emph{[A] light dawned on all those who study nature. They comprehended that reason has insight only into what it itself produces according to its own design; that it must take the lead with principles for its judgments according to constant laws and compel nature to answer its questions, rather than letting nature guide its movements by keeping reason, as it were, in leading-strings; for otherwise accidental observations, made according to no previously designed plan, can never connect up into a necessary law, which is yet what reason seeks and requires. Reason, in order to be taught by nature, must approach nature with its principles in one hand, according to which alone the agreement among appearances can count as laws, and, in the
other hand, the experiments thought out in accordance with these principles - yet in order to be instructed by nature not like a pupil, who has
recited to him whatever the teacher wants to say, but like an appointed judge who compels witnesses to answer the questions he puts to them.}
\end{quote}

The apprehension of these principles, this design of our own reason, may be said to constitute knowledge that is not empirical, but is discernible to us as a distinct form into which all empirical knowledge fits. To focus on this form rather than the objects that are observed within this form has been called `Kant's Copernican revolution'.

\begin{quote}
\emph{Up to now it has been assumed that all our cognition must conform to the objects; but all attempts to find out something about them \emph{\emph{a priori}} through concepts that would extend our cognition have, on this presupposition, come to nothing. Hence let us once try whether we do not get farther with the problems of metaphysics by assuming that the objects must conform to our cognition, which would agree better with the requested possibility of an \emph{a priori} cognition of them, which is to establish something about objects before they are given to us. This would be just like the first thoughts of Copernicus, who, when he did not make good progress in the explanation of the celestial motions if he assumed that the entire celestial host revolves around the observer, tried to see if he might not have greater success if he made the observer revolve and left the stars at rest.}
\end{quote}

I support this view wholeheartedly. In the preface I stressed that it is vain to try to build physical models that do not use our fundamental categories of perception as a basis, but instead try to invent another basis that either deny the existence of our categories of perception (like the flow of time), or let them follow as consequences. At best, we get an unnecessarily convoluted model. At worst, we get a self-contradictory model.

Kant stresses the lesser importance of the contents of perception by saying that it is given \emph{a posteriori}, in contrast to its form, which is given to us \emph{a priori}.

\begin{quote}
\emph{I call that in the appearance which corresponds to sensation its \emph{matter}, but that which allows the manifold of appearance to be intuited as ordered in certain relations a I call the \emph{form} of appearance. Since that within which the sensations can alone be ordered and placed in a certain form cannot itself be in turn sensation, the matter of all appearance is only given to us \emph{a posteriori}, but its form must all lie ready for it in the mind \emph{\emph{a priori}}, and can therefore be considered separately from all sensation.}
\end{quote}

Another way to put it is to say that the contents of perception is forever changing and possible to manipulate, whereas their form is absolute and given to us once and for all. Therefore they are the best handle to hold on to if we want to open the door to the underlying nature of the world.

In Kant's vocabulary, judgments that make use of the forms of perception but not their content are called \emph{synthetic a priori}. He contrasts such judgments with those that are \emph{analytic a priori}. The latter are defined by Kant as mere tautologies, where the conclusion is just a reformulation of the premise. The idea behind Kant's Copernican revolution can be expressed as the claim that there are true judgments that are \emph{synthetic a priori}, and that the corresponding knowledge is the sound metaphysical basis for all knowledge \emph{a posteriori}, that is, knowledge about matter, about the content of the world as we see it.

This is the line of thought that is used in the present study. The epistemic assumptions that we use as input can be seen as an attempt to capture those judgments \emph{synthetic a priori} that are relevant to a proper formulation of physics, seen as a symbolic model for the behavior of our sensations, of matter as it appears to us. Examples of such assumptions are intertwined dualism, the directed, sequential nature of time, the existence of distinct objects, the distinction between internal and relational attributes of these objects, the concepts of betweenness and straightness, and the possibility of object division. The last assumption can be seen as a prerequisite to set theory, since it gives meaning to the concept of a subset. Each of the daughter object in a division can be seen as a proper subset of the mother object.

This statement about a connection between set theory and the ability of physical objects to divide conforms with Kant's claim that mathematical statements are not \emph{analytic}, but \emph{synthetic a priori}. To arrive at a mathematical conclusion logic is not enough, we need an \emph{intuition}. He claims that this is true in arithmetics, as well as in other branches of mathematics, such as geometry.

\begin{quote}
\emph{Just as little is any principle of pure geometry analytic. That the straight line between two points is the shortest is a synthetic proposition. For my concept of the straight contains nothing of quantity, but only a quality. The concept of the shortest is therefore entirely additional to it, and cannot be extracted out of the concept of the straight line by any analysis. Help must here be gotten from intuition, by means of which alone the synthesis is possible.}
\end{quote}

To say that intuition or visualization is a necessary ingredient in all of mathematics is the same as to say that no mathematics can be conceived independently from the forms of our perceptions of the physical world. Mathematics is slave to physics, in a sense. We have adopted this perspective when we try to relate the orientability of space to the existence of parity-breaking weak interactions, which we claim give meaning to the subjective distinction between left and right. Without such a distinction it is impossible to \emph{intuition} the parity operation $x\rightarrow -x$. The mathematical concept of a vector space would never have been invented.

Kant himself uses a similar line of reasoning when he connects the concept of measured value, or \emph{magnitude}, to the existence of the form of perception we call time.

\begin{quote}
\emph{No one can define the concept of magnitude in general except by something like this: That it is the determination of a thing through which it can be thought how many units are posited in it. Only this how-many-times is grounded on successive repetition, thus on time and the synthesis (of the homogeneous) in it.}
\end{quote}

Kant views the properties of space and time as essential forms of perception, and thus as a basis of judgments \emph{synthetic a priori}.

\begin{quote}
\emph{Geometry is a science that determines the properties of space synthetically and yet \emph{\emph{a priori}}. What then must the representation of space be for such a cognition of it to be possible? It must originally be intuition; for from a mere concept no propositions can be drawn that go beyond the concept, which, however, happens in geometry [...]. But this intuition must be encountered in us \emph{\emph{a priori}}, i.e., prior to all perception of an object, thus it must be pure, not empirical intuition.}
\end{quote}

\begin{quote}
\emph{Time is a necessary representation that grounds all intuitions. In regard to appearances in general one cannot remove time, though one can very well take the appearances away from time. Time is therefore given \emph{\emph{a priori}}. In it alone is all actuality of appearances possible. The latter could all disappear, but time itself (as the universal condition of their possibility) cannot be removed.}
\end{quote}

As Kant sees it, time is a more fundamental form of perception than space.

\begin{quote}
\emph{Time is the \emph{a priori} formal condition of all appearances in general. Space, as the pure form of all outer intuitions, is limited as an \emph{\emph{a priori}} condition merely to outer intuitions.}
\end{quote}

By outer intuitions Kant means perceptions of objects external to ourselves. Internal perceptions such as fantasies does not have to have a spatial relationship, but must still be ordered temporally.

As discussed above, Kant regards the forms of perceptions, such as space and time, to be just that - forms for things as they appear to us. They have nothing to do with the things in themselves, or their relations.

\begin{quote}
\emph{Space represents no property at all of any things in themselves nor any relation of them to each other [...] We can accordingly speak of space, extended beings, and so on, only from the human standpoint. If we depart from the subjective condition under which alone we can acquire outer intuition, namely that through which we may be affected by objects, then the representation of space signifies nothing at all. This predicate is attributed to things only insofar as they appear to us, i.e., are objects of sensibility.}
\end{quote}

\begin{quote}
\emph{[Time] is only of objective validity in regard to appearances, because these are already things that we take as \emph{objects of our senses}; but it is no longer objective if one abstracts from the sensibility of our intuition, thus from that kind of representation that is peculiar to us, and speaks of \emph{things in general}. Time is therefore merely a subjective condition of our (human) intuition (which is always sensible, i.e., insofar as we are affected by objects), and in itself, outside the subject, is nothing.}
\end{quote}

In our vocabulary, we regard space and time to be attributes that relate objects, such as they appear to us. The more fundamental nature of time as compared to space is reflected in our treatement by the introduction of the discrete, sequential time in addition to the relational time that is part of space-time. We respect Kant's assertion that none of these attributes or forms refer to things in themselves, just to things as they appear to us as observing subjects. These appearences, including the appearences of our own bodies, is the only target of physics. The mathematical representation of physical states and physical law should use numbers and symbols that corresponds to forms and contents of perception only.

The fathers of the Copenhagen interpretation of quantum mechanics stressed this fact as much as I do, and as much as I think Kant would have done, had he been around in the 1920s. Nevertheless, I have not seen these physicists referring so enthusiastically to Kant as I have done above. For example, Werner Heisenberg offers a wonderful chapter in his book \emph{Physics and philosophy} \cite{heisenberg} that puts the concepts of modern science in the context of the concepts used throughout the history of philosophy. Heisenberg devotes considerable space to a discussion about Kant, but he does not seem to find any immediate parallells between Kant's concepts and ideas, and the concepts needed to interpret quantum mechanics. My impression is that Heisenberg did not quite understand all aspects of Kant's philosophy. Most importantly, trying to find a counterpart to the `thing in itself', Heisenberg suggests the abstract and pure mathematical representations of elementary particles as a candidate to the `thing in itself'. If I were to speak for Kant, I would say that his view is that physical modelling has nothing to do \emph{at all} with things in themselves, only with appearances.

If I were to point to the `thing itself' in my illustration of intertwined duality (Fig. \ref{Fig1}), I would say it is the paper on which the sketch is drawn, or something like that. We cannot identify it with the `currently unknowable' or `unknowable' according to Fig. \ref{Fig3}. The currently unknowable are things that can be known in the future by choosing the appropriate experimental mode of observation. The unknowable may be said to correspond to everything conceivable that is not part of the world, to the complement to the world, to the negation of the world. In contrast, the `thing in itself' has no such relation at all to the world as we see it, it cannot be conceived or described in the same terms, by the same categories.

Maybe the `thing in itself' can be identified with a philosophical incarnation of God. Every child knows that we run into contradictions if we try to describe a universal God with the same categories or attributes as we use to describe objects that are part of the world we perceive. We cannot relate God spatially to other objects, and we cannot relate him temporally to the sequence of worldy events. If we try, we have to ask ourselves meaningless questions such as "What did God do before he created the world?" Such forms of perception cannot be applied to God, just as they cannot be applied to the `thing in itself'.

One might say that what I have tried to do, in addition to what Kant did, is to establish a one-to-one correspondence between the forms of perception and the form of physical law, via the assumptions of `epistemic closure' and `epistemic invariance'. I have tried to make physics out of metaphysics. One parallell is undeniable: both the forms of perception and physical law are given to us once and for all, they cannot be changed at will. Also, the parallell between the contents of perception and the physical state is obviuos. Both can be changed at will, to some extent. Formulating my project in Kantian terms, one may say that I give arguments for the hypothesis that the form of physical law is knowledge that is \emph{synthetic a priori}; it is possible to derive it from the forms of perception. The derivation goes both ways, as I see it: we may equally well say that the forms of perception follow from the laws of physics.

\vspace{5mm}
\begin{center}
$\maltese$
\end{center}
\paragraph{}

Leaving Kant and returning to the fathers of the Copenhagen interpretation, they looked at quantum mechanics as a literal expression of the fact that physics concerns what can be known about things, rather than the thing in itself. The central role taken by the concept of knowledge in this interpretation of quantum mechancis means that subject and object are both indispensable aspects of the world. In the words of Bohr \cite{bohr6}:

\begin{quote}
\emph{[T]he finite magnitude of the quantum of action prevents an altogether sharp distinction being made between the phenomenon and the agency by which it is being observed.}
\end{quote}

\begin{quote}
\emph{We meet here in a new light the old truth that in our description of nature the purpose is not to disclose the real essence of the phenomena but only to track down, so far as it is possible, relations between the manifold aspects of our experience.}
\end{quote}

Both Bohr and Heisenberg emphasizes that we must use the subjective form of perception as a basis in every attempt to describe nature, as expressed in natural language. Bohr puts it like this:

\begin{quote}
\emph{[T]he recognition of the limitation of our forms of perception by no means implies that we can dispense with our customary ideas or their direct verbal expressions when reducing our sense impressions to order. No more is it likely that the fundamental concepts of the classical theories will ever become superfluous for the description of physical experience.}
\end{quote}

Heisenberg makes a similar point by saying that even if the concepts of science will change in the future, their ultimate foundation in our natural language, reflecting the way in which we perceive the world, will not \cite{heisenberg}:  

\begin{quote}
\emph{The general trend of human thinking in the nineteenth century had been toward an increasing confidence in the scientific method and in precise rational terms, and had led to a general scepticism with regard to those concepts of natural language which do not fit into the closed form of scientific thought - for instance, those of religion. Modern physics has in many ways increased this scepticism; but it has at the same time turned it against the overestimation of precise scientific concepts, against a too-optimistic view on progress in general, and finally against scepticism itself. The scepticism against precise scientific concepts does not mean that there should be a definite limitation for the application of rational thinking. On the contrary, one may say that the human ability to understand may be in a certain sense unlimited. But the existing scientific concepts cover always only a very limited part of reality, and the other part that has not yet been understood is infinite. Whenever we proceed from the known into the unknown we may hope to understand, but we may have to learn at the same time a new meaning of the word `understanding'. We know that any understanding must be based finally upon the natual language because it is only there that we can be certain to touch reality, and hence we must be sceptical about any scepticism with regard to this natural language and its essential concepts. Therefore, we may use these concepts as they have been used at all times. In this way modern physics has perhaps opened the door to a wider outlook on the relation between the human mind and reality.}
\end{quote}

The inevitable foundation in natural language of all attempts to describe the world, as emphasized by Bohr and Heisenberg, corresponds in our vocabulary to the idea that the basic categories of perception are also be the basic categories in any proper physical model. The fact that each sentence contains a \emph{subject} and a \emph{predicate} corresponds to the basic categories of knowledge given by the concept of \emph{objects} and their \emph{attributes}. To express knowledge in a sentence a \emph{subject} is not enough, we must say something about it using a \emph{predicate}. To define knowledge in a physical state it is not enough to refer to an \emph{object}, we must specify its properties using a set of \emph{attributes}.

The crucial role of the personal pronouns in natural language corresponds to the crucial role of the subjective aspect of the world assumed in this study. The symmetry between the subjective and objective aspects inherent in the model of intertwined duality is reflected, for example, in the fact that each personal pronoun comes in two cases: subject and object. The division of the subjective aspect of the world into individuals, and the fact that all knowledge has an individual root, is reflected in the fact that the personal pronouns are divided into first, second and third person. The fundamental nature of the past, the present and the future in the physical formalism presented here, corresponds to the fact that there are verb tenses in virtually all languages.

It is a common misunderstanding that the crucial role taken by the observer in the Copenhagen interpretation of quantum mechanics means that we give up on objectivity. However, objectivity can be present at different levels, and Heisenberg calls the level represented by a naive materialistic world view \emph{metaphysical realism} \cite{heisenberg}.

\begin{quote}
\emph{We "objectivate" a statement if we claim that its content does not depend on the conditions under which it can be verified. Practical realism assumes that there are statements that can be objectivated and that in fact the largest part of our experience in daily life consists of such statements. Dogmatic realism claims that there are no statements concerning the material world that cannot be objectivated. Practical realism has always been and will always be an essential part of natural science. Dogmatic realism, however, is, as we see it now, not a necessary condition for natural science. [...] Metaphysical realism goes one step further than dogmatic realism by saying that "the things really exist."}
\end{quote}

Wolfgang Pauli thinks along similar lines \cite{pauli3}.

\begin{quote}
\emph{I agree with Bohr in the opinion that the \emph{objectivity} of a scientific explanation of nature should be defined as liberally as possible: Every mode of looking at things which one can impart on others, which others having the necessary preliminary knowledge can understand and in turn apply, which we can talk about with others, shall be called objective. In tis sense all physical theories and laws are objective.}
\end{quote}

A more detailed discussion about his views on ontology is the following \cite{pauli1}.

\begin{quote}
\emph{In the new pattern of thought we do not assume any longer \emph{detached observer}, occurring in the idealizations of this classical type of theory, but an observer who by his indeterminable effects creates a new situation, theoretically described as a new state of the observed system. In this way every observation is a singling out of a particular factual result, here and now, from the theoretical possibilities, thereby making obvious the discontinuous aspect of the physical phenomena. Nevertheless, there remains still in the new kind of theory an \emph{objective reality}, inasmuch as these theories deny any possibility for the observer to influence the results of a measurement, once the experimental arrangement is chosen. Therefore particular qualities of an individual observer do not enter the conceptual framework of the theory. [...] In this wider sense the quantum-mechanical description of atomic phenomena is still an objective description, although the state of an object is not assumed any longer to remain independent of the way in which the possible sources of information about the object are irrevocably altered by obseration. The existence of such alterations reveals a new kind of wholeness in nature, unknown in classical physics, inasmuch as an attempt to subdivide a phenomenon defined by the whole experimental arrangement used for its observation creates an entirely new phenomenon.}
\end{quote}

Niels Bohr also discusses the `wholeness' in nature in terms of the indivisibility of quantum phenomena \cite{bohr3}.

\begin{quote}
\emph{[T]he whole experimental arrangement must be taken into account in a well-defined description of the phenomena. The indivisibility of quantum phenomena finds its consequent expression in the circumstance that every definable subdivision would require a change of the experimental arrangement with the appearance of new individual phenomena. Thus, the very foundation of a deterministic description has disappeared and the statistical character of the predictions is evidenced by the fact that in one and the same experimental arrangement there will in general appear observations corresponding to different individual processes.}
\end{quote}

Similarly \cite{bohr0}:

\begin{quote}
\emph{I advocated the application of the word \emph{phenomenon} exclusively to refer to the observations obtained under specified circumstances, including an account of the whole experimental arrangement. In such a terminology, the observational problem is free of any special intricacy since, in actual experiments, all observations are expressed by unambiguous statements referring, for instance, to the registration of the point at which an electron arrives at a photographic plate.}
\end{quote}

These considerations are taken into account in the formalism used in this study, since the physical state $S_{O}$ of the experimental setup corresponds to the macroscopic perception of this setup. Likewise, the set of alternative outcomes $\{S_{j}\}$ of the experiment corresponds to the set of possible macroscopic perceptions of the experimental setup after the measurement has been performed, for examples the set of distinguishable positions of a pointer in the measurement apparatus. We assign probabilities only to such macroscopic states, not to the states of the individual microscopic particles that we study in the experiment. We can use such deduced quasiobjects $\tilde{O}$ in order to express a reduced state $\check{S}_{O}$ where the microscopic composition of the experiment, including the observed specimen, is taken into accout. However, the naked perception, symbolized by $S_{O}$, is always the necessary and sufficent foundation for the physical description. In contrast, the states of the individual quasiobjects $\tilde{O}$, detached from the means of the observation that makes their deduction possible, are \emph{never} a sufficient foundation, just as Bohr and Pauli emphazises.

Bohr stresses the abstract nature of the elementary particles and other microscopic objects, just as we do when we call them quasiobjects \cite{bohr2}.

\begin{quote}
\emph{[I]t must be kept in mind that, according to the view taken above, radiation in free space as well as isolated material particles are abstractions, their properties on the quantum theory being definable and observable only through their interaction with other systems.}
\end{quote}

In this connection, it is interesting that Bohr \cite{bohr6} takes a similar view on photons as we do when we degrade them even further to bookkeeping devices called pseudoobjects.

\begin{quote}
\emph{In accordance with the classical electromagnetic conceptions we cannot, however, ascribe any proper material nature to light, since observation of light phenomena always depend on a transfer of energy and momentum to material particles. The tangible content of the idea of light quanta is limited, rather, to the account it enables us to make of the conservation of energy and momentum.}
\end{quote}

Given a mcroscopic state $S$, we define its evolution $u_{1}S(n)$ as the union of all states that can possibly be observed the next time instant $n+1$. It is easy to see similarities between this view and that of Pauli \cite{pauli5}.

\begin{quote}
\emph{[S]tatements in quantum mechanics are dealing only with possibilities, not with actualities. They have the form `This is not possible' or `Either this or that is possible', but they can never say `that will actually happen then and there'. The actual observation appears as an event outside the range of description by physical laws and brings forth in general a discontinuous selection out of the several possibilities foreseen by the statistical laws of the new theory.}
\end{quote}

The general discontinuous selection that Pauli talks about can be identified with a state reduction $u_{1}S(n)\rightarrow S(n+1)\subset u_{1}S(n)$. The fact that this process is fundamentally irreversible is emphasized by Heisenberg \cite{heisenberg}.

\begin{quote}
\emph{The observer has, rather, only the function of registering decisions, i.e., processes in space and time, and it does not matter whether the observer is an apparatus or a human being; but the registration, i.e., the transition from the "possible" to the "actual", is absolutely necessary here and cannot be omitted from the interpretation of quantum theory. At this point quantum theory is intrinsically connected with thermodynamics in so far as every act of observation is by its very nature an irreversible process; it is only through such irreversible processes that the formalism of quantum theory can be consistently connected with actual events in space and time. Again the irreversibility is - when projected into the mathematical representation of the phenomena - a consequence of the observer's incomplete knowledge of the system and in so far not completely "objective".}
\end{quote}

A state reduction means that the knowledge about the observed system increases. To avoid that our knowledge becomes more and more complete after repeated observations, there has to be a balancing loss of knowledge. Pauli expresses this fact as follows \cite{pauli2}.

\begin{quote}
\emph{[E]very experimental arrangement is accompanied by an undeterminable interaction between the measuring instrument and the system observed; as a result, any knowledge gained by an observation must be paid for by an irrevocable loss of some other knowledge. What knowledge is gained and what other knowledge is irrevocably lost, is left to the experimenter's free choice between mutually exclusive experimental arrangements. It is on this possibility of a free choice of mutually complementary experimental arrangements that the indeterministic character of natural laws postulated by quantum mechanics rests.}
\end{quote}

In our formalism we place the loss of knowledge in the evolution operator, meaning that $u_{1}S$ corresponds to a state of less knowledge than $S$ does, for a typical physical state $S$.

Heisenberg talks about the fundamental nature of the irreversibility in the temporal evolution in quantum mechanics, and compares it with the irreversibility in statistical mechanics. In the present treatment, we have tried to identify the `microscopic' irreversibility in quantum mechanics and the `macroscopic' irreversibility in statistical mechanics by defining entropy as the logarithm of the volume $V[S]$ of the same state $S$ that we act upon to define the `microscopic' evolution. The basic lines of thought are the same as the conventional ones, of course, as expressed, for instance, in this quote by Pauli \cite{pauli2}.

\begin{quote}
\emph{The first application of the calculus of probabilities in physics, which is fundamental for our understanding of the laws of nature, is the general statistical theory of heat, established by Boltzmann and Gibbs. This theory, as is well known, led necessarily to the interpretation of the entropy of a system as a function of its state, which, unlike the energy, depends on our \emph{knowledge} about the system. If this knowledge is the maximum knowledge which is consistent with the laws of nature in general (micro-state), the entropy is always null. On the other hand thermodynamic concepts are applicable to a system only when the knowledge of the initial state of the system is inexact; the entropy is then appropriately measured by the logarithm of a volume in phase space.}
\end{quote}

Pauli invented the exclusion principle and made significant contributions to the development of quantum field theory. Therefore it is interesting that he thought that we are still lacking an understanding of both these theoretical structures at a fundamental level. Concerning the exclusion principle, he writes the following \cite{pauli5}.

\begin{quote}
\emph{Already in my original paper I stressed the circumstance that I was unable to give a logical reason for the exclusion principle or to deduce it from more general assumptions. I had always the feeling and I still have it today, that this is a deficiency.}
\end{quote}

I think that the approach taken in this study gives a simple logical reason for Pauli's principle, as discussed in Section \ref{spinstatistics} in relation to Fig. \ref{Fig90}. The crucial point that enables understanding is that we do not consider the formalism of quantum mechanics or quantum field theory to be the basic layer of description. Instead we introduce the conceptually simple object state space $\emph{S}_{O}$ in which object states $S_{OO}$ can be represented. The statement that two objects have the same or overlapping states $S_{OO}$ lacks epistemic meaning. 

Regarding the concept of a \emph{field}, Pauli expresses the following concerns \cite{pauli1}:

\begin{quote}
\emph{While in the present theory there exists still a duality between the concepts of fields and of test bodies, I think that a new mathematical form of the physical law is required, which makes fields without test bodies not only physically but also logically impossible. It must also express properly the complementarity between the measurement of a field with an atomic object on the one hand, and the description of the same object as source of the field on the other hand. Indeed, these two possibilities should become automatically mutually exclusive as a result of a suitable form of the laws of nature.}
\end{quote}

The present formalism takes a step in the direction envisaged by Pauli, I think. Again, the introduction of the state $S$ is crucial. It relieves the wave function and fields from the burden to be the fundamental and perpetual carrier of the physical state. Just as we let the wave function be defined in specific experimental circumstances only, the same goes for the field. These matters are discussed in Section \ref{gaugeprinciple} in relation to Fig. \ref{Fig113}. More precisely, the field represents those aspects of the specimen that are not directly observed in the experiment, but influences the outcome of the observation. To define the field in such an experiment, we need \emph{a priori} knowledge about those objects that create the field (sources) as well as their influence on the part of the specimen that we observe (the test body).

It is clear from these considerations that the present formalism cannot constitute a quantum field theory, since we do not let any field (quantum or classical) survive the end of the experiment in which it is defined. From the epistemic perspective, perpetual quantum fields that penetrate the entire universe are not satisfying. To me, it is this unsound starting point that creates the difficulties with infinities that plauge these theories, at least at the formal level. Pauli expresses his concerns as follows \cite{pauli5}.

\begin{quote}
\emph{[T]he zero-point energy of the vacuum derived from the quantized field becomes infinite, a result which is directly connected with the fact that the system considered has an infinite number of degrees of freedom. It is clear that this zero-point energy has no physical reality, for instance it is not the source of a gravitational field. Formally it is easy to subtract constant infinite terms which are independent of the state considered and never change; nevertheless it seems to me that already this result is an indication that a fundamental change in the concepts underlying the present theory of quantized fields will be necessary.}
\end{quote}

Paul Dirac was as sceptic as Pauli when it came to quantum field theory \cite{dirac1}.

\begin{quote}
\emph{It seems clear that the present quantum mechanics is not in its final form. Some further changes will be needed, just about as drastic as the changes made in passing from Bohr's orbit theory to quantum mechanics. Some day a new quantum mechanics, a relativistic one, will be discovered, in which we will not have these infinities occurring at all.}
\end{quote}

I am aware that quantum field theory has developed since the days of Pauli and Dirac, where the invention of the renormalization group is the most important step. Many scientists seem satisfied with the present state of affairs. To me, the conceptual foundation of quantum field theory just does not look right. I suspect that this is the reason for the computational complexity and the need for various tricks to tame the infinities. Again, I have to stress that I do not master quantum field theory, so the reader may very well disregard my opinions.

The above remarks by Bohr, Dirac, Heisenberg and Pauli are more or less technical, unconventional at the most. However, at least Bohr, Heisenberg and Pauli did not hesitate to venture into speculations about deeeper matters such as the nature of life, the unconscious, and future transformations of physics that will push it further away from the classical picture. Bohr stresses that the existence of subjective experiences, the crucial quality of life, must be taken as a primary fact that cannot be reduced to something else \cite{bohr4}.

\begin{quote}
\emph{[T]he existence of life itself should be considered, both as regards its definition and observation, as a basic postulate of biology, not susceptible of further analysis, in the same way as the existence of the quantum of action, together with the ultimate atomicity of matter, forms the elementary basis of atomic physics. It will be seen that such a view-point is equally removed from the extreme doctrines of mechanism and vitalism. [...] it rejects as irrational all such attempts at introducing some kind of special biological laws inconsistent with well-established physical and chemical regularities, as have in our days been revived under the impression of the wonderful revelations of embryology regarding cell growth and division. In this connection it must be especially remembered that the possibility of avoiding any such inconsistency within the frame of complementarity is given by the very fact that no result of biological investigation can be unambiguously described otherwise than in terms of physics and chemistry, just as any account of experience even in atomic physics must ultimately rest on the use of the concepts indispensable for a conscious recording of sense impressions.}
\end{quote}

In the present approach we try to trace the apparent `white hole' in physics that lets us choose an experimental setup freely (without predefined probabilities for the different possible choices) to some activity in the brain that is unknowable in principle (Fig. \ref{Fig143}). This line of thought is similar to that expressed by Bohr \cite{bohr7}:

\begin{quote}
\emph{[T]he idea suggests itself that the minimal freedom we must allow the organism will be just large enough to permit it, so to say, to hide its ultimate secrets from us.}
\end{quote}

Also \cite{bohr5}:

\begin{quote}
\emph{[E]very experimental arrangement suitable for following the behavior of the atoms constituting an organism in as exhaustive a way as implied by the possibilities of physical observation and definition would be incompatible with the maintaining of the life of the organism. This would in fact be quite analogous to the circumstance that all observations obtained by experimental arrangements which allow of a space-time account of the behavior of the constituents of atoms and molecules stand in a complementary relation to those obtained under conditions permitting the study of the intrinsic stability of atomic structures so essential for the physical and chemical properties of matter.}
\end{quote}

We may connect the appearance in our minds of options, intentions and choices - and the impossibility to assign probabilities to these in advance - to the action of the unconscious. According to Pauli \cite{pauli4}:

\begin{quote}
\emph{This suggests a comparison between the inner process of sense-perception, or more generally every appearance of new content of consciousness, and observation in physics, insofar as physical measuring instruments can be regarded as technical extensions of the observer's sense organs. In the case of sense perception, however, new content of consiousness remains incorporated as a constituent part of the perceiving subject. Since the unconscious is not quantitatively measurable, and therefore not capable of mathematical description and since every extension of consciuousness ("bringing into consciousness") must by reaction alter the unconsciuous, we may expect a "problem of observation" in relation to the unconsciuous, which, while it presents analogies with that in atomic physics, nevertheless involves considerably greater difficulties.}
\end{quote}

As a final quote, I choose a view that Pauli expressed in 1954 on the current state of physics, and on the direction that he expects it to take in the future \cite{pauli1}.

\begin{quote}
\emph{There is a general agreement that present-day quantum mechanics leaves many fundamental questions unexplained, as for instance the atomistic character of electricity [...]. Besides that, no satisfactory interpretation is given of the characteristic variety of mass values and of the very different degrees of stability of the many particles, which in a very provisional way are called `elementary'. The limitations of the applicability of our present theory admitted, there is, naturally, a wide divergence of opinions about the direction of further development in future. Some physicists hope for the possibility of a return to the classical idea of the detached observer, whose effect on the observed system could always be eliminated by theoretically determinable corrections. Others, with whom I belong myself, have, on the contrary, hopes just in the opposite direction.}
\end{quote}

Among the unsolved questions raised by Pauli, I tentatively suggest that the `atomistic character of electricity' has something to do with the discreteness of the values of internal attributes that follows from the requirement of temporal identifiability of elementary particles in division processes (Section \ref{divideconserve}). The perspective I offer on the concept of rest mass might give a ray of hope when it comes to the derivation of the masses of elementary particles. However, I have done no attempt to make actual calculations. Therefore I have to remain humble in my claims.

In summary, my aim has been to continue along the path set out by Kant and the fathers of quantum mechanics. It is up to others to judge if I deviate from this path, or if I have misunderstood it from the outset.

\section{The dangers of metaphorical modelling}
In the preceding section, I quoted Pauli saying that he did not like the idea of fields without test bodies. Intuitively, we like to imagine an electromagnetic field as a wave that spreads, interacting with the matter it encounters along its way, causing diffraction and interference. However, in this study we have denied such a field any independent existence, saying that photons are just bookkeeping devices that are used to identify two objects $O_{1}$ and $O_{2}$ as possibly interacting. This means that attribtutes are carried from one object to the other, given the relevant conservation laws and the given ratio called `speed of light' between the distance between the objects and the time difference between the changes undergone by these objects.

Therefore, we should rather entertain the following picture. We introduce a family $C(\mathbf{r}_{4})$ of experimental contexts or arrangements such that the spatio-temporal distance $\mathbf{r}_{4}$ is varied between the source $O_{1}$ and the detector $O_{2}$ that possibly registers a photon emitted from the source. The detector corresponds to the test body, and we get a field $C(\mathbf{r}_{4})$ of experimental arrangements rather than an electromagnetic field.

As I see it, nothing happens between the emission and the absorption of a photon, since such hypothetical events are outside our knowledge by construction. The wiggly line shown in Figs. \ref{Fig98} and \ref{Fig98b}, indicating the exchange of a photon, is just a mental aid for the attribute accountant. We should not be carried away by such a picture, saying that there is `really' a particle traveling from $O_{1}$ to $O_{2}$, or that there is a field into which these objects are submerged. To construct physical models that \emph{depend} on such mental pictures is to fall in the trap of \emph{metaphorical modelling}, I think.

I am afraid that quantum field theory is an example of such a trap. Here, objects are seen as excitations of a field, making the field the fundamental entity on which the theory relies. To be able to form an image of what a quantum field is, we have to have solid state physics in mind. We picture a lattice or medium defined throughout space, in which the state of each lattice point can deviate from its equilibrium position, increasing the energy of the medium, just like an atom in a crystal may deviate from its equilibrium position in the lattice. The quantum aspect enters simply because the medium can be in a superposition of different states.

What we do, in effect, is to look at elementary particles \emph{as if} they were excitations in a lattice of other particles. But how do the particles in the lattice come about, then? The explanation bites its own tail and explains nothing. My argument is naive on purpose. Of course, nobody expects a quantum field to be a medium of particles, like a solid. But more generally speaking, the very notion of `excitation' requires that something is excited. As soon as we try to imagine an excitation, we cannot avoid the picture of an exitation in a collection of individual objects that form collective, wave-like patterns - like water waves, or ink dots on a paper forming wiggly lines.

Of course, these interpretational difficulties do not mean that quantum field theory has to be `wrong'. We may see it as a purely mathematical construction that makes it possible to extract predictions about the nature of the elementary particles and the way they interact. The involved mathematical objects and operations do not need to have any intuitive interpretation. It goes without saying, however, that such a perspective is at odds with the entire philosophy of this study. In particular, it goes against `epistemic minimalism', the idea that all mathematical objects and distinctions should correspond to objects and distinctions that are subjectively perceivable in principle. Looked at from this angle, quantum fields are as conceptually unsound as the aether. By excluding the existence of the aether, nature corrected the mistake that arose from the attempts in the nineteenth century to interpret the electromagnetic field. Maybe nature will soon correct our mistake when it comes to our attempt to interpret elementary particles in terms of excitations of quantum fields? Maybe the infinities that plague the theory, and the difficulties to include gravity, is a sign that we have set out on the wrong track from the beginning?

The basic problem with quantum field theory can be expressed as follows: We attempt to explain the fundamental layer of the objective world in terms of something that is not fundamental - a composite medium. Since we can make a subjective distinction between the fundamental and the non-fundamental layer, we should, from the epistemic perspective, use different kinds of models in the two cases; we should not use the metaphor of a phonon to explain the photon.

This mistake is analogous to that made in creation myths. The beginning of the world is explained in terms of a metaphor describing events happening in a world that already exists. For example, in Nordic mythology, the giant \emph{Ymer} was created in the void \emph{Ginnungagap} between the cold \emph{Nifelheim} and the hot \emph{Muspelheim} when the heat from Muspelheim melted the white frost from Nifelheim. Then the Gods \emph{Oden}, \emph{Vile} och \emph{Ve} killed Ymer and created the world from his body. In the modern version, Ginnungagap, Nifelheim and Muspelheim are replaced by quantum fields. The sudden hot wind from Muspelheim is replaced by a fluctuation from an unstable vacuum state \cite{krauss}. If we accept the traditional view on quantum fields we still have to ask ourselves in what sense these exist before the creation of the world. If we share Pauli's view that there cannot be any fields without test bodies, then the `explanation' loses its meaning. 

String theory suffers from the same problem as quantum field theory. A string must be interpreted as an extended object that can be conceptually divided into smaller, connected parts. Nevertheless, this non-fundamental object is used as a model for all the fundamental objects - the elementary particles. Looking at these particles as musical notes played on a string is a beautiful metaphor, but misleading as such, I think.

Just as for quantum field theory, we may avoid these difficulties by refraining from all attempts to interpret the string theory formalism. We may also try to find a better picture of what a string is. Several illustrations in this study resemble the world tubes traced out by closed strings, for instance Fig. \ref{Fig47}. In our interpretation, the surface of these tubes represents a boundary of knowledge. We cannot exclude any point inside the tubes as a state of the object we study, but we can exclude all the points outside the tube. Our tubes live in the state space $\mathcal{S}_{O}$ of objects, rather than in four-dimensional space-time. Naturally, the dimension of $\mathcal{S}_{O}$ is higher than four, since space-time can be seen as a subspace of $\mathcal{S}_{O}$. The surface of our tubes has co-dimension one in $\mathcal{S}_{O}$ rather than the dimension two that results if we see them as surfaces swept out by one-dimensional strings. Since the evolution of the state $S_{O}$ is not pointwise in terms of exact states $Z$, the boundary $\partial S$ can be looked upon as a membrane possessing tension, just like a string (Fig. \ref{Fig55}). The boundary of knowledge cannot be exactly pinpointed, as discussed in Section \ref{knowstate}. Therefore we should consider superpositions of different tube surfaces, just as we consider superpositions of different string states. However, our superpositions do not correspond to elements in a Hilbert space, since no probabilities or probability amplitudes can be associated to each possible tube surface. Again, the reason is that the tube surface cannot be exactly determined, so that no realizable alternative is associated with each possible surface.

\begin{figure}[tp]
\begin{center}
\includegraphics[width=80mm,clip=true]{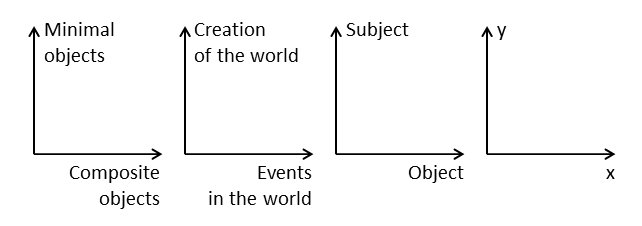}
\end{center}
\caption{Four examples of pairs of primary distinctions or `degrees of freedom'. We often try to `explain' the degree of freedom represented by the $y$-axis in terms of the degree of freedom represented on the $x$-axis. I call this strategy `metaphorical modelling', meaning that we treat processes on the $x$-axis as a metaphor for the appearance of the $y$-axis.}
\label{Fig153}
\end{figure}

Attempts to explain the subjective aspect of the world by saying that it `emerges' from the objective aspect might be fitted under the same umbrella of metaphorical modelling as the other examples that we have discussed. I have a hard time seeing what such an explanation could possibly mean. The distinction between these two aspects of the world must be considered fundamental, to be hard-wired into our faculties of perception.

We may compare the two ends of the distinction between the subjective and the objective to two independent degrees of freedom (Fig. \ref{Fig153}). One of them cannot be reduced to the other, or explained in terms of the other. We cannot account for the $y$-axis \emph{as if} it emerges from some processes taking place on the $x$-axis. We have to treat the existence of both the $x$- and the $y$-axis as a postulate if we want to use them both in our efforts to describe and understand the world. In the same way, we should not try to explain minimal objects in terms of composite objects, or the beginning of the world in terms of events taking place in a world that already exists. This is so even if we take the latter world to be a `meta-world' inside which the familiar world is created, where the meta-world may consist of quantum fields, or gods acting in mythological realms such as \emph{Ginnungagap}, \emph{Nifelheim} and \emph{Muspelheim}.

\section{Minimizing mysticism}

The traditional materialistic or `scientific' view of the world is that it is a world of objects that exists independent of our perception of it. The task for science is to describe it and understand its behavior. Concerning the subjective aspect of the world, the most common perspective is that it somehow `emerges' from the outside world. In either case, the objective aspect of the world is the fundamental aspect, and it is sufficent to understand this aspect in order to get a complete understanding of the world.

This world view rests on the implicit assumption that we can make inferences about the objective world in a way that is independent of the very same world. Figuratively speaking, the implicit idea is that our senses, logic, and mathematics `hover' above the objective world. For example, people often carelessly assume that we can form mathematical concepts taken from a Platonic world of ideas independent of the physical world.

In my own mind, this traditional materialistic world view is the archetype of mysticism, at the same time as it is inconsistent. It is mystical because it depends on a world whose existence we can never confirm. I mean here an objective world that can be described by the categories of our perceptions even if we do not perceive it. We may define mysticism as the belief in entities and worlds of which we have no everyday, tangible perception, and also the belief that these entities and worlds are necessary to account for the world in which we live. We may therefore say that materialists in the above sense are mystics by definition.

In addition, this world view is inconsistent for the following reason. We noticed the implicit assumption that it is possible to hover above the objective world, observe it and make statements about it, in a way that is not constrained by the world which we observe. This picture contradicts the view that the subjective aspect of the world `emerges' from the objective aspect, therefore being its slave.

In contrast, the intertwined dualism suggested in this study acknowledges both the subjective and the objective aspects of the world as fundamental ingredients in a complete world view. It is a `slim fit' in the sense that everything in the objective aspect has a root in the subjective aspect, and everything in the subjective aspect has a root in the objective aspect. It is therefore the most materialistic and least mystical world view that I can imagine, if we adopt consistency as a basic criterion. Epistemic closure is another term that I use to describe such a model.

It seems to me that the apparent inconsistency in the naive materialistic world view has the same root as the failure of all attempts to construct a purely formal basis for mathematics, as revealed by G\"{o}del. Namely, if we extend the power of the formal system (explanatory weight given to the objective aspect of the world) far enough, we finally reach a point where we bite a hole in our own system. We can prove that there are theorems that can be expressed within the system and are true, but cannot be proved within the system. In other words, the bold ansatz that formal systems are a sufficient basis for reason leads to the conclusion that there is a realm of independent mathematical truths `hovering' above the very same formal system. The situation is very similar to the one that I described above, where materialists implicitly adopt a view where we, as investigating subjects, hover above the objective world, being able to study every aspect of it, just like we study a picture in a book at arms length without being part of the picture ourselves. To be a bit blunt: if you believe that formal systems and computations are sufficient to understand reason, or if you believe that the objective aspect of the world is sufficient to understand the whole world, then you also believe in angels.

The G\"{o}delian problem arises because it is possible to formulate self-referential statements in strong enough formal systems. Analogously, self-reference is unavoidable in the naive materialistic world view, and the inconsistency arises because the adherents do not admit it, and does not properly take it into account. In contrast, in the world view suggested here, self-reference is built in from the beginning, by assumption, in the partial circularity inherent in the model of intertwined duality. As usual, it is a good idea to admit a problem, or just a fact, to be able to handle it.

In the world view presented here, the analogy between the foundations of mathematics and the `foundations' of nature arises very naturally, since mathematical reasoning is seen as a chain of physical object states $S_{O}$ that are governed by physical law just like any other series of events in our own body or in the outside world. The `white hole' in physical law allows nature to choose between states permitted by this physical law without any knowable selection rule. This means, for instance, that there may appear mathematical ideas without any knowable cause. The assumed fundamental distinction between proper an improper interpretations then kicks in, making it possible to judge whether the new idea is true or not. Expressed differently, the thinking of mathematicians is not completely restrained by the execution in the brain of deterministic or probabilistic computational schemes. Creativity cannot always be reduced to something else.

Many people would say that the attempts in this study to derive physics from philosophy, from pure thought, are misguided. We need to look at the world to learn anything about it, they say; all knowledge about the world is empirical. Kant opposed this view, and I agree with him. It is certainly true that we need empirical knowledge, that we cannot learn \emph{everything} about the world if we close our eyes and just think. But we can learn \emph{something}, at the most fundamental level. I would say that anyone who does not agree with this statement falls into the trap of believing in angels. In effect, such a person is saying that the workings of her thoughts does not reflect the workings of the world. Her thoughts are hovering above it and can move unrestrained by it. She believes that she is an angel herself. This is perfectly alright if she admits her mystic belief. But these people tend to say that they are materialists. This is simply not true. 

\backmatter

\end{document}